\documentclass[aps,prd,longbibliography,nofootinbib,square,sort&compress,10pt,fleqn,showpacs]{revtex4-1}
\usepackage[english]{babel}
\usepackage{amssymb}
\usepackage[fleqn]{amsmath}
\usepackage[T1]{fontenc}
\usepackage[x11names]{xcolor}
\usepackage{empheq}
\usepackage{color}
\allowdisplaybreaks
\usepackage{mathtools}
\usepackage{eucal}
\usepackage{graphicx}
\usepackage{float}
\usepackage{bookmark}
\usepackage{breakurl}
\usepackage{epsfig}
\usepackage{bm}
\usepackage[utf8x,latin1]{inputenc} 
\usepackage{emerald}
\usepackage{slantsc}
\usepackage{array}

\def\be{\begin{equation}}
\def\ee{\end{equation}}
\def\ba{\begin{eqnarray}}
\def\ea{\end{eqnarray}}
\def\bsu{\begin{subequations}}
\def\esu{\end{subequations}}

\def\B{{\rm B}}
\def\D{{\rm D}}

\def\a{\alpha}
\def\b{\beta}
\def\g{\gamma}     
\def\G{\Gamma}
\def\d{\delta}

\def\l{\lambda}
\def\m{\mu}
\def\n{\nu}

\def\r{\rho}
\def\s{\sigma}
\def\t{\tau}

\def\la{\label}
\def\pd{\partial}

\begin{document}
\title{Covariant Equations of Motion of Extended Bodies\\ with Arbitrary Mass and Spin Multipoles}
\author{Sergei M. Kopeikin}
\affiliation{Department of Physics \& Astronomy, University of Missouri, 322 Physics Bldg., Columbia, Missouri 65211, USA}
\email{E-mail: kopeikins@missouri.edu}

\begin{abstract}
Gravitational wave detectors allow to test general relativity and to study the internal structure and orbital dynamics of neutron stars and black holes in  inspiraling binary systems with a potentially unlimited rigor. Currently, analytic calculations of gravitational wave signal emitted by inspiralling compact binaries are based on the numerical integration of the asymptotic post-Newtonian expansions of the equations of motion in a pole-dipole approximation that includes masses and spins of the bodies composing the binary. Further progress in the accurate construction of gravitational-wave templates of the compact binaries strictly depends on our ability to significantly improve theoretical description of gravitational dynamics of extended bodies by taking into account the higher-order (quadrupole, octupole, etc.) multipoles in equations of motion of the bodies both in the radiative and conservative approximations of general relativity and other viable alternative theories of gravity. 

This paper employs the post-Newtonian approximations of a scalar-tensor theory of gravity along with the mathematical apparatus of the Cartesian symmetric trace-free tensors and the Blanchet-Damour multipole formalism to derive translational and rotational equations of motion of ${\mathbb N}$ extended bodies having arbitrary distribution of mass and velocity of matter. We assume that spacetime manifold can be covered globally by a single coordinate chart which asymptotically goes over to the Minkowskian coordinate chart at spatial infinity. We also introduce ${\mathbb N}$ local coordinate charts adapted to each body and covering a finite domain of space around the body. Gravitational field in the neighborhood of each body is parametrized by an infinite set of  mass and spin multipoles of the body as well as by the set of tidal gravitoelectric and gravitomagnetic multipoles of external ${\mathbb N}-1$ bodies. 
The origin of the local coordinates is set moving along the accelerated worldline of the center of mass of the corresponding body by an appropriate choice of the internal and external dipole moments of the gravitational field. Translational equations of motion of the body's center of mass and rotational equations of motion for its spin are derived by integrating microscopic equations of motion of body's matter and applying the method of asymptotic matching technique to splice together the post-Newtonian solutions of the field equations of the scalar-tensor theory of gravity for the metric tensor and scalar field obtained in the global and local coordinate charts. 

The asymptotic matching is also used for separating the post-Newtonian self-field effects from the external gravitational environment and constructing the effective background spacetime manifold. It allows us to present the equations of translational and rotational motion of each body in covariant form by making use of the Einstein principle of equivalence. This relaxes the slow-motion approximation and makes the covariant post-Newtonian equations of motion of extended bodies with weak self-gravity applicable for the case of relativistic speeds. Though the covariant equations of the first post-Newtonian order are still missing terms from the second post-Newtonian approximation they may be instrumental to get a glimpse of the last several orbital revolutions of stars in ultra-compact binary system just before merging. Our approach significantly generalizes the Mathisson-Papapetrou-Dixon covariant equations of motion with regard to the number of body's multipoles and the post-Newtonian terms having been taken into account. The equations of translational and rotational motion derived in the present paper include the entire infinite set of covariantly-defined mass and spin multipoles of the bodies. Thus, they can be used for much more accurate prediction of orbital dynamics of tidally deformed stars in inspiraling binary systems and construction of templates of gravitational waves at the merger stage of coalescing binary when the strong tidal distortions and gravitational coupling of higher-order mass and spin multipoles of the stars play a dominant role in the last few seconds of the binary life.   
\end{abstract}
\maketitle
\tableofcontents
\newpage

\section{Introduction}\la{intro}

Mathematical problem of derivation of relativistic equations of motion of extended bodies has been attracting theorists all the time since the discovery of general relativity. An enormous progress in solving this problem has been reached for the case of an isolated gravitating system consisting of spinning massive bodies in the, so-called, pole-dipole particle approximation \citep{Tagoshi_2001PhRvD,Wang_2007PhRvD,Marsat_2013CQGra} that was originally discussed by Mathisson \citep{mathisson_2010GReGr_1,mathisson_2010GReGr_2}, Papapetrou \citep{Papapetrou23101951} and Dixon \citep{dixon_1970_1,dixon_1970_2,dixon_1973GReGr,dixon_1974_3,dixon_1979} (see also papers of the other researchers \citep{Ohashi_2003PRD,steinhoff_2010PhRvD,Semerak_1999MNRAS,Kyrian_2007MNRAS} and references therein). These type of equations of motion is used for a comprehensive study of the nature of gravity through the monitoring orbital and rotational motion of bodies in the solar system \citep{sof89,kopeikin_2011book}, binary pulsars \citep{krawex_2009,weisberg_2010ApJ,damour_2009ASSL,damour_1998PhRvD}, and inspiralling compact binary systems made of neutron stars and/or black holes \citep{Reitze_2017PhyU}. A new branch of relativistic astrophysics -- gravitational wave astronomy\index{gravitational wave astronomy} can test general relativity in a strong field, fast-motion regime of coalescing binaries to unprecedented accuracy and probe the internal structure of neutron stars by measuring their Love numbers \citep{Hinderer_2008PhRvD,Poisson_2009PhRvD,Nagar_2009PhRvD,Raithel_2018ApJ,Yagi_2014PhRvD} through the gravitational response of their internal multipoles subject to the immense strength of the tidal gravitational field of inspiralling binary just before the merger \citep{Schutz_2018RSPTA}. Therefore, more advanced study of the dynamics of relativistic ${\mathbb N}$-body system is required to take into account gravitational perturbations generated by higher-order multipoles of extended bodies (quadrupole, octupole, etc.) that can significantly affect the orbital motion of the pole-dipole massive particles \citep{Blanchet_2002LRR,asada_2011,schaefer_2011mmgr,damour_2008IJMPA,Bini_PRD124034,Vines_Flanagan_PRD024046,Steinhoff_etal_PRD104028}. The study of these perturbations is also important for improving the solar system experimental tests of various gravity theories \citep{willLRR,Turyshev_2009PhyU} and for building more precise relativistic models of astronomical data processing \citep{Kopeikin_2000tmcs,koppolkor2006,Klioner_2010IAUS,Soffel_2017JGeod}. 

Over the last three decades most theoretical efforts in derivation of equations of motion were focused on solving two-body problem in general relativity in order to work out an exact analytic description of the higher-order post-Newtonian corrections beyond the quadrupole radiative approximation \citet{Landau1975} that would allow to construct sufficiently accurate waveforms of gravitational signal emitted by inspirlingthe  binary systems. One of the main obstacles in solving this problem is the self-interaction of gravitational field that strongly affects the orbital dynamics of  inspiraling  binaries through non-linearity of Einstein's field equations  \cite{Blanchet_2002LRR,Itoh_2009PhRvD,Damour_2016PhRvD93h4014D}. The non-linearity of gravitational field severely complicates derivation of equations of motion and computation of the waveform templates that are used for detecting gravitational wave signal by matched filtering technique and for estimating physical parameters of the binary system \citep{Saulson_2013}.  The non-linearity of the field equations leads to appearance of formally-divergent integrals in the post-Newtonian approximations \citep{ehlers_1976ApJ} that have to be regularized to prescribe them a unique and unambiguous finite value making physical sense. Major computational difficulty arises from using the Dirac delta-function as a source of gravitational field of point particles in curved spacetime \citep{poisson_2011}. Dirac's delta-function works well in a linear field theory like electrodynamics but it is not directly applicable for solving non-linear field equations in general relativity to account for the self-field effects of massive stars. This difficulty had been recognized by \citet{infeld_book} who pioneered the use of distributions in general relativity to replace the field singularities used in the original derivation of the Einstein-Infeld-Hoffmann (EIH) equations of motion \citep{eih}. In order to circumvent the mathematical difficulty arising from the usage of the delta-functions in the non-linear approximations of general relativity the Lorentz-invariant Hadamard "partie finie" method has been developed by French theorists \citep{bld1986,blanchet_2001JMP,blanchet_2004PhRvD,blanchet_2005PhRvD7}. It has been successfully used to regularize the divergent integrals up to the 3-d post-Newtonian approximation but faces certain limitations beyond it due to the presence of a specific pole in the quadrupole of the point-particle binary being intimately associated with the dimension of space and leading to ambiguities  \citep{Damour_2016PhRvD93h4014D}. Therefore, the Hadamard "partie finie" method was replaced with a more powerful method of dimensional regularization \citep{blanchet_2004PhRvD} to calculate equations of motion of point-like massive bodies in higher-order post-Newtonian approximations \citep{Marsat_2013CQGra,Marchand_2018PhRvD,Damour_2016PhRvD93h4014D}. There are other methods to calculate equations of motion of point-like particles in general relativity based on the matched asymptotic expansions \citep{das1,das2,gorbonos_2005CQGra}, the application of surface integral techniques like in the EIH approach \citep{eih,th_1985}, and the strong-field point-particle limit approach \citep{futamase_2007LRR,Itoh_2009PhRvD,asada_2011}.

It is well-understood that the point-like particle approximation is not enough for sufficiently accurate calculation of gravitational waveforms emitted by inspiraling compact binaries so that various types of mutual gravitational coupling of higher multipoles of moving bodies (spin, quadrupole, etc.) should be taken into account. Spin effects have been consistently tackled in a large number of papers \citep{Shibata_1993PhRvD,Rieth_1997CQGra,xu_1997PhRvD,Owen_1998PhRvD,Tagoshi_2001PhRvD,porto_2006,spin_Hamiltonian_schaefer,Wang_2007PhRvD,hergt_2008PhRvD,
tessmer_2013CQG,wang_2011PhRvDW,Marsat_2013CQGra} while only a few papers, e.g. \citep{xu_1997PhRvD,porto_2006}, attempted to compute the orbital post-Newtonian effects due to body's mass quadrupole demonstrating a substantial complexity of calculations. New generation of gravitational wave detectors will allow to  measure much more subtle effects of the multipolar coupling present in gravitational waveforms emitted by inspiralling compact binaries. Among them, especially promising from the fundamental point of view are the effects associated with the elastic properties of tidally-induced multipoles of neutron stars and black holes as they provide us with a direct experimental access to nuclear physics of condensed matter at ultra-high density of the neutron star's core and exploration of the true nature of astrophysical black holes.  Therefore, one of the challenging tasks for theorists working in gravitational wave astronomy is to derive equations of motion in relativistic ${\mathbb N}$-body problem with accounting for all effects of multipolar harmonics of extended bodies. This task is daunting and the progress in finding its solution is slow. Theoretical approach to resolve the primary difficulties in derivation of the equations of motion in isolated astronomical systems consisting of ${\mathbb N}$ extended bodies with arbitrary multipoles have been introduced in a series of papers by Brumberg and Kopeikin (BK) \citep{Kopejkin_1988CeMec,k89o,k89d,bk89,bk-nc} and further advanced by Damour, Soffel and Xu (DSX) \citep{dsx1,dsx2,dsx3,dsx4}. The two approaches are essentially similar but the advantage of the DSX formalism is the employment of the Blanchet-Damour (BD) multipoles of extended bodies which take into account the post-Newtonian corrections in the definition of the body's multipoles. The BD mass multipoles were introduced by \citet{bld} and the spin multipoles were devised by \citet{dyr2} -- see also \citep{di,blanchet_2005CQG}. The BD formulation of multipolar structure of gravitational field significantly improves the mathematical treatment of relativistic multipoles by \citet{thor} which suffers from appearance of divergent integrals from the Landau-Lifshitz pseudotensor of gravitational field \citep{Landau1975} entering the integral kernels. The BK-DSX formalism was adopted by the International Astronomical Union as a primary framework for dealing with the problems of relativistic celestial mechanics of the solar system \citep{iau2000,kopeikin_2011book}. \citet{racine_2005PhRvD,racine2013PhRvD} implemented it for a comprehensive study of the post-Newtonian dynamics of ${\mathbb N}$ extended, arbitrary-structured bodies and for derivation of their translational equations of motion with accounting for all mass and spin BD multipoles. However, \citet{racine_2005PhRvD} neither derived the post-Newtonian rotational equations of motion of the bodies nor they provided a covariant generalization of the equations of motion. 
 
In this paper we also use the BK-DSX formalism to derive translational and rotational equations of motion of ${\mathbb N}$ extended bodies in the post-Newtonian (PN) approximation of a scalar-tensor theory of gravity with a full account of arbitrary internal structure of the bodies which is mapped to the infinite set of the BD multipoles extended to the case of the scalar-tensor theory. Our mathematical approach deals explicitly with all integrals depending on the internal structure of the extended bodies and in this respect is different from the formalism applied by Racine-Vines-Flanagan (RVF) \citep{racine_2005PhRvD,racine2013PhRvD}. Besides the metric tensor, a scalar field is also a carrier of the long-range gravitational interaction in the scalar-tensor theory of gravity that brings about complications in computing the equations of motion. In particular, instead of two sets of general-relativistic BD multipoles we have to deal with an additional set of multipoles associated with the presence of the scalar field \citep{DamEsFar,kovl_2004,kopeikin_2011book}. We assume that the background value of the scalar field changes slowly that allows us to parameterize the scalar-tensor theory of gravity with two covariantly-defined parameters, $\b$ and $\g$, which correspond to the parameters of the parameterized post-Newtonian (PPN) formalism\index{PPN!formalism} \citep{willbook}. The $\b - \g$ parametrization of the equations of motion in ${\mathbb N}$-body problem is a powerful tool to test general relativity against the scalar-tensor theory of gravity in the solar system \citep{willbook,willLRR,nordtvedt_2001LNP}, in binary pulsars \citep{krawex_2009,damour_2009ASSL,wex2014} as well as with gravitational-wave detectors \citep{Will_2003CQG,Gair_2013LRR,Yunes_2013LRR} and pulsar-timing arrays \citep{Lee_2011AIPC,Yunes_2013LRR,Cornish_2018PhRvL}. The present paper significantly extends the result of papers \citep{racine_2005PhRvD,racine2013PhRvD} to the scalar-metric sector of gravitational physics, checks its consistency in Appendix \ref{appendixA}. Moreover, the present paper derives the post-Newtonian rotational equations for spins of massive bodies of ${\mathbb N}$-body system including all their multipoles.

Post-Newtonian dynamics of extended bodies \index{celestial mechanics!post-Newtonian} on curved spacetime manifold\index{manifold!spacetime} $M$ is known in literature as {\it relativistic celestial mechanics} -- the term coined by Victor Brumberg \citep{vab,brum}. Mathematical properties of the manifold $M$ are fully determined in general relativity by the metric tensor $g_{\a\b}$ which is found by solving Einstein's field equations. General-relativistic celestial mechanics admits a minimal number of fundamental constants characterizing geometry of curved spacetime -- the universal gravitational constant $G$ and the fundamental speed of gravity $c$ which is assumed to be equal the speed of light in vacuum \citep{Low_1999CQGra,Battista_2017IJMPA}. For experimental purposes \citet{willbook} denotes the fundamental speed in gravity sector as $c_g$ to distinguish it from the fundamental speed $c$ in matter sector of theory but he understands it in a rather narrow sense as the speed of weak gravitational waves propagating in radiative zone of an isolated gravitating system. On the other hand, \citet{Kopeikin_2004CQGra}  defines $c_g$ more generally as the fundamental speed that determines the rate of change of gravity field in both near and radiative zones. In the near zone $c_g$ defines the strength of gravitomagnetic field caused by rotational and/or translational motion of matter \citep{Kopeikin_2004CQGra,kopeikin_2011book,ciufolini_book}. Einstein postulated that in general relativity $c_g=c$ but this postulate along with general relativity itself, is a matter of experimental testing by radio-interferometry \citep{Kopeikin_2001ApJ,Fomalont_2003ApJ} or with gravitational-wave detectors \citep{Cornish_2017PhRvL}. The presence of additional (hypothetical) long-range fields coupled to gravity brings about other fundamental parameters of the scalar-tensor theory like $\b$ and $\g$ which are well-known in PPN formalism \citep{willbook}. The basic principles of the parameterized relativistic celestial mechanics \index{PPN} of extended bodies in scalar-tensor theory of gravity remain basically the same as in general relativity \citep{kopeikin_2011book,Battista_2017IJMPA}. 

Post-Newtonian celestial mechanics deals with an isolated gravitating ${\mathbb N}$-body system which theoretical concept cannot be fully understood without careful study of three aspects -- asymptotic structure of spacetime, approximation methods and equations of motion \citep{Ehlers_1980NYASA,frau_2004LRR} \footnote{The initial value problem is tightly related to the questions about origin and existence (stability) of an isolated gravitating system as well  \citep{ADM_paper,ehlers_1979,Rendall_1992JMP,Isenberg_2014,frau_2004LRR} but we don't elaborated on it in the present paper.}. 
In what follows, we adopt that spacetime is asymptotically-flat at infinity \citep{Adamo_2009,Faddeev_1982UFN,frau_2004LRR} and the post-Newtonian approximations (PNA) can be applied for solving the field equations. Strictly speaking, this assumption is not valid as our physical universe is described by Friedmann-Lema\^{i}tre-Robertson-Walker (FLRW) metric which is conformally-flat at infinity. Relativistic dynamics of extended bodies in FLRW universe requires development of the post-Friedmannian approximations for solving the field equations in case of an isolated gravitating system placed on the FLRW spacetime manifold \footnote{Notice that the term "post-Friedmannian" is used differently by various authors in cosmology \cite{Tegmark_2002PhRvD,Kasai_2007PThPh,CLIFTON_2012,Skordis_2015PhRvD}. We use this term in the sense used by \citet{bruni_2015PhRvD,bruni_2016PhRvD}.}. The post-Friedmannian approximation method is more general than the post-Newtonian approximations and includes additional small parameter that is the ratio of the characteristic length of the isolated gravitating system to the Hubble radius of the universe. Rigorous mathematical approach for doing the post-Friedmannian approximations is based on the field theory of the Lagrangian perturbations of pseudo-Riemannian manifolds \citep{Petrov_2017book} and it has been worked out in a series of our papers \citep{Ramirez_2002PhLB,Kopeikin_2013PhRvD,Kopeikin_2014AnPhy}. Relativistic celestial mechanics of an isolated gravitating systems in cosmology leads to a number of interesting predictions \citep{Kopeikin_2012PhRvD,galkop_2016PhRvD}. More comprehensive studies are required to fully incorporate various cosmological effects to the Bondi-Sachs formalism \citep{winicour_2016} that deals entirely with the gravitational waves in asymptotically-flat space time.

Equations of motion of ${\mathbb N}$-body system describe the time evolution of a set of independent variables in the configuration space of the system. These variables are integral characteristics of the continuous distribution of mass and current density of matter inside the bodies, and they are known as  mass and spin (or current) multipoles of gravitational field \citep{thor,bld,di}. Among them, mass-monopole, mass-dipole and spin-dipole of each body play a primary role in description of translational and rotational degrees of freedom. Higher-order multipoles of each body couples with the external gravitational field of other bodies of the isolated system and perturbs the evolution of the lower-order multipoles of the body in the configuration space.  Equations of motion are subdivided into three main categories corresponding to various degrees of freedom of the system configuration variables \citep{fockbook}. They are:
\begin{enumerate}
\item[I)] translational equations of motion of the linear momentum and the center of mass of each body,
\item[II)] rotational equations of motion\index{equations of motion!rotational} of the intrinsic angular momentum (spin) of each body,
\item[III)] evolutionary equations of the higher-order (quadrupole, etc.) multipoles of each body.
\end{enumerate}
Translational and rotational equations of motion are sufficient to describe the dynamics of pole-dipole massive particles which are physically equivalent to spherically-symmetric and rigidly-rotating bodies. Deeper understanding of celestial dynamics of arbitrary-structured extended bodies requires derivation of the evolutionary equations of the higher-order multipoles. Usually, a simplifying assumption of the rigid intrinsic rotation about the center of mass of each body is used for this purpose \citep{fockbook,brum,spyrou_1975ApJ,arminjon_2005PhRvD,Racine_2006CQG}. However, this assumption works only until one can neglect the tidal deformation of the body caused by the presence of other bodies in the system and, certainly, cannot be applied at the latest stages of a compact binary's inspiral before merger. It is worth noticing that  some authors refer to the translational and rotational equations of the linear momentum and spin of the bodies as to the laws of motion and precession \citep{Havas_1962PhRv,Ehlers_1980NYASA,th_1985,Zhang_1985PhRvD} relegating the term {\it equations of motion} to the center of mass and angular velocity of rotation of the bodies. We don't follow this terminology in the present paper.

The most works on the equations of motion of massive bodies have been done in some particular coordinate charts from which the most popular are the ADM \index{coordinates!ADM}\index{coordinates!harmonic}\index{ADM} and harmonic coordinates \citep{memmesh_2005PhRvD,hergt_2008PhRvD,schaefer_2011mmgr} \footnote{The ADM and harmonic coordinate charts are in general different structures but they can coincide under certain circumstances \citep{kopeikin_1999PhRvD}.}. However, the coordinate description of relativistic dynamics of ${\mathbb N}$-body system must have a universal physical meaning and predict the same dynamical effects irrespective of the choice of coordinates on spacetime manifold $M$. The best way to eliminate the appearance of possible spurious coordinate-dependent effects would be derivation of covariant equations of motion based entirely on the covariant definition of the configuration variables. To this end \citet{mathisson_2010GReGr_1,mathisson_2010GReGr_2}, \citet{Papapetrou23101951,pap1} and, especially, \citet{dixon_1970_1,dixon_1970_2,dixon_1974_3,dixon_1973GReGr,dixon_1979,dixon_2008,Dixon2015} had published a series of programmatic papers suggesting constructive steps toward the development of such fully-covariant algorithm for derivation of the set of equations of motion \footnote{See also \citep{Taub_1965,madore_1969}} known as Mathisson's {\it variational dynamics} or the Mathisson-Papapetrou-Dixon (MPD) formalism \citep{dixon_2008,Dixon2015}. However, the ambitious goal to make the MPD formalism independent of a specific theory of gravity and applicable to an arbitrary pseudo-Riemannian manifold created a number of hurdles that slowed down the advancement in developing the covariant dynamics of extended bodies. Nonetheless, continuing efforts to elaborate on the MPD theory had never stopped \citep{ehlers_1977GReGr,schattner_1979GReGr,Ohashi_2003PRD,steinhoff_2010PhRvD,Harte_2012,harte2015,dirk_2013PhLA,Obukhov_Puetzfeld2014,dirk_obukhov2014,dixon_2008}.  

In order to make the covariant MPD formalism connected to the more common coordinate-based derivations of the equations of motion of extended bodies the metric tensor $\bar g_{\a\b}$ of the effective background spacetime manifold $\bar{M}$ must be specified and Dixon's multipoles of the stress-energy skeleton \citep{dixon_1973GReGr,dixon_1979}\index{tensor!stress-energy!skeleton} have to be linked to the covariant definition of the BD multipoles of extended bodies. To find out this connection we tackle the problem of the covariant formulation of the equations of motion in a particular gauge associated with the class of conformal harmonic coordinates introduced by \citet{nutku_1969,nutku_1969ApJ}. Covariant formulation of the equations of motion is achieved at the final stage of our calculations by building the effective background manifold $\bar{M}$ and applying the Einstein equivalence principle for mapping the locally-defined BD multipoles to the arbitrary coordinates. This procedure has been proposed by \citet{Landau1975} and consistently developed and justified by \citet{th_1985}. It works perfect on torsionless manifolds\index{manifold!torsionless} with the affine connection\index{affine connection} being fully determined by the metric tensor. Its extension to the pseudo-Riemannian manifolds with torsion\index{torsion} and/or non-minimal coupling of matter to gravity requires further theoretical study which is not pursued in the present paper. Some steps forward in this direction have been made, for example, by \citet{yasskin_1980PhRvD,mao_2007PhRvD,march_2011GReGr,flanagan_2007PhRvD,hehl_2013PhLA} and \citet{dirk_2013PhRvD,dirk_2013PhLA}. 

Dynamics of matter in an isolated gravitating system consisting of ${\mathbb N}$ extended bodies is naturally split in two components -- the orbital motion of the center of mass of each body and the internal motion of matter with respect to the body's center of mass.
Therefore, the coordinate-based derivation of equations of motion of ${\mathbb N}$ extended bodies in the isolated gravitating system suggests a separation of the problem of motion in two parts: external and internal \citep{fockbook,Damour_1987book}. The external problem deals with the derivation of translational equations of bodies relative to each other while the internal problem provides the definition of physical multipoles of each body and translational equations of motion of the center of mass of the body with respect to the origin of the body-adapted local coordinates. The internal problem also gives us the evolutionary equations of the body's physical multipoles including the rotational equations of motion of body's spin. Solution of the external problem is rendered in a single global coordinate chart covering the entire manifold $M$. Solution of the internal problem is executed separately for each body in the body-adapted local coordinates. There are ${\mathbb N}$ local coordinate charts for ${\mathbb N}$ bodies making the atlas of the spacetime manifold. Mathematical construction of the global and local coordinates relies upon and is determined by the solutions of the field equations of the scalar-tensor theory of gravity. The coordinate-based approach to solving the problem of motion provides the most effective way for the unambiguous separation of the internal and external degrees of freedom of matter and for the definition of the internal multipoles of each body. Matching of the asymptotic expansions of the solutions of the field equations in the local and global coordinates allows to find out the structure of the coordinate transition functions on the manifold and to build the effective background metric $\bar g_{\a\b}$ on spacetime manifold $\bar{M}$ that is used for transforming the coordinate-dependent form of the equations of motion to the covariant one which can be compared with the MPD covariant equations of motion.  

The global coordinate chart is introduced for describing the orbital dynamics of the body's center of mass. It is not unique and is defined up to the group of diffeomorphisms which are consistent with the assumption that spacetime is asymptotically-flat at null infinity. This group is called the Bondi-Metzner-Sachs (BMS) group \citep{winicour_2016,Schmidt_1975GReGr} and it includes the Poincare transformations as a sub-group.  It means that in case of an isolated astronomical system\index{isolated astronomical system} embedded to the asymptotically-flat spacetime we can always introduce a non-rotating global coordinate chart with the origin located at the center of mass of the system such that at infinity: (1) the metric tensor approaches the Minkowski metric, $\eta_{\alpha\beta}$\index{Minkowski metric}, and (2) the global coordinates smoothly match the inertial coordinates of the Minkowski spacetime\index{spacetime!Minkowski}. The global coordinate chart is not sufficient for solving the problem of motion of extended bodies as it is not adequately adapted for the description of internal structure and motion of matter inside each body in the isolated ${\mathbb N}$-body system. This description is done more naturally in a local coordinate chart attached to each gravitating body as it allows us to exclude various spurious effects appearing in the global coordinates (like Lorentz contraction, geodetic precession, etc.) which have no relation to the motion of matter inside the body \citep{Kopejkin_1988CeMec,ashb2}. The body-adapted local coordinates replicate the inertial Lorentzian coordinates only in a limited domain of spacetime manifold $M$ inside a world tube around the body under consideration. Thus, a complete coordinate-based solution of the external and internal problems of celestial mechanics requires introduction of ${\mathbb N}+1$ coordinate charts -- one global and ${\mathbb N}$ local ones \citep{iau2000,Battista_2017IJMPA}. It agrees with the topological structure of manifold\index{manifold!topological structure} defined by a set of overlapping coordinate charts making the atlas of spacetime manifold\index{manifold!atlas} \cite{dfn,arno}. The equations of motion of the bodies are intimately connected to the differential structure\index{manifold!differential structure} of the manifold characterized by the metric tensor and its derivatives. It means that the mathematical presentations of the metric tensor in the local and global coordinates must be diffeomorphically equivalent that is the transition functions defining spacetime transformation from the local to global coordinates must map the components of the metric tensor of the internal problem of motion\index{problem of motion!internal} to those of the external\index{problem of motion!external} problem and vice verse. The principle of covariance\index{principle of relativity} is naturally satisfied when the law of transformation from the global to local coordinates is derived by matching the global and local asymptotic solutions of the field equations for the metric tensor. The coordinate transformation establishes a mutual functional relations between various geometric objects that appear in the solutions of the field equations, and determines the equation of motion of the origin of the local coordinates adapted to each body. The coordinate transformation are also employed to map the equations of motion of the center of mass of each body to the coordinate-free, covariant form.  

The brief content of our study is as follows. Next section \ref{notat} summarizes the main concepts and notations used in the present paper. In section \ref{stt} we discuss a scalar-tensor theory of gravity in application to the post-Newtonian celestial mechanics of ${\mathbb N}$-body system including the $\b--\g$ parametrization of the field equations, the small parameters, the post-Newtonian approximations and gauges. Parametrized post-Newtonian coordinate charts covering the entire spacetime manifold $M$ globally and in a local neighborhood of each body are set up in section \ref{prf}. They make up an atlas\index{manifold!atlas} of 
spacetime manifold. Geometrical properties of coordinates in relativity are characterized by the functional form of the metric tensor and its corresponding parameters - the internal and external multipoles of gravitational field - which are also introduced and explained in section \ref{prf} along with the multipolar structure of the scalar field. The local differential structure of spacetime manifold $M$ presumes that the functional forms of the metric tensor and scalar field given in different coordinates must smoothly match each other in the buffer regions where the coordinate charts overlap. The procedure of matching of the asymptotic expansions of the metric tensor and scalar field in the global and local coordinates is described in section \ref{pntb} that establishes: 1) the functional structure of the body-frame external multipoles of gravitational field in terms of the volume integrals taken from the distribution of mass density, matter current, pressure, etc., 2) defines the worldline $\mathcal{W}$ of the origin of the body-adapted local coordinates and yields the equation of its translational motion  with respect to the global coordinate chart, 3) defines the effective background metric, $\bar g_{\a\b}$, for each extended body that is used later on for derivation of the covariant equations of motion of the bodies. Section \ref{n4r6v} provides details of how the local coordinate chart adapted to each extended body is used for a detailed description of the body's own gravitational field inside and outside of the body and for definition of its mass, center of mass, linear and angular momentum (spin). This section also derives the equations of motion of body's center of mass moving along worldline ${\cal Z}$, and its spin in the body-adapted local coordinates. Translational equations of motion of body's center of mass in the global coordinates follow immediately after substituting the local equations of motion to the parametric description of the worldline $\mathcal{W}$ of the origin of the local coordinates with respect to the global coordinates. The parametric description of worldline ${\cal W}$ follows through the multipolar expansion of the external gravitational potentials in section \ref{c6a0n5} and that of the external multipoles in section \ref{111777}.
Section \ref{orbeom} derives the equations of translational motion of the worldline ${\cal Z}$ of the center of mass of each body in terms of the complete set of the Blanchet-Damour internal multipoles of the bodies comprising the ${\mathbb N}$-body system. 
Rotational equations of motion for spin of each body with the torque expressed in terms of the Blanchet-Damour multipoles, are derived in section \ref{sbdytvre0}. Finally, section 
\ref{ceom23} introduces the reader to the basic concepts of the Mathisson-Papapetrou-Dixon (MPD) variational dynamics and establishes a covariant form 
of the post-Newtonian translational and rotational equations of motion of extended bodies derived previously in the conformal harmonic coordinates in sections \ref{orbeom} and \ref{sbdytvre0}. 

The paper has four appendices. Appendix \ref{kjn34c} sets out auxiliary mathematical relationships for symmetric trace-free (STF) tensors. Appendix \ref{appendixA} compares our equations of translational motion from section \ref{c6a0n5} with similar equations derived by \citet{racine_2005PhRvD,racine2013PhRvD} and analyzes the reason for the seemingly different appearance of the equations. Appendix \ref{appndxon} explains the concept of Dixon's multipole moments of extended bodies and discusses their mathematical correspondence with the Blanchet-Damour multipole moments. Appendix \ref{appendixB} compares Dixon's covariant equations of translational and rotational motion of extended bodies with our covariant equations of motion from section \ref{ceom23}.    

\section{Primary Concepts and Mathematical Notations}\la{notat}

We consider an isolated gravitating system consisting of ${\mathbb N}$ extended bodies in the framework of a generic scalar-tensor theory of gravity. The bodies are indexed by either of three capital letters B, C, D from the Roman alphabet. Each of these indices takes values from 1 to ${\mathbb N}$. The bodies have arbitrary but physically-admissible distributions of mass, internal energy, pressure and velocity of matter which can depend on time. We exclude processes of the matter exchange between the bodies so that they interact between themselves only through the coupling to the gravity and/or scalar field force. We also exclude processes of nuclear transmutation of matter particles.  

It is now well-understood \citep{brum,racine_2005PhRvD,kopeikin_2011book,Soffel_2013book,Battista_2017IJMPA} that solution of the problem of motion of ${\mathbb N}$-body system requires introduction of one global coordinate chart, $x^\a$, covering the entire spacetime manifold and ${\mathbb N}$ local coordinate charts, $w^\a_\B$, adapted to each body B of the system. If there is no confusion with other bodies the sub-index B in the notation of the local coordinate chart of the body B is omitted.

Equations of scalar-tensor theory of gravity admit a class of conformal transformations of the metric tensor which allows to put the gravity field equations in two different forms which are referred to as the Einstein and Jordan frames respectively. The field equations in the Einstein frame makes the field equations looking exactly as Einstein's equations of general relativity with the scalar field entering solely the right-hand side of the field equations in the form of the stress-energy tensor. The metric tensor in the Einstein frame is coupled with the scalar field explicitly while the Ricci tensor is uncoupled from the scalar field. In the Jordan frame the situtaion is opposite -- the Ricci tensor couples with the scalar field explicitly while the metric tensor is uncoupled from the scalar field. It was debated for a while which frame -- Einstein or Jordan -- is physical \citep{Faraoni_1999ijtp,Bhandra_2007MPLA}. The answer is that all classical physical predictions are to be conformal-frame invariant \citep{Flanagan_2004cqg}. Therefore, the choice of the frame is a matter of mathematical convenience.  In the present paper we shall primarily work in the Jordan frame in which matter is minimally coupled to gravitational field like in general relativity.  

Let us single out a body B in the ${\mathbb N}$-body system and consider the metric tensor in the local, body-adapted coordinates. The metric outside the body is parametrized by two infinite sets of configuration parameters which are called the {\it internal} and {\it external} multipoles. The multipoles are purely spatial, 3-dimensional, symmetric trace-free (STF) Cartesian tensors \citep{thor,bld1986,Zschocke_2014} residing on the hypersurface ${\cal H}_{u_\B}$ of constant coordinate time $u_\B$ passing through the origin of the local coordinate chart, $w^\a_\B$. The internal multipoles characterize  gravitational field and internal structure of the body B itself and they are of two types -- the mass multipoles ${\cal M}^L_\B$, and the spin multipoles ${\cal S}^L_\B$ where the multi-index $L=i_1i_2\ldots i_l$ consists of a set of spatial indices with $l$ denoting the rank of the STF tensor $(l\ge 0)$. If there is no confusion, the index B of the internal multipoles is dropped off. There are also two types of external multipoles -- the gravitoelectric multipoles ${\cal Q}_L$, and the gravitomagnetic multipoles ${\cal C}_L$. The external multipoles with rank $l\ge 2$ characterize tidal gravitational field in the neighborhood of body B produced by other (external) bodies residing outside body B. Gravitoelectric dipole ${\cal Q}_i$ describes local acceleration of the origin of the local coordinates adapted to body B. Gravitomagnetic dipole ${\cal C}_i$ is the angular velocity of rotation of the spatial axes of the local coordinates. In what follows we set ${\cal C}_i=0$. The scalar field of the scalar-tensor theory of gravity has its own multipolar decomposition with the internal and external multipoles. The external multipoles of the scalar field are denoted as ${\cal P}_L$. The above-mentioned multipoles are called {\it canonical} as they are directly related to two degrees of freedom of vacuum gravitational field and one degree of freedom of the scalar field. The overall theory also admit the appearance of {\it non-canonical} STF multipoles in the process of derivation of the equations of motion. These multipoles are related to the gauge degrees of freedom and can be eliminated from the equations of motion by the appropriate choice in the definition of the canonical multipoles and the center of mass of body B.

Definitions of the {\it canonical} STF multipoles must be consistent with the differential structure of spacetime manifold $M$ determined by the solutions of the gravity field equations in the global and local coordinate charts. The consistency is achieved by applying the method of asymptotic matching of the external and internal solutions of the field equations that allows us to express the external multipoles, ${\cal Q}_L$ and ${\cal C}_L$, in terms of the internal multipoles, ${\cal M}^L_\B$ and ${\cal S}^L_\B$. The internal multipoles of an extended body B are defined by the integrals taken over body's volume from the correspondingly-chosen internal distribution of mass-energy inside the body. This distribution includes not only the internal characteristics of the body B (mass density, pressure, compression energy, etc.) but also the energy density of the tidal gravitational field produced by the external bodies.  

There are two important reference worldlines associated with the translational motion of each body B -- a worldline ${\cal W}$ of the origin of the body-adapted, local coordinates, $w^\a_\B$, and a worldline ${\cal Z}$ of the center of mass of the body. Equations of motion of the origin of the local coordinates are obtained by doing the asymptotic matching of the internal and external solutions of the field equations for the metric tensor. Equations of motion of the center of mass of the body are derived by integrating the macroscopic post-Newtonian equations of motion of matter which are consequence of the local law of conservation of the stress-energy tensor. The center of mass of each body is defined by the condition of vanishing of the internal mass dipole of the body in the multipolar expansion of the metric tensor in the Einstein frame, ${\cal I}^i_\B=0$. This definition imposes a constraint on the local acceleration ${\cal Q}_i$ that makes worldline ${\cal W}$ coinciding with ${\cal Z}$. It also eliminates the other extraneous ({\it non-canonical}) types of STF multipoles of gravitational field from the translational and rotational equations of motion.       

We use $G$ to denote the observed value of the universal gravitational constant and $c$ as a fundamental speed both in gravity and matter sectors of the theory. Every time, when there is no confusion about the system of units, we choose a geometrical system of units such that $G=c=1$ so that $G$ and $c$ don't appear in equations explicitly. We put a hat above any function that describes a contribution from the internal distribution of mass, velocity, etc. of body B in the local coordinates adapted to the body. A bar over any function denotes functions produced by the distributions of mass, velocity, etc. from the bodies being external with respect to body B. The bar also denotes the gravitational potentials entering the external multipoles as well as the metric tensor, ${\bar g}_{\a\b}$, of the effective background manifold, $\bar{M}$, that is used to construct covariant equations of motion of the bodies in section \ref{ceom23}.  

Primary mathematical symbols and notations used in the present paper are as follows:
\begin{itemize}
\item the capital Roman indices B,C,D label the extended bodies of ${\mathbb N}$-body system. Each of them takes values from the set $\{1,2,\ldots,N\}$,
\item the small Greek letters $\a,\b,\g,\ldots$ denote spacetime indices of tensors and run through values $0,1,2,3$, 
\item the small Roman indices $i,j,k,\ldots$ denote spatial tensor indices and take values $1,2,3$,
\item the capital Roman letters $L,K,N,S$ denote spatial tensor multi-indices, for example, $L\equiv \{i_1i_2\ldots i_l\}$, $N\equiv \{i_1i_2\ldots i_n\}$, $K-1\equiv \{i_1i_2\ldots i_{k-1}\}$, etc.,  
\item the Einstein summation rule is applied for repeated (dummy) indices and multi-indices, for example,  ${\cal P}^\a {\cal Q}_\a\equiv {\cal P}^0 {\cal Q}_0+{\cal P}^1 {\cal Q}_1+{\cal P}^2 {\cal Q}_2 + {\cal P}^3 {\cal Q}_3$, ${\cal P}^i {\cal Q}_i\equiv {\cal P}^1 {\cal Q}_1+{\cal P}^2 {\cal Q}_2 + {\cal P}^3 {\cal Q}_3$, ${\cal P}^L {\cal Q}_L={\cal P}^{i_1i_2\ldots i_l}{\cal Q}_{i_1i_2\ldots i_l}$, ${\cal P}^{K-1} {\cal Q}_{K-1}={\cal P}^{i_1i_2\ldots i_{k-1}}{\cal Q}_{i_1i_2\ldots i_{k-1}}$, etc.,
\item the Kronecker symbol $\delta_{ij}=\delta^{ij}=\delta^i_j=\delta^j_i$ in 3-dimensional space is a unit matrix
 \be\nonumber
\delta_{ij} \equiv \left\{ \begin{array}{cl}
         1 &\qquad\qquad \mbox{if $i=j$},\\
        0 & \qquad\qquad\mbox{if $i\not=j$},\end{array} \right.
\ee 
\item the Levi-Civita fully anti-symmetric symbol, $\varepsilon_{ijk}=\varepsilon^{ijk}$, in 3-dimensional space is defined as $\varepsilon_{123}=+1$, and
 \be\nonumber
\varepsilon_{ijk} \equiv \left\{ \begin{array}{cl}
         +1 &\qquad\qquad \mbox{if the set $\{i,j,k\}$ forms an even permutation},\\
        -1 & \qquad\qquad\mbox{if the set $\{i,j,k\}$ forms an odd permutation},\\
        0& \qquad\qquad\mbox{if, at least, two indices from the set $\{i,j,k\}$ coincide},\end{array} \right.
\ee
\item $E_{\a\b\g\d}$ is 4-dimensional generalization of the fully anti-symmetric, 3-dimensional Levi-Civita symbol,
\item $g_{\a\b}$ is a full metric of spacetime manifold $M$,
\item $\bar g_{\a\b}$ is the effective metric of the background spacetime manifold $\bar{M}$,
\item $\eta_{\a\b}={\rm diag}\{-1,+1,+1,+1\}$ is the Minkowski metric,
\item $h_{\a\b}$ is the metric perturbation of the Minkowski spacetime in the global coordinate chart,
\item $\hat h_{\a\b}$ is the metric perturbation of the Minkowski spacetime in the local coordinate chart of body B,
\item $w^\a_{\rm B}=(w^0_{\rm B},w^i_{\rm B})=(u_{\rm B},w^i_{\rm B})$ are the local coordinates adapted to a body B with $u_{\rm B}$ being the local coordinate time. Every time, when there is no confusion, we drop the sub-index B  from the notations of the local coordinates. Thus, by default $w^\a=(w^0,w^i)=(u,w^i)$ are the local coordinates adapted to body B with $u$ being the local coordinate time,
\item $x^\a=\{x^0,x^i\}=\{t,x^i\}$ are the global coordinates covering the entire spacetime manifold $M$ or $\bar{M}$. Notation for the manifold should not be confused with the mass internal monopole of body B which is denoted with ${\cal M}_\B$,
\item $\pd_\a=\pd/\pd x^\a$ is a partial derivative with respect to coordinate $x^\a$,
\item $\hat\pd_\a=\pd/\pd w^\a$ is a partial derivative with respect to the local coordinate $w^\a$,
\item shorthand notations for the multi-index partial derivatives with respect to coordinates $x^\a$ are: $\partial_L\equiv\partial_{i_1\ldots i_l}=\pd_{i_1}\pd_{i_2}...\pd_{i_l}$, $\partial_{L-1}\equiv\partial_{i_1\ldots i_{l-1}}$, $\partial_{pL-1}\equiv\partial_{pi_1...i_{l-1}}$, etc.,
\item shorthand notations for the multi-index partial derivatives with respect to coordinates $w^\a$ are: $\hat\partial_L\equiv\hat\partial_{i_1\ldots i_l}=\hat\pd_{i_1}\hat\pd_{i_2}...\hat\pd_{i_l}$, $\hat\partial_{L-1}\equiv\hat\partial_{i_1\ldots i_{l-1}}$, $\hat\partial_{pL-1}\equiv\hat\partial_{pi_1...i_{l-1}}$, etc.,
\item $\bar\nabla$ standing in front of a group of $p$ tensor indices denotes an operator of the covariant derivative of the $p-th$ order with respect to the background metric $\bar g_{\a\b}$, for example, $\bar\nabla_{\a_1\a_2...\a_p}=\bar\nabla_{\a_1}\bar\nabla_{\a_2}...\bar\nabla_{\a_p}$,
\item $\nabla$ standing in front of a group of $p$ tensor indices denotes a covariant derivative of the $p-th$ order with respect to the full metric $g_{\a\b}$, that is $\nabla_{\a_1\a_2...\a_p}=\nabla_{\a_1}\nabla_{\a_2}...\nabla_{\a_p}$,
\item $\frac{{\cal D}}{{\cal D}\tau}=\bar u^\a\bar\nabla_\a$ denotes a covariant derivative along vector $\bar u^\a$,
\item $\frac{{\cal D}_{\rm F}}{{\cal D}\tau}$ denotes a Fermi-Walker covariant derivative along vector $\bar u^\a$ \citep[Chapter 1, \S4]{syngebook},
\item tensor (Greek) indices of geometric objects on spacetime manifold ${M}$ are raised and lowered with the full metric $g_{\a\b}$,
\item tensor (Greek) indices of geometric objects on the effective background manifold $\bar{M}$ are raised and lowered with the background metric $\bar g_{\a\b}$,
\item tensor (Greek) indices of the metric tensor perturbation $h_{\a\b}$ are raised and lowered with the Minkowski metric $\eta_{\a\b}$,
\item the spatial (Roman) indices of geometric objects are raised and lowered with the Kronecker symbol $\delta^{ij}$. Effectively, it means that the position of the spatial indices - either superscript or subscript - does not matter,
\item a symbol of summation over {\it all} ${\mathbb N}$ bodies of ${\mathbb N}$-body system is denoted as $\sum\limits_\B\equiv\sum\limits_{\B=1}^N$, or $\sum\limits_{\rm C}\equiv\sum\limits_{{\rm C}=1}^N$, etc.,
\item the symbol of summation over ${\mathbb N}-1$ bodies of ${\mathbb N}$-body system excluding, let say body C, is $\sum\limits_{\B\not={\rm C}}\equiv\sum\limits_{\substack{\B=1\\\B\not={\rm C}}}^N$, 
\item the ordinary factorial is $l!=l(l-1)(l-2)...2\cdot 1$, 
\item the double factorial means 
\be\nonumber
l!! \equiv \left\{ \begin{array}{cl}
         l(l-2)(l-4)...4\cdot 2 &\qquad\qquad \mbox{if $l$ is even},\\
         &\\
        l(l-2)(l-4)...3\cdot 1 & \qquad\qquad\mbox{if $l$ is odd},\end{array} \right.
\ee
\item the round parentheses around a group of tensor indices denote full symmetrization, 
\be\nonumber
T_{(\a_1\a_2...\a_l)}=\frac1{l!}\sum_{\sigma\in S} T_{\sigma(\a_1)\sigma(\a_2)...\sigma(\a_l)}\;,
\ee
where $\sigma$ is a permutation of the set $S=\{\a_1,\a_2,...,\a_l\}$ 
\be\nonumber
\sigma=\left\{ \begin{array}{ccccc}
         \a_1&\a_2&\a_3&...&\a_l\\
        \sigma(\a_1)&\sigma(\a_2)&\sigma(\a_3)&...&\sigma(\a_l)\end{array} \right\}\;,
        \ee
        for example, $T_{(\a\b\g)}=\displaystyle\frac1{3!}\left(T_{\a\b\g}+T_{\b\g\a}+T_{\g\a\b}+T_{\b\a\g}+T_{\a\g\b}+T_{\g\b\a}\right)$, etc.\;,
\item the curled parentheses around a group of tensor indices denote {\it un-normalized} symmetrization over the smallest set of the index permutations, for example, $T_{\{\a}\delta_{\b\g\}}\equiv T_\a\delta_{\b\g}+T_\b\d_{\a\g}+T_\g\d_{\a\b}$, etc.,
\item the square parentheses around a pair of tensor indices denote anti-symmetrization, for example,
$T^{[\a\b]\g}=\displaystyle\frac12\left(T^{\a\b\g}-T^{\b\a\g}\right)$\;,etc.
\item the angular brackets around tensor indices denote a symmetric trace-free (STF) projection of tensor $T_L=T_{i_1i_2...i_l}$. The STF projection $T_{<L>}$ of tensor $T_L$ is constructed from its symmetric part, 
\be
S_L\equiv T_{(L)}=T_{(i_1i_2...i_l)}\;,
\ee
by subtracting all the permissible traces. This makes $T_{<L>}$ fully-symmetric and trace-free on all pairs of indices. The general formula for the STF projection is \citep{thor,bld1986}
\be\label{stfformula}
T_{<L>}\equiv\sum_{n=0}^{[l/2]}\frac{(-1)^n}{2^nn!}\frac{l!}{(l-2n)!}\frac{(2l-2n-1)!!}{(2l-1)!!}\delta_{(i_1i_2...}\delta_{i_{2n-1}i_{2n}}S_{i_{2n+1}...i_l)j_1j_1...j_nj_n}\;,
\ee
where $[l/2]$ is the largest integer less than or equal to $l/2$.
\item the STF spatial derivative is denoted by the angular parentheses embracing the STF indices, for example, $\pd_{<L>}\equiv\pd_{<i_1i_2...i_l>}$ or $\pd_{<K>}\equiv \pd_{<i_1i_2...i_k>}$\;,
\item the Christoffel symbols on spacetime manifold ${M}$ are: $\Gamma^\a_{\b\g}=\frac12 g^{\a\s}\left(\pd_\b g_{\g\s}+\pd_\g g_{\b\s}-\pd_\s g_{\b\g}\right)$\;,
\item the Christoffel symbols of the effective background manifold $\bar{M}$ are: $\bar\Gamma^\a_{\b\g}=\frac12 \bar g^{\a\s}\left(\pd_\b \bar g_{\g\s}+\pd_\g \bar g_{\b\s}-\pd_\s \bar g_{\b\g}\right)$\;,
\item the sign of the Riemann tensor on spacetime manifold ${M}$ is defined by convention (it is the same as in \citep{mtw}) 
\be\label{kk33cc25}
R_{\a\b\m\n}=\frac12\left(\pd_{\a\n} g_{\b\m}+\pd_{\b\m} g_{\a\n}-\pd_{\b\n} g_{\a\m}-\pd_{\a\m} g_{\b\n}\right)+ g_{\r\s}\left(\G^\r_{\a\n}\G^\s_{\b\m}-\G^\r_{\a\m}\G^\s_{\b\n}\right)\;,
\ee
\item the Riemann tensor of the effective background manifold $\bar{M}$ is 
\be\label{kk33xc}
\bar R_{\a\b\m\n}=\frac12\left(\pd_{\a\n} \bar g_{\b\m}+\pd_{\b\m} \bar g_{\a\n}-\pd_{\b\n} \bar g_{\a\m}-\pd_{\a\m} \bar g_{\b\n}\right)+ \bar g_{\r\s}\left(\bar\G^\r_{\a\n}\bar\G^\s_{\b\m}-\bar\G^\r_{\a\m}\bar\G^\s_{\b\n}\right)\;.
\ee
The sign conventions \eqref{kk33cc25} and \eqref{kk33xc} for the Riemann tensor are opposite to that from the Weinberg textbook \citep[Equation 6.6.2]{weinberg_book1972}. 
\end{itemize}
Other notations will be introduced and explained in the main text of the paper as they appear. Useful algebraic and differential identities of STF tensors are given in Appendix \ref{kjn34c} of the present paper.

\section{Scalar-Tensor Theory and Post-Newtonian Approximations}\index{scalar-tensor theory of gravity}\label{stt}

We consider an isolated ${\mathbb N}$-body system comprised of ${\mathbb N}$ extended bodies with non-singular interior described by the stress-energy tensor $T^{\a\b}$ of baryonic matter. The bodies have a localized matter support and are supposed to be isolated one from another in space in the sense that accretion, transfer and other fluxes of baryonic matter outside of the bodies are excluded.

Post-Newtonian celestial mechanics describes orbital and rotational motions of the bodies on a curved spacetime manifold ${M}$ defined by the metric tensor\index{metric tensor}, $g_{\a\b}$ obtained as a solution of the field equations of a metric-based theory of gravitation in the slow-motion\index{slow-motion approximation}\index{approximation!slow-motion} and weak-field\index{approximation!weak-field}\index{weak-field approximation} approximation. The class of viable metric theories of gravity, which can be employed for developing relativistic celestial mechanics, ranges from  general theory of relativity \citep{brum,Landau1975} to a scalar-vector-tensor theory of gravity proposed by \citet{Bekenstein2004} for describing orbital motion of galaxies in clusters at cosmological scale. It is not the goal of the present paper to review all these theories and we refer the reader to reviews by \citet{willLRR} and \citet{Turyshev_2009PhyU} for further details. 

We shall build the parametrized post-Newtonian celestial mechanics in the framework of a scalar-tensor theory of gravity introduced by \citet{1949Natur.164..637J,1959ZPhy..157..112J} and \citet{1956AcHPh.29..128F}, and independently re-discovered later by \citet{1961PhRv..124..925B,1962PhRv..125.2163D,1962PhRv..126.1875D}. The Jordan-Fierz-Brans-Dicke (JFBD) theory extends the Lagrangian\index{Lagrangian} of general relativity by introducing a long-range, nonlinear scalar field\index{scalar field} (or fields \cite{DamEsFar}) being minimally coupled to gravity. The presence of the scalar field\index{scalar field} causes deviation of the metric-based gravity theory from a pure geometric phenomenon. The scalar field effects are superimposed on gravitational effects of general relativity, thus, highlighting the geometric role of the metric tensor and making physical content of the theory richer. Recent discovery of the scalar Higgs boson\index{Higgs boson} at LHC\index{LHC} \citep{Dittmaier_2013PrPNP} and its possible connection to the effects of JFBD scalar field in gravitation and cosmology \citep{dehnen_1992IJTP} reinforce the significance of application of the scalar-tensor theory in relativistic astrophysics and celestial mechanics of isolated gravitating systems.

\subsection{Lagrangian and Field Equations}
\label{fieq}
Gravitational field in scalar-tensor theory of gravity is described by the metric tensor $g_{\alpha\beta}$ and a long-range scalar field $\Phi$ with non-linear self-interaction described by means of a coupling function $\omega(\Phi)$. Field equations in the Jordan frame of scalar-tensor theory are derived from the action \index{action} \cite{willbook}
\be\la{10.1}
S=-\frac{1}{16\pi}\int \Phi R\sqrt{-g}d^4x+\frac{1}{8\pi}\int L^\Phi\sqrt{-g}d^4x+\int L^{\rm M}\sqrt{-g}d^4x\;,
\ee
where $g={\rm det}[g_{\alpha\beta}]<0$ is the determinant of the metric tensor $g_{\alpha\beta}$, $R=g^{\a\b}R_{\a\b}$ is the Ricci scalar, $R_{\a\b}$ is the Ricci tensor,
\be
L^\Phi=\frac{\omega(\Phi)}{2\Phi}g^{\a\b}\pd_\a\Phi\pd_\b\Phi-V(\Phi)\;,
\ee
is the Lagrangian of the scalar field with $V(\Phi)$ being the potential of the scalar field, and $L^{\rm M}\equiv L(g_{\a\b},\psi)$ is the  Lagrangian of matter of ${\mathbb N}$-body system with $\psi$ denoting the dynamic variables characterizing the matter of the extended bodies comprising the system. We keep the self-coupling function $\omega(\Phi)$ of the scalar field unspecified for making covariant parametrization of possible deviations of the scalar-tensor theory from general relativity. Moreover, we assume the minimal coupling of the metric tensor $g_{\a\b}$ with matter variables $\psi$ without coupling to the scalar field $\Phi$. It explains why the Lagrangian $L^{\rm M}$ does not depend on the scalar field $\Phi$.  

The action (\ref{10.1}) is written in the Jordan frame\index{Jordan frame} in which the metric tensor $g_{\a\b}$ has a standard physical meaning of observable quantity used in the definitions of the proper time, the proper length, and in the geodesic equation of motion of test particles \citep{willbook}. Taking variational derivatives from the action \eqref{10.1} with respect to the metric tensor, we obtain gravitational field equations for the metric tensor,
\be\label{10.2aa}
R_{\mu\nu}-\frac12 g_{\m\n}R  =\frac1{\Phi}\left(\nabla_{\m\n}\Phi-g_{\m\n}\Box_g\Phi+T^\Phi_{\mu\nu}\right)+\frac{8\pi}{\Phi}T^{\rm M}_{\m\n}\;,
\ee
where, here and everywhere else, the operator $\nabla_\m$ denotes a covariant derivative\index{covariant derivative}\index{derivative!covariant} on the spacetime manifold with the metric $g_{\a\b}$, the g-box symbol 
\begin{equation}
  \label{covd}
  {\Box}_g\equiv g^{\mu\nu}\nabla_{\m\n}=g^{\m\n}\pd_{\m\n}-g^{\mu\nu}\Gamma^\alpha_{\mu\nu}\pd_\a\;,
\end{equation}
denotes the differential Laplace-Beltrami operator\index{Laplace-Beltrami operator} \citep{mtw,eisen} on manifold with metric $g_{\a\b}$,
$T^\Phi_{\mu\nu}$ and $T^M_{\mu\nu}$ are stress-energy tensors of scalar field and matter of ${\mathbb N}$-body system respectively. 
In particular, 
\be\la{ne5x6a}
T^\Phi_{\mu\nu}=\frac{\omega(\Phi)}{\Phi}\left(\pd_\m\Phi\pd_\n\Phi-\frac12 g_{\m\n}\pd^\a\Phi\pd_\a\Phi\right)+g_{\m\n}V(\Phi)\;,
\ee
and 
\begin{equation}
  \label{11.1}
  T^{\rm M}_{\m\n}=\rho\left(1+\Pi\right)u_\m u_\n+{\mathfrak{s}}_{\m\n}\;,
\end{equation}
where $\rho$ and $\Pi$ are the density\index{density of matter} and the specific internal energy\index{specific internal energy} of the baryonic matter, $u^\alpha=dx^\alpha/cd\tau$ is 4-velocity of the matter with $\tau$ being the proper time\index{proper time} along the worldline of matter's volume element, and ${\mathfrak{s}}^{\alpha\beta}$ is an arbitrary (but physically admissible) symmetric tensor of spatial stresses being orthogonal to the 4-velocity of matter
\begin{equation}
\label{pz1}
u^\alpha{\mathfrak{s}}_{\alpha\beta}=0.
\end{equation}
Equation (\ref{pz1}) means that the stress tensor has only spatial components in the frame co-moving with matter. 

Equation for the scalar field\index{scalar field} $\Phi$ is obtained by variation of action (\ref{10.1}) with respect to $\Phi$. After making use of a contracted form of (\ref{10.2aa}) it yields \cite{willbook}
\begin{equation}
\label{10.4as}
{\Box}_g\Phi = \frac1{3+2\omega(\Phi)} \bigg[8\pi T^{\rm M}-\frac{d\omega}{d\Phi}\pd^\a\Phi\pd_\a\Phi-2\Phi\frac{dV}{d\Phi}+4 V(\Phi)\bigg]\;,
\end{equation}
where $T^{\rm M}=g^{\a\b}T^{\rm M}_{\a\b}$ is the trace of the stress-energy tensor of matter which serves as a source of the scalar field\index{scalar field!source} along with its own kinetic (due to the self-coupling) and potential energies. 

Gravitational field and matter are tightly connected via the Bianchi identities of the field equations for the metric tensor \citep{Landau1975,mtw} which read
\be\la{bid6c}
\nabla_\n\bigg(R^{\m\n}-\frac12 g^{\m\n}R\bigg)\equiv 0\;.
\ee
The Bianchi identities make four out of ten components of the metric tensor fully independent. This freedom is usually fixed by picking up a specific gauge condition\index{gauge condition}, which imposes four constraints on the components of the metric tensor and/or its first derivatives. At the same time the Bianchi identity \eqref{bid6c} imposes four differential constraints on the stress-energy tensor of matter and scalar field which constitute microscopic equations of motion of matter \citep{Landau1975}. Due to the Bianchi identities \eqref{bid6c} the source of gravitational field standing in the right-hand side of \eqref{10.2aa} is also conserved. The law of conservation of this tensor is convenient to write down in the following form, 
\begin{equation}
  \label{h2a}
 8\pi \nabla_\n T_{\rm M}^{\m\n}=-\nabla_\n T_{\Phi}^{\m\n}+\frac{\nabla^\m\Phi}{2\Phi}\left(8\pi T_{\rm M}+ T_\Phi-3{\Box}_g\Phi\right)\;.
\end{equation}
After taking the covariant derivative from the stress-energy tensor of the scalar field \eqref{ne5x6a}, and making use of the scalar field equation \eqref{10.4as} we can check by direct calculation that the right-hand side of \eqref{h2a} vanishes. It yields the laws of conservation of the stress-energy tensor of baryonic matter of ${\mathbb N}$-body system, 
\begin{equation}
\label{h2}
\nabla_\n T_{\rm M}^{\m\n}=0\;.
\end{equation}
The conservation of the stress-energy leads to the (exact) equation of continuity\index{equation of continuity} 
\begin{equation}
  \label{pz2}
  \nabla_\a(\rho u^\alpha)=\frac{1}{\sqrt{-g}}\pd_\a(\rho\sqrt{-g} u^\alpha) =0\;,
\end{equation}
and to the thermodynamic law of conservation of energy that is expressed as a differential relation between the specific internal energy $\Pi$ and the stress tensor of matter
\begin{equation}
  \label{11.2}
  \rho u^{\alpha}\pd_\a\Pi+{\mathfrak{s}}^{\alpha\beta}\nabla_\a u_\b=0\;.
\end{equation}
These equations will be employed later on for solving the field equations and for derivation of equations of motion of the extended bodies.

\subsection{Post-Newtonian Approximations}\index{post-Newtonain approximations}\index{PNA}\label{pnap}

We shall assume that\index{scalar field!potential} potential $V(\Phi)$ of scalar field can be neglected in the following calulations. Discarding potential $V(\Phi)$ is justified from observational point of view in weak gravitational field (like in the solar system) as it does not reveal any measurable effect in orbital and rotational motion of celestial bodies on sufficiently long intervals of time \cite{willLRR,Turyshev_2009PhyU}. On the other hand, if the potential of scalar field is not identically nil, it may become important in astrophysical systems having strong gravitational field \index{gravitational field!strong} like compact binary neutron stars\index{neutron star} or black holes\index{black hole}, and its inclusion to the theory leads to important physical consequences \cite{DamEsFar,damesf1993}. 

Neglecting the scalar field potential simplifies the field equations \eqref{10.2aa} and \eqref{10.4as} and reduce them to the following form,
\ba\label{10.2}
R_{\mu\nu}-\frac12 g_{\m\n}R  &=&\frac1{\Phi}\left[8\pi T_{\m\n}+\frac{\omega(\Phi)}{\Phi}\left(\pd_\m\Phi\pd_\n\Phi-\frac12 g_{\m\n}\pd^\a\Phi\pd_\a\Phi\right)+\nabla_{\m\n}\Phi-g_{\m\n}\Box_g\Phi\right]\;,
\\
\label{10.4}
{\Box}_g\Phi &=& \frac1{3+2\omega(\Phi)} \left(8\pi T-\frac{d\omega}{d\Phi}\pd^\a\Phi\pd_\a\Phi\right)\;,
\ea
where we suppressed index ${\rm M}$ at the stress-energy tensor of the baryonic matter for simplicity: $T^{\m\n}\equiv T_{\rm M}^{\m\n}$ and $T\equiv T^\a{}_\a$.

Field equations (\ref{10.2}) and (\ref{10.4}) of the scalar-tensor theory of gravity represent a system of eleventh non-linear differential equations in partial derivatives. It is challenging to find their solution in the case of ${\mathbb N}$-body system made of extended bodies with sufficiently strong gravitational field which back reaction on the geometry of spacetime manifold\index{manifold!spacetime} cannot be neglected. Like in general relativity, an exact solution of this problem is not known and may not be available in analytic form. Hence, one has to resort to approximations to apply the analytic methods. Two basic methods have been worked out in asymptotically-flat spacetime -- the post-Minkowskian (PMA)\index{post-Minkowskian approximations}\index{approximations!post-Minkowskian}\index{PMA} and the post-Newtonian (PNA) approximations \index{PNA}\index{post-Newtonian approximations}\index{approximations!post-Newtonian}\citep{Damour_1987book,asada_2011,kopeikin_2011book}.  Post-Newtonian approximations are applicable in case when matter moves slowly and gravitational field\index{gravitational field!weak} is weak everywhere -- the conditions, which are satisfied, e.g., within the solar system. Post-Minkowskian approximations relax the requirement of the slow motion but the weak-field limitation remains. Strong field regime requires more involved techniques \citep{Itoh_2009PhRvD}. We use the method of the post-Newtonian approximations in this paper which is remarkably effective and consistent in describing gravitational field of isolated gravitating systems including binary pulsars containing dense neutron stars and binary black hole inspiraling toward a final merger \citep{Will_2011PNAS}.

Post-Newtonian approximation scheme suggests that the metric tensor can be expanded in the near zone\index{near zone} of ${\mathbb N}$-body system in powers with respect to the inverse powers of the fundamental speed $c$ \index{speed of gravity} \footnote{For historical reason the speed $c$ in all sectors of fundamental interactions is called "the speed of light" \citep{uzan_2005AmJPh}. It is clear that in gravity sector its physical meaning is the speed of gravity \citep{Low_1999CQGra,Kopeikin_2004CQGra}}. This expansion may be not analytic at higher post-Newtonian approximations in a certain class of coordinate charts including the harmonic coordinates\index{coordinates!harmonic} \citep{kake,bld1986}. Exact mathematical formulation of the basic axioms underlying the post-Newtonian expansion\index{post-Newtonian expansion!axioms} was given by Rendall \citep{rend}\index{Rendall}. Practically, it requires to have several small parameters characterizing ${\mathbb N}$-body system and the interior structure of the bodies. They are:
$\epsilon_{\rm i}\sim v_{\rm i}/c$, $\epsilon_{\rm e}\sim v_{\rm e}/c$, and $\eta_{\rm i}\sim U_{\rm i}/c^2$, $\eta_{\rm e}\sim U_{\rm e}/c^2$, where $v_{\rm i}$ is a characteristic internal velocity of motion of matter inside an extended body, $v_{\rm e}$ is a characteristic velocity of the relative motion of the bodies with respect to each other, $U_{\rm i}$ is the internal Newtonian gravitational potential inside each body, and $U_{\rm e}$ is the external Newtonian gravitational potential in the regions of space between the bodies. If we denote a characteristic radius of an extended body as $L$ and a characteristic distance between the bodies as $R$, the internal and external gravitational potentials will have the following estimates: $U_{\rm i}\simeq GM/L$ and $U_{\rm e}\simeq GM/R$, where $M$ is a characteristic mass of the body. Due to the virial theorem \index{virial theorem} of the Newtonian gravity \citep{Landau1975} the small parameters are not fully independent. Specifically, one has $\epsilon_e^2\sim\eta_e$ and
$\epsilon_{\rm i}^2\sim\eta_{\rm i}$ if the internal motions of matter inside the body are governed by the gravitational field of the body through macroscopic equations of motion. The slow-motion parameter $\epsilon_{\rm i}$ is not related to the weak-field parameter $\eta_{\rm i}$ in all other cases like  rotational motion of the body, convection of matter, sound waves, etc. Parameters $\epsilon_{\rm i}$ and $\epsilon_{\rm e}$ are the primary parameters in doing the post-Newtonian expansions of the solutions of the field equations for the metric tensor and scalar field. In what follows, we use a single notation $\epsilon$ to quantify the order of the parametric expansion in the post-Newtonian series.

Besides the small parameters $\epsilon$ and $\eta$, the post-Newtonian approximation utilizes two more small parameters: $\delta\sim L/R$ characterizing the dependence of body's gravitational field on its finite size $L$, and the asphericity parameter $\lambda\simeq\Delta L/L$ estimating how much the shape of the body under consideration deviates from sphere. These parameters appear in vacuum multipolar expansion of the metric tensor and scalar field. As the metric tensor has 10 algebraically-independent components we might expect appearance of 10 different types of tensor multipoles but only two types of them (mass and spin multipoles) are physically-significant because 8 types of the tensor multipoles are gauge-dependent and can be eliminated from the multipolar expansion of the metric tensor by using the gauge freedom of the theory  \citep{thor,bld,bld1986}. Multipolar expansion of the scalar field has naturally one type of the (scalar) multipoles which is fully independent of the choice of the metric gauge. The property of disappearance of 8 types of the tensor multipoles in the multipolar expansion of the metric tensor is known as the {\it effacing} principle \citep{Damour_1987book} which tells us that the only information about the internal structure of the body obtained from the measurement of its vacuum gravitational field, can be extracted from the {\it canonical} mass and spin multipoles of the body. It imposes certain limitations on our ability to get an unambiguous information about the distribution of mass, velocity, pressure, and other internal characteristics of the body, for example, \index{gravitational field!spherically-symmetric} gravitational field of an extended body having spherically-symmetric distribution of mass can not be distinguished from that of a massive point-like particle having the same mass due to the Birkhoff theorem that is valid in scalar-tensor theory of gravity as well as in general relativity \citep{faraoni_2018PRD}. 

In principle, translational and/or rotational equations of motion of extended bodies might depend on more than the two ({\it canonical} ) types of the multipoles of the bodies. This is because derivation of the equations of motion is based on integration of macroscopic equations of motion of matter over finite volumes of the bodies and it is not evident that the result of a such integration will not produce additional {\it non-canonical} types of the multipoles entering the gravitational force and/or torque exerted on each body. Had this  happened the parameter $\delta=L/R$ would appear in the post-Newtonian expansions even if the bodies comprising ${\mathbb N}$-body system were spherically-symmetric. Scrutiny theoretical study of the problem of motion in general relativity has shown that such non-canonical multipoles don't appear in the equations of motion of ${\mathbb N}$-body system and the internal structure of extended bodies is completely effaced up to 2.5PN approximation for spherically-symmetric bodies \citep{k85,Damour_1983grr,Damour_1987book,kovl_2008,mitchell_2007PhRvD} and up to 1PN approximation for arbitrary-structured bodies \citep{dsx2,kovl_2004,racine_2005PhRvD}. We demonstrate in the present paper that the effacing principle is also valid in scalar-tensor theory of gravity in 1PN approximation. The effacing of the internal structure and disappearance of the non-canonical multipoles of the bodies from equations of motion indicates that the equations can be extrapolated to the case of structureless bodies like black holes in compact binaries.

The multipoles of extended bodies have some {\it bare} values in case when the body is non-rotating and fully-isolated from external gravitational environment. The numerical value of the multipoles will deviate from the bare value if the body rotates and interacts gravitationally with other members of ${\mathbb N}$-body system as it brings about intrinsic deformations in the distribution of matter inside the body. The measured value of each multipole is a sum of its bare value and the induced deformations. The magnitude of the induced deformations depends on the parameters of elasticity of each body which are intrinsically related to the equation of state of the body's matter. These parameters are known as Love's numbers $\kappa_{{\rm n}l}$ where sub-index ${\rm n}=1,2,3$ indicates the physical type of the Love number and $l$ is the multipole number \citep{Zharkov_1978book,loven1,getino,Yip_2017}. Measurement of the Love numbers of neutron stars and black holes in compact inspiralling binaries is one of the main goals of gravitational wave astronomy \citep{Hinderer_2008PhRvD,Poisson_2009PhRvD,Nagar_2009PhRvD,Raithel_2018ApJ}. Generally speaking, the Love numbers $\kappa_{{\rm n}l}$ depend on the frequency of orbital harmonics and are different for each multipole \citep{Yagi_2014PhRvD}. Therefore, a complete study of the internal structure of neutron stars by means of the gravitational wave astronomy requires including all multipoles of the bodies to the translational and rotational equations of motion in order to get an exhaustive information about their internal physical characteristics -- equation of state, radius, distribution of mass density, etc.  The present paper accounts for all internal multipoles of the bodies.

\subsection{Post-Newtonian Expansions}

Post-Newtonian series are expansions of the metric tensor, scalar field and matter variables around their background values with respect to the small parameters introduced above. We denote $\Phi_0$ the background value of the scalar field\index{scalar field} $\Phi$ and assume that the dimensionless perturbation of the field, $\phi$, is small compared with $\Phi_0$. In cosmological case, $\Phi_0$ is not constant and changes subject to the Hubble expansion of the universe \citep{1993PhRvL..70.2217D}. The inverse value of the background scalar field is proportional to the universal gravitational constant $G\sim 1/\Phi_0$ as shown below in \eqref{10.6}. Therefore, the time variation of $\Phi_0$ causes a secular evolution of the universal gravitational constant $G=G_0+\dot G(t-t_0)$ as well as other PPN parameters of the scalar-tensor theory \citep{galkop_2016PhRvD}. The rate of the hypothetical secular variation of the universal gravitational constant has been measured by lunar laser ranging (LLR) and is negligibly small -- $\dot G/G_0=(7.1\pm 7.6)\times 10^{-14}$ yr$^{-1}$ \citep{LLR_2018CQGra}. Other techniques yield similar constraints \citep{WD_2011JCAP,Pitjeva_2013MNRAS,wex2014}. 
In this paper we consider the case of asymptotically-flat space time and treat $\Phi_0$ as constant.  We write {\it exact} decomposition
\begin{equation}
  \label{aa}
  \frac{\Phi}{\Phi_0}= 1+\phi\;,
\end{equation}
where $\phi$ is the dimensionless value of the scalar field $\Phi$ normalized to $\Phi_0$.

According to theoretical expectations \cite{1993PhRvL..70.2217D} and experimental limitation on PPN parameters \cite{willLRR,Turyshev_2009PhyU,LLR_2018CQGra}, the post-Newtonian perturbation $\phi$ of the scalar field has a very small magnitude, so that we can expand all quantities depending on the scalar field in a Maclaurin series with respect to $\phi$ using it as a small parameter in the expansion. In particular, the post-Newtonian decomposition of the coupling function $\omega(\Phi)$ can be written as
\begin{equation}
  \label{10.5}
  \omega(\Phi)=\omega_0+\omega'_0\phi+\mathcal{O}(\phi^2),
\end{equation}
where $\omega_0\equiv\omega(\Phi_0)$, $\omega'_0\equiv\left(d\omega/d\phi\right)_{\Phi=\Phi_0}$, and we impose the boundary condition on the scalar field\index{scalar field!boundary conditions} such that $\phi$ approaches zero as the distance from ${\mathbb N}$-body system approaches infinity -- see equations \eqref{12.3}, \eqref{12.3a}. The post-Newtonian expansion of the perturbation $\phi$ is given in the form
\be\la{b7c4}
\phi=\epsilon^2\phi^{\rm (2)}+\mathcal{O}(\epsilon^3)\;,
\ee
where the post-Newtonian correction ${\phi}^{\rm (2)}$ will be defined below, and the symbol $\mathcal{O}(\epsilon^3)$ indicate the expected magnitude of the residual terms. Notice that the linear term being proportional to $\epsilon$ does not appear in \eqref{b7c4} as it is incompatible with the field equations \eqref{10.4}.

The unperturbed value of the metric tensor $g_{\a\b}$ in asymptotically-flat spacetime is the Minkowski metric, $\eta_{\a\b}$. The metric tensor is expanded in the post-Newtonian series with respect to parameter $\epsilon$  around the Minkowski metric as follows
\begin{equation}
  \label{exp}
  g_{\alpha\beta}  =  \eta_{\alpha\beta}+\epsilon{h}^{\rm (1)}_{\alpha\beta}+ \epsilon^2{h}^{\rm (2)}_{\alpha\beta}+\epsilon^3{h}^{\rm (3)}_{\alpha\beta}+\epsilon^4{h}^{\rm (4)}_{\alpha\beta}+\mathcal{O}(\epsilon^5)\;.
\end{equation}
The generic post-Newtonian expansion of the metric tensor is not analytic with respect to parameter $\epsilon$ \citep{kake,bld1986,Damour_1987book}. However, the non-analytic (logarithmic) terms emerge only in higher post-Newtonian approximations and do not affect results of the present paper since we restrict ourselves with the first post-Newtonian approximation. Notice also that the linear, with respect to $\epsilon$, terms in the metric tensor expansion (\ref{exp}) do not originate from the field equations \eqref{10.2} and are pure coordinate-dependent effect. Hence, they can be eliminated by making an appropriate adjustment of the coordinate chart \citep{futamase_1983PhRvD,th_1985,kovl_2004}. If we kept them, they would make the coordinate grid non-orthogonal and rotating at classic (Newtonian) level. Reference frames with such properties are rarely used in astronomy and astrophysics. Therefore, we shall postulate that the linear term in expansion (\ref{exp}) is absent.

After eliminating the linear terms in the post-Newtonian expansion of the metric tensor and substituting the expansion to the field equations \eqref{10.2} we can check by inspection that various components of the metric tensor\index{metric tensor} and the scalar field have in the first post-Newtonian approximation the following form \citep{futamase_1983PhRvD}
\ba
  \label{11.4}
  g_{00} & = & -1+\epsilon^2{h}^{\rm (2)}_{00}+\epsilon^4{h}^{\rm (4)}_{00}+\mathcal{O}(\epsilon^6),\\
  \label{11.5}
  g_{0i} & = & \epsilon^3{h}^{\rm (3)}_{0i}+\mathcal{O}(\epsilon^5),\\
  \label{11.6}
  g_{ij} & = & \delta_{ij}+ \epsilon^2{h}^{\rm (2)}_{ij}+\mathcal{O}(\epsilon^4),
  \ea
where each term of the expansions will be defined and explained below. 
In order to simplify notations, we shall use the following abbreviations for the metric tensor perturbations:
\begin{equation} 
\label{not}
  h_{00}\equiv{h}^{\rm (2)}_{00}\;,\qquad  l_{00}\equiv{h}^{\rm (4)}_{00}\;,\qquad
  h_{0i}\equiv{h}^{\rm (3)}_{0i}\;,\qquad  h_{ij}\equiv{h}^{\rm (2)}_{ij}\;,\qquad
  h\equiv{h}^{\rm (2)}_{kk}\;.
\end{equation}

Post-Newtonian expansion of the metric tensor \eqref{11.4}-\eqref{11.6} introduces a corresponding expansion of the stress-energy tensor\index{stress-energy tensor!post-Newtonian expansion} of matter \eqref{11.1},
\begin{eqnarray}
  \label{11.8}
  T_{00} & = & {T}^{(0)}_{00}+\epsilon^2{T}^{(2)}_{00}+\mathcal{O}(\epsilon^4),\\
  \label{11.9}
  T_{0i} & = &\epsilon{T}^{(1)}_{0i}+\mathcal{O}(\epsilon^3),\\
  \label{11.10}
  T_{ij} & = & \epsilon^2{T}^{(2)}_{ij}+\mathcal{O}(\epsilon^4),
\end{eqnarray}
where 
\begin{eqnarray}
  \label{11.21}
  {T}^{(0)}_{00} & = & \rho^{\ast}\;,\\
  \label{11.23}
  {T}^{(1)}_{0i} & = & -\rho^{\ast}v^i\;,\\
  \label{11.24}
  {T}^{(2)}_{ij} & = & \rho^{\ast}v^iv^j+{\mathfrak{s}}^{ij}\;,\\
  \label{11.22}
  {T}^{(2)}_{00} & = & \rho^{\ast}\bigg(\frac{v^2}{2}+\Pi-h_{00}-\frac{h}{2}\bigg)\;,
\end{eqnarray}
$v^i=c u^i/u^0=dx^i/dt$ is 3-dimensional velocity of matter, and 
\begin{equation}
  \label{11.19}
  \rho^{\ast} \equiv \sqrt{-g}u^0\rho = \rho+\frac{\epsilon^2}{2}\rho(v^2+h)+\mathcal{O}(\epsilon^4)\;,
\end{equation}
is the invariant density of matter that is a useful mathematical variable in relativistic hydrodynamics \cite{fockbook} due to the exact law \eqref{pz2} of conservation of rest mass. This conservation law can be recast, following \eqref{pz2}, to the equation of continuity \cite{fockbook} 
\begin{equation}
  \label{11.20}
  \pd_t\rho^{\ast}+\pd_i(\rho^{\ast}v^i)=0\;,
\end{equation}
which has the {\it exact} Newtonian form in arbitrary coordinates. 
Since equation (\ref{11.20}) is exact it makes calculation of the total time derivative from a volume integral of arbitrary differentiable function $f(t,{\bm x})$ simple,
\begin{equation}
  \label{dmq}
  \frac{d}{dt}\int\limits_{{\cal V}_{\rm B}}\rho^{\ast}(t,\bm{x})f(t,\bm{x})d^3x=\int\limits_{{\cal V}_{\rm B}}\rho^{\ast}(t,\bm{x})\frac{df(t,\bm{x})}{dt}d^3x,
\end{equation}
where ${\cal V}_{\rm B}$ denotes a volume of body B, and the operator of the total time derivative\index{time derivative!total} is
\begin{equation}
  \label{qiw}
  \frac{d}{dt}=\frac{\partial}{\partial t}+v^i\frac{\partial}{\partial x^i}\;.
\end{equation}
In derivation of \eqref{dmq} we have taken into account that the boundary of the volume of body B can change as time progresses \citep{kopeikin_2011book} but there is no flux of baryonic matter through the boundary of the body. We also notice that equation (\ref{dmq}) is exact.

In what follows, we shall give up on the post-Newtonian expansion parameter $\epsilon$ in all subsequent equations because we work only in the first post-Newtonian approximation, and leaving out $\epsilon$ should not cause confusion. We also use the geometric system of units, $G=c=1$. Physical units like SI or CGS can be easily put back to our equations by making use of dimensional analysis \citep{Gibbings_2011}.  

\subsection{Conformal Harmonic Gauge}\index{harmonic gauge!conformal}\label{gfree}
The post-Newtonian field equations for the post-Newtonian components of the metric tensor and scalar field variables can be derived after substituting the post-Newtonian series of the previous section to the covariant equations (\ref{10.2}) and (\ref{10.4}), and arranging the terms in the expansion in the order of smallness with respect to parameter $\epsilon$. The post-Newtonian equations are covariant like the original field equations that is their form is independent of the choice of spacetime coordinates. Hence, their solutions are determined up to four arbitrary functions reflecting a freedom of coordinate transformations called the gauge freedom of the metric tensor\index{gauge freedom}. It is a common practice to limit the coordinate arbitrariness by imposing a gauge condition which limits the choice of coordinates on spacetime manifold\index{manifold}. The gauge condition does not fix the freedom in choosing coordinates completely -- a restricted class of coordinate transformations within the imposed gauge still remains. This class of transformations is called a {\it residual} gauge freedom\index{gauge freedom!residual} which plays an important role in theoretical formulation of relativistic dynamics of  ${\mathbb N}$-body system.

One of the most convenient gauge conditions in a scalar-tensor theory of gravity was proposed by Nutku\index{Nutku} \cite{nutku_1969,nutku_1969ApJ} as a generalization of the harmonic gauge of general relativity\index{harmonic gauge}
\begin{equation}
  \label{11.3}
  \pd_\n\left(\Phi\sqrt{-g}\;g^{\mu\nu}\right)=0\;.
\end{equation} 
The Nutku gauge condition\index{Nutku!gauge condition} (\ref{11.3}) is equivalent to the following condition imposed on the Christoffel symbols,
\begin{equation}
  \label{gau}
  g^{\mu\nu}\Gamma^\alpha_{\mu\nu} = g^{\a\b}\pd_\b\ln\Phi\;.
\end{equation}
Let us consider now the Laplace-Beltrami operator\index{Laplace-Beltrami operator} introduced above in (\ref{covd}) and write it down in the Nutku gauge in case of an arbitrary scalar function $F\equiv F(x^\alpha)$. It yields, 
\begin{equation}
\label{co}
\Box_gF\equiv g^{\a\b}\left(\pd_{\a\b}F-\pd_\a F\pd_\b\ln\Phi\right)\;.
\end{equation}
Any function $F$ that is subject to the homogeneous Laplace-Beltrami equation, $\Box_gF=0$, is called harmonic\index{harmonic function}. The Laplace-Beltrami operator \eqref{co} applied to each particular coordinate being considered as a scalar function $F=x^\a$, give us 
\be
{\Box}_g x^\alpha=-g^{\a\b}\pd_\b\ln\Phi\neq 0\;.
\ee
which means that the coordinates $x^\alpha$ are not harmonic functions on spacetime manifold in the Jordan frame\index{Jordan frame} and in the Nutku gauge. Nonetheless, such non-harmonic coordinates are more convenient in the scalar-tensor theory of gravity because they allow us to eliminate more coordinate-dependent terms from the field equations as compared with the harmonic gauge condition ${\Box}_g x^\alpha=0$ which is not equivalent to the Nutku gauge \eqref{11.3}. We call the class of
the coordinates satisfying the Nutku gauge (\ref{11.3}), the conformal harmonic coordinates \citep{kovl_2004}. As we have learned above, these coordinates are not harmonic in the Jordan frame but it can be shown that they are harmonic functions of spacetime manifold in the conformal Einstein frame with the metric $\tilde g_{\a\b}\equiv\Phi g_{\a\b}$. Indeed, in the Einstein frame, the Nutku gauge condition \eqref{gau} reads $\pd_\b\left(\sqrt{-\tilde g}\tilde g^{\a\b}\right)=0$, which is exactly the harmonic gauge condition.

The conformal harmonic coordinates have many properties similar to the harmonic coordinates in general relativity. Our preferences in choosing the conformal harmonic coordinates for constructing theory of motion of extended celestial bodies are justified by three factors: 
\begin{itemize}
\item[1)] the conformal harmonic coordinates become harmonic coordinates\index{coordinates!harmonic} in general relativity when the scalar field is switched off, $\Phi\rightarrow 0$, 
\item[2)] the conformal harmonic coordinates represent a natural generalization of the IAU 2000 resolutions \cite{iau2000} on relativistic reference frames from general relativity to scalar-tensor theory of gravity, 
\item[3)] the Nutku gauge condition (\ref{11.3}) significantly simplifies the field equations and facilitates finding their solutions like in case of the harmonic gauge in general relativity. 
\end{itemize}
Harmonic coordinates in the Jordan frame have been used by \citet{2000PhRvD..62b4019K} for constructing post-Newtonian reference frames in PPN formalism\index{PPN formalism}. The conformal harmonic coordinates were employed in our publications \citep{kovl_2004,kopeikin_2011book} for discussing relativistic celestial mechanics of the solar system. We shall also use the conformal harmonic coordinates in the present paper.

The gauge condition (\ref{gau}) does not fix the conformal harmonic coordinates uniquely. Let us change the coordinates
\begin{equation}
  \label{11.12a}
  x^\alpha\mapsto w^\alpha=w^\alpha(x^\alpha)\;,
\end{equation}
but keep the Nutku gauge condition (\ref{gau}) intact in the new coordinates. After applying the coordinate transformation \eqref{11.12a} to \eqref{gau} it is straightforward to show that the new conformal harmonic coordinates $w^\a$ must satisfy a homogeneous wave equation
\begin{equation}
  \label{11.b}
  g^{\mu\nu}(x^{\beta})\frac{\pd^2 w^\alpha}{\pd x^\m\pd x^\n}=0\;,
\end{equation}
which describes the residual gauge freedom\index{residual gauge freedom} in choosing the conformal harmonic coordinates that remains after imposing the Nutku gauge condition on the metric tensor. Equation \eqref{11.b}
has the infinite number of non-trivial solutions defining the entire set of the conformal harmonic coordinates on spacetime manifold. The residual gauge freedom in the scalar-tensor theory of gravity is similar to that existing in the harmonic gauge\index{harmonic gauge} of general relativity. We shall specify the set of the conformal harmonic coordinates used for derivation of equations of motion of celestial bodies in ${\mathbb N}$-body system, in section \ref{prf}.

\subsection{Post-Newtonian Field Equations}\label{pnfi}

Before writing down the field equations, it is worth noticing that the post-Newtonian approximation of the scalar-tensor theory of gravity with a variable coupling function $\omega(\Phi)$ has two parameters, $\omega_0$ and $\omega'_0$, characterizing deviation from general relativity. It is more convenient to bring these parameters to the standard form of PPN parameters, $\gamma$ and $\beta$, \cite{willbook}
\begin{eqnarray}
  \label{11.27}
  \gamma & = & \frac{\omega_0+1}{\omega_0+2}\;,\\
  \label{11.28}
  \beta& = & 1+ \frac{\omega'_0}{(2\omega_0+3)(2\omega_0+4)^2}\;.
\end{eqnarray}
General relativity is obtained as a limiting case of the scalar-tensor
theory when parameters $\gamma=\beta=1$ or $\omega_0\rightarrow\infty$. Notice that in order to get this limit convergent, the derivative of the coupling function, $\omega'_0$, must grow slower than $\omega_0^3$ as $\omega_0$ approaches infinity. Currently, there are no experimental data restricting the asymptotic behavior of $\omega'_0\sim\omega_0^3\beta$ which could help us to understand better the nature of the coupling function $\omega(\Phi)$. This makes the parameter $\beta$ a primary target for experimental study in the near-future gravitational experiments \cite{2008ASSL..349.....D,astrod_2009ExA,ciufolini_2008} including the advanced lunar laser ranging (LLR) \cite{2008AdSpR..42.1378K,2012NIMPA,Murthy_2013RPPh} and gravitational wave detectors \citep{Will_2003CQG}. The background scalar field $\Phi_0$ and the parameter of coupling $\omega_0$ determine the observed numerical value of the universal gravitational constant
\begin{equation}
  \label{10.6}
  G =\frac{2\omega_0+4}{2\omega_0+3 }\Phi_0^{-1}\;.
\end{equation}
Had the background value $\Phi_0$ of the scalar field been driven by cosmological evolution, the measured values of the universal gravitational constant $G$ and parameters $\beta$ and $\gamma$ would depend on time \citep{galkop_2016PhRvD}. Notice also that in the geometric system of units $G=1$, and equation \eqref{10.6} reads 
\begin{equation}
  \label{10.6axe}
  \Phi_0 =\frac{2\omega_0+4}{2\omega_0+3 }=\frac{2}{\g+1}\;,
\end{equation}
which allows us to express the background value $\Phi_0$ of the scalar field in terms of the PPN parameter $\g$. 

Let us now substitute the post-Newtonian expansions given by equations(\ref{11.4})--(\ref{11.10}) to the field equations (\ref{10.2}), (\ref{10.4}) and make use of the conformal harmonic gauge condition \eqref{11.3} in the first post-Newtonian approximation. It reads
\begin{eqnarray}
  \label{11.11}
  \pd_0\left(h_{kk}+h_{00}\right)+2(1-\g)\pd_0\varphi & = & 2\pd_j h_{0j}\;,\\
  \label{11.12}
  \pd_i\left(h_{kk}-h_{00}\right)+2(1-\g)\pd_i\varphi & = & 2\pd_j h_{ij};,
\end{eqnarray}
where, for the sake of simplifying the field equations, we have introduced a new notation of the post-Newtonian perturbation, ${\phi}^{(2)}$, of the scalar field, namely,
\begin{equation}
  \label{10.6aa}
  {\phi}^{(2)}\equiv(1-\gamma)\varphi\;. 
\end{equation}
It is worth noting that in the first post-Newtonian approximation the metric tensor component ${h}^{(4)}_{00}\equiv l_{00}$ does not enter \eqref{11.11}, \eqref{11.12} and should be taken into account only at the second post-Newtonian approximation which we don't consider in the present paper.

After making use of the stress-energy tensor (\ref{11.21})--(\ref{11.22}), definitions of the PPN parameters (\ref{11.27})--(\ref{11.28}) and \eqref{10.6axe}, one obtains the final form of the post-Newtonian field equations:
\begin{eqnarray}
  \label{11.29}
  \Box_\eta\varphi&=&-4\pi \rho^{\ast}\;,\\
  \label{nervc34}
  \Box_\eta h_{00} &=& -8\pi \gamma\rho^{\ast}\;,\\
    \label{11.32}
   \Box_\eta h_{ij} &=& -8\pi \gamma\rho^{\ast}\delta_{ij}\;,\\
   \label{11.31}
      \Box_\eta h_{0i} &=&\phantom{-}8\pi (1+\gamma)\rho^{\ast}v^i\;,\\
 \label{11.33}
  \Box_\eta l_{00} &=&  
   -8\pi\rho^{\ast}\left[(\gamma+\frac{1}{2})\,v^2+\Pi+\gamma\,\frac{{\mathfrak{s}}^{kk}}{\rho^{\ast}}-\frac{h_{kk}}{6}-(2\beta-\gamma-1)\varphi\right]- \frac12\Box_\eta\left[h_{00}^2+4(\beta-1)\varphi^2 \right] \;,
\end{eqnarray}
where the $\eta$-box symbol, $\Box_\eta\equiv\eta^{\mu\nu}\partial_\mu\partial_\nu$, is the D'Alembert (wave)\index{D'Alembert} operator \index{wave operator} of the Minkowski spacetime.
Equations (\ref{11.29})--(\ref{11.33}) are valid in the conformal harmonic coordinate charts defined by the gauge condition (\ref{11.3}) imposed on the components of the metric tensor. Their solution depend on the boundary conditions imposed on the metric tensor and the scalar field perturbations. In their own turn, the boundary conditions singled out a certain type of coordinate chart. We discuss the coordinate charts in next section. 

\section{Parametrized Post-Newtonian Coordinates}\label{prf}\index{coordinates!post-Newtonian!parametrized}

Standard textbooks on the post-Newtonian celestial mechanics \citep{fockbook,infeld_book,sof89,vab,ciufolini_book,kopeikin_2011book,Soffel_2013book} derives post-Newtonian equations of motion in a particular gauge to suppress the gauge-dependent effects and to bring the equations to a form which is suitable for finding analytic solutions and for computational applications like numerical orbital simulations, data processing, etc. The coordinate-based approach is also used for solving the field equations and deriving relativistic equations of motion of compact inspiralling binaries for the purposes of gravitational wave astronomy \citep{asada_2011,Baiotti_2016JPhCS,schaefer_2011mmgr,Blanchet_2002LRR}. The post-Newtonian equations admit a large freedom in making the gauge (coordinate) transformations on spacetime manifold as well as in the configuration space of the orbital parameters characterizing motion of bodies \citep{efroim_2003JMP,efroim_2004A&A}. Therefore, each single term taken in such post-Newtonian equations separately from the others makes no physical sense -- it can be always changed or even eliminated by making the post-Newtonian coordinate transformations. Only after the equations are solved and their solutions are substituted to observables, we can unambiguously discuss gravitational physics because the observables are invariantly defined. Therefore, a primary goal of the present paper is to derive the post-Newtonian equations of translational and rotational motion of arbitrary structured bodies in ${\mathbb N}$-body problem in a fully-covariant form. Nonetheless, coordinate-dependent form of equations of motion is more convenient for practical use in various applications. This is why we, first, derive the equations of motion in the conformal harmonic coordinates and, then, establish their correspondence to the covariant form of the equations of motion.  

Derivation of the covariant equations of motion of bodies from the field equations can be achieved directly by the methods of differential geometry like in the Mathisson-Papapetrou-Dixon (MPD) formalism. They can be compared with the coordinate-dependent form of the equations of motion by projecting the corresponding covariant quantities onto the coordinate basis but we use an alternative approach in the present paper. More specifically, we build a set of ${\mathbb N}$ local coordinate charts adapted to each body, derive equations of motion of each body in the local chart, and then, prolongate the coordinate-dependent description to the covariant form by making use of the Einstein principle of equivalence (EEP) applied on the effective background spacetime manifold $\bar{M}$ to the multipoles propagated along the accelerated worldline of the origin of the local coordinates. This procedure is equivalent to "comma-goes-to-semicolon" rule \citep[Chapter 16]{mtw} applied on the worldline of the origin of the local coordinates.  EEP effectively allows us to replace each spatial partial derivative $\hat\pd_i$ in the local coordinates with a covariant derivative $\bar\nabla_\a$ projected on the hypersurface being orthogonal to 4-velocity $\bar u^\a$ of the origin of the local coordinates. It also replaces each time derivative in the local coordinates with the Fermi-Walker covariant derivative of the Fermi-Walker transport -- see section \ref{n3cz52s} for more detail. 

Nonetheless, it is not guaranteed that taking the first post-Newtonian equations of motion and "covariantizing" them by making use "comma-goes-to-semicolon" rule, will automatically lead to results which are even formally valid in the fast-motion and thus, for binaries, strong-field regime. Each term in the "generalized" covariant equations of motion results from a corresponding term in the post-Newtonian equations of motion, which have themselves relied on the post-Newtonian field equations for their derivation. It is certainly conceivable and perhaps even likely, especially at sufficiently high orders in the multipole expansions, that there could exist higher-order non-linearities and higher-order time-derivative terms in some appropriate formally valid covariant equations of motion which would leave no imprint on the appropriately expanded post-Newtonian equations of motion. Such terms would then not be produced by the covariantization procedure as implemented in the present paper. The limits of application of the EEP to the derivation of the covariant equations of motion beyond the first post-Newtonian approximation requires additional study. 
  
Direct derivation of the covariant equations of motion of extended bodies having an arbitrary set of multipoles has been proposed in general relativity by \citet{mathisson_2010GReGr_1,mathisson_2010GReGr_2}, further developed by \citet{tulczyjew1,tulczyjew1_1962}, \citet{Papapetrou23101951,pap1,pap2}, \citet{Taub_1965}, \citet{madore_1969} and, especially, by \citet{dixon_1970_1,dixon_1970_2,dixon_1973GReGr,dixon_1974_3,dixon_1979} with some improvements made by \citet{ehlers_1977GReGr,schattner_1979GReGr} and \citet{Dixon2015}. Subsequent development of the MPD covariant approach  \citep{bini_2009GReGr,dirk_2013PhLA,Obukhov_Puetzfeld2014,helical_2012PhRvD} brought more progress to our understanding of the covariant nature of motion but it has not yet been elaborated to the extent that allows to apply the formalism in astrophysical work. The MPD approach operates on worldlines of the center-of-mass of the extended bodies which are considered as point-like particles endowed with an infinite set of Dixon's multipoles \citep{dixon_1973GReGr}. Such treatment of the extended bodies requires to replace the continuous stress-energy tensor of matter with a, so-called, stress-energy skeleton\index{tensor!stress-energy!skeleton} defined in terms of distributions \citep{shilov_1968}. The skeleton must lead to the same solution of the field equations and to the same equations of motion as the continuous stress-energy tensor. This identity has been checked in the linearized approximation of general relativity but it is not yet clear how to build the skeleton in the non-linear gravity regime that hampers extension of the MPD approach to astrophysical objects with strong gravity like neutron stars and black holes which equations of motion are currently derived by the matched asymptotic expansions technique \citep{das1,das2,gorbonos_2004JHEP,gorbonos_2005CQGra,futamase_2008PhRvD}. 

The MPD covariant approach to the problem of motion of ${\mathbb N}$-body system of extended bodies has an ambiguity concerning the most optimal definition of the center of mass of extended body. There are four competing mathematical definitions based on the, so-called, spin supplementary condition demanding the intrinsic angular momentum (spin) of the body to be orthogonal to either 4-velocity of the center of mass (Mathisson-Pirani condition) or to body's linear momentum (Tulczyjew-Dixon condition) or to some time-like vector (Newton-Wigner condition) or to the unit vector being tangent to the coordinate time axis (Corinaldesi-Papapetrou condition). Depending on the choice of the spin supplementary condition the MPD equations of motion take different forms leading to different solutions of the equations of motion which are intensively discussed in literature -- see, for example, \citep{Kyrian_2007MNRAS,helical_2012PhRvD,Lukes_2014PhRvD,Mikoczi_2017PhRvD,Costa_2018PhRvD} -- but there is no general agreement which solution corresponds to a real physical motion of the body. 

The above-mentioned problems with the MPD formalism convinced us to use more practical, coordinate-based route to the derivation of covariant equations of motion used along with the method of asymptotic matching of the solutions of the internal and external problems in ${\mathbb N}$-body problem and the Blanchet-Damour (BD) multipole formalism. The employment of a set of global and local coordinates is a necessary intermediate step in building the covariant theory of motion of extended bodies. Coordinates are necessary to give physically-meaningful definition of the BD multipoles of the bodies in the non-linear gravity regime, to unambiguously single out the center of mass of each body and its worldline and to separate the self-action force of each body \index{self-force} from the external gravitational force of the other bodies of ${\mathbb N}$-body system. The coordinate description is practically useful in astrophysics for computation of orbital motion of inspiralling binaries and in the relativistic celestial mechanics of the solar system \citep{kopeikin_2011book}. On the other hand, the coordinate description of the equations of motion can be easily converted to the covariant form as soon as the theory is completed. As we have learned above, discussion of the dynamics of ${\mathbb N}$-body problem requires introduction of one global and ${\mathbb N}$ local coordinate charts adapted to each body. Geometric properties of the coordinate charts as well as their kinematic and dynamic characteristics are defined by the boundary conditions imposed on the metric tensor and scalar field. 

\subsection{Global Coordinate Chart}\label{ssbf}\index{coordinates!post-Newtonian!global}

\subsubsection{Boundary Conditions}\label{zoki}

We consider an isolated system consisting of ${\mathbb N}$ extended bodies which are gravitationally bound, occupy a finite volume of space and there is no other matter outside it. Since there is no matter outside the system, the spacetime manifold with the metric tensor $g_{\alpha\beta}$ can be considered at infinity as asymptotically approaching to flat spacetime with the Minkowski metric $\eta_{\alpha\beta}=\mathrm{diag}(-1,+1,+1,+1)$. We further assume, in accordance with the post-Newtonian approximations, that there are no physical singularities on the manifold\index{manifold} like black holes\index{black hole}, wormholes\index{wormhole}, etc. among the bodies of the system, and that the bodies move slowly and gravitational field\index{gravitational field!weak} is weak everywhere.\index{coordinates!post-Newtonian!boundary conditions}

These founding assumptions allow us to cover the whole spacetime manifold with a global coordinate chart denoted as $x^\alpha=(x^0,x^i)$, where $x^0=t$ is the coordinate time and $x^i\equiv{\bm x}$ are the spatial coordinates. The global coordinates are used for describing orbital dynamics of the bodies, for calculating generation and propagation of gravitational waves emitted by the isolated system, and for formulating the global laws of conservation and conserved quantities \citep{Petrov_2017book}. The coordinate time, $t$, and spatial coordinates, $x^i$, have no immediate physical meaning in the regions of space where gravitational field is not negligible. However, when one approaches to infinity the global coordinates approximate the Lorentz coordinates of inertial observer in the Minkowski space. For this reason, one can interpret the coordinate time $t$ and the spatial coordinates $x^i$ respectively as the proper time and the proper distance measured by a set of the inertial observers located at rest at spatial infinity \citep{fockbook}. The global coordinates are not defined uniquely but up to a group of transformation preserving the asymptotic flatness of spacetime. Contrary to the original expectations this group of transformation is not 10-parametric Poincar\'e group but the infinite-dimensional BMS group which is isomorphic to the semi-direct product of the homogeneous Lorentz group with the Abelian group of super-translations \citep{frau_2004LRR}. The Poincar\'e group is a sub-group of the BMS group.

Precise mathematical description of properties of the global post-Newtonian coordinates can be given in terms of the metric tensor that is solution of the field equations \eqref{nervc34}--\eqref{11.33} with the boundary conditions\index{boundary} imposed at infinity. To formulate the boundary conditions, we introduce the metric perturbation 
\begin{equation}
  \label{mtp}
  h_{\alpha\beta}(t,\bm{x})\equiv g_{\alpha\beta}(t,\bm{x})-\eta_{\alpha\beta}\;,
\end{equation}
where $h_{\a\b}$ is the full post-Newtonian series defined in (\ref{exp}).
The global coordinates must match asymptotically with the inertial coordinates of the Minkowski spacetime which presumes that the products $rh_{\alpha\beta}$ and $r^2 h_{\alpha\beta,\gamma}$ where $r=|{\bm x}|$, are bounded at spatial infinity\index{spatial infinity}\index{infinity!spatial} \citep{fockbook,Damour_1983grr}, while at the future null infinity\index{future null infinity}\index{infinity!null future}
\begin{equation}
  \label{12.1}
  \lim_{\substack{r\rightarrow\infty\\t+r=\mathrm{const.}}}h_{\alpha\beta}(t,\bm{x})=0\;.
\end{equation}
Additional boundary condition must be imposed on the first derivatives of the metric tensor to exclude non-physical (advanced) radiative solutions associated with gravitational waves\index{gravitational waves} incoming to ${\mathbb N}$-body system from infinity. This condition is imposed because we have assumed that there are no sources of gravitational waves outside of the isolated ${\mathbb N}$-body system. It is formulated as follows \citep{fockbook,Damour_1983grr}
\begin{equation}
  \label{12.2}
  \lim_{\substack{r\rightarrow\infty\\t+r=\mathrm{const.}}}[\pd_r(rh_{\alpha\beta})+\pd_t(rh_{\alpha\beta})]=0\;,
\end{equation}
where $\pd_r$ and $\pd_t$ denote the partial derivatives with respect to radial coordinate $r$ and time $t$, respectively.
Though, the first post-Newtonian approximation does not include gravitational waves\index{gravitational waves}\index{gravitational wave}, the boundary condition (\ref{12.2}) tells us to choose the retarded solution of the field equation (\ref{nervc34})-(\ref{11.33}). 

Similarly, we impose the "no-incoming-radiation" conditions\index{boundary conditions!no-incoming radiation} on the perturbation $\varphi$ of the scalar field defined in (\ref{10.6aa}),
\begin{equation}
  \label{12.3}
  \lim_{\substack{r\rightarrow\infty\\t+r=\mathrm{const.}}}\varphi(t,\bm{x})=0\;,
\end{equation}
\begin{equation}
  \label{12.3a}
  \lim_{\substack{r\rightarrow\infty\\t+r=\mathrm{const.}}}[\pd_r(r\varphi)+\pd_t(r\varphi)]=0\;.
\end{equation}
These conditions eliminates the advanced radiative solution for the scalar field.

\subsubsection{Scalar Field}\la{n465vsd4}

Scalar field in the global coordinates is obtained as a solution of the field equation \eqref{11.29} with the no-incoming (scalar) radiation boundary conditions \eqref{12.3}, \eqref{12.3a}. This solution is a retarded potential
\be\la{n3cz4s3}
\varphi(t,{\bm x})= \int\limits_{{\mathbb R}^3}\frac{\rho^{\ast}(t-|\bm{x}-\bm{x}'|,\bm{x}')}{|\bm{x}-\bm{x}'|}d^3x'\;,
\ee
where the integration is performed over the entire space ${\mathbb R}^3$. The post-Newtonian expansion of the retarded potential is obtained by expanding the integrand in \eqref{n3cz4s3} around the instant of time $t$, and integrating each term of the expansion. In what follows, we need merely the first term of the expansion. Moreover, since the density of matter $\rho^{\ast}$ vanishes outside the bodies of ${\mathbb N}$-body system, the integration is carried out over only the volumes of the bodies, which yields
\be\label{12.5qqar}
  \varphi(t,\bm{x})  =  U(t,\bm{x})\;.
\ee
Here,
\be\la{asz9t2}
U(t,\bm{x})=\sum_{\rm C} U_{\rm C}(t,\bm{x})\;,
\ee
is a linear superposition of the Newtonian gravitational potentials $U_{\rm C}(t,\bm{x})$ of the bodies $(C=1,2,\ldots,N)$, and
\be
\label{12.10rtfz}
U_{\rm C}(t,\bm{x}) =\int\limits_{{\cal V}_{\rm C}}\frac{\rho^{\ast}(t,\bm{x}')}{|\bm{x}-\bm{x}'|}d^3x'\;,
  \ee  
where ${\cal V}_{\rm C}$ denotes the spatial volume occupied by the body C.

Subsequent derivation requires to single out one of the bodies, let say a body B, and split the scalar field in two parts -- internal and external,
\be\la{nx4z9d5}
U(t,\bm{x})=U_{\rm B}(t,\bm{x})+\bar U(t,\bm{x})\;,   
\ee  
where $U_{\rm B}$ denotes the internal gravitational potential produced by the body B alone, 
\be
U_{\rm B}(t,\bm{x}) =\int\limits_{{\cal V}_{\rm B}}\frac{\rho^{\ast}(t,\bm{x}')}{|\bm{x}-\bm{x}'|}d^3x'\;,
  \ee  
and 
\be\la{tb54vd}
\bar U(t,\bm{x})=\sum_{{\rm C}\not=\B} U_{\rm C}(t,\bm{x})\;,
\ee
denotes the external gravitational potential of all other bodies of ${\mathbb N}$-body system but the body B. 

\subsubsection{Metric Tensor}\la{mtsf183}
The metric tensor $g_{\alpha\beta}(t,{\bm x})$ in the global coordinates is obtained by solving the field equations (\ref{nervc34})--(\ref{11.33}) with the boundary conditions (\ref{12.1})--(\ref{12.2}). It yields \citep{kovl_2004,kopeikin_2011book}
\begin{eqnarray}
  \label{12.6}
  h_{00}(t,\bm{x}) & = & \phantom{-}2\,U(t,\bm{x})\;,\\
\label{12.7}
    h_{ij}(t,\bm{x})&= &\phantom{-}2\gamma\delta_{ij}U(t,\bm{x})\;,\\
  \label{12.8}
  h_{0i}(t,\bm{x})&=& -2(1+\gamma)\,U^i(t,\bm{x})\;,\\
   \label{12.9}
    l_{00}(t,\bm{x}) & = &\phantom{-} 2\Psi(t,\bm{x})
    -2\beta U^2(t,\bm{x})-\pd_{tt}\chi(t,\bm{x})\;,
\end{eqnarray}
where the operator $\pd_{tt}\equiv\pd^2/\pd t^2$, the post-Newtonian potential
\begin{equation}
  \label{12.9ex}
  \Psi(t,\bm{x}) \equiv \bigg(\gamma+\frac{1}{2}\bigg)\Psi_1(t,\bm{x})+(1-2\beta)\Psi_2(t,\bm{x})+\Psi_3(t,\bm{x})+\gamma\Psi_4(t,\bm{x})\;,
\end{equation}
and parameters $\g$ and $\b$ have been defined in (\ref{11.27}) and (\ref{11.28}) respectively.

Newtonian gravitational potential $U$ has been defined above in \eqref{asz9t2}. Post-Newtonian potentials $U^i, \chi$, $\Psi_n$ $(n=1,2,3,4)$ are linear combinations of the gravitational potentials produced by the bodies of ${\mathbb N}$-body system,
\begin{equation}
  \label{12.9a}
  U^i(t,\bm{x})=\sum_{\rm C} U^i_{\rm C}(t,\bm{x}),\qquad \Psi_n(t,\bm{x})=\sum_{\rm C}\Psi_{{\rm C}n}(t,\bm{x}),\qquad\chi(t,\bm{x})=\sum_{\rm C}\chi_{\rm C}(t,\bm{x})\;.
\end{equation}
Here, the summation index $C=1,2,\ldots,N$ numerates the bodies of $\mathbb{N}$-body system, and the gravitational potentials of body C are defined as integrals performed over a spatial volume ${\cal V}_{\rm C}$ occupied by the body's matter,
\begin{eqnarray}
\label{12.11}
  U^i_{\rm C}(t,\bm{x}) &=&\int\limits_{{\cal V}_{\rm C}}\frac{\rho^{\ast}(t,\bm{x}')v^i(t,\bm{x}')}{|\bm{x}-\bm{x}'|}d^3x'\;,
\\
  \label{12.13}
  \Psi_{{\rm C}1}(t,\bm{x})&=&\int\limits_{{\cal V}_{\rm C}}\frac{\rho^{\ast}(t,\bm{x}')v^2(t,\bm{x}')}{|\bm{x}-\bm{x}'|}d^3x'\;,
  \\
  \label{12.14a}
  \Psi_{{\rm C}2}(t,\bm{x})&=&\int\limits_{{\cal V}_{\rm C}}\frac{\rho^{\ast}(t,\bm{x}')U(t,\bm{x}')}{|\bm{x}-\bm{x}'|}d^3x'\;,
\\
  \label{12.15}
  \Psi_{{\rm C}3}(t,\bm{x})&=&\int\limits_{{\cal V}_{\rm C}}\frac{\rho^{\ast}(t,\bm{x}')\Pi(t,\bm{x}')}{|\bm{x}-\bm{x}'|}d^3x'\;,
\\
  \label{12.16}
  \Psi_{{\rm C}4}(t,\bm{x})&=&\int\limits_{{\cal V}_{\rm C}}\frac{{\mathfrak{s}}^{kk}(t,\bm{x}')}{|\bm{x}-\bm{x}'|}d^3x'\;,
\end{eqnarray}
where $v^i=v^i(t,{\bm x})$ is velocity of the element of matter located at time $t$ at a spatial point $x^i={\bm x}$ in the global coordinates, and $v^2=\d_{ij}v^iv^j$.  

Superpotential $\chi_{\rm C}$ is determined as a particular solution of the inhomogeneous Poisson equation\index{Poisson!equation}\index{equation!Poisson}
\begin{equation}
  \label{12.18}
  \triangle\chi_{\rm C}(t,{\bm x}) = -2U_{\rm C}(t,{\bm x})
\end{equation}
where $\triangle\equiv\d^{ij}\pd_i\pd_j$ is the Laplace operator in the Euclidean space. The source of the superpotential $\chi_{\rm C}$ is the Newtonian gravitational potential $U_{\rm C}$ that presents everywhere in a whole space. Nevertheless, because it falls off as $1/r$ at infinity, solution of the Poisson equation \eqref{12.18} has a compact support, and is given by an integral taken over the finite volume of body C  \citep{fockbook,willbook}
\be
  \label{12.12}
  \chi_{\rm C}(t,\bm{x}) = -\int\limits_{{\cal V}_{\rm C}}\rho^{\ast}(t,\bm{x}')|\bm{x}-\bm{x}'|d^3x'\;.
\ee 
It is useful to emphasize that all above-given volume integrals defining the metric tensor\index{tensor!metric} in the global coordinates, are taken on the space-like hypersurface ${\cal H}_t$ of constant coordinate time $t$. Changing the time coordinate does not change the functional form of the integrals but transforms the time hypersurface that makes the numerical value of the integrals different. This remark is important for understanding the post-Newtonian transformations and the technique of matched asymptotic expansions of the metric tensor and scalar field which we explain below in section \ref{pntb}. 

In what follows we single out a body B, and split all post-Newtonian potentials in two parts -- internal and external -- like we did above in \eqref{nx4z9d5} for the Newtonian gravitational potential
\ba\la{a1v3e}
U^i(t,\bm{x})&=&U^i_{\rm B}(t,\bm{x})+\bar U^i(t,\bm{x})\;, \\
\Psi(t,\bm{x})&=&\Psi_{\rm B}(t,\bm{x})+\bar\Psi(t,\bm{x})\;,\\
\chi(t,\bm{x})&=&\chi_{\rm B}(t,\bm{x})+\bar \chi(t,\bm{x})\;.
\ea  
Here, functions with sub-index B denote the internal potentials produced by the body B alone, 
\begin{eqnarray}
\label{12.11xx}
  U^i_{\rm B}(t,\bm{x}) &=&\int\limits_{{\cal V}_{\rm B}}\frac{\rho^{\ast}(t,\bm{x}')v^i(t,\bm{x}')}{|\bm{x}-\bm{x}'|}d^3x'\;,
\\
  \label{12.13xx}
  \Psi_{{\rm B}1}(t,\bm{x})&=&\int\limits_{{\cal V}_{\rm B}}\frac{\rho^{\ast}(t,\bm{x}')v^2(t,\bm{x}')}{|\bm{x}-\bm{x}'|}d^3x'\;,
  \\
  \label{12.14axx}
  \Psi_{{\rm B}2}(t,\bm{x})&=&\int\limits_{{\cal V}_{\rm B}}\frac{\rho^{\ast}(t,\bm{x}')U(t,\bm{x}')}{|\bm{x}-\bm{x}'|}d^3x'\;,
\\
  \label{12.15xx}
  \Psi_{{\rm B}3}(t,\bm{x})&=&\int\limits_{{\cal V}_{\rm B}}\frac{\rho^{\ast}(t,\bm{x}')\Pi(t,\bm{x}')}{|\bm{x}-\bm{x}'|}d^3x'\;,
\\
  \label{12.16xx}
  \Psi_{{\rm B}4}(t,\bm{x})&=&\int\limits_{{\cal V}_{\rm B}}\frac{{\mathfrak{s}}^{kk}(t,\bm{x}')}{|\bm{x}-\bm{x}'|}d^3x'\;,\\
  \label{12.18xx}
  \chi_\B(t,\bm{x}) &=& -\int\limits_{{\cal V}_{\rm B}}\rho^{\ast}(t,\bm{x}')|\bm{x}-\bm{x}'|d^3x'\;,
\end{eqnarray}
and functions covered with a bar, denote the external potentials,
\be\la{3d8b1a}
\bar{U}^i(t,{\bm x})=\sum\limits_{{\rm C}\not={\rm B}}U^i_{\rm C}(t,{\bm x})\;,\qquad
\bar{\Psi}(t,{\bm x})=\sum\limits_{{\rm C}\not={\rm B}}\Psi_{\rm C}(t,{\bm x})\;,\qquad
\bar{\chi}(t,{\bm x})=\sum\limits_{{\rm C}\not={\rm B}}\chi_{\rm C}(t,{\bm x})\;,
\ee
where potentials $U^i_{\rm C}$, $\Psi_{\rm C}$, $\chi_{\rm C}$ are given by integrals (\ref{12.11})--(\ref{12.16}) respectively. It is worth emphasizing \citep{Fichte_1950} that the integrand of integrals (\ref{12.14a}), \eqref{12.14axx} depends on the {\it total} gravitational potential $U$ of all bodies of ${\mathbb N}$-body system as defined in \eqref{asz9t2}.  It is also important to notice that the Newtonian gravitational potential $U(t,{\bm x})$ has a double camouflage in scalar-tensor theory of gravity. It appears in the solution \eqref{12.5qqar} of the field equation for scalar field $\varphi$, and, also, in (\ref{12.6}), (\ref{12.7}) describing perturbations of the metric tensor components $h_{00}$ and $h_{ij}$. It would be wrong, however, to interpret the metric tensor component $h_{00}=2U$, and the trace $h\equiv\d^{ij}h_{ij}=h_{kk}=6U$ like scalars - they can be expressed in terms of the scalar field $\varphi$ alone only in the global coordinates. By definition, the metric tensor perturbations, $h_{00}$ and $h_{kk}$, are transformed as tensors not as scalars. 

Mathematical description of orbital dynamics of extended bodies in ${\mathbb N}$-body system would be significantly simplified if we could keep position of the center of mass of ${\mathbb N}$-body system at the origin of the global coordinates for any instant of time. This condition suggests that the dipole, $\mathbb{D}^i$, of gravitational field of ${\mathbb N}$-body system in the multipolar expansion of $h_{00}(t,{\bm x})$ component of the metric tensor perturbation vanishes along with the dipole (linear momentum), $\mathbb{P}^i$, in the multipolar expansion of $h_{0i}$ component \citep{mtw}. This condition cannot be satisfied at higher post-Newtonian approximations due to the gravitational wave recoil which makes the system's center of mass moving with acceleration \citep{Fitchett_1983MNRAS}. Nonetheless, in the first and second post-Newtonian approximations the orbital dynamics of ${\mathbb N}$-body system is fully determined by the Lagrangian admitting ten conservation laws corresponding to ten infinitesimal generators of the Poincar\'e group preserving the invariance of the Lagrangian of ${\mathbb N}$-body problem \citep{fockbook,infeld_book,gk86,Damour_1985GReGr,Damour_1989MG5}. 
The post-Newtonian law of conservation of the total linear momentum, $\mathbb{P}^i$, allows to hold the center of mass of ${\mathbb N}$-body system always at the origin of the global coordinate chart \citep{kopeikin_2011book}.

\subsection{Local Coordinate Chart}\label{ems}\index{coordinates!post-Newtonian!local}

\subsubsection{Boundary Conditions}\label{pvz1}

We label the local coordinates adapted to body B by letters $w_{\rm B}^\alpha=(w_{\rm B}^0,w_{\rm B}^i)=(u_{\rm B}, w_{\rm B}^i)$ where $u_{\rm B}$ stands for the local coordinate time and $w_{\rm B}^i$ denote the spatial coordinates $({\rm B}=1,2,\dots,N)$. There are ${\mathbb N}$ local coordinate charts -- one for each body. In case when there is no confusion, we drop off the sub-index B in the notation of the local coordinates. Hence, by default the local coordinates adapted to body B will be denoted by $w^\a=(u,w^i)\equiv (u_\B, w_\B^i)$. The origin of the local coordinates adapted to body B moves along a reference worldline ${\cal W}$ which is chosen to be sufficiently close to the worldline $\cal Z$ of the center of mass\index{center of mass} of body B. Initially, the two worldlines are different but can be made identical after careful study of the problem of definition of the center of mass and its equations of motion relative to ${\cal W}$. This will be done in section \ref{n4r6v}. 

The local coordinates are used to describe the internal motion of matter inside the body, to define its center of mass, linear momentum, spin and the other, higher-order internal multipoles of body's gravitational field. The importance of the local coordinates for adequate mathematical description of relativistic dynamics of extended, self-gravitating massive bodies in ${\mathbb N}$-body system was emphasized by \citet{fockbook}. Concrete mathematical construction of the body-adapted, local coordinates was achieved in the post-Newtonian approximation by the technique of asymptotic matching in papers \citep{ashb2,Kopejkin_1988CeMec} -- for extended bodies, and in papers \citep{das2,DEath_1996book} -- for black holes. Later on, a more rigorous mathematical BK-DSX formalism of construction of the local coordinates has been elaborated in a series of publications \citep{bk89,bk-nc,dsx1,dsx2,dsx3} which led to the development and adoption of the IAU 2000 resolutions on general-relativistic reference frames in the solar system \citep{iau2000,kopeikin_2011book,Soffel_2013book}. Below we extend this formalism to the scalar-tensor theory of gravity.

Scalar field and metric tensor in the local coordinates adapted to body B are solutions of the field equations (\ref{11.29})--(\ref{11.32}) inside a bounded spatial domain enclosing worldline ${\cal Z}$ of the center of mass of body B and having radius spreading out to another nearest body from ${\mathbb N}$-body system. Thus, the right side of the inhomogeneous equations (\ref{11.29})--(\ref{11.33}) includes only matter of body B. In order to distinguish solutions of the field equations in the local coordinates from the corresponding solutions of the field equations in the global coordinates, we put a hat over functions of the local coordinates. Solution of the field equation for metric tensor or scalar field in the local coordinates is a linear combination of a particular solution of the inhomogeneous equation and a general solution of a homogeneous equation. The particular solution yields the internal gravitational field of body B alone while the general solution of the homogeneous equation pertains to the external field of other bodies C$\not=$B. The non-linear nature of the field equation \eqref{11.33} brings in mixed terms $l_{00}$ to the metric tensor perturbation describing a coupling between the first-order perturbations. 

The post-Newtonian solution of the scalar field equation (\ref{11.29}) in the local coordinates adapted to body B is written as a sum of two terms
\begin{equation}
  \label{1.1}
  \hat{\varphi}(u,\bm{w}) = \hat{\varphi}^{\mathrm{int}}(u,\bm{w})+\hat{\varphi}^{\mathrm{ext}}(u,\bm{w})\;,
\end{equation}
describing contributions of the internal matter of body B and external bodies C$\not=$B respectively. If we had no other bodies but the body B, the internal solution had to vanish at infinity. Hence, it obeys the boundary conditions similar to \eqref{12.3}, \eqref{12.3a}. The external solution must be regular at the origin of the local coordinates and diverges at infinity.  

Perturbation of the metric tensor in the local coordinates is denoted
\begin{equation}
  \label{nzim}
  \hat{h}_{\mu\nu}(u,\bm{w}) = \hat{g}_{\mu\nu}(u,\bm{w})-\eta_{\mu\nu}\;,
\end{equation}
where each component of $\hat{h}_{\mu\nu}$ is expanded in the post-Newtonian series similar to \eqref{11.4}--\eqref{11.6},
\ba
  \label{11.4www}
  \hat h_{00}(u,\bm{w}) & = & \epsilon^2\hat{h}^{\rm (2)}_{00}(u,\bm{w})+\epsilon^4\hat{h}^{\rm (4)}_{00}(u,\bm{w})+\mathcal{O}(\epsilon^6),\\
  \label{11.5www}
  \hat h_{0i}(u,\bm{w}) & = & \epsilon^3\hat{h}^{\rm (3)}_{0i}(u,\bm{w})+\mathcal{O}(\epsilon^5),\\
  \label{11.6www}
  \hat h_{ij}(u,\bm{w}) & = & \delta_{ij}+ \epsilon^2\hat{h}^{\rm (2)}_{ij}(u,\bm{w})+\mathcal{O}(\epsilon^4),
  \ea
and each term of the post-Newtonian series will be denoted  
\begin{equation} 
\label{notwww}
  \hat h_{00}\equiv\hat {h}^{\rm (2)}_{00}\;,\qquad  \hat l_{00}\equiv\hat{h}^{\rm (4)}_{00}\;,\qquad
  \hat h_{0i}\equiv\hat{h}^{\rm (3)}_{0i}\;,\qquad  \hat h_{ij}\equiv\hat{h}^{\rm (2)}_{ij}\;,\qquad
  \hat h\equiv\hat{h}^{\rm (2)}_{kk}\;.
\end{equation}

Post-Newtonian solution of the field equations (\ref{nervc34})--(\ref{11.33}) in the local coordinates is given as a sum of three terms \citep{th_1985}
\begin{equation}
  \label{1.2}
  \hat{h}_{\mu\nu}(u,\bm{w}) = \hat{h}^{\mathrm{int}}_{\mu\nu}(u,\bm{w}) +\hat{h}^{\mathrm{ext}}_{\mu\nu}(u,\bm{w}) +\hat{h}^{\mathrm{mix}}_{\mu\nu}(u,\bm{w})\;,
\end{equation}
where $\hat{h}^{\mathrm{int}}_{\mu\nu}$ describes gravitational field generated by the internal matter of body B, $\hat{h}^{\mathrm{ext}}_{\mu\nu}$ describes the tidal gravitational field produced by external bodies C$\not=$B, and the term $\hat{h}^{\mathrm{mix}}_{\mu\nu}$ is a contribution due to the non-linear coupling of the internal and external metric perturbations in the field equation (\ref{11.33}). In the first post-Newtonian approximation the coupling term $\hat{h}^{\mathrm{mix}}_{\mu\nu}$ appears only in $\hat l_{00}(u,\bm{w})$ component of the metric tensor perturbation. The body-frame field $\hat{h}^{\mathrm{int}}_{\mu\nu}(u,\bm{w})$ is the same as if the other bodies of ${\mathbb N}$-body system were absent. Therefore, it is defined by imposing the boundary conditions similar to \eqref{12.1}, \eqref{12.2}. Since the external metric perturbation $\hat{h}^{\mathrm{ext}}_{\mu\nu}(u,\bm{w})$ has a physical meaning of the tidal field caused by external bodies, it must be regular on the worldline ${\cal W}$ of the origin of the local coordinates. The coupling field $\hat{h}^{\mathrm{mix}}_{\mu\nu}(u,\bm{w})$ is obtained directly by finding a particular solution of the nonlinear part of the field equation \eqref{11.33}. Since the internal and external part of the metric tensor perturbation have been already specified, there is no need to impose a separate boundary condition on the coupling component of the metric tensor perturbation.   

The origin of the local coordinates moves along some, yet unspecified, worldline, $\cal W$, which will be determined later on by matching the solutions of the field equations obtained in the local and global coordinates in the buffer domain where the two coordinate charts overlap. For we are interested in derivation of equations of motion of the center of mass of each body, we wish to make the origin of the local coordinates coinciding with the center of mass of the body under consideration at any instant of time. This requires a precise post-Newtonian definition of the center of mass\index{center of mass}. Any deficiency in the definition of body's center of mass introduces to the equations of motion fictitious forces and torques\index{inertial force} that have no direct physical meaning. We prove in the present paper that the freedom in choosing the position of the center of mass is large enough to completely remove such fictitious forces and torques from the equations of motion of extended bodies in the scalar-tensor theory of gravity. 

We should also impose a limitation on the rotation of spatial axes of the local coordinates as they move along worldline ${\cal W}$. 
Spatial axes of the local coordinates\index{coordinates!non-rotating!kinematically} are called kinematical non-rotating if their spatial orientation does not change with respect to the spatial axes of the global coordinates at infinity as time goes on \citep{1989rfag.conf....1K,2004fuas.book.....K}. 
Dynamical non-rotating spatial coordinates\index{coordinates!non-rotating!dynamically} are defined by demanding that equations of motion of test particles in the local coordinates do not have the Coriolis\index{Coriolis force} and centrifugal forces\index{centripetal force} \citep{1989rfag.conf....1K}. Because ${\mathbb N}$-body system is isolated the spatial axes of the global coordinate do not rotate in any sense. On the other hand, the local coordinates are adapted to a single body B that is not fully isolated from external gravitational environment of other bodies of ${\mathbb N}$-body system. Therefore, we have to postulate whether the spatial axes of the local coordinates are non-rotating in kinematic or dynamic sense. For the sake of mathematical simplifications in writing solutions of the field equations it is more convenient to postulate that the spatial axes of the local coordinates are not rotating dynamically. Relativistic nature of gravitational interaction suggests that the spatial axes of the dynamically non-rotating local coordinates will be slowly rotating (precessing) in the kinematic sense with respect to the spatial axes of the global coordinates. Relativistic precession of the spatial axes of the local coordinates has a pure geometric origin and includes three physically-different terms that are called respectively de-Sitter (geodetic\index{geodetic precession}), Lense-Thirring (gravitomagnetic\index{gravitomagnetic precession}), and Thomas precession\index{Thomas precession} \citep{mtw}. Exact formula for matrix of the kinematic precession of spatial axes of the local coordinates is given below in equation \eqref{5.18}.

\subsubsection{Scalar Field: internal and external solutions}\label{qopsca}

In the local coordinates adapted to body B, the internal, $\hat{\varphi}^{\mathrm{int}}(u,\bm{w})$, and external, $\hat{\varphi}^{\mathrm{ext}}(u,\bm{w})$, parts of scalar field perturbation \eqref{1.1} have the following form,
\begin{eqnarray}
  \label{1.7}
  \hat{\varphi}^{\mathrm{int}}(u,\bm{w}) & = & \hat{U}_{\rm B}(u,{\bm w})\;,\\
  \label{1.7a}
  \hat{\varphi}^{\mathrm{ext}}(u,\bm{w})&=& \sum_{l=0}^{\infty}\frac{1}{l!} {\cal P}_Lw^L\;.
\end{eqnarray}
Here, the scalar field $\hat{\varphi}^{\mathrm{int}}(u,\bm{w})$ is a particular solution of inhomogeneous equation \eqref{11.29} with the right-hand side depending solely on the matter density $\rho^*$ of body B. It is expressed in terms of the Newtonian gravitational potential of body B, $\hat{U}_{\rm B}(u,{\bm w})$, that is defined below in equation (\ref{1.11}). The scalar field, $\hat{\varphi}^{\mathrm{ext}}(u,\bm{w})$, is a general solution of a homogeneous Laplace equation \eqref{11.29} without sources. As $\hat{\varphi}^{\mathrm{ext}}(u,\bm{w})$ must be regular at the origin of the local coordinates, the solution is given in the form of a Maclaurin series with respect to STF harmonic polynomials, $w^L\equiv w^{<i_1...i_l>}$, made out of the products of the spatial local coordinates $w^i$ and the Kronecker symbols $\d^{ij}$ -- see definition of STF tensor projection in \eqref{stfformula}. Coefficients of the expansion are scalar external multipoles, ${\cal P}_L\equiv{\cal P}_{<i_1...i_l>}(u)$, which are STF\index{STF} Cartesian tensors\index{tensor!Cartesian} in 3-dimensional Euclidean space that is tangent to hypersurface ${\cal H}_u$ of constant coordinate time $u$ taken at the origin of the local coordinates adapted to body B.   

\subsubsection{Metric Tensor: internal solution}\la{n2v5r}\index{metric tensor!internal solution}

The boundary conditions\index{boundary conditions} imposed on the internal solution $ \hat{h}_{\a\b}^{\mathrm{int}}$ for the metric tensor perturbation in the local coordinates adapted to body B are identical with those given in equations (\ref{12.1}), (\ref{12.2}). For this reason the internal solution has the same form as in the global coordinates but all functions now refers solely to body B. We obtain,
\begin{eqnarray}
  \label{1.8}
  \hat{h}_{00}^{\mathrm{int}}(u,\bm{w}) & = &\phantom{-} 2 \hat{U}_{\rm B}(u,\bm{w})\;,\\
  \label{1.9}
  \hat{h}_{0i}^{\mathrm{int}}(u,\bm{w}) & = & -2(1+\gamma)\hat{U}^i_{\rm B}(u,\bm{w})\;,\\
  \label{1.10}
  \hat h_{ij}^{\mathrm{int}}(u,\bm{w}) & = &\phantom{-} 2\gamma\delta_{ij}\hat{U}_{\rm B}(u,\bm{w})\;,\\
   \label{1.8aa}
    \hat{l}_{00}^{\mathrm{int}}(u,\bm{w}) & = &\phantom{-} 2 \hat{\Psi}_{\rm B}(u,\bm{w})
     - 2\beta\hat{U}^2_{\rm B}(u,\bm{w})-\pd_{uu}\hat{\chi}_{\rm B}(u,\bm{w})\;,
\end{eqnarray}
where the partial time derivative $\pd_{uu}\equiv\pd^2/\pd u^2$, 
\begin{eqnarray}
  \label{1.12a}
  \hat{\Psi}_{\rm B}(u,\bm{w}) & = & \bigg(\gamma+\frac{1}{2}\bigg)\hat{\Psi}_{{\rm B}1}(u,\bm{w}) +(1-2\beta)\hat{\Psi}_{{\rm B}2}(u,\bm{w})
  + \hat{\Psi}_{{\rm B}3}(u,\bm{w})+\gamma\hat{\Psi}_{{\rm B}4}(u,\bm{w})\;,
\end{eqnarray}
and index B indicates that the potential having this index is generated by matter of body B only. All the potentials are defined as integrals over volume ${\cal V}_{\rm B}$ occupied by matter of body B:
\begin{eqnarray}
  \label{1.11}
  \hat{U}_{\rm B}(u,\bm{w}) & = & \int\limits_{{\cal V}_{\rm B}}\frac{\rho^*(u,\bm{w}')}{|\bm{w}-\bm{w}'|}d^3w'\;,\\
  \la{qw4cx}
  \hat{U}^{i}_{\rm B}(u,\bm{w}) & = & \int\limits_{{\cal V}_{\rm B}}\frac{\rho^*(u,\bm{w}')\nu^i(u,\bm{w}')}{|\bm{w}-\bm{w}'|}d^3w'\;,
\\
  \label{1.14}
  \hat{\Psi}_{{\rm B}1}(u,\bm{w}) &=& \int\limits_{{\cal V}_{\rm B}}\frac{\rho^{\ast}(u,\bm{w}')\nu^2(u,\bm{w}')}{|\bm{w}-\bm{w}'|}d^3w'\;,
\\
  \label{1.15}
  \hat{\Psi}_{{\rm B}2}(u,\bm{w}) &=& \int\limits_{{\cal V}_{\rm B}}\frac{\rho^{\ast}(u,\bm{w}') \hat{U}_{\rm B}(u,\bm{w}')}{|\bm{w}-\bm{w}'|}d^3w'\;,
\\
  \label{1.16}
  \hat{\Psi}_{{\rm B}3}(u,\bm{w})& =& \int\limits_{{\cal V}_{\rm B}}\frac{\rho^{\ast}(u,\bm{w}')\Pi(u,\bm{w}')}{|\bm{w}-\bm{w}'|}d^3w'\;,
\\
  \label{1.17}
  \hat{\Psi}_{{\rm B}4}(u,\bm{w})& =&\int\limits_{{\cal V}_{\rm B}}\frac{{\mathfrak{s}}^{kk}(u,\bm{w}')}{|\bm{w}-\bm{w}'|}d^3w'\;,
  \\
    \label{1.13}
    \hat{\chi}_{\rm B}(u,\bm{w}) & =&  -\int\limits_{{\cal V}_{\rm B}}\rho^*(u,\bm{w}')|\bm{w}-\bm{w}'|d^3w',
\end{eqnarray}
$\nu^i=dw^i/du$ is the coordinate velocity of body's matter with respect to the origin of the local coordinates. Notice that the integrals \eqref{1.11}--\eqref{1.13} are taken over hypersurface ${\cal H}_u$ of coordinate time $u$ that is different from the hypersurface ${\cal H}_t$ of constant coordinate time $t$, which is used for spatial integration in equations \eqref{12.10rtfz}, (\ref{12.11})--(\ref{12.16}) defining gravitational potentials in the global coordinates $x^\a$. This is important for the post-Newtonian transformation of gravitational potentials as it requires to use a Lie transport of functions from hypersurface ${\cal H}_u$ to hypersurface ${\cal H}_t$ -- for more detail, see \cite[Section 5.2.3]{kopeikin_2011book}.

The internal potentials of the metric tensor in the local coordinates given by (\ref{1.8}) and (\ref{qw4cx}) are connected through the exact equation
\begin{equation}
  \label{1.19}
  \pd_u\hat{U}_{\rm B}(u,\bm{w}) +\pd_i\hat{U}^{i}_{\rm B}(u,\bm{w}) = 0\;,
\end{equation}
which is a direct consequence of the equation of continuity (\ref{11.20}) applied in the local coordinates.

\subsubsection{Metric Tensor: external solution}\label{mtex}\index{metric tensor!external solution}\la{mtextso}

Solution of the homogeneous field equations \eqref{nervc34}--\eqref{11.31} for the linearized metric tensor perturbation in the local coordinates adapted to body B yields the tidal gravitational field of external bodies of ${\mathbb N}$-body system in terms of the external STF multipoles \citep{kovl_2004,kopeikin_2011book}. The external solution is convergent at the origin of the local coordinates and its most general form is given by \citet{kovl_2004,kopeikin_2011book}
\begin{eqnarray}
  \label{1.8a}
  \hat{h}_{00}^{\mathrm{ext}}(u,\bm{w}) & = & 2\sum_{l=1}^{\infty}\frac{1}{l!}{\cal Q}_Lw^{L}\;,\\
  \label{1.8aaa}
  \hat h_{0i}^{\mathrm{ext}}(u,\bm{w}) & = & \sum_{l=2}^{\infty}\frac{l}{(l+1)!}\varepsilon_{ipq}{\cal C}_{pL-1}w^{qL-1}+\sum_{l=0}^{\infty}\frac{1}{l!}Z_{iL}w^{L} +\sum_{l=0}^{\infty}\frac{1}{l!}S_Lw^{iL},\\
  \label{1.10a}
  \hat h_{ij}^{\mathrm{ext}}(u,\bm{w}) & = & 2\delta_{ij}\sum_{l=1}^{\infty}\frac{1}{l!}A_Lw^{L}+\sum_{l=0}^{\infty}\frac{1}{l!}B_Lw^{ijL}
   +\sum_{l=1}^{\infty}\frac{1}{l!} \Big[ D_{iL-1}w^{jL-1}+\varepsilon_{ipq}E_{pL-1}w^{jqL-1} \Big]_{\mathrm{sym}(ij)}\\
  & & +\sum_{l=2}^{\infty}\frac{1}{l!} \Big[ F_{ijL-2}w^{L-2}+\varepsilon_{pq(i} G_{j)pL-2}w^{qL-2}\Big]\;,\nonumber
\end{eqnarray}
where $A_L$, $B_L$, etc., are STF Cartesian tensors defined on worldline ${\cal W}$ of the origin of the local coordinate, the symbol $\mathrm{sym}(ij)$ denotes symmetrization. 

Tensors  $A_L$, $B_L$, etc., are the {\it external} multipoles which depend on the coordinate time $u$ only, that is $A_L\equiv A_L(u)$, $B_L\equiv B_L(u)$, etc. Four gauge conditions (\ref{11.11}), (\ref{11.12}) imposed on the components (\ref{1.8a})--(\ref{1.10a}) of the metric tensor perturbations reveal that only 6 out of 10 external multipoles are algebraically-independent. This allows to eliminate four multipoles: $B_L,\,E_L,\,S_L,\,D_L$ from the local metric perturbation \citep{kovl_2004,kopeikin_2011book}. The remaining six multipoles: $A_L$, ${\cal C}_L$, $F_L$, $G_L$, ${\cal Q}_L$, $Z_L$ can be constrained by making use of the residual gauge freedom allowed by the differential equation (\ref{11.b}) that excludes four other multipoles -- $A_L$, $F_L$, $G_L$, $Z_L$ \citep{kovl_2004,kopeikin_2011book}. Finally, only two families of the external multipoles --  gravitoelectric multipoles ${\cal Q}_L$ and gravitomagnetic multipoles ${\cal C}_L$ -- have real physical meaning reflecting the existence of two degrees of freedom (polarization states) for the tidal gravitational field\index{gravitational field!tidal} of the metric tensor. 

After fixing the gauge freedom as indicated above, the external metric tensor\index{metric tensor!external} assumes in the local coordinates the following form
\begin{eqnarray}
  \label{1.24b}
  \hat{h}_{00}^{\mathrm{ext}}(u,\bm{w}) & = &2\sum_{l=1}^{\infty}\frac{1}{l!}{\cal Q}_Lw^{L},\\
  \label{1.25ba}
  \hat h_{0i}^{\mathrm{ext}}(u,\bm{w}) & = & \frac{1-\gamma}3\dot{\cal P} w^i
+\sum_{l=1}^{\infty}\frac{l+1}{(l+2)!}\varepsilon_{ipq}{\cal C}_{pL}w^{qL}
  +2\sum_{l=1}^{\infty}\frac{2l+1}{(2l+3)(l+1)!}\biggl[2\dot{\cal Q}_L+(\gamma-1)\dot{\cal P}_L\biggr]w^{iL}\;,
\\
  \label{1.26b}
  \hat h_{ij}^{\mathrm{ext}}(u,\bm{w}) & = & 2\delta_{ij}\sum_{l=1}^{\infty}\frac{1}{l!}\left[{\cal Q}_L+(\gamma-1){\cal P}_L\right]w^{L},
\end{eqnarray}
where the scalar external multipoles appear in the metric perturbations through the gauge conditions \eqref{11.11}, \eqref{11.12}, and a dot above the external multipoles denotes a total derivative with respect to time $u$. External dipole ${\cal Q}_i$ is acceleration of worldline ${\cal W}$ of the origin of the local frame adapted to body B with respect to a worldline of a freely falling particle, and monopole ${\cal P}$ is the value of the scalar field generated by external bodies ${\rm C}\not=\B$, taken at the origin of the local coordinates \citep{kopeikin_2011book}. It cannot be excluded from $\hat h^{\rm ext}_{0i}$ component by gauge transformation. On the other hand, the monopole ${\cal Q}$ in the metric perturbation is gauge-dependent and has been eliminated by re-scaling of the local coordinate time.  

The non-linear part $\hat l_{00}$ of the perturbation of the external metric tensor is determined as a particular solution of the field equation (\ref{11.33}) that yields \citep{kovl_2004} 
\begin{eqnarray}
  \label{1.24c}
  \hat l_{00}^{{\mathrm{ext}}}(u,\bm{w}) & = & -2\bigg(\sum_{l=1}^{\infty}\frac{1}{l!}{\cal Q}_Lw^{L}\bigg)^2 -2(\beta-1)\bigg(\sum_{l=1}^{\infty}\frac{1}{l!}{\cal P}_Lw^{L}\bigg)^2 +\sum_{l=1}^{\infty}\frac{1}{(2l+3)l!}\ddot{\cal Q}_Lw^{L}w^2\;,
\end{eqnarray}
where, here and everywhere else, a double dot above function denotes a second derivative with respect to time $u$. We have excluded the scalar field components ${\cal P}^2$  and ${\cal P}{\cal P}_i$ from the second term in the right-hand side of \eqref{1.24c} because ${\cal P}^2$ is removed by re-scaling of the local coordinate time while ${\cal P}{\cal P}^i$ is absorbed to, yet unknown, acceleration ${\cal Q}_i$, in (\ref{1.24b}). We might also decompose the product of two sums in (\ref{1.24c}) in algebraic sum of irreducible components and absorb the STF part of the decomposition to multipoles ${\cal Q}_L$ ($l\ge 2$). However, this way of writing solution \eqref{1.24c} complicates calculations and we don't implement it.

\subsubsection{Metric Tensor: the coupling component}\la{intermixt}

The coupling of the internal and external solutions of the linearized metric tensor perturbations is described by the mixed term $\hat{l}_{00}^{\mathrm{mix}}$. It is found as a particular solution of the inhomogeneous field equation (\ref{11.33}) with the right side taken as a product of the internal and external solutions found on the previous step of the post-Newtonian iterations. Solving (\ref{11.33}) yields
\begin{eqnarray}
  \label{1.24o}
  \hat{l}_{00}^{\mathrm{mix}}(u,\bm{w}) & = & -2\left\{\eta{\cal P}
 +2\sum_{l=1}^{\infty}\frac{1}{l!}\Big[{\cal Q}_L+(\beta-1){\cal P}_L\Big]w^{L}\right\}\hat{U}_{\rm B}(u,\bm{w})\\
  & & -2\sum_{l=1}^{\infty}\frac{1}{l!}\Big[{\cal Q}_L+2(\beta-1){\cal P}_L\Big]\;\int\limits_{{\cal V}_{\rm B}}\frac{\rho^{\ast}(u,\bm{w}')w'^L}{|\bm{w}-\bm{w}'|}d^3w'\;,\nonumber
\end{eqnarray}
where $\eta\equiv 4\beta-\gamma-3$ is called the Nordtvedt parameter \citep{willbook}, and ${\cal V}_{\rm B}$ denotes the volume of body B. The best experimental limitation on the numerical value of Nordtvedt's
parameter, $|\eta|<5\times 10^{-4}$, is known from lunar laser ranging (LLR) experiment \index{LLR}\cite{Hofman_2010}. Gravitational wave astronomy will improve its measurement by many orders of magnitude. Equation \eqref{1.24o} completes derivation of the metric tensor in the local coordinates in the post-Newtonian approximation.

\subsubsection{Body-Frame Internal Multipoles}\index{multipoles!internal}\label{mdloc}

Multipolar decomposition of the metric tensor of an isolated gravitating system\index{isolated gravitating system} residing in asymptotically-flat spacetime has been thoroughly studied by a number of researchers \citep{geroch_1970JMP,hansen_1974JMP,quevedo_1990ForPh,thor,gursel_1983GRG}. The most useful technique for the case of the post-Newtonian approximations has been worked out by \citet{bld,di,dyr2}. This technique has been extended to the case of a self-gravitating system embedded to a curved, non-asymptotically flat spacetime in general relativity \citep{th_1985,dsx1} and in scalar-tensor theory of gravity \citep{kovl_2004}, and is used in the present paper. 

A single body B from ${\mathbb N}$-body system interacts gravitationally with other bodies of the system and this interaction cannot be ignored in multipolar decomposition of gravitational field of the body. The presence of the external bodies brings about the interaction field (\ref{1.24o}) to the metric tensor in the local coordinates which energy density gives rise to contribution of gravitational field of the external fields to the definition of the internal multipoles of body B. It, first, looked like an ambiguity as it was unclear whether the contribution of the external fields has to be included to the definition of the body multipoles or not \citep{th_1985}. This issue was resolved in general relativity by \citet{dsx1} and in scalar-tensor theory of gravity by \citet{kovl_2004} who demonstrated that the contribution of the interaction field is to be included to the definition of body's internal multipoles in order to eliminate the {\it non-canonical} multipoles, ${\cal N}^L$ and ${\cal R}^L$ -- see \eqref{NL15} and \eqref{spin-85} -- originating from the non-linear part of the metric tensor perturbation (\ref{1.24o}), from the equations of motion of extended bodies. This effectively erases any dependence of the equations of motion on the internal structure of extended bodies and promotes application of the {\it effacing principle} \citep{Damour_1987book,kovl_2008} from spherically-symmetric bodies to all multipoles.

There are two families of the {\it canonical} internal multipoles in general relativity which are called mass and spin multipoles \citep{bld,bld1986,iau2000}. In scalar-tensor theory of gravity the mass multipoles are additionally split in two algebraically-independent families which are called {\it active} and {\it conformal} multipoles \cite{willbook}. The \textit{active} mass multipoles of a body B from ${\mathbb N}$-body system are defined by equation \citep{kovl_2004,kopeikin_2011book}
\begin{eqnarray}\label{1.31}
  {\cal M}^{L} & = & \int\limits_{{\cal V}_{\rm B}}\sigma(u,\bm{w})\left\{
  1-(2\beta-\gamma-1){\cal P}-\sum_{k=1}^{\infty}\frac{1}{k!}\Big[{\cal Q}_{K}+2(\beta-1) {\cal P}_{K}\Big]w^{<K>}\right\}w^{<L>}d^3w\\\nonumber
  &&+\frac{1}{(2l+3)}\left[\frac12\ddot{\cal N}^{<L>}-2(1+\gamma)\frac{2l+1}{l+1} \dot{\cal R}^{<L>} \right]
\end{eqnarray}
where the angular brackets around spatial indices denote STF Cartesian tensor \citep{thor,bld1986}, and 
\ba
  \label{NL15}
  \mathcal{N}^{L} &=& \int\limits_{{\cal V}_{\rm B}}\sigma(u,\bm{w})w^2w^{<L>}d^3w\;,\\
  \label{spin-85}
  {\cal R}^{L}&=&\int\limits_{{\cal V}_{\rm B}}\sigma^i(u,\bm{w}) w^{<iL>}d^3w\;,
\ea
are two additional {\it non-canonical}  sets of STF multipoles, ${\cal V}_{\rm B}$ is volume of body B over which the integration is performed. {\it Non-canonical} multipoles $\mathcal{N}^L$ generalize the second-order rotational moment of inertia of body B,
\begin{equation}
  \label{r5t1}
  \mathcal{N} = \int\limits_{{\cal V}_{\rm B}}\rho^{\ast}w^2d^3w\;,
\end{equation}
with respect to the origin of the local coordinates, and ${\cal R}^L$ are {\it non-canonical} multipoles associated with matter currents inside the body.  
The density $\sigma$ in \eqref{1.31} is called the \textit{active} mass density \citep{kovl_2004},
\begin{eqnarray}
  \label{pz3}
  \sigma(u,\bm{w}) & = & \rho^{\ast}(u,\bm{w})\left[1+(\gamma+\frac12)\nu^2(u,\bm{w})+\Pi(u,\bm{w})
  -  (2\beta-1)\hat{U}_{\rm B}(u,\bm{w})  \right]+\gamma{\mathfrak{s}}^{kk}(u,\bm{w})\;,
\end{eqnarray}
and the vector 
\be
  \label{pz6}
  \sigma^i(u,\bm{w})=\rho^{\ast}(u,\bm{w})\nu^i(u,\bm{w}),
\ee
is matter's current density. All integrals in (\ref{1.31})--\eqref{r5t1} are performed over hypersurface ${\cal H}_u$ of a constant coordinate time $u$.

The \textit{conformal} mass multipoles of the body B are defined as follows \citep{kovl_2004,kopeikin_2011book}
\begin{eqnarray}
  \label{1.34}
  {\cal I}^{L} & = &\int\limits_{{\cal V}_{\rm B}}\varrho(u,\bm{w})\bigg[1-(1-\gamma){\cal P}-\sum_{k=1}^{\infty}\frac{1}{k!}{\cal Q}_{K}w^{<K>}\bigg]w^{<L>}\,d^3w
  + \frac{1}{(2l+3)}\left[\frac12\ddot{\cal N}^{<L>}-4\frac{2l+1}{l+1} \dot{\cal R}^{<L>} \right],
\end{eqnarray}
where, again, the integration is performed over a hypersurface ${\cal H}_u$ of constant coordinate time $u$, and
\begin{equation}
  \label{pz5}
  \varrho = \rho^{\ast}(u,\bm{w})\left[1+\frac{3}{2}\nu^2(u,\bm{w})+\Pi(u,\bm{w})-\hat{U}_{\rm B}(u,\bm{w})\right]+{\mathfrak{s}}^{kk}(u,\bm{w})\;,
\end{equation}
is the \textit{conformal} mass density \index{mass density!conformal}of matter
which does not depend on the PPN parameters $\beta$ and $\gamma$ as contrasted to the definition \eqref{pz3} of the {\it active} mass density.

There is one more type of the multipoles called \textit{scalar} multipoles\index{scalar internal multipoles}\index{internal multipoles!scalar}, $\tilde{I}^L$. However, they are not independent, and relate to the \textit{active} and \textit{conformal} multipoles by simple formula \cite{kovl_2004}
\begin{equation}
  \label{wh1}
  \tilde{I}^{L} = 2{\cal M}^{<L>}-(1+\gamma){\cal I}^{<L>}\;.
\end{equation}

In addition to the gravitational mass multipoles, ${\cal M}^L$ and ${\cal I}^L$, there is a set of internal spin multipoles\index{multipoles!spin}\index{internal multipoles!spin}. In the Newtonian approximation they are defined by expression \citep{kovl_2004}
\begin{equation}
  \label{1.32}
  \mathcal{S}^{L} = \int\limits_{{\cal V}_{\rm B}}\varepsilon^{pq<i_l} w^{i_{l-1}...i_1>p}\sigma^q(u,{\bm w})d^3w,
\end{equation}
where matter's current density $\sigma^q$ has been defined in (\ref{pz6}). All multipoles of body B are functions of time $u$ only. They are the STF Cartesian tensors in the tangent Euclidean space attached to the worldline $\cal W$ of the origin of local coordinates adapted to body B. Definition \eqref{1.32} is sufficient for deriving the post-Newtonian translational equations of motion of the extended bodies in ${\mathbb N}$-body system. However, derivation of the post-Newtonian rotational equations of motion requires a post-Newtonian definition of the body's angular momentum (spin). We shall discuss it later in section \ref{x5wc7n}.

\section{Matched Asymptotic Expansions and Coordinate Transformations}\label{pntb}
\subsection{Basic Principles}\label{gzm}

Post-Newtonian transformations between the global and local coordinate charts are derived by the method of {\it matched asymptotic expansions} \citep{Lagerstrom_89}. It involves finding several different approximate solutions of the field equation, each of which is valid for a specific domain of space, and then combining these different solutions together in a buffer domain where all different solutions overlap, in order to obtain a single approximate solution. The technique of matched asymptotic expansions in general relativity was first implemented by \citet{demgrish_1974GRG} for deriving equations of motion of black holes in the Newtonian limit. D'Eath \citep{das2,DEath_1996book} significantly extended this technique to the next approximations of general relativity and it is now commonly used for derivation of equations of motion of black holes \citep{gorbonos_2004JHEP,gorbonos_2005CQGra,futamase_2008PhRvD}. Matching asymptotic expansions are indispensable in case of the singular perturbations of the field equations but the method turned out to be very effective also for derivation of equations of motion of extended bodies \citep{ashb2,Kopejkin_1988CeMec,bk-nc,dsx1,dsx2} and for constructing a post-Newtonian theory of reference frames in the solar system \citep{bk89,iau2000,kopeikin_2011book,Soffel_2013book}. 

In the present paper the independent dynamic field variables are scalar field and metric tensor which describe the asymptotic solutions of the field equations in the form of the post-Newtonian expansions which are valid in the spatial domains covered by the global or local coordinates. These solutions describe one and the same value of the dynamic variables in any type of coordinates which means that the solutions can be spliced in the spatial region where the coordinate charts overlap. The splicing relies upon the tensor transformation law applied to the post-Newtonian expansions of the metric tensor and scalar field.  The post-Newtonian transition functions entering the transformation establish the correspondence between the global and local coordinates. Coordinate distance from the origin of the local coordinates to the first singular points of the Jacobian of the transformation determines the domain of applicability of the local coordinates \citep{kopeikin_2011book}.

The matching procedure is organized as follows. We use conformal harmonic coordinates defined by the Nutku gauge condition (\ref{gau}). Transition functions\index{transition functions!of coordinate transformation}\index{coordinate transformation!transition functions} of the post-Newtonian coordinate transformation\index{coordinate transformation!post-Newtonian} are constrained by this condition and must obey differential equation (\ref{11.b}) describing the residual gauge freedom. Solutions of this homogeneous equation are to be continuously-differentiable functions that are regular at the origin of the local coordinates. These functions can be represented in the form of a Taylor series of the harmonic polynomials of the spatial local coordinates. Coefficients of the Taylor series are the STF Cartesian tensors defined on the worldline\index{worldline} $\cal W$ of the origin of the local coordinates. The transition functions are to be substituted to the matching equations describing the splicing of the internal and external solutions of the field equations in the global and local coordinates. Matching the asymptotic post-Newtonian expansions of the scalar field\index{scalar field} and the metric tensor allows us to fix all degrees of the residual gauge freedom in the final form of the post-Newtonian coordinate transformation and to determine a functional form of all external multipoles except for the external dipole ${\cal Q}_i$ which is not constrained by the matching conditions and must be found separately from the equations of motion of the center-of-mass of body B in the body-adapted local coordinates. 

Physically, the post-Newtonian transformation between coordinate times, $t$ and $u$, describes the Lorentz (velocity-dependent) and Einstein\index{Einstein!time dilation} (gravitational-field-dependent) time dilation\index{Lorentz!time dilation} associated with the different simultaneity\index{simultaneity} of events in the two coordinate charts \citep{Kopejkin_1988CeMec,ashb2}. It also includes an infinite series of the polynomial terms \citep{bk89,1990CeMDA..48...23B}. The post-Newtonian transformation between the spatial coordinates, $x^i$ and $w^i$, is a quadratic function of spatial coordinates. The linear part of the transformation includes the Lorentz and Einstein contractions of length as well as a matrix of rotation describing the post-Newtonian precession\index{precession!post-Newtonian} of the spatial axes of the local coordinates with respect to the global coordinates due to the translational and rotational motion of the bodies \citep{k85,Damour_1987book}.
The Lorentz length contraction\index{Lorentz! length contraction}\index{length contraction!Lorentz} takes into account the kinematic
aspects of the post-Newtonian transformation and depends on the relative velocity of motion of the local coordinates with respect to the global coordinates. The Einstein (gravitational) length contraction \index{Einstein! length contraction}\index{length contraction!Einstein} accounts for static effects of the scalar field and the metric tensor \citep{kovl_2004,kopeikin_2011book}. The quadratic part of the spatial transformation depends on the orbital acceleration of the local coordinates and accounts for the effects of the affine connection\index{affine connection} (the Christoffel symbols\index{Christoffel symbols}) of spacetime manifold\index{manifold}.

Let us now discuss the mathematical structure of the post-Newtonian transformation between the local coordinates, $w^\alpha=(w^0,w^i)=(u,\bm{w})$, and the global coordinates, $x^\alpha=(x^0,x^i)=(t,\bm{x})$ in more detail. This coordinate transformation must be compatible with the weak-field and slow-motion approximation used in the post-Newtonian expansions. Hence, the coordinate transformation is given as a post-Newtonian expansion:
\begin{eqnarray}
  \label{2.2}
  u & = & t +\xi^0(t,\bm{x}),\\
  \label{2.3}
  w^i & = & R_{\rm B}^{i}+\xi^i(t,\bm{x}),
\end{eqnarray}
where $\xi^0$ and $\xi^i$ are the post-Newtonian corrections to the Galilean transformation \index{Galilean transformation}, $u=t$, $R_{\rm B}^{i}\equiv x^i-x^i_{{ B}}(t)$, and $x^i_{\rm B}(t)$ is a spatial position of the origin of the local coordinates in the global coordinates. We denote velocity and acceleration of the origin of the local coordinates as $v_{\rm B}^i\equiv\dot{x}^i_{\rm B}$ and $a_{\rm B}^i\equiv\ddot{x}^i_{\rm B}$ respectively, where a dot above function denotes a derivative with respect to time $t$. At this step, we don't know yet equations for worldline $\cal W$ of the origin of the local coordinates adapted to body B nor for worldline $\cal Z$ of the body's center of mass. Therefore, it is natural to assume that originally the two worldlines, ${\cal W}$ and ${\cal Z}$, are different. Later on, we shall show that the two worldlines can be made identical by demanding the conservation of the linear momentum of body B. It can be always achieved by choosing the external dipole ${\cal Q}_i$ to compensate the non-inertial acceleration of the body's center-of-mass caused by tidal forces \citep{Kopejkin_1988CeMec,kovl_2004,kopeikin_2011book}. The presence of non-vanishing dipole ${\cal Q}_i$ in the local metric \eqref{1.24b} makes the local coordinates adapted to body B to be non-inertial. 

It is instructive to notice that the local coordinates used by \citet{th_1985} are inertial that is the origin of the Thorne-Hartle local coordinates moves along a geodesic worldline of the {\it effective} spacetime manifold $\bar{M}$ with metric, $\bar g_{\a\b}=\eta+\bar h_{\a\b}$, which is obtained from the original spacetime manifold ${M}$ with metric, $g_{\a\b}=\eta+h_{\a\b}$, by deleting from $h_{\a\b}$ the internal part of the metric $ h^{\rm int}_{\a\b}$.  In such local inertial coordinates the external dipole ${\cal Q}_i\equiv 0$ but the center of mass of body B does not move along geodesic in the most general case due to the tidal interaction of the internal multipoles ${\cal M}_L$ and ${\cal S}_L$ of the body with external gravitational field of other bodies. 

The asymptotic matching equations for independent dynamic variables -- the scalar field $\varphi$ and the metric tensor $g_{\m\n}$ -- are given by the laws of coordinate transformations of these geometric objects \citep{dfn}
\begin{eqnarray}
  \label{2.5}
  \varphi(t,\bm{x}) & = & \hat{\varphi}(u,\bm{w})\;,\\
  \label{2.6}
  g_{\mu\nu}(t,\bm{x}) & = & \hat{g}_{\alpha\beta}(u, \bm{w})\frac{\partial w^{\alpha}}{\partial x^{\mu}} \frac{\partial w^{\beta}}{\partial x^{\nu}}\;.
 \end{eqnarray}
Equations (\ref{2.5}), (\ref{2.6}) are valid in the spacetime region that is covered simultaneously by the local and global coordinates. Functions in the left-hand side of these equations are known and given in section \ref{mtsf183} as integrals from body's matter variables (density, pressure, etc.) performed over volumes of all bodies of ${\mathbb N}$-body system on hypersurface ${\cal H}_t$ of constant time $t$. The right-hand side of the matching equations contains, besides the known integrals from the matter variables of body B taken on hypersurface ${\cal H}_u$ of constant time $u$, yet unknown external multipoles, $P_L$, ${\cal Q}_L$, ${\cal C}_L$ of the external part of the metric tensor in the local coordinates and the transition functions $\xi^\a=(\xi^0,\xi^i)$ from the coordinate transformations (\ref{2.2}), (\ref{2.3}). We prove below that both the external multipoles and the transition functions can be determined by solving  matching equations (\ref{2.5}), (\ref{2.6}) that also yield equations of motion of the origin of the local coordinates, $x^i_\B=x^i_\B(t)$. Matching the post-Newtonian expansions of the metric tensor and scalar field does not yield equations of motion of the center of mass of body B. Additional procedure of integration of the microscopic equations of motion of matter of body B is required for this purpose to determine the motion of the center of mass of body B with respect to the origin of the local coordinates and to derive rotational equations of motion of body's spin. It is explained in section \ref{n4r6v}.

\subsection{Transition Functions}

A comprehensive description of the matching procedure establishing the correspondence between the global and local coordinates in ${\mathbb N}$-body problem is given in \citep{kovl_2004,kopeikin_2011book,Soffel_2013book}. Here, we summarize the main results of the matching.

Solving matching equations (\ref{2.5}), (\ref{2.6}) begins from $\hat g_{0i}$ component of the metric tensor perturbation in the local coordinates adapted to body B. This component does not contain 0.5 post-Newtonian term of the order of ${\cal O}(\epsilon)$ because we have chosen the spatial axes of the local coordinates dynamically non-rotating and orthogonal to worldline ${\cal W}$ of its origin at any instant of time. It eliminates the angular and linear velocity terms of the order of ${\cal O}(\epsilon)$ in $\hat g_{0i}$ and implies that function $\xi^0(t,\bm{x})$ in (\ref{2.2}) satisfies the following constraint \citep{kovl_2004,dsx1},
\begin{equation}
  \label{ha+}
  \pd_i\xi^0(t,{\bm x}) = -v_{\rm B}^i +\pd_i\kappa(t,\bm{x})\;,
\end{equation}
where $\kappa(t,\bm{x})$ is the post-Newtonian, yet unknown correction of the order of ${\cal O}(\epsilon^2)$.
Integration of the partial differential equation \eqref{ha+} yields,
\begin{equation}
  \label{2.8}
  \xi^0(t,\bm{x}) = \mathcal{A}(t)- v_{\rm B}^k R_{\rm B}^k+\kappa(t,\bm{x})\;,
\end{equation}
where $\mathcal{A}(t)$ is a constant of integration depending on time. 

At second step we use differential equation (\ref{11.b}) in order to find out the transition functions $\kappa$ from \eqref{2.8} and $\xi^i$ from \eqref{2.3}. We replace (\ref{2.8}) to (\ref{2.2}) and substitute it along with $w^i$ from (\ref{2.3}) in equation (\ref{11.b}) which yields two decoupled inhomogeneous Poisson equations\index{Poisson!equation}\index{equation!Poisson} for the post-Newtonian components of the transition functions,
\begin{eqnarray}
  \label{2.10}
  \triangle \kappa(t,\bm{x}) & = & 3v_{\rm B}^{k}a_{\rm B}^{k} +\ddot {\mathcal{A}}- \dot{a}^k_{\rm B}R_{\rm B}^{k} \;,\\
  \label{2.11}
  \triangle\xi^i(t,\bm{x}) & = & -a_{\rm B}^{i}\;,
\end{eqnarray}
where $\triangle\equiv\d^{ij}\pd_i\pd_j$ is the Laplace operator in the Euclidean space.
General solution of these elliptic-type equations must be regular at the origin of the local coordinates adapted to body B and consists of two parts -- a fundamental solution of the homogeneous Laplace equation\index{Laplace equation}\index{equation!Laplace} and a particular solution of the inhomogeneous Poisson equation\index{Poisson!equation}\index{equation!Poisson} \citep{dsx1,kovl_2004}
\begin{eqnarray}
  \label{2.12}
  \kappa & = & \left(\frac{1}{2}v_{\rm B}^{k}a_{\rm B}^{k}-\frac{1}{6}\ddot{\mathcal{A}}\right)R_{\rm B}^{2}-\frac{1}{10}\dot{a}^k_{\rm B}R_{\rm B}^{k}R_{\rm B}^{2}+\Xi(t,\bm{x})\;,\\
  \label{2.13}
  \xi^i & = & -\frac{1}{6} a_{\rm B}^{i} R_{\rm B}^{2}+\Xi^i(t,\bm{x})\;.
\end{eqnarray}
Here, functions $\Xi$ and $\Xi^i$ are the fundamental solutions of the homogeneous Laplace equation -- the harmonic polynomials with respect to the local spatial coordinates expressed in terms of the global coordinates, $w^i=R^i_\B+{\cal O}(\epsilon^2)$,
\begin{eqnarray}
  \label{eq1}
  \Xi(t,\bm{x}) & = & \sum_{l=0}^\infty\frac{1}{l!}\mathcal{B}^{L}R_{\rm B}^{<L>}\;,\\
  \label{eq2}
  \Xi^i(t,\bm{x}) & = & \sum_{l=1}^{\infty}\frac{1}{l!}\mathcal{D}^{iL}R_{\rm B}^{<L>} +\sum_{l=0}^{\infty}\frac{\varepsilon_{ipq}}{(l+1)!}\mathcal{F}^{pL}R_{\rm B}^{<qL>} +\sum_{l=0}^{\infty}\frac{1}{l!}\mathcal{E}^{L}R_{\rm B}^{<iL>}\;,
\end{eqnarray}
where the coefficients, $\mathcal{B}^{L}$, $\mathcal{D}^{L}$, $\mathcal{F}^{L}$ and $\mathcal{E}^{L}$ of the expansions are STF Cartesian tensors\index{Cartesian tensor} which should not be confused with the external multipoles entering the local metric tensor. These coefficients are defined on the worldline $\cal W$ of the origin of the local coordinates and depend only on time $t$ of the global coordinates. Explicit form of coefficients $\mathcal{B}^{L}$, $\mathcal{D}^{L}$, $\mathcal{F}^{L}$ is derived by substituting transitions functions $w^\a=(u,w^i)$ in the form of \eqref{2.2}, \eqref{2.3}, \eqref{2.8}, \eqref{2.12}--\eqref{eq2} to matching equations (\ref{2.5})--(\ref{2.6}) and solving them. This solution also determines the external multipoles and the equations of motion for the origin of the local coordinates - worldline ${\cal W}$. The overall procedure of solving the matching equations is rather long and technical and we don't describe it over here. The reader can found its comprehensive description in papers \citep{kovl_2004, xie_2010AcPSl} and in book \citep[Chapter 5]{kopeikin_2011book}. The matching solution is given in section \ref{u8b2p} below.

\subsection{Matching Solution}\la{u8b2p}
\subsubsection{Post-Newtonian Coordinate Transformation}\label{ffemb}

Parametrized post-Newtonian transformation between the local coordinates $w^\a$ adapted to body B and the global coordinates $x^\a$ is given by two equations \citep{Kopejkin_1988CeMec,kopeikin_2011book},
\begin{eqnarray}
  \label{5.12}
  u & = & t+\frac1{c^2}(\mathcal{A}-v^k_{\rm B}R_{\rm B}^k)+\frac1{c^4}\left[\mathcal{B}+\left(\frac{1}{3}v^k_{\rm B}a^k_{\rm B}-\frac{1}{6}\dot{\bar{U}}(t,{\bm x}_\B)- \frac{1}{10}\dot{a}^k_{\rm B}R^k_{\rm B}\right)R^2_{\rm B}+\sum_{l=1}^{\infty}\frac{1}{l!}\mathcal{B}^LR_{\rm B}^L\right]+\mathcal{O}(c^{-6})\;,\\
  \label{5.13}
  w^i & = & R^i_{\rm B}+\frac1{c^2}\left[\left(\frac{1}{2}v^i_{\rm B}v_{\rm B}^k+\delta^{ik}\gamma\bar{U}(t,{\bm x}_\B)+F^{ik}_\B\right)R_{\rm B}^k+ a^k_{\rm B}R^i_{\rm B}R^k_{\rm B}-\frac{1}{2}a^i_{\rm B}R_{\rm B}^2\right]+\mathcal{O}(c^{-4})\;,
\end{eqnarray}
where $R^i_{\rm B}=x^i-x^i_{\rm B}$
is the coordinate distance on the hypersurface ${\cal H}_t$ of constant time $t$ between the point of matching, $x^i$, and the origin of the local coordinates, $x^i_{\rm B}=x^i_{\rm B}(t)$, and we have shown in these equations the fundamental speed $c$ explicitly to attenuate the post-Newtonian order of different terms. 
 
Functions $\mathcal{A}$ and $\mathcal{B}$ depend on the global coordinate time $t$ and define transformation between the local time $u$ and the global coordinate time $t$ at the origin of the local coordinates. They obey the ordinary differential equations,
\begin{eqnarray}
  \label{5.14}
  \frac{d\mathcal{A}}{dt} & = & -\frac{1}{2}v_{\rm B}^2-\bar{U}(t,{\bm x}_\B)\;,\\
  \label{5.7a}
  \frac{d\mathcal{B}}{dt} & = & -\frac{1}{8}v^4_{\rm B}-\left(\gamma+\frac{1}{2}\right)v_{\rm B}^2\bar{U}(t,{\bm x}_\B)+\frac{1}{2}\bar{U}^2(t,{\bm x}_\B)+2(1+\gamma)v_{\rm B}^k\bar{U}^k(t,{\bm x}_\B) -\bar{\Psi}(t,{\bm x}_\B)+\frac{1}{2}\pd_{tt}\bar{\chi}(t,{\bm x}_\B)\;,
\end{eqnarray}
that describe the post-Newtonian transformation between time $u$ of the local coordinates and time $t$ of the global coordinates. The other functions entering (\ref{5.12}), (\ref{5.13}) are defined by algebraic relations 
\begin{eqnarray}
  \label{5.15}
  \mathcal{B}^i & = & 2(1+\gamma)\bar{U}^i(t,{\bm x}_\B)-(1+2\gamma)v^i_{\rm B}\bar{U}(t,{\bm x}_\B)-\frac{1}{2}v^i_{\rm B}v^2_{\rm B}\;,\\
  \label{5.16}
  \mathcal{B}^{ij} & = & 2(1+\gamma)\pd^{<i}\bar{U}^{j>}(t,{\bm x}_\B)-2(1+\gamma)v_{\rm B}^{<i}\pd^{j>}\bar{U}(t,{\bm x}_\B)+2a_{\rm B}^{<i}a_{\rm B}^{j>}\;,\\
  \label{5.17}
  \mathcal{B}^{iL} & = & 2(1+\gamma)\pd^{<L}\bar{U}^{i>}(t,{\bm x}_\B)-2(1+\gamma)v_{\rm B}^{<i}\pd^{L>}\bar{U}(t,{\bm x}_\B)\;, \qquad\qquad\qquad\qquad (l\ge 2),
\end{eqnarray}
where the angular brackets denote STF projection of indices, and the external (with respect to body B) potentials $\bar{U}$, $\bar{U}^i$, $\bar\Psi$, $\bar\chi$ are defined in \eqref{tb54vd}, \eqref{3d8b1a}. Notations $\bar{U}(t,{\bm x}_\B)$, $\bar{U}^i(t,{\bm x}_\B)$, $\bar\Psi(t,{\bm x}_\B)$ and $\bar\chi(t,{\bm x}_\B)$ mean that the potentials are taken at the origin of the local coordinates adapted to body B at instant of time $t$.

The skew-symmetric\index{anti-symmetric} rotational matrix\index{rotational matrix} ${F}^{ij}_\B$ is a solution of the ordinary differential equation 
\begin{equation}
  \label{5.18}
  \frac{d{F}^{ij}_\B}{dt}=2(1+\gamma)\pd^{[i}\bar{U}^{j]}(t,{\bm x}_\B)+(1+2\gamma)v_{\rm B}^{[i}\pd^{j]}\bar{U}(t,{\bm x}_\B)+v_{\rm B}^{[i}{\cal Q}^{j]}\;,
\end{equation}
describing the rate of the kinematic rotation of the spatial axes of the local coordinates adapted to body B with respect to the global coordinates \citep{Kopejkin_1988CeMec,kovl_2004}. Equation \eqref{5.18} has been derived here for arbitrary-structured bodies by the method of matched asymptotic expansions. The same equation was obtained independently for spinning test particle (gyroscope) through the Fermi-Walker transport of spin \citep[\S 40.7]{mtw}.
The first term in the right-hand side of (\ref{5.18}) describes the Lense-Thirring (gravitomagnetic) precession\index{precession!Lense-Thirring}\index{precession!gravitomagnetic} which is also called the dragging of inertial frames \citep{mtw,ciufolini_book}. The second term in the right-hand side of \eqref{5.18} describes the de-Sitter (geodetic) precession\index{precession!de-Sitter}\index{precession!geodetic}, and the third term describes the Thomas precession \index{precession!Thomas} depending on the local (non-geodesic) acceleration ${\cal Q}^i=\d^{ij}{\cal Q}_j$ of the origin of the local coordinates with respect to a geodesic worldline of a freely-falling test particle. In the scalar-tensor theory both the Lense-Thirring and de-Sitter precession depend on the PPN parameter $\gamma$ while the Thomas precession does not. The reason is that the Thomas precession is generically a special relativistic effect \citep{Fisher_1972AmJPh} that cannot depend on a particular version of an alternative theory of gravity. 

The Lense-Thirring and geodetic precession have been recently measured in Gravity Probe B gyroscope experiment \citep{GPB_2015CQGra} and by satellite laser ranging technique \citep{ciufolini_2012NewA,Ciufolini_2016EPJC}. Relativistic precession is an attractive mechanism for theoretical explanation of quasi-periodic oscillations (QPO) in the optical power density spectra of accreting black holes \citep{Veledina_2013ApJ}. It is also important to include relativistic precession of spins of stars in merging compact binaries for adequate prediction and analysis of gravitational waveforms emitted by the binaries \citep{buonanna_2003PhRvD,err_2006PhRvD,Gupta_2015CQGra,Harry_2016PhRvD}. 

\subsubsection{Body's Self-Action Force and Bootstrap Effect\index{bootstrap effect}}

Self-action force is a key concept in gravitational dynamics of extended bodies both in the Newtonian and relativistic gravity theories \citep{harte2008_2,Harte_2012,harte2015}. It is defined as the net action of the gravitational field generated by a single body from ${\mathbb N}$-body system on the body itself. The self-action force includes a conservative part and dissipative terms which are known as gravitational radiation-reaction force \citep{Detweiler_2011mmgr,Wald_2011mmgr,Pound_2015}. The self-action of the gravitational radiation appears for the first time at 1.5 PN approximation in scalar-tensor theory of gravity due to the emission of dipolar scalar field radiation \citep{willbook,Mirshekari_2013PhRvD87h4070M} and at 2.5 PN approximation in general relativity \citep{Chandra_1970ApJ,Damour_1983grr,gk_1983SvAL,gk86,schaefer_1985AnPhy} due to the emission of quadrupole gravitational waves by the moving bodies \citep{Landau1975,mtw}. Calculation of the radiation-reaction force beyond 2.5 post-Newtonian approximation is a challenging theoretical task \citep{poisson_2011,Detweiler_2011mmgr,Wald_2011mmgr,Pound_2015} which solution is of paramount importance for correct prediction of inspiral motion of compact binaries, especially in the extreme mass ratio limit \citep{Babak2015,wardell2015}.

\citet{chicone_2001PhLA} studied the origin of the self-action force by means of the mathematical theory of delay equations which include the field-retardation effects, and predicted that all of them must have runaway modes. It was shown that when retardation effects are small, the physically significant solutions belong to the so-called {\it slow manifold}\index{slow manifold}\index{manifold!slow} of the dynamic system which is identified with the attractor in the state space of the delay equation. It was also demonstrated via an example that when retardation effects are no longer small, the motion of the system exhibits bifurcation phenomena that are not contained in the local equations of motion. The bifurcation behavior of the solutions of the delay equations pointed out by \citet{chicone_2001PhLA} is absent in the conservative post-Newtonian approximations but has to be studied more attentively by analysts computing the gravitational waveforms of inspiral binary systems.

Radiation-reaction force does not prevent a sufficiently compact and non-spinning body from moving on a geodesic in a particularly chosen, regular effective external metric if a singular part of the full metric properly removed by regularization \citep{Detweiler_2003PRD}. Thus, the regular part of radiation-reaction force does not violate the Einstein principle of equivalence \cite{Whiting_2003IJMPD}. The singular part of the metric corresponds to the conservative part of the self-action force which apparently must obey the third Newton's law to get a vanishing net internal force, thus, preventing self-accelerated run-away motion of the body which we call a {\it bootstrap} effect. Bootstrapping can happen only in some non-conservative (non-viable) alternative theories of gravity \citep{willbook}. It does not occur in the first post-Newtonian approximation of scalar-tensor theory for arbitrary-structured bodies as one can see from matching equations \eqref{2.5}, \eqref{2.6} where all the terms depending on body's internal gravitational potentials mutually cancel out. The bootstrap effect is also absent in the second post-Newtonian approximation both in general relativity \citep{k85,mitchell_2007PhRvD} and in scalar-tensor theory of gravity \citep{Mirshekari_2013PhRvD87h4070M}. 

\subsubsection{World Line of the Origin of the Local Coordinates}

The origin of the local coordinates adapted to body B moves in spacetime along worldline ${\cal W}$. Matching equation \eqref{2.6} for the metric tensors in the local and global coordinates yields equations of translational motion of the origin of the local coordinates, $x^i_{\rm B}=x^i_{\rm B}(t)$, with respect to the global coordinates. It reads \citep[Equation 5.88]{kopeikin_2011book}
\begin{eqnarray}
  \label{5.8}
 a^i_{\rm B}& = & \pd^i\bar{U}(t,{\bm x}_\B)-{\cal Q}^i+F^{ij}_{\rm B}{\cal Q}_j
+\pd^i\bar{\Psi}(t,{\bm x}_\B)-\frac{1}{2}\pd_{tt}\pd^{i}{\bar\chi}(t,{\bm x}_\B)+2(1+\gamma)\dot{\bar{U}}^i(t,{\bm x}_\B)\\\nonumber
&&-2(1+\gamma)v^j_{\rm B}\pd^i\bar{U}^{j}(t,{\bm x}_\B)
 -(1+2\gamma)v^i_{\rm B}\dot{\bar{U}}(t,{\bm x}_\B)+(2-2\beta-\gamma)\bar{U}(t,{\bm x}_\B)\pd^i\bar{U}(t,{\bm x}_\B)\\\nonumber&&+(1+\gamma) v^2_{\rm B}\pd^i\bar{U}(t,{\bm x}_\B)
-\frac12 v^i_{\rm B}v^j_{\rm B} \pd^j\bar{U}(t,{\bm x}_\B)  -\frac12 v^i_{\rm B}v^j_{\rm B}a^j_{\rm B}-v^2_{\rm B}a^i_{\rm B}-(2+\gamma)a^i_{\rm B}\bar U(t,{\bm x}_\B)\;,    \nonumber
\end{eqnarray}
where a dot above function denotes a total derivative\index{time derivative!total} with respect to time $t$, $v^i_{\rm B}\equiv \dot{x}^i_{\rm B}$ and   
$a^i_{\rm B}\equiv\ddot{x}^i_{\rm B}$ are velocity and acceleration of the origin of the local coordinates relative to the global coordinates, ${\cal Q}^i=\d^{ij}{\cal Q}_j$ is a dipole term ($l=1$) in the external solution for $\hat{h}^{\rm ext}_{00}$ component of the metric tensor perturbation (\ref{1.8a}) which describes a local acceleration of the worldline ${\cal W}$. 

The right-hand side of \eqref{5.8} is a gravitational force per unit mass causing the coordinate acceleration $a^i_\B$ of the origin of the local coordinates of body B with respect to the global coordinates.  The force is explicitly expressed in terms of the external gravitational potentials, $\bar U$, $\bar U^i$, $\bar\Psi$, $\bar\chi$, and their time and/or spatial derivatives. It also depends on the external dipole, ${\cal Q}^i=\d^{ij}{\cal Q}_j$, which represents a local acceleration of worldline ${\cal W}$ with respect to a time-like geodesic on the effective spacetime manifold $\bar{M}$ which is explained in more detail in section \ref{po3v6}. Function ${\cal Q}^i$ does not depend on the choice of gauge condition and constitutes a part of definition of the state of motion of the origin of the local coordinates \citep{Ni_1978PRD}. Only after specification of ${\cal Q}^i$ as a function of time, formula (\ref{5.8}) becomes an ordinary differential equation which solution yields worldline $\cal W$ of the origin of the local coordinates as a known function of time ${x}^i_{\rm B}(t)$.

A trivial choice of the local acceleration, ${\cal Q}^i=0$, looks attractive as it immediately converts (\ref{5.8}) to a fully-determined differential equation. It is this choice that has been made, for example, by \citet{th_1985,dixon_1979} which means that worldline ${\cal W}$ of the origin of the local coordinates is a geodesic of the effective background manifold $\bar{M}$. However, this choice does not allow us to keep the origin of the local coordinates always at the center of mass of body B if the body has non-vanishing internal multipoles ${\cal M}^L$ and ${\cal S}^L$ which interact with the tidal field multipoles ${\cal Q}_L$ and ${\cal C}_L$ of the external bodies C$\not=$B from ${\mathbb N}$-body system. The interaction exerts a force on the body B and makes its center of mass moving along a non-geodesic worldline having ${\cal Q}_i\not=0$ \cite{th_1985,Kopejkin_1988CeMec}. Thus, worldline $\cal Z$ of the center of mass of body B is not geodesic in the most general case. If we want to retain the center of mass of body B at the origin of the body-adapted local coordinates at any instant of time, the acceleration ${\cal Q}^i$ must obey the equations of motion of body's center of mass with respect to the local coordinates. Derivation of this equation cannot be achieved by the method of matched asymptotic expansions and requires either integration of microscopic equations of matter over the volume of body B in the local coordinates \citep{dsx1,dsx2,kovl_2004} or finding asymptotes of the surface integrals in  buffer region of overlapping the local and global coordinates \citep{th_1985,asada_2011}. We deal with a regular distribution of matter inside the extended bodies and apply the technique of integration of the microscopic equations of motion to find the local accleration ${\cal Q}_i$ in section \ref{teofminlc}.

\subsubsection{Body-Frame External Multipoles}\la{ex5m3}

\paragraph{Scalar-Field Multipoles.} Matching determines the external (with respect to body B) tidal multipoles in terms of the partial derivatives from the gravitational potentials of external bodies  \cite{kovl_2004,kopeikin_2011book}. The external scalar field multipoles\index{scalar field!external multipoles} are obtained by solving \eqref{2.5}, and read
\begin{equation}
  \label{3.13}
  {\cal P}_L=\pd_L\bar{\varphi}(t,{\bm x}_\B)\;, \qquad (l\ge 0)
\end{equation}
where the external scalar field $\bar\varphi$ is expressed in terms of the external Newtonian potential $\bar{U}$ 
\be\la{scal123}
\bar{\varphi}(t,{\bm x})=\bar U(t,{\bm x})\;.
\ee
We remind that the scalar field perturbation $\varphi$ is coupled either with the factor $\gamma-1$ or $\b-1$, so that all physical effects of the scalar field are proportional to these factors and can be easily identified in the equations that follow. It should be noticed that the external scalar field monopole ${\cal P}$ $(l=0)$ and dipole ${\cal P}_i$ $(l=1)$ can not be removed from observable gravitational effects by rendering a coordinate transformation to a freely-falling frame because the scalar field is a true scalar. In other words, the gradient of scalar field is not equivalent to the inertial force caused by acceleration as it can not be eliminated by changing the state of motion of observer. It was the primary reason why Einstein\index{Einstein} abandoned a pure scalar field theory of gravity in favor of general relativity where gravitational field is identified with the components of the metric tensor, and, unlike a scalar field, can be removed by transformation to the local inertial frame. 

Rather remarkable, this difference in transformation properties between scalar field and metric tensor has no direct consequence for equivalence between inertial and gravitational masses of test bodies.  It was discovered \citep{dirk_obukhov_2015PhRvD} that the inertial and gravitational masses of massive test bodies remain equal in a wide class of scalar-tensor theories of gravity and the freely-falling test bodies move in the same way independently of their mass. This observation forces us to carefully discriminate between various formulations of the weak equivalence principle (WEP) in scalar-tensor theories. 

\paragraph{Gravitoelectric Multipoles.} External gravitoelectric multipoles\index{multipoles!external!gravitoelectric} ${\cal Q}_L\equiv {\cal Q}_{<i_1i_2\ldots i_l>}$ ($l\ge2$) are obtained by solving \eqref{2.6} and given by the following equation \citep[Equation 5.89]{kopeikin_2011book} \footnote{Be mindful of that the spatial indices are raised and lowered with the Kronecker symbol $\d^{ij}$ so that the position of the spatial indices does not matter.}
\begin{eqnarray}
  \label{5.9}
  {\cal Q}^L & = & \pd^{<L>}\bar{U}(t,{\bm x}_\B)\\
  & +& \pd^{<L>}\bar{\Psi}(t,{\bm x}_\B)-\frac{1}{2}\pd_{tt}\pd^{<L>}\bar{\chi}(t,{\bm x}_\B)+2(1+\gamma)\pd^{<L-1}\dot{\bar{U}}^{i_l>}(t,{\bm x}_\B)\nonumber\\
  & -& 2(1+\gamma)v_{\rm B}^j\pd^{<L>}\bar{U}^j(t,{\bm x}_\B)+(l-2\gamma-2)v_{\rm B}^{<i_l}\pd^{L-1>}\dot{\bar{U}}(t,{\bm x}_\B)\nonumber\\
  & +& (1+\gamma)v_{\rm B}^2\pd^{<L>}\bar{U}(t,{\bm x}_\B) -\frac{l}{2}v^j_{\rm B}v_{\rm B}^{<i_l}\pd^{L-1>j}\bar{U}(t,{\bm x}_\B)+(2-2\beta-l\gamma)\bar{U}(t,{\bm x}_\B)\pd^{<L>}\bar{U}(t,{\bm x}_\B)\nonumber\\
  & -&  (l^2-l+2\gamma+2)a_{\rm B}^{<i_l}\pd^{L-1>}\bar{U}(t,{\bm x}_\B)-lF^{j<i_l}_{\rm B}\pd^{L-1>}\bar{U}^j(t,{\bm x}_\B)+X^L\;\nonumber,\qquad\qquad(l\ge2)
\end{eqnarray}
where $X^L$ represents a contribution of the local inertial forces to the gravitoelectric multipole,
\begin{eqnarray}
  \label{1q4d}
  X^L \equiv \left\{ \begin{array}{cc}
         3a_{\rm B}^{<i_1}a_{\rm B}^{i_2>} &\qquad\qquad \mbox{if $l=2$};\\
         &\\
        0 & \qquad\qquad\mbox{if $l\geq3$}.\end{array} \right. 
\end{eqnarray}
We point out that in spite of the fact that the term $X^L$ appears in the expression \eqref{5.9} for the external multipoles, ${\cal Q}^L$, it is not a part of the curvature of spacetime manifold \citep{k89d,kovl_2004} and is exclusively associated with the local acceleration of worldline ${\cal W}$ of the origin of the body-adapted local coordinates. This is proved in section \ref{ne7c2x9}. 

\paragraph{Gravitomagnetic Multipoles.} External gravitomagnetic multipoles\index{multipoles!external!current-type} ${\cal C}_L\equiv {\cal C}_{<i_1i_2\ldots i_l>}$ for $l\ge2$ are also obtained by solving \eqref{2.6} and given by \citet[Equation 5.37]{xie_2010AcPSl} \footnote{Formula \eqref{3.29} corrects a typo in \citep[Equation 5.74]{kopeikin_2011book} for the external gravitomagnetic multipole ${\cal C}_L$.}
\begin{eqnarray}
  \label{3.29}
  \varepsilon_{ipk}{\cal C}_{pL} & = & 4(1+\gamma)\biggl[ v_{\rm B}^{[i}\pd^{k]<L>}\bar U(t,{\bm x}_\B) +\pd^{<L>[i}\bar{U}^{k]}(t,{\bm x}_\B)
   -\frac{l}{l+1}\delta^{<i_{l}[i}\pd^{k]L-1>}\dot{\bar{U}}(t,{\bm x}_\B)\biggr]\;,\qquad (l\ge1)
\end{eqnarray}
where the dot denotes the time derivative with respect to time $t$, the angular brackets denote STF symmetry with respect to multi-index $L=i_1,i_2,\ldots,i_l$, and the square brackets denote anti-symmetrization: $T^{[ij]}=(T^{ij}-T^{ji})/2$. The external multipoles ${\cal Q}_L$ and ${\cal C}_L$ are analogues of Dixon's multipoles $A_{\a_1...\a_l\m\n}$ and $B_{\a_1...\a_l\m\n}$ respectively -- see (\ref{om5g}) and (\ref{om6f}) below. We shall use the above-given expressions for the external multipoles in derivation of the equations of motion of extended bodies in next section. 

\section{Post-Newtonian Equations of Motion of an Extended Body in the Local Coordinates}\la{n4r6v}

Coordinate acceleration $a^i_\B$ of worldline $\cal W$ of the origin of the local coordinates adapted to body B with respect to the global coordinates is given by equation \eqref{5.8}. It depends on the local acceleration ${\cal Q}_i$ of the origin of the local coordinates with respect to a time-like geodesic of the effective background metric $\bar g_{\a\b}$. The acceleration ${\cal Q}_i$ cannot be determined by solving the matching equations \eqref{2.5}, \eqref{2.6}, and remains an arbitrary function of time. Center of mass of body B has not yet been defined but it certainly moves along worldline $\cal Z$ which is formally different from ${\cal W}$ in the most general case. However, we have enough freedom in choosing worldline ${\cal W}$ which we can use in order to make the two worldlines coincide. Mathematically, it means that the center of mass of body B remains at rest at the origin of the local coordinates adapted to body B as the body moves on spacetime manifold. This condition imposes a functional constraint on the local acceleration ${\cal Q}_i$ which converts the translational equations of motion \eqref{5.8} of the origin of the local coordinates to those for the center of mass of body B with respect to the global coordinates. In order to put the center of mass of body B to the origin of the local coordinates and to hold it in there, we have to know the translational equations of motion of the body's center of mass in the local coordinates adapted to the body.  

Derivation of translational equations of motion of the center of mass of body B in the local coordinates can be executed in three different ways, which are:
\begin{enumerate} 
\item the Fock-Papapetrou method of integration of microscopic equations of motion of matter over the body's volume \citep{fockbook,petrova,pap1,pap2,Papapetrou23101951} ;
\item the Mathisson-Dixon method of integration of skeleton of the stress-energy tensor of matter of body B given in terms of distributions \citep{mathisson_2010GReGr_1,mathisson_2010GReGr_2,dixon_1979} and amended with some regularization technique \citep{infeld_book,Damour_1983grr,poisson_2011};  
\item the Einstein-Infeld-Hoffmann (EIH) method of asymptotic surface integrals  \citep{eih,th_1985,asada_2011,racine_2005PhRvD,racine2013PhRvD}.
\end{enumerate}
The Mathisson-Dixon and EIH methods consider the extended bodies in ${\mathbb N}$-body system as singularities of gravitational field endowed with a set of the internal multipoles which represent the internal structure of the bodies. The multipoles in these approaches are not given in terms of volume integrals from a smooth distribution of matter inside the bodies but are merely functions of time given on worldline ${\cal Z}$ of each body's center of mass. On the other hand, the Fock-Papapetrou method operates with a continuous distribution of matter inside the bodies and defines the internal multipoles of the bodies in terms of the volume integrals like in section \ref{mdloc} of the present paper. It is assumed that the Mathisson-Dixon and EIH methods should give the same equations of motion for extended, arbitrary structured bodies as in the Fock-Papapetrou method. This is indeed true in case of pole-dipole particle approximation corresponding to rigidly rotating, spherically-symmetric bodies. However, this correspondence has been never checked for higher-order internal multipoles. We use the Fock-Papapetrou method of derivation of translational equations of motion of extended bodies having all mass and spin internal multipoles, and compare them with similar equations derived by \citet{racine_2005PhRvD,racine2013PhRvD} with EIH technique (see Appendix \ref{appendixA} and with the covariant equations derived by \citet{dixon_1979} (see Appendix \ref{appendixB}).  

In this section we define a center of mass and a linear momentum of body B, derive the post-Newtonian microscopic equations of motion of matter of the body in the local coordinates and, then, integrate them over the body's volume in order to get the post-Newtonian equations of motion of the linear momentum  and the center of mass of the body. As soon as the equations of motion for these quantities are established, the local acceleration ${\cal Q}_i$ is determined from the condition of vanishing of the linear momentum and the integral of the center of mass of the body which warrants that the center of mass of body B is always stay at the origin of the local coordinates. At the end of this section we give a post-Newtonian definition of the intrinsic angular momentum (spin) of body B and derive spin's rotational equations of motion in the local coordinates.

\subsection{Microscopic Equations of Motion of Matter}\label{pnem}

The microscopic post-Newtonian equations of motion of matter of body B include: 
\begin{itemize}
\item[1)] equation of continuity\index{equation of continuity}, 
\item[2)] thermodynamic equation relating the elastic energy, $\Pi=\Pi(u,{\bm w})$, to the stress tensor, ${\mathfrak{s}}_{\alpha\beta}={\mathfrak{s}}_{\a\b}(u,{\bm w})$, 
\item[3)] equation of conservation of the stress-energy tensor\index{conservation!stress-energy tensor}.
\end{itemize}
The equation of continuity\index{equation of continuity} of matter of body B in the body-adapted local coordinates $w^\a=(u,{\bm w})$ has the most simple form if we use the invariant density $\rho^{\ast}=\rho^{\ast}(u,{\bm w})$, defined in (\ref{11.19}). It reads
\begin{equation}
  \label{kp1}
  \frac{\pd\rho^{\ast}}{\pd u}+\frac{\pd\left(\rho^{\ast}\nu^i\right)}{\pd w^i}=0\;,
\end{equation} 
where $\nu^i=\nu^i(u,{\bm w})=dw^i/du$ is a coordinate velocity of matter in the local coordinates. Equation \eqref{kp1} is exact in any order of the post-Newtonian approximations like \eqref{11.20}. 

The thermodynamic equation relating the internal elastic energy, $\Pi$, and the stress tensor, ${\mathfrak{s}}_{\alpha\beta}$, of body B is required only in a linearized approximation where the stress-energy tensor is completely characterized by its spatial (stress) components ${\mathfrak{s}}_{ij}$. After making this substitution to the covariant equation (\ref{11.2}) we get the following thermodynamic equation in the local coordinates,
\begin{equation}
  \label{kp2}
  \rho^{\ast}\frac{d\Pi}{du}+{\mathfrak{s}}_{ij}\frac{\pd\nu^i}{\pd w^j} = 0\;,
\end{equation}
where the operator of the total time derivative\index{convective derivative}\index{derivative!convective}, $d/du\equiv\partial/\partial u+\nu^i\partial/\partial w^i$.

Covariant equation of conservation of the stress-energy tensor of matter of body B is \eqref{h2}. We need in the post-Newtonian approximation only the spatial component of this equation. Straightforward calculations with making use of the post-Newtonian components \eqref{11.21}--\eqref{11.22} of the stress-energy tensor of matter of body B yield the following form of the law of conservation \eqref{h2} in the local coordinates,
\ba
  \label{kp3}
   \rho^{\ast}\frac{d}{du}\bigg[\left(1+\frac{1}{2}\nu^2+\Pi+\frac{1}{2} \hat{h}_{00}+\frac13\hat{h}_{kk}\right)\nu^i+ \hat{h}_{0i}\bigg]&=&\frac{1}{2} \rho^{\ast}\frac{\partial (\hat{h}_{00}+\hat{l}_{00})}{\partial w^i}-\frac{\partial{\mathfrak{s}}_{ij}}{\partial w^j}\\
 & + &\rho^{\ast}\bigg[\frac{1}{4}(\nu^2+2\Pi+ \hat{h}_{00}) \frac{\partial \hat{h}_{00}}{\partial w^i}
  +\frac{1}{6}\nu^2\frac{\partial \hat{h}_{kk}}{\partial w^i}+\nu^k\frac{\partial \hat{h}_{0k}}{\partial w^i}\bigg]
  \nonumber\\
 &+&\frac{1}{2}\frac{\pd}{\pd w^j}\left[{\mathfrak{s}}_{ij}\left(\frac{\partial  \hat{h}_{00}}{\partial w^k}-\frac{1}{3}\frac{\partial \hat{h}_{kk}}{\partial w^k}\right)\right]+\frac{1}{6}{\mathfrak{s}}_{kk}\frac{\partial \hat{h}_{jj}}{\partial w^i}+\frac{\partial({\mathfrak{s}}_{ij}\nu^j)}{\partial u}\;,\nonumber
  \ea
where the metric tensor perturbations $\hat{h}_{00}$, $\hat{l}_{00}$, $\hat{h}_{0i}$, $\hat{h}_{ij}$ and $\hat{h}_{ii}$ in the local coordinates have been defined above in sections {\ref{n2v5r}} -- \ref{intermixt}.

\subsection{Post-Newtonian Mass of a Single Body }\label{pnmassdef}

There are two algebraically-independent definitions of the post-Newtonian mass in the scalar-tensor theory -- the {\it active} mass (Jordan's frame) and the {\it conformal} mass (Einstein's frame) which are defined respectively by equations (\ref{1.31}) and (\ref{1.34}) for multipolar index $l = 0$.
More specifically, the active mass \index{active mass}\index{mass!active}of body B is \cite{kovl_2004,kopeikin_2011book}
\begin{eqnarray}
  \label{activemass}
  \mathcal{M} 
  & = & \mathrm{M_{GR}}\Big[1+(1+\g-2\beta){\cal P}\Big] 
  + \frac{1}{6}(\gamma-1)\ddot{\mathcal{N}}-\frac{1}{2}\eta \int\limits_{{\cal V}_{\rm B}}\rho^{\ast}\hat{U}_{\rm B} d^3w
  - \sum_{l=1}^{\infty}\frac{1}{l!}\Big[(\gamma l+1){\cal Q}_L+2(\beta-1){\cal P}_L\Big]{\cal M}^{L}\;,
\end{eqnarray}
where ${\cal P}_L$, ${\cal Q}_L$ are the scalar field and gravitoelectric external multipoles given in \eqref{3.13} and \eqref{5.9} respectively, 
\begin{equation}
  \label{grmass}
  \mathrm{M_{GR}} = \int\limits_{{\cal V}_{\rm B}}\rho^{\ast}\bigg(1+\frac{1}{2}\nu^2+\Pi-\frac{1}{2}\hat{U}_{\rm B}\bigg)d^3w
\end{equation}
is a {\it bare} post-Newtonian mass of body B \citep{willbook}, ${\cal M}^{L}$ are {\it active}
 multipoles\index{multipoles!active} of the body defined in (\ref{1.31}), ${\cal N}$ is the rotational moment of inertia defined in \eqref{r5t1}, and $\ddot{\cal N}=d^2{\cal N}/du^2$ denote a second derivative of the moment of inertia with respect to time $u$.
 
Mass $\mathrm{M_{GR}}$ depends only on the internal distribution of mass, kinetic, thermal and gravitational energy densities of body B. It coincides with with the Tolman mass \index{Tolman mass}\index{mass!Tolman}\citep{PhysRev.35.875} of a single, isolated body residing in asymptotically-flat spacetime derived by volume integration of Tolman's superpotential \citep[Equation 1.4.32]{Petrov_2017book}. Had the body B been isolated, the mass $\mathrm{M_{GR}}$ would be conserved. However, in ${\mathbb N}$-body system gravitational interaction of body B with external bodies causes the body's tidal deformations which change the internal distribution of matter and shape of body B, thus, making ${\rm M_{GR}}$ dependent on time. The temporal change of  ${\rm M_{GR}}$ is governed by the ordinary differential equation \citep{dsx2,kopeikin_2011book}
\be\la{mch3c}
\dot{\mathrm M}_{\mathrm{GR}}=\sum\limits_{l=1}^\infty\frac1{l!}{\cal Q}_L\dot{\cal M}^{L}\;,
\ee
where the overdot denotes a derivative with respect to coordinate time $u$.

The conformal mass \index{conformal mass}\index{mass!conformal}of body B, $M\equiv {\cal I}$, is defined by equation \eqref{1.34} taken for $l=0$, and is \citep{kovl_2004,kopeikin_2011book}
\begin{eqnarray}
  \label{confmass}
  M 
  & = & \mathrm{M_{GR}}\left[1+(\gamma-1){\cal P}\right]-\sum_{l=1}^{\infty}\frac{l+1}{l!}{\cal Q}_L{\cal M}^{L}\;.
\end{eqnarray}
The conformal mass $M$ defines the inertial mass of a single body B in ${\mathbb N}$-body system as we shall demonstrate in section \ref{n2c4z21a}. 
In case of a single isolated body the last term in the right-hand side of \eqref{confmass} is absent but it appears in ${\mathbb N}$-body system (if the body under consideration is not spherically-symmetric) and can be interpreted in the spirit of Mach's principle \index{Mach's principle} stating that the body's inertial mass originates from its gravitational interaction with an external universe. Mach's idea is not completely right because the inertial mass of the body is primarily originating from the {\it bare} mass ${\rm M_{GR}}$ but it has a partial support as we cannot completely ignore the gravitational interaction of a single body with its external gravitational environment in the definition of the inertial mass of the body. This effect is important to take into account in inspiralling compact binaries as they are tidally distorted and, hence, the part of the inertial mass of each star associated with the very last term in \eqref{confmass} rapidly changes as the distance between them is decreasing. The overall time variation of the conformal mass $M$ is given by equation,
\be \label{x1674} 
\dot M=(\g-1)\left({\cal P}\sum\limits_{l=1}^\infty\frac1{l!}{\cal Q}_L\dot{\cal M}^{L}+\dot{\cal P}\mathrm{M_{GR}}\right)-\sum\limits_{l=1}^\infty\frac1{(l-1)!}\left({\cal Q}_L\dot{\cal M}^{L}+\frac{l+1}{l}\dot{\cal Q}_L{\cal M}^{L}\right)\;,
\ee
where we have made use of \eqref{mch3c}.

Relation between the active and conformal masses\index{active mass!relation to conformal mass} is obtained by comparing (\ref{activemass}) with (\ref{confmass}) 
\begin{eqnarray}
  \label{p3c2}
  M & = & \mathcal{M} + \frac{1}{2}\eta\int\limits_{{\cal V}_{\rm B}}\rho^{\ast}\hat{U}_{\rm B}d^3w -\frac{1}{6}(\gamma-1)\ddot{\mathcal{N}}
   +2(\beta-1)\bigg(\mathcal{M}{\cal P}+\sum_{l=1}^{\infty}\frac{1}{l!}{\cal P}_L{\cal M}^{L}\bigg)+(\gamma-1)\sum_{l=1}^{\infty}\frac{1}{(l-1)!}{\cal Q}_L{\cal M}^{L} \;,
\end{eqnarray}
where $\eta=4\beta-\gamma-3$ is called the Nordtvedt parameter \index{Nordtvedt!parameter} \citep{willbook}. We can see that the conformal mass $M$ of body B differs from its active mass ${\cal M}$. This fact was noticed by \citet{1962PhRv..126.1875D,1965AnPhy..31..235D} and \index{Nordtvedt} \citet{1973PhRvD...7.2347N,willbook} who found the integral term being proportional to the Nordtvedt parameter $\eta$ in the right-hand side of (\ref{p3c2}). The actual difference between the masses turns out to be more complicated and includes a term with the second time derivative of the rotational moment of inertia of the body as well as the tidal contributions originating from gravitational interaction of the body's internal multipoles with the external multipoles.  Had body B been completely isolated from the external gravitational field\index{gravitational field!external}, the difference between the active and conformal masses would be caused only by the Dicke-Nordtvedt self-gravity term depending on parameter $\eta$, and the second time derivative of the body's rotational moment of inertia due to, e.g., radial oscillations of the body. In case of ${\mathbb N}$-body system the gravitational field of $\mathbb{N}-1$ external bodies cannot be ignored in the definition of the post-Newtonian mass of a single body due to the gravitational coupling of the external and internal multipoles of the body. 

\subsection{Post-Newtonian Center of Mass and Linear Momentum of a Single Body}\la{o93c5}

Functional form of equations of motion of extended bodies in ${\mathbb N}$-body system depends crucially on the choice of the reference point inside body B that defines its center of mass. There is a large freedom in choosing definition of the center of mass beyond the Newtonian limit. Physically, any definition is allowed and makes a certain sense. However, the most optimal definition of the center of mass makes the equations of motion look simple and eliminates a number of spurious terms which would contaminate the equations of motion, like the {\it non-canonical} multipole moments ${\cal N}^L$ and ${\cal R}^L$ mentioned above, if the center of mass is not chosen properly. \citet{dsx1,dsx2} have shown that in general relativity the position of the center of mass of body B, which is a member of ${\mathbb N}$-body system, is the most optimally determined by picking up the zero value of the Blanchet-Damour mass dipole in the internal solution for the metric tensor perturbation. In scalar-tensor theory of gravity there are two possible definitions of the internal mass dipole depending on whether the Jordan or the Einstein frame is chosen for the multipole expansion of the metric tensor. The Jordan frame gives the {\it active} dipole moment ${\cal M}^i$, and the Einstein frame defines the {\it conformal} dipole ${\cal I}^i$. Before doing computations it is difficult to foresee which choice of the dipole is the best for positioning the center of mass of the body. Only after completing the derivation of the equations of motion it becomes clear that it is the {\it conformal} mass dipole yields the most optimal choice of the post-Newtonian center of mass of each body \citep{kovl_2004,kopeikin_2011book}. Physical reason for this is that the {\it conformal} dipole moment obeys the law of conservation of linear momentum, ${\mathfrak{p}}^i$, of each body in its own local coordinate chart while the post-Newtonian {\it active} dipole does not have such a property. 

Thus, we define the post-Newtonian center of mass of each body B by making use of the conformal definition (\ref{1.34}) of the internal multipoles of body B for a multipolar index $l = 1$. It yields 
\be\la{b2c4z9}
{\cal I}^i={\cal I}^i_{\rm b}+{\cal I}^i_{\rm c}\;,
\ee
where
\ba\la{brtv56b}
  {\cal I}^i_{\rm b} & = &\int\limits_{{\cal V}_{\rm B}}\varrho(u,\bm{w})\left[1-(1-\gamma){\cal P}-\sum_{l=1}^{\infty}\frac{1}{l!}{\cal Q}_Lw^{L}\right]w^i\,d^3w- \frac{2}{5}\left(3 \dot{\cal R}^i-\frac14\ddot{\cal N}^i \right)\;,
\ea
is the {\it bare} conformal dipole of body B, and ${\cal I}^i_{\rm c}$ is a {\it complementary} post-Newtonian translation that is introduced in order to have freedom in a residual adjustment of worldline ${\cal Z}$ of the center of mass of the body in the process of derivation of equations of motion. At this stage the translation ${\cal I}^i_{\rm c}$ is left undetermined. It will be specified later on -- see equations \eqref{m4g1x8p} and \eqref{d9n3v5}.  

The last two terms in the right-hand side of \eqref{brtv56b} can be written down more explicitly if we use a vector virial theorem, 
\be\la{jnw8j}
\frac25\left(3 \dot{\cal R}^i-\frac14\ddot{\cal N}^i\right) =\int\limits_{{\cal V}_{\rm B}}\left(\rho^*\nu^2+{\mathfrak{s}}_{kk}-\frac12\rho^*\hat U_{\rm B}\right)w^id^3w+\sum_{l=1}^\infty \frac{1}{(l-1)!}{\cal Q}_L{\cal M}^{iL}-\frac12\sum_{l=0}^\infty\frac{1}{(2l+3)l!} {\cal Q}_{iL}{\cal N}^L\;.
\ee
Replacing \eqref{jnw8j} to \eqref{brtv56b} brings the {\it bare} conformal dipole to the following form, 
\begin{eqnarray}
  \label{confdipole}
  {\cal I}^i_{\rm b}& = & \int\limits_{{\cal V}_{\rm B}} \rho^{\ast}(u,\bm{w}) \left[1+\frac{1}{2}\nu^2+\Pi-\frac{1}{2}\hat{U}_{\rm B}+(\gamma-1){\cal P}\right]w^id^3w
  -\sum_{l=1}^{\infty}\frac{l+1}{l!}{\cal Q}_L{\cal M}^{iL}
  - \frac{1}{2}\sum_{l=0}^{\infty}\frac{1}{(2l+3)l!}{\cal Q}_{iL}\mathcal{N}^L \;,
\end{eqnarray}
where the STF {\it non-canonical} multipole, $\mathcal{N}^L$, has been defined in \eqref{NL15}.

We will also need definition of the {\it active} dipole, ${\cal M}^i$, for it will appear in the equations of motion explicitly. Definition of the {\it active} mass dipole follows directly from the generic post-Newtonian formula for mass multipoles \eqref{1.31}  taken for $l=1$. After applying the virial theorem \eqref{jnw8j}, we find out that the {\it active} dipole, ${\cal M}^i$,  of body B relates to its {\it bare} conformal dipole, ${\cal I}^i_{\rm b}$ as follows,
\ba\la{nrtcvc67h}
{\cal M}^i&=&{\cal I}^i_{\rm b}+
(\gamma-1)\left(\frac{3}5\dot{\cal R}^i
-\frac{1}{10}\ddot{\cal N}^i\right) -
\frac{\eta}2\bigg(\int\limits_{{\cal V}_{\rm B}}\rho^*\hat U_{\rm B}w^id^3w+\sum_{l=0}^{\infty}\frac1{(2l+3)l!}{\cal Q}_{iL}{\cal N}^L\bigg)
\\\nonumber&-&
\sum_{l=1}^{\infty}\frac{(\gamma-1)l+2(\beta-1)}{l!}{\cal Q}_L{\cal
I}^{iL}-2(\beta-1)\left({\cal P}^k-{\cal Q}^k\right)\left({\cal M}^{ik}+\frac13\delta^{ik}{\cal N}\right)\;.
\ea
The volume integrals entering definitions \eqref{confdipole} and \eqref{nrtcvc67h} of the conformal and active dipoles of body B are performed over hypersurface ${\cal H}_u$ of constant time $u$. All other terms entering these definitions are taken on worldline ${\cal W}$ of the origin of the local coordinates adapted to the body, at the point of intersection of ${\cal W}$ with hypersurface ${\cal H}_u$. Therefore, the dipole is a function of time $u$ only.

The dipole defines a vector of displacement of the center of mass of body B from the origin of the local coordinates adapted to the body. If the origin of the local coordinates coincides with the center of mass of the body, the dipole vanishes. We draw attention of the reader that the post-Newtonian definition of the center of mass of body B depend (like in case of the post-Newtonian definition of body's mass) not only on the distribution of matter density, velocity and stresses inside the body but also on the terms describing the coupling of the internal and external multipoles. \citet{th_1985} were first who noticed the presence of such terms in the post-Newtonian definition of the center of mass (and other mass multipoles) but they did not provide their exact form that was found later by \citet{dsx1, dsx2} in general relativity and by \citet{kovl_2004} in scalar-tensor theory of gravity. We notice that dipole's definitions  \eqref{confdipole} and \eqref{nrtcvc67h} contain {\it non-canonical} multipoles, ${\cal R}^L$ and ${\cal N}^L$, which don't appear in the {\it canonical} multipole decomposition of the metric tensor perturbation in vacuum \citep{bld,bld1986,thor}. Comprehensive calculations of equations of motion of extended bodies by the Fock-Papapetrou method have revealed \citep{dsx1,dsx2,kovl_2004} that if the {\it non-canonical} multipoles ${\cal R}^L$ and ${\cal N}^L$ are removed from definition of the dipole, they appear explicitly in the equations of motion, thus, making them incompatible with the equations of motion in the Mathisson-Dixon or EIH approaches which cannot have the {\it non-canonical} multipoles, ${\cal R}^L$ and ${\cal N}^L$, at all. Therefore, it is natural to hold the {\it non-canonical} multipoles ${\cal R}^L$ and ${\cal N}^L$ in the definitions of the post-Newtonian mass, center of mass and mass multipoles ${\cal M}^L$ of body B.   

Definition \eqref{b2c4z9} of the conformal dipole of body B is used to define the position of its center of mass with respect to the origin of the local coordinates adapted to body B.  The center of mass, $w^i_{\rm cm}$, of the body is defined in its local coordinates by the overall value of its dipole,
\be\la{n5v1c8p}
Mw^i_{\rm cm}={\cal I}^i\;,
\ee
where $M$ is the post-Newtonian conformal mass of body B defined above in \eqref{confmass} . 
The post-Newtonian linear momentum ${\mathfrak{p}}^i$ of body B is defined as the first derivative of the dipole (\ref{b2c4z9}) with respect to the local time $u$,
\be\la{b5z0e}
{\mathfrak{p}}^i\equiv\dot {\cal I}^i(u)={\mathfrak{p}}^i_{\rm b}+\dot {\cal I}^i_{\rm c}\;,
\ee
where ${\mathfrak{p}}^i_{\rm b}\equiv\dot {\cal I}^i_{\rm b}$, and  the overdot denotes the time derivative with respect to $u$.
After taking the time derivative from the {\it bare} dipole \eqref{confdipole} and using the local equations of motion of matter (\ref{kp3}) to transform the integrand, we obtain \citep{kovl_2004},
\begin{eqnarray}
  \label{confmoment}
  {\mathfrak{p}}^i_{\rm b} 
  & = & \int\limits_{{\cal V}_{\rm B}}\rho^{\ast}\nu^i\bigg(1+\frac{1}{2}\nu^2+\Pi-\frac{1}{2}\hat{U}_{\rm B}\bigg)d^3w + \int\limits_{{\cal V}_{\rm B}}\left({\mathfrak{s}}_{ik}\nu^k-\frac{1}{2}\rho^{\ast}\hat{W}^i_{\rm B}\right)d^3w\\
  & &+ \frac{d}{du}\bigg[{\cal I}^i_{\rm c}-(1-\gamma){\cal P}{\cal M}^i-\sum_{l=1}^{\infty}\frac{l+1}{l!}{\cal Q}_L{\cal M}^{iL} -\frac{1}{2}\sum_{l=0}^{\infty}\frac{1}{(2l+3)l!}{\cal Q}_{iL}\mathcal{N}^{L}\bigg]\nonumber\\
  & &+ \sum_{l=1}^{\infty}\frac{1}{l!}\bigg[{\cal Q}_L\dot{{\cal M}}^{iL}+\frac{l}{2l+1}{\cal Q}_{iL-1}\dot{\mathcal{N}}^{L-1}-{\cal Q}_L\int\limits_{{\cal V}_{\rm B}}\rho^{\ast}\nu^iw^{L}d^3w\bigg]\nonumber\;,
\end{eqnarray}
where 
\begin{equation}
  \label{w3h7}
  \hat{W}^i_{\rm B} = \int\limits_{{\cal V}_{\rm B}}\frac{\rho^{\ast}(u,\bm{w}')\nu'^k(w^k-w'^k)(w^i-w'^i)}{|\bm{w}-\bm{w}'|^3}d^3w'\;,
\end{equation}
is a new internal potential of gravitational field of body B - c.f. \citep[Equation 4.32]{willbook}.

We remind now that the point $x^i_{\rm B}$ represents position of the origin of the local coordinates adapted to body B in the global coordinates taken at instant of time $t$. It moves along worldline ${\cal W}$ which we want to make identical to worldline ${\cal Z}$ of the center of mass of body B.
It can be achieved if we can retain the center of mass of body B at the origin of the local coordinates adapted to the body, that is to have for any instant of time, $w^i_{\rm cm}=0$. This condition means that both functions of time -- the conformal dipole ${\cal I}^i$ of the body and its linear momentum ${\mathfrak{p}}^i$ -- have to vanish,
\be\la{n5vz1o}
{\cal I}^i=0\;,\qquad\qquad {\mathfrak{p}}^i=0\;.
\ee
These constraints imposed on the conformal dipole and linear momentum of body B can be satisfied if, and only if, the local equation of motion of the center of mass of the body can be reduced to equation 
\begin{equation}
  \label{a0n5}
  \dot{\mathfrak{p}}^i(u) = \dot{\mathfrak{p}}^i_{\rm b}+\ddot{\cal I}^i_{\rm c}=0\;.
\end{equation}
It is remarkable that equation \eqref{a0n5} can be, indeed, fulfilled after making an appropriate choice of the external dipole ${\cal Q}_i$ that characterizes the acceleration of the origin of the local coordinates of body B with respect to a geodesic worldline of the effective external manifold $\bar{M}$. We prove this statement below in section \ref{teofminlc}.  

\subsection{Post-Newtonian Spin of a Single Body}\la{x5wc7n}

In the post-Newtonian approximation the spin multipoles of an extended body B appear in the multipolar decomposition of the metric tensor in the Newtonian form \eqref{1.32} where the body's spin corresponds to $l=1$. The Newtonian definition of spin is insufficient for derivation of the post-Newtonian equations of rotational motion and must be extended to include the post-Newtonian terms. The post-Newtonian definition of spin of a single body residing in asymptotically-flat spacetime can be extracted from the multipolar expansion of the metric tensor component $\hat{g}_{0i}(u,{\bm w})$ by taking into account terms of the post-post-Newtonian order \cite{dyr2}. The problem we face in the present paper is that we have to define the post-Newtonian spin of body B which is not residing in asymptotically-flat spacetime but is a member of ${\mathbb N}$-body system. We have also take into account the contribution of the scalar field as we work in scalar-tensor theory of gravity. 

Post-Newtonian definition of the spin can be extracted from the local law of conservation of the stress-energy complex $\Theta^{\m\n}$  
\be
\Theta^{\m\n}{}_{,\n}=0\;,
\ee
which is used for building definitions of conserved quantities in metric theories of gravity \citep{Petrov_2017book}. The stress-energy complex is not unique and is defined up to a term which divergence vanishes identically. One of the most convenient definitions of the symmetric stress-energy tensor in scalar-tensor theory of gravity was found by \citet{nutku_1969ApJ}. It generalizes the Landau-Lifshitz stress-energy complex \citep{Landau1975} and reads,
\be
\Theta^{\m\n}=-g(1+\phi)\left(T^{\m\n}+t^{\m\n}\right)\;,
\ee
where $g={\rm det}[g_{\m\n}]$, $\phi$ is the perturbation of the scalar field \eqref{aa}, $T^{\m\n}$ is the stress energy-tensor of matter, and $t^{\m\n}$ is an analog of the Landau-Lifshitz pseudo-tensor $t^{\m\n}_{\rm LL}$ of the gravitational field \citep{Landau1975}. The pseudo-tensor has been determined by \citet{nutku_1969ApJ} and reads
\be\la{m3c7z4}
t^{\m\n}=\frac1{16\pi}\left[\left(1+\phi^3\right)t^{\m\n}_{\rm LL}+\frac{2\omega(\phi)+3}{1+\phi}\left(\pd^\m\phi\pd^\n\phi-\frac12 g^{\m\n}\pd^\a\phi\pd_\a\phi\right)\right]\;.
\ee

Let us now introduce the post-Newtonian definition of a {\it bare} spin of body B in the local coordinates adapted to the body, as follows,
\be\label{spin-1}
{\cal S}^i_{\rm b} =
\int\limits_{{\mathbb R}^3}\varepsilon_{ijk}w^j\big[-\hat g(u,{\bm w})\big]\big[1+(\g-1)\hat\varphi(u,{\bm w})\big]\big[\hat T^{0k}(u,{\bm w})+\hat{t}^{0k}(u,{\bm w})\big]d^3w\,,
\ee
where $\varepsilon_{ijk}$ is 3-dimensional symbol of Levi-Civita and the integration is performed over the entire 3-dimensional space ${\mathbb R}^3$. Special attention should be paid to the variables entering definition \eqref{spin-1}. Namely, the scalar field perturbation $\hat\varphi$ is given by \eqref{1.1} and includes both external and internal parts, the stress-energy tensor $\hat T^{\mu\nu}$ depends solely on matter variables of body B as defined in equations \eqref{11.21}--\eqref{11.22} but it includes the overall - external and internal - post-Newtonian perturbations of the metric tensor \eqref{1.2} and scalar field \eqref{1.1}, while the Nutku pseudo-tensor $\hat{t}^{\mu\nu}$ introduced in \eqref{m3c7z4} depends only on the internal part of the post-Newtonian perturbations of the metric tensor \eqref{1.8}--\eqref{1.10} and scalar field \eqref{1.7}. These limitations introduced to the definition of spin of body B prevents appearance of divergent terms that could emerge from the integration of pseudo-tensor which is formally defined in the entire space ${\mathbb R}^3$.

Integrating by parts allows us to reduce \eqref{spin-1} to the integral over the volume ${\cal V}_{\rm B}$ of body B only. Expanding it in the post-Newtonian series yields explicit expression for the {\it bare} post-Newtonian spin of body B in the following form \citep{kovl_2004}
 \ba\label{spin-3}
 {\cal S}^i_{\rm b} &=&\displaystyle
\int\limits_{{\cal V}_{\rm B}}\rho^*\varepsilon_{ijk}w^j\nu^k\left[1+\frac12\nu^2+\Pi+(2\gamma+1)\hat U_{\rm B}+(1-\gamma){\cal P}\right]d^3w+\int\limits_{{\cal V}_{\rm B}}\varepsilon_{ijk}w^j{\mathfrak{s}}^{kp}\nu^pd^3w\\\nonumber
&+&\displaystyle\sum_{l=1}^{\infty}\frac1{l!}\bigg[3{\cal Q}_L+2(\gamma-1){\cal P}_L\bigg]\int\limits_{{\cal V}_{\rm B}}\rho^*\varepsilon_{ijk}w^j\nu^kw^Ld^3w-
\frac1{2}\int\limits_{{\cal V}_{\rm B}}\rho^*\varepsilon_{ijk}w^j\bigg[\hat
W^k_{\rm B} +(3+4\gamma)\hat U^k_{\rm B}\bigg]d^3w
\;,
\ea
where $\nu^i=dw^i/du$ is velocity of matter of body B in the local coordinates, the integration is over volume of body B, and vector potential $\hat W^k_{\rm B}$ is defined in (\ref{w3h7}). The reader can notice that the spin of body B which is a member of ${\mathbb N}$-body system, depends not only on the internal structure of the body but on the gravitational field of external bodies like in case of the internal mass multipoles.
We shall use definition (\ref{spin-3}) to derive the rotational equations of motion of body's spin below in this section and in section \ref{sbdytvre0}.

\subsection{Translational Equation of Motion of the Center of Mass of a Single Body}\la{teofminlc}

Translational equations of motion of the center of mass of body B with respect to the local coordinates $w^\a$ adapted to the body, are derived by Fock-Papapetrou method from the law of conservation \eqref{a0n5} of the total linear momentum ${\mathfrak{p}}^i$ of the body. In order to implement this law we have to find out the time derivative of the {\it bare} linear momentum, ${\mathfrak{p}}^i_{\rm b}$ of the body. To this end, we differentiate both sides of equation (\ref{confmoment}) one time with respect to the local coordinate time $u$, make use of the microscopic equations of motion (\ref{kp1})-(\ref{kp3}), and integrate by parts to re-arrange a number of terms. One obtains \citep{kopeikin_2011book,xie_2010AcPSl}
\begin{eqnarray}
  \label{Qi}
  \dot{\mathfrak{p}}^i_{\rm b} & = & \mathcal{M}{\cal Q}^i+\sum_{l=1}^{\infty}\frac{1}{l!}{\cal Q}_{iL}{\cal M}^{L}+\sum_{l=1}^{\infty}\frac{l}{(l+1)!}{\cal C}_{iL}\mathcal{S}^{L}\\
  & &- \sum_{l=1}^{\infty}\frac{1}{(l+1)!}\bigg[(l^2+l+4){\cal Q}_L+2(\gamma-1){\cal P}_L\bigg]\ddot{{\cal M}}^{iL}\nonumber\\
  & &- \sum_{l=1}^{\infty}\frac{2l+1}{(l+1)(l+1)!}\bigg[(l^2+2l+5)\dot{\cal Q}_L+2(\gamma-1)\dot{{\cal P}}_L\bigg]\dot{{\cal M}}^{iL}\nonumber\\
  & &- \sum_{l=1}^{\infty}\frac{2l+1}{(2l+3)(l+1)!}\bigg[(l^2+3l+6)\ddot{\cal Q}_L+2(\gamma-1)\ddot{{\cal P}}_L\bigg]{\cal M}^{iL}\nonumber\\
  & &- \sum_{l=1}^{\infty}\frac{1}{(l+1)!}\varepsilon_{ipq}\bigg[{\cal C}_{pL}\dot{{\cal M}}^{qL}+\frac{l+1}{l+2}\dot{\cal C}_{pL}{\cal M}^{qL}\bigg]\nonumber\\
  & &+ 2\sum_{l=0}^{\infty}\frac{l+1}{(l+2)!}\varepsilon_{ipq}\bigg[\Big(2{\cal Q}_{pL}+(\gamma-1){\cal P}_{pL}\Big)\dot{\mathcal{S}}^{qL}
  + \frac{l+1}{l+2}\Big(2\dot{\cal Q}_{pL}+(\gamma-1)\dot{{\cal P}}_{pL}\Big)\mathcal{S}^{qL}\bigg]\nonumber\\
  &&- ({\cal P}^i-{\cal Q}^i)\bigg[\frac{1}{2}\eta\int\limits_{{\cal V}_{\rm B}}\rho^{\ast}\hat{U}^{(\mathrm{B})}d^3w-\frac{1}{6}(\gamma-1)\ddot{\mathcal{N}}+\nonumber\\
  & &\phantom{- ({\cal P}^i-{\cal Q}^i)}+ 2(\beta-1)\bigg({\cal M}{\cal P}+\sum_{l=1}^{\infty}\frac{1}{l!}{\cal P}_L{\cal M}^{L}\bigg)+(\gamma-1)\sum_{l=1}^{\infty}\frac1{(l-1)!}{\cal Q}_L{\cal M}^{L}\bigg]\;,\nonumber
\end{eqnarray}
where the spatial indices are raised and lowered with the Kronecker symbol, the {\it active} mass multipoles ${\cal M}^L$ are defined in \eqref{1.31} and include the post-Newtonian corrections, the spin multipoles ${\cal S}^L$ are sufficient in the Newtonian limit \eqref{1.32}. We have not shown in \eqref{Qi} a number of terms which are directly proportional to the internal conformal dipole, ${\cal I}^i$, and the linear momentum, ${\mathfrak{p}}^i$, of body B because these terms vanish if the origin of the local coordinates coincides with the center of mass of body B under condition \eqref{n5vz1o} which we employ in the rest of the paper. The omitted dipole-dependent terms in \eqref{Qi} can be found in \citep[Equation 6.19]{kopeikin_2011book}. 

Equation \eqref{Qi} is the post-Newtonian generalization of the second Newton's law applied to body B and written down in the body-adapted local coordinates. Therefore, the right-hand side of (\ref{Qi}) is the net force exerted on body B. This force does not include the self-action force as the scalar-tensor theory of gravity belongs to the class of conservative theories \citep{willbook}. Formally, the self-action force terms appeared at different stages of the computation of the time derivative of the linear momentum but they all have mutually canceled out at the final expression \eqref{Qi}. The external force standing in the right-hand side of \eqref{Qi} consists of three parts:
\begin{enumerate}
\item the tidal gravitational force caused by the coupling of the internal {\it active} multipoles, ${\cal M}^L$, ${\cal S}^L$ of body B with the external multipoles ${\cal Q}_L$, ${\cal P}_L$, ${\cal C}_L$ for $l\ge 2$, 
\item the force of inertia consisting of ${\cal M}{\cal Q}_i$ and all other post-Newtonian terms being proportional to ${\cal Q}_i$, caused by the non-geodesic motion of the origin of the local coordinates adapted to body B;
\item the Dicke-Nordtvedt force that is proportional to the difference ${\cal P}^i-{\cal Q}^i$ as shown by the very last term in the right-hand side of \eqref{Qi}, caused by the violation of the strong principle of equivalence (SEP) in scalar-tensor theory of gravity. 
\end{enumerate}
In order to ensure vanishing of the total linear momentum of body B, $\dot {\mathfrak{p}}^i=0$, we shall choose the local acceleration ${\cal Q}_i$ to compensate all terms in the right-hand side of \eqref{Qi} along with the complementary term $\ddot{\cal I}^i_{\rm c}$ that is used for small residual adjustment of the acceleration. This choice eliminates the relative acceleration of the worldline $\cal Z$ of the center of mass of body B with respect to worldline $\cal W$ of the origin of the body-adapted local coordinates. In this locally-accelerated frame we can still have the center of mass of body B moving with respect to the origin of the local coordinates with constant velocity, but we impose further constraint \eqref{n5vz1o} to eliminate this rectilinear motion and to put the center of mass of body B at the origin of its own local coordinates. It makes worldlines ${\cal Z}$ and ${\cal W}$ identical. 

Solution of the law of conservation of the linear momentum \eqref{a0n5}, where $\dot {\mathfrak{p}}^i_{\rm b}$ is given by (\ref{Qi}), with respect to ${\cal Q}_i$ yields
\be\la{q6v4m}
{\cal Q}_i={\cal Q}_i^{\rm N}+{\cal Q}_i^{\rm pN}-\frac{\ddot {\cal I}^i_{\rm c}}{M}\;,
\ee
where the first term is the Newtonian part of acceleration, the second term is the post-Newtonian correction, and the third term is the complementary acceleration which allows us to make residual adjustments in the position of the center of mass of the body, if necessary. The residual freedom in choosing position of the center of mass of body B is fixed at the last steps of derivation of translational equations of motion -- see \eqref{m4g1x8p} and \eqref{d9n3v5}. 

The Newtonian and post-Newtonian counterparts of the local acceleration of body B are defined by the following equations,
\ba\la{b4cs2j}
M{\cal Q}_i^{\rm N}&=&
\left(M-{\cal M}\right){\cal P}_i-\sum_{l=1}^{\infty}\frac{1}{l!}{\cal Q}_{iL}{\cal M}^{L}\;,\\
\la{nrvug9}
M{\cal Q}_i^{\rm pN}&=&\sum_{l=1}^{\infty}\frac{1}{(l+1)!}\bigg[(l^2+l+4){\cal Q}_L+2(\gamma-1){\cal P}_L\bigg]\ddot{{\cal M}}^{iL}\\
  & +& \sum_{l=1}^{\infty}\frac{2l+1}{(l+1)(l+1)!}\bigg[(l^2+2l+5)\dot{\cal Q}_L+2(\gamma-1)\dot{{\cal P}}_L\bigg]\dot{{\cal M}}^{iL}\nonumber\\
  & +& \sum_{l=1}^{\infty}\frac{2l+1}{(2l+3)(l+1)!}\bigg[(l^2+3l+6)\ddot{\cal Q}_L+2(\gamma-1)
  \ddot{\mathcal{P}}_L\bigg]{\cal M}^{iL}\nonumber\\
& +& \sum_{l=1}^{\infty}\frac{1}{(l+1)!}\varepsilon_{ipq}\bigg[{\cal C}_{pL}\dot{{\cal M}}^{qL}+\frac{l+1}{l+2}\dot{\cal C}_{pL}{\cal M}^{qL}\bigg]-\sum_{l=1}^{\infty}\frac{l}{(l+1)!}{\cal C}_{iL}\mathcal{S}^{L}\nonumber\\
& -& 2\sum_{l=0}^{\infty}\frac{l+1}{(l+2)!}\varepsilon_{ipq}\bigg[\Big(2{\cal Q}_{pL}+(\gamma-1){\cal P}_{pL}\Big)\dot{\mathcal{S}}^{qL}
 + \frac{l+1}{l+2}\Big(2\dot{\cal Q}_{pL}+(\gamma-1)\dot{{\cal P}}_{pL}\Big)\mathcal{S}^{qL}\bigg]
\;,\nonumber
\ea
where $M$ and ${\cal M}$ are the {\it conformal}\index{conformal mass}\index{mass!conformal} and {\it active} \index{active mass}\index{mass!active}gravitational masses of body B. The two masses, $M$ and ${\cal M}$, are not equal according to (\ref{p3c2}). The difference between them plays a role of a scalar charge, $\mathfrak{q}\equiv {\cal M}-M$, of the scalar field $\phi$ which couples with the external dipole of the scalar field ${\cal P}^i=\bar U_{,i}$ and causes the Dicke-Nordtvedt anomalous acceleration, $\mathfrak{q}{\cal P}^i$, in \eqref{b4cs2j} \index{Dicke-Nordtvedt effect}\index{Dicke}\index{Nordtvedt}\citep{1962PhRv..126.1875D,1965AnPhy..31..235D,willbook}. In general relativity, $\mathfrak{q}=0$, and the Dicke-Nordtvedt acceleration in the right-hand side of (\ref{q6v4m}) vanishes. 

Equation (\ref{q6v4m}) is a condition for the fulfillment of the law of conservation of linear momentum (\ref{a0n5}) in local coordinates. It ensures that the worldline $\cal W$ of the origin of local coordinates does not accelerate with respect to the worldline $\cal Z$ of the center of mass of body B. Equation (\ref{q6v4m}) does not warranty, however, that $\cal W$ and $\cal Z$ coincides. The origin of the local coordinates still can move uniformly with respect to the center of mass of the body. To eliminate this uniform motion we impose condition, ${\mathfrak{p}}^i=0$. The freedom which remains is a constant relative displacement of the origin of the local coordinates with respect to the center of mass of the body. This constant displacement is removed by additional constraint imposed on the internal conformal dipole of the body, ${\cal I}^i=0$. This procedure results in the constraint \eqref{n5vz1o} and ensures that the worldlines $\cal W$ and $\cal Z$ coincide.  

Acceleration ${\cal Q}_i$ given in (\ref{q6v4m}) must be substituted to the equations of motion of the origin of the local coordinates (\ref{5.8}) to convert them to the translational equations of motion of the center of mass of body B in the global coordinates. These equations still contain the external gravitational potentials $\bar U$, $\bar\Psi$, $\bar U^i$, and $\bar\chi$ defined in \eqref{tb54vd}, \eqref{3d8b1a}, which are given in the form of integrals expressed in the global coordinates. These integrals should be explicitly expanded with respect to the internal multipoles\index{internal multipole} of the bodies of ${\mathbb N}$-body system in order to complete the theory. We shall conduct this computation in section \ref{c6a0n5} and derive translational equations of motion of extended bodies in ${\mathbb N}$-body system in terms of their internal multipoles as well as coordinates and velocities of their centers of mass. 

\subsection{Rotational Equations of Motion of Spin of a Single Body}\label{q5a9n6}

Rotational equations of motion of spin of an extended body are derived in the local coordinates by differentiating the {\it bare} spin of body B given  by Eq. (\ref{spin-3})
with respect to the local coordinate time $u$. After taking the time derivative and making use of the
microscopic equations of motion in the local coordinates given in section \ref{pnem}, we perform several
transformations in the integrand to reduce similar terms, integrate the contributions from partial derivatives by parts, and simplify the final result. After long and tedious calculation we obtain the following expression for the first time derivative of the bare spin of body B in the local coordinates adapted to the body \citep{kovl_2004}
\be\label{spin-4}
\frac{d {\cal S}^i_{\rm b}}{du} = {\cal T}^i_{\rm b} +{\cal T}_{\rm c}^i - \dot{\cal S}_{\rm c}^i\,,
\ee
where ${\cal T}^i_{\rm b}$ is the {\it bare} torque exerted on the body B due to the coupling of its internal multipoles with the external tidal multipoles, and ${\cal T}_{\rm c}^i$ is a post-Newtonian correction to the bare torque caused by the difference \eqref{nrtcvc67h} between the {\it active} and {\it conformal} dipoles of body B, while $\dot{\cal S}_{\rm c}^i\equiv d{\cal S}_{\rm c}^i/du$ and ${\cal S}_{\rm c}^i$ is a linear combination of terms which can be treated as a {\it complementary} contribution to the {\it bare} spin of the body.

Gravitational {\it bare} torque, ${\cal T}^i_{\rm b}$, and the other terms in the right-hand side of (\ref{spin-4}) read as follows \citep{kovl_2004},
\ba\label{spin-5} 
{\cal T}^i_{\rm b} &=&
\Bigl[1+(2\beta-\gamma-1){\cal P}\Bigr]\sum_{l=0}^{\infty}\frac1{l!}\varepsilon_{ijk}{\cal Q}_{kL}
{\cal M}^{jL}+\sum_{l=0}^{\infty}\frac{l+1}{(l+2)l!}\varepsilon_{ijk}{\cal C}_{kL}{\cal S}^{jL}\,,
\\\nonumber\\\label{spin-6}
{\cal T}_{\rm c}^i &=&
\varepsilon_{ijk}a^j_{\rm B}\left[(1-\gamma)\biggl(\frac{3}5\dot{\cal R}^k
-\frac{1}{10}\ddot{\cal N}^k\biggl) +
\frac{\eta}2\biggl(\int\limits_{{\cal V}_{\rm B}}\rho^*\hat U_{\rm B}w^kd^3w+\sum_{l=0}^{\infty}\frac1{(2l+3)l!}{\cal Q}_{kL}{\cal N}^L\biggl)
\right.\\\nonumber&+&\left.
\sum_{l=1}^{\infty}\frac{(\gamma-1)l+2(\beta-1)}{l!}{\cal Q}_L{\cal
M}^{kL}+2(\beta-1)a^p_{\rm B}\left({\cal M}^{kp}+\frac13\delta^{kp}{\cal N}\right)\right]\,,
\\\nonumber\\\label{spin-7}
{\cal S}_{\rm c}^i &=&
-\sum_{l=1}^{\infty}\frac{l}{(l+1)!}{\cal C}_L{\cal M}^{iL}
+\sum_{l=0}^{\infty}\frac{1}{(2l+3)l!}{\cal C}_{iL}{\cal
N}^{L}
\\\nonumber&+&
\sum_{l=0}^{\infty}\frac{1}{(2l+5)l!}\varepsilon_{ijk}\left[\frac12{\cal Q}_{kL}\dot{\cal
N}^{jL}- \frac{l+2(2\gamma+3)}{2(l+2)}\dot
{\cal Q}_{kL}{\cal N}^{jL}-\frac{2(1+\gamma)(2l+3)}{l+2}{\cal Q}_{kL}{\cal R}^{jL}\right]
\\\nonumber &+&
\frac{1-\gamma}5\varepsilon_{ijk}\bigl(3{\cal R}^ja^k_{\rm B}+{\cal
N}^j\dot a^i_{\rm B}\bigr)+ (\gamma-1){\cal P}{\cal S}_{\rm b}^i\,,
\ea
where the {\it non-canonical}  multipoles, ${\cal N}^L$ and ${\cal R}^L$ have been defined earlier in \eqref{NL15} and \eqref{spin-85} respectively, and in all post-Newtonian terms the global acceleration, $a_\B^i$, is interpreted as the difference between the dipole of the scalar field and the local acceleration, $a_{\rm B}^i={\cal P}_i-{\cal Q}_i$.

The bare torque, ${\cal T}^i_{\rm b}$, is caused by gravitational coupling of the internal and external multipoles of body B, and is rooted in general relativity. The complementary torque, ${\cal T}^i_{\rm c}$, is caused by the difference between the {\it conformal} and {\it active} dipoles of the body \eqref{nrtcvc67h} and exists only in the scalar-tensor theory. Indeed, by comparison of \eqref{spin-6} with \eqref{nrtcvc67h} we can see that 
\be\la{ne6v7l}
{\cal T}_{\rm c}^i=\varepsilon_{ijk}a_\B^j\left({\cal I}^k_{\rm b}-{\cal M}^k\right)=\varepsilon_{ijk}\left({\cal P}^j-{\cal Q}^j\right)\left({\cal I}^k_{\rm b}-{\cal M}^k\right)\;,
\ee
where ${\cal I}^i_{\rm b}$ is the bare {\it conformal} dipole \eqref{confdipole}, and ${\cal M}^i$ is the {\it active} dipole of body B respectively. Equation \eqref{ne6v7l} can be further transformed to yet another form by taking into account that the total {\it conformal} dipole \eqref{b2c4z9} vanishes, ${\cal I}^i=0$, due to our choice of the center of mass \eqref{n5vz1o}. After making use of this choice and implementing \eqref{b2c4z9}, the complementary torque takes on the following form,
\be\la{ne6v7lqq}
{\cal T}_{\rm c}^i=-\varepsilon_{ijk}\left({\cal P}^j-{\cal Q}^j\right){\cal M}^k-\varepsilon_{ijk}a_\B^j{\cal I}^k_{\rm c}\;,
\ee
where the complementary vector function ${\cal I}^k_{\rm c}$ is still arbitrary. It will be fixed later by condition \eqref{m4g1x8p}.

The complementary term $\dot{\cal S}_{\rm c}$ in \eqref{spin-4} is a total time derivative which is naturally combined with the {\it bare} spin, thus, forming the total spin of body B,
\be\label{spin-9}
{\cal S}^i \equiv {\cal S}^i_{\rm b} + {\cal S}_{\rm c}^i\;.
\ee
Defining the total torque in the local coordinates of body B by
\ba\label{b3ca8k}
{\cal T}^i&\equiv&{\cal T}^i_{\rm b} + {\cal T}_{\rm c}^i\\\nonumber
&=&\varepsilon_{ijk}\bigg[{\cal P}_k{\cal M}^j+ \sum_{l=1}^{\infty}\frac1{l!}{\cal Q}_{kL}{\cal M}^{jL} 
+a_\B^k {\cal I}^j_{\rm c}+(2\beta-\gamma-1){\cal P}\sum_{l=1}^{\infty}\frac1{l!}{\cal Q}_{kL}
{\cal M}^{jL}+\sum_{l=1}^{\infty}\frac{l+1}{(l+2)l!}{\cal C}_{kL}{\cal S}^{jL}\bigg]\;,
\ea
brings about the rotational equation of motion of spin of body $\B$ to its final form,
\be\label{spin-10}
\frac{d{\cal S}^i}{du} = {\cal T}^i \;,
\ee
which includes all Newtonian and post-Newtonian corrections. Derivation of the rotational equations of motion given in this section follows the approach proposed by \citet{dsx3} in general relativity and by \citet{kovl_2004} in scalar-tensor theory of gravity. 

\section{Multipolar Expansion of External Potentials in the Global Coordinates}\la{c6a0n5}

Equations of translational motion of each body B in the global coordinates are given in \eqref{5.8} where the local acceleration ${\cal Q}_i$ should be taken from \eqref{q6v4m}--\eqref{nrvug9}. However,
the external gravitational potentials of the body -- $\bar U$, $\bar\Psi$, $\bar U^i$, $\bar\chi$ -- defined in \eqref{tb54vd}, \eqref{3d8b1a} are represented in the form of volume integrals which have not yet been explicitly performed in terms of the configuration variables defining each body of ${\mathbb N}$-body system -- the internal multipoles, coordinates of the centers of mass and their velocities.  Computation of the integrals is rather straightforward and rendered by expanding an integrand in each integral defining the external potential, in a Taylor series around the point of the center of mass of body B with subsequent integration of the coefficients of the expansion over volume of body B. The resulting expansion of the external potentials is given in terms of the internal multipole moments of the bodies which are the integrals performed in the global coordinates, $x^\a$. Additional transformation of the internal multipoles from the global to the body-adapted local coordinates is required. This section describes the details of the overall procedure of the multipolar expansion of the external potentials which are used, then, in the translational equations of motion.

We have built the local coordinates, $w^\a=(u,w^i)\equiv(u_{\rm B},w^i_{\rm B})$, adapted to body B$\in\{1,2,...,N\}$ by the matched asymptotic expansion technique. We have suppressed the sub-index B in previous sections for all functions of the local coordinates adapted to body B to simplify notations. However, computations in this section involves the bodies of ${\mathbb N}$-body system which are external with respect to body B, and we need to distinguish the local coordinates built around each body C from those adapted to body B. Therefore, we shall use a sub-index C$\in\{1,2,...,N\}$ to explicitly label the local coordinates adapted to body C along with all configuration variables associated with it.

\subsection{Multipolar Expansion of Potential \texorpdfstring{$\bar U$}{\bar U}}\la{f8n3k}

The local coordinates adapted to body C are denoted $w^\a_{\rm C}=(u_{\rm C},w^i_{\rm C})$ and the sub-index C will appear explicitly in all computations associated with the body C. Post-Newtonian coordinate transformation between $w^\a_{\rm C}$ and the global coordinates $x^\a$ is identical to equations (\ref{5.12}), (\ref{5.13}) describing the transformation from the local coordinates adapted to body B to the global coordinates except that now we have to pin the label C to all quantities related to the local coordinates adapted to body C to distinguish them from the local coordinates adapted to body B. More specifically, the transformation reads
\begin{eqnarray}
  \label{5.12cc}
 u_{\rm C} & = & t+\frac1{c^2}\bigg(\mathcal{A}_{\rm C}-v^k_{\rm C}R_{\rm C}^k\bigg)+\frac1{c^4}\Bigg\{\bigg[\frac{1}{3}v^k_{\rm C}a^k_{\rm C}-\frac{1}{6}\dot{\bar{U}}_{\rm C}(t,{\bm x}_{\rm C})-\frac{1}{10}\dot{a}^k_{\rm C}R^k_{\rm C}\bigg]R_{\rm C}^2+\sum_{l=0}^{\infty}\frac{1}{l!}\mathcal{B}_{\rm C}^LR_{\rm C}^L\Bigg\}\;,\\
\label{5.13cc}
  w_{\rm C}^i & = & R^i_{\rm C}+\frac1{c^2}\Bigg[\bigg(\frac{1}{2}v^i_{\rm C}v_{\rm C}^k+D_{\rm C}^{ik}+F^{ik}_{\rm C}\bigg)R_{\rm C}^k+D_{\rm C}^{ijk}R_{\rm C}^jR_{\rm C}^k\Bigg]\;,
  \end{eqnarray}
where $R^i_{\rm C}=x^i-x^i_{\rm C}$, $x^i_{\rm C}=x^i_{\rm C}(t)$ marks the global spatial coordinates of the origin of the local coordinates adapted to body C, $v^i_{\rm C}=dx^i_{\rm C}/dt$ is velocity of the origin of the local coordinates of body C, $a^i_{\rm C}=dv^i_{\rm C}/dt$ is acceleration of the origin of the local coordinates, and we have made use of abbreviations,
\begin{eqnarray}
\label{w1w2}
 D_{\rm C}^{ik} &\equiv & \delta^{ik}\gamma\bar{U}_{\rm C}(t,{\bm x}_{\rm C})\;,\\
 D_{\rm C}^{ijk} &\equiv & \frac{1}{2}\left(a_{\rm C}^j\delta^{ik}+a_{\rm C}^k\delta^{ij}-a_{\rm C}^i\delta^{jk}\right)\;,
\end{eqnarray}
that allows us to shorten formula \eqref{5.13cc} and is also useful in the computations which follow. 
Equations for functions like ${\cal A}_{\rm C}={\cal A}_{\rm C}(t)$, $\mathcal{B}_{\rm C}^L=\mathcal{B}_{\rm C}^L(t)$, etc., in (\ref{5.12cc}), (\ref{5.13cc}) repeat the corresponding equations for ${\cal A}$, ${\cal B}^L$, etc., in section \ref{ffemb}, after attaching the sub-index C to all  functions in \eqref{5.14}--\eqref{5.18}. Notice that the potential $\bar U_{\rm C}(t,{\bm x}_{\rm C})$ in \eqref{w1w2} denotes the Newtonian gravitational potential of all massive bodies being external to body C,
\be\la{qu3f2i}
\bar U_{\rm C}(t,{\bm x})=\sum_{\B\not={\rm C}} U_\B(t,{\bm x})\;.
\ee
We emphasize that the instant of time $t$ that appears in \eqref{5.12cc} and which is also a time argument of all functions and functionals of body C in the global coordinate chart is the same as the instant of time $t$ for functions and functionals of body B. This is because we consider dynamics of the entire ${\mathbb N}$-body system as a continuous past-to-future diffeomorphism of spatial coordinates of the bodies taken on a hypersurface of simultaneity ${\cal H}_t$ which points have the same value of a single parameter -- time $t$. 

The multipolar expansions of the external gravitational potentials $\bar U$, $\bar{U}^i$, $\bar\Psi$, $\bar\chi$ of body B defined in \eqref{tb54vd}, \eqref{3d8b1a} are represented in the form of the multipolar expansions from a linear superposition of potentials $U_{\rm C}(t,{\bm x})$, ${U}^i_{\rm C}(t,{\bm x})$, $\Psi_{\rm C}(t,{\bm x})$, and $\chi_{\rm C}(t,{\bm x})$ correspondingly. Therefore, we focus on the multipolar expansions of the individual potentials. 

\begin{figure}
\hspace*{-2cm}
\includegraphics[scale=0.65]{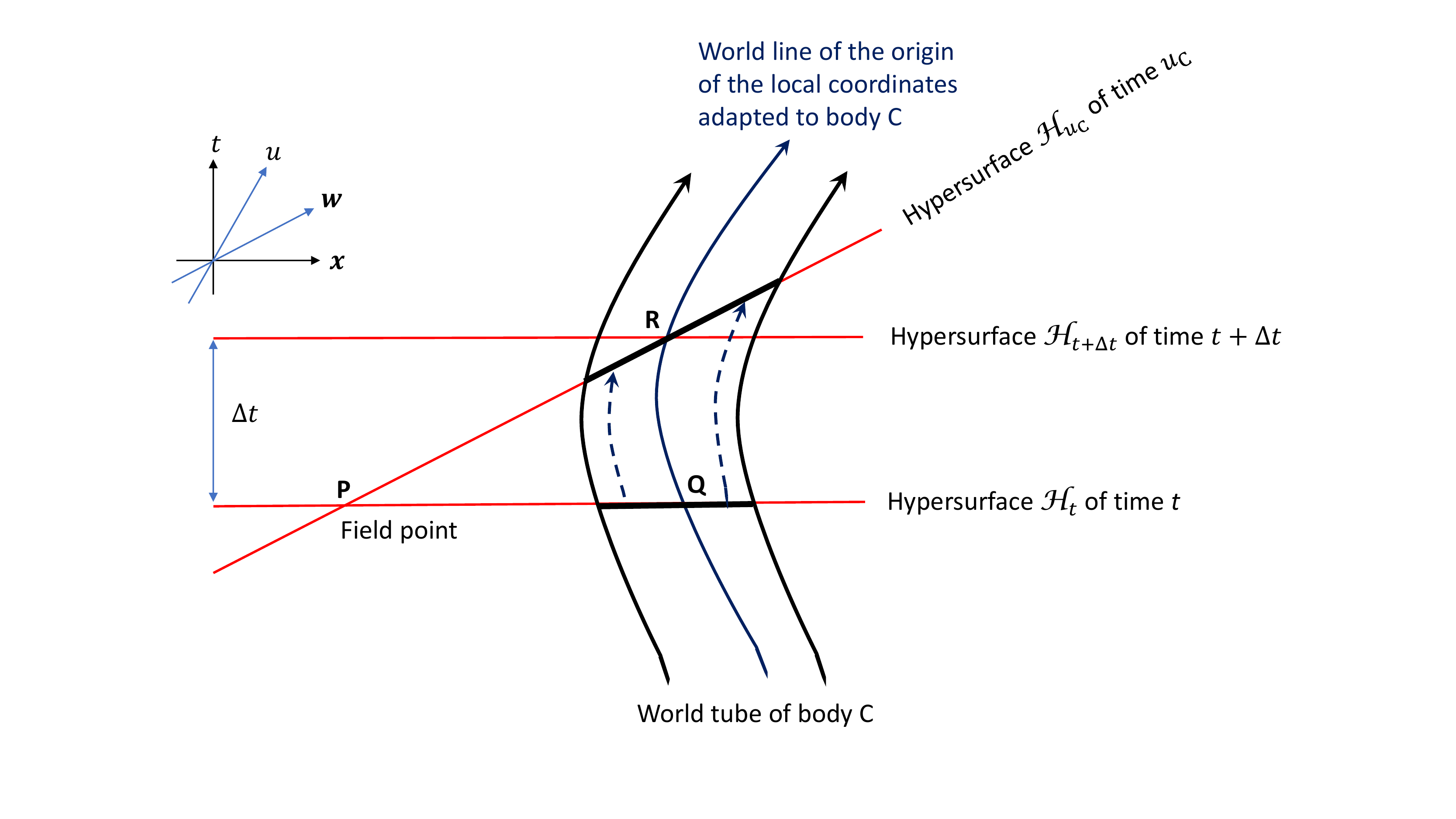}
\caption{Figure shows a world tube of matter of body C intersected by hypersurfaces of simultaneity in the global and local coordinates adapted to body C. Integration in the global coordinates goes over the hypersurface ${\cal H}_t$ of constant time $t$ passing through points P and Q. Integration in the local coordinates goes over the hypersurface ${\cal H}_{u_{\rm C}}$ of constant time $u_{\rm C}$ passing through points P and R. The two hypersurfaces intersect at the field point P having global coordinates $x^\a_P=\{t,{\bm x}\}$ and local coordinates $w^\a_P=\{u_{\rm C},{\bm w}_{\rm C}\}$. The points Q and R are lying on the world line ${\cal W}$ of the origin of the local coordinates adapted to body C. Lie transport of the elements of integration from ${\cal H}_t$ to ${\cal H}_{u_{\rm C}}$ is shown by dotted lines and carried out along world lines of matter particles forming the element of integration. Hypersurface ${\cal H}_{t+\Delta t}$ of constant time $t+\Delta t$ is passing through point R. Points Q and R have global coordinates $x^\a_Q=\{t,{\bm x}_{\rm C}(t)\}$ and 
$x^\a_R=\{t+\Delta t,{\bm x}_{\rm C}(t+\Delta t)\}$, respectively. Local coordinates of point R are $w^\a_R=\{u_{\rm C},0\}$. 
Time shift $\Delta t$ between hypersurfaces ${\cal H}_t$ and ${\cal H}_{t+\Delta t}$ is determined by the time transformation \eqref{5.12cc} applied to coordinates of two points, P and R which have the same value of the local time $u_{\rm C}$. It is given by $\Delta t= v^k_{\rm C} R^k_{\rm C}$.\label{fig1}}
\end{figure}

Potentials $U_{\rm C}(t,{\bm x})$, ${U}^i_{\rm C}(t,{\bm x})$, $\Psi_{\rm C}(t,{\bm x})$ are given in the global coordinates as integrals \eqref{12.10rtfz}, \eqref{12.11}--\eqref{12.16}  with a kernel, $|{\bm x}-{\bm x}'|^{-1}$, which is a Green function of the Laplace equation. This kernel\index{Laplace equation!kernel} is expanded into multipolar series as follows
\begin{eqnarray}
  \label{w1w5}
  \frac{1}{|\bm{x}-\bm{x}'|} & = & \frac{1}{|R_{\rm C}-R'_{\rm C}|}
   = \sum_{l=0}^{\infty}\frac{(-1)^l}{l!}R_{\rm C}'^{<L>}\pd_L\bigg(\frac{1}{R_{\rm C}}\bigg)\;,
\end{eqnarray}
where $R'^i_{\rm C}=x'^i-x^i_{\rm C}$ is the coordinate distance from the origin of the local coordinates $x^i_{\rm C}$ adapted to body C, $R^i_{\rm C}=x^i-x^i_{\rm C}$ is the coordinate distance from $x^i_{\rm C}$ to the field point, $R_C=\left(\d_{ij}R^i_{\rm C} R^j_{\rm C}\right)^{1/2}$, $\partial_L\equiv\partial_{i_1...i_l}$ denotes a partial derivative of $l$-th order with respect to spatial global coordinates where each $\partial_i=\partial/\partial x^i$, and the angular parentheses around indices indicate the STF projection, and the point $x'^i$ lies inside volume with radius $R'_C<R_C$ so that the series \eqref{w1w5} is convergent. Equation \eqref{w1w5} yields the multipolar expansion of the Newtonian potential of body C in the global coordinates as follows,
\be\la{mb4g0m}
U_{\rm C}(t,{\bm x})=\int\limits_{{\cal V}_{\rm C}}\frac{\rho^*(t,\bm{x}')}{|\bm{x}-\bm{x}'|}d^3x'=
 \sum_{l=0}^{\infty}\frac{(-1)^l}{l!}{\mathbb I}^{<L>}_{\rm C}\pd_L\bigg(\frac{1}{R_{\rm C}}\bigg)\;,
 \ee
where  
\be\label{mumo28}
{\mathbb I}^{L}_{\rm C}\equiv{\mathbb I}^{L}_{\rm C}(t)=\displaystyle\int\limits_{{\cal V}_{\rm C}}\rho^*(t,\bm{x}')R_{\rm C}'^{i_1}R_{\rm C}'^{i_2}...R_{\rm C}'^{i_l}d^3x'\;,
\ee
are the Newtonian mass moments computed in the global coordinates. We preserve the prime in the notation of the spatial coordinates $R'^i_{\rm C}=x'^i-x^i_{\rm C}$ that appear in the integrand of \eqref{mumo28} to prevent confusion of the point of integration $x'^i$ with the field point $x^i$.  
Symmetric multipoles ${\mathbb I}^{L}_{\rm C}$ have to be transformed from the global to local coordinates adapted to body C in order to express them in terms of the internal STF mass and spin multipoles defined in section \ref{mdloc}. The transformation procedure is somehow subtle and should be done with care as it involves not only a pointwise transformation of coordinates but a Lie transport of the integration points along worldlines of matter of body C \citep{bk-nc,kovl_2004} -- see Figure \ref{fig1}. 

It starts from the post-Newtonian transformation of radius-vector $R^i_{\rm C}=x^i-x^i_{\rm C}$ from the global to local coordinates $w_{\rm C}^i$ adapted to body C. This is achieved by applying the inverse coordinate transformation of \eqref{5.13cc}:
\be\la{ne6v12k}
 R_{\rm C}^i  =  w_{\rm C}^i-\frac1{c^2}\Bigg[\bigg(\frac{1}{2}v^i_{\rm C}v_{\rm C}^k+D_{\rm C}^{ik}+F_{\rm C}^{ik}\bigg)w_{\rm C}^k+D_{\rm C}^{ijk}w_{\rm C}^jw_{\rm C}^k\Bigg]\;.
 \ee
However, we actually need a post-Newtonian transformation not $R_{\rm C}^i$ but a radius-vector $R'^i_{\rm C}=x'^i-x^i_{\rm C}$ from the global to the local coordinates because it is $R'^i_{\rm C}$ which appears in the definition of ${\mathbb I}^{L}_{\rm C}$ in \eqref{mumo28} as a consequence of the Taylor expansion \eqref{mb4g0m}. This transformation is slightly different from (\ref{ne6v12k}) because in all integrals performed in the global coordinates the points $x^i$ and $x'^i$ are lying on hypersurface ${\cal H}_t$ of constant global coordinate time $t$, while the points $w^i_{\rm C}$ and $w'^i_{\rm C}$ are lying on hypersurfaces ${\cal H}_{u_{\rm C}}$ of constant local coordinate time $u_{\rm C}$ in all integrals defining the internal part of the metric tensor perturbation of body C. Hypersurface ${\cal H}_t$ differs from that ${\cal H}_{u_{\rm C}}$. Therefore, transformation of ${\mathbb I}^{<L>}_{\rm C}$ from the global to local coordinates must include not only the transformations between the coordinate points but also a Lie transport of the integration point with coordinates $x'^i$ from  hypersurface ${\cal H}_t$ to hypersurface ${\cal H}_{u_{\rm C}}$ performed along the time-like worldlines of matter of body C. The magnitude of the Lie transport of each point of integration depends on the size of spatial separation of the integration point $x'^i$ from the origin of the local coordinates adapted to body C, and is determined from the equation of time transformation \eqref{5.12cc}, and a condition that all points on the hypersurface ${\cal H}_{u_{\rm C}}$ have the same value of the local coordinate time $u_{\rm C}$ as the field point P in Fig. \ref{fig1}. The Lie transport\index{Lie transport} of the corresponding element of matter with coordinates  $x'^i$ is accompanied by the point-wise post-Newtonian transformation \eqref{5.12cc} applied to $x'^i$ and the resulting transform was worked out by \citet{bk-nc,kopejkin_1991INTSA} and its comprehensive explanation is given in full detail in our textbook \citep[Sections 5.2.3.1 and 6.3.2]{kopeikin_2011book}. It yields for the post-Newtonian Lie transform of the spatial coordinate $w'^i_{\rm C}$ the following result \citep[equation 6.56]{kopeikin_2011book},
\be
  \label{w1w3}
  R_{\rm C}'^i  =  w_{\rm C}'^i-\frac1{c^2}\Bigg[\bigg(\frac{1}{2}v^i_{\rm C}v_{\rm C}^k+D_{\rm C}^{ik}+F_{\rm C}^{ik}\bigg)w_{\rm C}'^k+D_{\rm C}^{ijk}w_{\rm C}'^jw_{\rm C}'^k+\nu_{\rm C}'^iv_{\rm C}^k\left(w_{\rm C}'^k-w_{\rm C}^k\right)\Bigg]\;,
\ee
where $\nu_{\rm C}'^i={v'}^i-v_{\rm C}^i$ is the relative velocity of matter of body C located at point $x'^i$ with respect to the origin of the local coordinates of the body, ${v'}^i=dx'^i/dt$, ${v}^i_{\rm C}=dx^i_{\rm C}/dt$. The difference between transformations \eqref{ne6v12k} and \eqref{w1w3} is in the presence of the very last term in \eqref{w1w3} which is due to the Lie transport of an element of integration from the hypersurface ${\cal H}_t$ of constant time $t$ to that ${\cal H}_{u_{\rm C}}$ of $u_{\rm C}$ along worldlines of matter some of which are shown by dotted lines in Fig. \ref{fig1}. This term brings about a seemingly different appearance of our translational equations of motion for the center of mass of each body as compared with translational equations of motion derived by \citet{racine_2005PhRvD} with corrections outlined in \citep{racine2013PhRvD}. This is a matter of choice of the hypersurface of integration ${\cal H}_{u_{\rm C}}$ in the local coordinates adapted to the body under consideration. We reconcile this issue in Appendix \ref{appendixA} -- see discussion following equation \eqref{v39b2x3}.  

Equation (\ref{w1w3}) allows us to transform ${\mathbb I}^{<L>}_{\rm C}\equiv {\mathbb I}^{<L>}_{\rm C}(t)$ from the global to local coordinates as follows \citep[equation 6.60]{kopeikin_2011book},
\begin{eqnarray}
  \label{glomult1}
  {\mathbb I}^{<L>}_{\rm C}& = &\mathfrak{I}_{\rm C}^{<L>}-\frac{l}{2}v_{\rm C}^kv_{\rm C}^{<i_l}\mathfrak{I}_{\rm C}^{L-1>k}+lF_{\rm C}^{k<i_l}\mathfrak{I}_{\rm C}^{L-1>k}
  - lD_{\rm C}^{k<i_l}\mathfrak{I}_{\rm C}^{L-1>k}
  \\
  &&
  -l\mathfrak{I}_{\rm C}^{jk<L-1}D_{\rm C}^{i_l>jk}-v_{\rm C}^k\dot{\mathfrak{I}}_{\rm C}^{k<L>}+v_{\rm C}^kR_{\rm C}^k\dot{\mathfrak{I}}_{\rm C}^{<L>}
   + v_{\rm C}^k\int\limits_{{\cal V}_{\rm C}} {\rho_{\rm C}^*}'{\nu}_{\rm C}^{\prime k}w_{\rm C}'^{<L>}d^3w_{\rm C}'\;,\nonumber
\end{eqnarray}
where a shorthand notation, ${\rho_{\rm C}^*}'\equiv \rho^*_{\rm C}(u_C,{\bm w}'_{\rm C})$, stands for the invariant density of matter at the integration point ${\bm w}'_{\rm C}$ in the local coordinates, the moments
\be\la{nev6x4}
\mathfrak{I}_{\rm C}^L\equiv \mathfrak{I}_{\rm C}(u_{\rm C})= \int\limits_{{\cal V}_{\rm C}} {\rho_{\rm C}^*}'w_{\rm C}'^{i_1}w_{\rm C}'^{i_2}\ldots w_{\rm C}'^{i_l}d^3w_{\rm C}'\;,
\ee
are {\it symmetric} moments of body C depending on the local time $u_{\rm C}$, and we have made use of the fact that the product of the mass density $\rho^*$ with 3-dimensional coordinate volume is Lie-invariant when transported from hypersurface ${\cal H}_t$ to hypersurface ${\cal H}_{u_{\rm C}}$ along worldlines of matter, that is $\rho^*_{\rm C}(t,{\bm x}')d^3x'=\rho^*_{\rm C}(u_{\rm C},{\bm w}'_{\rm C})d^3w'_{\rm C}$ \citep{kopeikin_2011book}. Notice that formula \eqref{glomult1} is not a pointwise transformation of the moments performed at the origin of the local coordinates adapted to body C because of the presence of the last but one term, $v_{\rm C}^kR_{\rm C}^k\dot{\mathfrak{I}}_{\rm C}^{<L>}$ which depends on the coordinate distance $R^k_{\rm C}$ from the origin of the local coordinates to the field point (point P in Fig. \ref{fig1}). At first glance, the appearance of this term may look strange as by definition \eqref{mumo28} the moments ${\mathbb I}^{<L>}_{\rm C}$ are solely functions of time $t$ alone. The reader should keep in mind that the moments $\mathfrak{I}_{\rm C}^L$ are functions of the local time $u_{\rm C}$ and, though both ${\mathbb I}^{<L>}_{\rm C}$ and $\mathfrak{I}_{\rm C}^L$ are functions pinned down to the origin of the local coordinates adapted to body C, they are taken at different points on the worldline ${\cal W}$ of the origin because the field point $(t,x^i)$ is considered as being fixed in the derivation of the transformation \eqref{glomult1}. Therefore, the transformation of the time arguments of the moments involves the time shift $\Delta t= v_{\rm C}^kR_{\rm C}^k$ of the moments along the worldline ${\cal W}$, which explains the origin of term $v_{\rm C}^kR_{\rm C}^k\dot{\mathfrak{I}}_{\rm C}^{k<L>}$ in \eqref{glomult1}. It is worth noticing that the term $v_{\rm C}^kR_{\rm C}^k\dot{\mathfrak{I}}_{\rm C}^{<L>}$ is not present in the transformation equations for multipole moments derived by \citet{racine_2005PhRvD} as they have computed the multipoles of each body C at the value of the local time $u_{\rm C}$ taken at the center of mass of body C which is different from our convention. This leads to the translational equations of motion which look different from ours by several terms. This apparent difference is not an indicator of mistake but, as we show in Appendix \ref{appendixA}, a matter of computational approach and conventions.

It should be emphasized that the moments $\mathfrak{I}_{\rm C}^L$ are \underline{\it not} the STF Cartesian tensors. Their STF projection is denoted as $\mathfrak{I}_{\rm C}^{<L>}$ and, in general, $\mathfrak{I}^L_{\rm C}\not=\mathfrak{I}^{<L>}_{\rm C}$. It  means that after contraction of any two indices in \eqref{nev6x4} we get the trace $\mathfrak{I}_{\rm C}^{kkL-2}\not=0$, and must be taken into account in subsequent calculations. The STF part of the Newtonian-like moments \eqref{nev6x4} is related to the STF post-Newtonian internal mass multipoles, ${\cal M}^L_{\rm C}\equiv{\cal M}^L_{\rm C}(u_{\rm C})$, of body C as follows \citep{xie_2010AcPSl},
\ba\la{bv34cw}
\mathfrak{I}_{\rm C}^{<L>}&=&{\cal M}_{\rm C}^{L}\bigg[1+(2\beta-\gamma-1){\cal P}_{\rm C}\bigg]
-\int\limits_{{\cal V}_{\rm C}} {\rho_{\rm C}^*}'\bigg[(\gamma+\frac12)\nu_{\rm C}'^2+\Pi_{\rm C}'+\gamma\frac{{\mathfrak{s}}_{\rm C}^{'kk}}{{\rho_{\rm C}^*}'}-(2\beta-1)\hat U_{\rm C}' \bigg]w_{\rm C}'^{<L>}d^3w_{\rm C}'\\
  &-&\frac{1}{2(2l+3)}\bigg[\ddot{\mathcal{N}}_{\rm C}^{<L>}-4(1+\gamma)\,\frac{2l+1}{l+1}\dot{\mathcal{R}}_{\rm C}^{<L>}\bigg]
   +\sum_{k=1}^{\infty}\frac{1}{k!}\bigg[{\cal Q}_{\rm C}^{K}+2(\beta-1) {\cal P}_{\rm C}^{K}\bigg]\int\limits_{{\cal V}_{\rm C}}{\rho_{\rm C}^*}'w_{\rm C}'^{<K>}w_{\rm C}'^{<L>}d^3w_{\rm C}'\nonumber\;, 
\ea
where a prime standing after function (like $\Pi_{\rm C}'$, etc.) in the integrand means that the function is taken at the point $w'^i_{\rm C}$, an overdot denotes a total time derivative with respect to the coordinate time $u_{\rm C}$ of the local coordinates adapted to body C,   
\be
{\cal P}^K_{\rm C}\equiv \sum_{\B\not={\rm C}}\pd_K U_\B(t,{\bm x}_{\rm C})\;,\qquad\qquad (k\ge 0)
\ee
are monopole and higher-order external multipoles of the scalar field generated by all bodies being external to body ${\rm C}$,
\be
{\cal Q}^K_{\rm C}\equiv \sum_{\B\not={\rm C}}\pd_K U_\B(t,{\bm x}_{\rm C})\qquad\qquad\qquad (k\ge 2)\;,
\ee
are higher-order gravitoelectric external multipoles of body C, and the local acceleration ${\cal Q}^i_{\rm C}$ is defined in \eqref{q6v4m} and must be referred to body C, the {\it non-canonical}  multipoles $ {\mathcal{N}}_{\rm C}^{L}$ and ${\mathcal{R}}_{\rm C}^{L}$ are defined by equations similar to \eqref{NL15}, \eqref{spin-85} where the integrals must be taken over a volume of body C,
\ba
  \label{w1w6}
  \mathcal{N}_{\rm C}^L &\equiv& \int_{{\cal V}_{\rm C}}\rho_{\rm C}^{\ast}(u_C,{\bm w}_{\rm C})w_{\rm C}^2w_{\rm C}^{<L>}d^3w_{\rm C}\;,
\\
  \label{w1w7}
  \mathcal{R}_{\rm C}^L &\equiv& \int_{{\cal V}_{\rm C}}\rho_{\rm C}^{\ast}(u_C,{\bm w}_{\rm C})\nu_{\rm C}^kw_{\rm C}^{<kL>}d^3w_{\rm C}\;.
\ea

Now, we replace expression \eqref{glomult1} for ${\mathbb I}^{<L>}_{\rm C}$  to multipolar expansion \eqref{mb4g0m} of the Newtonian potential of body C and use \eqref{bv34cw}. It results in,
\begin{eqnarray}
  \label{UC1}
  U_{\rm C}(t,\bm{x}) 
  & = & \sum_{l=0}^{\infty}\frac{(-1)^l}{l!}{\cal M}_{\rm C}^{L}\pd_L\bigg(\frac{1}{R_{\rm C}}\bigg)\bigg[1+(2\beta-\gamma-1){\cal P}_{\rm C}\bigg]\\
  &-& \sum_{l=0}^{\infty}\frac{(-1)^l}{l!}\pd_L\bigg(\frac{1}{R_{\rm C}}\bigg)\Bigg\{\int\limits_{{\cal V}_{\rm C}} {\rho_{\rm C}^*}'\bigg[(\gamma+\frac12)\nu_{\rm C}'^2+\Pi_{\rm C}'+\gamma\frac{{\mathfrak{s}}_{\rm C}^{'kk}}{{\rho_{\rm C}^*}'}-(2\beta-1)\hat U_{\rm C}' \bigg]w_{\rm C}'^{<L>}d^3w_{\rm C}'\nonumber\\
  &&+\frac{1}{2(2l+3)}\bigg[\ddot{\mathcal{N}}_{\rm C}^{<L>}-4(1+\gamma)\,\frac{2l+1}{l+1}\dot{\mathcal{R}}_{\rm C}^{<L>}\bigg]\nonumber\\
   &&-\sum_{k=1}^{\infty}\frac{1}{k!}\bigg[{\cal Q}_{\rm C}^{K}+2(\beta-1) {\cal P}_{\rm C}^{K}\bigg]\int\limits_{{\cal V}_{\rm C}}{\rho_{\rm C}^*}'w_{\rm C}'^{<K>}w_{\rm C}'^{<L>}d^3w_{\rm C}'\nonumber\\  
     &&+\frac{l}{2}v_{\rm C}^kv_{\rm C}^{<i_l}\mathfrak{I}_{\rm C}^{L-1>k}-lF_{\rm C}^{k<i_l}\mathfrak{I}_{\rm C}^{L-1>k}+lD_{\rm C}^{k<i_l}\mathfrak{I}_{\rm C}^{L-1>k}+l\mathfrak{I}_{\rm C}^{jk<L-1}D_{\rm C}^{i_l>jk}\nonumber\\
     && +v_{\rm C}^k\dot{\mathfrak{I}}_{\rm C}^{k<L>}-v_{\rm C}^kR_{\rm C}^k\dot{\mathfrak{I}}_{\rm C}^{<L>}- v_{\rm C}^k\int\limits_{{\cal V}_{\rm C}} {\rho_{\rm C}^*}'{\nu}_{\rm C}^{\,\prime k}w_{\rm C}'^{<L>}d^3w_{\rm C}'\Bigg\}\;.\nonumber
\end{eqnarray}
Neither the multipoles $\mathfrak{I}^L_{\rm C}$ nor the very last  integral in \eqref{glomult1} are the STF Cartesian tensors. Therefore, equation \eqref{UC1} must be further transformed to bring it to the form depending on the STF internal mass and spin multipoles, ${\cal M}^L_{\rm C}$ and ${\cal S}^L_{\rm C}$. This is achieved by making use of the following equations, 
\ba\la{nmb1}
v_{\rm C}^kv_{\rm C}^{<i_l}\mathfrak{I}_{\rm C}^{L-1>k}&=&v_{\rm C}^kv_{\rm C}^{<i_l}{\cal M}_{\rm C}^{L-1>k}+\frac{l-1}{2l-1}v_{\rm C}^{<i_1}v_{\rm C}^{i_2}\mathcal{N}_{\rm C}^{L-2>}\;, 
\\\la{nmb2}
D_{\rm C}^{k<i_l}\mathfrak{I}_{\rm C}^{L-1>k}&=& \gamma\bar U_{\rm C}(t,{\bm x}_{\rm C}){\cal M}_{\rm C}^L\;, 
\\\la{nmb3}
\mathfrak{I}_{\rm C}^{jk<L-1}D_{\rm C}^{i_l>jk}&=&a_{\rm C}^j{\cal M}_{\rm C}^{jL}-\frac{1}{2(2l+1)}a_{\rm C}^{<i_l}\mathcal{N}_{\rm C}^{L-1>}  \;,
\\\la{nmb4}
v_{\rm C}^k\dot{\mathfrak{I}}_{\rm C}^{k<L>}&=&v_{\rm C}^k\dot{{\cal M}}_{\rm C}^{kL} +\frac{l}{2l+1}v_{\rm C}^{<i_l}\dot{\mathcal{N}}_{\rm C}^{L-1>}  
\\\la{nmb5}
\int\limits_{{\cal V}_{\rm C}} {\rho_{\rm C}^*}'{\nu}_{\rm C}^{\,\prime k}w_{\rm C}'^{<L>}d^3w_{\rm C}'&=&\frac1{l+1}\dot{{\cal M}}_{\rm C}^{kL}+ \frac{l}{l+1}\varepsilon^{kp<i_l}{\mathcal{S}}_{\rm C}^{L-1>p}+\frac{2l-1}{2l+1}\delta^{k<i_l}\mathcal{R}_{\rm C}^{L-1>}\;,
\ea
where the overdot denotes a total time derivative with respect to coordinate time $u_{\rm C}$ of the local coordinates adapted to body C, and we have used everywhere in the post-Newtonian terms ${\mathfrak I}^{<L>}_{\rm C}={\cal M}^L_{\rm C}$ which is valid in the approximation under consideration.
Substituting \eqref{nmb1}--\eqref{nmb5} to equation \eqref{UC1} yields a multipolar post-Newtonian expansion of the Newtonian potential of body C given in terms of the internal {\it active} mass and spin multipoles of the body,  
\be\la{ustra1}
U_{\rm C}(t,\bm{x})=W_{\rm C}(t,{\bm x})+\Phi_{\rm C}(t,{\bm x})\;,
\ee
where
  \ba\label{je5v20}
  {W}_{\rm C}(t,\bm{x})
  & = & \sum_{l=0}^{\infty}\frac{(-1)^l}{l!}\pd_L\bigg(\frac{1}{R_{\rm C}}\bigg){\cal M}_{\rm C}^L\;,\\
  \la{hdv45x5}
 \Phi_{\rm C}(t,{\bm x}) &=&
 (2\beta-\gamma-1){\cal P}_{\rm C}\sum_{l=0}^{\infty}\frac{(-1)^l}{l!}\pd_L\bigg(\frac{1}{R_{\rm C}}\bigg){\cal M}_{\rm C}^L\\
 & -&\sum_{l=0}^{\infty}\frac{(-1)^l}{l!}\pd_L\bigg(\frac{1}{R_{\rm C}}\bigg)\Bigg\{\int\limits_{{\cal V}_{\rm C}} {\rho_{\rm C}^*}'\bigg[(\gamma+\frac12)\nu_{\rm C}'^2+\Pi_{\rm C}'+\gamma\frac{{\mathfrak{s}}_{\rm C}^{'kk}}{{\rho_{\rm C}^*}'}-(2\beta-1)\hat U_{\rm C}' \bigg]w_{\rm C}'^{<L>}d^3w_{\rm C}'\nonumber\\
  &&+\frac{1}{2(2l+3)}\bigg[\ddot{\mathcal{N}}_{\rm C}^L-4(1+\gamma)\,\frac{2l+1}{l+1}\dot{\mathcal{R}}_{\rm C}^L\bigg]\nonumber\\
   &&-\sum_{n=1}^{\infty}\frac{1}{n!}\bigg[{\cal Q}_{\rm C}^{N}+2(\beta-1) {\cal P}_{\rm C}^{N}\bigg]\int_{C}{\rho_{\rm C}^*}'w_{\rm C}'^{<N>}w_{\rm C}'^{<L>}d^3w_{\rm C}'\nonumber\\  
     &&+\frac{l}2 v_{\rm C}^kv_{\rm C}^{<i_l}{\cal M}_{\rm C}^{L-1>k}- lF_{\rm C}^{k<i_l}{\cal M}_{\rm C}^{L-1>k} +l\gamma\bar U(t,{\bm x}_{\rm C}){\cal M}_{\rm C}^L
     +la_{\rm C}^k{\cal M}_{\rm C}^{kL}+v_{\rm C}^k\dot{{\cal M}}_{\rm C}^{kL}-v_{\rm C}^kR_{\rm C}^k\dot{{\cal M}}^L\Bigg\}
     \nonumber\\
  &&+\sum_{l=0}^{\infty}\frac{(-1)^l}{(l+1)!}\pd_L\bigg(\frac{1}{R_{\rm C}}\bigg)v_{\rm C}^k\dot{{\cal M}}_{\rm C}^{kL}+\sum_{l=1}^{\infty}\frac{(-1)^ll}{(l+1)!}\varepsilon_{kpq}v_{\rm C}^k\pd_{qL-1}\bigg(\frac{1}{R_{\rm C}}\bigg)\mathcal{S}_{\rm C}^{pL-1}\nonumber\\
  &&+\sum_{l=1}^{\infty}\frac{(-1)^l(2l-1)}{(2l+1)l!}v_{\rm C}^k\pd_{kL-1}\bigg(\frac{1}{R_{\rm C}}\bigg)\mathcal{R}_{\rm C}^{L-1}- \frac12\sum_{l=0}^{\infty}\frac{(-1)^l}{(2l+3)l!}\pd_{pqL}\bigg(\frac{1}{R_{\rm C}}\bigg)v_{\rm C}^{<p}v_{\rm C}^q\mathcal{N}_{\rm C}^{L>} \nonumber\\
 &&-\frac12\sum_{l=0}^{\infty}\frac{(-1)^l}{(2l+3)l!}\pd_{pL}\bigg(\frac{1}{R_{\rm C}}\bigg)a_{\rm C}^{<p}\mathcal{N}_{\rm C}^{L>}+\sum_{l=0}^{\infty}\frac{(-1)^l}{(2l+3)l!}\pd_{pL}\bigg(\frac{1}{R_{\rm C}}\bigg)v_{\rm C}^{<p}\dot{\mathcal{N}}_{\rm C}^{L>}\;, \nonumber
 \end{eqnarray}
The reader can notice that \eqref{hdv45x5} includes explicitly a number of integrals depending on the intrinsic physical quantities of body C such as the internal velocity of matter $\nu^i_{\rm C}$, potential energy $\Pi_{\rm C}$, the stress tensor ${\mathfrak{s}}_{\rm C}^{ij}$, and self-gravity potential $\hat U_{\rm C}$, as well as the {\it non-canonical}   multipoles, $\mathcal{N}_{\rm C}^L$ and $\mathcal{R}_{\rm C}^L$. The appearance of such terms is not expected in the final equations of motion if the principle of effacing of the internal structure is valid. Indeed, subsequent calculations demonstrate that the multipolar expansions of other gravitational potentials also contain similar terms depending on the internal structure of body C which are mutually canceled out in the final form of the post-Newtonian equations of motion.

Multipolar expansion of the Newtonian potential was rather cumbersome because we had to take into account the post-Newtonian corrections to the definitions of the internal multipoles ${\cal M}^L$ and to implement the post-Newtonian transformation from the global to local coordinates. Multipolar expansions of other external potentials are less laborious as they show up only in the post-Newtonian terms in definition of the external gravitoelectric multipoles ${\cal Q}_L$. Thus, their multipolar expansions can be performed by operating merely with the Newtonian part of the coordinate transformations and taking the leading (Newtonian-order) terms in the definition of the {\it active} internal multipoles ${\cal M}^L$.

\subsection{Multipolar Expansion of Potential \texorpdfstring{$\bar U^i$}{\bar U^i}}\la{ui888}

The external vector potential $U_{\rm C}^i$ is defined in the global coordinates by equation \eqref{12.11} and depends on the velocity of matter $v^i$ of the body C taken with respect to the origin of the global coordinates. This velocity is a linear sum of two pieces, 
\be\la{n4448}
v^i=v^i_{\rm C}+\nu^i_{\rm C}\;,
\ee 
where $v^i_{\rm C}=dx^i_{\rm C}/dt$ is velocity of the origin of local coordinates adapted to body C with respect to the global coordinates, and $\nu^i_{\rm C}$ is velocity of matter of body C with respect to the origin of the local coordinates. After accounting for the linear decomposition of the velocity, the vector potential $U_{\rm C}^i$ is expanded in terms of the internal multipoles as follows,
\begin{eqnarray}
  \label{w1w8}
  U_{\rm C}^i(t,\bm{x})  =& & \int\limits_{{\cal V}_{\rm C}}\frac{\rho^*(t,\bm{x}')v'^i}{|\bm{x}-\bm{x}'|}d^3x'\\
   =& & \sum_{l=0}^{\infty}\frac{(-1)^l}{l!}\pd_L\bigg(\frac{1}{R_{\rm C}}\bigg){\cal M}_{\rm C}^{L}v_{\rm C}^i +\sum_{l=1}^{\infty}\frac{(-1)^l}{(l+1)!}\pd_L\bigg(\frac{1}{R_{\rm C}}\bigg)\dot{{\cal M}}_{\rm C}^{iL}\nonumber\\
  &+&\sum_{l=1}^{\infty}\frac{(-1)^ll}{(l+1)!}\varepsilon_{ipq}\pd_{qL-1}\bigg(\frac{1}{R_{\rm C}}\bigg)\mathcal{S}_{\rm C}^{pL-1}
 +\sum_{l=1}^{\infty}\frac{(-1)^l}{l!}\frac{2l-1}{2l+1}\pd_{iL-1}\bigg(\frac{1}{R_{\rm C}}\bigg)\mathcal{R}_{\rm C}^{L-1}\;,\nonumber
\end{eqnarray}
where ${\cal M}^L_{\rm C}$ and ${\cal S}^L_{\rm C}$  are the {\it canonical}  internal mass and spin multipoles of body C defined in \eqref{1.31} and \eqref{1.32} respectively, and ${\cal R}^L$ are the {\it non-canonical}  multipoles of body C defined in \eqref{spin-85}. 
\subsection{Multipolar Expansion of Potential \texorpdfstring{$\bar\Psi$}{\bar\Psi}}
Multipolar expansion of the external potential $\bar\Psi$ entering definition of the external tidal potential ${\cal Q}_L$ for body B, is a sum of gravitational potentials of the bodies being external with respect to body B,
\be
\bar\Psi(t,{\bm x})=\sum_{{\rm C}\not={\rm B}}\Psi_{\rm C}(t,{\bm x})\;,
\ee
where
\be\la{ctb5j}
\Psi_{\rm C}(t,{\bm x})\equiv \left(\gamma+\frac{1}{2}\right)\Psi_{{\rm C}1}(t,\bm{x})+(1-2\beta)\Psi_{{\rm C}2}(t,\bm{x})+\Psi_{{\rm C}3}(t,\bm{x})+\gamma\Psi_{{\rm C}4}(t,\bm{x})\;,
\ee
is a linear superposition of potentials $\Psi_{{\rm C}1}$, $\Psi_{{\rm C}2}$, $\Psi_{{\rm C}3}$, $\Psi_{{\rm C}4}$ defined in \eqref{12.13}--\eqref{12.16} respectively.

Potential $\Psi_{{\rm C}1}$ is a quadratic functional of matter's velocity with respect to the global coordinates. The square of the velocity is split in three peaces in accordance with decomposition \eqref{n4448},
\be\la{045v}
v^2=v^2_{\rm C}+2v^k_{\rm C}\nu^k_{\rm C}+\nu^2_{\rm C}\;.
\ee
Replacing $v^2$ with the right-hand side of \eqref{045v} in \eqref{12.13}, and performing multipolar decomposition of each integral with the help of \eqref{w1w5}, we obtain,
\begin{eqnarray} 
  \label{w1w9}
\Psi_{{\rm C}1}(t,\bm{x}) & = & \int\limits_{{\cal V}_{\rm C}}\frac{\rho^*(t,\bm{x}')v'^2}{|\bm{x}-\bm{x}'|}d^3x'\\
  & = & \sum_{l=0}^{\infty}\frac{(-1)^l}{l!}\pd_L\bigg(\frac{1}{R_{\rm C}}\bigg)\bigg({\cal M}_{\rm C}^{L}v_{\rm C}^2+\int\limits_{{\cal V}_{\rm C}}{\rho_{\rm C}^*}'{\nu}_{\rm C}^{\,\prime 2}w_{\rm C}'^{<L>}d^3w_{\rm C}'\bigg)\nonumber\\
  &+&2\sum_{l=1}^{\infty}\frac{(-1)^l}{(l+1)!}\pd_L\bigg(\frac{1}{R_{\rm C}}\bigg)v_{\rm C}^p\dot{{\cal M}}_{\rm C}^{pL}
 +2\sum_{l=1}^{\infty}\frac{(-1)^ll}{(l+1)!}v_{\rm C}^k\varepsilon_{kpq}\pd_{qL-1}\bigg(\frac{1}{R_{\rm C}}\bigg)\mathcal{S}_{\rm C}^{pL-1}\nonumber\\
  &+&2\sum_{l=1}^{\infty}\frac{(-1)^l}{l!}\frac{2l-1}{2l+1}v_{\rm C}^k\pd_{kL-1}\bigg(\frac{1}{R_{\rm C}}\bigg)\mathcal{R}_{\rm C}^{L-1}\;,\nonumber
\ea
where a prime after a function means that the function is taken at the integration point with coordinates $w'^i_{\rm C}$ in the local coordinates adapted to body C.

Potential $\Psi_{{\rm C}2}$ depends on the total Newtonian potential $U$ of all bodies in $\mathbb{N}$-body system. It is split in two pieces,
\be\label{c229}
U(t,{\bm x})=U_{\rm C}(t,{\bm x})+\bar U_{\rm C}(t,{\bm x})\;,
\ee
where $U_{\rm C}(t,{\bm x})$ is the Newtonian potential of body C, and $\bar U_{\rm C}(t,{\bm x})=\sum\limits_{\B\not={\rm C}}U_\B(t,{\bm x})$ is the Newtonian potential of all other bodies of $\mathbb{N}$-body system. Transformation of the Newtonian potential from the global to local coordinates of body C is sufficient in the Newtonian approximation: $U_{\rm C}(t,{\bm x})=U_{\rm C}(u_C,{\bm w}_{\rm C})$. The external Newtonian potential is decomposed in a Taylor series around the origin $x^i_{\rm C}$ of the local coordinates of body C, which is also transformed from the global to local coordinates,
\be\label{z33a1}
\bar U_{\rm C}(t,{\bm x})= \bar U_{\rm C}(t,{\bm x}_{\rm C})+\sum\limits_{k=1}^\infty\frac1{k!}\pd_K \bar U_{\rm C}(t,{\bm x}_{\rm C})w^K_{\rm C}\;,
\ee
where we have used notations
\be\label{22vv33d}
\bar U_{\rm C}(t,{\bm x}_{\rm C})\equiv\sum_{{\rm B}\not={\rm C}}U_{\rm B}(t,{\bm x}_{\rm C})\;,\qquad\qquad \pd_K\bar U_{\rm C}(t,{\bm x}_{\rm C})\equiv\lim_{{\bm x}\rightarrow{\bm x}_{\rm C}}\sum_{{\rm B}\not={\rm C}}\pd_{<i_1...i_k>} U_{\rm B}(t,{\bm x})\;.
\ee
Taking the above considerations into account, and performing calculations of integrals, we get a multipolar decomposition of potential $\Psi_{{\rm C}2}$ in the following form,
\ba
  \label{pc31}
\Psi_{{\rm C}2}(t,\bm{x}) & = & \int\limits_{{\cal V}_{\rm C}}\frac{\rho^*(t,\bm{x}')U(t,\bm{x}')}{|\bm{x}-\bm{x}'|}d^3x'\\
  & = & \sum_{l=0}^{\infty}\frac{(-1)^l}{l!}\pd_L\bigg(\frac{1}{R_{\rm C}}\bigg)\int\limits_{{\cal V}_{\rm C}}{\rho_{\rm C}^*}'U_{\rm C}'w_{\rm C}'^{<L>}d^3w_{\rm C}'\nonumber\\
& + & \sum_{l=0}^{\infty}\frac{(-1)^l}{l!}\pd_L\bigg(\frac{1}{R_{\rm C}}\bigg)\left[\bar U_{\rm C}(t,{\bm x}_{\rm C}){\cal M}^L_{\rm C}+\sum_{k=1}^\infty\frac1{k!}\pd_K\bar U_{\rm C}(t,{\bm x}_{\rm C})\int\limits_{{\cal V}_{\rm C}}{\rho_{\rm C}^*}'w_{\rm C}'^{<K>}w_{\rm C}'^{<L>}d^3w_{\rm C}'\right]\;.\nonumber
\ea

Multipolar decompositions of potentials $\Psi_{{\rm C}3}$, $\Psi_{{\rm C}4}$ are straightforward, and result in
\ba 
  \label{w1q2}
\Psi_{{\rm C}3}(t,\bm{x}) & = & \int\limits_{{\cal V}_{\rm C}}\frac{\rho^*(t,\bm{x}')\Pi(t,\bm{x}')}{|\bm{x}-\bm{x}'|}d^3x'= \sum_{l=0}^{\infty}\frac{(-1)^l}{l!}\pd_L\bigg(\frac{1}{R_{\rm C}}\bigg)\int\limits_{{\cal V}_{\rm C}}{\rho_{\rm C}^*}'\Pi_{\rm C}'w_{\rm C}'^{<L>}d^3w_{\rm C}'\;,\\\nonumber\\\nonumber\\
  \label{w1q3}
\Psi_{{\rm C}4}(t,\bm{x}) & = & \int\limits_{{\cal V}_{\rm C}}\frac{{\mathfrak{s}}^{kk}(t,\bm{x}')}{|\bm{x}-\bm{x}'|}d^3x'= \sum_{l=0}^{\infty}\frac{(-1)^l}{l!}\pd_L\bigg(\frac{1}{R_{\rm C}}\bigg)\int\limits_{{\cal V}_{\rm C}}{\mathfrak{s}}_{\rm C}'^{kk}w_{\rm C}'^{<L>}d^3w_{\rm C}'\;.
\end{eqnarray}

\subsection{Multipolar Expansion of Potential \texorpdfstring{$\bar\chi$}{\bar\chi}}

Multipolar expansion of external potential $\chi_{\rm C}(t,{\bm x})$ defined by equation \eqref{12.12}, is based on the multipolar expansion of coordinate distance $|\bm{x}-\bm{x}'|=|R_{\rm C}-R'_{\rm C}|$ that is a kernel of the integral in \eqref{12.12}, near the origin of the local coordinates that is the point with coordinates ${\bm x}_{\rm C}$. Taylor's expansion of the kernel $|\bm{x}-\bm{x}'|$ with respect to ${\bm x}'$ is given in terms of the Gegenbauer polynomials, ${\rm C}^{(-\frac12)}_l({\bm x})$, \citep[section 8.93]{gradryzh} and its STF expansion near the origin of the local coordinates of body B reads,
\begin{eqnarray}\la{12345}
|\bm{x}-\bm{x}'|=\sum_{l=0}^{\infty}\frac{(-1)^l}{l!}R_{\rm C}'^{<L>}\pd_LR_{\rm C}+
\sum_{l=0}^{\infty}\frac{(-1)^l}{(2l+3)l!}R_{\rm C}'^2R_{\rm C}'^{<L>}\pd_L\bigg(\frac{1}{R_{\rm C}}\bigg)\;.
\end{eqnarray}
Therefore, the multipolar expansion of external potential $\chi_{\rm C}(t,{\bm x})$ has the following form,
\ba\la{nb34v}
\chi_{\rm C}(t,{\bm x})&=&-\int\limits_{{\cal V}_{\rm C}}\rho^*(t,\bm{x}')|\bm{x}-\bm{x}'|d^3x'
=-\sum_{l=0}^{\infty}\frac{(-1)^l}{l!}\pd_LR_{\rm C}{\cal M}_{\rm C}^L-
\sum_{l=0}^{\infty}\frac{(-1)^l}{(2l+3)l!}\pd_L\bigg(\frac{1}{R_{\rm C}}\bigg)\mathcal{N}_{\rm C}^L\;,
\ea
which is a direct consequence of integration of \eqref{12345}. 

In what follows, we will need the multipolar expansion of the second partial derivative of the potential $\chi_{\rm C}$ with respect to the global coordinate time, $\pd_t^2\chi_{\rm C}(t,{\bm x})$, because it is this quantity that enters definition of the external gravitoelectric multipoles, ${\cal Q}_L$. The partial time derivative of $\chi_{\rm C}$ with respect to the global coordinate time, $t$, should be transformed  to the time derivative taken with respect to the local coordinate time $u_{\rm C}$ of body C which allow us to separate the internal, time-dependent physical processes inside body C, from the temporal changes caused by motion of body C with respect to the global coordinates. The law of transformation of the first time derivative is derived directly from the coordinate transformation \eqref{5.12cc}, \eqref{5.13cc} and is given by 
\be\la{udvc6}
\frac{\pd}{\pd t}=\frac{\pd}{\pd u_{\rm C}}\frac{\pd u_{\rm C}}{\pd t}+\frac{\pd}{\pd w^i}\frac{\pd w^i}{\pd t}=\frac{\pd}{\pd u_{\rm C}}-v_{\rm C}^k\frac{\pd}{\pd R_{\rm C}^k}\;,
\ee
where we have neglected all terms of the post-Newtonian order because they contribute only to the post-post-Newtonian approximation which we don't consider.
Applying \eqref{udvc6} one more time, we get for the second partial derivative,
\be\la{b4n0cw}
\frac{\pd^2}{\pd t^2}=\frac{\pd^2}{\pd u_{\rm C}^2}-2v_{\rm C}^k\frac{\pd^2}{\pd R^k_{\rm C}\pd u_{\rm C}}+v_{\rm C}^kv_{\rm C}^p\frac{\pd^2}{\pd R_{\rm C}^k\pd R_{\rm C}^p}-a_{\rm C}^k\frac{\pd}{\pd R_{\rm C}^k}\;.
\ee
Now, we employ \eqref{b4n0cw} to calculate the second time derivative from expansion \eqref{nb34v}. In doing this, we remind that the internal potentials, ${\cal M}_{\rm C}^L$ and $\mathcal{N}_{\rm C}^L$, are functions of the local coordinate time $u_{\rm C}$ only, and the partial derivative $\pd/\pd R_{\rm C}^i=\pd/\pd x^i\equiv\pd_i$. Therefore, taking the second time derivative from $\chi_{\rm C}$ results in,
\ba\hspace{-2cm}\la{ne9v4}
\frac{\pd^2\chi_{\rm C}}{\pd t^2}&=&-\sum_{l=0}^{\infty}\frac{(-1)^l}{l!}\left[\ddot{{\cal M}}_{\rm C}^L\pd_LR_{\rm C}-
 2\dot{{\cal M}}_{\rm C}^Lv_{\rm C}^k \pd_{kL}R_{\rm C}+{{\cal M}}_{\rm C}^Lv_{\rm C}^kv_{\rm C}^p\pd_{kpL}R_{\rm C}- {{\cal M}}_{\rm C}^La_{\rm C}^k\pd_{kL}R_{\rm C}\right]\\\nonumber
 &&- \sum_{l=0}^{\infty}\frac{(-1)^l}{(2l+3)l!}\left[\pd_L\left(\frac1{R_{\rm C}}\right)\ddot{\mathcal{N}}_{\rm C}^L-
 2v_{\rm C}^k \pd_{kL}\left(\frac1{R_{\rm C}}\right)\dot{\mathcal{N}}_{\rm C}^L+v_{\rm C}^kv_{\rm C}^p\pd_{kpL}\left(\frac1{R_{\rm C}}\right){\mathcal{N}}_{\rm C}^L- a_{\rm C}^k\pd_{kL}\left(\frac1{R_{\rm C}}\right){\mathcal{N}}_{\rm C}^L\right]\;,
 \ea
where we have discarded all post-Newtonian terms as they contribute only to the second post-Newtonian approximation which we don't consider, $v^i_{\rm C}= dx^i_{\rm C}/dt$ and $a^i_{\rm C}=dv^i_{\rm C}/dt$ are, respectively, velocity and acceleration of the origin of the local coordinates of body C with respect to the global coordinates. 

It is worth noticing that the partial derivatives from function $1/R_{\rm C}$ like $\pd_L(1/R_{\rm C})$, $\pd_{kL}(1/R_{\rm C})$, etc., are STF derivatives with respect to all indices. At the same time the partial derivatives from $R_{\rm C}$, like $\pd_L R_{\rm C}$, $\pd_{kL} R_{\rm C}$, etc., are not STF derivatives with respect to their indices, only that part of indices in the derivatives which is contracted with STF multipoles becomes symmetric and trace free. Transformation of the partial derivatives from $R_{\rm C}$ to their STF counterpart will be required in derivation of the equations of motion and is given below in \eqref{n4g2x0}.  

\section{Multipolar Expansion of External Multipoles in the Global Coordinates}\la{111777}
The external tidal  multipoles ${\cal P}_L$, ${\cal Q}_L$ and ${\cal C}_L$ of body B have been introduced in section \ref{ex5m3} in the form of the STF partial derivatives from the external potentials. We need explicit expressions of the external multipoles in terms of the multipolar series with respect to the internal multipoles of the extended bodies for calculating equations of motion of ${\mathbb N}$-body system in the global coordinates.
The present section provides this multipolar decomposition.

\subsection{Scalar-Field Multipoles \texorpdfstring{${\cal P}_L$}{{\cal P}_L}}\la{n328v}

Multipolar decomposition of the external scalar-field multipoles ${\cal P}_L$ of body B is obtained from \eqref{3.13} where the scalar field $\bar\varphi(t,{\bm x}_\B)=\bar W(t,{\bm x}_\B)$ of external bodies and, $\bar W(t,{\bm x}_\B)=\sum\limits_{{\rm C}\not=\B}W_C(t,{\bm x}_\B)$, is the external Newtonian potential. Multipolar decomposition of the potential $W_{\rm C}({\bm x}_B)$ of body C is given in \eqref{je5v20}. Making use of it, we get the external scalar-field multipoles
\be\la{rt3c5z3}
{\cal P}_L=\pd_L\bar W(t,{\bm x}_\B)=\sum\limits_{{\rm C}\not=\B}\sum_{n=0}^{\infty}\frac{(-1)^n}{n!}{\cal M}_{\rm C}^N\pd_{LN}\bigg(\frac{1}{R_{\rm C}}\bigg)_{{\bm x}={\bm x}_\B}\;,
\ee
where the STF index $N$ should not be confused with the number ${\mathbb N}$ of the extended bodies in ${\mathbb N}$-body system.
Expression \eqref{rt3c5z3} will be used later for calculating the post-Newtonian part of gravitational force depending on the external scalar-field multipoles.

\subsection{Gravitoelectric Multipoles \texorpdfstring{${\cal Q}_L$}{{\cal Q}_L}}\la{b3da53}
Gravitoelectric multipoles ${\cal Q}_L$ are defined by equation \eqref{5.9}. It is instructive to introduce potentials $\bar W$, $\bar V$ and $\bar V^i$ as linear combinations of potentials $W_{\rm C}$, $V_{\rm C}$, $V^i_{\rm C}$ of individual bodies from $\mathbb{N}$-body system,  
\be\la{n3c72b}
\bar{W}(t,{\bm x})\equiv\sum_{{\rm C}\not={\rm B}}{W}_{\rm C}(t,{\bm x})\;,\qquad\qquad
\bar V(t,{\bm x},l)\equiv\sum_{{\rm C}\not={\rm B}}V_{\rm C}(t,{\bm x},l)\;,\qquad\qquad \bar V_i(t,{\bm x},l)\equiv\sum_{{\rm C}\not={\rm B}}V_{\rm C}^i(t,{\bm x},l)\;,
\ee
where the scalar potential $W_{\rm C}$ has been defined earlier in \eqref{je5v20}, the scalar potential
\ba\la{nrrx4e}
V_{\rm C}(t,{\bm x},l)&\equiv&\Phi_{\rm C}(t,{\bm x})+\Psi_{\rm C}(t,{\bm x})-\frac12\pd_{tt}\chi_{\rm C}(t,{\bm x})-2(1+\gamma)v_{\rm B}^kU^k_{\rm C}(t,{\bm x})\\\nonumber&&+
(1+\gamma)v_{\rm B}^2U_{\rm C}(t,{\bm x})+(2-2\beta-l\gamma)\bar U(t,{\bm x}_\B) U_{\rm C}(t,{\bm x})\;,\ea
and the vector potential
\ba
\la{bex5b7}
V_{\rm C}^i(t,{\bm x},l)&\equiv&2(1+\gamma)\dot U_{\rm C}^i(t,{\bm x})+(l-2-2\gamma)v_{\rm B}^i\dot U_{\rm C}(t,{\bm x})-\frac{l}2v_{\rm B}^iv_{\rm B}^k\pd_kU_{\rm C}(t,{\bm x})\\\nonumber
&&-
(l^2-l+2+2\gamma)a_{\rm B}^iU_{\rm C}(t,{\bm x})-lF^{ki}_{\rm B}\pd_k U_{\rm C}(t,{\bm x})\;.
\ea
Notice that potentials $V_{\rm C}(t,{\bm x},l)$ and $V_{\rm C}^i(t,{\bm x},l)$ depend explicitly on the multipolar index $l$.
In terms of the new potentials the gravitoelectric multipole ${\cal Q}_L$ takes on a simpler expression,
\be\la{b4v9h2}
{\cal Q}_L=\pd_{<L>}\bar{W}(t,{\bm x}_\B)+\pd_{<L>}\bar V({\bm x}_{\rm B},l)+\pd_{<L-1}\bar V_{i_l>}({\bm x}_{\rm B},l)+X_{<L>}\;,\qquad\qquad (l\ge2)\;.
\ee

Multipolar expansion of potential ${W}_{\rm C}$ is given in \eqref{je5v20}. Multipolar expansion of two other gravitational potentials are obtained form the results of section \ref{111777}
\ba
\la{b3c7x5}
V_{\rm C}(t,{\bm x},l)&=& \sum_{n=0}^{\infty}\frac{(-1)^n}{n!}\pd_N\bigg(\frac{1}{R_{\rm C}}\bigg)
\left[(1+\gamma)v_{\rm B}^2+\left(\gamma+\frac12\right)v_{\rm C}^2\right]{\cal M}_{\rm C}^N\\
&+&\sum_{n=0}^{\infty}\frac{(-1)^n}{n!}\pd_N\bigg(\frac{1}{R_{\rm C}}\bigg)\left[(2-2\beta-l\gamma)\bar U(t,{\bm x}_\B)-\gamma(n+1)\bar U_{\rm C}(t,{\bm x}_{\rm C})\right]{\cal M}_{\rm C}^N\nonumber\\
 &-&\sum_{n=0}^{\infty}\frac{(-1)^n}{n!}\pd_N\bigg(\frac{1}{R_{\rm C}}\bigg)\bigg[ \frac{n}2 v_{\rm C}^pv_{\rm C}^{i_n}{\cal M}_{\rm C}^{pN-1}- nF_{\rm C}^{pi_n}{\cal M}_{\rm C}^{pN-1}
     +(n+1)a_{\rm C}^p{\cal M}_{\rm C}^{pN}+v_{\rm C}^p\dot{{\cal M}}_{\rm C}^{pN}-v_{\rm C}^pR_{\rm C}^p\dot{{\cal M}}_{\rm C}^N\bigg]\nonumber\\
     &+&\sum_{n=0}^{\infty}\frac{(-1)^n}{n!}\left[\frac12\ddot{{\cal M}}_{\rm C}^N\pd_NR_{\rm C}-
 \dot{{\cal M}}_{\rm C}^Nv_{\rm C}^p \pd_{pN}R_{\rm C}+\frac12{{\cal M}}_{\rm C}^Nv_{\rm C}^pv_{\rm C}^q\pd_{pqN}R_{\rm C}-\frac12 {{\cal M}}_{\rm C}^Na_{\rm C}^p\pd_{pN}R_{\rm C}\right]\nonumber\\
  &+&2(1+\gamma) \bigg[\sum_{n=1}^{\infty}\frac{(-1)^nn}{(n+1)!}\varepsilon_{kpq}\pd_{pN-1}\bigg(\frac{1}{R_{\rm C}}\bigg)\mathcal{S}_{\rm C}^{qN-1}v_{\B{\rm C}}^k
 -\sum_{n=0}^{\infty}\frac{(-1)^n}{(n+1)!}\pd_N\bigg(\frac{1}{R_{\rm C}}\bigg)\dot{{\cal M}}_{\rm C}^{pN}v_{\B{\rm C}}^p
 \nonumber\\
  &&\phantom{2(1+\gamma) \bigg[}-\sum_{n=0}^{\infty}\frac{(-1)^n}{n!}\pd_N\bigg(\frac{1}{R_{\rm C}}\bigg) v^p_{\rm B} v^p_{\rm C}{\cal M}_{\rm C}^{N}-
 \sum_{n=1}^{\infty}\frac{(-1)^n}{n!}\frac{2n-1}{2n+1}\pd_{pN-1}\bigg(\frac{1}{R_{\rm C}}\bigg)\mathcal{R}_{\rm C}^{N-1}v_{\B{\rm C}}^p
\nonumber\\
 &&\phantom{2(1+\gamma) \bigg[}-
  \sum_{n=1}^{\infty}\frac{(-1)^n}{n!}\frac{2n-1}{2n+1}\pd_{N-1}\bigg(\frac{1}{R_{\rm C}}\bigg)\dot{\mathcal{R}}_{\rm C}^{N-1}\bigg]\;,\nonumber 
  \ea
where 
\be
v^i_{\rm BC}\equiv v^i_{\rm B}-v^i_{\rm C}\;,
\ee
is the relative coordinate velocity between the bodies B and C, and the external potentials $\bar U$ and $\bar U_{\rm C}$ have been defined in \eqref{tb54vd} and {\eqref{22vv33d}.

Expression \eqref{bex5b7} for $V^i_{\rm C}$ contains the total time derivatives from the potentials taken on the worldline, $x^i_\B(t)$, of the origin of the local coordinates adapted to body B. It is expressed in terms of the partial time and spatial derivatives as follows,
\be\la{ner5c3}
\frac{d}{dt}=\frac{\pd}{\pd t}+v^i_{\rm B}\frac{\pd}{\pd x^i}\;,
\ee
where $v^i_{\rm B}=dx^i_\B/dt$ is velocity of the origin of the local coordinates adapted to body B with respect to the global coordinates.
The partial time derivative in \eqref{ner5c3} is taken with respect to the variables associated with each body C that is external to the body B. It is related to the partial time derivative taken with respect to the local coordinate time $u_{\rm C}$ of body C by equation \eqref{udvc6}. Hence, the total time derivative from the external potentials associated with body C taken on the worldline of body B reads, 
\be\la{wn9f1p}
\frac{d}{dt}=\frac{\pd}{\pd u_{\rm C}}+v^i_{\rm BC}\frac{\pd}{\pd x^i}\;.
\ee
where again $v^i_{\rm BC}\equiv v^i_{\rm B}-v^i_{\rm C}$ is the relative velocity between two bodies, B and C.
After employing \eqref{wn9f1p} for taking the total time derivatives in \eqref{bex5b7} and the multipolar expansions of other potentials entering definition of $\bar V^i_{\rm C}$, we get 
\ba\la{zf4k9}
V^i_{\rm C}(t,{\bm x},l)&=&2(1+\gamma)\Bigg[\sum_{n=1}^{\infty}\frac{(-1)^n}{(n+1)!}\pd_N\bigg(\frac{1}{R_{\rm C}}\bigg)\ddot{{\cal M}}_{\rm C}^{iN}+\sum_{n=0}^{\infty}\frac{(-1)^n}{n!}\pd_N\bigg(\frac{1}{R_{\rm C}}\bigg)\left(\dot{{\cal M}}_{\rm C}^{N}v_{\rm C}^i+{{\cal M}}_{\rm C}^{N}a_{\rm C}^i \right)\\
&&\phantom{++}-\sum_{n=1}^{\infty}\frac{(-1)^nn}{(n+1)!}\varepsilon_{ipq}\pd_{pN-1}\bigg(\frac{1}{R_{\rm C}}\bigg)\dot{\mathcal{S}}_{\rm C}^{qN-1}
-\sum_{n=1}^{\infty}\frac{(-1)^nn}{(n+1)!}\varepsilon_{ipq}\pd_{kpN-1}\bigg(\frac{1}{R_{\rm C}}\bigg){\mathcal{S}}_{\rm C}^{qN-1}v_{\B{\rm C}}^k
\nonumber\\
&&\phantom{++}+\sum_{n=1}^{\infty}\frac{(-1)^n}{(n+1)!}\pd_{pN}\bigg(\frac{1}{R_{\rm C}}\bigg)\dot{{\cal M}}_{\rm C}^{iN}v_{\B{\rm C}}^p+ \sum_{n=0}^{\infty}\frac{(-1)^n}{n!}\pd_{pN}\bigg(\frac{1}{R_{\rm C}}\bigg){{\cal M}}_{\rm C}^{N}v_{\B{\rm C}}^pv_{\rm C}^i 
\nonumber\\
&&\phantom{++}
+\sum_{n=1}^{\infty}\frac{(-1)^n}{n!}\frac{2n-1}{2n+1}\pd_{iN-1}\bigg(\frac{1}{R_{\rm C}}\bigg)\dot{\mathcal{R}}_{\rm C}^{N-1}
+\sum_{n=1}^{\infty}\frac{(-1)^n}{n!}\frac{2n-1}{2n+1}\pd_{ipN-1}\bigg(\frac{1}{R_{\rm C}}\bigg){\mathcal{R}}_{\rm C}^{N-1}v_{\B{\rm C}}^p\Bigg]\nonumber\\
&&\phantom{++}+(l-2-2\gamma)\Bigg[\sum_{n=0}^{\infty}\frac{(-1)^n}{n!}\pd_N\bigg(\frac{1}{R_{\rm C}}\bigg)\dot{{\cal M}}_{\rm C}^{N}v_{\rm B}^i
+\sum_{n=0}^{\infty}\frac{(-1)^n}{n!}\pd_{pN}\bigg(\frac{1}{R_{\rm C}}\bigg){{\cal M}}_{\rm C}^{N}v_{\B{\rm C}}^pv_{\rm B}^i\Bigg]\nonumber\\
&&\phantom{++}-(l^2-l+2+2\gamma)\sum_{n=0}^{\infty}\frac{(-1)^n}{n!}\pd_N\bigg(\frac{1}{R_{\rm C}}\bigg){{\cal M}}_{\rm C}^{N}a_{\rm B}^i-\frac{l}2\sum_{n=0}^{\infty}\frac{(-1)^n}{n!}\pd_{pN}\bigg(\frac{1}{R_{\rm C}}\bigg){{\cal M}}_{\rm C}^{N}v^p_{\rm B}v_{\rm B}^i
\nonumber\\
&&\phantom{++}-l\,F_{\rm B}^{ki}\pd_k\bar U(t,{\bm x}_\B)\nonumber\;.
 \ea    
Multipolar expansion of the external gravitoelectric multipole ${\cal Q}_L$ is obtained by substituting \eqref{b3c7x5} and \eqref{zf4k9} to definition \eqref{b4v9h2}. It is remarkable that each potential, $V(t,{\bm x},l)$ and $\bar V_i(t,{\bm x},l)$, entering \eqref{b4v9h2} depends separately on the {\it non-canonical} multipoles ${\cal R}^L$ and ${\cal N}^L$ but they are mutually canceled out in the linear combination $\pd_{<L>}\bar V(t,{\bm x},l)+\pd_{<L-1}\bar V_{i>}(t,{\bm x},l)$ so that the gravitoelectric multipoles ${\cal Q}_L$ depend exclusively on the {\it canonical} internal {\it active} mass, ${\cal M}^L_{\rm C}$, and spin, ${\cal S}^L_{\rm C}$, multipoles. We don't provide over here the explicit expression for the multipolar decomposition of ${\cal Q}_L$. It will be given below in section \ref{hdfr4c}.

\subsection{Gravitomagnetic Multipoles \texorpdfstring{${\cal C}_L$}{{\cal C}_L}}\la{44nn99}

Gravitomagnetic external multipoles, ${\cal C}_L$, have been defined in \eqref{3.29}. They represent a linear combination of the gravitomagnetic multipoles, $H^{ikL}_{\rm C}$, generated by all bodies of ${\mathbb N}$-body system which are external with respect to body C. More specifically, we re-formulate \eqref{3.29} as follows, 
\ba\la{he5v2z}
 \varepsilon_{ipk}{\cal C}_{pL} \equiv \bar H_{ikL}\;,
\ea
where 
\be\la{ny3v6z5}
\bar H_{ikL}=\sum_{{\rm C}\not=\B}H^{ikL}_{\rm C}(t,{\bm x}_\B)\;,
\ee
and $H^{ikL}_{\rm C}$ is a skew-symmetric tensor with respect to the first two indices and STF tensor with respect to the multi-index $L$, that is $H^{ikL}_{\rm C}\equiv H^{[ik]<L>}_{\rm C}$. The same property naturally holds for $\bar H_{ikL}$. 

For each body C equation \eqref{3.29} yields, 
 \ba\la{ne8v40}
 H^{ikL}_{\rm C}(t,{\bm x})& = & 4(1+\gamma)\bigg\{ v_{\rm B}^{[i}\pd^{k]L}_{\phantom{B}}U_{\rm C}(t,{\bm x}) +\pd^{L[i}_{\phantom{C}}{U}_{\rm C}^{k]}(t,{\bm x})\bigg\}\\
   &-&2(1+\g)\frac{l}{l+1}\left\{\delta^{i<i_{l}}\pd^{L-1>k}\dot{U}_{\rm C}(t,{\bm x})-\delta^{k<i_{l}}\pd^{L-1>i}\dot{U}_{\rm C}(t,{\bm x})\right\}\;.\nonumber
 \ea
According to definition \eqref{he5v2z} we have $\varepsilon_{ipk}{\cal C}_{pkL-1}\equiv 0$ due to the anti-symmetry of the Levi-Civita symbol and the STF symmetry of ${\cal C}_L$. It follows, then, that $H^{ik<kL-1>}_{\rm C}=0$ as well. This property can be confirmed by inspection after contracting the corresponding indices in the right-hand side of \eqref{ne8v40}, and remembering that according to equation of continuity \eqref{11.20}, we have in the global coordinates, $\pd_kU_{\rm C}^k+\pd_t U_{\rm C}=0$.  

 Multipolar expansion of $H^{ikL}_{\rm C}$ is obtained after making use of multipolar decomposition of potentials $U_{\rm C}$ and $U^i_{\rm C}$ given above in sections \ref{f8n3k}, \ref{ui888},
 \ba\la{ket3b7}
 H^{ikL}_{\rm C}(t,{\bm x})&=&2(1+\g)\sum_{n=0}^\infty\frac{(-1)^n}{n!}\Bigg[ v_{\rm BC}^{i}\pd^{kLN}\bigg(\frac1{R_{\rm C}}\bigg)- v_{\rm BC}^{k}\pd^{iLN}\bigg(\frac1{R_{\rm C}}\bigg)\Bigg]{\cal M}_{\rm C}^N\\\nonumber
	&-&2(1+\g)\sum_{n=0}^{\infty}\frac{(-1)^n}{(n+1)!}\Bigg[\dot{{\cal M}}_{\rm C}^{iN}\pd^{kLN}\bigg(\frac{1}{R_{\rm C}}\bigg)-\dot{{\cal M}}_{\rm C}^{kN}\pd^{iLN}\bigg(\frac{1}{R_{\rm C}}\bigg)\Bigg]\nonumber\\
 	&+&2(1+\g)\sum_{n=0}^{\infty}\frac{(-1)^n}{(n+2)n!}\Bigg[\varepsilon^{pqi}\pd^{kqLN}\bigg(\frac{1}{R_{\rm C}}\bigg)-\varepsilon^{pqk}\pd^{iqLN}\bigg(\frac{1}{R_{\rm C}}\bigg)\Bigg]\mathcal{S}_{\rm C}^{pN}
\nonumber\\\nonumber
	&-&2(1+\g)\sum_{n=0}^\infty\frac{(-1)^nl}{(l+1)n!}v^p_{\B{\rm C}}\Bigg[ \delta^{i<i_{l}}\pd^{L-1>Npk}\bigg(\frac1{R_{\rm C}}\bigg)- \delta^{k<i_{l}}\pd^{L-1>Npi}\bigg(\frac1{R_{\rm C}}\bigg)\Bigg]{\cal M}_{\rm C}^N\\\nonumber
	&-& 2(1+\g)\sum_{n=0}^\infty\frac{(-1)^nl}{(l+1)n!}\Bigg[\delta^{i<i_{l}}\pd^{L-1>Nk}\bigg(\frac1{R_{\rm C}}\bigg)-\delta^{k<i_{l}}\pd^{L-1>Ni}\bigg(\frac1{R_{\rm C}}\bigg)\Bigg] \dot{\cal M}_{\rm C}^N\;.
\ea
It is worth noticing that the {\it non-canonical} multipoles ${\cal R}_L$ which are present in the multipolar expansion \eqref{w1w8} of the external gravitomagnetic potential $\bar U^i_{\rm C}$, are canceled out in \eqref{ket3b7} after taking the skew-symmetric partial derivative, $\bar {U}^{[i,k]}$. Therefore, the external gravitomagnetic multipoles, ${\cal C}_L$, do not depend on the {\it non-canonical} multipoles ${\cal R}_L$.


\section{Translational Equations of Motion of Bodies in the Global Coordinates}\label{orbeom}

The aim of this section is to derive the post-Newtonian equations of translational motion of extended bodies in the global coordinates with taking into account all possible gravitational interactions taking place between mass and spin internal multipoles of the bodies in ${\mathbb N}$-body system. Our derivation is based on the Fock-Papapetrou method along with the matched asymptotic expansions technique and significantly extends the post-Newtonian equations of motion of extended bodies in gravitationally-bound systems beyond the pole-dipole approximation. A similar task was set forth and solved in the post-Newtonian approximation of general relativity by \citet{racine_2005PhRvD,racine2013PhRvD} who used the EIH technique of surface integration along with the post-Newtonian transformations of asymptotic expansions of the metric tensors and Blanchet-Damour multipole formalism. We shall compare our translational equations with those derived previously by \citet{racine_2005PhRvD,racine2013PhRvD} in Appendix \ref{appendixA}.  

\subsection{Computation of Gravitational Force}\la{cf34}
\subsubsection{Reduction of Similar Terms}\label{hdfr4c}
Translational equations of motion of the center of mass of body B in the global coordinates follow directly from the equations of motion (\ref{5.8}) of the origin of the local coordinates adapted to body B after making use of the specific value of the local acceleration ${\cal Q}_i$ defined in (\ref{q6v4m})--\eqref{nrvug9} and the multipolar decomposition of the external multipoles ${\cal P}_L$, ${\cal Q}_L$, ${\cal C}_L$ provided in section \ref{111777}. This makes the worldline ${\cal W}$ of the origin of the local coordinates of body B identical with the worldline ${\cal Z}$ of the body's center of mass.

It is instrumental to re-write the right-hand side of \eqref{5.8} in terms of the gravitational potentials $\bar V(t,{\bm x},l)$ and $\bar V_i(t,{\bm x},l)$ introduced above in \eqref{nrrx4e} and \eqref{bex5b7}. We have
\be\la{x5v0m4b}
a^i_{\rm B}=\pd_i\bar{W}(t,{\bm x}_\B)-{\cal Q}_i^{\rm N}+\pd_i\bar V(t,{\bm x}_{\rm B},1)+\bar V_i(t,{\bm x}_{\rm B},1)-{\cal Q}_i^{\rm pN}+\frac{\ddot {\cal I}^i_{\rm c}}{M_\B}-\frac12v_{\rm B}^iv_{\rm B}^ka^k_{\rm B}-F^{ik}_{\rm B}a^k_{\rm B}-v_{\rm B}^2a^i_{\rm B}+\g a^i_{\rm B}\bar U(t,{\bm x}_\B)\;,
\ee
where accelerations ${\cal Q}_i^{\rm N}$ and ${\cal Q}_i^{\rm pN}$ are determined by \eqref{b4cs2j} and \eqref{nrvug9}, the external gravitational potentials
\be
\bar V(t,{\bm x},1)\equiv\sum_{{\rm C}\not={\rm B}}V_{\rm C}(t,{\bm x},1)\;,\qquad\qquad \bar V_i(t,{\bm x},1)\equiv\sum_{{\rm C}\not={\rm B}}V^i_{\rm C}(t,{\bm x},1)\;,
\ee
and gravitational potentials of body C are given respectively by \eqref{nrrx4e} and \eqref{bex5b7} for the value of multipole index $l=1$,
\ba\la{vxrwcd7}
V_{\rm C}(t,{\bm x},1)&\equiv&
\Phi_{\rm C}(t,{\bm x})+\Psi_{\rm C}(t,{\bm x})-\frac12\pd_{tt}\chi_{\rm C}(t,{\bm x})-2(1+\gamma)v_{\rm B}^kU^k_{\rm C}(t,{\bm x})\\\nonumber&&+
(1+\gamma) v_{\rm B}^2U_{\rm C}(t,{\bm x})+(2-2\beta-\gamma)\bar U(t,{\bm x}_\B) U_{\rm C}(t,{\bm x})\;,\\\la{getvgi7}
V^i_{\rm C}(t,{\bm x},1)&\equiv&2(1+\gamma)\dot U_{\rm C}^i(t,{\bm x})-(1+2\gamma)v_{\rm B}^i\dot U_{\rm C}(t,{\bm x})-\frac12v_{\rm B}^iv_{\rm B}^k\pd_kU_{\rm C}(t,{\bm x})\\\nonumber
&&-2(1+\gamma)a^i_{\rm B}U_{\rm C}(t,{\bm x})-F^{ki}_{\rm B}\pd_k U_{\rm C}(t,{\bm x})\;.
\ea

Local acceleration ${\cal Q}_i^{\rm N}$ of the center of mass of body B is given by \eqref{b4cs2j} where the external gravitoelectric multipoles ${\cal Q}_L$ are defined in \eqref{b4v9h2} in terms of the derivatives from the potentials $\bar W(t,{\bm x})$, $\bar V(t,{\bm x},l)$ and $ \bar V_i(t,{\bm x},l)$. Taking into account in the definition \eqref{b4cs2j} of ${\cal Q}_i^{\rm N}$ that, according to \eqref{rt3c5z3}, the external scalar-field dipole ${\cal P}_i=\pd_i\bar W(t,{\bm x}_\B)$, we can reduce relativistic equation of motion \eqref{x5v0m4b} to the form of the second Newton's law,
\be\la{u4bvs3}
M_\B a^i_{\rm B}=F^i\;,
\ee
where $M_\B$ is the {\it conformal} mass of body B, and 
\ba\la{yckope}
F^i&=&\sum_{l=0}^\infty\frac1{l!}\pd_{<iL>}\bar{W}(t,{\bm x}_\B){\cal M}^L_\B\\\nonumber
&-&\sum_{l=0}^\infty\frac1{l!}\bigg\{\Big[v_{\rm B}^2-\g \bar{U}(t,{\bm x}_\B)\Big]\pd_{<iL>}\bar{U}(t,{\bm x}_\B){\cal M}^L_\B+\frac12v_{\rm B}^iv_{\rm B}^k\pd_{<kL>}\bar {U}(t,{\bm x}_\B) {\cal M}^L_\B\bigg\}\\\nonumber
&+&\sum_{l=0}^\infty\frac1{l!}\bigg\{\pd_{<iL>}\bar V(t,{\bm x}_\B,l+1){\cal M}^L_\B+ \pd_{<L}\bar V_{i>}(t,{\bm x}_\B,l+1){\cal M}^L_\B-M_\B {\cal Q}^{\rm pN}_i+\ddot{\cal I}^i_{\rm c}-F^{ik}_{\rm B}\pd_{<kL>}\bar{U}(t,{\bm x}_B){\cal M}^L_\B\bigg\}\;,
\ea
is a relativistic force exerted on body B by external bodies of $\mathbb{N}$-body system. It depends explicitly on the {\it active} internal multipoles, ${\cal M}^L_\B$ of body B, and we identify, here and everywhere else, the {\it active} mass ${\cal M}_\B$ of body B with a monopole value ($l=0$) of the {\it active} mass multipole of body B, that is ${\cal M}\equiv{\cal M}_\B$. It is instructive to emphasize that the Newtonian part of the force, given by the first term in the right-hand side of \eqref{yckope}, depends on the {\it active} dipole, ${\cal M}^i_\B$, of body B which does not vanish in scalar-tensor theory of gravity because position of the center of mass of body B is defined by the condition \eqref{n5vz1o} of vanishing of the {\it conformal} dipole moment, ${\cal I}^i_\B$ of body B \footnote{We remind that in general case, ${\cal M}^i_\B\not={\cal I}^i_\B$. The difference has a post-Newtonian order of magnitude.}.

Before proceeding to the explicit calculation of the gravitational force, we notice that there are some cancellations of similar terms in \eqref{yckope}. More specifically, 
\begin{itemize}
\item[--] The very last term in the third line of \eqref{b3c7x5} can be transformed to
\ba\la{m8d5v0}
\dot{\cal M}^N_{{\rm C}}v^p_{{\rm C}}R^p_{{\rm C}}\pd_{N}\bigg(\frac1{R_{\rm C}}\bigg) &=&\dot{\cal M}^{a_1...a_n}_{{\rm C}}v^p_{{\rm C}}\left[\pd_{<pa_1...a_n>}R_{{\rm C}}
-n\frac{2n-1}{2n+1}\d_{p<a_1}\pd_{a_2...a_n>}\bigg(\frac1{R_{\rm C}}\bigg)\right]\\\nonumber
&=&\dot{\cal M}^{a_1...a_n}_{{\rm C}}v^p_{{\rm C}}\left[\pd_{pa_1...a_n}R_{{\rm C}}-\frac{2}{2n+1}\d_{\{pa_1}\pd_{a_2...a_n\}}\bigg(\frac1{R_{\rm C}}\bigg)
-n\frac{2n-1}{2n+1}\d_{p<a_1}\pd_{a_2...a_n>}\bigg(\frac1{R_{\rm C}}\bigg)\right]\\\nonumber
&=&\dot{\cal M}^N_{{\rm C}}v^p_{{\rm C}}\pd_{pN}R_{{\rm C}}
-nv^p_{{\rm C}}\dot{\cal M}^{pN-1}_{{\rm C}}\pd_{N-1}\bigg(\frac1{R_{\rm C}}\bigg)\;.
\ea
The first and second terms in the very last line of \eqref{m8d5v0} cancel out, respectively, the second term in the forth line of \eqref{b3c7x5} and the forth term in the third line of \eqref{b3c7x5} which all depend on the time derivative $\dot{\cal M}^L_{\rm C}$.
\item[--] The very last term in \eqref{zf4k9} enters equation \eqref{yckope} in STF form $(l+1)F_\B^{k<i}\pd^{L>k}\bar U(t,{\bm x}_\B)$ which can be decomposed with the help of peeling formula \eqref{hh33e5z} separating the STF index $i$ from that $L$, so that we get 
\be\la{j3v7g5}
\sum\limits_{l=0}^\infty\frac{l+1}{l!}F_\B^{k<i}\pd^{L>k}\bar{U}(t,{\bm x}_\B){\cal M}_\B^L=\sum\limits_{l=0}^\infty\frac{1}{l!}F_\B^{ki}\pd_{kL}\bar{U}(t,{\bm x}_\B){\cal M}_\B^L+\sum\limits_{l=0}^\infty\frac{1}{l!}F_\B^{kp}\pd_{ikL}\bar U(t,{\bm x}_\B){\cal M}_\B^{pL}\;.
\ee
The first term in the right-hand side of \eqref{j3v7g5} cancels the very last (precessional) term in \eqref{yckope}. 
\item[--] Each potential $V(t,{\bm x},l+1)$ and $\bar V_i(t,{\bm x},l+1)$ entering \eqref{yckope} depends on the {\it non-canonical} multipoles ${\cal R}^L$ and ${\cal N}^L$ but they are mutually canceled out in the linear combination $\pd_{<iL>}\bar V_(t,{\bm x},l+1)+\pd_{<L}\bar V_{i>}(t,{\bm x},l+1)$ so that the right side of the translational equations of motion \eqref{yckope} depends only on the {\it active} internal mass and spin multipoles  ${\cal M}^L_{\rm C}$ and ${\cal S}^L_{\rm C}$ of the bodies. 
\item[--] Finally, we notice that the post-Newtonian term, $X_L$, which is a part of ${\cal Q}_L$, does not appear in \eqref{yckope}. The term $X_L$ would appear in \eqref{yckope} only in the form of the quadrupole-dipole coupling, $X_{ip}{\cal M}^p_\B$, as a consequence of its definition \eqref{1q4d}. However, with sufficient accuracy the {\it active} mass dipole ${\cal M}^p_\B={\cal I}^p_\B=0$ in the post-Newtonian approximation due to the choice of the center of mass \eqref{n5vz1o}.   
\end{itemize}

The above-mentioned cancellations simplify \eqref{yckope} and recast it to 
\ba\la{sr5v6h}
F^i&=&\sum_{l=0}^\infty\frac1{l!}\pd_{<iL>}\bar{W}(t,{\bm x}_\B){\cal M}^L_\B-\bigg(v_\B^2\d^{ik}+\frac12v_{\rm B}^iv_{\rm B}^k\bigg)\sum_{l=0}^\infty\frac1{l!}\pd_{<kL>}\bar{U}(t,{\bm x}_\B) {\cal M}^L_\B\\\nonumber
&-&F^{pk}_{\rm B}\sum_{l=0}^\infty\frac1{l!}\pd_{<ipL>}\bar{U}(t,{\bm x}_\B){\cal M}^{kL}_\B+\sum_{l=0}^\infty\frac1{l!}\pd_{<iL>}\bar\Omega(t,{\bm x}_\B,l){\cal M}^L_\B\\\nonumber
&+& \sum_{l=0}^\infty\frac1{l!}\pd_{<L}\bar \Omega_{i>}(t,{\bm x}_\B,l){\cal M}^L_\B-M_\B {\cal Q}^{\rm pN}_i+\ddot{M}^i_{\rm c}\;,
\ea
where the external potentials
\be\la{popka34}
\bar {W}(t,{\bm x})\equiv\sum_{{\rm C}\not={\rm B}}{W}_{\rm C}(t,{\bm x})\;,\qquad\qquad \bar \Omega(t,{\bm x},l)\equiv\sum_{{\rm C}\not={\rm B}}\Omega_{\rm C}(t,{\bm x},l)\;,\qquad\qquad \bar \Omega_i(t,{\bm x},l)\equiv\sum_{{\rm C}\not={\rm B}}\Omega^i_{\rm C}(t,{\bm x},l)\;,
\ee
represent the linear superposition of gravitational potentials $W_{\rm C}$, $\Omega_{\rm C}$ and $\Omega^i_{\rm C}$ generated by body C$\not=$B.
Multipolar expansion of potential ${W}_{\rm C}(t,{\bm x})$ is given in \eqref{je5v20}. The new potentials $\Omega_{\rm C}(t,{\bm x},l)$ and $\Omega^i_{\rm C}(t,{\bm x},l)$ are modifications of $V_{\rm C}(t,{\bm x},l+1)$ and $V^i_{\rm C}(t,{\bm x},l+1)$ respectively after taking into account the above-mentioned cancellations of similar terms in \eqref{yckope}. They read, 
\ba\la{om3gfx6}
\Omega_{\rm C}(t,{\bm x},l)&=&
\sum_{n=0}^{\infty}\frac{(-1)^n}{n!}\pd_N\bigg(\frac{1}{R_{\rm C}}\bigg)
\left[\Big(\gamma+1\Big) v_{\rm B}^2+\Big(\gamma+\frac12\Big)v_{\rm C}^2\right]{\cal M}_{\rm C}^N\\
&+&\sum_{n=0}^{\infty}\frac{(-1)^n}{n!}\pd_N\bigg(\frac{1}{R_{\rm C}}\bigg)\left[(2-2\beta-l\gamma)\bar U(t,{\bm x}_\B)-\gamma(n+1)\bar U_{\rm C}(t,{\bm x}_{\rm C})\right]{\cal M}_{\rm C}^N\nonumber\\
&+&\sum_{n=0}^{\infty}\frac{(-1)^n}{n!}\pd_{pN}\bigg(\frac{1}{R_{\rm C}}\bigg)\bigg( \frac{1}2 v_{\rm C}^pv_{\rm C}^{k}{\cal M}_{\rm C}^{kN}- F_{\rm C}^{kp}{\cal M}_{\rm C}^{kN}\bigg)
     -\sum_{n=0}^{\infty}\frac{(-1)^n(n+1)}{n!}\pd_N\bigg(\frac{1}{R_{\rm C}}\bigg)a_{\rm C}^p{\cal M}_{\rm C}^{pN}\nonumber\\
     &+&\frac12\sum_{n=0}^{\infty}\frac{(-1)^n}{n!}\bigg[\ddot{{\cal M}}_{\rm C}^N\pd_NR_{\rm C}+{{\cal M}}_{\rm C}^Nv_{\rm C}^pv_{\rm C}^q\pd_{pq<N>}R_{\rm C}- {{\cal M}}_{\rm C}^Na_{\rm C}^p\pd_{p<N>}R_{\rm C}\bigg]\nonumber\\ 
&-&2(1+\gamma) \bigg[\sum_{n=0}^{\infty}\frac{(-1)^n}{(n+2)n!}\varepsilon_{kpq}\pd_{pN}\bigg(\frac{1}{R_{\rm C}}\bigg)\mathcal{S}_{\rm C}^{qN}v_{\B{\rm C}}^k
 +\sum_{n=0}^{\infty}\frac{(-1)^n}{(n+1)!}\pd_N\bigg(\frac{1}{R_{\rm C}}\bigg)\dot{{\cal M}}_{\rm C}^{pN}v_{\B{\rm C}}^p
 \nonumber\\
  &&\phantom{.........}+\sum_{n=0}^{\infty}\frac{(-1)^n}{n!}\pd_N\bigg(\frac{1}{R_{\rm C}}\bigg) v^p_{\rm B} v^p_{\rm C}{\cal M}_{\rm C}^{N}\bigg]\;,\nonumber\\\nonumber\\\nonumber\\
\la{omi84v6}
\Omega^i_{\rm C}(t,{\bm x},l)&=&2(1+\gamma)\Bigg[\sum_{n=1}^{\infty}\frac{(-1)^n}{(n+1)!}\pd_N\bigg(\frac{1}{R_{\rm C}}\bigg)\ddot{{\cal M}}_{\rm C}^{iN}+\sum_{n=0}^{\infty}\frac{(-1)^n}{n!}\pd_N\bigg(\frac{1}{R_{\rm C}}\bigg)\left(\dot{{\cal M}}_{\rm C}^{N}v_{\rm C}^i+{{\cal M}}_{\rm C}^{N}a_{\rm C}^i \right)\\
&&\phantom{++}+\sum_{n=0}^{\infty}\frac{(-1)^n}{(n+2)n!}\varepsilon_{ipq}\pd_{pN}\bigg(\frac{1}{R_{\rm C}}\bigg)\dot{\mathcal{S}}_{\rm C}^{qN}
+\sum_{n=0}^{\infty}\frac{(-1)^n}{(n+2)n!}\varepsilon_{ipq}\pd_{kpN}\bigg(\frac{1}{R_{\rm C}}\bigg){\mathcal{S}}_{\rm C}^{qN}v_{\B{\rm C}}^k
\nonumber\\
&&\phantom{++}+\sum_{n=1}^{\infty}\frac{(-1)^n}{(n+1)!}\pd_{pN}\bigg(\frac{1}{R_{\rm C}}\bigg)\dot{{\cal M}}_{\rm C}^{iN}v_{\B{\rm C}}^p+ \sum_{n=0}^{\infty}\frac{(-1)^n}{n!}\pd_{pN}\bigg(\frac{1}{R_{\rm C}}\bigg){{\cal M}}_{\rm C}^{N}v_{\B{\rm C}}^pv_{\rm C}^i \bigg]
\nonumber\\
&&\phantom{++}+(l-1-2\gamma)\Bigg[\sum_{n=0}^{\infty}\frac{(-1)^n}{n!}\pd_N\bigg(\frac{1}{R_{\rm C}}\bigg)\dot{{\cal M}}_{\rm C}^{N}v_{\rm B}^i
+\sum_{n=0}^{\infty}\frac{(-1)^n}{n!}\pd_{pN}\bigg(\frac{1}{R_{\rm C}}\bigg){{\cal M}}_{\rm C}^{N}v_{\B{\rm C}}^pv_{\rm B}^i\Bigg]\nonumber\\
&&\phantom{++}-\frac{l+1}2\sum_{n=0}^{\infty}\frac{(-1)^n}{n!}\pd_{pN}\bigg(\frac{1}{R_{\rm C}}\bigg){{\cal M}}_{\rm C}^{N}v^p_{\rm B}v_{\rm B}^i
-(l^2+l+2+2\gamma)\sum_{n=0}^{\infty}\frac{(-1)^n}{n!}\pd_N\bigg(\frac{1}{R_{\rm C}}\bigg){{\cal M}}_{\rm C}^{N}a_{\rm B}^i\nonumber\;.
 \ea
Notice that both potentials $\Omega_{\rm C}(t,{\bm x},l)$ and $\Omega^i_{\rm C}(t,{\bm x},l)$ depend on the multipolar index $l$ explicitly which should be taken into account when rendering summation in \eqref{yckope}.
m
\subsubsection{STF Derivatives from the Scalar Potentials \texorpdfstring{$\bar W$}{\bar W} and \texorpdfstring{$\bar U$}{\bar U}}\la{aaaaa1}

The force contains the STF derivatives from the scalar potentials $\bar W$ and $\bar U$ that appear in the first line of \eqref{sr5v6h}. The derivatives are computed with the help of expansions \eqref{je5v20} and observation that we can equate $U_{\rm C}$ to $W_{\rm C}$ in the post-Newtonian terms. Accounting for the fact that the partial derivative of any order from the inverse distance, $R^{-1}_{\rm C}$, is automatically STF Cartesian tensor because this function is a fundamental solution of the Laplace equation, we get
\ba\la{bumb3d4}
\sum_{l=0}^\infty\frac1{l!}\pd_{<iL>}\bar{W}(t,{\bm x}){\cal M}^L_\B&=&\sum_{l=0}^\infty\sum_{n=0}^{\infty}\frac{(-1)^n}{l!n!}\pd_{iLN}\bigg(\frac{1}{R_{\rm C}}\bigg){\cal M}_\B^L{\cal M}_{\rm C}^N\;,\\
\sum_{l=0}^\infty\frac1{l!}\pd_{<kL>}\bar{U}(t,{\bm x}) {\cal M}^L_\B&=&\sum_{l=0}^\infty\sum_{n=0}^{\infty}\frac{(-1)^n}{l!n!}\pd_{kLN}\bigg(\frac{1}{R_{\rm C}}\bigg){\cal M}_\B^L{\cal M}_{\rm C}^N\;,\\
\sum_{l=0}^\infty\frac1{l!}\pd_{<ipL>}\bar{U}(t,{\bm x}){\cal M}^{kL}_\B&=&\sum_{l=0}^\infty\sum_{n=0}^{\infty}\frac{(-1)^n}{l!n!}\pd_{ipLN}\bigg(\frac{1}{R_{\rm C}}\bigg){\cal M}_\B^{kL}{\cal M}_{\rm C}^N\;,
\ea 
where we have dropped the angular (STF) brackets around spatial indices of the partial derivatives from $R^{-1}_{\rm C}$ as they are redundant because the partial and STF derivatives of $R^{-1}_{\rm C}$ are identical, $\pd_{iLN}R^{-1}_{\rm C}=\pd_{<iLN>}R^{-1}_{\rm C}$, etc.
  
\subsubsection{STF Derivatives from the Scalar Potential \texorpdfstring{$\bar\Omega$}{\bar\Omega}}\la{aaaaa2}

Computation of the STF partial derivative from $\bar\Omega$ in the second line of equation \eqref{sr5v6h} for the force $F^i$, involves taking the partial derivatives from the coordinate distance $R_{\rm C}$. We already know that all the partial derivatives taken from the inverse distance, $R^{-1}_{\rm C}$, are automatically STF derivatives in the sense that $\pd_L R^{-1}_{\rm C}=\pd_{<L>} R^{-1}_{\rm C}$ for any number $l$ of indices. On the other hand, the partial derivatives from $R_{\rm C}$ are not the STF derivatives, that is $\pd_L R_{\rm C}\not=\pd_{<L>} R_{\rm C}$. The partial derivatives from $R_{\rm C}$ enter the forth line of formula \eqref{om3gfx6} for $\Omega_{\rm C}(t,{\bm x},l)$, and additional partial derivatives from $R_{\rm C}$ are taken in \eqref{sr5v6h} in the form of $\pd_{<iL>}\bar\Omega(t,{\bm x},l)$. The derivatives from $R_{\rm C}$ have to be converted to the STF partial derivatives in order to represents all terms in the equations of motion as expansions with respect to the STF Cartesian tensors. This is achieved by making use of the following complementary relation allowing to transform a partial derivative of order $p$ from $R_{\rm C}$ to its STF counterpart \citep[Equation A21{\it b}]{bld1986}: 
\be\la{n4g2x0}
\pd_{a_1a_2\ldots a_p}R_{\rm C}=\pd_{<a_1a_2\ldots a_p>}R_{{\rm C}}+\frac2{2p-1}\d_{\{a_1a_2}\pd_{a_3\ldots a_p\}}\bigg(\frac1{R_{{\rm C}}}\bigg)\;,
\ee
where the curl brackets around tensor indices denote a full symmetrization with respect to the smallest set of permutations $(1,2,\ldots,p)$ of the indices. 

Let us consider a transformation of the partial derivatives from $R_{\rm C}$ to the STF derivatives more explicitly. The first term in $\pd_{<iL>}\bar\Omega$ with the partial derivatives from $R_{\rm C}$ is proportional to ${\cal M}^L_{\B}\ddot{\cal M}^N_{{\rm C}}\pd_{<iL>N}R_{\rm C}$. It is converted to the STF derivative by applying \eqref{n4g2x0} in two steps. First, we use \eqref{n4g2x0} in reverse order,  
\ba\la{ger6c5}
{\cal M}^L_{\B}\ddot{\cal M}^N_{{\rm C}}\pd_{<iL>N}R_{\rm C}
&=&{\cal M}^L_{\B}\ddot{\cal M}^N_{{\rm C}}\left[\pd_{iLN}R_{\rm C}-\frac2{2l+1}\d_{\{ia_1}\pd_{a_2\ldots a_l\}b_1\ldots b_n}\bigg(\frac1{R_{{\rm C}}}\bigg)\right]\;,
\ea
with the purpose to get the symmetric partial derivative $\pd_{iLN}R_{\rm C}$ from the partial derivative $\pd_{<iL>N}R_{\rm C}$ which contains a mixture of the STF and symmetric derivatives. Second, we apply \eqref{n4g2x0} in direct order for converting the symmetric derivative $\pd_{iLN}R_{\rm C}$ to its STF counterpart,
\ba\la{in3v6}
{\cal M}^L_{\B}\ddot{\cal M}^N_{{\rm C}}\pd_{iLN}R_{\rm C}
&=&{\cal M}^L_{\B}\ddot{\cal M}^N_{{\rm C}}\left[\pd_{<iLN>}R_{{\rm C}}+\frac2{2l+2n+1}\d_{\{ia_1}\pd_{a_2\ldots a_lb_1\ldots b_n\}}\bigg(\frac1{R_{{\rm C}}}\bigg)\right]\;.
\ea
Expanding the symmetric permutation symbol in the second term of \eqref{in3v6} to a corresponding number of terms and remembering that the Laplacian from $R^{-1}_{\rm C}$ vanishes, $\Delta R^{-1}_{\rm C}=0$,  we eventually get,
\ba\la{in3v6qq}
{\cal M}^L_{\B}\ddot{\cal M}^N_{{\rm C}}\pd_{<iL>N}R_{\rm C}
&=&{\cal M}^L_{\B}\ddot{\cal M}^N_{{\rm C}}\pd_{<iLN>}R_{{\rm C}}
+\left[\frac{2l}{2l+2n+1}-\frac{2l}{2l+1}\right]{\cal M}^{iL-1}_{\B}\ddot{\cal M}^N_{{\rm C}}\pd_{<NL-1>}\bigg(\frac1{R_{{\rm C}}}\bigg)\\\nonumber
&&+\frac{2n}{2l+2n+1}{\cal M}^{L}_{\B}\ddot{\cal M}^{iN-1}_{{\rm C}}\pd_{<LN-1>}\bigg(\frac1{R_{{\rm C}}}\bigg)
\\\nonumber&&
+\frac{2ln}{2l+2n+1}{\cal M}^{pL-1}_{\B}\ddot{\cal M}^{pN-1}_{{\rm C}}\pd_{<iL-1N-1>}\bigg(\frac1{R_{{\rm C}}}\bigg)
\;.
\ea
Proceeding in a similar way, we get for two other partial derivatives from $R_C$,
\ba
\la{vet5v7}
{\cal M}^L_{\B}{\cal M}^N_{{\rm C}}v^p_{{\rm C}}v^q_{{\rm C}}\pd_{<iL>pqN}R_{\rm C}
&=&{\cal M}^L_{\B}{\cal M}^N_{{\rm C}}v^p_{{\rm C}}v^q_{{\rm C}}\pd_{<ipqLN>}R_{{\rm C}}
+\frac{2}{2l+2n+5}{\cal M}^{L}_{\B}{\cal M}^N_{{\rm C}}v^2_{{\rm C}}\pd_{<iLN>}\bigg(\frac1{R_{{\rm C}}}\bigg)
\\\nonumber
&&+\frac{4}{2l+2n+5}{\cal M}^{L}_{\B}{\cal M}^{N}_{{\rm C}}v^i_{{\rm C}}v^p_{{\rm C}}\pd_{<pLN>}\bigg(\frac1{R_{{\rm C}}}\bigg)
\\\nonumber
&&+\left[\frac{2l}{2l+2n+5}-\frac{2l}{2l+1}\right]{\cal M}^{iL-1}_{\B}{\cal M}^N_{{\rm C}}v^p_{{\rm C}}v^q_{{\rm C}}\pd_{<pqNL-1>}\bigg(\frac1{R_{{\rm C}}}\bigg)
\\\nonumber
&&+\frac{2n}{2l+2n+5}{\cal M}^{L}_{\B}{\cal M}^{iN-1}_{{\rm C}}v^p_{{\rm C}}v^q_{{\rm C}}\pd_{<pqLN-1>}\bigg(\frac1{R_{{\rm C}}}\bigg)
\\\nonumber
&&+\frac{4l}{2l+2n+5}{\cal M}^{qL-1}_{\B}{\cal M}^N_{{\rm C}}v^p_{{\rm C}}v^q_{{\rm C}}\pd_{<ipNL-1>}\bigg(\frac1{R_{{\rm C}}}\bigg)
\\\nonumber
&&+\frac{4n}{2l+2n+5}{\cal M}^{qN-1}_{{\rm C}}{\cal M}^L_{\B}v^p_{{\rm C}}v^q_{{\rm C}}\pd_{<ipLN-1>}\bigg(\frac1{R_{{\rm C}}}\bigg)
\\\nonumber
&&+
\frac{2ln}{2l+2n+5}{\cal M}^{kL-1}_{\B}{\cal M}^{kN-1}_{{\rm C}}v^p_{{\rm C}}v^q_{{\rm C}}\pd_{<ipqL-1N-1>}\bigg(\frac1{R_{{\rm C}}}\bigg)\;,
\\\la{mw9n7}
{\cal M}^L_{\B}{\cal M}^N_{{\rm C}}a^p_{{\rm C}}\pd_{<iL>pN}R_{\rm C}
&=&{\cal M}^L_{\B}{\cal M}^N_{{\rm C}}a^p_{{\rm C}}\pd_{<ipLN>}R_{{\rm C}}+\frac{2}{2l+2n+3}a^i_{{\rm C}}{\cal M}^{L}_{\B}{\cal M}^N_{{\rm C}}\pd_{<LN>}\bigg(\frac1{R_{{\rm C}}}\bigg)\\\nonumber
&&+\left[\frac{2l}{2l+2n+3}-\frac{2l}{2l+1}\right]{\cal M}^{iL-1}_{\B}{\cal M}^N_{{\rm C}}a^p_{{\rm C}}\pd_{<pNL-1>}\bigg(\frac1{R_{{\rm C}}}\bigg)\\\nonumber&&+\frac{2n}{2l+2n+3}{\cal M}^{L}_{\B}{\cal M}^{iN-1}_{{\rm C}}a^p_{{\rm C}}\pd_{<pLN-1>}\bigg(\frac1{R_{{\rm C}}}\bigg)\\\nonumber
&&+\frac{2l}{2l+2n+3}a^p_{{\rm C}}{\cal M}^{pL-1}_{\B}{\cal M}^N_{{\rm C}}\pd_{<iNL-1>}\bigg(\frac1{R_{{\rm C}}}\bigg)\\\nonumber&&+\frac{2n}{2l+2n+3}a^p_{{\rm C}}{\cal M}^{pN-1}_{{\rm C}}{\cal M}^L_{\B}\pd_{<iLN-1>}\bigg(\frac1{R_{{\rm C}}}\bigg)\\\nonumber
&&+
\frac{2ln}{2l+2n+3}{\cal M}^{qL-1}_{\B}{\cal M}^{qN-1}_{{\rm C}}a^p_{{\rm C}}\pd_{<ipL-1N-1>}\bigg(\frac1{R_{{\rm C}}}\bigg)\;.
\ea
Employing these relations to present all terms in  $\pd_{<iL>}\bar\Omega$ in the STF form, we compute its contribution to the force \eqref{sr5v6h} as follows 
\ba
\la{uv3f6l}
\sum_{l=0}^\infty\frac1{l!}\pd_{<iL>}\Omega_{\rm C}(t,{\bm x},l){\cal M}^L_\B&=&\sum_{l=0}^\infty\sum_{n=0}^{\infty}\frac{(-1)^n}{l!n!}\pd_{iLN}\bigg(\frac{1}{R_{\rm C}}\bigg)
\left[(1+\g) v_{\rm BC}^2-\frac12\frac{2l+2n+3}{2l+2n+5}v_{\rm C}^2\right]{\cal M}_\B^L{\cal M}_{\rm C}^N
\\
&+&\sum_{l=0}^\infty\sum_{n=0}^{\infty}\frac{(-1)^n}{l!n!}\pd_{iLN}\bigg(\frac{1}{R_{\rm C}}\bigg)\left[(2-2\beta-l\gamma)\bar U(t,{\bm x}_\B)-\gamma(n+1)\bar U_{\rm C}(t,{\bm x}_{\rm C})\right]{\cal M}_\B^L{\cal M}_{\rm C}^N\nonumber
\\
&+&\sum_{l=0}^\infty\sum_{n=0}^{\infty}\frac{(-1)^n}{l!n!}\pd_{ipLN}\bigg(\frac{1}{R_{\rm C}}\bigg)\bigg( \frac{1}2 v_{\rm C}^pv_{\rm C}^{k}- F_{\rm C}^{kp}\bigg){\cal M}_\B^L{\cal M}_{\rm C}^{kN}\nonumber
\\
&-&\sum_{l=0}^\infty\sum_{n=0}^{\infty}\frac{(-1)^n}{l!n!}(n+1)\pd_{iLN}\bigg(\frac{1}{R_{\rm C}}\bigg)a_{\rm C}^p{\cal M}_\B^L{\cal M}_{\rm C}^{pN}\nonumber
\\
&+&\sum_{l=0}^\infty\sum_{n=0}^{\infty}\frac{(-1)^n}{l!n!}\left[\frac1{2}\ddot{{\cal M}}_{\rm C}^N\pd_{<iLN>}R_{\rm C}+\frac1{2}{{\cal M}}_{\rm C}^Nv_{\rm C}^pv_{\rm C}^q\pd_{<ipqLN>}R_{\rm C}- \frac1{2}{{\cal M}}_{\rm C}^Na_{\rm C}^p\pd_{<ipLN>}R_{\rm C}\right]{\cal M}_\B^L\nonumber
\\
     &+&\sum_{l=0}^\infty\sum_{n=0}^{\infty}\frac{(-1)^n}{l!n!}\left[\frac{1}{2l+2n+3} -\frac1{2l+3}\right]\pd_{LN} \bigg(\frac{1}{R_{\rm C}}\bigg){\cal M}_\B^{iL}\ddot{\cal M}_{\rm C}^N\nonumber
     \\ 
&-&\sum_{l=0}^\infty\sum_{n=0}^{\infty}\frac{(-1)^n}{l!n!}\frac{1}{2l+2n+3} \pd_{LN} \bigg(\frac{1}{R_{\rm C}}\bigg){\cal M}_\B^{L}\ddot{\cal M}_{\rm C}^{iN} \nonumber
     \\ 
	&-&\sum_{l=0}^\infty\sum_{n=0}^{\infty}\frac{(-1)^n}{l!n!}\frac{1}{2l+2n+5} \pd_{iLN} \bigg(\frac{1}{R_{\rm C}}\bigg){\cal M}_\B^{pL}\ddot{\cal M}_{\rm C}^{pN} \nonumber
     \\ 
&+&\sum_{l=0}^\infty\sum_{n=0}^{\infty}\frac{(-1)^n}{l!n!}\frac{2}{2l+2n+5} \pd_{pLN} \bigg(\frac{1}{R_{\rm C}}\bigg){\cal M}_\B^{L}{\cal M}_{\rm C}^{N}v^p_{\rm C} v^i_{\rm C} \nonumber
     \\      
 &+&\sum_{l=0}^\infty\sum_{n=0}^{\infty}\frac{(-1)^n}{l!n!}\left[\frac{1}{2l+2n+7}-\frac1{2l+3}\right] \pd_{pqLN} \bigg(\frac{1}{R_{\rm C}}\bigg){\cal M}_\B^{iL}{\cal M}_{\rm C}^N v_{\rm C}^p v_{\rm C}^q \nonumber\\ 
&-&\sum_{l=0}^\infty\sum_{n=0}^{\infty}\frac{(-1)^n}{l!n!}\frac{1}{2l+2n+7} \pd_{pqLN} \bigg(\frac{1}{R_{\rm C}}\bigg){\cal M}_\B^{L}{\cal M}_{\rm C}^{iN}v_{\rm C}^p v_{\rm C}^q \nonumber\\ 
 &+&\sum_{l=0}^\infty\sum_{n=0}^{\infty}\frac{(-1)^n}{l!n!}\frac{2}{2l+2n+7} \pd_{ipLN} \bigg(\frac{1}{R_{\rm C}}\bigg)\left({\cal M}_\B^{qL}{\cal M}_{\rm C}^N-{\cal M}_\B^{L}{\cal M}_{\rm C}^{qN}\right)v_{\rm C}^p v_{\rm C}^q \nonumber\\ 
 &-&\sum_{l=0}^\infty\sum_{n=0}^{\infty}\frac{(-1)^n}{l!n!}\frac{1}{2l+2n+9} \pd_{ipqLN} \bigg(\frac{1}{R_{\rm C}}\bigg){\cal M}_\B^{kL}{\cal M}_{\rm C}^{kN}v_{\rm C}^p v_{\rm C}^q \nonumber\\ 
 &-&\sum_{l=0}^\infty\sum_{n=0}^{\infty}\frac{(-1)^n}{l!n!}\frac{1}{2l+2n+3} \pd_{LN} \bigg(\frac{1}{R_{\rm C}}\bigg){\cal M}_\B^{L}{\cal M}_{\rm C}^{N}a_{\rm C}^i \nonumber\\ 
 &-&\sum_{l=0}^\infty\sum_{n=0}^{\infty}\frac{(-1)^n}{l!n!}\left[\frac{1}{2l+2n+5}-\frac1{2l+3}\right] \pd_{pLN} \bigg(\frac{1}{R_{\rm C}}\bigg){\cal M}_\B^{iL}{\cal M}_{\rm C}^N a_{\rm C}^p  \nonumber\\ 
&+&\sum_{l=0}^\infty\sum_{n=0}^{\infty}\frac{(-1)^n}{l!n!}\frac{1}{2l+2n+5} \pd_{pLN} \bigg(\frac{1}{R_{\rm C}}\bigg){\cal M}_\B^{L}{\cal M}_{\rm C}^{iN}a_{\rm C}^p  \nonumber\\ 
 &-&\sum_{l=0}^\infty\sum_{n=0}^{\infty}\frac{(-1)^n}{l!n!}\frac{1}{2l+2n+5} \pd_{iLN} \bigg(\frac{1}{R_{\rm C}}\bigg)\left({\cal M}_\B^{pL}{\cal M}_{\rm C}^N-{\cal M}_\B^{L}{\cal M}_{\rm C}^{pN}\right)a_{\rm C}^p  \nonumber\\ 
 &+&\sum_{l=0}^\infty\sum_{n=0}^{\infty}\frac{(-1)^n}{l!n!}\frac{1}{2l+2n+7} \pd_{ipLN} \bigg(\frac{1}{R_{\rm C}}\bigg){\cal M}_\B^{qL}{\cal M}_{\rm C}^{qN}a_{\rm C}^p  \nonumber\\ 
  &-&2(1+\gamma)\sum_{l=0}^\infty\sum_{n=0}^{\infty}\frac{(-1)^n}{l!n!}\frac1{n+2}\varepsilon_{kpq}\pd_{ipLN}\bigg(\frac{1}{R_{\rm C}}\bigg){\cal M}_\B^L\mathcal{S}_{\rm C}^{qN}v_{\B{\rm C}}^k\nonumber\\
 &-&2(1+\gamma)\sum_{l=0}^\infty\sum_{n=0}^{\infty}\frac{(-1)^n}{l!n!}\frac1{n+1}\pd_{iLN}\bigg(\frac{1}{R_{\rm C}}\bigg){\cal M}_\B^L\dot{{\cal M}}_{\rm C}^{pN}v_{\B{\rm C}}^p
 \nonumber\;,
\ea
where, in the fifth line of this long formula, we keep the angular brackets around indices of the spatial derivatives from $R_{\rm C}$ to make clear that these are the STF partial derivatives from $R_{\rm C}$.

\subsubsection{STF Derivatives from the Vector Potential \texorpdfstring{$\bar\Omega^i$}{\bar\Omega^i}}\la{aaaaa3}

Our next step is the computation of the STF derivative $\pd_{<L}\bar\Omega_{i>}(t,{\bm x},l)$ that appears in the third line of equation \eqref{sr5v6h} for force $F^i$. Calculation of this term is based on application of the index peeling-off formula \eqref{hh33e5z} which yields, 
\ba\la{me7v20}
\sum_{l=0}^\infty\frac1{l!}\pd_{<L}\bar\Omega_{i>}(t,{\bm x},l){\cal M}_\B^L&=&\sum_{l=0}^\infty\frac1{(l+1)!}\pd_{<L>}\bar \Omega_{i}(t,{\bm x},l){\cal M}_\B^L+\sum_{l=0}^\infty\frac{l}{(l+1)!}\pd_{i<L-1}\bar \Omega_{p>}(t,{\bm x},l){\cal M}_\B^{pL-1}\\\nonumber
&-&\sum_{l=0}^\infty\frac1{(l+1)!}\frac{2l}{2l+1}\pd_{p<L-1>}\bar \Omega_{p}(t,{\bm x},l){\cal M}_\B^{iL-1}\;,
\ea
that helps to disentangle one index from the remaining STF indices and simplifies computation of the partial derivative. 
Vector potential $\bar\Omega_i$ is given by the last term in \eqref{popka34} as a linear superposition of vector-potentials $\Omega^i_{\rm C}$ of bodies with index C$\not=$B that are external with respect to body B. Applying \eqref{me7v20} to the individual $\Omega^i_{\rm C}$ defined in \eqref{omi84v6}, we obtain the first term in the right-hand side of \eqref{me7v20},
\ba
\la{hye5b3}
\sum_{l=0}^\infty\frac1{(l+1)!}\pd_{L}\Omega^i_{\rm C}(t,{\bm x},l){\cal M}_\B^L&=&2(1+\gamma)\sum_{l=0}^\infty\sum_{n=0}^{\infty}\frac{(-1)^n}{(l+1)!n!}\frac1{n+1}\pd_{LN}\bigg(\frac{1}{R_{\rm C}}\bigg){\cal M}_\B^L\ddot{{\cal M}}_{\rm C}^{iN}\\\nonumber
&+&2(1+\g)\sum_{l=0}^\infty\sum_{n=0}^{\infty}\frac{(-1)^n}{(l+1)!n!}\pd_{LN}\bigg(\frac{1}{R_{\rm C}}\bigg)\left(\dot{{\cal M}}_{\rm C}^{N}v_{\rm C}^i+{{\cal M}}_{\rm C}^{N}a_{\rm C}^i \right){\cal M}_\B^L
\\\nonumber
&+&2(1+\g)\sum_{l=0}^\infty\sum_{n=0}^{\infty}\frac{(-1)^n}{(l+1)!n!}\frac1{n+2}\varepsilon_{ipq}\pd_{pLN}\bigg(\frac{1}{R_{\rm C}}\bigg){\cal M}_\B^L\dot{\mathcal{S}}_{\rm C}^{qN}\\\nonumber
&+&2(1+\g)\sum_{l=0}^\infty\sum_{n=0}^{\infty}\frac{(-1)^n}{(l+1)!n!}\frac1{n+2}\varepsilon_{ipq}\pd_{kpLN}\bigg(\frac{1}{R_{\rm C}}\bigg){\cal M}_\B^L{\mathcal{S}}_{\rm C}^{qN}v_{\B{\rm C}}^k
\\\nonumber
&+&2(1+\g)\sum_{l=0}^\infty\sum_{n=1}^{\infty}\frac{(-1)^n}{(l+1)!n!}\frac1{n+1}\pd_{pLN}\bigg(\frac{1}{R_{\rm C}}\bigg){\cal M}_\B^L\dot{{\cal M}}_{\rm C}^{iN}v_{\B{\rm C}}^p\\\nonumber
&+&2(1+\g)\sum_{l=0}^\infty \sum_{n=0}^{\infty}\frac{(-1)^n}{(l+1)!n!}\pd_{pLN}\bigg(\frac{1}{R_{\rm C}}\bigg){\cal M}_\B^L{{\cal M}}_{\rm C}^{N}v_{\B{\rm C}}^pv_{\rm C}^i 
\\\nonumber
&+&\sum_{l=0}^\infty\sum_{n=0}^{\infty}\frac{(-1)^n}{(l+1)!n!}\Big(l-1-2\gamma\Big)\pd_{LN}\bigg(\frac{1}{R_{\rm C}}\bigg){\cal M}_\B^L\dot{{\cal M}}_{\rm C}^{N}v_{\rm B}^i\\\nonumber
&+&\sum_{l=0}^\infty\sum_{n=0}^{\infty}\frac{(-1)^n}{(l+1)!n!}\Big(l-1-2\gamma\Big)\pd_{pLN}\bigg(\frac{1}{R_{\rm C}}\bigg){\cal M}_\B^L{{\cal M}}_{\rm C}^{N}v_{\B{\rm C}}^pv_{\rm B}^i\\\nonumber
&-&\frac{1}2\sum_{l=0}^\infty\sum_{n=0}^{\infty}\frac{(-1)^n}{l!n!}\pd_{pLN}\bigg(\frac{1}{R_{\rm C}}\bigg){\cal M}_\B^L{{\cal M}}_{\rm C}^{N}v^p_{\rm B}v_{\rm B}^i\\\nonumber
&-&\sum_{l=0}^\infty\sum_{n=0}^{\infty}\frac{(-1)^n}{(l+1)!n!}\Big(l^2+l+2+2\gamma\Big)\pd_{LN}\bigg(\frac{1}{R_{\rm C}}\bigg){\cal M}_\B^L{{\cal M}}_{\rm C}^{N}a_{\rm B}^i\;.
\ea
In order to compute the second and third terms in the right-hand side of \eqref{me7v20} it is useful to reformulate them by changing the index of summation, $l\rightarrow l+1$, which also replaces STF index $L-1\rightarrow L$. This procedure is convenient for reduction of similar terms in the final equation for the force which consists of the contributions of many separate pieces. We have,
\ba\label{92vs43}
\sum_{l=0}^\infty\frac{l}{(l+1)!}\pd_{iL-1}\bar \Omega_{p}(t,{\bm x},l){\cal M}_\B^{pL-1}&=&\sum_{l=0}^\infty\frac{1}{l!(l+2)}\pd_{iL}\bar \Omega_{p}(t,{\bm x},l+1){\cal M}_\B^{pL}
\\\label{b4v1c7}
\sum_{l=0}^\infty\frac1{(l+1)!}\frac{2l}{2l+1}\pd_{pL-1}\bar \Omega_{p}(t,{\bm x},l){\cal M}_\B^{iL-1}&=&\sum_{l=0}^\infty\frac{1}{(l+2)!}\frac{2(l+1)}{2l+3}\pd_{pL}\bar \Omega_{p}(t,{\bm x},l+1){\cal M}_\B^{iL}\;,
\ea
and the STF derivatives are
\ba
\la{yrtr3}
\sum_{l=0}^\infty\frac{1}{l!(l+2)}\pd_{iL}\Omega^p_{\rm C}(t,{\bm x},l+1){\cal M}_\B^{pL}&=&2(1+\gamma)\sum_{l=0}^\infty\sum_{n=0}^{\infty}\frac{(-1)^n}{l!(n+1)!}\frac1{l+2}\pd_{iLN}\bigg(\frac{1}{R_{\rm C}}\bigg){\cal M}_\B^{pL}\ddot{{\cal M}}_{\rm C}^{pN}\\\nonumber
&+&2(1+\g)\sum_{l=0}^\infty\sum_{n=0}^{\infty}\frac{(-1)^n}{l!n!}\frac1{l+2}\pd_{iLN}\bigg(\frac{1}{R_{\rm C}}\bigg)\left(\dot{{\cal M}}_{\rm C}^{N}v_{\rm C}^p+{{\cal M}}_{\rm C}^{N}a_{\rm C}^p \right){\cal M}_\B^{pL}
\\\nonumber
&+&2(1+\g)\sum_{l=0}^\infty\sum_{n=0}^{\infty}\frac{(-1)^n(n+1)(l+1)}{(l+2)!(n+2)!}\varepsilon_{pkq}\pd_{ikLN}\bigg(\frac{1}{R_{\rm C}}\bigg){\cal M}_\B^{pL}\dot{\mathcal{S}}_{\rm C}^{qN}\\\nonumber
&+&2(1+\g)\sum_{l=0}^\infty\sum_{n=0}^{\infty}\frac{(-1)^n(n+1)(l+1)}{(l+2)!(n+2)!}\varepsilon_{pkq}\pd_{ijkLN}\bigg(\frac{1}{R_{\rm C}}\bigg){\cal M}_\B^{pL}{\mathcal{S}}_{\rm C}^{qN}v_{\B{\rm C}}^j
\\\nonumber
&+&2(1+\g)\sum_{l=0}^\infty\sum_{n=0}^{\infty}\frac{(-1)^n}{l!(n+1)!}\frac1{l+2}\pd_{ijLN}\bigg(\frac{1}{R_{\rm C}}\bigg){\cal M}_\B^{pL}\dot{{\cal M}}_{\rm C}^{pN}v_{\B{\rm C}}^j\\\nonumber
&+&2(1+\g)\sum_{l=0}^\infty \sum_{n=0}^{\infty}\frac{(-1)^n}{l!n!}\frac1{l+2}\pd_{ijLN}\bigg(\frac{1}{R_{\rm C}}\bigg){\cal M}_\B^{pL}{{\cal M}}_{\rm C}^{N}v_{\B{\rm C}}^jv_{\rm C}^p 
\\\nonumber
&+&\sum_{l=0}^\infty\sum_{n=0}^{\infty}\frac{(-1)^n}{l!n!}\frac{l-2\gamma}{l+2}\pd_{iLN}\bigg(\frac{1}{R_{\rm C}}\bigg){\cal M}_\B^{pL}\dot{{\cal M}}_{\rm C}^{N}v_{\rm B}^p\\\nonumber
&+&\sum_{l=0}^\infty\sum_{n=0}^{\infty}\frac{(-1)^n}{l!n!}\frac{l-2\gamma}{l+2}\pd_{ijLN}\bigg(\frac{1}{R_{\rm C}}\bigg){\cal M}_\B^{pL}{{\cal M}}_{\rm C}^{N}v_{\B{\rm C}}^jv_{\rm B}^p\\\nonumber
&-&\frac{1}2\sum_{l=0}^\infty\sum_{n=0}^{\infty}\frac{(-1)^n}{l!n!}\pd_{ijLN}\bigg(\frac{1}{R_{\rm C}}\bigg){\cal M}_\B^{pL}{{\cal M}}_{\rm C}^{N}v^j_{\rm B}v_{\rm B}^p\\\nonumber
&-&\sum_{l=0}^\infty\sum_{n=0}^{\infty}\frac{(-1)^n}{l!n!}\frac{l^2+3l+4+2\gamma}{l+2}\pd_{iLN}\bigg(\frac{1}{R_{\rm C}}\bigg){\cal M}_\B^{pL}{{\cal M}}_{\rm C}^{N}a_{\rm B}^p\;,
 \\\nonumber\\\nonumber
 \ea
\ba\hspace{-1cm}
\la{u5bz4o}
\sum_{l=0}^\infty\frac1{(l+2)!}\frac{2(l+1)}{2l+3}\pd_{pL}\Omega^p_{\rm C}(t,{\bm x},l+1){\cal M}_\B^{iL}&=&-\sum_{l=0}^\infty\sum_{n=0}^{\infty}\frac{4(1+\gamma)(-1)^n}{(l+2)!n!}\frac{l+1}{2l+3}\pd_{LN}\bigg(\frac{1}{R_{\rm C}}\bigg){\cal M}_\B^{iL}\ddot{{\cal M}}_{\rm C}^{N}\\\nonumber
&+&\sum_{l=0}^\infty\sum_{n=0}^{\infty}\frac{4(1+\gamma)(-1)^n}{(l+2)!n!}\frac{l+1}{2l+3}\pd_{pLN}\bigg(\frac{1}{R_{\rm C}}\bigg)\left(\dot{{\cal M}}_{\rm C}^{N}v_{\rm C}^p+{{\cal M}}_{\rm C}^{N}a_{\rm C}^p \right){\cal M}_\B^{iL}
\\\nonumber
&-&\sum_{l=0}^\infty\sum_{n=0}^{\infty}\frac{4(1+\gamma)(-1)^n}{(l+2)!n!}\frac{l+1}{2l+3}\pd_{jLN}\bigg(\frac{1}{R_{\rm C}}\bigg){\cal M}_\B^{iL}\dot{{\cal M}}_{\rm C}^{N}v_{\B{\rm C}}^j\\\nonumber
&+&\sum_{l=0}^\infty \sum_{n=0}^{\infty}\frac{4(1+\gamma)(-1)^n}{(l+2)!n!}\frac{l+1}{2l+3}\pd_{pjLN}\bigg(\frac{1}{R_{\rm C}}\bigg){\cal M}_\B^{iL}{{\cal M}}_{\rm C}^{N}v_{\B{\rm C}}^jv_{\rm C}^p 
\\\nonumber
&+&\sum_{l=0}^\infty\sum_{n=0}^{\infty}\frac{2(l-2\gamma)(-1)^n}{(l+2)!n!}\frac{l+1}{2l+3}\pd_{pLN}\bigg(\frac{1}{R_{\rm C}}\bigg){\cal M}_\B^{iL}\dot{{\cal M}}_{\rm C}^{N}v_{\rm B}^p\\\nonumber
&+&\sum_{l=0}^\infty\sum_{n=0}^{\infty}\frac{2(l-2\gamma)(-1)^n}{(l+2)!n!}\frac{l+1}{2l+3}\pd_{pjLN}\bigg(\frac{1}{R_{\rm C}}\bigg){\cal M}_\B^{iL}{{\cal M}}_{\rm C}^{N}v_{\B{\rm C}}^jv_{\rm B}^p\\\nonumber
&-&\sum_{l=0}^\infty\sum_{n=0}^{\infty}\frac{(-1)^n}{l!n!}\frac{1}{2l+3}\pd_{pjLN}\bigg(\frac{1}{R_{\rm C}}\bigg){\cal M}_\B^{iL}{{\cal M}}_{\rm C}^{N}v^j_{\rm B}v_{\rm B}^p\\\nonumber
&-&\sum_{l=0}^\infty\sum_{n=0}^{\infty}\frac{(-1)^n(2l^2+6l+8+4\gamma)}{(l+2)!n!}\frac{l+1}{2l+3}\pd_{pLN}\bigg(\frac{1}{R_{\rm C}}\bigg){\cal M}_\B^{iL}{{\cal M}}_{\rm C}^{N}a_{\rm B}^p\;,
 \ea

\subsubsection{Post-Newtonian Local Acceleration \texorpdfstring{${\cal Q}_i^{\rm pN}$}{{\cal Q}_i^{\rm pN}}}\la{aaaaa4}
 
We also need to express the post-Newtonian part ${\cal Q}_i^{\rm pN}$ of the local acceleration \eqref{nrvug9} of body B explicitly as a function of the STF mass and spin internal multipoles of all bodies in $\mathbb{N}$-body system. For completing this task we, first of all, transform the terms in the first three lines of expression \eqref{nrvug9} for ${\cal Q}_i^{\rm pN}$ by making use of the fact that the external gravitoelectric multipoles ${\cal Q}_L={\cal P}_L$ for $l\ge 2$ in all post-Newtonian terms. After accounting for this equality, equation \eqref{nrvug9} can be reshuffled to the following form, 
\ba\la{ger9b3}
M_\B {\cal Q}_i^{\rm pN}&=&3\left({\cal Q}_k\ddot{\cal M}_{\B}^{ik}+2\dot {\cal Q}_k\dot{\cal M}_{\B}^{ik}+\ddot {\cal Q}_k{\cal M}_{\B}^{ik}\right)+(\g-1)\left({\cal P}_k\ddot{\cal M}_{\B}^{ik}+\frac32\dot {\cal P}_k\dot{\cal M}_{\B}^{ik}+\frac35\ddot {\cal P}_k{\cal M}_{\B}^{ik}\right)\\\nonumber
&+&\sum_{l=2}^{\infty}\frac{l^2+l+2+2\gamma}{(l+1)!}{\cal P}_L \ddot{{\cal M}}_\B^{iL}
   + \sum_{l=2}^{\infty}\frac{2l+1}{l+1}\frac{l^2+2l+3+2\gamma}{(l+1)!}\dot{{\cal P}}_L\dot{{\cal M}}_\B^{iL}\\
  & +& \sum_{l=2}^{\infty}\frac{2l+1}{2l+3}\frac{l^2+3l+4+2\gamma}{(l+1)!}\ddot{{\cal P}}_L{\cal M}_\B^{iL}-\sum_{l=1}^{\infty}\frac{l}{(l+1)!}{\cal C}_{iL}\mathcal{S}_\B^{L}\nonumber\\
& +& \sum_{l=1}^{\infty}\frac{1}{(l+1)!}\varepsilon_{ipq}\bigg[{\cal C}_{pL}\dot{{\cal M}}^{<qL>}+\frac{l+1}{l+2}\dot{\cal C}_{pL}{\cal M}^{<qL>}\bigg]\nonumber\\
& -& 2(1+\g)\sum_{l=1}^{\infty}\frac{l+1}{(l+2)!}\varepsilon_{ipq}\bigg[{\cal P}_{pL}\dot{\mathcal{S}}_\B^{qL}
 + \frac{l+1}{l+2}\dot{{\cal P}}_{pL}\mathcal{S}_\B^{qL}\bigg]\nonumber\\
&-& \frac{1}{2}\varepsilon_{ikq}\bigg[(4{\cal Q}_k+2(\gamma-1){\cal P}_k)\dot{\mathcal{S}}_\B^q+(2\dot{\cal Q}_k+(\gamma-1)\dot{{\cal P}}_k)\mathcal{S}_\B^q\bigg]
\;.\nonumber
\ea
At the second step of the computation, we take advantage of equation of motion \eqref{5.8} in the global coordinates to replace the local acceleration ${\cal Q}_i$ everywhere in \eqref{ger9b3} with its global counterpart $a^i_\B$. The Newtonian approximation is sufficient, ${\cal Q}_i=\pd_i\bar U(t,{\bm x}_\B)-a^i_\B={\cal P}^i-a^i_\B$, where we employed $\pd_i\bar U(t,{\bm x}_\B)={\cal P}_i$ in accordance with definition of the external scalar-field multipoles  provided in section \ref{ex5m3}. Proceeding in this way, we obtain  
\ba\la{g9b3q1}
M_\B {\cal Q}_i^{\rm pN}&=&
\sum_{l=0}^{\infty}\frac{l^2+l+2+2\gamma}{(l+1)!}{\cal P}_L \ddot{{\cal M}}_\B^{iL}
   + \sum_{l=0}^{\infty}\frac{2l+1}{l+1}\frac{l^2+2l+3+2\gamma}{(l+1)!}\dot{{\cal P}}_L\dot{{\cal M}}_\B^{iL}\\
  & +& \sum_{l=0}^{\infty}\frac{2l+1}{2l+3}\frac{l^2+3l+4+2\gamma}{(l+1)!}\ddot{{\cal P}}_L{\cal M}_\B^{iL}-\sum_{l=0}^{\infty}\frac{l}{(l+1)!}{\cal C}_{iL}\mathcal{S}_\B^{L}\nonumber\\
& +& \sum_{l=0}^{\infty}\frac{1}{(l+1)!}\varepsilon_{ipq}\bigg[{\cal C}_{pL}\dot{{\cal M}}_\B^{qL}+\frac{l+1}{l+2}\dot{\cal C}_{pL}{\cal M}_\B^{qL}\bigg]\nonumber\\
& -& 2(1+\g)\sum_{l=0}^{\infty}\frac{l+1}{(l+2)!}\varepsilon_{ipq}\bigg[{\cal P}_{pL}\dot{\mathcal{S}}_\B^{qL}
 + \frac{l+1}{l+2}\dot{{\cal P}}_{pL}\mathcal{S}_\B^{qL}\bigg]\nonumber\\
&+&\varepsilon_{ikq}\bigg(2a^k_\B\dot{\mathcal{S}}_\B^q+\dot{a}^k_\B\mathcal{S}_\B^q\bigg)-3\left(a_\B^k\ddot{\cal M}_{\B}^{ik}+2\dot{a}_\B^k\dot{\cal M}_{\B}^{ik}+\ddot{a}_\B^k{\cal M}_{\B}^{ik}\right)\nonumber
\;,
\ea
where we have formally extended summation to value $l=0$ in all sums by taking into account that in terms of the post-Newtonian order of magnitude the {\it active} dipole of each body vanishes, ${\cal M}^i_\B=0$. 

The external multipoles, ${\cal P}_L$ and ${\cal C}_L$ are expressed in terms of the external gravitational potentials, $\bar U$ and $\bar U^i$ of the body B with the help of \eqref{3.13}, \eqref{scal123} and \eqref{3.29} respectively. Particular attention should be paid to the term ${\cal C}_{iL}{\cal S}_\B^L$. After a few algebraic transformations it becomes,
\be
{\cal C}_{iL}{\cal S}_\B^L=\d_{pq}{\cal C}_{ipL-1}{\cal S}_\B^{qL-1}=\frac12\varepsilon_{jpk}\varepsilon_{jqk}{\cal C}_{piL-1}{\cal S}_\B^{qL-1}=\frac12\varepsilon_{jqk}\bar H_{jk<iL-1>}{\cal S}^{qL-1}\;,
\ee
where, at the last step, we have used \eqref{he5v2z}.
After substituting $\bar H_{jk<iL-1>}$ from \eqref{ny3v6z5}, \eqref{ne8v40} to the above expression and noticing that contraction of any two indices in STF multipole ${\cal S}_\B^L$ vanishes, we get
\be
{\cal C}_{iL}{\cal S}_\B^L=2(1+\g)\varepsilon_{jqk}\left[v^j_\B\pd_{ikL-1}\bar U(t,{\bm x}_\B)-\pd_{ikL-1}\bar U^j(t,{\bm x}_\B)\right]{\cal S}^{qL-1}_\B+\frac{2(1+\g)}{l+1}\varepsilon_{ijq}\pd_{jL-1}\dot{\bar U}(t,{\bm x}_\B){\cal S}^{qL-1}_\B\;.
\ee
After implementing this and other replacements of the external multipoles in \eqref{g9b3q1} with the corresponding external global potentials, and reducing similar terms, the post-Newtonian local acceleration takes on the following form,
\ba\la{c5s0k2f}
M_\B {\cal Q}_i^{\rm pN}&=&\bar\Xi_{\rm C}^i(t,{\bm x}_\B)
+\varepsilon_{ikq}\left(2a^k_\B\dot{\mathcal{S}}_\B^q+\dot{a}^k_\B\mathcal{S}_\B^q\right)-3\left(a_\B^k\ddot{\cal M}_{\B}^{ik}+2\dot{a}_\B^k\dot{\cal M}_{\B}^{ik}+\ddot{a}_\B^k{\cal M}_{\B}^{ik}\right)\;,
\ea
where the first term 
\be
\bar\Xi_{\rm C}^i(t,{\bm x})=\sum_{{\rm C}\not=\B}\Xi_{\rm C}^i(t,{\bm x})\;,
\ee
is a linear superposition of vectors
\ba\la{6c0ea7}
\Xi_{\rm C}^i(t,{\bm x})&=&
\sum_{l=0}^{\infty}\frac{l^2+l+2+2\gamma}{(l+1)!}\pd_L U_{\rm C}(t,{\bm x}) \ddot{{\cal M}}_\B^{iL}\\\nonumber
   &+& \sum_{l=0}^{\infty}\frac{2l^2+3l+3+2\gamma}{(l+1)!}\pd_L\dot{ U}_{\rm C}(t,{\bm x})\dot{{\cal M}}_\B^{iL}\\
  & +& \sum_{l=0}^{\infty}\frac{2l^3 +9l^2+12l+8+4\gamma}{(l+2)(2l+3)l!}\pd_L\ddot{ U}_{\rm C}(t,{\bm x}){\cal M}_\B^{iL}\nonumber\\
 & +&4(1+\g) \sum_{l=0}^{\infty}\frac{1}{(l+1)!}\bigg[v_\B^{[i}\pd^{k]L} U_{\rm C}(t,{\bm x}) +\pd^{L[i} U_{\rm C}^{k]}(t,{\bm x}) \bigg]\dot{{\cal M}}_\B^{kL}\nonumber\\
& +&4(1+\g) \sum_{l=0}^{\infty}\frac{1}{(l+2)l!}\bigg[a_\B^{[i}\pd^{k]L} U_{\rm C}(t,{\bm x})+v_\B^{[i}\pd^{k]L}\dot{ U}_{\rm C}(t,{\bm x}) +\pd^{L[i}\dot{ U}_{\rm C}^{k]}(t,{\bm x}) \bigg]{{\cal M}}_\B^{kL}\nonumber\\
&+&2(1+\g)\sum_{l=0}^{\infty}\frac{l+1}{(l+2)!}\varepsilon_{jkq}\left[v^j_\B\pd_{ikL}U_{\rm C}(t,{\bm x})-\pd_{ikL}\bar U^j_{\rm C}(t,{\bm x})\right]{\cal S}^{qL}_\B\nonumber\\
&-&2(1+\g)\sum_{l=0}^{\infty}\frac{l+1}{(l+2)!}\varepsilon_{ikq}\pd_{kL} \dot U_{\rm C}(t,{\bm x}){\cal S}_\B^{qL} \nonumber\\
&  -&2(1+\g)\sum_{l=0}^{\infty}\frac{l+1}{(l+2)!}\varepsilon_{ikq}\pd_{kL} U_{\rm C}(t,{\bm x})\dot{\cal S}_\B^{qL} \nonumber
\;.
\ea
Finally, after making use of multipolar decompositions of potentials $U_{\rm C}=W_{\rm C}$ and $U_{\rm C}^i$ given in \eqref{je5v20} and \eqref{w1w8} respectively, vector $\Xi^i_{\rm C}$ becomes
\ba\la{d4x01b}
\Xi_{\rm C}^i(t,{\bm x})&=&
\sum_{l=0}^{\infty}\sum_{n=0}^{\infty}\frac{(-1)^n}{n!}\frac{l^2+l+2+2\gamma}{(l+1)!}\pd_{LN} \bigg(\frac1{R_{\rm C}}\bigg) \ddot{{\cal M}}_\B^{iL}{\cal M}_{\rm C}^N
\\\nonumber
   &+& \sum_{l=0}^{\infty}\sum_{n=0}^{\infty}\frac{(-1)^n}{n!}\frac{2l^2+3l+3+2\gamma}{(l+1)!}\pd_{LN}\bigg(\frac1{R_{\rm C}}\bigg)\dot{{\cal M}}_\B^{iL}\dot{\cal M}_{\rm C}^N
   \\\nonumber
   &+& \sum_{l=0}^{\infty}\sum_{n=0}^{\infty}\frac{(-1)^n}{n!}\frac{2l^2+3l+3+2\gamma}{(l+1)!}\pd_{pLN}\bigg(\frac1{R_{\rm C}}\bigg)\dot{{\cal M}}_\B^{iL}{\cal M}_{\rm C}^Nv^p_{\B{\rm C}}
   \\
  & +& \sum_{l=0}^{\infty}\sum_{n=0}^{\infty}\frac{(-1)^n}{n!}\frac{2l^3 +9l^2+12l+8+4\gamma}{(l+2)(2l+3)l!}\pd_{LN}\bigg(\frac1{R_{\rm C}}\bigg){\cal M}_\B^{iL}\ddot{\cal M}_{\rm C}^N\nonumber
  \\
  & +& \sum_{l=0}^{\infty}\sum_{n=0}^{\infty}\frac{(-1)^n}{n!}\frac{4l^3 +18l^2+24l+16+8\gamma}{(l+2)(2l+3)l!}\pd_{pLN}\bigg(\frac1{R_{\rm C}}\bigg){\cal M}_\B^{iL}\dot{\cal M}_{\rm C}^Nv^p_{\B{\rm C}}\nonumber
  \\ 
 & +& \sum_{l=0}^{\infty}\sum_{n=0}^{\infty}\frac{(-1)^n}{n!}\frac{2l^3 +9l^2+12l+8+4\gamma}{(l+2)(2l+3)l!}\pd_{pLN}\bigg(\frac1{R_{\rm C}}\bigg){\cal M}_\B^{iL}{\cal M}_{\rm C}^Na^p_{\B{\rm C}}\nonumber
 \\   
 & +& \sum_{l=0}^{\infty}\sum_{n=0}^{\infty}\frac{(-1)^n}{n!}\frac{2l^3 +9l^2+12l+8+4\gamma}{(l+2)(2l+3)l!}\pd_{pqLN}\bigg(\frac1{R_{\rm C}}\bigg){\cal M}_\B^{iL}{\cal M}_{\rm C}^N
  v^p_{\B{\rm C}}v^q_{\B{\rm C}}\nonumber
  \\    
  &+&2(1+\g) \sum_{l=0}^{\infty}\sum_{n=0}^{\infty}\frac{(-1)^n}{(l+1)!n!}
 \bigg[\pd_{pLN}\bigg(\frac1{R_{\rm C}}\bigg)\dot{\cal M}_\B^{pL}{\cal M}_{\rm C}^N v_{\B{\rm C}}^i
 -\pd_{iLN}\bigg(\frac1{R_{\rm C}}\bigg)\dot{\cal M}_\B^{pL}{\cal M}_{\rm C}^Nv_{\B{\rm C}}^p\bigg]\nonumber
  \\  
  &+&2(1+\g) \sum_{l=0}^{\infty}\sum_{n=0}^{\infty}\frac{(-1)^n}{(l+1)!(n+1)!}
 \bigg[\pd_{iLN}\bigg(\frac1{R_{\rm C}}\bigg)\dot{\cal M}_\B^{pL}\dot{\cal M}_{\rm C}^{pN}-\pd_{pLN}\bigg(\frac1{R_{\rm C}}\bigg)\dot{\cal M}_\B^{pL}\dot{\cal M}_{\rm C}^{iN} 
\bigg]\nonumber
  \\   
&+&2(1+\g) \sum_{l=0}^{\infty}\sum_{n=0}^{\infty}\frac{(-1)^n(n+1)}{(l+1)!(n+2)!}\bigg[\varepsilon_{kpq}\pd_{ipLN}\bigg(\frac1{R_{\rm C}}\bigg)\dot{\cal M}_\B^{kL}{\cal S}_{\rm C}^{qN} -\varepsilon_{ipq}\pd_{kpLN}\bigg(\frac1{R_{\rm C}}\bigg)\dot{\cal M}_\B^{kL}{\cal S}_{\rm C}^{qN}       \bigg]  \nonumber
  \\   
&+&2(1+\g) \sum_{l=0}^{\infty}\sum_{n=0}^{\infty}\frac{(-1)^n}{n!}\frac{(l+1)}{(l+2)!}
 \bigg[\pd_{pLN}\bigg(\frac1{R_{\rm C}}\bigg){\cal M}_\B^{pL}{\cal M}_{\rm C}^N a_{\B{\rm C}}^i
 -\pd_{iLN}\bigg(\frac1{R_{\rm C}}\bigg){\cal M}_\B^{pL}{\cal M}_{\rm C}^Na_{\B{\rm C}}^p\bigg]\nonumber
  \\  
  &+&2(1+\g) \sum_{l=0}^{\infty}\sum_{n=0}^{\infty}\frac{(-1)^n}{n!}\frac{(l+1)}{(l+2)!}
 \bigg[\pd_{pLN}\bigg(\frac1{R_{\rm C}}\bigg){\cal M}_\B^{pL}\dot{\cal M}_{\rm C}^N v_{\B{\rm C}}^i
 -\pd_{iLN}\bigg(\frac1{R_{\rm C}}\bigg){\cal M}_\B^{pL}\dot{\cal M}_{\rm C}^Nv_{\B{\rm C}}^p\bigg]\nonumber
  \\  
 &+&2(1+\g) \sum_{l=0}^{\infty}\sum_{n=0}^{\infty}\frac{(-1)^n}{n!}\frac{(l+1)}{(l+2)!}
 \bigg[\pd_{kpLN}\bigg(\frac1{R_{\rm C}}\bigg){\cal M}_\B^{pL}{\cal M}_{\rm C}^N v_{\B{\rm C}}^iv_{\B{\rm C}}^k
 -\pd_{ikLN}\bigg(\frac1{R_{\rm C}}\bigg){\cal M}_\B^{pL}{\cal M}_{\rm C}^Nv_{\B{\rm C}}^pv_{\B{\rm C}}^k\bigg]\nonumber
  \\  
 &+&2(1+\g) \sum_{l=0}^{\infty}\sum_{n=0}^{\infty}\frac{(-1)^n}{(n+1)!}\frac{(l+1)}{(l+2)!}\bigg[\pd_{iLN}\bigg(\frac1{R_{\rm C}}\bigg){\cal M}_\B^{pL}\ddot{\cal M}_{\rm C}^{pN}-\pd_{pLN}\bigg(\frac1{R_{\rm C}}\bigg){\cal M}_\B^{pL}\ddot{\cal M}_{\rm C}^{iN} \bigg]
 \nonumber\\
  &+&2(1+\g) \sum_{l=0}^{\infty}\sum_{n=0}^{\infty}\frac{(-1)^n}{(n+1)!}\frac{(l+1)}{(l+2)!}\bigg[\pd_{ikLN}\bigg(\frac1{R_{\rm C}}\bigg){\cal M}_\B^{pL}\dot{\cal M}_{\rm C}^{pN}v^k_{\B{\rm C}}-\pd_{pkLN}\bigg(\frac1{R_{\rm C}}\bigg){\cal M}_\B^{pL}\dot{\cal M}_{\rm C}^{iN}v^k_{\B{\rm C}}\bigg]
 \nonumber\\ 
 &+&2(1+\g) \sum_{l=0}^{\infty}\sum_{n=0}^{\infty}\frac{(-1)^n(n+1)}{(n+2)!}\frac{l+1}{(l+2)!}\bigg[\varepsilon_{kpq}\pd_{ipLN}\bigg(\frac1{R_{\rm C}}\bigg){\cal M}_\B^{kL}\dot{\cal S}_{\rm C}^{qN} -\varepsilon_{ipq}\pd_{kpLN}\bigg(\frac1{R_{\rm C}}\bigg){\cal M}_\B^{kL}\dot{\cal S}_{\rm C}^{qN}       \bigg]  \nonumber
  \\ 
   &+&2(1+\g) \sum_{l=0}^{\infty}\sum_{n=0}^{\infty}\frac{(-1)^n(n+1)}{(n+2)!}\frac{l+1}{(l+2)!}\bigg[\varepsilon_{kpq}\pd_{ijpLN}\bigg(\frac1{R_{\rm C}}\bigg){\cal M}_\B^{kL}{\cal S}_{\rm C}^{qN}v^j_{\B{\rm C}} -\varepsilon_{ipq}\pd_{jkpLN}\bigg(\frac1{R_{\rm C}}\bigg){\cal M}_\B^{kL}{\cal S}_{\rm C}^{qN} v^j_{\B{\rm C}}      \bigg]  \nonumber
  \\           
   &-&2(1+\g) \sum_{l=0}^{\infty}\sum_{n=0}^{\infty}\frac{(-1)^n}{n!}\frac{l+1}{(l+2)!}\varepsilon_{kpq}\pd_{ikLN}\bigg(\frac1{R_{\rm C}}\bigg){\cal S}_\B^{qL}{\cal M}_{\rm C}^{N}v^p_{\B{\rm C}}\nonumber
  \\
  &+&2(1+\g) \sum_{l=0}^{\infty}\sum_{n=0}^{\infty}\frac{(-1)^n}{(n+1)!}\frac{l+1}{(l+2)!}\varepsilon_{kpq}\pd_{ikLN}\bigg(\frac1{R_{\rm C}}\bigg){\cal S}_\B^{qL}\dot{\cal M}_{\rm C}^{pN}\nonumber
  \\
     &+&2(1+\g) \sum_{l=0}^{\infty}\sum_{n=0}^{\infty}\frac{(-1)^n (n+1)}{(n+2)!}\frac{l+1}{(l+2)!}\pd_{ikqLN}\bigg(\frac1{R_{\rm C}}\bigg){\cal S}_\B^{qL}{\cal S}_{\rm C}^{kN}\nonumber
  \\
&-&2(1+\g) \sum_{l=0}^{\infty}\sum_{n=0}^{\infty}\frac{(-1)^n}{n!}\frac{l+1}{(l+2)!}\varepsilon_{ikq}\pd_{kLN}\bigg(\frac1{R_{\rm C}}\bigg){\cal S}_\B^{qL}\dot{\cal M}_{\rm C}^{N}\nonumber
\\
&-&2(1+\g) \sum_{l=0}^{\infty}\sum_{n=0}^{\infty}\frac{(-1)^n}{n!}\frac{l+1}{(l+2)!}\varepsilon_{ikq}\pd_{kpLN}\bigg(\frac1{R_{\rm C}}\bigg){\cal S}_\B^{qL}{\cal M}_{\rm C}^{N}v^p_{\B{\rm C}}\nonumber
\\
 &-&2(1+\g) \sum_{l=0}^{\infty}\sum_{n=0}^{\infty}\frac{(-1)^n}{n!}\frac{l+1}{(l+2)!}\varepsilon_{ikq}\pd_{kLN}\bigg(\frac1{R_{\rm C}}\bigg)\dot{\cal S}_\B^{qL}{\cal M}_{\rm C}^{N}\nonumber
\;.
\ea   
\subsubsection{Complementary Vector Function \texorpdfstring{${\cal I}^i_{\rm c}$}{{\cal I}^i_{\rm c}}. Adjustment of the Center of Mass.}

We notice that the last three terms in \eqref{c5s0k2f} represent a second time derivative from the product, $a_\B^k{\cal M}_{\B}^{ik}$. These terms can be eliminating from the net force \eqref{sr5v6h} by choosing the {\it complementary} vector function ${\cal I}^i_{\rm c}$ in definition \eqref{b2c4z9} of the center of mass of body B, as follows
\be
\la{m4g1x8p}
{\cal I}^i_{\rm c}=-3a_\B^k{\cal M}_{\B}^{ik}\;.
\ee
This choice slightly simplifies equations of translational motion and makes a small adjustment of the worldline ${\cal Z}$ of the center of mass of body B as compared with the choice ${\cal I}^i_{\rm c}=0$ which was used, for example, in \citep{racine_2005PhRvD,racine2013PhRvD}. 

The terms which are proportional to spin ${\cal S}^i_\B$ of body B, in the right-hand side of \eqref{c5s0k2f} do not represent a second time derivative and will be left in the equations of motion. In principle, we can always group some terms in the net force \eqref{sr5v6h} to form a second time derivative that can be eliminated from the force. This procedure can make sense for simplifying the translational equations of motion of body B. However, it brings additional terms to the rotational equations of motion for body's spin and, in overall, may be not so effective. Therefore, we don't implement it beyond applying equation \eqref{m4g1x8p}.  

\subsection{Explicit Formula for Gravitational Force}\la{n2c4z21a}

After making adjustment \eqref{m4g1x8p} of the worldline of the center of mass of body B, translational equations of motion \eqref{u4bvs3} take on the following form
\begin{equation}
  \label{MBaB1}  
M^{\phantom{i}}_{\rm B}a^i_{\rm B} = F^i_{\rm N} + F^i_{\rm pN}\;.
\end{equation}
where $M_{\rm B}$ is the inertial ({\it conformal}) mass\index{mass!inertial} of the body, and the net gravitational force $F^i$ is split in two components: $F^i_{\rm N}$ is the Newtonian gravitational force, and $F^i_{\rm pN}$ is the post-Newtonian gravitational force. The force components read,
\ba\la{huj3d6}
 F^i_{\rm N}&=&\sum_{l=0}^\infty\frac1{l!}\pd_{<iL>}\bar{W}(t,{\bm x}_\B){\cal M}^L_\B\;,\\
 \la{poni56}
 F^i_{\rm pN}&=&\sum_{l=0}^\infty\frac1{l!}\pd_{<iL>}\bar\Omega(t,{\bm x}_\B){\cal M}^L_\B+ \sum_{l=0}^\infty\frac1{l!}\pd_{<L}\bar \Omega_{i>}(t,{\bm x}_\B){\cal M}^L_\B-\bar\Xi^i(t,{\bm x}_\B)\\\nonumber
&&-\bigg(v_\B^2\d^{ik}+\frac12v_{\rm B}^iv_{\rm B}^k\bigg)\sum_{l=0}^\infty\frac1{l!}\pd_{<kL>}\bar{U}(t,{\bm x}_\B) {\cal M}^L_\B-F^{pk}_{\rm B}\sum_{l=0}^\infty\frac1{l!}\pd_{<ipL>}\bar{U}(t,{\bm x}_\B){\cal M}^{kL}_\B\\\nonumber
&&-M^{-1}_\B\varepsilon_{ikq}\left(2\sum_{l=0}^\infty\frac1{l!}\pd_{<kL>}\bar U(t,{\bm x}_\B){\cal M}^L_\B\dot{\mathcal{S}}_\B^q+\sum_{l=0}^\infty\frac1{l!}\pd_{<kL>}\dot{\bar U}(t,{\bm x}_\B){\cal M}^L_\B{\mathcal{S}}_\B^q+\sum_{l=0}^\infty\frac1{l!}\pd_{<kL>}{\bar U}(t,{\bm x}_\B)\dot{\cal M}^L_\B{\mathcal{S}}_\B^q\right)\;,
\ea
where the very last two terms with spin multipoles originate from the middle group of the spin-dependent terms in \eqref{c5s0k2f} after replacing acceleration $a^i_\B$ with the Newtonian equations of motion of body B.

Computation of the explicit form of the force is now achieved by substituting to \eqref{huj3d6}, \eqref{poni56} the STF derivatives of gravitational potentials obtained in sections \ref{aaaaa1}--\ref{aaaaa4}, and employing relations \eqref{zz33aa22}, \eqref{mumu45x} for computations of partial derivatives from $R_{\rm C}=|{\bm x}-{\bm x}_{\rm C}|$,
\ba\label{acser243}
\pd_{<L>}R_{\B{\rm C}}^{-1}     \equiv\lim_{{\bm x}\rightarrow{\bm x}_{\rm B}}\pd_{<L>}R_{{\rm C}}^{-1} &=&(-1)^l(2l-1)!!\frac{R^{<L>}_{\B{\rm C}}}{R^{2l+1}_{\B{\rm C}}}\;,\\\label{nrvx31w}
\pd_{<L>}R_{\B{\rm C}}    \equiv\lim_{{\bm x}\rightarrow{\bm x}_{\rm B}} \pd_{<L>}R_{{\rm C}} &=&(-1)^{l+1}(2l-3)!!\frac{R^{<L>}_{\B{\rm C}}}{R^{2l-1}_{\B{\rm C}}}\;,
\ea
which are taken at point ${\bm x}_\B$ -- the center of mass of body B. It is worth noticing that $\pd_{<L>}R_{\B{\rm C}}^{-1}=\pd_{L}R_{\B{\rm C}}^{-1}$ due to the fact that function $R^{-1}_{\rm C}$ is a fundamental solution of the Laplace equation, $\triangle R^{-1}_{\rm C}=0$, everywhere but the point $x^i=x^i_{\rm C}$.

\subsubsection{Newtonian Force}

The total Newtonian gravitational force, $F^i_{\rm N}$, is given by a linear superposition of gravitational forces exerted on the body B by all other bodies of ${\mathbb N}$-body system. Using \eqref{n3c72b}, \eqref{bumb3d4} and taking the partial derivative in \eqref{huj3d6} with the help of \eqref{acser243}, we get
\begin{equation}
  \label{w1q5}
  F^i_{\rm N}  =\sum_{C\neq B}\sum_{l=0}^{\infty}\sum_{n=0}^{\infty}\frac{(-1)^n}{l!n!}{\cal M}_{\rm B}^{L}{\cal M}_{\rm C}^{N}\pd_{iLN}R_{\B{\rm C}}^{-1}=
  -\sum_{C\neq B}\sum_{l=0}^{\infty}\sum_{n=0}^{\infty}\frac{(-1)^l(2l+2n+1)!!}{l!n!}\frac{{\cal M}_{\rm B}^{L}{\cal M}_{\rm C}^{N}}{R_{\rm BC}^{2l+2n+3}}R_{\rm BC}^{<iLN>}\;,
\end{equation}
where ${\cal M}_{\rm B}^{L}={\cal M}_{\rm B}^{<a_1...a_l>}$ are active STF\index{STF} multipoles of body B, ${\cal M}_{\rm C}^{N}={\cal M}_{\rm C}^{<b_1...b_n>}$ are active STF multipoles of the external body C,  $R_{\rm BC}=|{\bm R}_{\rm BC}|=\left(\d_{ij}R^i_{\rm BC}R^j_{\rm BC}\right)^{1/2}$, 
\be\la{9b2c5}
R^i_{\rm BC}\equiv x^i_{\rm B}-x^i_{\rm C}=x^i_{\rm B}(t)-x^i_{\rm C}(t)\;,
\ee
is the coordinate distance between the centers of mass of the bodies, $R^{<iLN>}_{\rm BC}=R^{<ia_1...a_lb_1...b_n>}_{\rm BC}$, and the repeated indices mean the Einstein summation rule. 

We draw attention to the reader that the coordinates of the centers of mass of all bodies are computed at the same instant of global time $t$ that is $x^i_{\rm B}=x^i_{\rm B}(t)$, $x^i_{\rm C}=x^i_{\rm C}(t)$, and so on.  On the other hand, each body STF multipole is a function of the coordinate time of the corresponding local coordinates adapted to the body. According to the procedure of derivation of the equations of motion adopted in the present paper, the numerical values of all local coordinate times are computed at the center of mass of body B which acceleration $a^i_{\rm B}$ enters the left-hand side of \eqref{MBaB1}. In other words, we have ${\cal M}_{\rm B}^{L}\equiv{\cal M}_{\rm B}^L(u^*_{\rm B})$ and ${\cal M}_{\rm C}^{L}\equiv{\cal M}_{\rm C}^L(u^*_{\rm C})$ (and similar convention is applied to the spin multipoles) where the local times
\ba
\label{jk9wss5f}
u^*_{\rm B}&=&u_{\rm B}|_{{\bm x}={\bm x}_{\rm B}}=t+\frac1{c^2}{\cal A}_{\rm B}(t)+{\cal O}\left(\frac1{c^4}\right)\;,\\
\label{acz521sa}
u^*_{\rm C}&=&u_{\rm C}|_{{\bm x}={\bm x}_{\rm B}}=t+\frac1{c^2}\left[{\cal A}_{\rm C}(t)-v^k_{\rm C}(t)R^k_{\rm BC}\right]+{\cal O}\left(\frac1{c^4}\right)\;,
\ea
where time dilation functions ${\cal A}_{\rm B}$ and ${\cal A}_{\rm C}$ are defined by solutions of the ordinary differential equations 
\ba
\label{byc29x4}
\frac{d\mathcal{A}_{\rm B}}{dt} & = & -\frac{1}{2}v_{\rm B}^2(t)-\bar{U}_{\rm B}(t,{\bm x}_\B)\;\\
\label{moeb6732c}
\frac{d\mathcal{A}_{\rm C}}{dt} & = & -\frac{1}{2}v_{\rm C}^2(t)-\bar{U}_{\rm C}(t,{\bm x}_{\rm C})\;,
\ea
which constitute a part of the coordinate transformation between the local and global coordinates of the corresponding massive body. The mass $M_{\rm B}$ in the left-hand side of equation \eqref{MBaB1} is computed at the time $u^*_{\rm B}$ defined above in \eqref{jk9wss5f}.

We emphasize that the Newtonian gravitational force (\ref{w1q5}) in scalar-tensor theory of gravity depends on the {\it active} multipoles which include the post-Newtonian corrections as shown in \eqref{1.31}. The force also has a post-Newtonian contribution from the {\it active}  mass dipole ${\cal M}^i$ of the bodies (terms with $l=1$ and $n=1$) which do not vanish because the center of mass of the body is defined by the condition of vanishing conformal mass dipole, ${\cal I}^i=0$ of each body. The {\it active} dipole ${\cal M}^i\not={\cal I}^i$ according to \eqref{nrtcvc67h}. 

Additional notice is that the inertial mass, $M_{\rm B}$, in the left side of (\ref{MBaB1}) is the {\it conformal} mass \eqref{confmass} of body B  while the gravitational force in the right side of (\ref{MBaB1}) depends on the {\it active} mass ${\cal M}_{\rm B}$ -- see \eqref{activemass} -- of body B and the {\it active} masses of other bodies, which corresponds, for example, to the terms with $l=0$ in the right-hand side of (\ref{w1q5}). The {\it active} and {\it conformal} masses do not coincide as follows from (\ref{p3c2}). It violates the strong principle of equivalence\index{principle of equivalence!strong} in scalar-tensor theory of gravity \citep{1962PhRv..125.2163D,1973PhRvD...7.2347N,willbook}.

\subsubsection{Post-Newtonian Force}
The post-Newtonian gravitational force can be represented in the form of a linear superposition of STF partial derivatives taken from functions $R^{-1}_{\B{\rm C}}$ and $R_{\B{\rm C}}$,
\ba\la{eee3s}
F^i_{\rm pN}&=&\frac12\sum_{{\rm C}\not=\B}\sum_{l=0}^\infty\sum_{n=0}^{\infty}\frac{(-1)^n}{l!n!}{\cal M}_\B^L\Big[\ddot{{\cal M}}_{\rm C}^N\pd_{<iLN>}- {{\cal M}}_{\rm C}^Na_{\rm C}^p\pd_{<ipLN>}+{{\cal M}}_{\rm C}^Nv_{\rm C}^pv_{\rm C}^q\pd_{<ipqLN>}\Big]R_{\rm BC}\\\nonumber
&&+
\sum_{{\rm C}\not=\B}\sum_{l=0}^{\infty}\sum_{n=0}^{\infty}\frac{(-1)^n}{l!n!}\bigg[\Big(\a^{iLN}_{\rm F}+\beta^{iLN}_{\rm F}\Big)\pd_{<LN>}+\Big(\a^{ipLN}_{\rm F}+\beta^{ipLN}_{\rm F}\Big)\pd_{<pLN>}+\a^{ipqLN}_{\rm F}\pd_{<pqLN>}\\\nonumber
&&+\Big(\a^{LN}_{\rm F}+\beta^{LN}_{\rm F}+\gamma^{LN}_{\rm F}\Big)\pd_{<iLN>}+\Big(\mu^{pLN}_{\rm F}+\nu^{pLN}_{\rm F}+\rho^{pLN}_{\rm F}\Big)\pd_{<ipLN>}+\sigma^{pqLN}_{\rm F}\pd_{<ipqLN>}\Big]R_{\B{\rm C}}^{-1}\\\nonumber
\;,
\ea
where all partial derivatives are understood in the sense of equations \eqref{acser243}, \eqref{nrvx31w} and the coefficients of the differential operator are
\ba\label{oq7v3d}
\a^{iLN}_{\rm F}&=&\bigg[v^i_\B-2(1+\g)v^i_{\B{\rm C}}\bigg]{\cal M}^{L}_\B\dot{\cal M}^{N}_{\rm C}+\left[\frac{2(1+\g)}{n+1}-\frac1{2l+2n+3}\right]{\cal M}^{L}_\B\ddot{\cal M}^{iN}_{\rm C}\\\nonumber
&&+2(1+\g)\left[\frac1{n+1}\dot{\cal M}^{L}_\B\dot{\cal M}^{iN}_{\rm C} -\dot{\cal M}^{L}_\B{\cal M}^N_{\rm C} v^i_{\B{\rm C}}\right]-\frac{2l^2+3l+3+2\g}{l+1} \dot{\cal M}^{iL}_\B\dot{\cal M}^N_{\rm C}\\\nonumber
&&-\frac1{2l+3}\left[(l+2)(2l+1)+\frac{2n}{2l+2n+3}\right] {\cal M}^{iL}_\B\ddot{\cal M}^N_{\rm C} -\frac{l^2+l+2+2\g}{l+1} \ddot{\cal M}^{iL}_\B{\cal M}^N_{\rm C}\;,\\
\la{bet1111}
\beta^{iLN}_{\rm F}&=&\left[\Big(2+2\g-\frac1{2l+2n+3}\Big)a^i_{\rm C}-\Big(l+2+2\g\Big)a^i_\B\right]{\cal M}^L_\B{\cal M}^N_{\rm C}\;,\\\nonumber
\\\la{xxx5d6}
\a^{ipLN}_{\rm F}&=&\left[\frac2{2l+2n+5}v^i_{\rm C} v^p_{\rm C}-2(1+\g)v^i_{\B{\rm C}} v^p_{\B{\rm C}}-v^i_\B v^p_{\rm C}-\frac2{M_\B}\varepsilon_{ipq}\dot{\cal S}^q_\B\right]{\cal M}^{L}_\B{\cal M}^{N}_{\rm C}\\\nonumber
&&+\frac{2(1+\g)}{n+1}{\cal M}^{L}_\B\dot{\cal M}^{iN}_{\rm C} v^p_{\B{\rm C}}+2\left[\frac{(l+2)(2l+1)}{2l+3}v^p_{{\rm C}}-(l+1)v^p_\B\right]{\cal M}^{iL}_\B\dot{\cal M}^{N}_{\rm C}\\\nonumber
&&-\frac{2l^2+3l+3+2\g}{l+1}\dot{\cal M}^{iL}_\B{\cal M}^{N}_{\rm C} v^p_{\B{\rm C}}-\frac1{M_\B}\varepsilon_{ipq}\left({\cal M}^L_\B\dot{\cal M}^N_{\rm C}{\cal S}^{q}_\B+\dot{\cal M}^L_\B{\cal M}^N_{\rm C}{\cal S}^{q}_\B\right)\\\nonumber
&&+\frac{2(1+\g)}{n+2}\varepsilon_{ipq}\left({\cal M}^L_\B\dot{\cal S}^{qN}_{\rm C}+\dot{\cal M}^L_\B{\cal S}^{qN}_{\rm C}\right)+\frac{2(1+\g)}{l+2}\varepsilon_{ipq}\left(\dot{\cal S}^{qN}_\B{\cal M}^L_{\rm C}+{\cal S}^{qN}_\B\dot{\cal M}^L_{\rm C}\right)\;,\\
\la{bet2222}
\beta^{ipLN}_{\rm F}&=&\left[\Big(l+1-\frac1{2l+2n+5}\Big)a^p_{\rm C}-la^p_\B\right]{\cal M}^{iL}_\B{\cal M}^{N}_{\rm C}+\frac{1}{2l+2n+5}a^p_{\rm C}{\cal M}^{L}_\B{\cal M}^{iN}_{\rm C}\;,
\\\nonumber\\
\la{puk34w}
\a^{ipqLN}_{\rm F}&=&\frac1{2l+2n+7}\Big({\cal M}^{iL}_\B{\cal M}^N_{\rm C}-{\cal M}^{L}_\B{\cal M}^{iN}_{\rm C}\Big)v^p_{\rm C} v^q_{\rm C}-(l+1){\cal M}^{iL}_\B{\cal M}^N_{\rm C} v^p_{\B{\rm C}} v^q_{\B{\rm C}}\\\nonumber
&&+\frac{2(1+\g)}{l+2}\varepsilon^{ipk}{\cal S}^{kL}_\B{\cal M}^N_{\rm C} v^q_{\B{\rm C}}+\frac{2(1+\g)}{n+2}\varepsilon^{ipk}{\cal M}^L_\B{\cal S}^{kN}_{\rm C} v^q_{\B{\rm C}}-\frac1{M_\B}\varepsilon^{ipk}{\cal M}^L_\B {\cal M}^N_{\rm C}{\cal S}^k_{\rm C} v^q_{\B{\rm C}}\;,
\\\nonumber\\
\la{no23c}
\alpha^{LN}_{\rm F}&=&\left[(1+\g)v^2_{\B{\rm C}}-\frac12\frac{2l+2n+3}{2l+2n+5}v^2_{\rm C}-v^2_\B\right]{\cal M}^L_\B{\cal M}^N_{\rm C}+{\cal M}^{kL}_\B\dot{\cal M}^N_{\rm C} v^k_\B-\frac{2(1+\g)}{n+1}{\cal M}^{L}_\B\dot{\cal M}^{kN}_{\rm C} v^k_{\B{\rm C}}\\\nonumber
&&-\frac1{2l+2n+5}{\cal M}^{kL}_\B\ddot{\cal M}^{kN}_{\rm C}+\frac{2(1+\g)}{l+1}\left( \dot{\cal M}^{kL}_\B{\cal M}^{N}_{\rm C} v^k_{\B{\rm C}}-\frac1{n+1} \dot{\cal M}^{kL}_\B\dot{\cal M}^{kN}_{\rm C}\right)\;,\\
\la{btg42x}
\beta^{LN}_{\rm F}&=&-(l+1){\cal M}^{kL}_\B{\cal M}^{N}_{\rm C} a^k_\B-(n+1){\cal M}^{L}_\B{\cal M}^{kN}_{\rm C} a^k_{\rm C}-\frac1{2l+2n+5}\Big({\cal M}^{kL}_\B{\cal M}^{N}_{\rm C}-{\cal M}^{L}_\B{\cal M}^{kN}_{\rm C}\Big)a^k_{\rm C}\;,\\
\gamma^{LN}_{\rm F}&=&\bigg[(2-2\beta-l\g)\bar U(t,{\bm x}_\B)-\g(n+1)\bar U(t,{\bm x}_{\rm C})\bigg]{\cal M}^L_\B{\cal M}^N_{\rm C}\;,\\\nonumber\\
\la{vert3s}
\mu^{pLN}_{\rm F}&=&\frac12\frac{2l+2n+3}{2l+2n+7}{\cal M}^{L}_\B{\cal M}^{kN}_{\rm C} v^p_{\rm C} v^k_{\rm C}+\frac{2(1+\g)}{n+2}\varepsilon_{pkq}{\cal M}^{L}_\B{\cal S}^{qN}_{\rm C}  v^k_{\B{\rm C}}\\\nonumber
&& +\bigg(v^k_\B v^p_{\B{\rm C}}-\frac12 v^k_\B v^p_\B+\frac{2}{2l+2n+7} v^p_{\rm C} v^k_{\rm C}\bigg){\cal M}^{kL}_\B{\cal M}^{N}_{\rm C} \\\nonumber
&&\frac{2(1+\g)}{(l+1)(n+2)}\varepsilon_{pkq}\dot{\cal M}^{kL}_\B{\cal S}^{qN}_{\rm C}-\frac{2(1+\g)}{l+2}\varepsilon_{pkq}{\cal S}^{kL}_\B\bigg({\cal M}^{N}_{\rm C} v^q_{\B{\rm C}}  -\frac1{n+1}\dot{\cal M}^{qN}_{\rm C}\bigg)\;,\\
\la{nv2c6a}
\nu^{pLN}_{\rm F}&=&\frac1{2l+2n+7}{\cal M}^{kL}_\B{\cal M}^{kN}_{\rm C}  a^p_{\rm C} \;,\\
\label{zw3k4b}
\rho^{pLN}_{\rm F}&=&-F^{kp}_{\rm C}{\cal M}^L_\B{\cal M}^{kN}_{\rm C}-F^{pk}_\B{\cal M}^{kL}_\B{\cal M}^{N}_{\rm C}\;,\\\nonumber\\
\la{bey6c4}
\sigma^{pqLN}_{\rm F}&=&-\frac1{2l+2n+9}{\cal M}^{kL}_\B{\cal M}^{kN}_{\rm C} v^p_{\rm C} v^q_{\rm C}-\frac{2(1+\g)}{(n+2)(l+2)}{\cal S}^{pL}_\B{\cal S}^{qN}_{\rm C}\;.
\ea
The coefficients \eqref{oq7v3d}--\eqref{bey6c4} depend on the {\it active} mass and spin multipoles of the bodies of $\mathbb{N}$-body system and their time derivatives. They also depend on velocities of the centers of mass and their accelerations with respect to the origin of the global coordinates. Coefficient \eqref{zw3k4b} describes dependence of the force on the matrix of relativistic precession for each body which is a solution of the equation of relativistic precession \eqref{5.18}. Post-Newtonian force for arbitrary-structured extended bodies with accounting for all mass and spin multipoles of the bodies has been derived in general relativity by \citet{racine_2005PhRvD,racine2013PhRvD}. We compare their result with our expression \eqref{eee3s} for the force in Appendix \ref{appendixA}. 

\subsection{Reduced Post-Newtonian Force}

The post-Newtonian force \eqref{eee3s} depends explicitly on the coordinate accelerations $a^i_\B$ and $a^i_{\rm C}$ of the centers of mass of extended bodies. In case, when velocities of bodies are  significantly smaller than the fundamental speed $c$, we can use the Newtonian equations of motion of bodies, $M_{\B}a^i_\B=F^i_{\rm N}$, in order to replace the accelerations, $a^i_{\rm B}$, with the explicit form of the Newtonian force, $F^i_{\rm N}$, taken from \eqref{w1q5}. It gives us the reduced post-Newtonian force which depends on three types of interaction between multipoles of the extended bodies in $\mathbb{N}$-body system which are due to mass-mass, spin-mass and spin-spin gravitational couplings. In order to set in order the different types of the multipole-multipole interactions, which enter different coefficients \eqref{oq7v3d}--\eqref{bey6c4}, we split the post-Newtonian gravitational force in three main constituents,
\be\la{pNf5s}
F^i_{\rm pN}=F^i_{\rm M}+F^i_{\rm S}+F^i_{\rm P}\;,
\ee
where $F^i_{\rm M}$ is the force caused by the gravitational interaction between the mass multipoles of extended bodies, $F^i_{\rm S}$ is the force caused by the spin-mass and spin-spin multipole interactions, and the force $F^i_{\rm P}$ is due to the relativistic precession of the body-adapted local coordinates with respect to the spatial axes of the global coordinates. We describe the structure of each of the three components of \eqref{pNf5s} below.

\subsubsection{Mass Multipole Coupling Force}
The mass multipole coupling force $F^i_{\rm M}$ consists of a number of terms describing mutual gravitational interaction between the mass multipoles of two, three, and four bodies comprising ${\mathbb N}$-body system. Besides, the force includes terms depending on the first and second time-derivatives of the mass multipoles as well. The force can be represented as a sum of vectorial components, 
\ba\la{1aaa1}
F^i_{\rm M}&=&
F^i_{{\cal M}{\cal M}} +F^i_{{\cal M}\dot{{\cal M}}} +F^i_{{\cal M}\ddot{{\cal M}}}
+F^i_{\dot{{\cal M}}\dot{{\cal M}}}+F^i_{{\cal M}{\cal M}{\cal M}}+
F^i_{{\cal M}{\cal M}{\cal M}{\cal M}}\;,
\ea
where each particular term in the right-hand-side of \eqref{1aaa1} is labeled in correspondence with the number of the mass multipoles and/or their time derivatives participating in the multipole-to-multipole coupling. Specific expressions for different terms in (\ref{1aaa1}) are given in the form of products of the coupling coefficients $ A^{LN}_{{\cal M}{\cal M}}$, $A^{LN}_{{\cal M}\dot{{\cal M}}}$, etc.,  with the explicit expressions of STF derivatives \eqref{acser243}} and \eqref{nrvx31w}. The components of the mass-mass multipole coupling force, $F_{\rm M}$, are as follows:
\begin{eqnarray}
  \label{w1q8}
  F^i_{{\cal M}{\cal M}}
  & = & \sum_{C\neq B}\sum_{l=0}^{\infty}\sum_{n=0}^{\infty} \Bigg[ 
    A^{LN}_{{\cal M}{\cal M}}\frac{R_{\rm BC}^{<iLN>}}{R_{\rm BC}^{2l+2n+3}}
    +A^{ijLN}_{{\cal M}{\cal M}}\frac{R_{\rm BC}^{<jLN>}}{R_{\rm BC}^{2l+2n+3}}
   +A^{jLN}_{{\cal M}{\cal M}}\frac{R_{\rm BC}^{<ijLN>}}{R_{\rm BC}^{2l+2n+5}}
     \\
  & &\phantom{\sum_{C\neq B}\sum_{l=0}^{\infty}\sum_{n=0}^{\infty}}+A^{ijpLN}_{{\cal M}{\cal M}}\frac{R_{\rm BC}^{<jpLN>}}{R_{\rm BC}^{2l+2n+5}} +B^{jpLN}_{{\cal M}{\cal M}}\frac{R_{\rm BC}^{<ijpLN>}}{R_{\rm BC}^{2l+2n+5}}
  +{\cal C}^{jpLN}_{{\cal M}{\cal M}}\frac{R_{\rm BC}^{<ijpLN>}}{R_{\rm BC}^{2l+2n+7}}\Bigg]\;,\nonumber
\\
  \label{w1q9}
  F^i_{{\cal M}\dot{{\cal M}}}
  & = & \sum_{C\neq B}\sum_{l=0}^{\infty}\sum_{n=0}^{\infty} \Bigg[ 
  A^{iLN}_{{\cal M}\dot{{\cal M}}}\frac{R_{\rm BC}^{<LN>}}{R_{\rm BC}^{2l+2n+1}}
  +A^{LN}_{{\cal M}\dot{{\cal M}}}\frac{R_{\rm BC}^{<iLN>}}{R_{\rm BC}^{2l+2n+3}} +A^{ijLN}_{{\cal M}\dot{{\cal M}}}\frac{R_{\rm BC}^{<jLN>}}{R_{\rm BC}^{2l+2n+3}} \\\nonumber
    & &\phantom{ \sum_{C\neq B}\sum_{l=0}^{\infty}\sum_{n=0}^{\infty}}
 +B^{jLN}_{{\cal M}\dot{{\cal M}}}\frac{R_{\rm BC}^{<ijLN>}}{R_{\rm BC}^{2l+2n+3}}
  +{\cal C}^{jLN}_{{\cal M}\dot{{\cal M}}}\frac{R_{\rm BC}^{<ijLN>}}{R_{\rm BC}^{2l+2n+5}}\Bigg]\;,
\\
  \label{q1w2}
  F^i_{{\cal M}\ddot{{\cal M}}}
   &=&  \sum_{C\neq B}\sum_{l=0}^{\infty}\sum_{n=0}^{\infty} \Bigg[ A^{iLN}_{{\cal M}\ddot{{\cal M}}}\frac{R_{\rm BC}^{<LN>}}{R_{\rm BC}^{2l+2n+1}}
 +A^{LN}_{{\cal M}\ddot{{\cal M}}}\frac{R_{\rm BC}^{<iLN>}}{R_{\rm BC}^{2l+2n+1}}
+B^{LN}_{{\cal M}\ddot{{\cal M}}}\frac{R_{\rm BC}^{<iLN>}}{R_{\rm BC}^{2l+2n+3}}\Bigg]\;,
\\
  \label{q1w4}
  F^i_{\dot{{\cal M}}\dot{{\cal M}}}
   &=& \sum_{C\neq B}\sum_{l=0}^{\infty}\sum_{n=0}^{\infty} \Bigg[ 
   A^{iLN}_{\dot{{\cal M}}\dot{{\cal M}}}\frac{R_{\rm BC}^{<LN>}}{R_{\rm BC}^{2l+2n+1}}
   +A^{LN}_{\dot{{\cal M}}\dot{{\cal M}}}\frac{R_{\rm BC}^{<iLN>}}{R_{\rm BC}^{2l+2n+3}}\Bigg]\;,
\\
  \label{q1w9}
  F^i_{{\cal M}{\cal M}{\cal M}} 
  & = &  \sum_{C\neq B}\sum_{D\neq C}\sum_{l=0}^{\infty}\sum_{n=0}^{\infty}\sum_{k=0}^{\infty} A^{LNK}_{{\cal M}{\cal M}{\cal M}}\frac{R_{\rm BC}^{<iLN>}R_{\rm CD}^{<K>}}{R_{\rm BC}^{2l+2n+3}R_{\rm CD}^{2k+1}}\\\nonumber
  &+&\sum_{C\neq B}\sum_{D\neq B}\sum_{l=0}^{\infty}\sum_{n=0}^{\infty}\sum_{k=0}^{\infty}B^{LNK}_{{\cal M}{\cal M}{\cal M}}\frac{R_{\rm BC}^{<iLN>}R_{\rm BD}^{<K>}}{R_{\rm BC}^{2l+2n+3}R_{\rm BD}^{2k+1}}\;,
  \\
 \label{z1q3}
  F^i_{{\cal M}{\cal M}{\cal M}{\cal M}}&=&
  \sum_{C\neq B}\sum_{D\neq C}\sum_{l=0}^{\infty}\sum_{n=0}^{\infty}\sum_{k=0}^{\infty}\sum_{s=0}^{\infty}\Bigg[
  A^{LNSK}_{{\cal M}{\cal M}{\cal M}{\cal M}}\frac{R_{\rm BC}^{<LN>}R_{\rm CD}^{<iKS>}}{R_{\rm BC}^{2l+2n+1}R_{\rm CD}^{2k+2s+3}}\\
  & &\phantom{\sum_{C\neq B}\sum_{D\neq C}\sum_{l=0}^{\infty}\sum_{n=0}^{\infty}\sum_{k=0}^{\infty}\sum_{s=0}^{\infty}} +B^{LNSK}_{{\cal M}{\cal M}{\cal M}{\cal M}}\frac{R_{\rm BC}^{<ijLN>}R_{\rm CD}^{<jKS>}}{R_{\rm BC}^{2l+2n+3}R_{\rm CD}^{2k+2s+3}}
  +{\cal C}^{LNSK}_{{\cal M}{\cal M}{\cal M}{\cal M}}\frac{R_{\rm BC}^{<ijLN>}R_{\rm CD}^{<jKS>}}{R_{\rm BC}^{2l+2n+5}R_{\rm CD}^{2k+2s+3}}
  \nonumber\\
  & &\phantom{\sum_{C\neq B}\sum_{D\neq C}\sum_{l=0}^{\infty}\sum_{n=0}^{\infty}\sum_{k=0}^{\infty}\sum_{s=0}^{\infty}} +\bigg(A^{jLNSK}_{{\cal M}{\cal M}{\cal M}{\cal M}}+B^{jLNSK}_{{\cal M}{\cal M}{\cal M}{\cal M}}\bigg)\frac{R_{\rm BC}^{<iLN>}R_{\rm CD}^{<jKS>}}{R_{\rm BC}^{2l+2n+3}R_{\rm CD}^{2k+2s+3}}\nonumber\\
  &&\phantom{\sum_{C\neq B}\sum_{D\neq C}\sum_{l=0}^{\infty}\sum_{n=0}^{\infty}\sum_{k=0}^{\infty}\sum_{s=0}^{\infty}}+\bigg({\cal C}^{iLNSK}_{{\cal M}{\cal M}{\cal M}{\cal M}}+D^{iLNSK}_{{\cal M}{\cal M}{\cal M}{\cal M}}\bigg)\frac{R_{\rm BC}^{<jLN>}R_{\rm CD}^{<jKS>}}{R_{\rm BC}^{2l+2n+3}R_{\rm CD}^{2k+2s+3}}
  \Bigg]\nonumber\\
  &+& \sum_{C\neq B}\sum_{D\neq B}\sum_{l=0}^{\infty}\sum_{n=0}^{\infty}\sum_{k=0}^{\infty}\sum_{s=0}^{\infty}\Bigg[
  D^{LNSK}_{{\cal M}{\cal M}{\cal M}{\cal M}}\frac{R_{\rm BC}^{<LN>}R_{\rm BD}^{<iKS>}}{R_{\rm BC}^{2l+2n+1}R_{\rm BD}^{2k+2s+3}}\nonumber\\
  & & \phantom{\sum_{C\neq B}\sum_{D\neq C}\sum_{l=0}^{\infty}\sum_{n=0}^{\infty}\sum_{k=0}^{\infty}\sum_{s=0}^{\infty}}+E^{jLNSK}_{{\cal M}{\cal M}{\cal M}{\cal M}}\frac{R_{\rm BC}^{<iLN>}R_{\rm BD}^{<jKS>}}{R_{\rm BC}^{2l+2n+3}R_{\rm BD}^{2k+2s+3}}
  +H^{iLNSK}_{{\cal M}{\cal M}{\cal M}{\cal M}}\frac{R_{\rm BC}^{<jLN>}R_{\rm BD}^{<jKS>}}{R_{\rm BC}^{2l+2n+3}R_{\rm BD}^{2k+2s+3}}\Bigg]\;,\nonumber
\end{eqnarray}
where the coupling coefficients $A^{LN}_{{\cal M}{\cal M}}$, $A^{ijLN}_{{\cal M}{\cal M}}$, etc. are given by
\begin{eqnarray}
\hspace{-2cm}
  \label{t1q6}
  A^{LN}_{{\cal M}{\cal M}} & = & \frac{(-1)^l(2l+2n+1)!!}{l!n!}\bigg[2(\gamma+1)v_{\rm B}^kv_{\rm C}^k-\gamma v_{\rm B}^2 -\bigg(\g+\frac12+\frac{1}{2l+2n+5}\bigg)v_{\rm C}^{2}\bigg]{\cal M}_{\rm B}^{L}{\cal M}_{\rm C}^{N}\;,
\\
  \label{t4w1}
  A^{LN}_{{\cal M}\dot{{\cal M}}} & = & \frac{(-1)^l(2l+2n+1)!!}{l!n!}\bigg[
2v_{\rm BC}^{k}\bigg(\frac{\gamma+1}{n+1}{\cal M}_{\rm B}^{L}\dot{{\cal M}}_{\rm C}^{kN}-\frac{\gamma+1}{l+1}\dot{{\cal M}}_{\rm B}^{kL}{\cal M}_{\rm C}^{N}\bigg)
   +\bigg(\frac{2\g+1}{l+2}v_{\rm C}^k-v_{\rm B}^{k}\bigg){\cal M}_{\rm B}^{kL}\dot{{\cal M}}_{\rm C}^{N}\bigg]\;,
\\
  \label{e1q6}
  A^{LN}_{\dot{{\cal M}}\dot{{\cal M}}} & = & 2(\gamma+1)\frac{(-1)^l(2l+2n+1)!!}{(l+1)!(n+1)!}\dot{{\cal M}}_{\rm B}^{kL}\dot{{\cal M}}_{\rm C}^{kN},
\\
  \label{t8w1}
  A^{LN}_{{\cal M}\ddot{{\cal M}}} & = & \frac{(-1)^{l}(2l+2n-1)!!}{2l!n!}{\cal M}_{\rm B}^{L}\ddot{{\cal M}}_{\rm C}^{N},
\\
  \label{t1q8}
  A^{iLN}_{{\cal M}{\cal M}} & = & \frac{(-1)^l(2l+2n+3)!!}{l!n!}\bigg[\frac12\frac{2l+2n+3}{2l+2n+7}v_{\rm C}^{i}v_{\rm C}^{p}{\cal M}_{\rm B}^{L}{\cal M}_{\rm C}^{pN}
  \\
  & &\phantom{ \frac{(-1)^l(2l+2n+3)!!}{l!n!}} +\bigg(\frac{1}{2}v_{\rm B}^{i}v_{\rm B}^{p}-v_{C}^{i}v_{\rm B}^{p}+\frac{2}{2l+2n+7}v_{\rm C}^{i}v_{\rm C}^{p}\bigg){\cal M}_{\rm B}^{pL}{\cal M}_{\rm C}^{N}\bigg],\nonumber
\\
  \label{t3w1}
  A^{iLN}_{{\cal M}\dot{{\cal M}}} & = & \frac{(-1)^l(2l+2n-1)!!}{l!n!}
  \bigg\{\bigg[v^i_\B-2(\gamma+1)v_{\rm BC}^{i}\bigg]{\cal M}_{\rm B}^{L}\dot{{\cal M}}_{\rm C}^{N}-2(\gamma+1)\dot{{\cal M}}_{\rm B}^{L}{\cal M}_{\rm C}^{N}v_{\rm BC}^{i}\bigg\}\;,
\\
  \label{e1q5}
  A^{iLN}_{\dot{{\cal M}}\dot{{\cal M}}} &=& \frac{(-1)^l(2l+2n-1)!!}{l!n!}\bigg[\frac{2(\gamma+1)}{n+1}\dot{{\cal M}}_{\rm B}^{L}\dot{{\cal M}}_{\rm C}^{iN}-\frac{2l^2+3l+2\gamma+3}{l+1}\dot{{\cal M}}_{\rm B}^{iL}\dot{{\cal M}}_{\rm C}^{N}\bigg],
\\
  \label{t9w1}
  A^{iLN}_{{\cal M}\ddot{{\cal M}}} & = & \frac{(-1)^l(2l+2n-1)!!}{l!n!}\Bigg\{\bigg[\frac{2(\g+1)}{n+1}-\frac{1}{2l+2n+3}\bigg]{\cal M}_{\rm B}^{L}\ddot{{\cal M}}_{\rm C}^{iN}\\\nonumber
  & &\phantom{\frac{(-1)^l(2l+2n-1)!!}{l!n!}} +\bigg[\frac{1}{2l+2n+3}-\frac{(l+2)(2l+1)}{(2l+3)}\bigg]{\cal M}_{\rm B}^{iL}\ddot{{\cal M}}_{\rm C}^{N}-\frac{l^2+l+2\gamma+2}{l+1}\ddot{{\cal M}}_{\rm B}^{iL}{\cal M}_{\rm C}^{N}\Bigg\}\;,
\\
  \label{t1q7}
  A^{ijLN}_{{\cal M}{\cal M}}& = &\frac{(-1)^l(2l+2n+1)!!}{l!n!}\bigg[v_{\rm B}^{i}v_{C}^j-\frac{2}{2l+2n+5}v_{\rm C}^{i} v_{\rm C}^j+2(\gamma+1)v_{\rm BC}^{i}v_{\rm BC}^{j}\bigg]{\cal M}_{\rm B}^{L}{\cal M}_{\rm C}^{N}\;,
\ea
\ba
  \label{t5w1}
  A^{ijLN}_{{\cal M}\dot{{\cal M}}} & = & \frac{(-1)^{l}(2l+2n+1)!!}{l!n!}\bigg\{
  2\bigg[(l+1)v_{\rm BC}^j+\frac{1}{2l+3}v^j_{\rm C}\bigg]{\cal M}_{\rm B}^{iL}\dot{{\cal M}}_{\rm C}^{N}\\\nonumber
  & &\phantom{ \frac{(-1)^l(2l+2n+1)!!}{l!n!}} -\frac{2(\gamma+1)}{n+1}{\cal M}_{\rm B}^{L}\dot{{\cal M}}_{\rm C}^{iN}v_{\rm BC}^j +\frac{2l^2+3l+2\gamma+3}{l+1}\dot{{\cal M}}_{\rm B}^{iL}{\cal M}_{\rm C}^{N}v_{\rm BC}^j\bigg\}\;,
\\
  \label{t1q9}
  A^{ijpLN}_{{\cal M}{\cal M}} & = & \frac{(-1)^{l}(2l+2n+3)!!}{l!n!}\bigg\{\frac{1}{2l+2n+7}\bigg[{\cal M}_{\rm B}^{iL}{\cal M}_{\rm C}^{N}-{\cal M}_{\rm B}^{L}{\cal M}_{\rm C}^{iN}\bigg]v_{\rm C}^{j}v_{\rm C}^{p}\\
  & &\phantom{ \frac{(-1)^l(2l+2n+3)!!}{l!n!}} -\frac{1}{2l+3}\bigg[2v_{\rm B}^{j}v_{\rm B}^{p}-3v_{\rm B}^{j}v_{\rm C}^p+(l+2)(2l+1)v_{\rm BC}^{j}v_{\rm BC}^{p}\bigg]{\cal M}_{\rm B}^{iL}{\cal M}_{\rm C}^{N}\bigg\}\;,\nonumber
\\
  \label{e1a5}
  A^{LNK}_{{\cal M}{\cal M}{\cal M}} & = &\frac{(-1)^{l+k}(2l+2n+1)!!(2k-1)!!}{l!n!k!}\bigg[\g(n+1)\bigg]{\cal M}_{\rm B}^{L}{\cal M}_{\rm C}^{N}{\cal M}_D^{K},
\\
  \label{e1a6}
  A^{LNSK}_{{\cal M}{\cal M}{\cal M}{\cal M}} & = & -\frac{(-1)^{l+s}(2l+2n-1)!!(2k+2s+1)!!}{l!n!k!s!}\bigg[2(\gamma+1)-\frac1{2l+2n+3}\bigg]\frac{{\cal M}_{\rm B}^{L}{\cal M}_{\rm C}^{N}{\cal M}_{\rm C}^{S}{\cal M}_D^{K}}{\mathcal{M}_{\rm C}},
\\
  \label{e1a9}
  A^{iLNSK}_{{\cal M}{\cal M}{\cal M}{\cal M}} & = & -\frac{(-1)^{l+s}(2l+2n+1)!!(2k+2s+1)!!}{l!n!k!s!}\frac1{2l+2n+5}\frac{{\cal M}_{\rm B}^{iL}{\cal M}_{\rm C}^{N}{\cal M}_{\rm C}^{S}{\cal M}_D^{K}}{\mathcal{M}_{\rm C}}\;,
\\
  \label{e1q1}
  B^{LN}_{{\cal M}\ddot{{\cal M}}}& = &\frac{(-1)^{l}(2l+2n+1)!!}{(2l+2n+5)l!n!}{\cal M}_{\rm B}^{kL}\ddot{{\cal M}}_{\rm C}^{kN},
\\
  \label{t6w1}
  B^{iLN}_{{\cal M}\dot{{\cal M}}} & = & \frac{(-1)^{l}(2l+2n+1)!!}{l!n!}{\cal M}_{\rm B}^{L}\dot{{\cal M}}_{\rm C}^{N}v_{\rm C}^i,
\\
  \label{t1w1}
  B^{ijLN}_{{\cal M}{\cal M}}& = &\frac{(-1)^{l}(2l+2n+3)!!}{2l!n!}{\cal M}_{\rm B}^{L}{\cal M}_{\rm C}^{N}v_{\rm C}^{i}v_{\rm C}^{j},
\\
  \label{z1a5}
  B^{LNK}_{{\cal M}{\cal M}{\cal M}} & = & \frac{(-1)^{l+k}(2l+2n+1)!!(2k-1)!!}{l!n!k!}\bigg[\gamma l+2(\beta-1)\bigg]{\cal M}_{\rm B}^{L}{\cal M}_{\rm C}^{N}{\cal M}_D^{K},
\\
  \label{e1a7}
  B^{LNSK}_{{\cal M}{\cal M}{\cal M}{\cal M}} & = &- \frac{(-1)^{l+s}(2l+2n+1)!!(2k+2s+1)!!}{l!n!k!s!}\frac{{\cal M}_{\rm B}^{L}{\cal M}_{\rm C}^{N}{\cal M}_{\rm C}^{S}{\cal M}_D^{K}}{2\mathcal{M}_{\rm C}}\;,
\\
  \label{z1a2}
  B^{iLNSK}_{{\cal M}{\cal M}{\cal M}{\cal M}} & = & \frac{(-1)^{l+s}(2l+2n+1)!!(2k+2s+1)!!}{l!n!k!s!}\bigg(n+1+\frac1{2l+2n+5}\bigg)\frac{{\cal M}_{\rm B}^{L}{\cal M}_{\rm C}^{iN}{\cal M}_{\rm C}^{S}{\cal M}_D^{K}}{\mathcal{M}_{\rm C}},
  \\
  \label{t2w1}
  {\cal C}^{ijLN}_{{\cal M}{\cal M}}& = &\frac{(-1)^{l}(2l+2n+5)!!}{l!n!(2l+2n+9)}{\cal M}_{\rm B}^{pL}{\cal M}_{\rm C}^{pN}v_{\rm C}^{i}v_{\rm C}^{j},
\\
  \label{z1a3}
  {\cal C}^{LNSK}_{{\cal M}{\cal M}{\cal M}{\cal M}} & = & \frac{(-1)^{l+s}(2l+2n+3)!!(2k+2s+1)!!}{l!n!k!s!}\frac1{2l+2n+7}\frac{{\cal M}_{\rm B}^{pL}{\cal M}_{\rm C}^{pN}{\cal M}_{\rm C}^{S}{\cal M}_D^{K}}{\mathcal{M}_{\rm C}}\;,
\\
  \label{z1a1}
  {\cal C}^{iLNSK}_{{\cal M}{\cal M}{\cal M}{\cal M}} & = & \frac{(-1)^{l+s}(2l+2n+1)!!(2k+2s+1)!!}{l!n!k!s!}\frac1{2l+2n+5}\frac{{\cal M}_{\rm B}^{L}{\cal M}_{\rm C}^{iN}{\cal M}_{\rm C}^{S}{\cal M}_D^{K}}{\mathcal{M}_{\rm C}}\;,
\\
  \label{z1a6}
  D^{LNSK}_{{\cal M}{\cal M}{\cal M}{\cal M}} & = & \frac{(-1)^{l+s}(2l+2n-1)!!(2k+2s+1)!!}{l!n!k!s!}\bigg[ l+2(\gamma+1)\bigg]\frac{{\cal M}_{\rm B}^{L}{\cal M}_{\rm B}^{S}{\cal M}_{\rm C}^{N}{\cal M}_D^{K}}{\mathcal{M}_{\rm B}}\;,
\\
  \label{e1a8}
  D^{iLNSK}_{{\cal M}{\cal M}{\cal M}{\cal M}} & = & \frac{(-1)^{l+s}(2l+2n+1)!!(2k+2s+1)!!}{l!n!k!s!}\bigg[\frac{(l+2)(2l+1)}{2l+3}-\frac1{2l+2n+5}\bigg]
  \frac{{\cal M}_{\rm B}^{iL}{\cal M}_{\rm C}^{N}{\cal M}_{\rm C}^{S}{\cal M}_D^{K}}{\mathcal{M}_{\rm C}},
\\
  \label{z1a7}
  E^{iLNSK}_{{\cal M}{\cal M}{\cal M}{\cal M}} & = & -\frac{(-1)^{l+s}(2l+2n+1)!!(2k+2s+1)!!}{l!n!k!s!}\big(l+1\big)\frac{{\cal M}_{\rm B}^{iL}{\cal M}_{\rm B}^{S}{\cal M}_{\rm C}^{N}{\cal M}_D^{K}}{\mathcal{M}_{\rm B}}\;,
\\
  \label{z1a8}
 H^{iLNSK}_{{\cal M}{\cal M}{\cal M}{\cal M}} & = & -\frac{(-1)^{l+s}(2l+2n+1)!!(2k+2s+1)!!}{(l-1)!n!k!s!}\frac{{\cal M}_{\rm B}^{iL}{\cal M}_{\rm B}^{S}{\cal M}_{\rm C}^{N}{\cal M}_D^{K}}{\mathcal{M}_{\rm B}}\;,
\ea

\subsubsection{Spin Multipole Coupling Force}
The spin multipole post-Newtonian force entering the translational equations of motion has the following structure,
\ba
 \label{FS1}
  F^i_{\rm S} & = &  F^i_{\mathcal{S}{\cal M}}+F^i_{\dot{\mathcal{S}}{\cal M}} +F^i_{\mathcal{S}\dot{{\cal M}}}+F^i_{\mathcal{S}\mathcal{S}}+F^i_{s{\cal M}{\cal M}}+F^i_{s{\cal M}\dot{{\cal M}}}+F^i_{\dot{s}{\cal M}{\cal M}},
\ea
where each component of the force is expressed in terms of the corresponding coupling coefficients $A^{pLN}_{\cal SI}$, $A^{ipLN}_{\cal \dot SI}$, etc., and the STF Cartesian tensors made out of the tensor products of the relative coordinate distances \eqref{9b2c5} between the bodies. Forces $F^i_{\mathcal{S}{\cal M}}$, $F^i_{\dot{\mathcal{S}}{\cal M}}$ and $F^i_{\mathcal{S}\dot{{\cal M}}}$ describe gravitational interaction between the spin and mass multipoles of the bodies.  The force $F^i_{\mathcal{S}\mathcal{S}}$ describes the spin-spin multipole interaction between the bodies. It generalizes to higher multipoles the known spin-spin gravitational force of interaction between spins of rigidly rotating, spherically-symmetric bodies given by \citet[page 275, equation 19]{vab}, and \citet[equation 54]{Barker_Oconnell_1975PhRvD}. The last three terms in the right-hand side of \eqref{FS1} labeled with a small Roman letter {\it s} take their origin from the last three terms in \eqref{poni56}. They describe gravitational interaction of spin of body B and its first time derivative with the mass multipoles of other bodies. 

The spin coupling force components are 
\begin{eqnarray}
  \label{z1q8}
  F^i_{\mathcal{S}{\cal M}} &=&\sum_{C\neq B}\sum_{l=0}^{\infty}\sum_{n=0}^{\infty}\Bigg[
  A^{pLN}_{\mathcal{S}{\cal M}}\frac{R_{\rm BC}^{<ipLN>}}{R_{\rm BC}^{2l+2n+5}}
  +A^{ipqLN}_{\mathcal{S}{\cal M}}\frac{R_{\rm BC}^{<pqLN>}}{R_{\rm BC}^{2l+2n+5}}\Bigg]\;,
\\
  \label{t1q1}
  F^i_{\mathcal{S}\dot{{\cal M}}} &=& \sum_{C\neq B}\sum_{l=0}^{\infty}\sum_{n=0}^{\infty}\Bigg[
  A^{ipLN}_{\mathcal{S}\dot{{\cal M}}}\frac{R_{\rm BC}^{<pLN>}}{R_{\rm BC}^{2l+2n+3}}
  +A^{pLN}_{\mathcal{S}\dot{{\cal M}}}\frac{R_{\rm BC}^{<ipLN>}}{R_{\rm BC}^{2l+2n+5}}\Bigg]\;,
\\
  \label{z1q9}
  F^i_{\dot{\mathcal{S}}{\cal M}} &=&  \sum_{C\neq B}\sum_{l=0}^{\infty}\sum_{n=0}^{\infty}A^{ipLN}_{\dot{\mathcal{S}}{\cal M}}\frac{R_{\rm BC}^{<pLN>}}{R_{\rm BC}^{2l+2n+3}}\;,
\\
  \label{t1q2}
  F^i_{\mathcal{S}\mathcal{S}}
  &=&  \sum_{C\neq B}\sum_{l=0}^{\infty}\sum_{n=0}^{\infty}A^{pqLN}_{\mathcal{S}\mathcal{S}}\frac{R_{\rm BC}^{<ipqLN>}}{R_{\rm BC}^{2l+2n+7}}\;,
  \\
  \label{t1q3}
  F^i_{s{\cal M}{\cal M}}  &=& \sum_{C\neq B}\sum_{l=0}^{\infty}\sum_{n=0}^{\infty}
  A^{ipqLN}_{s{\cal M}{\cal M}}\frac{R_{\rm BC}^{<pqLN>}}{R_{\rm BC}^{2l+2n+5}}\;,
\\
  \label{t1q3a}
  F^i_{s{\cal M}\dot{{\cal M}}}  &=& \sum_{C\neq B}\sum_{l=0}^{\infty}\sum_{n=0}^{\infty}
  A^{ipLN}_{s{\cal M}\dot{{\cal M}}}\frac{R_{\rm BC}^{<pLN>}}{R_{\rm BC}^{2l+2n+3}}
  \;,
\\
  \label{t1q4}
  F^i_{\dot{s}{\cal M}{\cal M}} &=&  \sum_{C\neq B}\sum_{l=0}^{\infty}\sum_{n=0}^{\infty}
  A^{ipLN}_{\dot{s}{\cal M}{\cal M}}\frac{R_{\rm BC}^{<pLN>}}{R_{\rm BC}^{2l+2n+3}}\;,
\end{eqnarray}
where the coupling coefficients entering the various members of the spin coupling force are  
\begin{eqnarray}
  \label{x1q7}
  A^{pLN}_{\mathcal{S}{\cal M}} &=& 
    2(1+\gamma)\frac{(-1)^l(2l+2n+3)!!}{l!n!}\varepsilon_{kpq}v_{\rm BC}^{q}\bigg[\frac{\mathcal{S}_{\rm B}^{kL}{\cal M}_{\rm C}^{N}}{l+2}+\frac{\mathcal{S}_{\rm C}^{kN}{\cal M}_{\rm B}^{L}}{n+2}\bigg]\;,\\
  \label{x1q8}
  A^{ipqLN}_{\mathcal{S}{\cal M}} &=& 
  2(1+\gamma)\frac{(-1)^l(2l+2n+3)!!}{l!n!}\varepsilon_{ipk}v_{\rm BC}^q\bigg[\frac{\mathcal{S}_{\rm B}^{kL}{\cal M}_{\rm C}^{N}}{l+2}+\frac{\mathcal{S}_{\rm C}^{kN}{\cal M}_{\rm B}^{L}}{n+2}\bigg]\;,
\\
  \label{as11}
  A^{pLN}_{\mathcal{S}\dot{{\cal M}}} &=& 
  -2(1+\gamma)\frac{(-1)^l(2l+2n+3)!!}{l!n!}\varepsilon_{kpq}\bigg[\frac{\mathcal{S}_{\rm B}^{kL}\dot{{\cal M}}_{\rm C}^{qN}}{(l+2)(n+1)}+
  \frac{\mathcal{S}_{\rm C}^{qN}\dot{{\cal M}}_{\rm B}^{kL}}{(l+1)(n+2)}\bigg]\;,
\\
  \label{diffRacine2005}
  A^{ipLN}_{\mathcal{S}\dot{{\cal M}}} & = & -2(\gamma+1)\frac{(-1)^l(2l+2n+1)!!}{l!n!}\varepsilon_{ipq}\bigg[\frac{\mathcal{S}_{\rm B}^{qN}\dot{{\cal M}}_{\rm C}^{L}}{(l+2)}
 +\frac{\dot{{\cal M}}_{\rm B}^{L}\mathcal{S}_{\rm C}^{qN}}{n+2}\bigg]\;,
\\
  \label{x1q9}
  A^{ipLN}_{\dot{\mathcal{S}}{\cal M}}  &=&  -2(1+\gamma)\frac{(-1)^l(2l+2n+1)!!}{l!n!}\varepsilon_{ipq}\bigg[\frac{\dot{\mathcal{S}}_{\rm B}^{qN}{\cal M}_{\rm C}^{L}}{l+2}+\frac{{\cal M}_{\rm B}^{L}\dot{\mathcal{S}}_{\rm C}^{qN}}{n+2}\bigg]\;,
\\
  \label{as12}
  A^{pqLN}_{\mathcal{S}\mathcal{S}} & = & 2(1+\gamma)\frac{(-1)^l(2l+2n+5)!!}{l!n!(l+2)(n+2)}\mathcal{S}_{\rm B}^{pL}\mathcal{S}_{\rm C}^{qN}\;,
\\
  \label{as14}
  A^{ipqLN}_{s{\cal M}{\cal M}} & = & \frac{(-1)^l(2l+2n+3)!!}{l!n!}\varepsilon_{ikp}\frac{\mathcal{S}_{\rm C}^k}{\mathcal{M}_{\rm B}}{\cal M}_{\rm B}^{L}{\cal M}_{\rm C}^{N}v_{\rm BC}^q\;,
\\
  \label{as13}
  A^{ipLN}_{s{\cal M}\dot{{\cal M}}} & = & \frac{(-1)^l(2l+2n+1)!!}{l!n!}\varepsilon_{ipq}\frac{\mathcal{S}_{\rm B}^q}{\mathcal{M}_{\rm B}}\left({\cal M}_{\rm B}^{L}\dot{{\cal M}}_{\rm C}^{N}+\dot{{\cal M}}_{\rm B}^{L}{\cal M}_{\rm C}^{N}\right)\;,
\\
  \label{as15}
  A^{ipLN}_{\dot{s}{\cal M}{\cal M}}   & = &  2\frac{(-1)^l(2l+2n+1)!!}{l!n!}\varepsilon_{ipq}\frac{\dot{\mathcal{S}}_{\rm B}^q}{\mathcal{M}_{\rm B}}{\cal M}_{\rm B}^{L}{\cal M}_{\rm C}^{N}\;.
\end{eqnarray}
\subsubsection{Precession Multipole Coupling Force}
Finally, the force caused by the relativistic precession of spatial axes of the local coordinates adapted to each body is 
\begin{eqnarray}
  \label{t1q5}
  F^i_{\rm P} & = & \sum_{C\neq B}\sum_{l=0}^{\infty}\sum_{n=0}^{\infty}\frac{(-1)^l(2l+2n+3)!!}{l!n!}\bigg[
 F^{pk}_\B{\cal M}^{kL}_\B{\cal M}^{N}_{\rm C}+F^{kp}_{\rm C}{\cal M}^L_\B{\cal M}^{kN}_{\rm C}\bigg]\frac{R_{\rm BC}^{<ikLN>}}{R_{\rm BC}^{2l+2n+5}}\;.
\end{eqnarray}
This completes derivation of the translational equations of motion of extended bodies in the global coordinates.

\subsection{Comments}
The post-Newtonian force in translational equations of motion has been calculated in this paper for the system of ${\mathbb N}$ extended bodies with an arbitrary internal structure, shape and density distribution. It includes the Newtonian and post-Newtonian forces due to the gravitational coupling between all internal mass and spin multipoles of extended bodies in ${\mathbb N}$-body system. The force \eqref{w1q8}, denoted as $F^i_{\cal MM}$, converges in monopole approximation to Einstein-Infeld-Hoffman (EIH)\index{equation!EIH}\index{EIH!equations of motion} equations of motion \citep{eih,fockbook,petrova} of point-like particles. The force (\ref{z1q8}), denoted as $F^i_{\cal SM}$, yields the correct analytic expression for the Lense-Thirring\index{Lense-Thirring} (gravitomagnetic\index{gravitomagnetic force}) force due to the gravitational coupling of body's intrinsic spin to orbital angular momentum of the body \citep{vab,ciufolini_book}. The force \eqref{t1q2}, denoted as $F^i_{\cal SS}$, is reduced to the known spin-spin coupling force  \citep{vab,Barker_Oconnell_1975PhRvD,Barker_Oconnell_1976PhRvD,Barker_Oconnell1987JMP} when higher-order multipoles $(l\ge 1)$ are neglected.

Calculation of the post-Newtonian force in quadrupole approximation ($l=2$) were completed by \citet{xu_1997PhRvD} in general relativity. Their result disagrees  by a sufficiently large number of terms with our expression for the post-Newtonian force (\ref{pNf5s}) in the quadrupole approximation. We could not identify the mathematical reason of this disagreement which origin has yet to be clarified. On the other hand, the complete post-Newtonian force for the quadrupole and all other higher-order multipoles taken into account, derived in general relativity by Racine\index{Racine}, Vine and Flanagan\index{Flanagan} (RVF) \citep{racine_2005PhRvD,racine2013PhRvD} by means of a different mathematical technique \citep{th_1985,futamase_1985PhRvD,futamase_2007LRR,asada_2011} nicely coincides (in case of the PPN parameters $\g=\b=1$) with our expression \eqref{pNf5s} in spite of different appearance of a few extra terms. Mathematical origin of this discrepancy is due to the different convention in the definition of time moments at which the numerical value of the body multipoles are to be computed on their world lines. This is explained in more detail in Appendix \ref{appendixA}. 

In particular, the term that had been missed in \citep[Equation 6.12c]{racine_2005PhRvD} and recovered in \citep[Equation 1.1]{racine2013PhRvD}, is given by our coupling coefficient $B^{LNSK}_{{\cal M}{\cal M}{\cal M}{\cal M}}$ in equation (\ref{e1a7}) which enters our expression (\ref{z1q3}) for the post-Newtonian force component $F^i_{{\cal M}{\cal M}{\cal M}{\cal M}}$. Notice also that we give our coupling coefficients for the expansion of force while \citet{racine_2005PhRvD} provide their coupling coefficients for acceleration of body B. Therefore, our tensor coupling coefficients must be divided by the inertial mass  $M_{\rm B}$ of body B in order to get the RVF coefficients. It is also worth noticing that, contrary to our choice of {\it dynamically} non-rotating local coordinates, \citet{racine_2005PhRvD} had chosen the body-adapted local coordinates as being {\it kinematically} non-rotating with respect to the global coordinates. For this reason the force \eqref{t1q5} caused by the relativistic precession of the local frame is absent in the RVF equations of motion. The present paper generalizes translational equations of motion derived by \citet{racine_2005PhRvD,racine2013PhRvD} to the realm of scalar-tensor theory of gravity parametrized with two covariantly-defined parameters\index{PPN parameters}, $\b$ and $\g$. This generalization is important for testing scalar-tensor theories of gravity with gravitational wave detectors and for developing more comprehensive experiments within the solar system.

It is instructive to better understand the correspondence between the post-Newtonian force (\ref{pNf5s}) for spherically-symmetric bodies and the EIH force \citep{eih}. The EIH equations of motion are traditionally viewed as equations of motion of point-like test particles which are modeled as non-rotating solid spheres having spherically-symmetric distribution of mass. The post-Newtonian force (\ref{pNf5s}) depends on the STF internal multipoles and it is reduced to the EIH force if we neglect all STF multipoles except of monopole ($l=0$) that corresponds to the relativistic (Tolman) mass of the body \citep{Tolman_book,Landau1975,mtw} if the body is fully isolated from the external gravitational environment preventing its tidal deformations. However, the spherically-symmetric distribution of matter does not ensure vanishing internal multipoles of the body. Indeed, the post-Newtonian definition of the mass multipoles (\ref{1.31}) includes the terms depending on volume integral, ${\cal Q}_K\int_{{\cal V}_{\rm B}}\sigma w^{<K>}w^{<L>}d^3w$, which does not vanish after integration over the unit sphere making the post-Newtonian force depending on the rotational moments of inertia of the spherically-symmetric bodies. Thus, the post-Newtonian force of interaction between rigid, spherically-symmetric bodies in ${\mathbb N}$-body system is not completely reduced to the EIH force but includes additional terms depending on the size of extended bodies. It makes clear that spherical bodies of finite size do not move like massive point particles and the effacing principle is violated \citep{kovl_2008}.     

Finite-size post-Newtonian effects in general-relativistic equations of motion of spherically-symmetric bodies were discussed previously by \citet{vab}, \citet{spyrou_1975ApJ,spyrou_1978GReGr,spyrou_1979GRG,spyrou_1981GRG},  \citet{caporali_1981_1,caporali_1981_2}, \citet{dallas_1977CeMec},\citet{vincent_1986CeMec} and, more recently, by \citet{arminjon_2005PhRvD}. The post-Newtonian correction to the EIH force obtained by these authors depends on the second-order rotational moments of inertia ${\cal N}$ defined in \eqref{r5t1}. We have shown in \citep[Section 6.3.4]{kopeikin_2011book} that this correction is not physical and represents a spurious, coordinate-dependent effect which can be removed by adjusting position of the center of mass and transforming body's quadrupole moment from the global to the body-adapted local coordinates. This fact was also noticed by \citet{nordtvedt_1994PhRvD}. Nonetheless, the post-Newtonian force of interaction between spherically-symmetric bodies can depend on the rotational moments of inertia of the second order in scalar-tensor theory of gravity -- see \citep[equation 6.85]{kopeikin_2011book}.  

\section{Rotational Equations of Motion of Spin in the Global Coordinates}\la{sbdytvre0}

Translational equations of motion of the centers of mass of extended, arbitrary structured bodies are not sufficient to describe gravitational dynamics of $\mathbb{N}$-body system. This is because the translational equations depend on the mass and spin multipoles of all bodies which are complicated functions of time. Therefore, they must be complemented with equations describing temporal evolution of the multipoles in order to close the system of differential equations for the configuration variables characterizing dynamics of $\mathbb{N}$-body system. Derivation of the complete system of the evolution equations for configuration variables is a daunting task as it includes among other issues, solution of the post-Newtonian problem of the elastic response of an extended body to the tidal perturbations caused by the presence of external bodies and calculation of rotational deformations of the body due to its rotation. Calculation of the tidal and rotational responses requires a corresponding development of the post-Newtonian theory of elastic deformations of extended, self-gravitating bodies \citep{xws1,xws2,xwsk} with its further dissemination to treat more subtle effects of viscosity and multi-layer structure of stars in astrophysical systems emitting gravitational waves. The overall task seems to be very complicated and will be discussed somewhere else. The present paper centers on the developing of equation of temporal evolution of the most important configuration variable in gravitational dynamics of $\mathbb{N}$-body system -- the intrinsic angular momentum  or spin of the bodies. Spin is closely related to three rotational degrees of freedom of a rigidly rotating extended body characterized by the vector of angular velocity. Therefore, we call the equation of temporal evolution for spin as rotational equations of motion.

Rotational equations of motion of spin of body B in the body-adapted local coordinates, $w^\alpha=(u,{\bm w})$, have been already derived in section \ref{q5a9n6}. The rotational equations of motion are parameterized with the local coordinate time $u_\B$ of the body-adpated coordinates and describe the force precession of body's spin, ${\cal S}^i_\B$, caused by gravitational coupling of the internal mass and spin multipoles of body B with the external multipoles. In its own turn, the body-adapated local frame is subject to the Fermi-Walker transport \cite{mtw} describing the relativistic precession of the spatial axes of the local coordinates with respect to the global coordinates in accordance with equation (\ref{5.18}). It is convenient from computational point of view to transform the rotational equations of motion of each body from the local to global coordinates to parameterize them with a single parameter - the global coordinate time $t$ and to include the Fermi-Walker transport to the evolution equation of spin. Moreover, we want to express all external multipoles in the rotational equations in the form of explicit functions of the global coordinates and multipole moments of the bodies. This procedure will formulate the rotational equations of motion in terms of the same set of configuration variables as that in the translational equations of motion of bodies.  

Let us define the spin components of body B measured with respect to the global coordinates as $S^i$. They are related to the spin components, ${\cal S}^i$, measured with respect to the body-adapted local coordinates by means of a post-Newtonian rotational transformation,
\be
S^i={\cal S}^j\left(\d^{ij}-F^{ij}_\B\right)\;,
\ee
where $F^{ij}_\B$ is the matrix of the Fermi-Walker precession of the local coordinates of body B. Then, rotational equations of motion of spin $S^i$ in the global coordinates are
\be \label{aa2653}
\frac{dS^i}{dt}=\frac{d{\cal S}^i}{du}\frac{du}{dt}-\frac{dF^{ij}_\B}{dt}{\cal S}^j-F^{ij}_\B\frac{dS^j}{du}\;,
\ee
where all derivatives are taken along the worldline ${\cal Z}$ of the center of mass of body B. Using equations \eqref{5.14}, \eqref{5.18}, \eqref{spin-10} for computing the time derivatives in \eqref{aa2653}, we get the rotational equations of spin of body B in the form,
\be \label{y775c12}
\frac{dS^i}{dt}=T^i
\ee 
where the spin $S^i$ is considered now as a function of time $t$ that is $S^i=S^i(u)|_{u=t}$. The total torque $T^i=T^i_{\rm B}+T^i_{\rm FW}$ is a linear combination of a torque $T^i_{\rm B}$ caused by the gravitational interaction of the internal multipoles of body B with the external multipoles, and a torque $T^i_{\rm FW}$ stemming from the Fermi-Walker precession,
\ba \label{y775c}
T^i_{\rm B}&=&\left\{1+\frac12 v^2_\B-\bar{U}(t,{\bm x}_\B) \right\} {\cal T}^i-F^{ij}_\B{\cal T}^j\;,\\
\label{y775c1}
T^i_{\rm FW}&=&\left\{v_{\rm B}^{[i}a^{j]}_\B-2(1+\gamma)\pd^{[i}\bar{U}^{j]}(t,{\bm x}_\B)-2(1+\gamma)v_{\rm B}^{[i}\pd^{j]}\bar{U}(t,{\bm x}_\B)\right\}S^j_\B\;.
\ea
Torque ${\cal T}^i$ in \eqref{y775c} has been introduced earlier in \eqref{b3ca8k}.
The next step in derivation of the rotational equations of motion is to compute the torque in the right-hand side of \eqref{y775c12} in an explicit analytic form as a function of common configuration variables -- the global coordinates of the center of mass of the bodies and their internal mass and spin multipole moments.  

\subsection{Computation of Torque}

Torque $T^i_{\rm B}$ in \eqref{y775c} is proportional to torque ${\cal T}^i$ given by equation \eqref{b3ca8k} that is computed by accounting for \eqref{n5vz1o}, \eqref{ne6v7l} and \eqref{b4v9h2}. It yields,
\ba\la{b5h1l8}
T^i_{\rm B}&=&\varepsilon_{ijk}\sum_{l=0}^\infty\frac1{l!}\left[1+\frac12 v^2_\B+(2\b-\g-2)\bar{U}(t,{\bm x}_\B)\right]\pd_{<kL>}\bar{W}(t,{\bm x}_\B){\cal M}^{jL}_\B\\\nonumber
&+&\varepsilon_{ijk}\sum_{l=0}^\infty\frac1{l!}\bigg[\pd_{<kL>}\bar V(t,{\bm x}_\B,l+1){\cal M}^{jL}_\B+\pd_{<L}\bar V_{k>}(t,{\bm x}_\B,l+1){\cal M}^{jL}_\B+\frac{l+1}{l+2}{\cal C}_{kL}{\cal S}^{jL}_\B\bigg]\\\nonumber
&-&
\varepsilon_{jpk}F^{ij}_\B\sum_{l=0}^\infty\frac1{l!}\pd_{<kL>}\bar{W}(t,{\bm x}_\B){\cal M}^{pL}_\B+\varepsilon_{ijk}a_\B^k\Big(3 a^p_\B{\cal M}^{jp}_\B+{\cal I}^j_{\rm c}\Big)\;,
\ea
where we have taken into account that the {\it active} dipole moment ${\cal M}^i_\B$ can be neglected in the post-Newtonian terms.
Gravitational potentials $\bar V(t,{\bm x}_\B,l+1)$ and $\bar V^i(t,{\bm x}_\B,l+1)$ are defined in \eqref{n3c72b} as sums taken over all bodies of $\mathbb{N}$-body system from potentials $V_{\rm C}$ and $V^i_{\rm C}$ given in \eqref{nrrx4e}, \eqref{bex5b7} along with \eqref{b3c7x5} and \eqref{zf4k9}. The linear sum of the STF derivatives from potentials  $\bar V(t,{\bm x}_\B,l+1)$ and $\bar V^i(t,{\bm x}_\B,l+1)$ that appear in \eqref{b5h1l8} does not contain the {\it non-canonical} potentials ${\cal R}^L$ and ${\cal N}^L$ which are mutually canceled out. We also notice that the acceleration-dependent terms in the third line of \eqref{b5h1l8} actually vanish because of the adjustment of the position of the center of mass of body B given by the {\it complementary} dipole function $I_{\rm c}^i$ defined in \eqref{m4g1x8p}. After summing up all terms in \eqref{b5h1l8} and accounting for the index peeling-off formula \eqref{me7v20}, we reduce the torque to a simpler form
\ba\la{on4c1f}
T^i_{\rm B}&=&\varepsilon_{ijk}\sum_{l=0}^\infty\frac1{l!}\left[1+\frac12 v^2_\B+2(\b-\g-1)\bar{U}(t,{\bm x}_\B)\right]\pd_{<kL>}\bar{W}(t,{\bm x}_\B){\cal M}^{jL}_\B
\\\nonumber
&+&\varepsilon_{ijk}\sum_{l=0}^\infty\frac1{l!}\pd_{<kL>}\bar\Omega(t,{\bm x}_\B,l){\cal M}^{jL}_\B+\varepsilon_{ijk}\sum_{l=0}^\infty\frac1{(l+1)!}\pd_{L}\bar\Omega_{k}(t,{\bm x}_\B,l){\cal M}^{jL}_\B\\\nonumber
&+&\varepsilon_{ijk}\sum_{l=0}^\infty\frac{l}{(l+1)!}\pd_{kL-1}\bar\Omega_{p}(t,{\bm x}_\B,l){\cal M}^{pjL-1}_\B
+\sum_{l=0}^\infty\frac{l+1}{(l+2)l!}\bar H_{jiL}{\cal S}^{jL}_\B\\
&-&\varepsilon_{jpk}F^{ij}_\B\sum_{l=0}^\infty\frac1{l!}\pd_{<kL>}\bar{W}(t,{\bm x}_\B){\cal M}^{pL}_\B
-\varepsilon_{ijk}\sum_{l=0}^\infty\frac1{l!}\left[F^{pk}_\B\pd_{<pL>}\bar{W}(t,{\bm x}_\B){\cal M}^{jL}_\B+F^{qp}_\B\pd_{<kqL>}\bar{W}(t,{\bm x}_\B){\cal M}^{jpL}_\B\right]\;,\nonumber
\ea
where the potentials $\bar W$, $\bar\Omega$, $\bar\Omega_k$ are given in \eqref{popka34}--\eqref{omi84v6} and tensor $\bar H_{jiL}$ is explained in \eqref{he5v2z}--\eqref{ket3b7}. The STF derivatives from $\bar W$, $\bar\Omega$ and $\bar\Omega_k$ have been computed in \eqref{bumb3d4}, \eqref{uv3f6l}, and \eqref{me7v20}--\eqref{u5bz4o} respectively. 

The torque depends on the contraction of the STF derivatives of the potentials with the Levi-Civita symbol $\varepsilon_{ijk}$. For computational convenience of the reader we provide their exact form below in order to facilitate tracking down the process of the computation. Because each of the barred potential is a linear superposition of the corresponding potentials of each body labeled with a letter C, we write down the corresponding formulas of contraction of the Levi-Civita symbol with the STF derivatives for the single-body potentials. Contraction of the derivatives from potentials $W_{\rm C}$ and $\Omega_{\rm C}$ with the Levi-Civita symbol are,
\ba\la{ff22uu}
\varepsilon_{ijk}\sum_{l=0}^\infty\frac1{l!}\pd_{<kL>}{W}_{\rm C}(t,{\bm x}){\cal M}^{jL}_\B&=&\varepsilon_{ijk}\sum_{l=0}^\infty\sum_{n=0}^{\infty}\frac{(-1)^n}{l!n!}\pd_{kLN}\bigg(\frac{1}{R_{\rm C}}\bigg){\cal M}^{jL}_\B{\cal M}_{\rm C}^N\;,
\\
\la{lllw6}
\varepsilon_{ijk}\sum_{l=0}^\infty\frac1{l!}\pd_{<kL>}\Omega_{\rm C}(t,{\bm x},l){\cal M}^{jL}_\B&=&
\ea
\ba
\nonumber\hspace{+2cm}
&&\varepsilon_{ijk}\sum_{l=0}^\infty\sum_{n=0}^{\infty}\frac{(-1)^n}{l!n!}\pd_{kLN}\bigg(\frac{1}{R_{\rm C}}\bigg)
\left[(1+\g) v_{\rm BC}^2-\frac12\frac{2l+2n+3}{2l+2n+5}v_{\rm C}^2\right]{\cal M}^{jL}_\B{\cal M}_{\rm C}^N
\\
&+&\varepsilon_{ijk}\sum_{l=0}^\infty\sum_{n=0}^{\infty}\frac{(-1)^n}{l!n!}\pd_{kLN}\bigg(\frac{1}{R_{\rm C}}\bigg)\left[(2-2\beta-l\gamma)\bar U(t,{\bm x}_\B)-\gamma(n+1)\bar U(t,{\bm x}_{\rm C})\right]{\cal M}^{jL}_\B{\cal M}_{\rm C}^N\nonumber
\\
&+&\varepsilon_{ijk}\sum_{l=0}^\infty\sum_{n=0}^{\infty}\frac{(-1)^n}{l!n!}\pd_{kpLN}\bigg(\frac{1}{R_{\rm C}}\bigg)\bigg( \frac{1}2 v_{\rm C}^pv_{\rm C}^{q}- F_{\rm C}^{qp}\bigg){\cal M}^{jL}_\B{\cal M}_{\rm C}^{qN}\nonumber
\\
&+&\varepsilon_{ijk}\sum_{l=0}^\infty\sum_{n=0}^{\infty}\frac{(-1)^n}{l!n!}(n+1)\pd_{kLN}\bigg(\frac{1}{R_{\rm C}}\bigg)a_{\rm C}^p{\cal M}^{jL}_\B{\cal M}_{\rm C}^{pN}\nonumber
\\
&+&\varepsilon_{ijk}\sum_{l=0}^\infty\sum_{n=0}^{\infty}\frac{(-1)^n}{2l!n!}\left[\ddot{{\cal M}}_{\rm C}^N\pd_{<kLN>}R_{\rm C}+{{\cal M}}_{\rm C}^Nv_{\rm C}^pv_{\rm C}^q\pd_{<kpqLN>}R_{\rm C}- {{\cal M}}_{\rm C}^Na_{\rm C}^p\pd_{<kpLN>}R_{\rm C}\right]{\cal M}^{jL}_\B\nonumber
\\
     &-&\varepsilon_{ijk}\sum_{l=0}^\infty\sum_{n=0}^{\infty}\frac{(-1)^n}{l!n!}\frac{1}{2l+2n+3} \pd_{LN} \bigg(\frac{1}{R_{\rm C}}\bigg){\cal M}_\B^{jL}\ddot{\cal M}_{\rm C}^{kN} \nonumber
     \\ 
     &-&\varepsilon_{ijk}\sum_{l=0}^\infty\sum_{n=0}^{\infty}\frac{(-1)^n}{l!n!}\frac{1}{2l+2n+5} \pd_{kLN} \bigg(\frac{1}{R_{\rm C}}\bigg){\cal M}_\B^{jpL}\ddot{\cal M}_{\rm C}^{pN} \nonumber
     \\ 
&+&\varepsilon_{ijk}\sum_{l=0}^\infty\sum_{n=0}^{\infty}\frac{(-1)^n}{l!n!}\frac{2}{2l+2n+5} \pd_{pLN} \bigg(\frac{1}{R_{\rm C}}\bigg){\cal M}_\B^{jL}{\cal M}_{\rm C}^{N}v^p_{\rm C} v^k_{\rm C} \nonumber
     \\      
 &-&\varepsilon_{ijk}\sum_{l=0}^\infty\sum_{n=0}^{\infty}\frac{(-1)^n}{l!n!}\frac{1}{2l+2n+7} \pd_{pqLN} \bigg(\frac{1}{R_{\rm C}}\bigg){\cal M}_\B^{jL}{\cal M}_{\rm C}^{kN}v_{\rm C}^p v_{\rm C}^q \nonumber\\ 
 &+&\varepsilon_{ijk}\sum_{l=0}^\infty\sum_{n=0}^{\infty}\frac{(-1)^n}{l!n!}\frac{2}{2l+2n+7} \pd_{kpLN} \bigg(\frac{1}{R_{\rm C}}\bigg)\left({\cal M}_\B^{jqL}{\cal M}_{\rm C}^N-{\cal M}_\B^{jL}{\cal M}_{\rm C}^{qN}\right)v_{\rm C}^p v_{\rm C}^q \nonumber\\ 
 &-&\varepsilon_{ijk}\sum_{l=0}^\infty\sum_{n=0}^{\infty}\frac{(-1)^n}{l!n!}\frac{1}{2l+2n+9} \pd_{kpqLN} \bigg(\frac{1}{R_{\rm C}}\bigg){\cal M}_\B^{jmL}{\cal M}_{\rm C}^{mN}v_{\rm C}^p v_{\rm C}^q \nonumber\\ 
 &-&\varepsilon_{ijk}\sum_{l=0}^\infty\sum_{n=0}^{\infty}\frac{(-1)^n}{l!n!}\frac{1}{2l+2n+3} \pd_{LN} \bigg(\frac{1}{R_{\rm C}}\bigg){\cal M}_\B^{jL}{\cal M}_{\rm C}^{N}a_{\rm C}^k \nonumber\\ 
 &+&\varepsilon_{ijk}\sum_{l=0}^\infty\sum_{n=0}^{\infty}\frac{(-1)^n}{l!n!}\frac{1}{2l+2n+5} \pd_{pLN} \bigg(\frac{1}{R_{\rm C}}\bigg){\cal M}_\B^{jL}{\cal M}_{\rm C}^{kN}a_{\rm C}^p  \nonumber\\ 
 &-&\varepsilon_{ijk}\sum_{l=0}^\infty\sum_{n=0}^{\infty}\frac{(-1)^n}{l!n!}\frac{1}{2l+2n+5} \pd_{kLN} \bigg(\frac{1}{R_{\rm C}}\bigg)\left({\cal M}_\B^{jpL}{\cal M}_{\rm C}^N-{\cal M}_\B^{jL}{\cal M}_{\rm C}^{pN}\right)a_{\rm C}^p  \nonumber\\ 
 &+&\varepsilon_{ijk}\sum_{l=0}^\infty\sum_{n=0}^{\infty}\frac{(-1)^n}{l!n!}\frac{1}{2l+2n+7} \pd_{kpLN} \bigg(\frac{1}{R_{\rm C}}\bigg){\cal M}_\B^{jqL}{\cal M}_{\rm C}^{qN}a_{\rm C}^p  \nonumber\\ 
 &-&2(1+\gamma)\varepsilon_{ijk}\sum_{l=0}^\infty\sum_{n=0}^{\infty}\frac{(-1)^n}{l!n!}\frac1{n+1}\pd_{kLN}\bigg(\frac{1}{R_{\rm C}}\bigg){\cal M}^{jL}_\B\dot{{\cal M}}_{\rm C}^{pN}v_{\B{\rm C}}^p
 \nonumber\\
   &-&2(1+\gamma)\varepsilon_{ijk}\sum_{l=0}^\infty\sum_{n=0}^{\infty}\frac{(-1)^n}{l!n!}\frac1{n+2}\varepsilon_{mpq}\pd_{kpLN}\bigg(\frac{1}{R_{\rm C}}\bigg){\cal M}^{jL}_\B\mathcal{S}_{\rm C}^{qN}v_{\B{\rm C}}^m\nonumber\;.
\ea
The very last term in the right-hand side of \eqref{lllw6} contains a product of two Levi-Civita symbols which can be expressed as a linear combination of the Kronecker delta-symbols \citep[Exercise 3.13]{mtw},
 \ba\la{bbb34}
\varepsilon_{ijk}\varepsilon_{mpq}&\equiv&\left\| \begin{array}{ccc}
         \d_{im}&\d_{ip}&\d_{iq}\\
         \d_{jm}&\d_{jp}&\d_{jq}\\
        \d_{km}&\d_{kp}&\d_{kq}\end{array} \right\|
=\d_{im}\d_{jp}\d_{kq}+\d_{ip}\d_{jq}\d_{km}+\d_{iq}\d_{jm}\d_{kp}-\d_{jm}\d_{ip}\d_{kq}-\d_{jp}\d_{iq}\d_{km}-\d_{jq}\d_{im}\d_{kp}\;.
\ea
It allows us to recast the term with two Levi-Civita symbols to a more transparent form
\ba\la{pmmm76}
\varepsilon_{ijk}\varepsilon_{mpq}\pd_{kpLN}\bigg(\frac{1}{R_{\rm C}}\bigg){\cal M}^{jL}_\B\mathcal{S}_{\rm C}^{qN}v_{\B{\rm C}}^m&=&2\pd_{ipLN}\bigg(\frac1{R_{\rm C}}\bigg){\cal M}^{qL}_\B{\cal S}_{\rm C}^{N[q}v_{\B{\rm C}}^{p]}-2\pd_{pqLN}\bigg(\frac1{R_{\rm C}}\bigg){\cal M}^{qL}_\B{\cal S}_{\rm C}^{N[i}v_{\B{\rm C}}^{p]}\;,
\ea

The two terms in \eqref{on4c1f} depending on the contraction of the Levi-Civita symbol with the STF derivatives of the vector potential $\Omega^i_{\rm C}$ are 
\ba
\la{hye5b3qq}\hspace{-0.5cm}
\varepsilon_{ijk}\sum_{l=0}^\infty\frac1{(l+1)!}\pd_{L}\Omega^k_{\rm C}(t,{\bm x},l){\cal M}_\B^{jL}&=&2(1+\gamma)\varepsilon_{ijk}\sum_{l=0}^\infty\sum_{n=0}^{\infty}\frac{(-1)^n}{(l+1)!n!}\frac1{n+1}\pd_{LN}\bigg(\frac{1}{R_{\rm C}}\bigg){\cal M}_\B^{jL}\ddot{{\cal M}}_{\rm C}^{kN}\\\nonumber
&+&2(1+\g)\varepsilon_{ijk}\sum_{l=0}^\infty\sum_{n=0}^{\infty}\frac{(-1)^n}{(l+1)!n!}\pd_{LN}\bigg(\frac{1}{R_{\rm C}}\bigg)\left(\dot{{\cal M}}_{\rm C}^{N}v_{\rm C}^k+{{\cal M}}_{\rm C}^{N}a_{\rm C}^k \right){\cal M}_\B^{jL}
\\\nonumber
&+&2(1+\g)\varepsilon_{ijk}\sum_{l=0}^\infty\sum_{n=0}^{\infty}\frac{(-1)^n}{(l+1)!n!}\frac1{n+2}\varepsilon_{kpq}\pd_{pLN}\bigg(\frac{1}{R_{\rm C}}\bigg){\cal M}_\B^{jL}\dot{\mathcal{S}}_{\rm C}^{qN}\\\nonumber
&+&2(1+\g)\varepsilon_{ijk}\sum_{l=0}^\infty\sum_{n=0}^{\infty}\frac{(-1)^n}{(l+1)!n!}\frac1{n+2}\varepsilon_{kpq}\pd_{mpLN}\bigg(\frac{1}{R_{\rm C}}\bigg){\cal M}_\B^{jL}{\mathcal{S}}_{\rm C}^{qN}v_{\B{\rm C}}^m
\\\nonumber
&+&2(1+\g)\varepsilon_{ijk}\sum_{l=0}^\infty\sum_{n=1}^{\infty}\frac{(-1)^n}{(l+1)!n!}\frac1{n+1}\pd_{pLN}\bigg(\frac{1}{R_{\rm C}}\bigg){\cal M}_\B^{jL}\dot{{\cal M}}_{\rm C}^{kN}v_{\B{\rm C}}^p\\\nonumber
&+&2(1+\g)\varepsilon_{ijk}\sum_{l=0}^\infty \sum_{n=0}^{\infty}\frac{(-1)^n}{(l+1)!n!}\pd_{pLN}\bigg(\frac{1}{R_{\rm C}}\bigg){\cal M}_\B^{jL}{{\cal M}}_{\rm C}^{N}v_{\B{\rm C}}^pv_{\rm C}^k 
\\\nonumber
&+&\varepsilon_{ijk}\sum_{l=0}^\infty\sum_{n=0}^{\infty}\frac{(-1)^n}{(l+1)!n!}\Big(l-1-2\gamma\Big)\pd_{LN}\bigg(\frac{1}{R_{\rm C}}\bigg){\cal M}_\B^{jL}\dot{{\cal M}}_{\rm C}^{N}v_{\rm B}^k\\\nonumber
&+&\varepsilon_{ijk}\sum_{l=0}^\infty\sum_{n=0}^{\infty}\frac{(-1)^n}{(l+1)!n!}\Big(l-1-2\gamma\Big)\pd_{pLN}\bigg(\frac{1}{R_{\rm C}}\bigg){\cal M}_\B^{jL}{{\cal M}}_{\rm C}^{N}v_{\B{\rm C}}^pv_{\rm B}^k\\\nonumber
&-&\frac{1}2\varepsilon_{ijk}\sum_{l=0}^\infty\sum_{n=0}^{\infty}\frac{(-1)^n}{l!n!}\pd_{pLN}\bigg(\frac{1}{R_{\rm C}}\bigg){\cal M}_\B^{jL}{{\cal M}}_{\rm C}^{N}v^p_{\rm B}v_{\rm B}^k\\\nonumber
&-&\varepsilon_{ijk}\sum_{l=0}^\infty\sum_{n=0}^{\infty}\frac{(-1)^n}{(l+1)!n!}\Big(l^2+l+2+2\gamma\Big)\pd_{LN}\bigg(\frac{1}{R_{\rm C}}\bigg){\cal M}_\B^{jL}{{\cal M}}_{\rm C}^{N}a_{\rm B}^k\;,
 \\\nonumber\\\nonumber\
 \ea
\ba
\la{yrtr3qq}
\varepsilon_{ijk}\sum_{l=0}^\infty\frac{l}{(l+1)!}\pd_{kL-1}\Omega^p_{\rm C}(t,{\bm x},l){\cal M}_\B^{jpL-1}&=&\ea
\ba
\nonumber
\hspace{+4cm}
&&2(1+\gamma)\varepsilon_{ijk}\sum_{l=0}^\infty\sum_{n=0}^{\infty}\frac{(-1)^n}{l!(n+1)!}\frac1{l+2}\pd_{kLN}\bigg(\frac{1}{R_{\rm C}}\bigg){\cal M}_\B^{jpL}\ddot{{\cal M}}_{\rm C}^{pN}\\\nonumber
&+&2(1+\g)\varepsilon_{ijk}\sum_{l=0}^\infty\sum_{n=0}^{\infty}\frac{(-1)^n}{l!n!}\frac1{l+2}\pd_{kLN}\bigg(\frac{1}{R_{\rm C}}\bigg)\left(\dot{{\cal M}}_{\rm C}^{N}v_{\rm C}^p+{{\cal M}}_{\rm C}^{N}a_{\rm C}^p \right){\cal M}_\B^{jpL}
\\\nonumber
&+&2(1+\g)\varepsilon_{ijk}\sum_{l=0}^\infty\sum_{n=0}^{\infty}\frac{(-1)^n(n+1)(l+1)}{(l+2)!(n+2)!}\varepsilon_{pmq}\pd_{kmLN}\bigg(\frac{1}{R_{\rm C}}\bigg){\cal M}_\B^{jpL}\dot{\mathcal{S}}_{\rm C}^{qN}\\\nonumber
&+&2(1+\g)\varepsilon_{ijk}\sum_{l=0}^\infty\sum_{n=0}^{\infty}\frac{(-1)^n(n+1)(l+1)}{(l+2)!(n+2)!}\varepsilon_{pmq}\pd_{kbmLN}\bigg(\frac{1}{R_{\rm C}}\bigg){\cal M}_\B^{jpL}{\mathcal{S}}_{\rm C}^{qN}v_{\B{\rm C}}^b
\\\nonumber
&+&2(1+\g)\varepsilon_{ijk}\sum_{l=0}^\infty\sum_{n=0}^{\infty}\frac{(-1)^n}{l!(n+1)!}\frac1{l+2}\pd_{kmLN}\bigg(\frac{1}{R_{\rm C}}\bigg){\cal M}_\B^{jpL}\dot{{\cal M}}_{\rm C}^{pN}v_{\B{\rm C}}^m\\\nonumber
&+&2(1+\g)\varepsilon_{ijk}\sum_{l=0}^\infty \sum_{n=0}^{\infty}\frac{(-1)^n}{l!n!}\frac1{l+2}\pd_{kmLN}\bigg(\frac{1}{R_{\rm C}}\bigg){\cal M}_\B^{jpL}{{\cal M}}_{\rm C}^{N}v_{\B{\rm C}}^mv_{\rm C}^p 
\\\nonumber
&+&\varepsilon_{ijk}\sum_{l=0}^\infty\sum_{n=0}^{\infty}\frac{(-1)^n}{l!n!}\frac{l-2\gamma}{l+2}\pd_{kLN}\bigg(\frac{1}{R_{\rm C}}\bigg){\cal M}_\B^{jpL}\dot{{\cal M}}_{\rm C}^{N}v_{\rm B}^p\\\nonumber
&+&\varepsilon_{ijk}\sum_{l=0}^\infty\sum_{n=0}^{\infty}\frac{(-1)^n}{l!n!}\frac{l-2\gamma}{l+2}\pd_{kmLN}\bigg(\frac{1}{R_{\rm C}}\bigg){\cal M}_\B^{jpL}{{\cal M}}_{\rm C}^{N}v_{\B{\rm C}}^mv_{\rm B}^p\\\nonumber
&-&\frac{1}2\varepsilon_{ijk}\sum_{l=0}^\infty\sum_{n=0}^{\infty}\frac{(-1)^n}{l!n!}\pd_{kmLN}\bigg(\frac{1}{R_{\rm C}}\bigg){\cal M}_\B^{jpL}{{\cal M}}_{\rm C}^{N}v^m_{\rm B}v_{\rm B}^p\\\nonumber
&-&\varepsilon_{ijk}\sum_{l=0}^\infty\sum_{n=0}^{\infty}\frac{(-1)^n}{l!n!}\frac{l^2+3l+4+2\gamma}{l+2}\pd_{kLN}\bigg(\frac{1}{R_{\rm C}}\bigg){\cal M}_\B^{jpL}{{\cal M}}_{\rm C}^{N}a_{\rm B}^p\;,
 \\\nonumber
 \ea
Again, we use \eqref{bbb34} in order to simplify those terms in \eqref{hye5b3qq}, \eqref{yrtr3qq} which contain the product of two Levi-Civita symbols. More specifically, the two terms in equation \eqref{hye5b3qq} are simplified to  
\ba\la{qq2z4x21}
\varepsilon_{ijk}\varepsilon_{kpq}\pd_{pLN}\bigg(\frac{1}{R_{\rm C}}\bigg){\cal M}^{jL}_\B\dot{\mathcal{S}}_{\rm C}^{qN}&=&2\pd_{iLN}\bigg(\frac1{R_{\rm C}}\bigg){\cal M}^{qL}_\B\dot{\cal S}_{\rm C}^{qN}-2\pd_{pLN}\bigg(\frac1{R_{\rm C}}\bigg){\cal M}^{pL}_\B\dot{\cal S}_{\rm C}^{iN}\;,\\
\la{v3c19n3}
\varepsilon_{ijk}\varepsilon_{kpq}\pd_{mpLN}\bigg(\frac{1}{R_{\rm C}}\bigg){\cal M}^{jL}_\B\mathcal{S}_{\rm C}^{qN}v_{\B{\rm C}}^m&=&2\pd_{ipLN}\bigg(\frac1{R_{\rm C}}\bigg){\cal M}^{qL}_\B{\cal S}_{\rm C}^{qN}v_{\B{\rm C}}^{p}-2\pd_{pqLN}\bigg(\frac1{R_{\rm C}}\bigg){\cal M}^{pL}_\B{\cal S}_{\rm C}^{iN}v_{\B{\rm C}}^{q}\;,
\ea
and the two other terms in \eqref{yrtr3qq} are
\ba\la{o5b2x8}
\varepsilon_{ijk}\varepsilon_{pmq}\pd_{kmLN}\bigg(\frac{1}{R_{\rm C}}\bigg){\cal M}^{jpL}_\B\dot{\mathcal{S}}_{\rm C}^{qN}&=&2\pd_{jkLN}\bigg(\frac1{R_{\rm C}}\bigg){\cal M}^{jL[i}_\B\dot{\cal S}_{\rm C}^{k]N}+\pd_{ikLN}\bigg(\frac1{R_{\rm C}}\bigg){\cal M}^{qkL}_\B\dot{\cal S}_{\rm C}^{qN}\;,\\
\la{rt2c4l6}
\varepsilon_{ijk}\varepsilon_{pmq}\pd_{kbmLN}\bigg(\frac{1}{R_{\rm C}}\bigg){\cal M}^{jpL}_\B{\mathcal{S}}_{\rm C}^{qN}v_{\B{\rm C}}^b&=&2\pd_{jkpLN}\bigg(\frac1{R_{\rm C}}\bigg){\cal M}^{jL[i}_\B{\cal S}_{\rm C}^{k]N}v_{\B{\rm C}}^p+\pd_{ikpLN}\bigg(\frac1{R_{\rm C}}\bigg){\cal M}^{qkL}_\B{\cal S}_{\rm C}^{qN}v_{\B{\rm C}}^p\;,
\ea

Multipolar expansion of the term in \eqref{on4c1f} containing the product of the STF derivative of $H_{\rm C}^{jiL}$ with the spin multipoles, reads
\ba\la{n1z9v5s8}
\sum_{l=0}^\infty\frac{l+1}{(l+2)l!} H_{\rm C}^{jiL}{\cal S}^{jL}_\B
&=&4(1+\g)\sum_{l=0}^\infty\sum_{n=0}^\infty\frac{(-1)^n}{n!}\frac{l+1}{(l+2)l!} v_{\rm BC}^{[j}\pd^{i]LN}\bigg(\frac1{R_{\rm C}}\bigg){\cal M}_{\rm C}^N{\cal S}^{jL}_\B\\\nonumber
&-&4(1+\g)\sum_{l=0}^\infty\sum_{n=0}^{\infty}\frac{(-1)^n}{(n+1)!}\frac{l+1}{(l+2)l!}\dot{{\cal M}}_{\rm C}^{N[j}\pd^{i]LN}\bigg(\frac{1}{R_{\rm C}}\bigg){\cal S}^{jL}_\B\nonumber\\
 &+&4(1+\g)\sum_{l=0}^\infty\sum_{n=0}^{\infty}\frac{(-1)^n}{(n+2)n!}\frac{l+1}{(l+2)l!}\varepsilon^{pq[j}\pd^{i]qLN}\bigg(\frac{1}{R_{\rm C}}\bigg)\mathcal{S}_{\rm C}^{pN}{\cal S}^{jL}_\B
\nonumber\\\nonumber
  &+&2(1+\g)\sum_{l=0}^\infty\sum_{n=0}^\infty\frac{(-1)^n}{n!}\frac{1}{(l+3)l!}\bigg[v^p_{\B{\rm C}} \pd^{pjNL}\bigg(\frac1{R_{\rm C}}\bigg){\cal M}_{\rm C}^N +\pd^{jNL}\bigg(\frac1{R_{\rm C}}\bigg)\dot{\cal M}_{\rm C}^N\bigg]{\cal S}^{ijL}_\B 
\ea

\subsection{Explicit Formula for Torque}

The total torque $T^i$ governing precession of spin of body B in the global coordinates is given in the right-hand side of the rotational equations of motion \eqref{y775c12} as a sum of two terms, $T^i_{\rm B}+T^i_{\rm FW}$, where the pure gravitational torque, $T^i_{\rm B}$, has been defined in \eqref{y775c} and \eqref{on4c1f} and the Fermi-Walker torque, $T^i_{\rm FW}$, is given in \eqref{y775c1}. After substituting equations \eqref{ff22uu}--\eqref{n1z9v5s8} into \eqref{on4c1f} and reducing similar terms the gravitational torque can be represented as a sum of the Newtonian and post-Newtonian terms, $T^i_{\rm B}=T^i_{\rm N}+T^i_{\rm pN}$. Hence, the total torque $T^i$ is given by
\be\la{mevzrw235}
T^i=T^i_{\rm N}+T^i_{\rm pN}+T^i_{\rm FW}\;,
\ee
where $T^i_{\rm N}$ is the Newtonian part of the torque, $T^i_{\rm pN}$ is its post-Newtonian counterpart, and $T^i_{\rm FW}$ is the Fermi-Walker torque. We provide explicit multipolar expressions for the gravitational torque in subsections \ref{nt9123} and \ref{nt9124} below. Explicit multipolar expansion of the Fermi-Walker torque is given in subsection \ref{nt9125}. 

\subsubsection{Newtonian Torque}\label{nt9123}

The Newtonian torque, $T^i_{\rm N}$, is defined by the very first term in equation \eqref{on4c1f},
\be\label{gg44cc22}
T^i_{\rm N}=\varepsilon_{ijk}\sum_{l=0}^\infty\frac1{l!}\pd_{<kL>}\bar W(t,{\bm x}_\B){\cal M}^{jL}_\B=\varepsilon_{ijk}\sum_{{\rm C}\not=\B}\sum_{l=0}^\infty\frac1{l!}\pd_{<kL>}W_{\rm C}(t,{\bm x}_\B){\cal M}^{jL}_\B\;
\ee
where $\pd_{<kL>}W_{\rm C}(t,{\bm x}_\B)=\lim_{{\bm x}\rightarrow{\bm x}_\B}\pd_{<kL>}W_{\rm C}(t,{\bm x})$, and multipolar expansion of gravitational potential $W_{\rm C}(t,{\bm x})$ has been defined in \eqref{je5v20}. After taking the partial STF derivatives from the potential $W_{\rm C}$, the Newtonian torque takes on the following explicit form,
\be\la{dc5se1}
T^i_{\rm N}= \varepsilon_{ijk}\sum_{{\rm C}\not=\B}\sum_{l=0}^\infty\sum_{n=0}^\infty\frac{(-1)^n}{l!n!}{\cal M}^{jL}_\B{\cal M}_{\rm C}^N  \pd_{<kLN>}R^{-1}_{\B{\rm C}}\;.
\ee
Applying \eqref{acser243} yields the Newtonian torque in its final form,
\be\la{m5btz4s}
T^i_{\rm N}=
-\varepsilon_{ijk}\sum_{{\rm C}\not=\B}\sum_{l=0}^\infty\sum_{n=0}^\infty\frac{(-1)^l(2l+2n+1)!!}{l!n!}{\cal M}^{jL}_\B{\cal M}^N_{\rm C}\frac{R_{\B{\rm C}}^{<kLN>}}{R_{\B{\rm C}}^{2l+2n+3}}\;,
\ee
where, here and everywhere else, all multipoles of body B are taken at the time $u^*_{\rm B}$ given by \eqref{jk9wss5f}, and all multipoles of body C$\not=$B are taken at time $u^*_{\rm C}$ given by \eqref{acz521sa}. Formula of the multipolar expansion for the Newtonian torque has been also derived by \citet{Racine_2006CQG} in general relativity. Torque \eqref{m5btz4s} depends on the {\it active} mass multipoles in the right-hand side of this equation and generalizes the results of \citep{Racine_2006CQG} to scalar-tensor theory of gravity. Equation \eqref{m5btz4s} reduces to the expression derived by \citet{Racine_2006CQG} in case of the PPN parameters $\b=\g=1$.  

We draw attention of the reader to the fact that the {\it active} multipoles in \eqref{m5btz4s} are defined with taking into account all post-Newtonian contributions from the stress-energy tensor of the extended bodies in accordance with their definition \eqref{1.31}. It is also worth noticing that the {\it active} dipole ${\cal M}^i_\B$ of each body is explicitly included to the right-hand side of the Newtonian torque \eqref{m5btz4s} as it does not vanish because the center of mass of each body B is defined by the condition of vanishing {\it conformal} dipole, ${\cal I}^i_\B=0$, in accordance with \eqref{n5vz1o}. It means that in contrast to general theory of relativity (c.f. \citep[Equation 91]{Racine_2006CQG}), the dipole-monopole gravitational torque that is the term with $l=0$, $n=0$ in \eqref{m5btz4s}, is present in the scalar-tensor theory of gravity even if the origin of the local coordinates is fixed exactly at the center of mass of the body. The dipole-monopole torque in the rotational equation of motion of spin causes an {\it anomalous} precession of body's spin as compared with general relativity. The anomalous precession of the spin is caused by the difference between the {\it active}, ${\cal M}^i_\B$, and {\it conformal}, ${\cal I}^i_\B$, dipole moments of the body B in scalar-tensor theory of gravity. This resembles the Dicke-Nordtvedt effect of violation of strong principle of equivalence in translational motion of the bodies, which is caused by the difference between {\it active}, ${\cal M}_\B$, and {\it conformal}, $M_\B$, masses of the body, to the case of rotational motion of the bodies. Measurement of the anomalous pole-dipole torque can help to set a direct experimental limitation on the PPN parameter $\beta$ which is currently measured only indirectly through the measurement of the Nordtvedt parameter $\eta=4\b-\g-3$, primarily by LLR technique \citep{LLR_2008ASSL,LLR_2010A&A,LLR_2018CQGra}, after subtracting the best numerical estimate of the parameter $\g$ obtained, for example, from the measurement of gravitational bending of light \citep{Fomalont_2009ApJ,Bertotti_2003Natur,Kopeikin_2007PhLA}.     

\subsubsection{Post-Newtonian Torque}\label{nt9124}

Multipolar expansion of the post-Newtonian gravitational torque, $T^i_{\rm pN}$, can be represented in the form of a linear operator from the STF partial derivatives with respect to spatial coordinates similarly to the presentation of the post-Newtonian force in the translational equations of motion,
\ba\la{we534h1}
T^i_{\rm pN}&=&\sum_{{\rm C}\not=\B}\sum_{l=0}^{\infty}\sum_{n=0}^{\infty}\frac{(-1)^n}{l!n!}\bigg[\Big(\a^{iLN}_{\rm T}+\beta^{iLN}_{\rm T}\Big)\pd_{<LN>}+\Big(\a^{ipLN}_{\rm T}+\beta^{ipLN}_{\rm T}+\gamma^{ipLN}_{\rm T}\Big)\pd_{<pLN>}\\\nonumber
&&\phantom{\sum_{{\rm C}\not=\B}\sum_{l=0}^{\infty}\sum_{n=0}^{\infty}\frac{(-1)^n}{l!n!}\bigg[}
+\Big(\a^{ipqLN}_{\rm T}+\beta^{ipqLN}_{\rm T}\Big)\pd_{<pqLN>}+\a^{ikpqLN}_{\rm T}\pd_{<kpqLN>}\bigg]R_{\rm C}^{-1}\\\nonumber
&+&\sum_{{\rm C}\not=\B}\sum_{l=0}^{\infty}\sum_{n=0}^{\infty}\frac{(-1)^n}{l!n!}\bigg[\a^{LN}_{\rm T}\pd_{<iLN>}+\mu^{pLN}_{\rm T}\pd_{<ipLN>}+\sigma^{pqLN}_{\rm T}\pd_{<ipqLN>}\Big]R_{\rm C}^{-1}\\\nonumber
&+&\frac12\varepsilon_{ijk}\sum_{{\rm C}\not=\B}\sum_{l=0}^\infty\sum_{n=0}^{\infty}\frac{(-1)^n}{l!n!}{\cal M}_\B^{jL}\Big[\ddot{{\cal M}}_{\rm C}^N\pd_{<kLN>}- {{\cal M}}_{\rm C}^Na_{\rm C}^p\pd_{<kpLN>}+{{\cal M}}_{\rm C}^Nv_{\rm C}^pv_{\rm C}^q\pd_{<kpqLN>}\Big]R_{\rm C}\;,
\ea
where the STF derivatives from $R^{-1}_{\B{\rm C}}$ and $R_{\B{\rm C}}$ are understood in the sense of equations \eqref{acser243}, \eqref{nrvx31w}. The coefficients of operator \eqref{we534h1} are
\ba
\la{m1x7z0n4}
\a_{\rm T}^{iLN}&=&\varepsilon_{ipk}\bigg[\frac{2(1+\g)}{(l+1)(n+1)}-\frac1{2l+2n+3}
\bigg]{\cal M}_\B^{pL}\ddot{\cal M}_{\rm C}^{kN}+\varepsilon_{ipk}\bigg[v^k_\B-\frac{2(1+\g)}{l+1}v^k_{\B{\rm C}}\bigg]
{\cal M}_\B^{pL}\dot{\cal M}_{\rm C}^{N}\;,\\
\la{e9v1z7m}
\beta_{\rm T}^{iLN}&=&\varepsilon_{ikp}\bigg[\Big(\frac1{2l+2n+3}
-2\frac{1+\g}{l+1}\Big)a^k_{\rm C}+\Big(l+2\frac{1+\g}{l+1}\Big)a^k_{\rm B}\bigg]{\cal M}_\B^{pL}{\cal M}_{\rm C}^{N}\;,
\\\label{bb33vv1}
\a_{\rm T}^{ipLN}&=&\varepsilon_{ikp}\bigg[\frac12 v_\B^2+(1+\g)v^2_{\B{\rm C}}-\frac12\frac{2l+2n+3}{2l+2n+5}v_{\rm C}^2\bigg]{\cal M}_\B^{kL}{\cal M}_{\rm C}^{N}-\varepsilon_{jkp}F^{ij}_\B{\cal M}^{kL}_\B{\cal M}_{\rm C}^{N} \\\nonumber
&&+\varepsilon_{ikq}\bigg[v^p_{\B{\rm C}}v^q_B-\frac12 v^p_B v^q_B-\frac{2(1+\g)}{l+1}v^p_{\B{\rm C}}v^q_{\B{\rm C}}+\frac2{2l+2n+5}v_{\rm C}^pv_{\rm C}^q-F^{pq}_\B\bigg]{\cal M}^{kL}_\B{\cal M}_{\rm C}^{N}\\\nonumber
&&+\varepsilon_{ikp}\bigg[\frac{2(1+\g)}{(l+2)(n+1)}-\frac1{2l+2n+5}\bigg]{\cal M}_\B^{kqL}\ddot{\cal M}_{\rm C}^{qN}
+\frac{2(1+\g)}{n+1}\bigg[\frac{\varepsilon_{ikq}}{l+1}v^p_{\B{\rm C}}-\varepsilon_{ikp}v_{\B{\rm C}}^q\bigg]{\cal M}^{kL}_\B\dot{\cal M}_{\rm C}^{qN}\\\nonumber
&&+\varepsilon_{ikp}\bigg[v^q_\B-\frac{2(1+\g)}{l+2}v^q_{\rm BC}\bigg]{\cal M}_\B^{kqL}\dot{\cal M}_{\rm C}^{N}-2(1+\g)\frac{l+1}{l+2}\bigg[{\cal S}_\B^{pL}{\cal M}_{\rm C}^{N}v^i_{\B{\rm C}}-\frac1{n+1}{\cal S}_\B^{pL}\dot{\cal M}_{\rm C}^{iN}\bigg]\\\nonumber
&&+\frac{2(1+\g)}{l+3}{\cal S}_\B^{ipL}\dot{\cal M}_{\rm C}^{N}
-\frac{4(1+\g)}{(l+1)(n+2)}{\cal M}_\B^{pL}\dot{\cal S}_{\rm C}^{iN}\;,\\\nonumber
\\
\beta_{\rm T}^{ipLN}&=&\varepsilon_{ikp}\bigg[n+1+\frac1{2l+2n+5}\bigg]{\cal M}_\B^{kL}{\cal M}_{\rm C}^{qN}a_{\rm C}^q
+\frac1{2l+2n+5}\varepsilon_{ikq}{\cal M}_\B^{kL}{\cal M}_{\rm C}^{qN}a_{\rm C}^p\\\nonumber
&&+\varepsilon_{ikp}\bigg[\frac{2(1+\g)}{l+2}a^q_{\rm C}-\frac1{2l+2n+5}a_{\rm C}^q-\frac{l^2+3l+4+2\g}{l+2}a^q_\B\bigg]{\cal M}_\B^{kqL}{\cal M}_{\rm C}^{N}\;,
\\
\gamma_{\rm T}^{ipLN}&=&-\gamma\varepsilon_{ikp}\bigg[(l+2)\bar U(t,{\bm x}_\B)+(n+1))\bar U(t,{\bm x}_{\rm C})\bigg]{\cal M}^{kL}_\B{\cal M}^N_{\rm C}\;,
\\\label{bb33vv2}
\alpha_{\rm T}^{ipqLN}&=&\varepsilon_{ijq}\bigg[\frac12-\frac2{2l+2n+7}\bigg] {\cal M}^{jL}_\B{\cal M}^{kN}_{\rm C} v^k_{\rm C} v^p_{\rm C}-\frac1{2l+2n+7}\varepsilon_{ijk}{\cal M}^{jL}_\B{\cal M}^{kN}_{\rm C} v^p_{\rm C} v^q_{\rm C}\\\nonumber
&&+\varepsilon_{ijq}\bigg[-\frac12 v^k_{\B} v^p_\B +v^p_{\B{\rm C}} v^k_\B-\frac{2(1+\g)}{l+2}v^p_{\B{\rm C}}v^k_{\B{\rm C}}+\frac2{2l+2n+7} v^k_{\rm C} v^p_{\rm C}\bigg]{\cal M}^{jkL}_\B{\cal M}^{N}_{\rm C}\\\nonumber
&&+\frac{2(1+\g)}{(l+2)(n+1)}\varepsilon_{ijq}{\cal M}^{jkL}_\B\dot{\cal M}^{kN}_{\rm C} v^p_{\B{\rm C}}-\varepsilon_{ijq}\bigg(F^{kp}_{\rm C}{\cal M}^{jL}_\B{\cal M}^{kN}_{\rm C}+F^{pk}_\B{\cal M}^{jkL}_\B{\cal M}^N_{\rm C}\bigg)\\\nonumber
&&+\frac{2(1+\g)}{n+2}\bigg[\frac{l-1}{l+1}{\cal M}^{pL}_\B{\cal S}^{iN}_{\rm C} v^q_{\B{\rm C}}-{\cal M}^{pL}_\B{\cal S}^{qN}_{\rm C} v^i_{\B{\rm C}}\bigg]+\frac{2(1+\g)}{l+3}{\cal S}^{ipL}_\B{\cal M}^N_{\rm C} v^q_{\B{\rm C}}
\\\nonumber
&&
+\frac{2(1+\g)}{(l+2)(n+2)}\bigg[{\cal M}^{ipL}_\B\dot{\cal S}^{qN}_{\rm C} -{\cal M}^{pqL}_\B\dot{\cal S}^{iN}_{\rm C}-(l+1)\varepsilon_{ijq}{\cal S}^{pL}_\B{\cal S}^{jN}_{\rm C}\bigg]\;,\\\nonumber
\\
\beta^{ipqLN}_{\rm T}&=&\frac1{2l+2n+7}\varepsilon_{ijq}{\cal M}^{jkL}_\B{\cal M}^{kN}_{\rm C} a^p_{\rm C}\;,
\\
\alpha_{\rm T}^{ikpqLN}&=&\frac{2(1+\g)}{(l+2)(n+2)}\bigg[{\cal M}_\B^{iqL}{\cal S}_{\rm C}^{kN}v^p_{\B{\rm C}}-{\cal M}_\B^{qkL}{\cal S}_{\rm C}^{iN}v^p_{\B{\rm C}}\bigg]-\frac1{2l+2n+9}\varepsilon_{ijk}{\cal M}^{jnL}_\B{\cal M}^{nN}_{\rm C} v^p_{\rm C} v^q_{\rm C}\;,
\\
\alpha^{LN}_{\rm T}&=&2(1+\g)\bigg[\frac{l+1}{l+2}{\cal S}^{jL}_\B {\cal M}^N_{\rm C} v^j_{\B{\rm C}}-\frac{l+1}{(l+2)(n+1)}{\cal S}^{jL}_\B \dot{\cal M}^{jN}_{\rm C}+
\frac{2}{(l+1)(n+2)}{\cal M}_\B^{qL}\dot{\cal S}_{\rm C}^{qN}\bigg]\;,
\\\label{ac3es5q}
\mu^{pLN}_{\rm T}&=&\frac{2(1+\g)}{n+2}\bigg[\frac{l+1}{l+2}\varepsilon_{qpk}{\cal S}^{kN}_\B {\cal S}^{qL}_{\rm C} 
+{\cal M}^{kL}_\B{\cal S}^{pN}_{\rm C} v^k_{\B{\rm C}}-\frac{l-1}{l+1}{\cal M}^{kL}_\B{\cal S}^{kN}_{\rm C} v^p_{\B{\rm C}}
+\frac{1}{l+2}{\cal M}^{kpL}_\B\dot{\cal S}^{kN}_{\rm C}\bigg] \;,\\\label{jjuu33vva}
\sigma^{pqLN}_{\rm T}&=&\frac{2(1+\g)}{(l+2)(n+2)}{\cal M}_\B^{kpL}{\cal S}_{\rm C}^{kN}v^q_{\B{\rm C}}
\ea

\subsubsection{Fermi-Walker Torque}\label{nt9125}

The Fermi-Walker torque \eqref{y775c1} can be easily calculated by making use of equations \eqref{UC1}, \eqref{w1w8} and replacing acceleration of the center of mass $a^i_\B=F^i_{\rm N}/M_\B$, where the Newtonian force $F^i_{\rm N}$ is shown in \eqref{w1q5}. Taking the STF derivatives from the corresponding expressions we get
\ba\label{sss42w}
T^i_{\rm FW}&=&2(1+\g)\sum_{C\not=\B}\sum_{l=0}^\infty\frac{(-1)^l}{l!}{\cal S}^j_\B\bigg({\cal M}^L_{\rm C} v^{[i}_{\B{\rm C}}\pd^{j]L}_{\phantom{C}}+\frac1{l+1}\dot{\cal M}^{L[i}_{\rm C}\pd^{j]L}_{\phantom{C}}-\frac1{l+2}{\cal S}^{pL}_{\rm C}\varepsilon^{pq[i}\pd^{j]L}\bigg)R^{-1}_{\B{\rm C}}\\\nonumber
&+&\frac1{M_\B}\sum_{C\not=\B}\sum_{l=0}^\infty\sum_{n=0}^\infty\frac{(-1)^n}{l!n!}{\cal M}^L_\B{\cal M}^N_{\rm C}{\cal S}^j_\B v^{[i}_\B\pd^{j]LN}R^{-1}_{\B{\rm C}}\;.
\ea
Taking the STF derivatives from $R^{-1}_{\B{\rm C}}$ defined in \eqref{acser243} we obtain the multipolar expansion of the Fermi-Walker torque,
\ba\label{sss42waa}
T^i_{\rm FW}&=&-2(1+\g)\sum_{C\not=\B}\sum_{l=0}^\infty\frac{(2l+1)!!}{l!}\bigg({\cal M}^L_{\rm C} v^{[i}_{\B{\rm C}}R_{\B{\rm C}}^{<j]L>}+\frac1{l+1}\dot{\cal M}^{L[i}_{\rm C} R_{\B{\rm C}}^{<j]L>}-\frac1{l+2}{\cal S}^{pL}_{\rm C}\varepsilon^{pq[i}R_{\B{\rm C}}^{<j]L>}\bigg)\frac{{\cal S}^j_\B}{ R^{2l+3}_{\B{\rm C}}}\\\nonumber
&&-\frac1{M_\B}\sum_{C\not=\B}\sum_{l=0}^\infty\sum_{n=0}^\infty\frac{(-1)^l(2l+2n+1)}{l!n!}\frac{v^{[i}_\B R_{\B{\rm C}}^{<j]LN>}}{R_{\B{\rm C}}^{2l+2n+3}}{\cal M}^L_\B{\cal M}^N_{\rm C}{\cal S}^j_\B \;.
\ea

\subsection{Reduced post-Newtonian Torque}

It is instructive to represent the post-Newtonian torque $T^i_{\rm pN}$ in yet another form by splitting up coefficients \eqref{m1x7z0n4}--\eqref{jjuu33vva} into various terms describing different types of gravitational coupling between the internal multipoles of extended bodies like mass-mass, mass-spin, spin-spin multipole interaction as well as the geometric coupling due to the Fermi-Walker precession. This requires to reduce the coefficients depending on the acceleration $a^i_{\rm B}$ of the center of mass of body B by making use of the Newtonian equations of translational motion, $M_{\B}a^i_\B=F^i_{\rm N}$, with the explicit form of the Newtonian force $F^i_{\rm N}$ given in \eqref{w1q5}. We perform this procedure and split the post-Newtonian torque in three main constituents,
\be\la{evzr2bti}
T^i_{pN}=T^i_{\rm M}+T^i_{\rm S}+T^i_{\rm P}\;,
\ee
where $T^i_{\rm M}$ is caused by the gravitational coupling between the mass multipoles of extended bodies, $T^i_{\rm S}$ describes gravitational interaction between the spin and mass multipoles, and $T^i_{\rm P}$ originates from the Fermi-Walker precession of the spatial axes of the body-adapted local coordinates. Specific expressions for each terms in the right-hand side of \eqref{evzr2bti} are given below.

\subsubsection{Mass Multipole Coupling Torque}
The mass-mass multipole coupling torque $T^i_{\rm M}$ consists of various terms describing two-, three-, and four-body gravitational interactions between the internal mass multipoles of the bodies comprising ${\mathbb N}$-body system. The torque depends on the interaction between the first and second time-derivatives of the mass multipoles as well. It has the following schematic structure,  
\ba\la{ba8wv41x}
T^i_{\rm M}&=&
T^i_{{\cal M}{\cal M}} +T^i_{{\cal M}\dot{{\cal M}}} +T^i_{{\cal M}\ddot{{\cal M}}}+T^i_{{\cal M}{\cal M}{\cal M}}+T^i_{{\cal M}{\cal M}{\cal M}{\cal M}}\;,
\ea
where each particular term denotes the number of the gravitationally-coupled multipoles. Specific expressions for different terms in (\ref{ba8wv41x}) are given below in terms of the coordinate distances (\ref{9b2c5}) between the bodies
and the corresponding coupling coefficients ${\cal K}^{LN}_{\cal II}, {\cal K}^{iLN}_{{\cal M}\dot{\cal M}}, {\cal K}^{iLN}_{\cal I\ddot I},$ etc., which are shown explicitly in equations \eqref{n6bac4sc}--\eqref{as7v3c0m}. The torque components read
\begin{eqnarray}
  \label{firsttorq}
  T^i_{{\cal M}{\cal M}}
  & = & \sum_{C\neq B}\sum_{l=0}^{\infty}\sum_{n=0}^{\infty} \Bigg[ 
    {\cal K}^{ipLN}_{{\cal M}{\cal M}}\frac{R_{\rm BC}^{<pLN>}}{R_{\rm BC}^{2l+2n+3}}
   +{\cal K}^{ipqLN}_{{\cal M}{\cal M}}\frac{R_{\rm BC}^{<pqLN>}}{R_{\rm BC}^{2l+2n+5}}\Bigg]\\\nonumber
  &+&\sum_{C\neq B}\sum_{l=0}^{\infty}\sum_{n=0}^{\infty} \Bigg[{\cal K}^{ikpqLN}_{{\cal M}{\cal M}}\frac{R_{\rm BC}^{<kpqLN>}}{R_{\rm BC}^{2l+2n+7}}+{\cal L}^{ikpqLN}_{{\cal M}{\cal M}}\frac{R_{\rm BC}^{<kpqLN>}}{R_{\rm BC}^{2l+2n+5}}\Bigg]\,,
\\
  \label{unqocn46c}
  T^i_{{\cal M}\dot{{\cal M}}}
  & = & \sum_{C\neq B}\sum_{l=0}^{\infty}\sum_{n=0}^{\infty} \Bigg[ 
  {\cal K}^{iLN}_{{\cal M}\dot{{\cal M}}}\frac{R_{\rm BC}^{<LN>}}{R_{\rm BC}^{2l+2n+1}}
  +{\cal K}^{ipLN}_{{\cal M}\dot{{\cal M}}}\frac{R_{\rm BC}^{<pLN>}}{R_{\rm BC}^{2l+2n+3}} 
  +{\cal K}^{ipqLN}_{{\cal M}\dot{{\cal M}}}\frac{R_{\rm BC}^{<pqLN>}}{R_{\rm BC}^{2l+2n+5}}\Bigg]\;,
\\
\label{un2c4vq}
T^i_{{\cal M}\ddot{{\cal M}}}
&=&  \sum_{C\neq B}\sum_{l=0}^{\infty}\sum_{n=0}^{\infty} \Bigg[{\cal K}^{iLN}_{{\cal M}\ddot{{\cal M}}}\frac{R_{\rm BC}^{<LN>}}{R_{\rm BC}^{2l+2n+1}}
+{\cal K}^{ipLN}_{{\cal M}\ddot{{\cal M}}}\frac{R_{\rm BC}^{<pLN>}}{R_{\rm BC}^{2l+2n+3}}+{\cal L}^{ipLN}_{{\cal M}\ddot{{\cal M}}}\frac{R_{\rm BC}^{<pLN>}}{R_{\rm BC}^{2l+2n+1}}\Bigg]\;,
\\
  \label{we43c45}
  T^i_{{\cal M}{\cal M}{\cal M}} 
  & = &  \sum_{C\neq B}\sum_{D\neq C}\sum_{l=0}^{\infty}\sum_{n=0}^{\infty}\sum_{k=0}^{\infty} {\cal K}^{ipLNK}_{{\cal M}{\cal M}{\cal M}}\frac{R_{\rm BC}^{<pLN>}R_{\rm CD}^{<K>}}{R_{\rm BC}^{2l+2n+3}R_{\rm CD}^{2k+1}}\\\nonumber
  &+&\sum_{C\neq B}\sum_{D\neq B}\sum_{l=0}^{\infty}\sum_{n=0}^{\infty}\sum_{k=0}^{\infty}{\cal L}^{ipLNK}_{{\cal M}{\cal M}{\cal M}}\frac{R_{\rm BC}^{<pLN>}R_{\rm BD}^{<K>}}{R_{\rm BC}^{2l+2n+3}R_{\rm BD}^{2k+1}}\;,
  \\
 \label{lasttorq}
  T^i_{{\cal M}{\cal M}{\cal M}{\cal M}}&=&
  \sum_{C\neq B}\sum_{D\neq C}\sum_{l=0}^{\infty}\sum_{n=0}^{\infty}\sum_{k=0}^{\infty}\sum_{s=0}^{\infty}\Bigg[
  {\cal K}^{ipLNSK}_{{\cal M}{\cal M}{\cal M}{\cal M}}\frac{R_{\rm BC}^{<LN>}R_{\rm CD}^{<pKS>}}{R_{\rm BC}^{2l+2n+1}R_{\rm CD}^{2k+2s+3}}
  +{\cal K}^{ipqLNSK}_{{\cal M}{\cal M}{\cal M}{\cal M}}\frac{R_{\rm BC}^{<pLN>}R_{\rm CD}^{<qKS>}}{R_{\rm BC}^{2l+2n+3}R_{\rm CD}^{2k+2s+3}}\\\nonumber
  &&+{\cal K}^{iLNSK}_{{\cal M}{\cal M}{\cal M}{\cal M}}\frac{R_{\rm BC}^{<pLN>}R_{\rm CD}^{<pKS>}}{R_{\rm BC}^{2l+2n+3}R_{\rm CD}^{2k+2s+3}}+{\cal L}^{ipLNSK}_{{\cal M}{\cal M}{\cal M}{\cal M}}\frac{R_{\rm BC}^{<pqLN>}R_{\rm CD}^{<qKS>}}{R_{\rm BC}^{2l+2n+5}R_{\rm CD}^{2k+2s+3}}
  +{\cal M}^{ipLNSK}_{{\cal M}{\cal M}{\cal M}{\cal M}}\frac{R_{\rm BC}^{<pqLN>}R_{\rm CD}^{<qKS>}}{R_{\rm BC}^{2l+2n+3}R_{\rm CD}^{2k+2s+3}}\Bigg]\nonumber\\
  &+&\sum_{C\neq B}\sum_{D\neq B}\sum_{l=0}^{\infty}\sum_{n=0}^{\infty}\sum_{k=0}^{\infty}\sum_{s=0}^{\infty}\Bigg[{\cal N}^{ipLNSK}_{{\cal M}{\cal M}{\cal M}{\cal M}}\frac{R_{\rm BC}^{<LN>}R_{\rm BD}^{<pKS>}}{R_{\rm BC}^{2l+2n+1}R_{\rm BD}^{2k+2s+3}}+{\cal N}^{ipqLNSK}_{{\cal M}{\cal M}{\cal M}{\cal M}}\frac{R_{\rm BC}^{<pLN>}R_{\rm BD}^{<qKS>}}{R_{\rm BC}^{2l+2n+3}R_{\rm BD}^{2k+2s+3}}\Bigg]\;.\nonumber
\end{eqnarray}
The coupling coefficients of the mass-mass multipole interaction that appear in \eqref{firsttorq}--\eqref{lasttorq} are:
\ba\la{n6bac4sc}
{\cal K}^{ipLN}_{{\cal M}{\cal M}}&=&\frac{(-1)^l(2l+2n+1)!!}{l!n!}\Bigg\{\varepsilon_{ipq}\bigg[\frac12 v^2_\B+(1+\g)v^2_{\B{\rm C}}-\frac12\frac{2l+2n+3}{2l+2n+5}v_{\rm C}^2\bigg]{\cal M}_\B^{qL}{\cal M}_{\rm C}^{N}
\\\nonumber
&&+\varepsilon_{ikq}\bigg[v^k_B v^p_{\B{\rm C}}-\frac12  v^k_B v^p_B-\frac{2(1+\g)}{l+1}v^k_{\B{\rm C}}v^p_{\B{\rm C}}+\frac2{2l+2n+5}v_{\rm C}^k v_{\rm C}^p\bigg]{\cal M}^{qL}_\B{\cal M}_{\rm C}^{N}\Bigg\} \;,\\
{\cal K}^{ipqLN}_{{\cal M}{\cal M}}&=&\frac{(-1)^l(2l+2n+3)!!}{l!n!}\times\\\nonumber
&&\Bigg\{\varepsilon_{ijq}\bigg[\frac12-\frac2{2l+2n+7}\bigg] {\cal M}^{jL}_\B{\cal M}^{kN}_{\rm C} v^k_{\rm C} v^p_{\rm C}-\frac1{2l+2n+7}\varepsilon_{ijk}{\cal M}^{jL}_\B{\cal M}^{kN}_{\rm C} v^p_{\rm C} v^q_{\rm C}\\\nonumber
&&+\varepsilon_{ijq}\bigg[-\frac12 v^k_{\B} v^p_\B +v^p_{\B{\rm C}} v^k_\B-\frac{2(1+\g)}{l+2}v^k_{\B{\rm C}}v^p_{\B{\rm C}}+\frac2{2l+2n+7} v^k_{\rm C} v^p_{\rm C}\bigg]{\cal M}^{jkL}_\B{\cal M}^{N}_{\rm C}  \Bigg\}\;,\\
{\cal K}^{ikpqLN}_{{\cal M}{\cal M}}&=& \frac{(-1)^l}{l!n!}\frac{(2l+2n+5)!!}{2l+2n+9}\varepsilon_{ijk}{\cal M}_\B^{jaL}{\cal M}_{\rm C}^{aN}v^p_{\rm C} v_{\rm C}^q\;, \\
{\cal L}^{ikpqLN}_{{\cal M}{\cal M}}&=& \frac{(-1)^l(2l+2n+3)!!}{2l!n!}\varepsilon_{ijk}{\cal M}_\B^{jL}{\cal M}_{\rm C}^{N}v^p_{\rm C} v_{\rm C}^q \;,   \\
{\cal K}^{iLN}_{{\cal M}\dot{\cal M}}&=& \frac{(-1)^l(2l+2n-1)!!}{l!n!}\varepsilon_{ijk}\bigg[v^k_\B-\frac{2(1+\g)}{l+1}v^k_{\B{\rm C}}\bigg]
{\cal M}_\B^{jL}\dot{\cal M}_{\rm C}^{N}\;, \\
{\cal K}^{ipLN}_{{\cal M}\dot{\cal M}}&=&\frac{(-1)^l(2l+2n+1)!!}{l!n!}\Bigg\{
\frac{2(1+\g)}{n+1}\bigg[\frac{1}{l+1}\varepsilon_{ikq}v^p_{\B{\rm C}}-\varepsilon_{ipq}v_{\B{\rm C}}^k\bigg]{\cal M}^{qL}_\B\dot{\cal M}_{\rm C}^{kN}\\\nonumber
&&-\varepsilon_{ikp}\bigg[v^q_{\rm B}-\frac{2(1+\g)}{l+2}v^q_{\rm BC}\bigg]{\cal M}_\B^{kqL}\dot{\cal M}_{\rm C}^{N}\Bigg\} \\
{\cal K}^{ipqLN}_{{\cal M}\dot{\cal M}}&=&\frac{(-1)^l(2l+2n+3)!!}{l!(n+1)!}\bigg[\frac{2(1+\g)}{l+2}\bigg]\varepsilon_{ijq}{\cal M}^{jkL}_\B\dot{\cal M}^{kN}_{\rm C} v^p_{\B{\rm C}}\;,\\
{\cal K}^{iLN}_{{\cal M}\ddot{\cal M}}&=& \frac{(-1)^l(2l+2n-1)!!}{l!n!}\varepsilon_{ipk}\bigg[\frac{2(1+\g)}{(l+1)(n+1)}-\frac1{2l+2n+3}
\bigg]{\cal M}_\B^{pL}\ddot{\cal M}_{\rm C}^{kN}\;,\\
{\cal K}^{ipLN}_{{\cal M}\ddot{\cal M}}&=&\frac{(-1)^l(2l+2n+1)!!}{l!n!}\varepsilon_{ipq}
\bigg[\frac{2(1+\g)}{(l+2)(n+1)}+\frac1{2l+2n+5}\bigg]{\cal M}_\B^{kqL}\ddot{\cal M}_{\rm C}^{kN}\;,  \\
{\cal L}^{ipLN}_{{\cal M}\ddot{\cal M}}&=& \frac{(-1)^l(2l+2n-1)!!}{2l!n!}\varepsilon_{ijp}{\cal M}_\B^{jL}\ddot{\cal M}_{\rm C}^N\;, \\
{\cal K}^{ipLNK}_{{\cal M}{\cal M}{\cal M}}&=&  \frac{(-1)^{l+k}(2l+2n+1)!!(2k-1)!!}{l!n!k!}\bigg[\g(n+1)\bigg]\varepsilon_{ijp}{\cal M}_B^{jL}{\cal M}_{\rm C}^N{\cal M}_D^K\;, \\
{\cal L}^{ipLNK}_{{\cal M}{\cal M}{\cal M}}&=& \frac{(-1)^{l+k}(2l+2n+1)!!(2k-1)!!}{l!n!k!}\bigg[\g(l+2)\bigg]\varepsilon_{ijp}{\cal M}_B^{jL}{\cal M}_{\rm C}^N{\cal M}_D^K\;, \\
{\cal K}^{ipLNSK}_{{\cal M}{\cal M}{\cal M}{\cal M}}&=&\frac{(-1)^{l+s}(2l+2n-1)!!(2s+2k+1)!!}{l!n!s!k!} \varepsilon_{ijp}\bigg[\frac1{2l+2n+3}-2\frac{1+\g}{l+1}\bigg]\frac{{\cal M}_\B^{jL}{\cal M}_{\rm C}^{N}{\cal M}^S_{\rm C}{\cal M}^K_\D}{{\cal M}_{\rm C}}\;, \\
{\cal N}^{ipLNSK}_{{\cal M}{\cal M}{\cal M}{\cal M}}&=&\frac{(-1)^{l+s}(2l+2n-1)!!(2s+2k+1)!!}{l!n!s!k!} \varepsilon_{ijp}\bigg[l+2\frac{1+\g}{l+1}\bigg]\frac{{\cal M}_\B^{jL}{\cal M}_{\rm C}^{N}{\cal M}^S_{\rm B}{\cal M}^K_\D}{{\cal M}_{\rm B}}\;, \\
{\cal K}^{ipqLNSK}_{{\cal M}{\cal M}{\cal M}{\cal M}}&=& \frac{(-1)^{l+s}(2l+2n+1)!!(2s+2k+1)!!}{l!n!s!k!}\varepsilon_{ijp}\Bigg\{\bigg[n+1+\frac1{2l+2n+5}\bigg]
\frac{{\cal M}_\B^{jL}{\cal M}_{\rm C}^{qN}{\cal M}^S_{\rm C}{\cal M}^K_\D}{{\cal M}_{\rm C}} \\\nonumber
&&  +\varepsilon_{ijp}\bigg[\frac{2(1+\g)}{l+2}-\frac1{2l+2n+5}\bigg]
\frac{{\cal M}_\B^{jqL}{\cal M}_{\rm C}^{N}{\cal M}^S_{\rm C}{\cal M}^K_\D}{{\cal M}_{\rm C}}  \Bigg\}\;,\\
{\cal K}^{iLNSK}_{{\cal M}{\cal M}{\cal M}{\cal M}}&=& \frac{(-1)^{l+s}(2l+2n+1)!!(2s+2k+1)!!}{l!n!s!k!(2l+2n+5)}\varepsilon_{ijq}\frac{{\cal M}_\B^{jL}{\cal M}_{\rm C}^{qN}{\cal M}^S_{\rm C}{\cal M}^K_\D}{{\cal M}_{\rm C}}\;,  \\
{\cal L}^{ipLNSK}_{{\cal M}{\cal M}{\cal M}{\cal M}}&=&  \frac{(-1)^{l+s+1}(2l+2n+3)!!(2s+2k+1)!!}{l!n!s!k!(2l+2n+7)}\varepsilon_{ijp}\frac{{\cal M}_\B^{jkL}{\cal M}_{\rm C}^{kN}{\cal M}^S_{\rm C}{\cal M}^K_\D}{{\cal M}_{\rm C}}\;,  \\
{\cal M}^{ipLNSK}_{{\cal M}{\cal M}{\cal M}{\cal M}}&=&\frac{(-1)^{l+s+1}(2l+2n+1)!!(2s+2k+1)!!}{2l!n!s!k!}\varepsilon_{ijp}\frac{{\cal M}_\B^{jL}{\cal M}_{\rm C}^{N}{\cal M}^S_{\rm C}{\cal M}^K_\D}{{\cal M}_{\rm C}}\;,  
\\\la{as7v3c0m}
{\cal N}^{ipqLNSK}_{{\cal M}{\cal M}{\cal M}{\cal M}}&=& \frac{(-1)^{l+s+1}(2l+2n+1)!!(2s+2k+1)!!}{l!n!s!k!}\frac{l^2+3l+4+2\g}{l+2}\varepsilon_{ijp}
\frac{{\cal M}_\B^{jqL}{\cal M}_{\rm C}^{N}{\cal M}^S_\B{\cal M}^K_\D}{{\cal M}_\B} \;.
\ea

\subsubsection{Spin Multipole Coupling Torque}

The post-Newtonian torque describing the spin-mass and spin-spin coupling between the internal multipoles of the extended bodies consists of four terms,
\ba
 \label{ty4bs9v2}
  T^i_{\rm S} & = &  T^i_{\mathcal{S}{\cal M}}+T^i_{\dot{\mathcal{S}}{\cal M}} +T^i_{\mathcal{S}\dot{\cal M}}+T^i_{\mathcal{S}\mathcal{S}}\;,
\ea
where each component of the torque is expressed in terms of the corresponding coupling coeffcients ${\cal K}_{\cal SI}$, ${\cal K}_{\cal \dot SI}$, etc. 
The components of the spin multipole coupling torque are 
\begin{eqnarray}
  \label{t1zap1}
  T^i_{\mathcal{S}{\cal M}} &=&\sum_{C\neq B}\sum_{l=0}^{\infty}\sum_{n=0}^{\infty}\Bigg[
  {\cal K}^{ipLN}_{\mathcal{S}{\cal M}}\frac{R_{\rm BC}^{<pLN>}}{R_{\rm BC}^{2l+2n+3}}
  +{\cal K}^{ipqLN}_{\mathcal{S}{\cal M}}\frac{R_{\rm BC}^{<pqLN>}}{R_{\rm BC}^{2l+2n+5}}+{\cal K}^{ikpqLN}_{\mathcal{S}{\cal M}}\frac{R_{\rm BC}^{<kpqLN>}}{R_{\rm BC}^{2l+2n+7}}\\\nonumber
  &&+{\cal K}^{LN}_{\mathcal{S}{\cal M}}\frac{R_{\rm BC}^{<iLN>}}{R_{\rm BC}^{2l+2n+3}}+{\cal K}^{pLN}_{\mathcal{S}{\cal M}}\frac{R_{\rm BC}^{<ipLN>}}{R_{\rm BC}^{2l+2n+5}}+{\cal L}^{pqLN}_{\mathcal{S}{\cal M}}\frac{R_{\rm BC}^{<ipqLN>}}{R_{\rm BC}^{2l+2n+7}}\Bigg]\;,
\\
  \label{t1zap2}
  T^i_{\mathcal{S}\dot{{\cal M}}} &=& \sum_{C\neq B}\sum_{l=0}^{\infty}\sum_{n=0}^{\infty}\Bigg[
  {\cal K}^{ipLN}_{\mathcal{S}\dot{{\cal M}}}\frac{R_{\rm BC}^{<pLN>}}{R_{\rm BC}^{2l+2n+3}}
  +{\cal K}^{LN}_{\mathcal{S}\dot{{\cal M}}}\frac{R_{\rm BC}^{<iLN>}}{R_{\rm BC}^{2l+2n+3}}\Bigg],
\\
  \label{t1zap3}
  T^i_{\dot{\mathcal{S}}{\cal M}} &=&  \sum_{C\neq B}\sum_{l=0}^{\infty}\sum_{n=0}^{\infty}\Bigg[{\cal K}^{ipLN}_{\dot{\mathcal{S}}{\cal M}}\frac{R_{\rm BC}^{<pLN>}}{R_{\rm BC}^{2l+2n+3}}+{\cal K}^{ipqLN}_{\dot{\mathcal{S}}{\cal M}}\frac{R_{\rm BC}^{<pqLN>}}{R_{\rm BC}^{2l+2n+5}}+{\cal K}^{LN}_{\dot{\mathcal{S}}{\cal M}}\frac{R_{\rm BC}^{<iLN>}}{R_{\rm BC}^{2l+2n+3}}+{\cal K}^{pLN}_{\dot{\mathcal{S}}{\cal M}}\frac{R_{\rm BC}^{<ipLN>}}{R_{\rm BC}^{2l+2n+5}}\Bigg]\;,
\\
  \label{t1zap4}
  T^i_{\mathcal{S}\mathcal{S}}
  &=&  \sum_{C\neq B}\sum_{l=0}^{\infty}\sum_{n=0}^{\infty}\Bigg[{\cal K}^{ipqLN}_{\mathcal{S}\mathcal{S}}\frac{R_{\rm BC}^{<pqLN>}}{R_{\rm BC}^{2l+2n+5}}+{\cal K}^{pLN}_{\mathcal{S}\mathcal{S}}\frac{R_{\rm BC}^{<ipLN>}}{R_{\rm BC}^{2l+2n+5}}\Bigg]\;.
\end{eqnarray}
The coupling coefficients that appear in \eqref{t1zap1}--\eqref{t1zap4} are,
\ba
{\cal K}^{ipLN}_{\mathcal{S}{\cal M}} &=&2(1+\g)\frac{(-1)^{l}(2l+2n+1)!!}{l!n!}\frac{l+1}{l+2}{\cal S}_\B^{pL}{\cal M}_{\rm C}^N v_{\B{\rm C}}^i\;,  \\
{\cal K}^{ipqLN}_{\mathcal{S}{\cal M}}&=& 2(1+\g)\frac{(-1)^{l}(2l+2n+3)!!}{l!n!}\times\\\nonumber
&&\phantom{+++++++++} \bigg[\frac1{n+2}\bigg(\frac{l-1}{l+1}{\cal M}_\B^{pL}{\cal S}_{\rm C}^{iN}v^q_{\B{\rm C}}-{\cal M}_\B^{pL}{\cal S}_{\rm C}^{qN}v^i_{\B{\rm C}}\bigg)+\frac1{l+3}{\cal S}_\B^{ipL}{\cal M}_{\rm C}^N v^q_{\B{\rm C}}\bigg],\\
{\cal K}^{ikpqLN}_{\mathcal{S}{\cal M}}&=& 2(1+\g)\frac{(-1)^{l}(2l+2n+5)!!}{l!n!(l+2)(n+2)}\bigg[{\cal M}_\B^{qkL}{\cal S}_{\rm C}^{iN}-{\cal M}_\B^{qiL}{\cal S}_{\rm C}^{kN}\bigg]v^p_{\B{\rm C}} \;,  \\
{\cal K}^{LN}_{\mathcal{S}{\cal M}}&=& 2(1+\g)\frac{(-1)^{l+1}(2l+2n+1)!!}{l!n!} \frac{l+1}{l+2}{\cal S}_\B^{jL}{\cal M}_{\rm C}^N v^j_{\B{\rm C}}\;, \\
{\cal K}^{pLN}_{\mathcal{S}{\cal M}}&=& 2(1+\g)\frac{(-1)^{l}(2l+2n+3)!!}{l!n!(n+2)} \bigg[{\cal M}^{kL}_\B{\cal S}_{\rm C}^{pN} v^k_{\B{\rm C}}-\frac{l-1}{l+1}{\cal M}^{kL}_\B{\cal S}_{\rm C}^{kN} v^p_{\B{\rm C}}\bigg] \;, \\
{\cal L}^{pqLN}_{\mathcal{S}{\cal M}}&=& 2(1+\g)\frac{(-1)^{l+1}(2l+2n+5)!!}{l!n!(l+2)(n+2)}{\cal M}^{kpL}_\B{\cal S}_{\rm C}^{kN} v^q_{\B{\rm C}}\;, \\
{\cal K}^{ipLN}_{\mathcal{S}\dot{{\cal M}}}&=&2(1+\g)\frac{(-1)^{l+1}(2l+2n+1)!!}{l!n!}\bigg[\frac{l+1}{(l+2)(n+1)}{\cal S}_\B^{pL}\dot{\cal M}^{iN}_{\rm C}-\frac1{l+3}{\cal S}_\B^{ipL}\dot{\cal M}^{N}_{\rm C}\bigg] \;,  \\
{\cal K}^{LN}_{\mathcal{S}\dot{{\cal M}}}&=&2(1+\g)\frac{(-1)^{l}(2l+2n+1)!!}{l!(n+1)!} \frac{l+1}{l+2} {\cal S}_\B^{jL}\dot{\cal M}^{jL}_{\rm C}\;,  \\
{\cal K}^{ipLN}_{\dot{\mathcal{S}}{\cal M}}&=&4(1+\g)\frac{(-1)^{l}(2l+2n+1)!!}{(l+1)!n!(n+2)}{\cal M}^{pL}_\B\dot{\cal S}^{iN}_{\rm C}\;,   \\
{\cal K}^{ipqLN}_{\dot{\mathcal{S}}{\cal M}}&=&2(1+\g)\frac{(-1)^{l}(2l+2n+3)!!}{l!n!(l+2)(n+2)}\bigg[{\cal M}^{ipL}_\B\dot{\cal S}^{qN}_{\rm C}-{\cal M}^{pqL}_\B\dot{\cal S}^{iN}_{\rm C}\bigg]\;,   \\
{\cal K}^{LN}_{\dot{\mathcal{S}}{\cal M}}&=& 4(1+\g)\frac{(-1)^{l+1}(2l+2n+1)!!}{(l+1)!n!(n+2)} {\cal M}^{jL}_\B\dot{\cal S}^{jN}_{\rm C}\;, \\
{\cal K}^{pLN}_{\dot{\mathcal{S}}{\cal M}}&=& 2(1+\g)\frac{(-1)^{l}(2l+2n+3)!!}{l!n!(l+2)(n+2)} {\cal M}^{jpL}_\B\dot{\cal S}^{jN}_{\rm C}\;, \\
{\cal K}^{ipqLN}_{\mathcal{S}\mathcal{S}}&=& 2(1+\g)\frac{(-1)^{l}(2l+2n+3)!!}{l!n!(n+2)}\frac{l+1}{l+2} \varepsilon_{iqj}{\cal S}_\B^{pL}{\cal S}_{\rm C}^{jN}\;, \\
{\cal K}^{pLN}_{\mathcal{S}\mathcal{S}}&=&2(1+\g)\frac{(-1)^{l}(2l+2n+3)!!}{l!n!(n+2)}\frac{l+1}{l+2} \varepsilon_{pkq}{\cal S}_\B^{kL}{\cal S}_{\rm C}^{qN}\;.
\ea

\subsubsection{Precession-Multipole Coupling Torque}
The Fermi-Walker precession causes a spatial rotation of each body-adapted local coordinates with respect to the distant observers at spatial infinity which is interpreted in the global coordinates as torque $T^i_{\rm P}$ caused by the geometric coupling of the matrix of relativistic precession to the internal mass multipoles of extended bodies. Picking up the precessional terms in the coupling coefficients $\a^{ipLN}_{\rm T}$ and $\a^{ipqLN}_{\rm T}$ in \eqref{bb33vv1} and \eqref{bb33vv2}, we get for the torque
\ba\label{tb56c41}
T^i_{\rm P}&=&\sum_{C\neq B}\sum_{l=0}^{\infty}\sum_{n=0}^{\infty}\frac{(-1)^l(2l+2n+1)!!}{l!n!}\bigg(\varepsilon_{kpq}F^{iq}_\B+\varepsilon_{ikq}F^{pq}_\B\bigg){\cal M}^{kL}_\B{\cal M}^N_{\rm C} \frac{R^{<pLN>}_{\B{\rm C}}}{R^{2l+2n+3}_{\B{\rm C}}} \\\nonumber
&-&\sum_{C\neq B}\sum_{l=0}^{\infty}\sum_{n=0}^{\infty}\frac{(-1)^l(2l+2n+3)!!}{l!n!}\varepsilon_{ijq}\bigg(F^{kp}_{\rm C}{\cal M}^{jL}_\B{\cal M}^{kN}_{\rm C}+F^{pk}_\B{\cal M}^{jkL}_\B{\cal M}^N_{\rm C}\bigg)\frac{R^{<pqLN>}_{\B{\rm C}}}{R^{2l+2n+5}_{\B{\rm C}}} \;.
\ea 

\citet{Racine_2006CQG} analyzed spin evolution equations for a wide class of extended bodies and gave a surface integral derivation of the leading-order evolution equations for the spin of a relativistic body interacting with other bodies. He expanded the spin evolution equations in the multipolar series but was unable to obtain the torque beyond the Newtonian formula \eqref{m5btz4s}. The present section significantly extends the result of paper \citep{Racine_2006CQG} and provides the multipolar expansion of the torque in the post-Newtonian approximation which has been never published before.
  
\section{Covariant Equations of Motion of Extended Bodies with All Multipoles}\la{ceom23}

This section formulates the translational and rotational equations of motion derived in the previous sections, in the covariant form in the spirit of the "covariantization" approach worked out by \citet{th_1985} who followed earlier developments outlined in \citep{Landau1975,mtw}. The covariantization procedure allows us to relax the slow-motion limitation of the first post-Newtonian approximation as the covariant equations of motion are apparently Lorentz-invariant and are applicable at both slow- and ultra-relativistic speeds. However, it should be understood that such covariant equations are still missing gravity-field potentials from the second, and higher-order post-Newtonian approximations and their application is limited by the weak-field, first post-Newtonian approximation. Nonetheless, the covariant equations of motion derived in this section may be instrumental in order to get a glimpse of the relativistic dynamics of very last several orbits of inspiralling binary system emitting gravitational waves before the bodies in the binary merge. 

Before discussing our own formalism we introduce the reader to the theory of Mathisson-Papapetrou-Dixon (MPD) equations of motion of extended bodies with higher-order multipoles that is considered as one of the most comprehensive and rigorous approaches for solving the fundamental problem of derivation of equations of motion of extended bodies in general relativity \citep{dixon_1979,dixon_2008,Dixon2015} and in the affine-metric theories of gravity \citep{dirk_2013PhLA,dirk_obukhov2014}. The original MPD theory has been developed mainly in the test-body approximation and had a number of other issues which made the domain of its astrophysical application fairly limited \citep{th_1985,Pound_2015}. In order to circumvent this issue, Harte \citep{harte2008_1,harte2008_2,harte2010,Harte_2012,harte2015} has developed a solid theoretical platform for stretching out the domain of applicability of the MPD theory to extended bodies with a strong self-gravity field. The concrete results obtained in this section, are fully consistent with the basic principles of Harte's general formalism and confirm validity of its predictions in the framework of the post-Newtonian dynamics of extended self-gravitating bodies possessing the entire collection of mass and spin multipoles.        

\subsection{The Mathisson Variational Dynamics}\la{tv3ef}

The goal to build a covariant post-Newtonian theory of motion of extended bodies and to find out the relativistic corrections to the equations of motion of a point-like particle which account for {\it all} multipoles characterizing the interior structure of the extended bodies was put forward by Mathisson\index{Mathisson} \citep{mathisson_2010GReGr_1,mathisson_2010GReGr_2} and further explored by \citet{Taub_1965}, \citet{tulczyjew1,tulczyjew1_1962}, and \citet{madore_1969}. However, the most significant advance in tackling this problem was achieved by Dixon\index{Dixon} \citep{dixon_1970_1,dixon_1970_2,dixon_1974_3,dixon_1973GReGr,dixon_1979} who elaborated on mathematically rigorous derivation of multipolar covariant equations of motion of extended bodies from the microscopic law of conservation of matter,  
\be\la{wk10}
\nabla_\a T^{\a\b}=0\;,
\ee
where $\nabla_\a$ denotes a covariant derivative on spacetime manifold ${M}$ with metric $g_{\a\b}$, and $T^{a\b}$ is the stress-energy tensor\index{tensor!stress-energy} of matter composing the extended bodies. Mathisson has dubbed this approach to the derivation of covariant equations of motion as {\it variational dynamics} \citep{mathisson_2010GReGr_1}. Comprehensive reviews of the historical development and current status of the variational dynamics can be found in papers by \citet{sauer_2000,dixon_2008,Dixon2015}. 

Dixon has significantly improved the Mathisson variational dynamics by employing a novel method of integration of the linear connection \index{linear connection} in general relativity\index{general relativity} as well as other innovations which allowed him to advance the original Mathisson's theory of variational dynamics. 
The generic mathematical technique used by Dixon\index{Dixon} to achieve this goal was the formalism of two-point world function,\index{world function} $\sigma(z,x)$, and its partial derivatives (called sometimes bi-tensors\index{bi-tensor}) introduced by Synge \citep{syngebook}, the distributional theory of multipoles stemmed from the theory of generalized functions \citep{Gelfand_1964,shilov_1968}, and the horizontal and vertical (or Ehresmann's \citep{Kolar_1993}) covariant derivatives\index{covariant derivative!horizontal}\index{covariant derivative!vertical} of two-point tensors defined on a vector bundle\index{vector bundle} formed by the direct product of the reference time-like worldline ${\cal Z}$ and a space-like hypersurface consisting of geodesics emitted at each instant of time from point $z$ on ${\cal Z}$ in all directions being orthogonal to ${\cal Z}$. 

An extended body in Dixon's approach is idealized as a time-like world tube\index{world tube} filled with continuous matter which stress-energy tensor $T^{\a\b}$ vanishes\index{stress-energy tensor} outside the tube. By making use of the bi-tensor propagators, $K^\a{}_\m\equiv K^\a{}_\m(z,x) $ and $H^\a{}_\m\equiv H^\a{}_\m(z,x)$, composed out of the inverse matrices of the first-order partial derivatives of the world function $\sigma(z,x)$ with respect to $z$ and $x$, Dixon defined the total linear momentum\index{linear momentum}, ${\mathfrak p}^\a\equiv {\mathfrak p}^\a(z)$, and the total angular momentum\index{spin}, $S^{\a\b}\equiv S^{\a\b}(z)$, of the extended body by integrals over a space-like hypersurface $\Sigma$, \citep[Equations 66--67]{dixon_1979}
\ba\la{wk11}
{\mathfrak p}^\a&\equiv&\int\limits_\Sigma K^\a{}_\m T^{\m\n}\sqrt{- g}d\Sigma_\n\;,\\
\la{wk12}
S^{\a\b}&\equiv&-2\int\limits_\Sigma X^{[\a}H^{\b]}{}_\m T^{\m\n}\sqrt{- g}d\Sigma_\n\;,
\ea
where $z\equiv z^\a(\tau)$ is a reference worldline $\mathcal{Z}$ of a representative point that is associated with the center of mass\index{center of mass} of the body with $\tau$ being the proper time on this worldline, vector 
\be\label{wk12777}
X^\a=-{g}^{\a\b}(z)\frac{\pd\sigma(z,x)}{\pd z^\b}\;,
\ee
is tangent to a geodesic emitted from the point $z$ and passing through point $x$. The oriented element of integration on the hypersurface, 
\ba\label{pd6s}
d\Sigma_\a=\frac1{3!}E_{\a\m\n\s} dX^\m\wedge dX^\n\wedge dX^\s\;,
\ea
where $E_{\a\m\n\s}$ is 4-dimensional, fully anti-symmetric symbol of Levi-Chivita, and the symbol $\wedge$ denotes the wedge product \citep[\S 3.5]{mtw} of the 1-forms $dX^\a$. Notice that Dixon's definition \eqref{wk12} of $S^{\a\b}$ yields (after a duality transformation) spin of the body that has an opposite sign as compared to our definition \eqref{spin-3} of spin.

It is further assumed in Dixon's formalism that the linear momentum, ${\mathfrak p}^\a$, is proportional to the {\it dynamic} velocity, ${\mathfrak n}^\a$, of the body \citep[Equation 83]{dixon_1979}
\be\label{q13m} 
{\mathfrak p}^\a\equiv M{\mathfrak n}^\a\;,
\ee
where $M=M(\tau)$ is the total mass of the body which, in general, can depend on time. The {\it dynamic} velocity is a unit vector, ${\mathfrak n}_\a {\mathfrak n}^\a=-1$. The {\it kinematic} 4-velocity of the body moving along worldline $\mathcal{Z}$ is tangent to this worldline, $ u^\a=dz^\a/d\tau$. It relates to the {\it dynamic} 4-velocity by condition, ${\mathfrak n}_\a { u}^\a=-1$, while the normalization condition of the {\it kinematic} 4-velocity is ${ u}_\a { u}^\a=-1$. Notice that in the most general case the dynamic and kinematic velocities are not equal due to the gravitational interaction between the bodies of $\mathbb{N}$-body system -- see \citep[Equation 88]{dixon_1979} and \citep{ehlers_1977GReGr} for more detail. 

Dixon defines the mass dipole, $m^\a=m^\a(z,\Sigma)$, of the body \citep[Equations 78]{dixon_1979},
\be 
\label{q14m}
m^\a\equiv S^{\a\b}{\mathfrak n}_\b\;,
\ee
and chooses the worldline $z=z^\a(\tau)$ of the center of mass of the body by condition, $m^\a=0$ This condition is equivalent due to (\ref{q13m}) and \eqref{q14m}, to  
\be\la{q13am}
{\mathfrak p}_\b S^{\a\b}=0\;,
\ee
which is known as Dixon's supplementary condition \index{Dixon's supplementary condition}\citep[Equation 81]{dixon_1979}. 

Dixon builds the body-adapted, local coordinates at each point $z$ on worldline ${\cal Z}$ as a set of the Riemann normal coordinates \citep[Chapter III, \S 7]{Schouten_book} denoted by $X^\a$ with the time coordinate $X^0$ along a time-like geodesic in the direction of the {\it dynamic} velocity ${\mathfrak n}^\a$, and the spatial coordinates $X^i=\{X^1,X^2,X^3\}$ lying on the hypersurface $\Sigma=\Sigma(z)$ consisting of all space-like geodesics passing through $z$ orthogonal to the unit vector ${\mathfrak n}^\a$ so that,
\be\label{be4y6}
{\mathfrak n}_\a X^\a=0\;.
\ee
It is important to understand that the Fermi normal coordinates (FNC) of observer moving along time-like geodesic do not coincide with the Riemann normal coordinates (RNC) used by Dixon \citep{dixon_1979,dixon_2008}. The FNC are constructed under condition that the Christoffel symbols vanish at {\it every} point along the geodesic \citep[Chapter III, \S 8]{Schouten_book} while the Christoffel symbols of the RNC vanish only at  a single event on spacetime manifold. The correspondence between the RNC and the FNC is discussed, for example, in \citep[Chapter 5]{poissonwill_book}, \citep{nesterov_1999CQG} and generalization of the FNC for the case of accelerated and locally-rotating observers is given in \citep[\S 13.6]{mtw} and \citep{Ni_1978PRD}. The present paper uses the conformal-harmonic gauge \eqref{11.3} to build the body-adapted local coordinates which coincide with the FNC of accelerated observer only in the linearized approximation of the Taylor expansion of the metric tensor with respect to the spatial coordinates around the worldline of the observer. 

Further development of the variational dynamics requires a clear separation of the matter and field variables in the solution of the full Einstein's field equations. This problem has not been solved in the MPD approach explicitly \footnote{In the present paper the separation of the matter and field variables in the metric tensor is achieved by means of the matched asymptotic expansion technique.}. It was replaced with the solution of a simpler problem of the separation of the matter and field variables in the equations of motion \eqref{wk10} by introducing a symmetric tensor distribution $\hat T^{\m\n}$ known as the stress-energy {\it skeleton} of the body \citep{mathisson_2010GReGr_1,mathisson_2010GReGr_2,dixon_1979}. Effectively, it means that the variational dynamics of each body is described on the effective background manifold $\bar{M}$ that is equivalent to the full manifold $M$ from which the self-field effects of the body have been removed. We denote the geometric quantities and fields defined on the effective background manifold with a bar above the corresponding object. Mathematical construction of the effective background manifold in our formalism is given below in section \ref{po3v6}.

\citet[Equation 140]{dixon_1979} defined high-order multipoles of an extended body in the normal Riemann coordinates, $X^\a$, by means of a tensor integral   
\be\label{q12m} 
I^{\a_1...\a_l\m\n}(z)=\int X^{\a_1}...X^{\a_l}\hat T^{\m\n}(z,X)\sqrt{-\bar g(z)}DX\;,\qquad\qquad (l\ge 2)
\ee
where $X^\a\equiv X^\a(z,x)$ is the same vector as in (\ref{wk12777}), $\hat T^{\m\n}$ is the stress-energy {\it skeleton} of the body, and the integration is performed over the tangent space of the point $z$ with the volume element of integration $DX=dX^0\wedge dX^1\wedge dX^2\wedge dX^3$. Definition (\ref{q12m}) implies the following symmetries,
\be\la{2v3j}
I^{\a_1...\a_l\m\n}=I^{(\a_1...\a_l)(\m\n)}\;,
\ee
where the round parentheses around the tensor indices denote a full symmetrization\index{symmetrization}. Microscopic equation of motion \eqref{wk10} also  tells us that
\be\label{gr7x3q}
I^{(\a_1...\a_l\m)\n}=0\;,
\ee
and a similar relation holds after exchanging indices $\m$ and $\n$ due to symmetry \eqref{2v3j}. Dixon's multipoles have a number of interesting symmetries which are discussed in \citep{dixon_1974_3,Dixon2015} and summarized in Appendix \ref{appndxon} of the present paper. Appendix \ref{oonn3388} compares the Dixon multipoles \eqref{q12m} with the Blanchet-Damour multipoles \eqref{1.31}, \eqref{1.32} and establishes a relationship between them in the post-Newtonian approximation of general relativity when the effects of the hypothetical scalar field are ignored.

\citet{dixon_1979} presented a number of theoretical arguments suggesting that the covariant equations of motion of the extended body have the following covariant form \citep[Equations 4.9--4.10]{dixon_1973GReGr}
\ba\la{q15m}
\frac{{\cal D}{\mathfrak p}_\a}{{\cal D}\tau}&=&\frac12 {\bar u}^\b S^{\m\n}\bar R_{\m\n\b\a}+\frac12\sum\limits_{l=2}^{\infty}\frac1{l!}\bar\nabla_{\a} A_{\b_1...\b_l\m\n}I^{\b_1...\b_l\m\n}
\\\la{q16m}
\frac{{\cal D}S^{\a\b}}{{\cal D}s}&=&2{\mathfrak p}^{[\a}{\bar u}^{\b]}+\sum\limits_{l=1}^{\infty}\frac{1}{l!}{B}_{\g_1...\g_l\s\m\n}\bar g^{\s[\a}I^{\b]\g_1...\g_l\m\n}\;,
\ea
where ${\cal D}/{\cal D}\tau\equiv {\bar u}^\a\bar\nabla_\a$
is the covariant derivative taken along the reference line $z=z(\tau),$ the moments $I^{\a_1...\a_l\m\n}$ are defined in \eqref{q12m}, $A_{\b_1...\b_l\m\n}$ and ${B}_{\g_1...\g_l\s\m\n}$ are the symmetric tensors computed at point $z$, and the bar above any tensor indicates that it belongs to the background spacetime manifold $\bar{M}$. 

\citet{th_1985} call body's multipoles $I^{\a_1...\a_l\m\n}$ the {\it internal} multipoles. Tensors $A_{\b_1...\b_l\m\n}$ and $B_{\g_1...\g_l\m\n\s}$ are called the {\it external} multipoles of the background spacetime. The external multipoles are the {\it normal} tensors in the sense of \citet{Veblen_1923}. They are reduced to the repeated partial derivatives of the metric tensor, $\bar g_{\m\n}$, and the Christoffel symbols, $\bar\Gamma_{\s\m\n}$, in the Riemann normal coordinates taken at the origin of the coordinate $X=0$ (corresponding to the point $z$ in coordinates $x^\a$) \citep{dixon_1979,Schouten_book},
\ba\label{om5g}
A_{\b_1...\b_l\m\n}&=&\lim_{X\rightarrow 0}\pd_{\b_1...\b_l}\bar g_{\m\n}(X)\;,\\
\label{om6f}
B_{\b_1...\b_l\s\m\n}&=&2\lim_{X\rightarrow 0}\pd_{\b_1...\b_l}\Gamma_{\s\m\n}(X)\\\nonumber
&=&\lim_{X\rightarrow 0}\left[\pd_{\b_1...\b_l\s}\bar g_{\m\n}(X)+\pd_{\b_1...\b_l\m}\bar g_{\n\s}(X)-\pd_{\b_1...\b_l\n}\bar g_{\s\m}(X)\right]\;.
\ea
In arbitrary coordinates $x^\a$, the normal tensors are expressed in terms of the Riemann tensor, $\bar R^\a{}_{\m\b\n}$, and its covariant derivatives \citep[Chapter III, \S 7]{Schouten_book}. More specifically, if the terms being quadratic with respect to the Riemann tensor are neglected, the external Dixon multipoles read,
\ba\label{zxc3d}
A_{\b_1...\b_l\m\n}&=&2\frac{l-1}{l+1}\bar\nabla_{(\b_1...\b_{l-2}}\bar R_{|\m|\b_{l-1}\b_l)\n}\;,\\\label{ydrv3}
B_{\b_1...\b_l\s\m\n}&=&\frac{2l}{l+2}\left[\bar\nabla_{(\b_1...\b_{l-1}}\bar R_{|\m|\s\b_l)\n}+\bar\nabla_{(\b_1...\b_{l-1}}\bar R_{|\s|\m\b_l)\n}-\bar\nabla_{(\b_1...\b_{l-1}}\bar R_{|\s|\n\b_l)\m}\right]
\ea  
where the vertical bars around an index means that it is excluded from the symmetrization denoted by the round parentheses. Notice that each term with the Riemann tensor in \eqref{zxc3d}, \eqref{ydrv3} is symmetric with respect to the first and forth indices of the Riemann tensor. This tells us that $A_{\b_1...\b_l\m\n}=A_{(\b_1...\b_l)(\m\n)}$ and $B_{\g_1...\g_l\s\m\n}=B_{(\g_1...\g_l)(\s\m)\n}$ in accordance with the symmetries of \eqref{om5g}, \eqref{om6f}.

Substituting these expressions to \eqref{q15m}, \eqref{q16m} yields the Dixon equations of motion in the following form, 
\ba\la{q15ms}
\frac{{\cal D} {\mathfrak p}_\a}{{\cal D}\tau}&=&\frac12 {\bar u}^\b S^{\m\n}\bar R_{\m\n\b\a}+\sum\limits_{l=2}^{\infty}\frac{l-1}{(l+1)!} \bar\nabla_{\a(\b_1...\b_{l-2}}\bar R_{|\m|\b_{l-1}\b_l)\n} J^{\b_1...\b_{l-1}\m\b_l\n}\;,
\\\la{q16mc}
\frac{{\cal D} S^{\a\b}}{{\cal D}\tau}&=&2{\mathfrak p}^{[\a}{\bar u}^{\b]}+2\sum\limits_{l=1}^{\infty}\frac{l(l+1)}{(l+2)!}\bar\nabla_{(\g_1...\g_{l-1}}\bar R_{|\m|\s\g_l)\n}\bar g^{\s[\a} J^{\b]\g_1...\g_{l-1}\m\g_l\n}\;,
\ea
where 
\be\la{wk12b}
J^{\a_1...\a_{p}\l\m\s\n}\equiv I^{\a_1...\a_{p}[\l[\s\m]\n]}\;,
\ee
denotes the internal multipoles with a skew symmetry with respect to two pairs of indices, $[\l\m]$ and $[\s\n]$. The Dixon  $I$ and $J$  multipoles are compared in Appendix \ref{appndxon} of the present paper. Comparison of Dixon's equations of motion \eqref{q15ms}, \eqref{q16mc} with our covariant equations is given in Appendix \ref{appendixB}. 

Mathematical elegance and apparently covariant nature of the variational dynamics has been attracting researchers to work on improving various aspects of derivation of the MPD equations of motion  \citep{beig_1967CMaPh,schattner_1979GReGr,ehlers_1977GReGr,Ehlers_1980NYASA,bailey_1980AnPhy,steinhoff_2010PhRvD,dirk_obukhov2014,Obukhov_Puetzfeld2014,Pound_2015}. From astrophysical point of view Dixon's formalism is viewed as being of considerable importance for the modeling the gravitational waves emitted by the extreme mass-ratio inspirals (EMRIs) which are binary black holes consisting of a supermassive black hole and a stellar mass black hole. EMRIs form a key science goal for the planned space based gravitational wave observatory LISA and the equations of motion of the black holes in those systems must be known with unprecedented accuracy \citep{Schutz_2018RSPTA,Babak2015}. Nonetheless, in spite of the power of Dixon's mathematical apparatus, there are several issues which make the MPD theory of the variational dynamics yet unsuitable for relativistic celestial mechanics, astrophysics and gravitational wave astronomy which have been pointed out by Dixon himself \citep{dixon_1979} and by \citet{th_1985}. 

The main problem is that the variational dynamics is too generic and does not engage any particular theory of gravity. It tacitly assumes that some valid theory of gravity is chosen, gravitational field equations are solved, and the metric tensor is known. However, the field equations and the equations of motion of matter are closely tied up -- matter generates gravity while gravity governs motion of matter. Due to this coupling the definition of the center of mass, linear momentum, spin, and other body's internal multipoles depend on the metric tensor which, in its own turn, depends on the multipoles through the non-linearity of the field equations. It complicates the problem of interpretation of the gravitational stress-energy skeleton in the non-linear regime of gravitational field and makes the MPD equations \eqref{q15m}, \eqref{q16m} valid solely in the linearized approximation of general relativity. For the same reason it is difficult to evaluate the residual terms in the existing derivations of the MPD equations and their multipolar extensions. One more serious difficulty relates to the lack of prescription for separation of self-gravity effects of moving body from the external gravitational environment. The MPD equations of motion are valid on the background effective manifold $\bar{M}$ but its exact mathematical formulation remains unclear in the framework of the variational dynamics alone \citep{Pound_2015}. Because of these shortcomings the MPD variational dynamics has not been commonly used in real astrophysical applications in spite that it is sometime claimed as a "standard theory" of the equations of motion of massive bodies in relativistic gravity \citep{bini_2009GReGr}. 

In order to complete the MPD approach to variational dynamics and make it applicable in astrophysics several critical ingredients have to be added. More specifically, what we need are:
\begin{enumerate}
\item the procedure of unambiguous characterization and determination of the gravitational self-force and self-torque exerted by the body on itself, and the proof that they are actually vanishing;
\item the procedure of building the effective background spacetime manifold $\bar{M}$ with the background metric $\bar g_{\a\b}$ used to describe the motion of the body which is a member of ${\mathbb N}$-body system;
\item the precise algorithm for calculating the body's internal multipoles (\ref{q12m}) and their connection to the gravitational field of the body;
\item the relationship between the Blanchet-Damour mass and spin body's multipoles, ${\cal M}^{\a_1...\a_l}$ and ${\cal S}^{\a_1...\a_l}$, the Dixon internal multipoles (\ref{q12m}) and the gravitational stress-energy skeleton.
\end{enumerate}

In this section we implement the formalism of derivation of covariant equations of motion of massive bodies proposed by \citet{th_1985} which yields a complete set of the covariant equations of translational and rotational motion. It relies upon the construction of the effective background manifold $\bar{M}$ by solving the field equations of scalar-tensor theory of gravity and applying the asymptotic matching technique which separates the self-field effects from the external gravitational environment, defines all external multipoles and establishes the local equations of motion of the body in the body-adapted local coordinates. The body's internal multipoles are defined in the conformal harmonic gauge by solving the field equations in the body-adapted local coordinates as proposed by \citet{bld}. The covariant equations of motion follow immediately from the local equations of motion by applying the Einstein equivalence principle \citep{th_1985}. We compare our covariant equations of motion, derived in this section, with the MPD equations in Appendix \ref{appendixB}. 

\subsection{The Effective Background Manifold}\la{po3v6}      
Equations of translational motion \eqref{MBaB1} of an extended body B in the global coordinate chart depend on an infinite set of configuration variables -- the internal mass and spin multipoles of the body, ${\cal M}^L_\B$ and ${\cal S}^L_\B$, and the external gravitoelectric and gravitomagnetic multipoles - ${\cal Q}_L$ and ${\cal C}_L$ -- all are pinned down to the worldline ${\cal Z}$ of the center of mass of the body. The same equations in the local coordinate chart adapted to the body B are given by \eqref{Qi} after applying the law of conservation of the linear momentum of the body \eqref{a0n5}. 
These equations in two different coordinate charts are interconnected by the spacetime coordinate transformation \eqref{5.12}, \eqref{5.13} -- the proof is given below in subsection \ref{o8b4m1}. It points out that the equations of motion derived in the local coordinates can be lifted to the generic covariant form by making use of the Einstein equivalence principle applied to body B that can be treated as a massive particle endowed with the internal multipoles ${\cal M}^L_\B$ and ${\cal S}^L_\B$, and moving along the worldline ${\cal Z}$ on the effective background spacetime manifold $\bar{M}$ which properties are characterized by the external multipoles ${\cal Q}_L$ and ${\cal C}_L$ that presumably depends on the curvature tensor on $\bar{M}$ and its covariant derivatives. The covariant form of the equations is independent of a particular realization of the conformal-harmonic coordinates but we hold on the gauge conditions \eqref{11.3} to prevent the appearance of gauge-dependent, nonphysical multipoles of gravitational field in the equations of motion. 

The power of our approach to the covariant equations of motion is that the effective background manifold $\bar{M}$ for each body B is not postulated or introduced ad hoc. It is constructed by solving the field equations in the local and global charts and separating the field variables -- scalar field and metric tensor perturbations -- in the internal and external parts. The separation is fairly straightforward in the local chart. The internal part of the metric tensor, $\hat h^{\rm int}_{\a\b}$ and scalar field $\hat\varphi^{\rm int}$, are determined by matter of body B and is expanded in the multipolar series outside the body which are singular at the origin of the body-adapted local coordinates. The external part of the metric tensor $\hat h^{\rm ext}_{\a\b}$ and scalar field $\hat\varphi^{\rm ext}$ are solutions of vacuum field equations and, hence, are regular at the origin of the local chart. There are also internal-external coupling component $\hat l^{\rm int}_{00}$ of the metric tensor perturbation but it is a non-linear functional of the internal solution and its multipolar series is also singular at the origin of the local chart of body B.

The effective background manifold is regular at the origin of the local coordinates and its geometry is entirely determined by the external part of the metric tensor, $\bar g_{\a\b}=\eta_{\a\b}+h^{\rm ext}_{\a\b}$. This is fully consistent with the result of matching of the asymptotic expansions of the metric tensor and scalar field in the global and local coordinates described in section \ref{pntb}. All terms which multipolar expansions are singular at the origin of the local chart are canceled out identically in the matching equations \eqref{2.5}, \eqref{2.6}. This establishes a one-to-one correspondence between the external metric perturbation $h^{\rm ext}_{\a\b}$ in the local chart and its counterpart in the global coordinate chart which is uniquely defined by the external gravitational potentials $\bar U, \bar U^i, \bar\Psi, \bar\chi$ given in (\ref{3d8b1a}). In the rest of this section we demonstrate that translational equations of motion of body B are equations of perturbed time-like geodesic of a massive particle on the effective background manifold with the metric $\bar g_{\a\b}$. The particle has mass $M=M_\B$ and internal multipoles ${\cal M}^L={\cal M}^L_\B$ and ${\cal S}^L={\cal S}^L_\B$. The perturbation of the geodesic is the local acceleration ${\cal Q}_i$ caused by the interaction of particle's multipoles with the external gravitoelectric and gravitomagnetic multipoles, ${\cal Q}_L$ and ${\cal C}_L$, which are fully expressed in terms of the covariant derivatives of the Riemann tensor, $\bar R_{\a\b\m\n}$ and scalar field $\bar\varphi$ of the background manifold. Covariant equations of rotational motion of the body spin are described by the Fermi-Walker transport with the external torques caused by the coupling of the internal and external multipoles of the body.

The effective background metric $\bar g_{\a\b}$ is given in the global coordinates by the following equations, 
(cf. \citep{th_1985}),
\ba\la{v5x8n}
\bar g_{00}(t,{\bm x})&=&-1+2\bar U(t,{\bm x})+2\left[\bar\Psi(t,{\bm x})-\beta\bar U^2(t,{\bm x})-\frac12\pd_{tt}\bar\chi(t,{\bm x})\right]\;,\\
\la{v5x8na}
\bar g_{0i}(t,{\bm x})&=&-2(1+\g)\bar U^i(t,{\bm x})\;,\\
\la{v5x8nb}
\bar g_{ij}(t,{\bm x})&=&\delta_{ij}+2\g\delta_{ij}\bar U(t,{\bm x})\;,
\ea
where the potentials in the right-hand side of (\ref{v5x8n})--(\ref{v5x8nb}) are defined in \eqref{tb54vd}, \eqref{3d8b1a} as functions of the global coordinates $x^\a=(t,{\bm x})$. The background metric in arbitrary coordinates can be obtained from (\ref{v5x8n})--(\ref{v5x8nb}) by performing a corresponding coordinate transformation. It is worth emphasizing that the effective metric $\bar g_{\a\b}$ is constructed for each body of the $\mathbb{N}$-body system separately and is a function of the external gravitational potentials which depend on which body is chosen. It means that we have a collection of $\mathbb{N}$ effective manifolds $\bar{M}$ -- one for each extended body. Another prominent point to draw attention of the reader is the fact that the effective metric of the extended body B depends on the gravitational field of the body itself through the non-linear interaction term $\Psi_{\rm  C 2}$ in the potential $\bar\Psi$ - see \eqref{12.14a} and its multipolar expansion \eqref{pc31}. This dependence of the background metric tensor on the gravitational field of the body itself is known as the {\it back-action} effect of gravitational field \citep{th_1985,ashb2}. It was first noticed by \citet{Fichte_1950} who pointed out that derivation of the post-Newtonian equations of motion of ${\mathbb N}$ bodies of comparable masses, given in the first edition of the "Classical Theory of Fields" by Landau and Lifshitz, is erroneous as they missed the {\it back-action} term in the effective metric. This error was corrected and did not appear in the subsequent editions of the Landau-Lifshitz textbook \citep{Landau1975}.    

The background metric, $\bar g_{\a\b}$, is a starting point of the covariant development of the equations of motion. It has the Christoffel symbols
\be\la{qq2wkw}
\bar\G^\a_{\m\n}=\frac12\bar g^{\a\b}\left(\pd_\n\bar g_{\b\m}+\pd_\m\bar g_{\b\n}-\pd_\b\bar g_{\m\n}\right)\;,
\ee
which can be directly calculated in the global coordinates, $x^\a$, by taking partial derivatives from the metric components (\ref{v5x8n})--(\ref{v5x8nb}). In what follows, we shall make use of a covariant derivative\index{covariant derivative}\index{derivative!covariant} defined on the background manifold $\bar{M}$ with the help of the Christoffel symbols\index{Christoffel symbols!background} $\bar\G^\a_{\m\n}$. The covariant derivative on the background manifold\index{manifold!background}, $\bar{M}$, is denoted $\bar\nabla_\a$ in order to distinguish it from the covariant derivative defined on the original spacetime manifold, ${M}$, denoted $\nabla_\a$. For example, the covariant derivative of vector field $V^\a$ is defined on the background manifold by the following equation
\be\la{r5v6h}
\bar\nabla_\b V^\a=\pd_\b V^\a+\bar\G^\a_{\m\b}V^\m\;,
\ee 
which is naturally extended to tensor fields of arbitrary type and rank in a standard way \citep{kopeikin_2011book}.
It is straightforward to define other geometric objects on the background manifold like the Riemann tensor \eqref{kk33xc}, 
\be\la{b6c2j}
\bar R^\a{}_{\m\b\n}=\pd_\b\bar\G^\a_{\m\n}-\pd_\n\bar\G^\a_{\m\b}+\bar\G^\a_{\s\b}\bar\G^\s_{\m\n}-\bar\G^\a_{\s\n}\bar\G^\s_{\m\b}\;,
\ee
and its contractions -- the Ricci tensor $\bar R_{\m\n}=\bar R^\a{}_{\m\a\n}$, and the Ricci scalar $\bar R=\bar g^{\m\n}\bar R_{\m\n}$. Tensor indices on the background manifold\index{manifold!background} are raised and lowered with the help of the metric $\bar g_{\a\b}$. 

The background metric tensor ${\bar {\rm g}}_{\a\b}(u,{\bm w})$ in the local coordinates $w^\a=(u,w^i)$ adapted to body B is given by 
\be\la{k8b2d}
{\bar {\rm g}}_{\a\b}(u,{\bm w})=\eta_{\a\b}+\hat h^{\rm ext}_{\a\b}(u,{\bm w})\;,
\ee
where the perturbation, $\hat h^{\rm ext}_{\a\b}$, is given by the polynomial expansions \eqref{1.24b}--\eqref{1.26b} of the external gravitational field with respect to the local spatial coordinates.
Notice that at the origin of the local coordinates, where $w^i=0$, the background metric $\bar g_{\a\b}$ is reduced to the Minkowski metric $\eta_{\a\b}$. It means that on the effective background manifold $\bar{M}$ the coordinate time $u$ is identical to the proper time $\t$ measured on the worldline ${\cal W}$ of the origin of the local coordinates adapted to body B,
\be\la{2n8}
\t=u\;.
\ee 

Post-Newtonian transformation from the global to local coordinates,  $w^\a=w^\a(x^\b)$, has been provided in section \ref{u8b2p}. It smoothly matches the two forms of the background metric, $\bar g_{\a\b}(t,{\bm x})$ and ${\bar {\rm g}}_{\a\b}(u,{\bm w})$ on the background manifold $\bar{M}$ in the sense that
\be\label{ww11cc77}
\bar g_{\m\n}(t,{\bm x})= {\bar{\rm g}}_{\a\b}(u,{\bm w})\frac{\pd w^\a}{\pd x^\m}\frac{\pd w^\b}{\pd x^\n}\;.
\ee
This should be compared with the law of transformation \eqref{2.6} applied to the full metric $g_{\a\b}$ on spacetime manifold ${M}$ which includes besides the external part also the internal and internal-external coupling components of the metric tensor perturbations but they are mutually canceled out in \eqref{2.6} leaving only the external terms, thus, converting \eqref{2.6} to \eqref{ww11cc77} without making any additional assumptions about the structure of the effective background manifold. The cancellation of the internal and internal-external components of the metric tensor perturbations in \eqref{2.6} is a manifestation of the {\it effacing} principle \citep{kovl_2008} that excludes the internal structure of body B from the definition of the effective background manifold $\bar{M}$ used for description of motion of the body \citep{Battista_2017IJMPA}. Compatibility of equations \eqref{2.6} and \eqref{ww11cc77} confirms that the internal and external problems of the relativistic celestial mechanics in $\mathbb{N}$-body system are completely decoupled regardless of the structure of the extended bodies and can be extrapolated to compact astrophysical objects like neutron stars and black holes.

In what follows, we will need a matrix of transformation\index{matrix of transformation} taken on the worldline of the origin of the local coordinates,
\be\la{p0c3b5}
\Lambda^\a{}_\b\equiv\Lambda^\a{}_\b(\tau)=\lim_{{\bm x}\rightarrow{\bm x}_\B}\frac{\pd w^\a}{\pd x^\b}\;.
\ee
The components of this matrix can be easily computed from equations of coordinate transformation \eqref{5.12}, \eqref{5.13} and its complete post-Newtonian form is shown in \citep[Section 5.1.3]{kopeikin_2011book}. With an accuracy being sufficient for derivation of the covariant post-Newtonian equations of motion in the present paper, it reads
\ba\la{1as1a}
\Lambda^0{}_0&=&1+\frac12v^2_{\rm B}-\bar U(t,{\bm x}_\B)\;,\\\la{1as1b}
\Lambda^0{}_i&=&-v^i_{\rm B}(1+\frac12v^2_{\rm B})+2(1+\g)\bar U^i(t,{\bm x}_\B)-(1+2\g)v^i_{\rm B}\bar U(t,{\bm x}_\B)\;,\\\la{1as1c}
\Lambda^i{}_0&=&-v^i_{\rm B}\left[1+\frac12v^2_{\rm B}+\g\bar U(t,{\bm x}_\B)\right]-F^{ij}_\B v^j_{\rm B}\;,\\\la{1as1d}
\Lambda^i{}_j&=&\d^{ij}\left[1+\g\bar U(t,{\bm x}_\B)\right]+\frac12v^i_{\rm B}v^j_{\rm B}+F^{ij}_\B\;,
\ea
where $F^{ij}_\B$ is the skew-symmetric matrix of the Fermi-Walker precession\index{precession!relativistic} of the spatial axes of the local frame adapted to body B, with respect to the global coordinates -- see \eqref{5.18}. 

We will also need the inverse matrix of transformation between the local and global coordinates taken on the worldline $\cal W$ of the origin of the local coordinates. We shall denote this matrix as
\be\la{p0s8}
\Omega^\a{}_\b\equiv{\Omega}^\a{}_\b(\tau)=\lim_{{\bm w}\rightarrow 0}\frac{\pd x^\a}{\pd w^\b}\;.
\ee
In accordance with the definition of the inverse matrix\index{inverse matrix} we have
\be\la{m9b1z}
\Lambda^\a{}_\b\Omega^\b{}_\g=\d^\a_\g\;,\qquad\qquad \Omega^\a{}_\b\Lambda^\b{}_\g=\d^\a_\g\;.
\ee
 Solving \eqref{m9b1z} with respect to the components of $\Omega^\a{}_\b$, we get
\ba\la{1a2sa}
\Omega^0{}_0&=&1+\frac12v^2_{\rm B}+\bar U(t,{\bm x}_\B)\;,\\\la{1as1b2}
\Omega^0{}_i&=&v^i_{\rm B}(1+\frac12v^2_{\rm B})+F^{ij}_\B v^j_{\rm B}-2(1+\g)\bar U^i(t,{\bm x}_\B)+(2+\g)v^i_{\rm B}\bar U(t,{\bm x}_\B)\;,\\\la{1as1c2}
\Omega^i{}_0&=&v^i_{\rm B}\left[1+\frac12v^2_{\rm B}+\bar U(t,{\bm x}_\B)\right]\;,\\\la{1as1d2}
\Omega^i{}_j&=&\d^{ij}\left[1-\g\bar U(t,{\bm x}_\B)\right]+\frac12v^i_{\rm B}v^j_{\rm B}-F^{ij}_\B\;,
\ea
As we shall see below, the matrices $\Lambda^\a{}_\b$ and $\Omega^\a{}_\b$ are instrumental in lifting the geometric objects pinned down to the worldline ${\cal W}$ and residing on 3-dimensional hypersurface ${\cal H}_u$ of constant time $u$ of the tangent space to the background manifold, from ${\cal H}_u$ up to 4-dimensional spacetime manifold $\bar{M}$.   

In order to arrive to the covariant formulation of the translational and rotational equations of motion we take the equations of motion derived in the local coordinates of body B, and prolongate them to the 4-dimensional, covariant form with the help of the transformation matrices and replacing the partial derivatives with the covariant ones. This is in accordance with the Einstein principle of equivalence which establishes a correspondence between spacetime manifold and its tangent space \citep{mtw}. It turns out that, eventually, all direct and inverse transformation matrices cancel out due to \eqref{m9b1z} and the equations acquire a final, covariant 4-dimensional form without any reference to the original coordinate charts that were used in the intermediate transformations. In what follows, we carry out this type of calculations.

\subsection{Geodesic Worldline and 4-Force on the Background Manifold}\la{o8b4m1}

Our algorithm of derivation of equations of motion defines the center of mass of body B by equating the conformal dipole of the body to zero, ${\cal I}^i=0$. The linear momentum, ${\mathfrak{p}}^i$ also vanishes ${\mathfrak{p}}^i=d{\cal I}^i/du=0$, as explained in section \ref{o93c5}. We have shown that these two conditions can be always satisfied by choosing the appropriate value \eqref{q6v4m}--\eqref{nrvug9} of the local acceleration, ${\cal Q}_i$, of the origin of the local coordinates adapted to body B in such a way that the worldline ${\cal W}$ of the origin of the local coordinates coincides with the worldline ${\cal Z}$ of the center of mass of the body. This specific choice of ${\cal Q}_i$ converts the equations of motion of the origin of the local coordinates of body B \eqref{5.8} to the equations of motion of its center of mass in the global coordinates. Below we prove that this equation can be interpreted on the background manifold $\bar{M}$ as the equation of time-like geodesic of a massive particle with the conformal mass, $M=M_\B$, of body B that is perturbed by the force of inertia produced by the local acceleration ${\cal Q}_i$ of the origin of the local coordinates. This is in concordance with the {\it effacing} principle \citep{Damour_1987book,kovl_2008,Battista_2017IJMPA} which determines dynamics in general relativity and scalar-tensor theory of gravity and suggests that the laws governing the motion of self-interacting masses are structurally identical to the laws governing the motion of test bodies \citep{harte2015}.

Let us introduce a 4-velocity ${\bar u}^\a$ of the center of mass of body B. In the global coordinates, $x^\a$, the worldline $\cal Z$ of the body's center of mass is described parametrically by $x^0_{\rm B}=t$, and $x^i_{\rm B}(t)$. The 4-velocity is defined by
\be\la{n8c3d}
{\bar u}^\a=\frac{dx^\a_{\rm B}}{d\t}\;,
\ee
where $\t$ is the proper time along the worldline $\cal Z$. The increment $d\t$ of the proper time is related to the increments $dx^\a$ of the global coordinates by equation, 
\be\la{n5b7x}
d\t^2=-\bar{g}_{\a\b}dx^\a dx^\b\;,
\ee
which tells us that the 4-velocity (\ref{n8c3d}) is normalized to unity, ${\bar u}_\a {\bar u}^\a=\bar g_{\a\b}{\bar u}^\a {\bar u}^\b=-1$. In the local coordinates the worldline ${\cal Z}$ is given by equations, $w^\a=(\t,\,w^i=0)$, and the 4-velocity has components $\bar{\rm u}^\a=(1,0,0,0)$. In the global coordinates the components of the 4-velocity are, ${u}^\a=\left(dt/d\t,\,dx^i_{\rm B}/d\t\right)$, which yields 3-dimensional velocity of the body's center of mass, $v^i_\B={\bar u}^i/{\bar u}^0=dx^i_{\rm B}/dt$.  Components of the 4-velocity are transformed from the local to global coordinates in accordance to the transformation equation, ${\bar u}^\a=\Omega^\a{}_\b{\bar{\rm u}}^\b$, which points out that in the global coordinates ${\bar u}^\a=\Omega^\a{}_0$. On the other hand, a covector of 4-velocity obeys the transformation equation, ${\bar u}_\a=\Lambda^\b{}_\a{\bar{\rm u}}_\b$, where ${\bar{\rm u}}_\a=(-1,0,0,0)$ are components of the covector of 4-velocity in the local coordinates. Thus, in the global coordinates ${\bar u}_\a=-\Lambda^0{}_\a$. The above presentation of the components of 4-velocity in terms of the matrices of transformation along with equation \eqref{m9b1z} makes it evident that 4-velocity is subject to two reciprocal conditions of orthogonality,
\be\la{brtv68h}
\Lambda^i{}_\a {\bar u}^\a=0\;, \qquad\qquad {\bar u}_\a\Omega^\a{}_i=0\;.
\ee
Equations \eqref{brtv68h} will be used later on in the procedure of lifting the spatial components of the internal and external multipoles to the covariant form.

In the covariant description of the equations of motion, an extended body B from ${\mathbb N}$-body system is treated as a particle having a conformal mass, $M=M_\B$, the {\it active} mass ${\cal M}\equiv{\cal M}_\B$, the {\it active} mass multipoles ${\cal M}^L\equiv{\cal M}^L_\B$, and the {\it active} spin multipoles ${\cal S}^L\equiv{\cal S}^L_\B$ attached to the particle, in other words, to the center of mass of the body. This set of the internal multipoles fully characterizes the internal structure of the body. The multipoles, in general, depend on time including the mass of the body which is not constant due to the temporal change of the multipoles (\ref{mch3c}) caused by tidal interaction. The mass and spin multipoles are fully determined by their spatial components in the body-adapted local coordinates in terms of integrals from the stress-energy distribution of matter through the solution of the field equations -- see sections \ref{mdloc}. Covariant generalization of the multipoles from the spatial to spacetime components is provided by the condition of orthogonality of the multipoles to the 4-velocity ${\bar u}^\a$ of the center of mass of the body as explained below in section \ref{n4v7a9}.

We postulate that the covariant definition of the linear momentum of the body is 
\be\label{zowv34as}
{\mathfrak{p}}^\a=M{\bar u}^\a\;,
\ee 
where ${\mathfrak{p}}^\a$ is a covariant generalization of 3-dimensional linear momentum ${\mathfrak{p}}^i$ of body B introduced in \eqref{b5z0e} where, for the time being, we do not specify the {\it complementary} part $\dot{\cal I}^\a_{\rm c}$. 
We are looking for the covariant translational equations of motion of body B in the following form
\begin{equation}
\label{geodes}
\frac{{\cal D} \mathfrak{p}^\a}{{\cal D}\tau}\equiv {\bar u}^\b\bar{\nabla}_\b {\mathfrak{p}}^\a=\frac{d{\mathfrak{p}}^\a}{d\tau}+\bar\Gamma^\a_{\mu\nu}{\mathfrak{p}}^\mu {\bar u}^\nu=F^{\a},
\end{equation}
where $F^\a$ is a 4-force that causes the worldline $\cal Z$ of the center of mass of the body to deviate from the geodesic worldline of the background manifold $\bar{M}$. We introduce this force to equation (\ref{geodes}) because the body's center of mass experiences a local acceleration ${\cal Q}_i$ given by (\ref{q6v4m}) which means that it is not in a state of a free fall and does not move on geodesic of the background manifold. In order to establish the mathematical form of the force $F^\a$ it is more convenient to re-write \eqref{geodes} in terms of 4-acceleration $a^\a\equiv {\cal D}{\bar u}^\a/{\cal D\tau}={\bar u}^\b \bar\nabla_\b {\bar u}^\a$
\be\la{betv67h}
M\left(\frac{d{\bar u}^\a}{d\tau}+\bar\Gamma^\a_{\mu\nu}{\bar u}^\mu {\bar u}^\nu\right)=F^{\a}-\dot M{\bar u}^\a\;,
\ee
where $\dot M$ is given in (\ref{x1674}). 

In what follows, it is more convenient to operate with a 4-force per unit mass defined by $f^{\a}\equiv F^\a/M$. Equation of motion \eqref{betv67h} is reduced to
\be\la{betv88h}
\frac{d{\bar u}^\a}{d\tau}+\bar\Gamma^\a_{\mu\nu}{\bar u}^\mu {\bar u}^\nu=f^{\a}-\frac{\dot M}{M}{\bar u}^\a\;,
\ee
The force $f^\a$ is orthogonal to 4-velocity, ${u}_\a f^\a=0$ as a consequence of \eqref{geodes} and the 4-velocity normalization condition. Hence, in the global coordinates the time component of the force is related to its spatial components as follows, $f_0=-v^i_{\rm B}f_i$. The condition of the orthogonality also yields the contravariant time component of the force in terms of its spatial components,
\be\la{q8h7m}
f^0=-\frac1{\bar g_{00}}\bar g_{ij}v^i_{\rm B}f^j\;.
\ee 
Our task is to prove that the covariant equation of motion (\ref{betv88h}) is exactly the same as the equation of motion (\ref{5.8}) of the center of mass of body B derived in the global coordinates that was obtained by asymptotic matching of the external and internal solutions of the field equations.
To this end we re-parameterize equation (\ref{betv88h}) by coordinate time $t$ instead of the proper time $\tau$, which yields
\ba\la{b4c8d}
a^i_{\rm B}&=&-\bar\G^i_{00}-2\bar\G^i_{0p}v^p_{\rm B}-\bar\G^i_{pq}v^p_{\rm B}v^q_{\rm B}
+\left(\bar\G^0_{00}+2\bar\G^0_{0p}v^p_{\rm B}+\bar\G^0_{pq}v^p_{\rm B}v^q_{\rm B}\right)v^i_{\rm B}+\left(f^i-f^0v^i_{\rm B}\right)\left(\frac{d\tau}{dt}\right)^2\;,
\ea
where $v^i_{\rm B}=dx^i_{\rm B}/dt$ and $a^i_{\rm B}=dv^i_{\rm B}/dt$ are the coordinate velocity and acceleration of the body's center of mass with respect to the global coordinates. 

We calculate the Christoffel symbols, $\bar\G^\a_{\m\n}$, the derivative $d\tau/dt$, substitute them to (\ref{b4c8d}) along with (\ref{q8h7m}), and retain only the Newtonian and post-Newtonian terms. Equation \eqref{b4c8d} takes on the following form
\begin{eqnarray}
  \label{aBi1}
  a_{\rm B}^i & = & \pd^i\bar{U}(t,{\bm x}_\B)+\pd^i\bar{\Psi}(t,{\bm x}_\B)-\frac{1}{2}\pd_{tt}\pd^i\bar{\chi}(t,{\bm x}_\B)+2(\gamma+1)\dot{\bar{U}}^{i}(t,{\bm x}_\B)\\
  & & -2(\gamma+1)v_{\rm B}^j\pd^i\bar{U}^j(t,{\bm x}_\B) -(2\gamma+1) v_{\rm B}^i\dot{\bar{U}}(t,{\bm x}_\B)\nonumber\\
  & & -2(\beta+\gamma)\bar{U}(t,{\bm x}_\B)\pd^i\bar{U}(t,{\bm x}_\B) +\gamma v_{\rm B}^2 \pd^i\bar{U}(t,{\bm x}_\B) - v_{\rm B}^iv_{\rm B}^j\pd^j\bar{U}(t,{\bm x}_\B)\nonumber\\\nonumber
  & & +f^i-v_{\rm B}^iv^k_{\rm B}f^k-\left[2\bar{U}(t,{\bm x}_\B)+v_{\rm B}^2\right]f^i\;.
\end{eqnarray}
This equation exactly matches translational equation of motion (\ref{5.8}) if we make the following identification of the spatial components $f^i$ of the force per unit mass with the local acceleration ${\cal Q}^i$,
\be\la{c5f6}
f^i\equiv -{\cal Q}^i-\frac12v^i_{\rm B}v^j_{\rm B}{\cal Q}_j+F^{ij}_\B {\cal Q}_j+\g\bar U(t,{\bm x}_\B){\cal Q}^i\;,
\ee 
By simple inspection we reveal that the right-hand side of the post-Newtonian force (\ref{c5f6}) can be written down in a covariant form 
\be\la{f5t8}
f^\a= -\bar g^{\a\b}\Lambda^{i}{}_\b {\cal Q}_{i}=\bar g^{\a\b}{\cal Q}_\b=-{\cal Q}^\a\;,
\ee
where $\Lambda^{i}{}_\b$ is given above in (\ref{1as1a})-(\ref{1as1d}), and ${\cal Q}_i$ is a vector of 4-acceleration in the local coordinates. The quantity ${\cal Q}_\a=\Lambda^{i}{}_\a {\cal Q}_{i}$ defines the covariant form of the local acceleration in the global coordinates with ${\cal Q}_\a$ being orthogonal to 4-velocity, ${\bar u}^\a {\cal Q}_\a=0$, which is a direct consequence of the condition \eqref{brtv68h}.  Explicit form of ${\cal Q}_i$ in the local coordinates is given in (\ref{q6v4m}) and should be used in (\ref{f5t8}) along with the covariant form of the external -- ${\cal Q}_L$, ${\cal C}_L$, ${\cal P}_L$ and internal -- ${\cal M}^L$, $\mathcal{S}^L$ multipoles in order to get $f^\a=-\bar g^{a\b}{\cal Q}_\b$. The covariant form of the multipoles is a matter of discussion in next subsection.

\subsection{Four-dimensional Form of Multipoles}\la{n4v7a9}
\subsubsection{Internal Multipoles}

The mathematical procedure that was used in construction of the local coordinates adapted to an extended body B in ${\mathbb N}$-body system indicates that all type of multipoles are defined at the origin of the local coordinates as the STF Cartesian tensors having only spatial components with their time components being identically nil. It means that the multipoles are projections of 4-dimensional tensors on hyperplane passing through the origin of the local coordinates orthogonally to 4-velocity ${\bar u}^\a$ of the worldline ${\cal Z}$ of the center of mass of the body. The 4-dimensional form of the internal multipoles can be established by making use of the law of transformation from the local to global coordinates,
\be\la{x6v4n}
{\cal M}^{\a_1...\a_l}\equiv\Omega^{\a_1}{}_{i_1}...\Omega^{\a_l}{}_{i_l}{\cal M}^{i_1i_2...i_l}\;,\qquad\qquad
{\cal S}^{\a_1...\a_l}\equiv\Omega^{\a_1}{}_{i_1}...\Omega^{\a_l}{}_{i_l}{\cal S}^{i_1i_2...i_l}\;,
\ee
where the matrix of transformation $\Omega^{\a}{}_{i}$ is given in \eqref{1a2sa}--\eqref{1as1d2}. Transforming 3-dimensional STF Cartesian tensors to 4-dimensional form does not change the property of the tensors to be symmetric and trace-free in the sense that we have for any pair of spacetime (Greek) indices
\be\label{gt4zx}
\bar g_{\a_1\a_2}{\cal M}^{\a_1\a_2...\a_l}=0\;,\qquad\qquad \bar g_{\a_1\a_2}{\cal S}^{\a_1\a_2...\a_l}=0\;.
\ee
The 4-dimensional form \eqref{x6v4n} of the multipoles along with equation \eqref{brtv68h} confirms that the multipoles are orthogonal to 4-velocity, that is
\be\la{m1zz1}
{\bar u}_{\a_1} {\cal M}^{\a_1...\a_l}=0\;,\qquad\qquad
{\bar u}_{\a_1} {\cal S}^{\a_1...\a_l}=0\;,
\ee
and due to the symmetry of the internal multipoles, equation (\ref{m1zz1}) is valid to each index. 

Notice that the matrix of transformation\index{matrix of transformation} (\ref{p0s8}) has been used in making up the contravariant components of the multipoles (\ref{x6v4n}) which are tensors of type $\genfrac[]{0pt}{2}{l}{0}$. Tensor components of the multipoles, ${\cal M}_{\a_1...\a_l}$ and ${\cal S}_{\a_1...\a_l}$, which are of the type $\genfrac[]{0pt}{2}{0}{l}$ are obtained by lowering each index of ${\cal M}^{\a_1...\a_l}$ and ${\cal S}^{\a_1...\a_l}$  respectively with the help of the background metric tensor $\bar g_{\a\b}$. It is worth emphasizing that we have introduced 4-dimensional definitions of the internal multipoles as tensors of type $\genfrac[]{0pt}{2}{l}{0}$ on the ground of transformation equations $\eqref{x6v4n}$ because we defined the spatial components of ${\cal M}^{i_1...i_l}$ and ${\cal S}^{i_1...i_l}$ as integrals \eqref{1.31} and \eqref{1.32} taken from the STF products of the components of 3-dimensional coordinate $w^i$ which behaves as a vector under the linear coordinate transformations. Another reason to use the contravariant components ${\cal M}^{i_1...i_l}$ and ${\cal S}^{i_1...i_l}$ as a starting point for their 4-dimensional prolongation is that the internal multipoles are the coefficients of the Cartesian tensors of type $\genfrac[]{0pt}{2}{l}{0}$  in the Taylor expansions \eqref{je5v20}, \eqref{hdv45x5} and \eqref{w1w8} of the gravitational potentials $U_\B(t,{\bm x})$ and $U^i_\B(t,{\bm x})$ with respect to the components of the partial derivatives $\pd_{i_1...i_l} r^{-1}_\B$ which are considered as the STF Cartesian tensors of type $\genfrac[]{0pt}{2}{0}{l}$.   

\subsubsection{External Multipoles}\la{ne7c2x9}
The external multipoles, ${\cal P}_{i_1...i_l}$, ${\cal Q}_{i_1...i_l}$ and ${\cal C}_{i_1...i_l}$,  have been defined at the origin of the local coordinates of body B by external solutions of the field equations for the metric tensor  and scalar field in such a way that they are purely spatial STF Cartesian tensors of type $\genfrac[]{0pt}{2}{0}{l}$ - see section \ref{mtextso}. It means that 4-dimensional tensor extensions of the external multipoles must be orthogonal to 4-velocity of the origin of the local coordinates which is, by construction, identical to 4-velocity $\bar u^\a$ of the worldline ${\cal Z}$ of the center of mass of the body B,
\be\la{m1zz2}
{\bar u}^{\a_1} {\cal Q}_{\a_1\a_2...\a_l}=0\;,\qquad {\bar u}^{\a_1} {\cal P}_{\a_1\a_2...\a_l}=0\;,\qquad
{\bar u}^{\a_1} {\cal C}_{\a_1\a_2...\a_l}=0\;.
\ee
These orthogonality conditions suggests that the 4-dimensional components of the external multipoles are obtained from their 3-dimensional counterparts by making use of the matrix of transformation (\ref{p0c3b5}) which yields
\be\la{x6v4na}
{\cal Q}_{\a_1...\a_l}\equiv\Lambda^{i_1}{}_{\a_1}...\Lambda^{i_l}{}_{\a_l}{\cal Q}_{i_1...i_l}\;,\qquad\quad
{\cal C}_{\a_1...\a_l}\equiv\Lambda^{i_1}{}_{\a_1}...\Lambda^{i_l}{}_{\a_l}{\cal C}_{i_1...i_l}\;,\qquad\quad
{\cal P}_{\a_1...\a_l}\equiv\Lambda^{i_1}{}_{\a_1}...\Lambda^{i_l}{}_{\a_l}{\cal P}_{i_1...i_l}\;.
\ee
We have used in here the matrix of transformation (\ref{p0c3b5}) because the external multipoles are defined originally as tensor coefficients of the  Taylor expansions of the external potentials $\bar U$, $\bar\Psi$, etc., which are expressed in terms of partial derivatives from these potentials and behave under coordinate transformations like tensors of type $\genfrac[]{0pt}{2}{0}{l}$. 
Definitions \eqref{x6v4na} and the properties of the matrices of transformation suggest that 4-dimensional tensors ${\cal Q}_{\a_1...\a_l}$, ${\cal C}_{\a_1...\a_l}$ and ${\cal P}_{\a_1...\a_l}$ are STF tensors in the sense of \eqref{gt4zx} that is $\bar g^{\a_1\a_2}{\cal Q}_{\a_1...\a_l}=0$, etc.

It is known that in general relativity the external multipoles, ${\cal Q}_{i_1...i_l}$ and ${\cal C}_{i_1...i_l}$ are defined in the local coordinates by partial derivatives of the Riemann tensor, $\bar R^\a{}_{\m\b\n}$, of the background metric (\ref{k8b2d}) taken at the origin of the local coordinates \citep{th_1985,1986PhRvD..34.3617S,zhang_1986PhRvD,poisson_2011}. This definition remains valid with some modification in the scalar-tensor theory of gravity which is explained below. 
The external multipoles\index{scalar field!external multipoles}, ${\cal P}_{i_1...i_l}$, of the scalar field are not related in any way to the Riemann tensor because they depend merely on derivatives of the background scalar field $\bar\varphi$.

As we show below, the 4-dimensional tensor formulation of the external multipoles is achieved by contracting the Riemann tensor with vectors of 4-velocity, ${\bar u}^\a$, and taking the covariant derivatives $\bar\nabla_\a$ projected on the hyperplane being orthogonal to the 4-velocity. The projection is fulfilled with the help of the operator of projection,
\be\la{n6g0s}
\pi^{\alpha}_{\beta}\equiv \delta^{\alpha}_{\beta}+{\bar u}^{\alpha}{\bar u}_{\beta}\;, \qquad\quad\pi^{\a\b}=\bar g^{\a\b}+{\bar u}^\a {\bar u}^\b\;,\qquad\quad\pi_{\a\b}=\bar g_{\a\b}+{\bar u}_\a {\bar u}_\b\;,  
\ee
The operator of projection satisfies the following relations: $\pi^\alpha_\g\pi^\g_\beta=\pi^\alpha_\beta$, $\pi^{\a\b}=\bar g^{\a\g}\pi^\b_\g$, $\pi_{\a\b}=\bar g_{\a\g}\pi^\g_\b$, and $\pi^\a_\a=3$. The latter property points out that $\pi^{\alpha}_{\beta}$ has only three algebraically-independent components which are reduced to the Kronecker symbol when $\pi^\a_\b$ is computed in the local coordinates of body B, that is in the local coordinates $\pi^0_0=0\;,\pi^i_0=\pi^0_i=0\;,\pi^i_j=\d^i_j$. In other words, the projection operator is a 3-dimensional Kronecker symbol $\d^i_j$ lifted up to 4-dimensional effective background manifold $\bar{M}$. We notice that the operator of the projection has some additional algebraic properties. Namely,
\be\la{nrv7b3}
\pi^\a_\b\Lambda^i{}_\a=\Lambda^i{}_\b\;,\qquad\qquad \pi^\b_\a\Omega^\a{}_i=\Omega^\b{}_i\;,
\ee
that are in accordance with the condition of orthogonality \eqref{brtv68h}. They point out that $\pi^{\alpha}_{\beta}$ can be also represented as a product of two reciprocal transformation matrices,
\be
\pi^{\alpha}_{\beta}=\Omega^\a{}_i\Lambda^i{}_\b\;.
\ee

The projection operator is required to extend 3-dimensional spatial derivatives of geometric objects to their 4-dimensional counterparts. Indeed, in the local coordinates the external multipoles are purely spatial Cartesian tensors which are expressed in terms of the partial spatial derivatives of the external perturbations of the metric tensor and/or scalar field. It means that the extension of a spatial partial derivative to its 4-dimensional form must preserve its orthogonality to the 4-velocity ${\bar u}^\a$ of the worldline ${\cal Z}$ which is achieved by coupling the spatial derivatives with the projection operator. For example, 4-dimensional STF form of the external STF scalar multipole ${\cal P}_L\equiv{\cal P}_{i_1...i_l}={\cal P}_{<i_1...i_l>}$ introduced in (\ref{3.13}) in terms of the spatial derivatives of the external scalar field, reads
\ba\la{o6b1o9}
{\cal P}_{\a_1...\a_l}&=&\Lambda^{i_1}{}_{\a_1}...\Lambda^{i_l}{}_{\a_l}{\cal P}_{i_1...i_l}=\Lambda^{<i_1}{}_{\a_1}...\Lambda^{i_l>}{}_{\a_l}\bar\nabla_{<i_1...i_l>}\bar\varphi
 =\Lambda^{<i_1}{}_{\a_1}...\Lambda^{i_l>}{}_{\a_l}\Omega^{\b_1}{}_{<i_1}...\Omega^{\b_l}{}_{i_l>}\bar\nabla_{\b_1...\b_l}\bar\varphi\\\nonumber &=&
 {\pi}^{\beta_1}_{<\alpha_1}\cdots {\pi}^{\beta_l}_{\alpha_l>}\bar\nabla_{\beta_1\cdots\beta_l}\bar\varphi\;,
\ea
where $\bar\varphi$ is the background scalar field perturbation, and the angular brackets around Greek indices indicate 4-dimensional generalization of 3-dimensional STF tensor defined earlier in \eqref{stfformula}. Extending 3-dimensional Kronecker symbol and other 3-tensors to 4-dimensional form we get, 
\be\label{stf4form}
T_{<\a_1...\a_l>}\equiv\sum_{n=0}^{[l/2]}\frac{(-1)^n}{2^nn!}\frac{l!}{(l-2n)!}\frac{(2l-2n-1)!!}{(2l-1)!!}\pi_{(\a_1\a_2...}\pi_{\a_{2n-1}\a_{2n}}S_{\a_{2n+1}...\a_l)\b_1\g_1...\b_n\g_n}\pi^{\b_1\g_1}...\pi^{\b_n\g_n}\;.
\ee 
We also notice that the projection operator can be effectively used to rise and/or to lower 4-dimensional (Greek) indices of the internal and external multipoles like the metric tensor $\bar g_{\a\b}$. This is because all multipoles are orthogonal to the 4-velocity ${\bar u}^\a$. Thus, for example, ${\cal Q}_{\a\b}\bar g^{\b\g}={\cal Q}_{\a\b}\pi^{\b\g}={\cal Q}_\a{}^\g$, etc.

The external multipoles ${\cal Q}_{\a_1...\a_l}$ and ${\cal C}_{\a_1...\a_l}$ are directly connected to the Riemann tensor of the background manifold and its covariant derivatives. In order to establish this connection we work in the local coordinates and employ a covariant definition of the Riemann tensor \eqref{kk33xc} of the background manifold where the background metric tensor in the local coordinates is  
\be\label{mm33xx}
\bar{\rm g}_{\a\b}=\eta_{\a\b}+\hat h_{\a\b}^{\rm ext}(u,{\bm w})+\hat l_{\a\b}^{\rm ext}(u,{\bm w})\;,
\ee
with the perturbations $\hat h_{\a\b}^{\rm ext}$ and $\hat l_{\a\b}^{\rm ext}$ defined in \eqref{1.24b}--\eqref{1.24c} respectively. 
The products of the connections entering \eqref{kk33xc} at the post-Newtonian level of approximation requires the following components of the Christoffel symbols
\be\label{kk22ss}
\bar\Gamma^i_{00}=\bar\Gamma^0_{0i}=-\frac12 \pd_i\hat h_{00}^{\rm ext}\;,\qquad\qquad \bar\Gamma^i_{jk}= \frac12\left(\pd_j\hat h_{ik}^{\rm ext}+\pd_k\hat h_{ik}^{\rm ext}-\pd_i\hat h_{jk}^{\rm ext}\right)\;.
\ee
Substituting \eqref{mm33xx} and \eqref{kk22ss} to \eqref{kk33xc} and taking into account all post-Newtonian terms we get the STF part of the Riemann tensor component $[\bar R_{0i0j}]^{\rm STF}\equiv \bar R_{0<i|0|j>}$ in the following form,
\ba\la{riem634}
 \left[\bar R_{0i0j}\right]^{\rm STF}&=&-D_{<ij>}+3D_{<i}D_{j>}+2DD_{<ij>}
 \\\nonumber&&
 +2(\g-1)D_{<i}H_{j>}+2(\b-1)\Bigl[H_{<i}H_{j>}+\left(H-{\cal P}\right)H_{<ij>}\Bigr]\\\nonumber
&&+2\sum_{l=0}^\infty\frac{(l-1)(l+1)}{(2l+5)(l+2)!}\ddot {\cal Q}_{L<i}w_{j>L}+(\g-1)\sum_{l=0}^\infty\frac{(2l+1)(l+1)}{(2l+5)(l+2)!}\ddot {\cal P}_{L<i}w_{j>L}\\\nonumber
&&-\frac12\sum_{l=0}^\infty\frac{l+7}{(2l+7)(l+3)l!}\ddot {\cal Q}_{<ij>L}w^{L}w^2-(\g-1)\sum_{l=0}^\infty\frac{1}{(2l+7)(l+3)l!}\ddot {\cal P}_{<ij>L}w^{L}w^2\\\nonumber
&&+\sum_{l=0}^\infty\frac{l+1}{(l+2)!}\varepsilon_{pq<i}\dot {\cal C}_{j>pL}w^{qL}\;,
\ea
where we have discarded all terms of the post-post-Newtonian order and
introduced the shorthand notations 
\ba
D&\equiv&D(u,{\bm w})= \sum_{k=1}^\infty \frac1{k!}{\cal Q}_{K}(u)w^K\;,\\
H&\equiv&H(u,{\bm w})= \sum_{k=0}^\infty \frac1{k!}{\cal P}_{K}(u)w^K\;,\\
D_{i_1...i_l}&\equiv&D_{i_1...i_l}(u,{\bm w})=  \partial_{i_1...i_l}D=\sum_{k=0}^\infty \frac1{k!}{\cal Q}_{i_1...i_lK}(u)w^K\;,\\
H_{i_1...i_l}&\equiv&H_{i_1...i_l}(u,{\bm w})=\partial_{i_1...i_l}H=\sum_{k=0}^\infty \frac1{k!}{\cal P}_{i_1...i_lK}(u)w^K\;.
\ea
Notice that at the origin of the local coordinates where $w^i=0$, we have $D(u,0)=0$, $H(u,0)={\cal P}$, $D_{i_1...i_p}(u,0)={\cal Q}_{i_1...i_p}$ and $H_{i_1...i_p}(u,0)={\cal P}_{i_1...i_p}$. Therefore, at the origin of the local coordinates, that is on the worldline ${\cal Z}$, the value of the STF Riemann tensor \eqref{riem634} is simplified to
\be\la{evy53c}
\left[\bar R_{0i0j}\right]^{\rm STF}_{\cal Z}=-{\cal Q}_{<ij>}
+3{\cal Q}_{<i}{\cal Q}_{j>}+2(\g-1){\cal Q}_{<i}{\cal P}_{j>}+2(\b-1){\cal P}_{<i}{\cal P}_{j>}\;.
\ee
This relationship establishes the connection between the external mass quadrupole ${\cal Q}_{ij}$ and the STF Riemann tensor. The reader should notice that \eqref{evy53c} includes terms depending on acceleration ${\cal Q}_i$ of the worldline of the center of mass of body B. This may look strange as the curvature of spacetime (the Riemann tensor) does not depend on the choice of the worldline of the local coordinates.  Indeed, it can be verified that the acceleration-dependent terms in \eqref{evy53c} are mutually canceled out with the similar terms coming out of the explicit expression for ${\cal Q}_{ij}$ taken from \eqref{5.9}, and obtained by the asymptotic matching technique.  

Relationship between the STF covariant derivative of $l$-th order from the Riemann tensor and the external gravitoelectric multipole of the same order is derived by taking covariant derivatives $l$ times from both sides of \eqref{riem634}. Covariant derivative of the order $l$ from the Riemann tensor is a linear operator on the background manifold that involves the products of the Christoffel symbols and the covariant derivatives of the order $l-1$ from the Riemann tensor. They can be calculated by iterations starting from $l=1$. Straightforward but tedious calculation shows that at the post-Newtonian level of approximation the covariant derivative of the order $l-2$ combined with the Riemann tensor to STF tensor of the order $l$, reads,
\ba\la{u5b1x}
\left[\bar\nabla_{i_1...i_{l-2}}\bar{R}_{0i_{l-1}0i_l}\right]^{\rm STF}=
\left[\pd_{i_1...i_{l-2}}\bar{R}_{0i_{l-1}0i_l}\right]^{\rm STF}&&\\\nonumber
+2\sum_{k=0}^{l-3}(k+1)\pd_{<i_1...i_{l-k-3}}\left[D_{i_{l-k-2}...i_{l-1}}D_{i_l>}\right]
&+&2(\g-1)\sum_{k=0}^{l-3}(k+2)\pd_{<i_1...i_{l-k-3}}\left[D_{i_{l-k-2}...i_{l-1}}H_{i_l>}\right]\;.
\ea
Applying the Leibniz rule of differentiation to the product of two functions \citep[Equation 0.42]{gradryzh}  standing in the right-hand side of \eqref{u5b1x}, we obtain a more simple expression,
\ba
\left[\bar\nabla_{i_1...i_{l-2}}\bar{R}_{0i_{l-1}0i_l}\right]^{\rm STF}&=&
\left[\pd_{i_1...i_{l-2}}\bar{R}_{0i_{l-1}0i_l}\right]^{\rm STF}
+2\sum_{k=0}^{l-3}\sum_{s=0}^{k}\frac{(l-k-2)k!}{s!(k-s)!}D_{<i_1...i_{s+1}}D_{i_{s+2}...i_l>}\\\nonumber
&+&2(\g-1)\sum_{k=0}^{l-3}\sum_{s=0}^{k}\frac{(l-k-1)k!}{s!(k-s)!}H_{<i_1...i_{s+1}}D_{i_{s+2}...i_l>}\;.
\ea
The $l-2$-th order partial derivatives from terms $D_{<i}D_{j>}$, $DD_{<ij>}$, etc., entering $\left[\pd_{i_1...i_{l-2}}\bar R_{0i_{l-1}0i_l}\right]^{\rm STF}$, are also calculated with the help of the Leibniz rule, yielding 
\ba
\pd_{<i_1...i_{l-2}}\left[D_{i_{l-1}}D_{i_l>}\right]&=&\sum_{k=0}^{l-2}\frac{(l-2)!}{k!(l-k-2)!}D_{<i_1...i_{k+1}}D_{i_{k+2}...i_{l}>}\;,\\
\pd_{<i_1...i_{l-2}}\left[D_{i_{l-1}i_l>}D\right]&=&
\sum_{k=1}^{l-2}\frac{(l-2)!}{k!(l-k-2)!}D_{<i_1...i_k}D_{i_{k+1}...i_l>}\;,\\
\pd_{<i_1...i_{l-2}}\left[D_{i_{l-1}}H_{i_l>}\right]&=&\sum_{k=0}^{l-2}\frac{(l-2)!}{k!(l-k-2)!}D_{<i_1...i_{k+1}}H_{i_{k+2}...i_{l}>}\;,\\
\pd_{<i_1...i_{l-2}}\left[H_{i_{l-1}}H_{i_l>}\right]&=&\sum_{k=0}^{l-2}\frac{(l-2)!}{k!(l-k-2)!}H_{<i_1...i_{k+1}}H_{i_{k+2}...i_{l}>}\;,\\
\pd_{<i_1...i_{l-2}}\left[D_{i_{l-1}i_l>}(H-{\cal P})\right]&=&
\sum_{k=1}^{l-2}\frac{(l-2)!}{k!(l-k-2)!}H_{<i_1...i_k}H_{i_{k+1}...i_l>}\;.
\ea

Actually, we need the covariant derivatives of the STF part of the Riemann tensor only at the origin of the local coordinates adapted to body B. Therefore, after taking the STF covariant derivatives from the Riemann tensor we take the value of the local spatial coordinates $w^i=0$, which eliminates all terms depending on the time derivatives of the external multipoles in the right-hand side of \eqref{riem634} for the STF part of the Riemann tensor. Hence, the STF covariant derivative of the Riemann tensor taken on the worldline of the center of mass of body B reads,
\ba\la{vx4a6c}
\left[\bar\nabla_{i_1...i_{l-2}}\bar{R}_{0i_{l-1}0i_l}\right]^{\rm STF}_{\cal Z}=
-{\cal Q}_{<i_1...i_l>}+
3\sum_{k=0}^{l-2}\frac{(l-2)!}{k!(l-k-2)!}{\cal Q}_{<i_1...i_{k+1}}{\cal Q}_{i_{k+2}...i_{l}>}\phantom{+++++++++}&&\\\nonumber
+2\bigg[\sum_{k=1}^{l-2}\frac{(l-2)!}{k!(l-k-2)!}{\cal Q}_{<i_1...i_k}{\cal Q}_{i_{k+1}...i_{l}>}+\sum_{k=0}^{l-3}\sum_{s=0}^{k}\frac{(l-k-2)k!}{s!(k-s)!}{\cal Q}_{<i_1...i_{s+1}}{\cal Q}_{i_{s+2}...i_l>}\bigg]&&\\\nonumber
+2(\g-1)\bigg[\sum_{k=1}^{l-2}\frac{(l-2)!}{k!(l-k-2)!}{\cal Q}_{<i_1...i_k}{\cal P}_{i_{k+1}...i_{l}>}+\sum_{k=0}^{l-3}\sum_{s=0}^{k}\frac{(l-k-1)k!}{s!(k-s)!}{\cal P}_{<i_1...i_{s+1}}{\cal Q}_{i_{s+2}...i_l>}\bigg]&&\\\nonumber
+2(\b-1)\bigg[\sum_{k=1}^{l-2}\frac{(l-2)!}{k!(l-k-2)!}{\cal P}_{<i_1...i_k}{\cal P}_{i_{k+1}...i_{l}>}+\sum_{k=0}^{l-2}\frac{(l-2)!}{k!(l-k-2)!}{\cal P}_{<i_1...i_{k+1}}{\cal P}_{i_{k+2}...i_{l}>}\bigg]&&\;.
\ea
 
It is rather straightforward now to convert \eqref{vx4a6c} to 4-dimensional form valid in arbitrary coordinates on the effective manifold $\bar{M}$ by making use of the transformation matrices and the operator of projection as it was explained above. We introduce a new notation for the covariant STF derivative of the Riemann tensor taken on the worldline ${\cal Z}$,
\be
  \label{sd11}
{\cal E}_{\a_1...\a_l}  \equiv {\pi}^{\beta_1}_{<\alpha_1}\pi^{\b_2}_{\a_2}.... {\pi}^{\beta_l}_{\alpha_l>}\left[\bar\nabla_{\beta_1...\beta_{l-2}}\bar R_{\mu\beta_{l-1}\beta_l\nu} {u}^{\mu}{u}^{\nu}\right]^{\rm STF}_{\cal Z}\;,
\ee
and use it for transformation of \eqref{vx4a6c} to arbitrary coordinates. It yields a covariant expression for the external gravitoelectric multipoles ${\cal Q}_{\a_1...\a_l}$ in terms of the STF covariant derivatives from the Riemann tensor,
\begin{eqnarray}
\label{we69}
{\cal Q}_{\a_1...\a_l} & = &  {\cal E}_{<\a_1...\a_l>}
+3\sum_{k=0}^{l-2}\frac{(l-2)!}{k!(l-k-2)!}{\cal E}_{<\a_1...\a_{k+1}}{\cal E}_{\a_{k+2}...\a_{l}>}\\\nonumber
&+&2\bigg[\sum_{k=1}^{l-2}\frac{(l-2)!}{k!(l-k-2)!}{\cal E}_{<\a_1...\a_k}{\cal E}_{\a_{k+1}...\a_{l}>}+\sum_{k=0}^{l-3}\sum_{s=0}^{k}\frac{(l-k-2)k!}{s!(k-s)!}{\cal E}_{<\a_1...\a_{s+1}}{\cal E}_{\a_{s+2}...\a_l>}\bigg]\\\nonumber
&+&2(\g-1)\bigg[\sum_{k=1}^{l-2}\frac{(l-2)!}{k!(l-k-2)!}{\cal E}_{<\a_1...\a_k}{\Phi}_{\a_{k+1}...\a_{l}>}+\sum_{k=0}^{l-3}\sum_{s=0}^{k}\frac{(l-k-1)k!}{s!(k-s)!}{\Phi}_{<\a_1...\a_{s+1}}{\cal E}_{\a_{s+2}...\a_l>}\bigg]\\\nonumber
&+&2(\b-1)\bigg[\sum_{k=1}^{l-2}\frac{(l-2)!}{k!(l-k-2)!}{\Phi}_{<\a_1...\a_k}{\Phi}_{\a_{k+1}...\a_{l}>}+\sum_{k=0}^{l-2}\frac{(l-2)!}{k!(l-k-2)!}{\Phi}_{<\a_1...\a_{k+1}}{\Phi}_{\a_{k+2}...\a_{l}>}\bigg]\;,
\ea
where 
we have made identification: ${\cal E}_a\equiv{\cal Q}_\a$. At this stage of calculation, it is worth noticing that 4-acceleration of the center of mass of body B, $a_\a\equiv {\bar u}^\b \bar\nabla_\b {\bar u}^\a$, is not exactly equal to ${\cal E}_\a$ because of a term depending on the time derivative of body's mass, $\dot M$, in the right-hand side of \eqref{betv67h}. Only in case when the mass is conserved, $a^\a={\cal E}^\a$.

Similar, but less tedious procedure allows us to calculate 4-dimensional form of the external gravitomagnetic multipoles ${\cal C}_{\a_1...\a_l}$ in terms of the STF covariant derivative of the Riemann tensor. We get,
\be
\label{we70}
{\cal C}_{\a_1...\a_l}  \equiv 
 {\pi}^{\beta_1}_{<\alpha_1}\pi^{\b_2}_{\a_2}...{\pi}^{\b_l}_{\alpha_l>}\left[\bar\nabla_{\beta_1...\beta_{l-2}}\bar R_{\s\m\n\beta_{l-1}}\varepsilon_{\b_l}{}^{\s\m}{\bar u}^\n\right]^{\rm STF}_{\cal Z}\;.
\ee
where we have utilized 3-dimensional covariant tensor of Levi-Civita $\varepsilon_{\a\beta\gamma}$ which is a projection of 4-dimensional, fully-antisymmetric Levi-Civita symbol $E_{\a\mu\nu\r}$ \index{Levi-Civita!symbol} \citep[\S 3.5]{mtw} on the hyperplane being orthogonal to 4-velocity ${\bar u}^\a$,
\be\label{vareps67}
\varepsilon_{\a\beta\gamma}\equiv (-\bar g)^{1/2}{\bar u}^\m\pi^\n_\a \pi^\r_\b  \pi^\s_\g  E_{\m\n\r\s}  \;.
\ee 
It can be checked by inspection that in the global coordinates the right-hand sides of (\ref{we69}) and (\ref{we70}) are reduced to ${\cal Q}_L$ and ${\cal C}_L$ respectively as it must be. 

4-dimensional definitions of the external multipoles given in this section allow us to transform products of the multipoles given in the local coordinates to their covariant counterparts, for example, ${\cal Q}_{L}{\cal M}^L\equiv {\cal Q}_{i_1...i_l}{\cal M}^{i_1...i_l}={\cal Q}_{\a_1...\a_l}{\cal M}^{\a_1...\a_l}$, etc. In all such products the matrices of transformation cancel out giving rise to covariant expressions being independent of a particular choice of coordinates.

\subsection{Covariant Translational Equations of Motion}\la{n3cz52s}

A generic form of the covariant translational equations of motion have been formulated in \eqref{betv67h}. Substituting to these equations the force $F^\a=-M{\cal Q}^\a$ where ${\cal Q}^\a$ was introduced in \eqref{f5t8}, yields 
\begin{eqnarray}
  \label{ubpu;b}
 M\frac{{\cal D} {\bar u}^\m}{{\cal D}\tau}&=&F^\mu-\dot M {\bar u}^\a \;,
\end{eqnarray}
where the force
\be\la{g6xc2c7}
F^\mu=F^\m_{\mathfrak{q}}+{F}^\m_{\cal Q}+{F}^\m_{\cal C}+{F}^\m_{\cal P}\;,
\ee
and the second term in the right-hand side of \eqref{ubpu;b} is due to the non-conservation of mass (\ref{x1674}) having the following covariant form 
\ba\la{x1674qq}
\dot M&=&(\g-1)\left({\cal P}\sum\limits_{l=1}^\infty\frac1{l!}{\cal Q}_{\a_1...\a_l}\frac{{\cal D}_{\rm F}{\cal M}^{\a_1...\a_l}}{{\cal D}\tau}+\frac{{\cal D}_{\rm F}{\cal P}}{{\cal D}\tau}{\cal M}\right)\\\nonumber
&-&\sum\limits_{l=1}^\infty\frac1{(l-1)!}{\cal Q}_{\a_1...\a_l}\frac{{\cal D}_{\rm F}{\cal M}^{\a_1...\a_l}}{{\cal D}\tau}-\sum\limits_{l=1}^\infty\frac{l+1}{l!}{\cal M}^{\a_1...\a_l}\frac{{\cal D}_{\rm F}{\cal E}_{\a_1...\a_l}}{{\cal D}\tau}\;,
\ea
where we have used the covariant Fermi-Walker derivative of the multipole moments which is a covariant generalization of the total time derivative in the local coordinates. The Fermi-Walker derivative is explained in more detail at the end of this section -- see equation \eqref{p2b8r3}.

Gravitational force $F^\mu$ in the right-hand side of \eqref{ubpu;b} is the 4-dimensional extension of 3-dimensional force \eqref{c5f6} with the local 4-acceleration ${\cal Q}_i$ defined in \eqref{q6v4m} where the complementary function ${\cal I}^i_{\rm c}$ is chosen as follows
${\cal I}^i_{\rm c}=3{\cal Q}_k{\cal M}^{ik}$, or, in 4-dimensional form
\be\la{d9n3v5}
{\cal I}^\a_{\rm c}=3{\cal Q}_\b{\cal M}^{\a\b}\;.
\ee
This form of ${\cal I}^\a_{\rm c}$ eliminates the terms depending on the local acceleration $Q_\a$ coupled with the quadrupole moment ${\cal M}^{\a\b}$ of the body from the force $F^\a$, and delivers a covariant definition of the center of mass of body B. It is similar but not exactly equal to the choice \eqref{m4g1x8p} of this function in the global coordinates. 

The first term in the right side of (\ref{g6xc2c7}) describes the Dicke-Nordtvedt anomalous force caused by the violation of the {\it strong} principle of equivalence (SEP)\index{SEP}\index{principle of equivalence!strong}\index{principle of equivalence!violation}  
\be\la{x8g2m9}
F^\a_{\mathfrak{q}}= \mathfrak{q}{\cal P}^\a\;,
\ee
where
\be\la{x1675}{\cal P}^\a={\pi}^{\a\b}\bar\nabla_\b\bar\varphi\;,
\ee
is an external scalar-field dipole \index{scalar field!dipole} and $\mathfrak{q}\equiv{\cal M}-M$ is 
the difference between the {\it active} -- ${\cal M}$, and {\it conformal} -- $M$, masses of body B. The quantity $\mathfrak{q}$ can be interpreted as an effective scalar charge of body B interacting with the external scalar field\index{scalar field!charge} and causing the body to accelerate with respect to a body having negligible self-gravity but the same set of internal multipole moments. The anomalous scalar-field gravitational force $F^\m_{\mathfrak{q}}$ was predicted by Dicke \index{Dicke} and its effect in three body system (Earth-Moon-Sun) was studied by Nordtvedt in the framework of PPN formalism \index{Nordtvedt effect} \citep[\S 8.1]{willbook}.  Explicit expression for the scalar charge $\mathfrak{q}$ is obtained from (\ref{p3c2}) and reads
\ba\label{a32d58}
\mathfrak{q}&=& \frac{1}{2}\eta\int\limits_{{\cal V}_{\rm B}}\rho^{\ast}\hat{U}_{\rm B}d^3w -\frac{1}{6}(\gamma-1)\frac{{\cal D}_{\rm F}^2{\cal N}}{{\cal D}\tau^2}
  +2(\beta-1)\mathcal{M}{\cal P}\\\nonumber
  &&+2(\beta-1)\sum_{l=1}^{\infty}\frac{1}{l!}{\cal P}_{\a_1...\a_l}{\cal M}^{\a_1...\a_l}+(\gamma-1)\sum_{l=1}^{\infty}\frac{1}{(l-1)!}{\cal Q}_{\a_1...\a_l}{\cal M}^{\a_1...\a_l}\;.
\ea
The first and second terms in the right-hand side of (\ref{a32d58}) compose a {\it bare} part of the scalar charge being proportional to self-gravitational energy of the body and the second time derivative of the body's moment of inertia ${\cal N}$. Standard treatment of the Nordtvedt effect \citep[\S 8.1]{willbook} takes into account only the very first term in the right-hand side of \eqref{a32d58}  which is proportional to the Nordtvedt parameter $\eta$ assuming that the time derivative of the moment of inertia is either negligibly small or that its average value vanishes for periodic motions and/or stationary rotation of celestial bodies. This assumption may be sufficient in case of slow-motion and weak gravitational field approximation. However, it is not true in strongly gravitating ${\mathbb N}$-body systems like coalescing binary neutron stars and/or black holes. The remaining terms in the right-hand side of \eqref{a32d58} describe gravitational coupling of the internal multipoles of body B and external multipoles of gravitational field. The dominant term, $2(\beta-1)\mathcal{M}{\cal P}$, is usually included to the Einstein-Infeld-Hoffmann force \citep[Equation 6.82]{kopeikin_2011book} and is not treated as a part of the Nordtvedt effect. The coupling terms depending on high-order multipoles in \eqref{a32d58} are fairly small in the solar system and have never been taken into account so far. Nonetheless, they become large at the latest stage of evolution of coalescing binary systems and can be used for more deep testing of scalar-tensor theory of gravity by gravitational wave detectors. 

The other components of the 4-dimensional force standing in the right-hand side of (\ref{ubpu;b}) describe gravitational interaction between the internal multipoles of body B and the external multipoles. We have,
\begin{eqnarray}
  \label{sd15a}
  F^{\mu}_{\cal Q} & = & \sum_{l=1}^{\infty}\frac{1}{l!}\bar g^{\mu\nu}{\cal Q}_{\nu\a_1...\a_l}{\cal M}^{\a_1...\a_l}
   -\sum_{l=2}^{\infty}\frac{l^2+l+4}{(l+1)!}{\cal Q}_{\a_1...\a_l}\frac{{\cal D}^2_{\rm F}{\cal M}^{\mu\a_1...\a_l}}{{\cal D}\tau^2}\\
  & & -\sum_{l=2}^{\infty}\frac{2l+1}{l+1}\frac{l^2+2l+5}{(l+1)!}\frac{{\cal D}_{\rm F}{\cal Q}_{\a_1...\a_l}}{{\cal D}\tau}\frac{{\cal D}_{\rm F}{\cal M}^{\mu\a_1...\a_l}}{{\cal D}\tau}
   -\sum_{l=2}^{\infty}\frac{2l+1}{2l+3}\frac{l^2+3l+6}{(l+1)!}{\cal M}^{\mu\a_1...\a_l}\frac{{\cal D}^2_{\rm F}{\cal Q}_{\a_1...\a_l}}{{\cal D}\tau^2}\nonumber\\
 &&+4\sum_{l=1}^{\infty}\frac{l+1}{(l+2)!}\varepsilon^{\mu\rho}{}_{\sigma}{\cal Q}_{\rho\a_1...\a_l}\frac{{\cal D}_{\rm F}{\mathcal{S}}^{\sigma\a_1...\a_l}}{{\cal D}\tau}
 +4\sum_{l=1}^{\infty}\frac{l+1}{l+2}\frac{l+1}{(l+2)!}\varepsilon^{\mu\rho}{}_{\sigma}\mathcal{S}^{\sigma\a_1...\a_l}\frac{{\cal D}_{\rm F} {\cal Q}_{\rho\a_1...\a_l}}{{\cal D}\tau}\nonumber\\
&& -\frac2{\cal M}\sum_{l=1}^{\infty}\frac{1}{l!}\varepsilon^{\mu\rho}{}_{\sigma}{\cal Q}_{\rho\a_1...\a_l}{\cal M}^{\a_1...\a_l}\frac{{\cal D}_{\rm F}{\mathcal{S}}^{\sigma}}{{\cal D}\tau}
 -\frac1{\cal M}\sum_{l=1}^{\infty}\frac{1}{l!}\varepsilon^{\mu\rho}{}_{\sigma}\mathcal{S}^{\sigma}\frac{{\cal D}_{\rm F}}{{\cal D}\tau}\bigg( {\cal Q}_{\rho\a_1...\a_l}{\cal M}^{\a_1...\a_l}\bigg)\nonumber\\ 
  \label{sd16}
  F^{\mu}_{\cal C}     &= & \sum_{l=1}^{\infty}\frac{l}{(l+1)!}\bar g^{\mu\nu}{\cal C}_{\nu\a_1...\a_l}\mathcal{S}^{\a_1...\a_l}\\
  &&-\sum_{l=1}^{\infty}\frac{1}{(l+1)!}\varepsilon^{\mu\rho}{}_\sigma\left[{\cal C}_{\rho\a_1...\a_l}\frac{{\cal D}_{\rm F}{\cal M}^{\sigma\a_1...\a_l}}{{\cal D}\tau}+\frac{l+1}{l+2}{\cal M}^{\sigma\a_1...\a_l}\frac{{\cal D}_{\rm F}{\cal C}_{\rho\a_1...\a_l}}{{\cal D}\tau} 
  \right]\nonumber\\
  \label{sd17a}
  F^\m_{\cal P}&=& 2(1-\g)\bigg[
   \sum_{l=1}^{\infty}\frac{1}{(l+1)!}{\cal P}_{\a_1...\a_l}\frac{{\cal D}^2_{\rm F}{\cal M}^{\mu\a_1...\a_l}}{{\cal D}\tau^2}\\\nonumber
   &&+\sum_{l=1}^{\infty}\frac{2l+1}{l+1}\frac{1}{(l+1)!}\frac{{\cal D}_{\rm F}{\cal P}_{\a_1...\a_l}}{{\cal D}\tau}\frac{{\cal D}_{\rm F}{\cal M}^{\mu\a_1...\a_l}}{{\cal D}\tau}\\\nonumber
   &&+\sum_{l=1}^{\infty}\frac{2l+1}{2l+3}\frac{1}{(l+1)!}{\cal M}^{\mu\a_1...\a_l}\frac{{\cal D}^2_{\rm F}{\cal P}_{\a_1...\a_l}}{{\cal D}\tau^2}\nonumber\\\nonumber
 &&-\sum_{l=0}^{\infty}\frac{l+1}{(l+2)!}\varepsilon^{\mu\rho}{}_{\sigma}{\cal P}_{\rho\a_1...\a_l}\frac{{\cal D}_{\rm F}{\mathcal{S}}^{\sigma\a_1...\a_l}}{{\cal D}\tau}\\\nonumber
 &&-\sum_{l=0}^{\infty}\frac{l+1}{l+2}\frac{l+1}{(l+2)!}\varepsilon^{\mu\rho}{}_{\sigma}\mathcal{S}^{\sigma\a_1...\a_l}\frac{{\cal D}_{\rm F} {\cal P}_{\rho\a_1...\a_l}}{{\cal D}\tau}\bigg]\nonumber
\end{eqnarray}

Time derivatives of the internal and external multipoles of body B in the local coordinates are taken at the fixed value of the spatial coordinates, $w^i=0$, that is at the origin of the local coordinates. The multipoles are STF Cartesian tensors which are orthogonal to 4-velocity of worldline ${\cal Z}$ representing the motion of the origin of the local coordinates which coincides with the center of mass of body B. This worldline is not a geodesic on the effective background manifold $\bar{M}$ but is accelerating with the local acceleration $Q_\a$. Therefore, the time derivative of the multipoles   
corresponds to the Fermi-Walker covariant derivative -- denoted as ${\cal D}_{\rm F}/{\cal D}\tau$ -- on the background manifold\index{manifold!background} taken along the direction of the 4-velocity vector ${\bar u}^\a$ with accounting for the Fermi-Walker transport \citep[Chapter 1, \S4]{syngebook}. For example, the first time derivative taken from 3-dimensional internal multipole $\dot{\cal M}^L\equiv\dot{\cal M}^{i_1i_2...i_l}$ in the local coordinates is mapped to the 4-dimensional Fermi-Walker covariant derivative as follows, 
\be\la{p2b8r3}
\dot{\cal M}^L\mapsto\frac{{\cal D}_{\rm F}{\cal M}^{\a_1\a_2...\a_l}}{{\cal D}\tau}\equiv\frac{{\cal D}{\cal M}^{\a_1\a_2...\a_l}}{{\cal D}\tau}+l{\cal Q}_\b{u}^{<\a_1} {\cal M}^{\a_2...\a_l>\b}\;,
\ee
where ${\cal D}{\cal M}^{<\a_1\a_2...\a_l>}/{\cal D}\tau\equiv {\bar u}^\beta\bar\nabla_\b{\cal M}^{<\a_1\a_2...\a_l>}$ is a standard covariant derivative of tensor ${\cal M}^{<\a_1\a_2...\a_l>}$, and ${\cal Q}^\a$ is 4-acceleration of the origin of the local coordinates. In a similar way, the second time derivative from 3-dimensional internal multipole, $\ddot{\cal M}^L\equiv\ddot{\cal M}^{i_1i_2...i_l}$, can be mapped to the 4-dimensional Fermi-Walker covariant derivative of the second order by applying the rule \eqref{p2b8r3} two times,
\ba\la{huc4}
\ddot{\cal M}^L\mapsto\frac{{\cal D}^2_{\rm F} {\cal M}^{\a_1\a_2...\a_l}}{{\cal D}\tau^2}&\equiv& \frac{{\cal D}^2 {\cal M}^{\a_1\a_2...\a_l}}{{\cal D}\tau^2}
+2l{\cal Q}_\b{u}^{<\a_1}\frac{{\cal D} {\cal M}^{\a_2...\a_l>\b}}{{\cal D}\tau}\\\nonumber
&+&l\frac{{\cal D}{\cal Q}_\b}{{\cal D}\tau}{u}^{<\a_1} {\cal M}^{\a_2...\a_l>\b}
+l{\cal Q}_\b{\cal Q}^{<\a_1} {\cal M}^{\a_2...\a_l>\b}+l^2{\cal Q}_\b{\cal Q}_\g{u}^{<\a_1}{u}^{\a_2} {\cal M}^{\a_3...\a_l>\b\g}
\;,
\ea
where ${\cal D}{\cal Q}^\a/{\cal D}\tau={\bar u}^\b\bar\nabla_\b {\cal Q}^\a$ is the covariant derivative of the 4-acceleration of the origin of the local frame taken along the direction of its 4-velocity.

Comparison of our covariant equations \eqref{ubpu;b}--\eqref{sd17a} of translational motion of the center of mass of body B with the corresponding equation \eqref{q15ms} derived by \citet{dixon_1979} will be done in Appendix \ref{ndj45xc}.

\subsection{Covariant Rotational Equations of Motion}
Covariant rotational equations of motion generalize 3-dimensional form \eqref{b3ca8k}, \eqref{spin-10} of the rotational equations for spin of body B which is a member of $\mathbb{N}$-body system, to a 4-dimensional, coordinate-independent form. Spin is a vector that is orthogonal to 4-velocity of the worldline ${\cal Z}$ of the center of mass of body B and carried out along this worldline according to the Fermi-Walker transportation rule. The covariant form of \eqref{b3ca8k} is based on the Fermi-Walker derivative, and reads 
\be\label{ui4vsr}
\frac{{\cal D}_{\rm F}{\cal S}^\mu}{{\cal D}\tau}={\cal T}^\mu\;,
\ee
or more explicitly,
\be\label{ui6s41}
\frac{{\cal D}{\cal S}^\mu}{{\cal D}\tau}={\cal T}^\mu-\left({\cal S}^\b {\cal Q}_\b\right) {\bar u}^{\mu}\;,
\ee
where the second term in the right-hand side is due to the fact that the Fermi-Walker transport is executed along the accelerated worldline ${\cal Z}$ of the center of mass of body B, the torque ${\cal T}^\mu$ is a covariant generalizations of 3-torque \eqref{b3ca8k}, and the center-of-mass condition \eqref{d9n3v5} has been implemented. We have, 
\ba\label{ac5s03v}
{\cal T}^\mu&=&
-\varepsilon^{\mu\rho}{}_\sigma\bigg[{\cal P}_\rho{\cal M}^\sigma+3\left({\cal P}_\rho-{\cal Q}_\rho\right) {\cal Q}_\b{\cal M}^{\sigma\b}
+(2\beta-\gamma-1){\cal P}\sum_{l=1}^{\infty}\frac1{l!}{\cal Q}_{\rho\a_1...\a_l}
{\cal M}^{\sigma\a_1...\a_l}\bigg]\\\nonumber
&&-\varepsilon^{\mu\rho}{}_\sigma\sum_{l=1}^{\infty}\frac1{l!}{\cal Q}_{\rho\a_1...\a_l}{\cal M}^{\sigma\a_1...\a_l} 
-\varepsilon^{\mu\rho}{}_\sigma\sum_{l=1}^{\infty}\frac{l+1}{(l+2)l!}{\cal C}_{\rho\a_1...\a_l}{\cal S}^{\sigma\a_1...\a_l}\;,
\ea
where the external multipole moments ${\cal Q}_{\a_1...\a_l}$ and ${\cal C}_{\a_1...\a_l}$ are expressed in terms of the Riemann tensor of the background manifold in accordance with equations \eqref{we69} and \eqref{we70} respectively. Acceleration $Q^\a=-F^\a/M$, where the force $F^\a$ is taken from \eqref{g6xc2c7}, and ${\cal P}_\a$ is defined in \eqref{x1675}.
It should be noticed that the terms entering the first line of the right-hand side of \eqref{ac5s03v} are present only in the scalar-tensor theory of gravity while the last two terms are the genuine general-relativistic components of the torque caused by the presence of the tidal gravitoelectric and gravitomagnetic fields respectively. 

Comparison of our equation \eqref{ui6s41} for evolution of spin of body B with the corresponding equation \eqref{q16mc} derived by \citet{dixon_1979} will be done in Appendix \ref{iopn3e4}.

\section*{Acknowledgments}
This work started off about two years ago during my stay as a visiting scholar at Shanghai Astronomical Observatory which generous support and hospitality are highly appreciated. A part of this work was reported at 656-th WE-Heraeus-Seminar on "Fundamental Physics in Space" in Bremen, Germany and is available online \url{https://www.zarm.uni-bremen.de/fps2017/pdf/Vortraege/Kopeikin_Covariant_Equations_of_Motion.pdf}. I am grateful to Dr. M. List and Dr. C. L\"ammerzahl for inviting me to the Seminar and for providing travel support. I am thankful to W.G. Dixon, \'Eanna Flanagan, Dirk Puetzfeld and Yuri N. Obukhov for valuable conversations and clarification of some subtle issues of their own approaches to the problem of derivation of equations of motion. Critical remarks and constructive comments of an anonymous referee were instrumental for significant improvement of the manuscript and reconciliation with the results of the present paper with the translational equations of motion obtained by \citet{racine_2005PhRvD,racine2013PhRvD}.   
\newpage

\appendix
\section{Auxiliary Mathematical Properties of STF Tensors}\label{kjn34c}

Definition of the symmetric trace-free (STF) Cartesian tensor was introduced by Pirani \citep{Pirani1964} and is given in equation \ref{stfformula} of the present paper. Here, we provide the reader with a number of auxiliary algebraic and differential identities involving STF tensors that were instrumental for doing our computations. 

Perhaps one of the most important algebraic identities of STF tensors is the {\it index-peeling} formula  that helps to separate a single index from the rest of other STF indices in STF tensor. Let us demonstrate how this formula is applied in case of a product of vector with STF tensor. We denote two STF tensors as $T_L\equiv T_{<L>}$ and $R_L\equiv R_{<L>}$, and let $V_i$ be an arbitrary covector. The index-peeling formula reads \citep[Equation 2.14]{di} 
\ba\label{hh33e5z}
V_{<i}T_{L>}&=&\frac1{l+1}V_i T_L+\frac{l}{l+1}T_{i<L-1}V_{i_l>}-\frac{2l}{(l+1)(2l+1)}V_kT_{k<L-1}\delta_{i_l>i}\;.
\ea
The index-peeling formula can be applied to two or more indices by successive iterations.

The index-peeling formula \eqref{hh33e5z} is directly extended from covector $V_i$ to tensors. For example, by replacing $V_i\mapsto \d_{ij}$ in \eqref{hh33e5z}, and reducing similar terms we can get the following identities \citep{racine_2005PhRvD}
\ba\label{jj33xx22}
T_{i<L}\d_{j>j}&=&\frac{2l+3}{2l+1}T_{iL}\;,\\\label{vxc3ed}
T_{j<L}\d_{j>i}&=&\frac{1}{(l+1)(2l+1)}T_{iL}\;.
\ea
Replacing $V_i\mapsto R_{Li}$ in \eqref{hh33e5z} yields\citep[Equation 4.26]{dsx2}
\be\label{ggwwcc}
R_{L<i}T_{L>}=\frac{1}{(l+1)(2l+1)}R_{iL}T_L\;.
\ee
Two other useful formulas are for a product of the unit vectors $n^i=x^i/r$, where $r=\left(\d_{ij}x^ix^j\right)^{1/2}$. They are \citep[Equations A22a and A23]{bld1986}
\ba\label{yy33d1}
n^{<iL>}&=&n^in^{<L>}-\frac{l}{2l+1}\d^{i<i_l}n^{L-1>}\;,\\
n^in^{<iL>}&=&\frac{l+1}{2l+1}n^{<L>}\;.
\ea
Differential identities for the STF partial derivatives from the radial distance $r$ are \citep[Equations A32 and A34]{bld1986}
\ba\label{zz33aa22}
\pd_{<L>}r^{-1}&=&\pd_L r^{-1}=(-1)^l(2l-1)!!\frac{n^{<L>}}{r^{l+1}}\;,\\\label{mumu45x}
\pd_{<L>}r&=&(-1)^{l+1}(2l-3)!!\frac{n^{<L>}}{r^{l-1}}\;.
\ea
A partial spatial derivative from STF tensor $n^{<L>}$ is \citep[Equation A24]{bld1986}
\be
r\pd_in^{<L>}=(l+1)n^in^{<L>}-(2l+1)n^{<iL>}\;.
\ee
Other useful algebraic and differential identities for STF tensors are given in papers \citep{thor,bld1986,di,dsx2,racine_2005PhRvD,Dixon_2013}.

\section{Comparison with the Racine-Vines-Flanagan Equations of Motion}\label{appendixA}
Translational equations of motion for arbitrary structured bodies have been derived by \citet{racine_2005PhRvD} with a corrigendum published in  \citep{racine2013PhRvD}. Definitions of the internal multipoles of body B in those papers are the same as in the present paper. The Racine-Vines-Flanagan (RVF) equations of motion are given in \citep[equations 6.11--6.16]{racine_2005PhRvD} and, besides directly computed terms, contain four terms depending on STF Cartesian tensor function $\hat P^{\rm C}_K=\hat P^{\rm C}_{<K>}$ \citep[equation 6.16]{racine_2005PhRvD} which is \footnote{Notice that we use indices B and C to label the bodies of $\mathbb{N}$-body system while \citet{racine_2005PhRvD} use an index B instead of C, and an index A instead of B. We prefer to use our index notations to facilitate the comparison of the equations of motion. Relabeling the RVF equations is achieved with the simple replacements of the body's indices: $\B\rightarrow{\rm C}$ and ${\rm A}\rightarrow\B$.} 
\be\label{h6dv4az}
\hat P_{\rm C}^K\equiv \ddot{\cal M}^K_{\rm C}+2k\dot{\cal M}^{<K-1}_{\rm C} v_{\rm C}^{i_k>}+k(k-1){\cal M}^{<K-2}_{\rm C} v^{i_{k-1}}_{\rm C} v^{i_k>}_{\rm C}\;.
\ee 
Function $\hat P_{\rm C}^K$ enters equations (6.13a), (6.13b), and (6.13g) in \citep{racine_2005PhRvD}. The terms with $\hat P_{\rm C}^K$ must be developed explicitly in order to combine it similar terms in other parts of the RVF equations of motion. 

It is more convenient to develop the products of $\hat P_{\rm C}^K$ with the STF combinations of a unit vector, $n^i_{\rm CB}=R^i_{\rm CB}/R_{\rm CB}$, where $R^i_{\rm CB}=x^i_{\rm C}-x^i_{\rm B}$ is the coordinate distance between centers of mass of bodies B and C. The RVF equations of motion depend on four such combinations which have not been shown in \citep{racine_2005PhRvD,racine2013PhRvD} so that we present them explicitly. Two of them are products, $n^{<iKL>}_{\rm CB}{\cal M}_{\rm B}^{jL}\hat P_{\rm C}^{jK}$ and $n^{<KL>}_{\rm CB}{\cal M}_{\rm B}^{L}\hat P_{\rm C}^{iK}$, which appear in the first and second terms in the right-hand side of equation (6.12a) in \citep{racine_2005PhRvD}. In order to compute these terms we successively apply the index-peeling formula \eqref{hh33e5z} two times to separate the index of velocity of body B in $\hat P_{\rm C}^{iK}$ from the STF multi-indices and, then, render contraction of the multi-indices. It yields,  
\ba \label{gg35s}\hspace{-2cm}
n^{<iKL>}_{\rm CB}{\cal M}_{\rm B}^{jL}\hat P_{\rm C}^{jK} &=&n^{<iKL>}_{\rm CB}{\cal M}_{\rm B}^{jL}\left(\ddot{\cal M}_{\rm C}^{jK}+2v^j_{\rm C}\dot{\cal M}_{\rm C}^{K}\right)+2kn^{<ijLK-1>}_{\rm CB}\\\nonumber
&\times&\left[v^j_{\rm C}{\cal M}_{\rm B}^{pL}\left(\dot{\cal M}_{\rm C}^{pK-1}+ v^p_{\rm C}{\cal M}_{\rm C}^{K-1}\right)-\frac1{2k+1}\left(2v^p_{\rm C}\dot{\cal M}_{\rm C}^{pK-1}+v^2_{\rm C}{\cal M}_{\rm C}^{K-1}\right){\cal M}_{\rm B}^{jL}\right]\\\nonumber
&+&k(k-1)n^{<ijpLK-2>}_{\rm CB} v^p_{\rm C}{\cal M}_{\rm C}^{qK-2}\left(v^j_{\rm C}{\cal M}_{\rm B}^{qL}-\frac{4}{2k+1} v^q_{\rm C}{\cal M}_{\rm B}^{jL}\right)\;,\\
\label{ve5x3q}
n^{<KL>}_{\rm CB}{\cal M}_{\rm B}^{L}\hat P_{\rm C}^{iK} &=&n^{<KL>}_{\rm CB}{\cal M}_{\rm B}^{L}\left(\ddot{\cal M}_{\rm C}^{iK}+2v^i_{\rm C}\dot{\cal M}_{\rm C}^{K}\right)+2k{\cal M}_{\rm B}^{L}\\\nonumber
&\times&\left[n^{<jLK-1>}_{\rm CB}v^j_{\rm C}\left(\dot{\cal M}_{\rm C}^{pK-1}+ v^i_{\rm C}{\cal M}_{\rm C}^{K-1}\right)-\frac{1}{2k+1}n^{<iLK-1>}_{\rm CB}\left(2v^p_{\rm C}\dot{\cal M}_{\rm C}^{pK-1}+v^2_{\rm C}{\cal M}_{\rm C}^{K-1}\right)\right]\\\nonumber
&+&k(k-1) v^j_{\rm C} v^p_{\rm C}{\cal M}_{\rm B}^{L}\left(n^{<jpLK-2>}_{\rm CB}{\cal M}_{\rm C}^{iK-2}-\frac4{2k+1}n^{<ijLK-2>}_{\rm CB}{\cal M}_{\rm C}^{pK-2}\right)\;.
\ea
There are two other terms in the RVF equations of motion which contain combinations, $n^{<KL>}_{\rm CB}{\cal M}_{\rm B}^{iL}\hat P_{\rm C}^{K}$ and $n^{<iKL>}_{\rm CB}{\cal M}_{\rm B}^{L}\hat P_{\rm C}^{K}$, in the second and seventh terms of the right-hand side of equation (6.12a) in \citep{racine_2005PhRvD}. These terms are easy to deal with. Straightforward application of \eqref{h6dv4az} and contraction of multi-indices yield,
\ba \label{nnxx55za}
n^{<KL>}_{\rm CB}{\cal M}_{\rm B}^{iL}\hat P_{\rm C}^{K}&=&{\cal M}_{\rm B}^{iL}\left[n^{<KL>}_{\rm CB}\ddot{\cal M}^K_{\rm C}+2kn^{<jLK-1>}_{\rm CB}\dot{\cal M}^{K-1}_{\rm C} v_{\rm C}^{j}+k(k-1)n^{<jpLK-2>}_{\rm CB}{\cal M}^{K-2}_{\rm C} v^{j}_{\rm C} v^{p}_{\rm C}\right] \;,\\\label{v45xzc2}
n^{<iKL>}_{\rm CB}{\cal M}_{\rm B}^{L}\hat P_{\rm C}^{K}&=&{\cal M}_{\rm B}^{L}\left[ n^{<iKL>}_{\rm CB}\ddot{\cal M}^K_{\rm C}+2kn^{<ijLK-1>}_{\rm CB}\dot{\cal M}^{K-1}_{\rm C} v_{\rm C}^{j}+k(k-1)n^{<ijpLK-2>}_{\rm CB}{\cal M}^{K-2}_{\rm C} v^{j}_{\rm C} v^{p}_{\rm C} \right]\;.
\ea

Substituting \eqref{gg35s}--\eqref{v45xzc2} to equations (6.13a), (6.13b), (6.13g) of the paper \citep{racine_2005PhRvD}, and making use of \eqref{acser243}, \eqref{nrvx31w} from the present paper in the inverse order, allow us to write down the RVF equations of motion given in \citep[equation 6.11]{racine_2005PhRvD} with typos fixed in \citep{racine2013PhRvD}, as follows
\be\label{h7dc35x1}
M_B a^i_B={\mathfrak F}^i_{\rm N}+{\mathfrak F}^i_{\rm pN}\;,
\ee
where $M_\B$ is the inertial (relativistic mass) of body B, $a^i_\B=d^2x^i_\B/dt^2$ is acceleration of the center of mass of body B, ${\mathfrak F}^i_{\rm N}$ is the Newtonian force, and ${\mathfrak F}^i_{\rm pN}$ is the post-Newtonian force. After taking into account our equations \eqref{gg35s}--\eqref{v45xzc2} the RVF forces can be written down similar to our equations \eqref{w1q5} and \eqref{eee3s} in the form of the partial derivative operator,
\ba\label{old5c2m}
{\mathfrak F}^i_{\rm N}&=&\sum_{C\neq B}\sum_{l=0}^{\infty}\sum_{n=0}^{\infty}\frac{(-1)^n}{l!n!}{\cal M}_{\rm B}^{L}(\tau_{\rm B}){\cal M}_{\rm C}^{N}(\tau_{\rm C})\pd_{iLN}R_{\B{\rm C}}^{-1}\;,\\
\la{RVFeee3s}
{\mathfrak F}^i_{\rm pN}&=&\frac12\sum_{{\rm C}\not=\B}\sum_{l=0}^\infty\sum_{n=0}^{\infty}\frac{(-1)^n}{l!n!}{\cal M}_\B^L\Big[\ddot{{\cal M}}_{\rm C}^N\pd_{<iLN>}-\left(2\dot{\cal M}^N_{{\rm C}}v^p_{{\rm C}}+ {{\cal M}}_{\rm C}^Na_{\rm C}^p\right)\pd_{<ipLN>}+{{\cal M}}_{\rm C}^Nv_{\rm C}^pv_{\rm C}^q\pd_{<ipqLN>}\Big]R_{\rm BC}\\\nonumber
&&+
\sum_{{\rm C}\not=\B}\sum_{l=0}^{\infty}\sum_{n=0}^{\infty}\frac{(-1)^n}{l!n!}\bigg[\Big(\a^{iLN}_{\rm RVF}+\beta^{iLN}_{\rm RVF}\Big)\pd_{<LN>}+\Big(\a^{ipLN}_{\rm RVF}+\beta^{ipLN}_{\rm RVF}\Big)\pd_{<pLN>}+\a^{ipqLN}_{\rm RVF}\pd_{<pqLN>}\\\nonumber
&&+\Big(\a^{LN}_{\rm RVF}+\beta^{LN}_{\rm RVF}+\gamma^{LN}_{\rm RVF}\Big)\pd_{<iLN>}+\Big(\mu^{pLN}_{\rm RVF}+\nu^{pLN}_{\rm RVF}+\rho^{pLN}_{\rm RVF}\Big)\pd_{<ipLN>}+\sigma^{pqLN}_{\rm RVF}\pd_{<ipqLN>}\Big]R_{\B{\rm C}}^{-1}\\\nonumber
&&+3\left(a_\B^k\ddot{\cal M}_{\B}^{ik}+2\dot{a}_\B^k\dot{\cal M}_{\B}^{ik}+\ddot{a}_\B^k{\cal M}_{\B}^{ik}\right)
\;,
\ea
where all partial derivatives are understood in the sense of equations \eqref{acser243}, \eqref{nrvx31w}. We have explicitly indicated the time arguments of the multipoles in the expression for the Newtonian force \eqref{old5c2m} which, according to \citep[Eq. 5.9]{racine_2005PhRvD}, are the proper times of the bodies taken on their world lines at the points of intersection with hypersurface ${\cal H}_t$ of constant coordinate time $t$ of the global coordinate chart (c.f. \eqref{jk9wss5f}--\eqref{acz521sa}),
\ba
\label{0d5c2xb}
\tau_{\rm B}&=&u_{\rm B}|_{{\bm x}={\bm x}_{\rm B}}=t+\frac1{c^2}{\cal A}_{\rm B}(t)+{\cal O}\left(\frac1{c^4}\right)\;,\\
\label{ze18va4}
\tau_{\rm C}&=&u_{\rm C}|_{{\bm x}={\bm x}_{\rm C}}=t+\frac1{c^2}{\cal A}_{\rm C}(t)+{\cal O}\left(\frac1{c^4}\right)\;,
\ea
where time dilation functions ${\cal A}_{\rm B}$ and ${\cal A}_{\rm C}$ are defined by solutions of the ordinary differential equations \eqref{byc29x4} and
\eqref{moeb6732c}.
 
The coefficients in the RVF post-Newtonian force \eqref{RVFeee3s} can be directly compared to those in our equation \eqref{eee3s} where we have to take $\b=\g=1$ in order to bring it to general-relativistic form. The comparison is tedious but rather straightforward. It results in  
\ba
\a^{iLN}_{\rm RVF}&=&\a^{iLN}_{\rm F}-\frac2{2l+2n+3}v^i_{\rm C}{\cal M}^{L}_\B\dot{\cal M}^{N}_{\rm C}\;,\\
\a^{ipLN}_{\rm RVF}&=&\a^{ipLN}_{\rm F}+\frac2{2l+2n+5}v^p_{\rm C}{\cal M}^{L}_\B\dot{\cal M}^{iN}_{\rm C}
+\left(\frac{2}{2l+3}-\frac{2}{2l+2n+5}\right)v^p_{\rm C}{\cal M}^{iL}_\B\dot{\cal M}^{N}_{\rm C}\;,\\
\a^{LN}_{\rm RVF}&=& \a^{LN}_{\rm F}-\frac{2}{2l+2n+5} v^k_{\rm C}{\cal M}^{kL}_\B\dot{\cal M}^N_{\rm C}-\frac{2l+2n+3}{2l+2n+5}v^k_{\rm C}{\cal M}^{L}_\B\dot{\cal M}^{kN}_{\rm C}\;,\\
\mu^{pLN}_{\rm RVF}&=&\mu^{pLN}_{\rm F}+\frac2{2l+2n+7}v^p_{\rm C}{\cal M}^{kL}_\B\dot{\cal M}^{kN}_{\rm C} \;,
\ea
and all other remaining coefficients in \eqref{RVFeee3s} and \eqref{eee3s} are identical for $\b=\g=1$, except for $\rho^{pLN}_{\rm RVF}=0$. The reason for vanishing $\rho^{pLN}_{\rm RVF}$ is that the local coordinate system adapted to body B has been chosen by \citet{racine_2005PhRvD} as {\it kinematically} non-rotating with respect to the spatial axes of the global coordinates while we operate with {\it dynamically} non-rotating local frame of body B. Kinematically non-rotating local frame is not carried out along the worldline of the body's center of mass in accordance with the Fermi-Walker transportation rule. It means that particles of matter moving with respect to the body must experience the centrifugal and Coriolis forces in this frame. These forces become sufficiently large at the latest stages of evolution of inspiralling compact binaries and affect computation of templates of gravitational waveforms. This effect is, however, purely coordinate-dependent and can be removed by choosing {\it dynamically} non-rotating local frame adapted to body B which is our choice.

Now, we notice a useful formula
\be
\label{yb3ca2qq}
\sum_{l=0}^\infty\sum_{n=0}^\infty\frac{(-1)^n}{l!n!}{\cal M}^L_\B\dot{\cal M}^N_{{\rm C}}v^p_{{\rm C}}R^p_{{\rm C}}\pd_{iLN}\bigg(\frac1{R_{\rm C}}\bigg) =
\sum_{l=0}^\infty\sum_{n=0}^\infty\frac{(-1)^n}{l!n!}{\cal M}^L_\B\left[\dot{\cal M}^N_{{\rm C}}v^p_{{\rm C}}\pd_{<iL>pN}R_{{\rm C}}
+v^p_{{\rm C}}\dot{\cal M}^{pN}_{{\rm C}}\pd_{iLN}\bigg(\frac1{R_{\rm C}}\bigg)\right]\;,
\ee
which expansion in terms of the STF derivatives is as follows,
\ba
\label{yb3ca2}
\sum_{l=0}^\infty\sum_{n=0}^\infty\frac{(-1)^n}{l!n!}{\cal M}^L_\B\dot{\cal M}^N_{{\rm C}}v^p_{{\rm C}}R^p_{{\rm C}}\pd_{iLN}\bigg(\frac1{R_{\rm C}}\bigg)
&=&\sum_{l=0}^\infty\sum_{n=0}^\infty\frac{(-1)^n}{l!n!}\bigg\{{\cal M}^L_{\B}\dot{\cal M}^N_{{\rm C}}v^p_{{\rm C}}\pd_{<ipLN>}R_{{\rm C}}\\\nonumber
&&+\frac{2}{2l+2n+3}v^i_{{\rm C}}{\cal M}^{L}_{\B}\dot{\cal M}^N_{{\rm C}}\pd_{<LN>}\bigg(\frac1{R_{{\rm C}}}\bigg)\\\nonumber
&&+\left(\frac{2}{2l+2n+5}-\frac{2}{2l+3}\right){\cal M}^{iL}_{\B}\dot{\cal M}^N_{{\rm C}}v^p_{{\rm C}}\pd_{<pLN>}\bigg(\frac1{R_{{\rm C}}}\bigg)\\\nonumber
&&-\frac{2}{2l+2n+5}{\cal M}^{L}_{\B}\dot{\cal M}^{iN}_{{\rm C}}v^p_{{\rm C}}\pd_{<pLN>}\bigg(\frac1{R_{{\rm C}}}\bigg)\\\nonumber
&&+\left(\frac{2}{2l+2n+5}v^p_{{\rm C}}{\cal M}^{pL}_{\B}\dot{\cal M}^N_{{\rm C}}+v^p_{{\rm C}}\dot{\cal M}^{pN}_{{\rm C}}\right)\pd_{<iLN>}\bigg(\frac1{R_{{\rm C}}}\bigg)\\\nonumber&&-\frac{2}{2l+2n+5}v^p_{{\rm C}}\dot{\cal M}^{pN}_{{\rm C}}{\cal M}^L_{\B}\pd_{<iLN>}\bigg(\frac1{R_{{\rm C}}}\bigg)\\\nonumber
&&-
\frac{2}{2l+2n+7}{\cal M}^{qL}_{\B}\dot{\cal M}^{qN}_{{\rm C}}v^p_{{\rm C}}\pd_{<ipLN>}\bigg(\frac1{R_{{\rm C}}}\bigg)\bigg\}\;.
\ea
Derivation of \eqref{yb3ca2} is based on application of \eqref{m8d5v0} and transformation  \eqref{mw9n7} where replacements, $a^i_{\rm C}\rightarrow v^i_{\rm C}$ and ${\cal M}^L_{\rm C}\rightarrow \dot{\cal M}^L_{\rm C}$ must be done in all terms. Employing \eqref{yb3ca2} in \eqref{RVFeee3s} we find out that the RVF post-Newtonian force ${\mathfrak F}^i_{\rm pN}$ relates to our post-Newtonian force \eqref{eee3s} in a fairly simple way,
\ba\label{v39b2x3}
{\mathfrak F}^i_{\rm pN}&=&F^i_{\rm pN}+3\left(a_\B^k\ddot{\cal M}_{\B}^{ik}+2\dot{a}_\B^k\dot{\cal M}_{\B}^{ik}+\ddot{a}_\B^k{\cal M}_{\B}^{ik}\right)\\\nonumber
&-&\sum_{n=0}^\infty\frac{(-1)^n}{l!n!}\rho^{pLN}_{\rm F}\pd_{ipLN}\bigg(\frac1{R_{\B{\rm C}}}\bigg)
-\sum_{{\rm C}\not=\B}\sum_{l=0}^\infty\sum_{n=0}^\infty\frac{(-1)^n}{l!n!}{\cal M}^L_\B\dot{\cal M}^N_{{\rm C}}v^p_{{\rm C}}R^p_{\B{\rm C}}\pd_{iLN}\bigg(\frac1{R_{\B{\rm C}}}\bigg)\;.
\ea
The first three acceleration-dependent terms in the right-hand side of \eqref{v39b2x3} following $F^i_{\rm pN}$, are identical to those in our equation \eqref{c5s0k2f}. Hence, these terms are due to the different choice of the center of mass of body B in \citep{racine_2005PhRvD} corresponding to the complementary function, ${\cal I}^i_{\rm c}=0$, in the definition of the center of mass of body B as compared to the choice adopted for this function in equation \eqref{m4g1x8p} of the present paper. The next term in the right-hand side of \eqref{v39b2x3} depends on coefficient $\rho^{pLN}_{\rm F}$ given in \eqref{zw3k4b}. This coefficient defines the relativistic transport of the multipoles adapted to body B, along the worldline of the body's center of mass. Our convention is that the local frame is carried out in accordance with the Fermi-Walker transportation law while \citet[section 5F]{racine_2005PhRvD} decided to make the local frame non-rotating with respect to the spatial axes of the global coordinates. This difference is a matter of choosing either {\it kinematical} or {\it dynamical} definition of the rotation of the body-adapted local frame and is easy to reconcile. 

The very last term in the right-hand side of \eqref{v39b2x3} is due to the different time arguments of the multipoles ${\cal M}_C$ taken at slightly different points on the worldline of body C. Indeed, by comparing \eqref{jk9wss5f} with \eqref{0d5c2xb} and \eqref{acz521sa} with \eqref{ze18va4}, we conclude that the time arguments of the multipoles ${\cal M}_{\rm B}$ of body B are identical, $\tau_{\rm B}=u^*_{\rm B}$, while the time arguments of multipoles of body C are shifted one with respect to another, $\tau_{\rm C}=u^*_{\rm C}+v^k_{\rm C}(t)R^k_{\rm BC}$. Looking back to Fig. \ref{fig1} we can say that the multipoles ${\cal M}_{\rm B}$ of body B are taken at point P while the multipoles ${\cal M}_C$ of body C are taken at point R in our approach and at the point Q in the paper by \citet{racine_2005PhRvD}. This observation allows us to connect the RVF Newtonian force \eqref{old5c2m} with our Newtonian force \eqref{w1q5} by taking the Taylor expansion of the multipoles ${\cal M}_{\rm C}$. It yields
\be\label{ked9v2x0}
{\mathfrak F}^i_{\rm N}=F^i_{\rm N}+\sum_{{\rm C}\not=\B}\sum_{l=0}^\infty\sum_{n=0}^\infty\frac{(-1)^n}{l!n!}{\cal M}^L_\B\dot{\cal M}^N_{{\rm C}}v^p_{{\rm C}}R^p_{\B{\rm C}}\pd_{iLN}\bigg(\frac1{R_{\B{\rm C}}}\bigg)\;.
\ee
The last term in the right-hand side of \eqref{ked9v2x0} exactly cancels the very last term in \eqref{v39b2x3} after substituting \eqref{v39b2x3} and \eqref{ked9v2x0} to the total force in the right-hand side of \eqref{h7dc35x1}. This makes it clear that our translational equations of motion are essentially the same as those derived by \citet{racine_2005PhRvD,racine2013PhRvD} except of several terms which are a matter of slightly different conventions adopted to define the center of mass of the bodies and rotation of the spatial axes of the body-adapted local frame. It is remarkable that the agreement is achieved in spite of using different mathematical technique based on the Fock-Papapetrou-Chandrasekhar approach \citep{fockbook,pap1,pap2,1969ApJ...158...55C,Chandra_1970ApJ} to the derivation of equations of motion of extended bodies in ${\mathbb N}$-body system made of matter with continuous stress-energy tensor. Finally, we bring attention of the reader to the fact that our equations of translational motion are more economic than that given in \citep{racine_2005PhRvD,racine2013PhRvD} in the sense that the post-Newtonian force $F^i_{\rm pN}$ in our approach have been reduced to the form \eqref{pNf5s} containing lesser number of terms than the corresponding force ${\mathfrak F}^i_{\rm pN}$ in \citep{racine_2005PhRvD,racine2013PhRvD}. It might be more effective to implement our form of the equations of motion with quadrupole and higher-order multipoles to the numerical integration of the orbital evolution of tidally-deformed neutron star binaries and prediction of gravitational wave signal from the mergers -- see, for example, \citep{Bini_PRD124034,Vines_Flanagan_PRD024046,Steinhoff_etal_PRD104028}.

\section{The Dixon multipole moments}\label{appndxon}

\citet{dixon_1979} has defined internal multipoles of an extended body B in the normal Riemann coordinates, $X^\a$, by means of a tensor integral \eqref{q12m}
\be\label{q12ttt} 
I^{\a_1...\a_l\m\n}(z)=\int X^{\a_1}...X^{\a_l}\hat T^{\m\n}(z,X)\sqrt{-\bar g(z)}DX\;,\qquad\qquad (\;l\ge 2\;)
\ee
where $\hat T^{\m\n}$ is the stress-energy {\it skeleton} of the body, the integration is performed over the tangent 4-dimensional space to background manifold $\bar{M}$ at point $z$ taken on a reference worldline ${\cal Z}$, and the volume element of integration $DX=dX^0\wedge dX^1\wedge dX^2\wedge dX^3$. The reason for appearance of the {skeleton} $\hat T^{\m\n}$ in \eqref{q12ttt} instead of the regular stress-energy tensor $T^{\m\n}$ was to incorporate the self-field effects of gravitational field of the body to the definition of the higher-order multipoles \footnote{The influence of the self-field effects on multipoles was studied by \citet{thor} and \citet{bld,dyr2} with different techniques.}. According to \citep{dixon_1979}, the skeleton $\hat T^{\m\n}(z,x)$ is a distribution \citep{shilov_1968} defined on the worldline ${\cal Z}$\index{generalized function} in such a way that it contains a complete information about the body but is entirely independent of the geometry of the surrounding spacetime to which the body is embedded. The skeleton is lying on the hyperplane made out of vectors $X^\a$ which are orthogonal to the vector of dynamic velocity ${\mathfrak n}^\a$. It gives the following constraint, 
\be\label{2x4a3z11} 
({\mathfrak n}_\a X^\a)X^{[\l}\hat T^{\m][\n}X^{\s]}=0\;,
\ee
which points out that the skeleton distribution is concentrated on the hyperplane ${\mathfrak n}_\a X^\a=0$.

Definition (\ref{q12ttt}) suggests that the Dixon multipole moments have the following symmetries,
\be\la{o39v3j}
I^{\a_1...\a_l\m\n}=I^{(\a_1...\a_l)(\m\n)}\;,
\ee
where the round parentheses around the tensor indices denote a full symmetrization\index{symmetrization}. In addition to \eqref{o39v3j} there are more symmetries of the Dixon multipoles due to the one-to-one mapping of the microscopic equation of motion \eqref{wk10} to a similar equation for the stress-energy {skeleton} \citep{dixon_1979} 
\be\label{on3c45}
\bar\nabla_{\n} \hat T^{\m\n}(z,X)=0\;. 
\ee
Multiplying \eqref{on3c45} with $X^{\a_1}...X^{\a_l}X^{\a_{l+1}}$, integrating over 4-dimensional volume and taking into account that $\hat T^{\m\n}$ vanishes outside hyperplane ${\mathfrak n}_\a X^\a=0$, yields \citep[Equation 143]{dixon_1979},
\be\label{gr7xews}
I^{(\a_1...\a_l\m)\n}=0\;,
\ee
and a similar relation holds after exchanging indices $\m$ and $\n$ due to symmetry \eqref{o39v3j}.
The number of algebraically independent components of $I^{\a_1...\a_l\m\n}$ obeying \eqref{o39v3j} is $N_1(l)=C^{l+3}_3\times C^5_3$ where $C^p_q=\frac{p!}{q!(p-q)!}$ is a binomial coefficient. Constraints \eqref{gr7xews} reduce the number of the algebraically independent components of the multipoles $I^{\a_1...\a_l\m\n}$ by $N_2(l)=C^{l+4}_3\times C^4_3$ making the number of linearly independent components of $I^{\a_1...\a_l\m\n}$ equal to
$N_3(l)=N_1(l)-N_2(l)=(l+3)(l+2)(l-1)$.

The multipoles $I^{\a_1...\a_l\m\n}$ are coupled to the Riemann tensor\index{Riemann tensor} $\bar R^\a{}_{\m\b\n}$ characterizing the curvature\index{curvature} of the effective background spacetime. Therefore, they can be replaced with a more suitable set of {\it reduced} moments $J^{\a_1...\a_l\l\m\n\r}$ which are defined by the following formulas \citep{dixon_1979,dixon_1973GReGr}
\be\la{wk12dcr}
J^{\a_1...\a_p\lambda\m\s\n}\equiv I^{\a_1...\a_p[\lambda[\s\m]\n]}\;,
\ee
where the square parentheses around the tensor indices denote a full anti-symmetrization\index{anti-symmetrization}, and the nested square brackets in (\ref{wk12dcr}) denote the anti-simmetrization on pairs of indices $[\l,\m]$ and $[\n,\r]$ independently. Definition \eqref{wk12dcr} tells us that
tensor $J^{\a_1...\a_p\lambda\m\s\n}$ is fully symmetric with respect to the first $p$ indices and is skew-symmetric with respect to the pairs of indices $\lambda,\mu$ and $\s,\n$,
\be
J^{\a_1...\a_p\lambda\m\s\n}=J^{(\a_1...\a_p)[\lambda\m][\s\n]}\;.
\ee  
Among other properties of $J^{\a_1...\a_p\lambda\m\s\n}$ we have
\be
J^{\a_1...\a_p\l[\m\s\n]}=0\;,\qquad\qquad
J^{\a_1...[\a_p\l\m]\s\n}=0\;,
\ee
which are consequences of the definition \eqref{wk12dcr}, and 
\be \label{kerxv34}
{\mathfrak n}_{\a_1} J^{\a_1...\a_p\l\m\s\n}=0\;,
\ee
that is the condition of orthogonality following from the constraint \eqref{2x4a3z11}.

Equation (\ref{wk12dcr}) can be transformed to another form. For this we write down the anti-symmetric part of \eqref{wk12dcr} explicitly as a combination of four terms, change notations of indices $\{\a_1...\a_p\m\n\}\rightarrow\{\a_1...\a_{l-2}\a_{l-1}\a_l\}$, and make a full symmetrization with respect to the set of indices $\{\a_1...\a_l\}$. It gives,
\be\label{bwc3aj}
J^{(\a_1...\a_{l-1}|\m|\a_l)\n}=\frac14\left[I^{(\a_1...\a_{l-1}\a_l)\m\n}-I^{(\a_1...\a_{l-2}|\m|\a_{l-1}\a_l)\n}-I^{(\a_1...\a_{l-2}\a_{l-1}|\n\m|\a_l)}+I^{(\a_1...\a_{l-2}|\m\n|\a_{l-1}\a_l)}\right]\;,
\ee
where the indices enclosed to vertical bars are excluded from symmetrization. Remembering that each of the $I$ moments is separately symmetric with respect to the first $l$ and the last two indices we can be recast \eqref{bwc3aj} to the following form,
\be
J^{(\a_1...\a_{l-1}|\m|\a_l)\n}=\frac14\left[I^{(\a_1...\a_{l-1}\a_l)\m\n}-I^{(\m(\a_1...\a_{l-1})\a_l)\n}-I^{(\n(\a_1...\a_{l-1}\a_l)\m}+I^{(\m\n(\a_1...\a_{l-2})\a_{l-1}\a_l)}\right]\;.
\ee
We now use the constrain \eqref{gr7xews} and notice that
\be
I^{(\a_1...\a_{l-1}\a_l\m)\n}=\frac{1}{l+1}\left[I^{\a_1...\a_{l-1}\a_l\m\n}+lI^{(\m(\a_1...\a_{l-1})\a_l)\n}\right]=0\;,
\ee
which gives
\be\label{m4bvz}
I^{(\m(\a_1...\a_{l-1})\a_l)\n}=-\frac{1}{l} I^{\a_1...\a_{l-1}\a_l\m\n}\;,
\ee
and, because of the symmetry with respect to indices $\m$ and $\n$,
\be\label{h3fs5a}
I^{(\n(\a_1...\a_{l-1})\a_l)\m}=-\frac{1}{l} I^{\a_1...\a_{l-1}\a_l\m\n}\;.
\ee
We also have
\ba
I^{(\a_1...\a_{l-1}\a_l\m\n)}&=&\frac{2!l!}{(l+2)!}\\\nonumber
&\times&\left[I^{\a_1...\a_{l-1}\a_l\m\n}+l I^{(\m(\a_1...\a_{l-1})\a_l)\n}+l I^{(\n(\a_1...\a_{l-1})\a_l)\m}+\frac{l(l-1)}{2}I^{(\m\n(\a_1...\a_{l-2})\a_{l-1}\a_l)}\right]=0\;,
\ea
which yields
\be\label{jerc2}
I^{(\m\n(\a_1...\a_{l-2})\a_{l-1}\a_l)}=\frac{2}{l(l-1)}I^{\a_1...\a_{l-1}\a_l\m\n}\;.
\ee
Replacing \eqref{m4bvz}, \eqref{h3fs5a} and \eqref{jerc2} to \eqref{bwc3aj} yields
\be\label{x8d4ak} 
J^{(\a_1...\a_{l-1}|\m|\a_l)\n}=\frac14\frac{l+1}{l-1}I^{\a_1...\a_l\m\n}\;,
\ee
that shows the algebraic equivalence between the symmetrised $J^{(\a_1...\a_{l-1}|\m|\a_l)\n}$ and $I^{\a_1...\a_l\m\n}$ multipole moments for $l\ge 2$. Due to the orthogonality condition \eqref{kerxv34} we conclude that 
\be \label{juevz5}
{\mathfrak n}_{\a_1}I^{\a_1...\a_l\m\n}=0\;,
\ee
for the first $l$ indices of $I^{\a_1...\a_l\m\n}$. The number of these conditions is the same as the number of components of tensor $I^{\a_1...\a_{l-1}\m\n}$ that is $N_3(l-1)=(l+2)(l+1)(l-2)$. It reduces the number of linearly independent components of $I^{\a_1...\a_l\m\n}$ to $N=N_3(l)-N_3(l-1)=(l+2)(3l-1)$ \citep{dixon_1979,Dixon2015}.


\section{Comparison with Mathisson-Papapetrou-Dixon Equations of Motion}\label{appendixB}
\subsection{Comparison of Dixon's and Blanchet-Damour multipole moments}\label{oonn3388}
Before comparing our covariant equations of motion \eqref{ubpu;b}, \eqref{ui6s41} with analogous equations \eqref{q15ms}, \eqref{q16mc} derived by \citet{dixon_1979} in the MPD formalism, we need to establish the correspondence between the Dixon multipole moments $I^{\a_1...\a_l\m\n}$ and the STF mass and spin multipoles ${\cal M}^{\a_1...\a_l}$ and ${\cal S}^{\a_1...\a_l}$ that are used in the present paper. 
To this end we notice that the original definition \eqref{q12ttt} of multipoles $I^{\a_1...\a_l\m\n}$ contains the time components, $X^0$, of vector $X^\a$ which are nonphysical as they cannot be measured by a local observer with dynamic velocity ${\mathfrak n}^\a$ at point $z$ on the reference worldline ${\cal Z}$. Only those components of $I^{\a_1...\a_l\m\n}$ which are orthogonal to ${\mathfrak n}^\a$ can be measured. This explains the physical meaning of the orthogonality condition \eqref{juevz5}. 

It is reasonable to introduce a new notation for the physically-meaningful components of Dixon's multipoles, 
\be\label{p3gz7}
{\cal J}^{\a_1...\a_l\m\n}=P^{\a_1}_{\b_1}...P^{\a_l}_{\b_l}\int_\Sigma X^{\b_1}...X^{\b_l}\hat T^{\m\n}(z,X)\sqrt{-\bar g(z)}d\Sigma\;,   \qquad\qquad (\;l\ge 2\;)\;,
\ee
where the integration is performed in 4-dimensional spacetime over the hypersurface $\Sigma$ passing through the point $z$ with the element of integration $d\Sigma={\mathfrak n}^\a d\Sigma_\a$, and 
\be 
P^{\a}_{\b}=\d^{\a}_{\b}+{\mathfrak n}^\a {\mathfrak n}_\b\;,
\ee
is the operator of projection on the hypersurface $\Sigma$ making all vectors $X^\a$ in \eqref{p3gz7} orthogonal to ${\mathfrak n}^\a$. The multipoles ${\cal J}^{\a_1...\a_l\m\n}$ have the same symmetries \eqref{o39v3j}, \eqref{gr7xews} as $I^{\a_1...\a_l\m\n}$, 
\ba \label{mev49}
{\cal J}^{\a_1...\a_l\m\n}&=&{\cal J}^{(\a_1...\a_l)(\m\n)}\;,\\\label{bstw5v}
{\cal J}^{(\a_1...\a_l\m)\n}&=&0\;,
\ea
while the orthogonality condition \eqref{juevz5} is identically satisfied and is no longer considered as an additional constraint. The projection operator is idempotent \citep{idempotence} that is
\be\label{kkk9}
P^\a_\g P^\g_\b=P^\a_\b\;,
\ee
which makes only 3 out of 4 components of $X^\a$ linearly-independent in \eqref{p3gz7}. On the other hand, the indices $\m$ and $\n$ in ${\cal J}^{\a_1...\a_l\m\n}$ still take values from the set $\{0,1,2,3\}$. Thus, equation \eqref{mev49} tells us that the number of components of ${\cal J}^{\a_1...\a_l\m\n}$ is $C^{l+2}_2\times C^5_3=5(l+2)(l+1)$ while the number of constraints \eqref{bstw5v} is $C^{l+3}_2\times C^4_3=2(l+3)(l+2)$. It gives the number of the algebraically-independent components of ${\cal J}^{\a_1...\a_l\m\n}$ equal to $N=(l+2)(3l-1)$ which exactly coincides with the number of algebraically-independent components of Dixon's multipoles $I^{\a_1...\a_l\m\n}$. 

Picking up the local Riemann coordinates in such a way that $X^0$ component of vector $X^\a$ is directed along the dynamic velocity ${\mathfrak n}^\a$ and three other components $X^i=\{X^1,X^2,X^3\}$ are lying in the hypersurface $\Sigma$, yields skeleton's structure, 
\be\label{hhrr6c3}
\hat T^{\m\n}(z,X)=\int_{-\infty}^{+\infty}\d(X^0)\hat T^{\m\n}_\perp(X^i)dX^0\;,
\ee
where $\d(X^0)$ is Dirac's delta-function and the distribution $\hat T^{\m\n}_\perp\in\Sigma$.  Substituting \eqref{hhrr6c3} to \eqref{p3gz7} and taking into account that in these coordinates $DX=dX^0d\Sigma$, we obtain that Dixon's multipoles $I^{\a_1...\a_l\m\n}={\cal J}^{\a_1...\a_l\m\n}$ and, due to the tensor nature of the multipoles, this equality is retained in arbitrary coordinates.   

Exact nature of the distribution $\hat T^{\m\n}_\perp(X^i)$ in full general relativity is not yet known due to the non-linearity of the Einstein equations. Nonetheless, the Dirac delta-function is a reasonable candidate being sufficient to work in the post-Newtonian approximation with a corresponding regularization techniques \citep{blanchet_2001JMP}. For the purpose of the present paper it is sufficient to assume that in arbitrary coordinates the stress-energy skeleton \eqref{hhrr6c3} has the following structure \citep{Ohashi_2003PRD,steinhoff_2010PhRvD,Pound_2015} 
\be\label{bb22sz}
\hat T^{\m\n}(z,x)=\sum_{l=0}^\infty \int_{-\infty}^{+\infty}\bar\nabla_{\a_1...\a_l}\bigg[{\mathsf t}^{\a_1...\a_l\m\n}(z)\frac{\d_4\left(x-z\right)}{\sqrt{-\bar g(z)}}\bigg]\frac{ds}{\sqrt{-\bar g_{\m\n}(z) {\mathfrak n}^\m  {\mathfrak n}^\n}}\;,
\ee
where $s$ is an affine parameter along the geodesic in direction of the dynamic velocity ${\mathfrak n}^\a$, $\d_4\left(x-z\right)\equiv \d_4\left[x^\a-z^\a(s)\right]$ is 4-dimensional Dirac's delta-function, ${\mathsf t}^{\a_1...\a_l\m\n}$ are generalized multipole moments defined on the worldline ${\cal Z}$ that are orthogonal to ${\mathfrak n}^\a$ in the first $l$ indices (${\mathfrak n}_{\a_1}{\mathsf t}^{\a_1...\a_l\m\n}=0$), and $\bar\nabla_{\a_1...\a_l}\equiv \bar\nabla_{\a_1}...\bar\nabla_{\a_l}$ is a covariant derivative of the order $l$ taken with respect to the argument $x\equiv x^\a$ of the Dirac delta-function on the background manifold. Notice that expression \eqref{bb22sz} is a simplification of the original Mathisson theory \citep{mathisson_2010GReGr_1,mathisson_2010GReGr_1} proposed by \citet{tulczyjew1}.  \citet{dixon_1979} did not specify the nature of the singularity entering definition \eqref{bb22sz} assuming that Dirac's delta-function is solely valid in the pole-dipole approximation while a more general type of distribution is required in the definition of the stress-energy {skeleton} for high-order multipoles. The Dirac delta-function is widely adopted in computations of equations of motion of relativistic binary systems \citep{schaefer_2011mmgr,spin_Hamiltonian_schaefer,Blanchet_2002LRR} amended with corresponding regularization techniques to deal with the singularities in the non-linear approximations of general relativity \citep{Damour_1987book,blanchet_2004PhRvD,blanchet_2005PhRvD7,Dixon_2013}.

The generalized multipoles ${\mathsf t}^{\a_1...\a_l\m\n}$ are used to derive the explicit form of the MPD equations of motion in terms of the linear momentum ${\mathfrak p}^\a$, angular momentum $S^{\a\b}$ and Dixon's multipole moments $I^{\a_1...\a_l\m\n}$ as demonstrated by \citet{mathisson_2010GReGr_1,mathisson_2010GReGr_2}, \citet{pap1,pap2}, \citet{dixon_1979} and other researchers   \citep{bini_2009GReGr,dirk_obukhov_2015PhRvD,steinhoff_2010PhRvD,dirk_obukhov2014,Ohashi_2003PRD}.  It turns out that the generalized multipoles ${\mathsf t}^{\a_1...\a_l\m\n}$ are effectively equivalent to the body multipoles, ${\cal J}^{\a_1...\a_l\m\n}$. Indeed, replacing the stress-energy skeleton \eqref{bb22sz} to \eqref{q12ttt}, transforming the most general coordinates $x^\a$ in \eqref{bb22sz} to the local Riemannian coordinates $X^\a$,  and taking the covariant derivatives yield 
\be\label{ncv4d}
{\cal J}^{\a_1...\a_l\m\n}=P^{\a_1}_{\b_1}...P^{\a_l}_{\b_l}\sum_{n=0}^\infty {\mathsf t}^{\g_1...\g_p\m\n}\int X^{\b_1}...X^{\b_l}\frac{\pd^n \d_4(X)}{\pd X^{\g_1}...\pd X^{\g_n}}DX\;.
\ee
Integrating by parts, taking the partial derivatives from $X^\a$, and accounting for the integral properties of delta-function \citep{shilov_1968}, we conclude 
\be\label{jj239}
{\cal J}^{\a_1...\a_l\m\n}=(-1)^ll!{\mathsf t}^{\a_1...\a_l\m\n}\;.
\ee
 
To proceed further on, we shall assume that the dynamic velocity ${\mathfrak n}^\a$ is equal to the kinematic velocity ${\bar u}^\a$. This assumption is consistent with Dixon's mathematical development and agrees with our covariant definition \eqref{zowv34as} of the linear momentum of an extended body moving on the background spacetime manifold. It also allows us to employ
the results obtained previously by \citet{Ohashi_2003PRD}, to retrieve  a covariant expression for the generalized multipoles ${\mathsf t}^{\a_1...\a_l\m\n}$ of the gravitational skeleton $\hat T^{\m\n}$ from the multipolar expansion of the metric tensor of a single body. We have derived the generalized multipoles of the stress-energy skeleton from \citep[Equation 3.1]{Ohashi_2003PRD} after reconciling the sign conventions of the metric tensor perturbation and the normalization coefficients of multipoles adopted in \citep{Ohashi_2003PRD} with those adopted by \citet[Equation 2.32]{bld1986} which we also use in the present paper. The generalized moments of the stress-energy {skeleton} read,  
\ba\label{gg55zz99} 
{\mathsf t}^{\a_1...\a_l\m\n}&=&\frac{(-1)^l}{l!}\left[{\bar u}^\m {\bar u}^\n{\cal M}^{\a_1...\a_l}+\frac{2}{l+1}{\bar u}^{(\m}\dot{\cal M}^{\n)\a_1...\a_l}+\frac{1}{(l+1)(l+2)}\ddot{\cal M}^{\m\n\a_1...\a_l}\right]\\\nonumber
&-&\frac{(-1)^l}{l!}\left[\frac{2l}{l+1}{\bar u}^{(\m}{\varepsilon}_\b{}^{\n)<\a_1}{\cal S}^{\a_2...\a_l>\b}+\frac{2}{l+2}\varepsilon_\b{}^{<\a_1(\m}\dot{\cal S}^{\n)\a_2...\a_l>\b}\right]\;,
\ea
where the dot above functions denotes the Fermi-Walker covariant derivative \eqref{p2b8r3} and \eqref{huc4}.
Comparing \eqref{gg55zz99} with \eqref{jj239} we obtain the relationship between the Dixon internal multipoles and the mass and spin multipoles used in the present paper,
\ba\label{ney1c} 
{\cal J}^{\a_1...\a_l\m\n}&=&{\bar u}^\m {\bar u}^\n{\cal M}^{\a_1...\a_l}+\frac{2}{l+1}{\bar u}^{(\m}\dot{\cal M}^{\n)\a_1...\a_l}+\frac{1}{(l+1)(l+2)}\ddot{\cal M}^{\m\n\a_1...\a_l}\\\nonumber
&-&\frac{2l}{l+1}{\bar u}^{(\m}{\varepsilon}_\b{}^{\n)<\a_1}{\cal S}^{\a_2...\a_l>\b}-\frac{2}{l+2}\varepsilon_\b{}^{<\a_1(\m}\dot{\cal S}^{\n)\a_2...\a_l>\b}\;.
\ea

We still have to take into account the identity \eqref{bstw5v} in order to eliminate linearly-dependent components of ${\cal J}^{\a_1...\a_l\m\n}$. The most easy way is to take the double skew-symmetric part with respect to the last four indices as shown in equation \eqref{wk12dcr}. It yields 
\be\label{cbxczw}
I^{\a_1...\a_l\m\n}\equiv {\cal J}^{\a_1...[a_{l-1}[\a_l\m]\n]}=4\left\{{\cal M}^{<\a_1...[\a_{l-1}[\a_l>}{ u}^{\m]}{u}^{\n]}+\frac{l}{l+1}{\cal S}^{\b<\a_1...[\a_{l-1}}{ u}^{(\m]}{\varepsilon}^{\a_l>\n)}{}_\b\right\}\;,
\ee
where we have taken into account that in calculating the skew-symmetric part of 4-velocity $u^\mu$ with a purely spatial tensor we have, for example, 
\be
{\cal M}^{\a_1...\a_{l-1}[\a_l}{\bar u}^{\m]}=\pi^{\a_l}_{\b_l}{\cal M}^{\a_1...\a_{l-1}[\b_l}{\bar u}^{\m]}=\frac12{\cal M}^{\a_1...\a_{l-1}\a_l}{\bar u}^{\m}\;,
\ee
and so on.
Relation between Dixon's $J$ and $I$ multipole moments has been defined in \eqref{x8d4ak}. Substituting expression \eqref{cbxczw} for the Dixon multipoles $I$ in the right-hand side of \eqref{x8d4ak} provides a correspondence between the symmetrized Dixon multipoles $J$ and the Blanchet-Damour mass and spin multipoles in the following form 
\be\label{ut3cw}
J^{(\a_1...\a_{l-1}|\m|\a_l)\n}=\frac{l+1}{l-1}\left[{\cal M}^{<\a_1...[\a_{l-1}[\a_l>}{\bar u}^{\m]}{\bar u}^{\n]}+\frac{l}{l+1}{\cal S}^{\b<\a_1...[\a_{l-1}}{\bar u}^{(\m]}{\varepsilon}^{\a_l>\n)}{}_\b\right]\;,
\ee
where the anti-symmetrization goes over the pair of indices $[\a_{l-1}\m]$ and $[\a_l\n]$. Contracting both sides of \eqref{ut3cw} with 4-velocity allows us to express the Blanchet-Damour mass and spin multipoles in terms of projections of the Dixon multipoles onto 4-velocity of the center of mass of the body. More specifically, we have
\ba\label{sssssss5}
{\cal M}^{\a_1...\a_l}&=&4\frac{l-1}{l+1}J^{<\a_1...\a_{l-1}|\m|\a_l>\n}{\bar u}_\mu{\bar u}_\nu\;,\qquad\qquad (\;l\ge2\;)\\
\label{bbbb4ds}
{\cal S}^{\a_1...\a_l}&=&2\frac{l-1}{l}J^{<\a_1...\a_{l-1}|\m\n\sigma|}\varepsilon^{\a_l>}{}_{\mu\nu}{\bar u}_\sigma\qquad\qquad (\;l\ge2\;)\;.
\ea
It is worth emphasizing that in this section we work in the framework of general relativity. Therefore, all internal mass and spin multipoles, ${\cal M}^{\a_1...\a_l}$ and ${\cal S}^{\a_1...\a_l}$, have only general-relativistic value with vanishing scalar field contribution. In particular, the mass dipole, ${\cal M}^i=0$, due to the choice of the origin of the local coordinates at the center of mass of the body.

\subsection{Comparison of translational equations of motion}\label{ndj45xc}
In order to compare our translational equations of motion \eqref{ubpu;b} with Dixon's equation \eqref{q15ms} we need to symmetrize the covariant derivatives in the right-hand side of \eqref{q15ms}. It is achieved with the help of the following algebraic transformation, 
\ba\label{yu564x}
\bar\nabla_{\a(\b_1...\b_{l-2}}R_{|\m|\b_{l-1}\b_l)\n} J^{\b_1...\b_{l-1}\m\b_l\n}&=&\bar\nabla_{(\a\b_1...\b_{l-2}}R_{|\m|\b_{l-1}\b_l)\n} J^{\b_1...\b_{l-1}\m\b_l\n}\\\nonumber
&+&
\frac{2}{l+1}\bar\nabla_{\n(\b_1...\b_{l-2}}R_{|\m|\b_{l-1}\b_l)\a} J^{\b_1...\b_{l-1}\m\b_l\n}+{\cal O}(R^2)\;,
\ea
where the residual terms are proportional to the square of the Riemann tensor, and have been discarded. These quadratic-in-curvature terms are important for the post-Newtonian equations of motion but complicate the equations which follow and, hence, will be omitted every time when they appear. Substituting \eqref{ut3cw} to the right-hand side of \eqref{yu564x} yields
\be\label{bye62c}
\bar\nabla_{\a(\b_1...\b_{l-2}}R_{|\m|\b_{l-1}\b_l)\n} J^{\b_1...\b_{l-1}\m\b_l\n}=\frac{l+1}{l-1}\left[{\cal E}_{\a\b_1...\b_l}{\cal M}^{\b_1...\b_l}+\frac{l}{l+1}{\cal C}_{\a\b_1...\b_l}{\cal S}^{\b_1...\b_l} \right]+{\cal O}(R^2)\;,
\ee
where the external multipole moments ${\cal E}_{\a_1...\a_l}$ and ${\cal C}_{\a_1...\a_l}$ have been defined in \eqref{sd11} and \eqref{we70} respectively.
Substituting \eqref{bye62c} to the right-hand side of \eqref{q15ms} recasts it to
\be\label{hwcz41}
\frac{{\cal D} {\mathfrak p}_\a}{{\cal D}\tau}=\frac12 {\bar u}^\b S^{\m\n}\bar R_{\m\n\b\a}+\sum\limits_{l=2}^{\infty}\frac{1}{l!} \left[{\cal E}_{\a\b_1...\b_l}{\cal M}^{\b_1...\b_l}+\frac{l}{l+1}{\cal C}_{\a\b_1...\b_l}{\cal S}^{\b_1...\b_l} \right]+{\cal O}(R^2)\;.
\ee
The very first term in the right-hand side depending on $S^{\a\b}$,  can be incorporated to the sum over the spin moments by making use of the duality relation between body's intrinsic spin ${\cal S}^\a$ and spin-tensor \footnote{The minus sign in \eqref{ju3v8m} appears because Dixon's definition \eqref{wk12} of $S^{\a\b}$ has an opposite sign as compared to our definition \eqref{spin-3} of spin ${\cal S}^\a$. } $S^{\a\b}$ 
\be\label{ju3v8m}
 S^{\m\n}=-\varepsilon^{\m\n}{}_\a{\cal S}^\a\;,
 \ee
 where the Levi-Civita tensor $\varepsilon_{\a\b\g}$ has been defined above in \eqref{vareps67}. It yields
 \be  
 {\bar u}^\b S^{\m\n}\bar R_{\m\n\b\a}={\cal C}_{\a\b}{\cal S}^\b\;,
 \ee
where ${\cal C}_{\a\b}$ is given by \eqref{we70} for $l=2$. Making use of \eqref{ju3v8m}  allows to rewrite \eqref{hwcz41} in the final form
\be\label{uvnet467}
\frac{{\cal D} {\mathfrak p}_\a}{{\cal D}\tau}=\sum\limits_{l=2}^{\infty}\frac{1}{l!} {\cal E}_{\a\b_1...\b_l}{\cal M}^{\b_1...\b_l}+\sum\limits_{l=1}^{\infty} \frac{l}{(l+1)!}{\cal C}_{\a\b_1...\b_l}{\cal S}^{\b_1...\b_l}+{\cal O}(R^2)\;.
\ee

Thus, Dixon's equation of translational motion \eqref{q15ms} given in terms of Dixon's internal multipoles and Veblen's tensor extensions of the Riemann tensor are brought to the form \eqref{uvnet467} given in terms of the gravitoelectric, ${\cal E}_{\a\b_1...\b_l}$, and gravitomagnetic, ${\cal C}_{\a\b_1...\b_l}$, external multipoles as well as mass, ${\cal M}^{\b_1...\b_l}$ and spin, ${\cal S}^{\b_1...\b_l}$ internal multipoles. Comparing with the complete covariant form of the translational equations of motion \eqref{ubpu;b}--\eqref{sd17a} taken for the case of general relativity one can see that Dixon's equation reproduces only two terms in the complete expression for the post-Newtonian force, more specifically -- the very first term of the post-Newtonian force $F^\a_{\cal Q}$ in \eqref{sd15a} and that of $F^\a_{\cal C}$ in \eqref{sd16}. The terms which are missed in the Dixon's translational equations of motion but are present in our equations \eqref{ubpu;b}--\eqref{sd17a} include the quadratic-in-curvature terms through \eqref{we69} and the terms which depend on the time derivatives of multipoles both external and internal ones. The terms with the time derivatives of the multipoles must be present in the equations of motion but they have been omitted by Dixon as he has taken into account only his $J$ multipoles while, in fact, all components of the Dixon's $I$ multipoles must be taken into account. Independent derivation of the translational equations of motion by \citet{racine_2005PhRvD,racine2013PhRvD} with different mathematical technique corroborates our conclusions about the missing terms in Dixon's translational equations of motion \eqref{q15ms}. It does not mean that the MPD formalism is erroneous. It merely indicates that much more work is required to take into account all the missing contributions to the post-Newtonian translational equations of motion derived in the framework of the Mathisson variational dynamics.

\subsection{Comparison of rotational equations of motion}\label{iopn3e4}

Dixon's equations of rotational motion are given by equation \eqref{q16mc}. The first term in the right-hand side of this equation vanishes in our approach because the linear momentum of the body $\mathfrak{p}^\a$ is chosen to be parallel to 4-velocity $\bar u^\a$ of the center of mass of body B. We express the spin of the body ${\cal S}^\a$ in terms of the spin tensor $S^{\lambda\sigma}$ by inverting \eqref{ju3v8m},
\be \label{lec27b}
{\cal S}^\a=-\frac12\varepsilon^\a{}_{\lambda\s}S^{\lambda\s}\;.
\ee 
Taking a covariant derivative from both sides of \eqref{lec27b} and replacing the covariant derivative from $S^{\b\g}$ with the terms from the right side of \eqref{q16mc} yields
\be \la{op4vx3ww}
\frac{{\cal D}{\cal S}^\a}{{\cal D}\tau}=-\varepsilon^\a{}_{\lambda\s}\sum\limits_{l=1}^{\infty}\frac{1}{l!}\nabla_{(\b_1...\b_{l-1}}\bar R_{|\m|\r\b_l)\n}g^{\r\lambda}
\left[{\cal M}^{\s\b_1...\b_{l-1}\b_l}{\bar u}^{\m}{\bar u}^{\n}+\frac{l+1}{l+2}{\cal S}^{\s\g\b_1...\b_{l-1}}{\bar u}^{\m}{\varepsilon}^{\b_l\n}{}_\g\right]\;,
\ee
where we have also used \eqref{ut3cw} to replace the Dixon internal multipole moments with the Blanchet-Damour mass and spin multipoles. Now, we employ the covariant definitions \eqref{sd11} and \eqref{we70} of the gravitoelectric and gravitomagnetic external multipoles in \eqref{op4vx3ww} that takes on the following form,
\be\label{xcz5ds}
\frac{{\cal D}{\cal S}^\a}{{\cal D}\tau}=-\varepsilon^{\a\lambda}{}_\s\sum\limits_{l=1}^{\infty}\frac{1}{l!}
\left[{\cal E}_{\lambda\b_1...\b_l}{\cal M}^{\s\b_1...\b_l}+\frac{l+1}{l+2}{\cal C}_{\lambda\b_1...\b_l}{\cal S}^{\s\b_1...\b_l}\right]\;.
\ee

Now, we can compare Dixon's equation of rotational motion \eqref{xcz5ds} with our equation \eqref{ui4vsr} where only general-relativistic terms in the torque \eqref{ac5s03v} must be retained. These terms are making up the second line in \eqref{ac5s03v} and they are in a perfect agreement with Dixon's torque in the right-hand side of \eqref{xcz5ds}. The difference between \eqref{xcz5ds} and \eqref{ui6s41} is in the presence of the very last term in the right-hand side of \eqref{ui6s41} as compared with \eqref{xcz5ds}. This term is associated with the Fermi-Walker transport of spin along an accelerated worldline of the body center of mass. The absence of this term in Dixon's rotational equation of motion \eqref{xcz5ds} tells us that the reference world line ${\cal W}$ of the origin of the normal Riemann coordinates used by \citet{dixon_1979,Dixon2015} for computation of his own results, is a time-like geodesic which, in the most general case, does not coincide with the worldline ${\cal Z}$ of the body center of mass because of the gravitational interaction of the internal moments of the body with the external gravitoelectric and gravitomagnetic multipoles.

\bibliographystyle{unsrtnat}
\bibliography{EQM_references}

\begin{thebibliography}{302}
\providecommand{\natexlab}[1]{#1}
\providecommand{\url}[1]{\texttt{#1}}
\expandafter\ifx\csname urlstyle\endcsname\relax
  \providecommand{\doi}[1]{doi: #1}\else
  \providecommand{\doi}{doi: \begingroup \urlstyle{rm}\Url}\fi

\bibitem[{Tagoshi} et~al.(2001){Tagoshi}, {Ohashi}, and
  {Owen}]{Tagoshi_2001PhRvD}
H.~{Tagoshi}, A.~{Ohashi}, and B.~J. {Owen}.
\newblock {Gravitational field and equations of motion of spinning compact
  binaries to 2.5 post-Newtonian order}.
\newblock \emph{\prd}, 63\penalty0 (4):\penalty0 044006, February 2001.
\newblock \doi{10.1103/PhysRevD.63.044006}.

\bibitem[{Wang} and {Will}(2007)]{Wang_2007PhRvD}
H.~{Wang} and C.~M. {Will}.
\newblock {Post-Newtonian gravitational radiation and equations of motion via
  direct integration of the relaxed Einstein equations. IV. Radiation reaction
  for binary systems with spin-spin coupling}.
\newblock \emph{\prd}, 75\penalty0 (6):\penalty0 064017, March 2007.
\newblock \doi{10.1103/PhysRevD.75.064017}.

\bibitem[{Marsat} et~al.(2013){Marsat}, {Boh{\'e}}, {Faye}, and
  {Blanchet}]{Marsat_2013CQGra}
S.~{Marsat}, A.~{Boh{\'e}}, G.~{Faye}, and L.~{Blanchet}.
\newblock {Next-to-next-to-leading order spin-orbit effects in the equations of
  motion of compact binary systems}.
\newblock \emph{Classical and Quantum Gravity}, 30\penalty0 (5):\penalty0
  055007, March 2013.
\newblock \doi{10.1088/0264-9381/30/5/055007}.

\bibitem[{Mathisson}(2010{\natexlab{a}})]{mathisson_2010GReGr_1}
M.~{Mathisson}.
\newblock {Republication of: The mechanics of matter particles in general
  relativity}.
\newblock \emph{General Relativity and Gravitation}, 42:\penalty0 989--1010,
  April 2010{\natexlab{a}}.
\newblock \doi{10.1007/s10714-010-0938-z}.

\bibitem[{Mathisson}(2010{\natexlab{b}})]{mathisson_2010GReGr_2}
M.~{Mathisson}.
\newblock {Republication of: New mechanics of material systems}.
\newblock \emph{General Relativity and Gravitation}, 42:\penalty0 1011--1048,
  April 2010{\natexlab{b}}.
\newblock \doi{10.1007/s10714-010-0939-y}.

\bibitem[{Papapetrou}(1951{\natexlab{a}})]{Papapetrou23101951}
A.~{Papapetrou}.
\newblock Spinning test-particles in general relativity. i.
\newblock \emph{Proceedings of the Royal Society of London Series A},
  209:\penalty0 248--258, October 1951{\natexlab{a}}.
\newblock \doi{10.1098/rspa.1951.0200}.

\bibitem[{Dixon}(1970{\natexlab{a}})]{dixon_1970_1}
W.~G. {Dixon}.
\newblock {Dynamics of Extended Bodies in General Relativity. I. Momentum and
  Angular Momentum}.
\newblock \emph{Royal Society of London Proceedings Series A}, 314:\penalty0
  499--527, January 1970{\natexlab{a}}.
\newblock \doi{10.1098/rspa.1970.0020}.

\bibitem[{Dixon}(1970{\natexlab{b}})]{dixon_1970_2}
W.~G. {Dixon}.
\newblock {Dynamics of Extended Bodies in General Relativity. II. Moments of
  the Charge-Current Vector}.
\newblock \emph{Royal Society of London Proceedings Series A}, 319:\penalty0
  509--547, November 1970{\natexlab{b}}.
\newblock \doi{10.1098/rspa.1970.0191}.

\bibitem[{Dixon}(1973)]{dixon_1973GReGr}
W.~G. {Dixon}.
\newblock {The definition of multipole moments for extended bodies}.
\newblock \emph{General Relativity and Gravitation}, 4:\penalty0 199--209, June
  1973.
\newblock \doi{10.1007/BF02412488}.

\bibitem[{Dixon}(1974)]{dixon_1974_3}
W.~G. {Dixon}.
\newblock {Dynamics of Extended Bodies in General Relativity. III. Equations of
  Motion}.
\newblock \emph{Royal Society of London Philosophical Transactions Series A},
  277:\penalty0 59--119, August 1974.
\newblock \doi{10.1098/rsta.1974.0046}.

\bibitem[{Dixon}(1979)]{dixon_1979}
W.~G. {Dixon}.
\newblock {Extended bodies in general relativity: their description and
  motion.}
\newblock In J.~{Ehlers}, editor, \emph{Isolated Gravitating Systems in General
  Relativity}, pages 156--219, Amsterdam, 1979. North-Holland.

\bibitem[{Ohashi}(2003)]{Ohashi_2003PRD}
A.~{Ohashi}.
\newblock {Multipole particle in relativity}.
\newblock \emph{\prd}, 68\penalty0 (4):\penalty0 044009, August 2003.
\newblock \doi{10.1103/PhysRevD.68.044009}.

\bibitem[{Steinhoff} and {Puetzfeld}(2010)]{steinhoff_2010PhRvD}
J.~{Steinhoff} and D.~{Puetzfeld}.
\newblock {Multipolar equations of motion for extended test bodies in general
  relativity}.
\newblock \emph{\prd}, 81\penalty0 (4):\penalty0 044019, February 2010.
\newblock \doi{10.1103/PhysRevD.81.044019}.

\bibitem[{Semer{\'a}k}(1999)]{Semerak_1999MNRAS}
O.~{Semer{\'a}k}.
\newblock {Spinning test particles in a Kerr field - I}.
\newblock \emph{\mnras}, 308:\penalty0 863--875, September 1999.
\newblock \doi{10.1046/j.1365-8711.1999.02754.x}.

\bibitem[{Kyrian} and {Semer{\'a}k}(2007)]{Kyrian_2007MNRAS}
K.~{Kyrian} and O.~{Semer{\'a}k}.
\newblock {Spinning test particles in a Kerr field - II}.
\newblock \emph{\mnras}, 382:\penalty0 1922--1932, December 2007.
\newblock \doi{10.1111/j.1365-2966.2007.12502.x}.

\bibitem[{Soffel}(1989)]{sof89}
M.~H. {Soffel}.
\newblock \emph{{Relativity in Astrometry, Celestial Mechanics and Geodesy}}.
\newblock Springer, Berlin, 1989.

\bibitem[{Kopeikin} et~al.(2011){Kopeikin}, {Efroimsky}, and
  {Kaplan}]{kopeikin_2011book}
S.~{Kopeikin}, M.~{Efroimsky}, and G.~{Kaplan}.
\newblock \emph{Relativistic Celestial Mechanics of the Solar System}.
\newblock Wiley, Weinheim, September 2011.

\bibitem[{Kramer} and {Wex}(2009)]{krawex_2009}
M.~{Kramer} and N.~{Wex}.
\newblock The double pulsar system: a unique laboratory for gravity.
\newblock \emph{Classical and Quantum Gravity}, 26\penalty0 (7):\penalty0
  073001, 2009.
\newblock URL \url{http://stacks.iop.org/0264-9381/26/i=7/a=073001}.

\bibitem[{Weisberg} et~al.(2010){Weisberg}, {Nice}, and
  {Taylor}]{weisberg_2010ApJ}
J.~M. {Weisberg}, D.~J. {Nice}, and J.~H. {Taylor}.
\newblock Timing measurements of the relativistic binary pulsar {PSR}
  {B}1913+16.
\newblock \emph{\apj}, 722:\penalty0 1030--1034, oct 2010.
\newblock \doi{10.1088/0004-637X/722/2/1030}.

\bibitem[{Damour}(2009)]{damour_2009ASSL}
T.~{Damour}.
\newblock Binary systems as test-beds of gravity theories.
\newblock In M.~Colpi, P.~Casella, V.~Gorini, U.~Moschella, and A.~Possenti,
  editors, \emph{Physics of Relativistic Objects in Compact Binaries: From
  Birth to Coalescence}, pages 1--41. Springer Netherlands, Dordrecht, 2009.
\newblock ISBN 978-1-4020-9264-0.
\newblock \doi{10.1007/978-1-4020-9264-0_1}.
\newblock URL \url{https://doi.org/10.1007/978-1-4020-9264-0_1}.

\bibitem[{Damour} and {Esposito-Far{\`e}se}(1998)]{damour_1998PhRvD}
T.~{Damour} and G.~{Esposito-Far{\`e}se}.
\newblock {Gravitational-wave versus binary-pulsar tests of strong-field
  gravity}.
\newblock \emph{\prd}, 58\penalty0 (4):\penalty0 042001, August 1998.
\newblock \doi{10.1103/PhysRevD.58.042001}.

\bibitem[{Reitze}(2017)]{Reitze_2017PhyU}
D.~H. {Reitze}.
\newblock {First detections of gravitational waves emitted from binary black
  hole mergers}.
\newblock \emph{Physics Uspekhi}, 60:\penalty0 823, November 2017.
\newblock \doi{10.3367/UFNe.2016.11.038176}.

\bibitem[{Flanagan} and {Hinderer}(2008)]{Hinderer_2008PhRvD}
{\'E}.~{\'E}. {Flanagan} and T.~{Hinderer}.
\newblock {Constraining neutron-star tidal Love numbers with gravitational-wave
  detectors}.
\newblock \emph{\prd}, 77\penalty0 (2):\penalty0 021502, January 2008.
\newblock \doi{10.1103/PhysRevD.77.021502}.

\bibitem[{Binnington} and {Poisson}(2009)]{Poisson_2009PhRvD}
T.~{Binnington} and E.~{Poisson}.
\newblock {Relativistic theory of tidal {L}ove numbers}.
\newblock \emph{\prd}, 80\penalty0 (8):\penalty0 084018, October 2009.
\newblock \doi{10.1103/PhysRevD.80.084018}.

\bibitem[{Damour} and {Nagar}(2009)]{Nagar_2009PhRvD}
T.~{Damour} and A.~{Nagar}.
\newblock {Relativistic tidal properties of neutron stars}.
\newblock \emph{\prd}, 80\penalty0 (8):\penalty0 084035, October 2009.
\newblock \doi{10.1103/PhysRevD.80.084035}.

\bibitem[{Raithel} et~al.(2018){Raithel}, {{\"O}zel}, and
  {Psaltis}]{Raithel_2018ApJ}
C.~A. {Raithel}, F.~{{\"O}zel}, and D.~{Psaltis}.
\newblock Tidal deformability from {GW}170817 as a direct probe of the neutron
  star radius.
\newblock \emph{\apjl}, 857:\penalty0 L23, April 2018.
\newblock \doi{10.3847/2041-8213/aabcbf}.

\bibitem[{Yagi}(2014)]{Yagi_2014PhRvD}
K.~{Yagi}.
\newblock Multipole {L}ove relations.
\newblock \emph{\prd}, 89\penalty0 (4):\penalty0 043011, February 2014.
\newblock \doi{10.1103/PhysRevD.89.043011}.

\bibitem[{Schutz}(2018)]{Schutz_2018RSPTA}
B.~F. {Schutz}.
\newblock {Gravitational-wave astronomy: delivering on the promises}.
\newblock \emph{Philosophical Transactions of the Royal Society of London
  Series A}, 376:\penalty0 20170279, May 2018.
\newblock \doi{10.1098/rsta.2017.0279}.

\bibitem[{Blanchet}(2002)]{Blanchet_2002LRR}
L.~{Blanchet}.
\newblock Gravitational radiation from post-{N}ewtonian sources and
  inspiralling compact binaries.
\newblock \emph{Living Reviews in Relativity}, 5:\penalty0 3, April 2002.
\newblock \doi{https://doi.org/10.12942/lrr-2002-3}.

\bibitem[{Asada} et~al.(2011){Asada}, {Futamase}, and {Hogan}]{asada_2011}
H.~{Asada}, T.~{Futamase}, and P.~{Hogan}.
\newblock \emph{Equations of Motion in General Relativity}.
\newblock Oxford University Press, New York, 2011.

\bibitem[{Sch{\"a}fer}(2011)]{schaefer_2011mmgr}
G.~{Sch{\"a}fer}.
\newblock Post-{N}ewtonian methods: {A}nalytic results on the binary problem.
\newblock In L.~{Blanchet}, A.~{Spallicci}, and B.~{Whiting}, editors,
  \emph{Mass and Motion in General Relativity. Fundamental Theories of
  Physics}, volume 162, pages 167--210. Berlin, Springer, 2011.
\newblock \doi{10.1007/978-90-481-3015-3_6}.

\bibitem[{Damour}(2008)]{damour_2008IJMPA}
T.~{Damour}.
\newblock Introductory lectures on the effective one body formalism.
\newblock \emph{International Journal of Modern Physics A}, 23:\penalty0
  1130--1148, 2008.
\newblock \doi{10.1142/S0217751X08039992}.

\bibitem[{Bini} et~al.(2012){Bini}, {Damour}, and {Faye}]{Bini_PRD124034}
D.~{Bini}, T.~{Damour}, and G.~{Faye}.
\newblock Effective action approach to higher-order relativistic tidal
  interactions in binary systems and their effective one body description.
\newblock \emph{\prd}, 85:\penalty0 124034, Jun 2012.
\newblock \doi{10.1103/PhysRevD.85.124034}.
\newblock URL \url{https://link.aps.org/doi/10.1103/PhysRevD.85.124034}.

\bibitem[{Vines} and {Flanaga}n(2013)]{Vines_Flanagan_PRD024046}
J.~E. {Vines} and \'E.~\'E. {Flanaga}n.
\newblock First-post-{N}ewtonian quadrupole tidal interactions in binary
  systems.
\newblock \emph{\prd}, 88:\penalty0 024046, Jul 2013.
\newblock \doi{10.1103/PhysRevD.88.024046}.
\newblock URL \url{https://link.aps.org/doi/10.1103/PhysRevD.88.024046}.

\bibitem[{Steinhoff} et~al.(2016){Steinhoff}, {Hinderer}, {Buonanno}, and
  {Taracchini}]{Steinhoff_etal_PRD104028}
J.~{Steinhoff}, T.~{Hinderer}, A.~{Buonanno}, and A.~{Taracchini}.
\newblock Dynamical tides in general relativity: {E}ffective action and
  effective-one-body {H}amiltonian.
\newblock \emph{\prd}, 94:\penalty0 104028, Nov 2016.
\newblock \doi{10.1103/PhysRevD.94.104028}.
\newblock URL \url{https://link.aps.org/doi/10.1103/PhysRevD.94.104028}.

\bibitem[{Will}(2006)]{willLRR}
C.~M. {Will}.
\newblock The confrontation between general relativity and experiment.
\newblock \emph{Living Reviews in Relativity}, 9:\penalty0 3 (cited on May 22,
  2018), March 2006.

\bibitem[{Turyshev}(2009)]{Turyshev_2009PhyU}
S.~G. {Turyshev}.
\newblock {Experimental tests of general relativity: recent progress and future
  directions}.
\newblock \emph{Physics Uspekhi}, 52:\penalty0 1--27, January 2009.
\newblock \doi{10.3367/UFNe.0179.200901a.0003}.

\bibitem[{Kopeikin} and {Gwinn}(2000)]{Kopeikin_2000tmcs}
S.~{Kopeikin} and C.~{Gwinn}.
\newblock {Sub-Microarcsecond Astrometry and New Horizons in Relativistic
  Gravitational Physics}.
\newblock In K.~J. {Johnston}, D.~D. {McCarthy}, B.~J. {Luzum}, and G.~H.
  {Kaplan}, editors, \emph{IAU Colloq. 180: Towards Models and Constants for
  Sub-Microarcsecond Astrometry}, pages 303--307, Washington DC, 2000. U.S.
  Naval Observatory.

\bibitem[{Kopeikin} et~al.(2006){Kopeikin}, {Korobkov}, and
  {Polnarev}]{koppolkor2006}
S.~{Kopeikin}, P.~{Korobkov}, and A.~{Polnarev}.
\newblock {Propagation of light in the field of stationary and radiative
  gravitational multipoles}.
\newblock \emph{Classical and Quantum Gravity}, 23:\penalty0 4299--4322, July
  2006.
\newblock \doi{10.1088/0264-9381/23/13/001}.

\bibitem[{Klioner} et~al.(2010){Klioner}, {Seidelmann}, and
  {Soffel}]{Klioner_2010IAUS}
S.~A. {Klioner}, P.~K. {Seidelmann}, and M.~H. {Soffel}.
\newblock \emph{{{\rm (Eds.)} Relativity in Fundamental Astronomy: Dynamics,
  Reference Frames, and Data Analysis}}.
\newblock Proceedings of the IAU Symposium 261. Cambridge University Press,
  Cambridge, January 2010.

\bibitem[{Soffel} et~al.(2017){Soffel}, {Kopeikin}, and
  {Han}]{Soffel_2017JGeod}
M.~{Soffel}, S.~{Kopeikin}, and W.-B. {Han}.
\newblock {Advanced relativistic VLBI model for geodesy}.
\newblock \emph{Journal of Geodesy}, 91:\penalty0 783--801, July 2017.
\newblock \doi{10.1007/s00190-016-0956-z}.

\bibitem[{Landau} and {Lifshitz}(1975)]{Landau1975}
L.~D. {Landau} and E.~M. {Lifshitz}.
\newblock \emph{The classical theory of fields}.
\newblock Pergamon Press, Oxford, 1975.

\bibitem[{Itoh}(2009)]{Itoh_2009PhRvD}
Y.~{Itoh}.
\newblock {Third-and-a-half order post-Newtonian equations of motion for
  relativistic compact binaries using the strong field point particle limit}.
\newblock \emph{\prd}, 80\penalty0 (12):\penalty0 124003, December 2009.
\newblock \doi{10.1103/PhysRevD.80.124003}.

\bibitem[{Damour} et~al.(2016){Damour}, {Jaranowski}, and
  {Sch{\"a}fer}]{Damour_2016PhRvD93h4014D}
T.~{Damour}, P.~{Jaranowski}, and G.~{Sch{\"a}fer}.
\newblock {Conservative dynamics of two-body systems at the fourth
  post-Newtonian approximation of general relativity}.
\newblock \emph{\prd}, 93\penalty0 (8):\penalty0 084014, April 2016.
\newblock \doi{10.1103/PhysRevD.93.084014}.

\bibitem[{Saulson}(2013)]{Saulson_2013}
P.~R. {Saulson}.
\newblock {Gravitational wave detection: Principles and practice}.
\newblock \emph{Comptes Rendus Physique}, 14:\penalty0 288--305, April 2013.
\newblock \doi{10.1016/j.crhy.2013.01.007}.

\bibitem[{Ehlers} et~al.(1976){Ehlers}, {Rosenblum}, {Goldberg}, and
  {Havas}]{ehlers_1976ApJ}
J.~{Ehlers}, A.~{Rosenblum}, J.~N. {Goldberg}, and P.~{Havas}.
\newblock {Comments on gravitational radiation damping and energy loss in
  binary systems}.
\newblock \emph{\apjl}, 208:\penalty0 L77--L81, September 1976.
\newblock \doi{10.1086/182236}.

\bibitem[{Poisson} et~al.(2011){Poisson}, {Pound}, and {Vega}]{poisson_2011}
E.~{Poisson}, A.~{Pound}, and I.~{Vega}.
\newblock The motion of point particles in curved spacetime.
\newblock \emph{Living Reviews in Relativity}, 14\penalty0 (7), 2011.
\newblock \doi{10.12942/lrr-2011-7}.
\newblock URL \url{http://www.livingreviews.org/lrr-2011-7}.

\bibitem[{Infled} and {Plebanski}(1960)]{infeld_book}
L.~{Infled} and J.~{Plebanski}.
\newblock \emph{{Motion and Relativity}}.
\newblock Pergamon Press, New York, 1960.

\bibitem[{Einstein} et~al.(1938){Einstein}, {Infeld}, and {Hoffmann}]{eih}
A.~{Einstein}, L.~{Infeld}, and B.~{Hoffmann}.
\newblock The gravitational equations and the problem of motion.
\newblock \emph{The Annals of Mathematics}, 39\penalty0 (1):\penalty0 65--100,
  1938.

\bibitem[{Blanchet} and {Damour}(1986)]{bld1986}
L.~{Blanchet} and T.~{Damour}.
\newblock {Radiative gravitational fields in general relativity. I - General
  structure of the field outside the source}.
\newblock \emph{Royal Society of London Philosophical Transactions Series A},
  320:\penalty0 379--430, December 1986.

\bibitem[{Blanchet} and {Faye}(2001)]{blanchet_2001JMP}
L.~{Blanchet} and G.~{Faye}.
\newblock {Lorentzian regularization and the problem of point-like particles in
  general relativity}.
\newblock \emph{Journal of Mathematical Physics}, 42:\penalty0 4391--4418,
  September 2001.
\newblock \doi{10.1063/1.1384864}.

\bibitem[{Blanchet} et~al.(2004){Blanchet}, {Damour}, and
  {Esposito-Far{\`e}se}]{blanchet_2004PhRvD}
L.~{Blanchet}, T.~{Damour}, and G.~{Esposito-Far{\`e}se}.
\newblock {Dimensional regularization of the third post-Newtonian dynamics of
  point particles in harmonic coordinates}.
\newblock \emph{\prd}, 69\penalty0 (12):\penalty0 124007, June 2004.
\newblock \doi{10.1103/PhysRevD.69.124007}.

\bibitem[{Blanchet} and {Iyer}(2005)]{blanchet_2005PhRvD7}
L.~{Blanchet} and B.~R. {Iyer}.
\newblock {Hadamard regularization of the third post-Newtonian gravitational
  wave generation of two point masses}.
\newblock \emph{\prd}, 71\penalty0 (2):\penalty0 024004, January 2005.
\newblock \doi{10.1103/PhysRevD.71.024004}.

\bibitem[{Marchand} et~al.(2018){Marchand}, {Bernard}, {Blanchet}, and
  {Faye}]{Marchand_2018PhRvD}
T.~{Marchand}, L.~{Bernard}, L.~{Blanchet}, and G.~{Faye}.
\newblock {Ambiguity-free completion of the equations of motion of compact
  binary systems at the fourth post-Newtonian order}.
\newblock \emph{\prd}, 97\penalty0 (4):\penalty0 044023, February 2018.
\newblock \doi{10.1103/PhysRevD.97.044023}.

\bibitem[{D'Eath}(1975{\natexlab{a}})]{das1}
P.~D. {D'Eath}.
\newblock {Dynamics of a small black hole in a background universe}.
\newblock \emph{\prd}, 11:\penalty0 1387--1403, March 1975{\natexlab{a}}.

\bibitem[{D'Eath}(1975{\natexlab{b}})]{das2}
P.~D. {D'Eath}.
\newblock {Interaction of two black holes in the slow-motion limit}.
\newblock \emph{\prd}, 12:\penalty0 2183--2199, October 1975{\natexlab{b}}.

\bibitem[{Gorbonos} and {Kol}(2005)]{gorbonos_2005CQGra}
D.~{Gorbonos} and B.~{Kol}.
\newblock {Matched asymptotic expansion for caged black holes: regularization
  of the post-Newtonian order}.
\newblock \emph{Classical and Quantum Gravity}, 22:\penalty0 3935--3959,
  October 2005.
\newblock \doi{10.1088/0264-9381/22/19/009}.

\bibitem[{Thorne} and {Hartle}(1985)]{th_1985}
K.~S. {Thorne} and J.~B. {Hartle}.
\newblock {Laws of motion and precession for black holes and other bodies}.
\newblock \emph{\prd}, 31:\penalty0 1815--1837, April 1985.

\bibitem[{Futamase} and {Itoh}(2007)]{futamase_2007LRR}
T.~{Futamase} and Y.~{Itoh}.
\newblock The post-{N}ewtonian approximation for relativistic compact binaries.
\newblock \emph{Living Reviews in Relativity}, 10:\penalty0 2, March 2007.
\newblock \doi{10.12942/lrr-2007-2}.

\bibitem[{Shibata}(1993)]{Shibata_1993PhRvD}
M.~{Shibata}.
\newblock {Gravitational waves induced by a particle orbiting around a rotating
  black hole: Spin-orbit interaction effect}.
\newblock \emph{\prd}, 48:\penalty0 663--666, July 1993.
\newblock \doi{10.1103/PhysRevD.48.663}.

\bibitem[{Rieth} and {Sch{\"a}fer}(1997)]{Rieth_1997CQGra}
R.~{Rieth} and G.~{Sch{\"a}fer}.
\newblock {Spin and tail effects in the gravitational-wave emission of compact
  binaries}.
\newblock \emph{Classical and Quantum Gravity}, 14:\penalty0 2357--2380, August
  1997.
\newblock \doi{10.1088/0264-9381/14/8/029}.

\bibitem[{Xu} et~al.(1997){Xu}, {Wu}, and {Sch{\"a}fer}]{xu_1997PhRvD}
C.~{Xu}, X.~{Wu}, and G.~{Sch{\"a}fer}.
\newblock {Binary systems with monopole, spin, and quadrupole moments}.
\newblock \emph{\prd}, 55:\penalty0 528--539, January 1997.
\newblock \doi{10.1103/PhysRevD.55.528}.

\bibitem[{Owen} et~al.(1998){Owen}, {Tagoshi}, and {Ohashi}]{Owen_1998PhRvD}
B.~J. {Owen}, H.~{Tagoshi}, and A.~{Ohashi}.
\newblock {Nonprecessional spin-orbit effects on gravitational waves from
  inspiraling compact binaries to second post-Newtonian order}.
\newblock \emph{\prd}, 57:\penalty0 6168--6175, May 1998.
\newblock \doi{10.1103/PhysRevD.57.6168}.

\bibitem[{Porto}(2006)]{porto_2006}
R.~A. {Porto}.
\newblock Post-{N}ewtonian corrections to the motion of spinning bodies in
  nonrelativistic general relativity.
\newblock \emph{Phys. Rev. D}, 73:\penalty0 104031, May 2006.
\newblock \doi{10.1103/PhysRevD.73.104031}.
\newblock URL \url{https://link.aps.org/doi/10.1103/PhysRevD.73.104031}.

\bibitem[{Steinhoff} et~al.(2008){Steinhoff}, {Sch\"afer}, and
  {Hergt}]{spin_Hamiltonian_schaefer}
J.~{Steinhoff}, G.~{Sch\"afer}, and S.~{Hergt}.
\newblock {ADM} canonical formalism for gravitating spinning objects.
\newblock \emph{\prd}, 77:\penalty0 104018, May 2008.
\newblock \doi{10.1103/PhysRevD.77.104018}.
\newblock URL \url{https://link.aps.org/doi/10.1103/PhysRevD.77.104018}.

\bibitem[{Hergt} and {Sch{\"a}fer}(2008)]{hergt_2008PhRvD}
S.~{Hergt} and G.~{Sch{\"a}fer}.
\newblock {Higher-order-in-spin interaction Hamiltonians for binary black holes
  from source terms of Kerr geometry in approximate ADM coordinates}.
\newblock \emph{\prd}, 77\penalty0 (10):\penalty0 104001, May 2008.
\newblock \doi{10.1103/PhysRevD.77.104001}.

\bibitem[{Tessmer} et~al.(2013){Tessmer}, {Hartung}, and
  {Sch{\"a}fer}]{tessmer_2013CQG}
M.~{Tessmer}, J.~{Hartung}, and G.~{Sch{\"a}fer}.
\newblock {Aligned spins: orbital elements, decaying orbits, and last stable
  circular orbit to high post-Newtonian orders}.
\newblock \emph{Classical and Quantum Gravity}, 30\penalty0 (1):\penalty0
  015007, January 2013.
\newblock \doi{10.1088/0264-9381/30/1/015007}.

\bibitem[{Wang} et~al.(2011){Wang}, {Steinhoff}, {Zeng}, and
  {Sch{\"a}fer}]{wang_2011PhRvDW}
H.~{Wang}, J.~{Steinhoff}, J.~{Zeng}, and G.~{Sch{\"a}fer}.
\newblock {Leading-order spin-orbit and spin(1)-spin(2) radiation-reaction
  Hamiltonians}.
\newblock \emph{\prd}, 84\penalty0 (12):\penalty0 124005, December 2011.
\newblock \doi{10.1103/PhysRevD.84.124005}.

\bibitem[{Kopejkin}(1988)]{Kopejkin_1988CeMec}
S.~M. {Kopejkin}.
\newblock {Celestial coordinate reference systems in curved space-time}.
\newblock \emph{Celestial Mechanics}, 44:\penalty0 87--115, March 1988.
\newblock \doi{10.1007/BF01230709}.

\bibitem[{Kopeikin}(1989{\natexlab{a}})]{k89o}
S.~M. {Kopeikin}.
\newblock Relativistic frames of reference in the solar system.
\newblock \emph{Soviet Astronomy}, 33:\penalty0 550--555, October
  1989{\natexlab{a}}.

\bibitem[{Kopeikin}(1989{\natexlab{b}})]{k89d}
S.~M. {Kopeikin}.
\newblock Asymptotic matching of gravitational fields in the solar system.
\newblock \emph{Soviet Astronomy}, 33:\penalty0 665--672, December
  1989{\natexlab{b}}.

\bibitem[{Brumberg} and {Kopejkin}(1989{\natexlab{a}})]{bk89}
V.~A. {Brumberg} and S.~M. {Kopejkin}.
\newblock {Relativistic theory of celestial reference frames}.
\newblock In J.~{Kovalevsky}, I.~I. {Mueller}, and B.~{Kolaczek}, editors,
  \emph{{Reference Frames in Astronomy and Geophysics}}, volume 154, pages
  115--141, Amsderdam, 1989{\natexlab{a}}. Astrophysics and Space Science
  Library, Kluwer.

\bibitem[{Brumberg} and {Kopejkin}(1989{\natexlab{b}})]{bk-nc}
V.~A. {Brumberg} and S.~M. {Kopejkin}.
\newblock {Relativistic Reference Systems and Motion of Test Bodies in the
  Vicinity of the Earth}.
\newblock \emph{Nuovo Cimento B Serie}, 103:\penalty0 63--98,
  1989{\natexlab{b}}.

\bibitem[{Damour} et~al.(1991){Damour}, {Soffel}, and {Xu}]{dsx1}
T.~{Damour}, M.~{Soffel}, and C.~{Xu}.
\newblock {General-relativistic celestial mechanics. I. Method and definition
  of reference systems}.
\newblock \emph{\prd}, 43:\penalty0 3273--3307, May 1991.

\bibitem[{Damour} et~al.(1992){Damour}, {Soffel}, and {Xu}]{dsx2}
T.~{Damour}, M.~{Soffel}, and C.~{Xu}.
\newblock {General-relativistic celestial mechanics. II. Translational
  equations of motion}.
\newblock \emph{\prd}, 45:\penalty0 1017--1044, February 1992.

\bibitem[Damour et~al.(1993)Damour, Soffel, and Xu]{dsx3}
T.~Damour, M.~Soffel, and C.~Xu.
\newblock {General-relativistic celestial mechanics. III. Rotational equations
  of motion}.
\newblock \emph{Phys. Rev. D}, 47:\penalty0 3124--3135, Apr 1993.
\newblock \doi{10.1103/PhysRevD.47.3124}.
\newblock URL \url{http://link.aps.org/doi/10.1103/PhysRevD.47.3124}.

\bibitem[{Damour} et~al.(1994){Damour}, {Soffel}, and {Xu}]{dsx4}
T.~{Damour}, M.~{Soffel}, and C.~{Xu}.
\newblock {General-relativistic celestial mechanics. IV. Theory of satellite
  motion}.
\newblock \emph{\prd}, 49:\penalty0 618--635, January 1994.

\bibitem[{Blanchet} and {Damour}(1989)]{bld}
L.~{Blanchet} and T.~{Damour}.
\newblock Post-{N}ewtonian generation of gravitational waves.
\newblock \emph{Ann. Inst. H. Poincar\'e}, 50\penalty0 (4):\penalty0 337--408,
  1989.

\bibitem[{Damour} and {Iyer}(1991{\natexlab{a}})]{dyr2}
T.~{Damour} and B.~R. {Iyer}.
\newblock {Post-newtonian generation of gravitational waves. II. The spin
  moments.}
\newblock \emph{Annales de l'I.~H.~P., section A}, 54\penalty0 (2):\penalty0
  115--164, 1991{\natexlab{a}}.
\newblock \doi{10.1103/PhysRevD.43.3259}.
\newblock URL \url{http://www.numdam.org/article/AIHPA_1991__54_2_115_0.pdf}.

\bibitem[{Damour} and {Iyer}(1991{\natexlab{b}})]{di}
T.~{Damour} and B.~R. {Iyer}.
\newblock {Multipole analysis for electromagnetism and linearized gravity with
  irreducible Cartesian tensors}.
\newblock \emph{\prd}, 43:\penalty0 3259--3272, May 1991{\natexlab{b}}.
\newblock \doi{10.1103/PhysRevD.43.3259}.

\bibitem[{Blanchet} et~al.(2005){Blanchet}, {Damour}, and
  {Iyer}]{blanchet_2005CQG}
L.~{Blanchet}, T.~{Damour}, and B.~R. {Iyer}.
\newblock {Surface-integral expressions for the multipole moments of
  post-Newtonian sources and the boosted Schwarzschild solution}.
\newblock \emph{Classical and Quantum Gravity}, 22:\penalty0 155--181, January
  2005.
\newblock \doi{10.1088/0264-9381/22/1/011}.

\bibitem[{Thorne}(1980)]{thor}
K.~S. {Thorne}.
\newblock {Multipole expansions of gravitational radiation}.
\newblock \emph{Rev. Mod. Phys.}, 52:\penalty0 299--340, April 1980.

\bibitem[{Soffel} et~al.(2003){Soffel}, {Klioner}, {Petit}, {Wolf}, {Kopeikin},
  {Bretagnon}, {Brumberg}, {Capitaine}, {Damour}, {Fukushima}, {Guinot},
  {Huang}, {Lindegren}, {Ma}, {Nordtvedt}, {Ries}, {Seidelmann},
  {Vokrouhlick{\'y}}, {Will}, and {Xu}]{iau2000}
M.~{Soffel}, S.~A. {Klioner}, G.~{Petit}, P.~{Wolf}, S.~M. {Kopeikin},
  P.~{Bretagnon}, V.~A. {Brumberg}, N.~{Capitaine}, T.~{Damour},
  T.~{Fukushima}, B.~{Guinot}, T.-Y. {Huang}, L.~{Lindegren}, C.~{Ma},
  K.~{Nordtvedt}, J.~C. {Ries}, P.~K. {Seidelmann}, D.~{Vokrouhlick{\'y}},
  C.~M. {Will}, and C.~{Xu}.
\newblock {The IAU 2000 Resolutions for Astrometry, Celestial Mechanics, and
  Metrology in the Relativistic Framework: Explanatory Supplement}.
\newblock \emph{Astron. J.}, 126:\penalty0 2687--2706, December 2003.
\newblock \doi{10.1086/378162}.

\bibitem[{Racine} and {Flanagan}(2005)]{racine_2005PhRvD}
{\'E}.~{Racine} and {\'E}.~{\'E}. {Flanagan}.
\newblock {Post-1-Newtonian equations of motion for systems of arbitrarily
  structured bodies}.
\newblock \emph{\prd}, 71\penalty0 (4):\penalty0 044010, February 2005.
\newblock \doi{10.1103/PhysRevD.71.044010}.

\bibitem[{Racine} et~al.(2013){Racine}, {Vines}, and
  {Flanagan}]{racine2013PhRvD}
{\'E}.~{Racine}, J.~E. {Vines}, and {\'E}.~{\'E}. {Flanagan}.
\newblock {Erratum: Post-1-Newtonian equations of motion for systems of
  arbitrarily structured bodies [Phys. Rev. D 71, 044010 (2005)]}.
\newblock \emph{\prd}, 88\penalty0 (8):\penalty0 089903, October 2013.
\newblock \doi{10.1103/PhysRevD.88.089903}.

\bibitem[{Damour} and {Esposito-Farese}(1992)]{DamEsFar}
T.~{Damour} and G.~{Esposito-Farese}.
\newblock {Tensor-multi-scalar theories of gravitation }.
\newblock \emph{Classical and Quantum Gravity}, 9:\penalty0 2093--2176,
  September 1992.

\bibitem[{Kopeikin} and {Vlasov}(2004)]{kovl_2004}
S.~{Kopeikin} and I.~{Vlasov}.
\newblock {Parametrized post-Newtonian theory of reference frames, multipolar
  expansions and equations of motion in the N-body problem}.
\newblock \emph{Phys. Rep.}, 400:\penalty0 209--318, November 2004.
\newblock \doi{10.1016/j.physrep.2004.08.004}.

\bibitem[{Will}(1993)]{willbook}
C.~M. {Will}.
\newblock \emph{{Theory and Experiment in Gravitational Physics}}.
\newblock Cambridge University Press, Cambridge, March 1993.

\bibitem[{Nordtvedt}(2001)]{nordtvedt_2001LNP}
K.~{Nordtvedt}.
\newblock Lunar laser ranging -- a comprehensive probe of the post-{N}ewtonian
  long range interaction.
\newblock In C.~{L{\"a}mmerzahl}, C.~W.~F. {Everitt}, and F.~W. {Hehl},
  editors, \emph{Gyros, Clocks, Interferometers ...: Testing Relativistic
  Gravity in Space}, volume 562 of \emph{Lecture Notes in Physics}, pages
  317--329, Berlin, 2001. Springer.

\bibitem[{Wex}(2014)]{wex2014}
N.~{Wex}.
\newblock {Testing Relativistic Gravity with Radio Pulsars}.
\newblock In S.~M. {Kopeikin}, editor, \emph{{Frontiers in Relativistic
  Celestial Mechanics. Vol. 2. Applications and Experiments}}, pages 39--102,
  Berlin, 2014. De Gruyter.

\bibitem[{Will}(2003)]{Will_2003CQG}
C.~M. {Will}.
\newblock {Testing gravity using space gravitational-wave detectors}.
\newblock \emph{Classical and Quantum Gravity}, 20:\penalty0 S219--S225, May
  2003.
\newblock \doi{10.1088/0264-9381/20/10/325}.

\bibitem[{Gair} et~al.(2013){Gair}, {Vallisneri}, {Larson}, and
  {Baker}]{Gair_2013LRR}
J.~R. {Gair}, M.~{Vallisneri}, S.~L. {Larson}, and J.~G. {Baker}.
\newblock Testing general relativity with low-frequency, space-based
  gravitational-wave detectors.
\newblock \emph{Living Reviews in Relativity}, 16:\penalty0 7, September 2013.
\newblock \doi{10.12942/lrr-2013-7}.

\bibitem[{Yunes} and {Siemens}(2013)]{Yunes_2013LRR}
N.~{Yunes} and X.~{Siemens}.
\newblock Gravitational-wave tests of general relativity with ground-based
  detectors and pulsar-timing arrays.
\newblock \emph{Living Reviews in Relativity}, 16:\penalty0 9, November 2013.
\newblock \doi{10.12942/lrr-2013-9}.

\bibitem[{Lee}(2011)]{Lee_2011AIPC}
K.~J. {Lee}.
\newblock Testing gravity theories in the radiative regime using pulsar timing
  arrays.
\newblock \emph{AIP Conference Proceedings}, 1357:\penalty0 73--76, August
  2011.
\newblock \doi{10.1063/1.3615081}.

\bibitem[{Cornish} et~al.(2018){Cornish}, {O'Beirne}, {Taylor}, and
  {Yunes}]{Cornish_2018PhRvL}
N.~J. {Cornish}, L.~{O'Beirne}, S.~R. {Taylor}, and N.~{Yunes}.
\newblock Constraining alternative theories of gravity using pulsar timing
  arrays.
\newblock \emph{Physical Review Letters}, 120\penalty0 (18):\penalty0 181101,
  May 2018.
\newblock \doi{10.1103/PhysRevLett.120.181101}.

\bibitem[{Brumberg}(1972)]{vab}
V.~A. {Brumberg}.
\newblock \emph{{Relativistic Celestial Mechanics}}.
\newblock Nauka, Moscow (in Russian), 1972.

\bibitem[{Brumberg}(1991)]{brum}
V.~A. {Brumberg}.
\newblock \emph{{Essential Relativistic Celestial Mechanics}}.
\newblock Adam Hilger, New York, 1991.

\bibitem[{Low}(1999)]{Low_1999CQGra}
R.~J. {Low}.
\newblock {Speed limits in general relativity}.
\newblock \emph{Classical and Quantum Gravity}, 16:\penalty0 543--549, February
  1999.
\newblock \doi{10.1088/0264-9381/16/2/016}.

\bibitem[{Battista} et~al.(2017){Battista}, {Esposito}, and
  {Dell'Agnello}]{Battista_2017IJMPA}
E.~{Battista}, G.~{Esposito}, and S.~{Dell'Agnello}.
\newblock {On the foundations of general relativistic celestial mechanics}.
\newblock \emph{International Journal of Modern Physics A}, 32:\penalty0
  1730022-335, September 2017.
\newblock \doi{10.1142/S0217751X17300228}.

\bibitem[{Kopeikin}(2004)]{Kopeikin_2004CQGra}
S.~M. {Kopeikin}.
\newblock {The speed of gravity in general relativity and theoretical
  interpretation of the Jovian deflection experiment}.
\newblock \emph{Classical and Quantum Gravity}, 21:\penalty0 3251--3286, July
  2004.
\newblock \doi{10.1088/0264-9381/21/13/010}.

\bibitem[{Ciufolini} and {Wheeler}(1995)]{ciufolini_book}
I.~{Ciufolini} and J.~A. {Wheeler}.
\newblock \emph{{Gravitation and Inertia}}.
\newblock Princeton University Press, Princeton, 1995.

\bibitem[{Kopeikin}(2001)]{Kopeikin_2001ApJ}
S.~M. {Kopeikin}.
\newblock Testing the relativistic effect of the propagation of gravity by
  {Very Long Baseline Interferometry}.
\newblock \emph{\apjl}, 556:\penalty0 L1--L5, July 2001.
\newblock \doi{10.1086/322872}.

\bibitem[{Fomalont} and {Kopeikin}(2003)]{Fomalont_2003ApJ}
E.~B. {Fomalont} and S.~M. {Kopeikin}.
\newblock The measurement of the light deflection from {Jupiter: E}xperimental
  results.
\newblock \emph{\apj}, 598:\penalty0 704--711, November 2003.
\newblock \doi{10.1086/378785}.

\bibitem[{Cornish} et~al.(2017){Cornish}, {Blas}, and
  {Nardini}]{Cornish_2017PhRvL}
N.~{Cornish}, D.~{Blas}, and G.~{Nardini}.
\newblock Bounding the speed of gravity with gravitational wave observations.
\newblock \emph{Physical Review Letters}, 119\penalty0 (16):\penalty0 161102,
  October 2017.
\newblock \doi{10.1103/PhysRevLett.119.161102}.

\bibitem[{Ehlers}(1980)]{Ehlers_1980NYASA}
J.~{Ehlers}.
\newblock {Isolated systems in general relativity}.
\newblock \emph{Annals of the New York Academy of Sciences}, 336:\penalty0
  279--294, February 1980.
\newblock \doi{10.1111/j.1749-6632.1980.tb15936.x}.

\bibitem[{Frauendiener}(2004)]{frau_2004LRR}
J.~{Frauendiener}.
\newblock Conformal infinity.
\newblock \emph{Living Reviews in Relativity}, 7:\penalty0 1, February 2004.
\newblock \doi{10.12942/lrr-2004-1}.

\bibitem[{Arnowitt} et~al.(1962){Arnowitt}, {Deser}, and {Misner}]{ADM_paper}
R.~{Arnowitt}, S.~{Deser}, and C.~W. {Misner}.
\newblock {The dynamics of general relativity}.
\newblock In L.~{Witten}, editor, \emph{Gravitation: an introduction to current
  research}, pages 227--264, New York, 1962. Wiley.

\bibitem[{Ehlers}(1979)]{ehlers_1979}
J.~{Ehlers}.
\newblock \emph{Isolated Gravitating Systems in General Relativity}.
\newblock North-Holland, Amsterdam, 1979.

\bibitem[{Rendall}(1992{\natexlab{a}})]{Rendall_1992JMP}
A.~D. {Rendall}.
\newblock {The initial value problem for a class of general relativistic fluid
  bodies}.
\newblock \emph{Journal of Mathematical Physics}, 33:\penalty0 1047--1053,
  March 1992{\natexlab{a}}.
\newblock \doi{10.1063/1.529766}.

\bibitem[{Isenberg}(2014)]{Isenberg_2014}
J.~{Isenberg}.
\newblock The initial value problem in general relativity.
\newblock In A.~{Ashtekar} and V.~{Petkov}, editors, \emph{Springer Handbook of
  Spacetime}, page 303. Springer, Berlin, 2014.
\newblock \doi{10.1007/978-3-642-41992-8_16}.

\bibitem[Adamo et~al.(2009)Adamo, Kozameh, and Newman]{Adamo_2009}
T.~M. Adamo, C.~Kozameh, and E.~T. Newman.
\newblock Null geodesic congruences, asymptotically-flat spacetimes and their
  physical interpretation.
\newblock \emph{Living Reviews in Relativity}, 12\penalty0 (1):\penalty0 6, Sep
  2009.
\newblock ISSN 1433-8351.
\newblock \doi{10.12942/lrr-2009-6}.
\newblock URL \url{https://doi.org/10.12942/lrr-2009-6}.

\bibitem[{Faddeev}(1982)]{Faddeev_1982UFN}
L.~D. {Faddeev}.
\newblock {The energy problem in Einstein's theory of gravitation /Dedicated to
  the memory of V. A. Fock/}.
\newblock \emph{Uspekhi Fizicheskikh Nauk}, 136:\penalty0 435--457, March 1982.

\bibitem[{Tegmark}(2002)]{Tegmark_2002PhRvD}
M.~{Tegmark}.
\newblock {Measuring the metric: A parametrized post-Friedmannian approach to
  the cosmic dark energy problem}.
\newblock \emph{\prd}, 66\penalty0 (10):\penalty0 103507, November 2002.
\newblock \doi{10.1103/PhysRevD.66.103507}.

\bibitem[{Kasai}(2007)]{Kasai_2007PThPh}
M.~{Kasai}.
\newblock {Apparent Acceleration through Large-Scale Inhomogeneities ---
  Post-Friedmannian Effects of Inhomogeneities on the Luminosity Distance ---}.
\newblock \emph{Progress of Theoretical Physics}, 117:\penalty0 1067--1075,
  June 2007.
\newblock \doi{10.1143/PTP.117.1067}.

\bibitem[{Clifton} et~al.(2012){Clifton}, {Ferreira}, {Padilla}, and
  {Skordis}]{CLIFTON_2012}
T.~{Clifton}, P.~G. {Ferreira}, A.~{Padilla}, and C.~{Skordis}.
\newblock Modified gravity and cosmology.
\newblock \emph{Physics Reports}, 513\penalty0 (1):\penalty0 1 -- 189, 2012.
\newblock ISSN 0370-1573.
\newblock \doi{https://doi.org/10.1016/j.physrep.2012.01.001}.
\newblock URL
  \url{http://www.sciencedirect.com/science/article/pii/S0370157312000105}.
\newblock Modified Gravity and Cosmology.

\bibitem[{Skordis} et~al.(2015){Skordis}, {Pourtsidou}, and
  {Copeland}]{Skordis_2015PhRvD}
C.~{Skordis}, A.~{Pourtsidou}, and E.~J. {Copeland}.
\newblock {Parametrized post-Friedmannian framework for interacting dark energy
  theories}.
\newblock \emph{\prd}, 91\penalty0 (8):\penalty0 083537, April 2015.
\newblock \doi{10.1103/PhysRevD.91.083537}.

\bibitem[{Milillo} et~al.(2015){Milillo}, {Bertacca}, {Bruni}, and
  {Maselli}]{bruni_2015PhRvD}
I.~{Milillo}, D.~{Bertacca}, M.~{Bruni}, and A.~{Maselli}.
\newblock {Missing link: A nonlinear post-Friedmann framework for small and
  large scales}.
\newblock \emph{\prd}, 92\penalty0 (2):\penalty0 023519, July 2015.
\newblock \doi{10.1103/PhysRevD.92.023519}.

\bibitem[{Rampf} et~al.(2016){Rampf}, {Villa}, {Bertacca}, and
  {Bruni}]{bruni_2016PhRvD}
C.~{Rampf}, E.~{Villa}, D.~{Bertacca}, and M.~{Bruni}.
\newblock {Lagrangian theory for cosmic structure formation with vorticity:
  Newtonian and post-Friedmann approximations}.
\newblock \emph{\prd}, 94\penalty0 (8):\penalty0 083515, October 2016.
\newblock \doi{10.1103/PhysRevD.94.083515}.

\bibitem[{Petrov} et~al.(2017){Petrov}, {Kopeikin}, {Lompay}, and
  {Tekin}]{Petrov_2017book}
A.~N. {Petrov}, S.~M. {Kopeikin}, R.~R. {Lompay}, and B.~{Tekin}.
\newblock \emph{Metric Theories of Gravity: {P}erturbations and Conservation
  Laws}.
\newblock De Gruyter, Berlin, April 2017.

\bibitem[{Ram{\'{\i}}rez} and {Kopeikin}(2002)]{Ramirez_2002PhLB}
J.~{Ram{\'{\i}}rez} and S.~{Kopeikin}.
\newblock {A decoupled system of hyperbolic equations for linearized
  cosmological perturbations}.
\newblock \emph{Physics Letters B}, 532:\penalty0 1--7, April 2002.
\newblock \doi{10.1016/S0370-2693(02)01471-5}.

\bibitem[{Kopeikin} and {Petrov}(2013)]{Kopeikin_2013PhRvD}
S.~M. {Kopeikin} and A.~N. {Petrov}.
\newblock {Post-Newtonian celestial dynamics in cosmology: Field equations}.
\newblock \emph{\prd}, 87\penalty0 (4):\penalty0 044029, February 2013.
\newblock \doi{10.1103/PhysRevD.87.044029}.

\bibitem[{Kopeikin} and {Petrov}(2014)]{Kopeikin_2014AnPhy}
S.~M. {Kopeikin} and A.~N. {Petrov}.
\newblock {Dynamic field theory and equations of motion in cosmology}.
\newblock \emph{Annals of Physics}, 350:\penalty0 379--440, November 2014.
\newblock \doi{10.1016/j.aop.2014.07.029}.

\bibitem[{Kopeikin}(2012)]{Kopeikin_2012PhRvD}
S.~M. {Kopeikin}.
\newblock {Celestial ephemerides in an expanding universe}.
\newblock \emph{\prd}, 86\penalty0 (6):\penalty0 064004, September 2012.
\newblock \doi{10.1103/PhysRevD.86.064004}.

\bibitem[{Galiautdinov} and {Kopeikin}(2016)]{galkop_2016PhRvD}
A.~{Galiautdinov} and S.~M. {Kopeikin}.
\newblock {Post-Newtonian celestial mechanics in scalar-tensor cosmology}.
\newblock \emph{\prd}, 94\penalty0 (4):\penalty0 044015, August 2016.
\newblock \doi{10.1103/PhysRevD.94.044015}.

\bibitem[{M{\"a}dler} and {Winicour}(2016)]{winicour_2016}
T.~{M{\"a}dler} and J.~{Winicour}.
\newblock {Bondi-Sachs Formalism}.
\newblock \emph{Scholarpedia}, 11:\penalty0 33528, 2016.
\newblock URL \url{http://dx.doi.org/10.4249/scholarpedia.335281}.

\bibitem[{Fock}(1959)]{fockbook}
V.~A. {Fock}.
\newblock \emph{{The Theory of Space, Time and Gravitation}}.
\newblock Pergamon Press, New York, February 1959.

\bibitem[{Spyrou}(1975)]{spyrou_1975ApJ}
N.~{Spyrou}.
\newblock {The N-body problem in general relativity}.
\newblock \emph{\apj}, 197:\penalty0 725--743, May 1975.
\newblock \doi{10.1086/153562}.

\bibitem[{Arminjon}(2005)]{arminjon_2005PhRvD}
M.~{Arminjon}.
\newblock {Equations of motion according to the asymptotic post-Newtonian
  scheme for general relativity in the harmonic gauge}.
\newblock \emph{\prd}, 72\penalty0 (8):\penalty0 084002, October 2005.
\newblock \doi{10.1103/PhysRevD.72.084002}.

\bibitem[{Racine}(2006)]{Racine_2006CQG}
{\'E}.~{Racine}.
\newblock {Spin and energy evolution equations for a wide class of extended
  bodies}.
\newblock \emph{Classical and Quantum Gravity}, 23:\penalty0 373--390, January
  2006.
\newblock \doi{10.1088/0264-9381/23/2/007}.

\bibitem[{Havas} and {Goldberg}(1962)]{Havas_1962PhRv}
P.~{Havas} and J.~N. {Goldberg}.
\newblock Lorentz-invariant equations of motion of point masses in the general
  theory of relativity.
\newblock \emph{Physical Review}, 128:\penalty0 398--414, October 1962.
\newblock \doi{10.1103/PhysRev.128.398}.

\bibitem[{Zhang}(1985)]{Zhang_1985PhRvD}
X.-H. {Zhang}.
\newblock {Higher-order corrections to the laws of motion and precession for
  black holes and other bodies}.
\newblock \emph{\prd}, 31:\penalty0 3130--3134, June 1985.
\newblock \doi{10.1103/PhysRevD.31.3130}.

\bibitem[{Memmesheimer} and {Sch{\"a}fer}(2005)]{memmesh_2005PhRvD}
R.-M. {Memmesheimer} and G.~{Sch{\"a}fer}.
\newblock {Third post-Newtonian constrained canonical dynamics for binary point
  masses in harmonic coordinates}.
\newblock \emph{\prd}, 71\penalty0 (4):\penalty0 044021, February 2005.
\newblock \doi{10.1103/PhysRevD.71.044021}.

\bibitem[{Kopeikin} et~al.(1999){Kopeikin}, {Sch{\"a}fer}, {Gwinn}, and
  {Eubanks}]{kopeikin_1999PhRvD}
S.~M. {Kopeikin}, G.~{Sch{\"a}fer}, C.~R. {Gwinn}, and T.~M. {Eubanks}.
\newblock {Astrometric and timing effects of gravitational waves from localized
  sources}.
\newblock \emph{\prd}, 59\penalty0 (8):\penalty0 084023, April 1999.
\newblock \doi{10.1103/PhysRevD.59.084023}.

\bibitem[{Papapetrou}(1951{\natexlab{b}})]{pap1}
A.~{Papapetrou}.
\newblock Equations of motion in general relativity.
\newblock \emph{Proceedings of the Physical Society A}, 64:\penalty0 57--75,
  January 1951{\natexlab{b}}.

\bibitem[{Dixon}(2008)]{dixon_2008}
W.~G. {Dixon}.
\newblock Mathisson's "{New Mechanics}": Its aims and realisation.
\newblock \emph{Acta Physica Polonica B Proceedings Supplement}, 1:\penalty0
  27--54, 2008.
\newblock URL
  \url{http://www.actaphys.uj.edu.pl/fulltext?series=Sup&vol=1&page=27}.

\bibitem[Dixon(2015)]{Dixon2015}
W.~G. Dixon.
\newblock {The New Mechanics of Myron Mathisson and Its Subsequent
  Development}.
\newblock In D.~Puetzfeld, C.~L{\"a}mmerzahl, and B.~Schutz, editors,
  \emph{Equations of Motion in Relativistic Gravity}, pages 1--66, Cham, 2015.
  Springer.
\newblock ISBN 978-3-319-18335-0.
\newblock \doi{10.1007/978-3-319-18335-0_1}.
\newblock URL \url{https://doi.org/10.1007/978-3-319-18335-0_1}.

\bibitem[{Taub}(1965)]{Taub_1965}
A.~H. {Taub}.
\newblock {The motion of multipoles in general relativity}.
\newblock In G.~Barb\`era, editor, \emph{IV Centenario Della Nascita di Galileo
  Galilei, 1564-1964}, pages 100--118, Firenze, 1965. Pubblicazioni del
  Comitato Nazionale per le Manifestazioni Celebrative.

\bibitem[{Madore}(1969)]{madore_1969}
J.~{Madore}.
\newblock {The equations of motion of an extended body in general relativity}.
\newblock \emph{Annales de l'I.H.P. Physique th\'eorique}, 11\penalty0
  (2):\penalty0 221--237, 1969.
\newblock \doi{10.1103/PhysRevD.80.084018}.
\newblock URL \url{http://www.numdam.org/article/AIHPA_1969__11_2_221_0.pdf}.

\bibitem[{Ehlers} and {Rudolph}(1977)]{ehlers_1977GReGr}
J.~{Ehlers} and E.~{Rudolph}.
\newblock {Dynamics of extended bodies in general relativity center-of-mass
  description and quasirigidity}.
\newblock \emph{General Relativity and Gravitation}, 8:\penalty0 197--217,
  March 1977.
\newblock \doi{10.1007/BF00763547}.

\bibitem[{Schattner}(1979)]{schattner_1979GReGr}
R.~{Schattner}.
\newblock {The center of mass in general relativity}.
\newblock \emph{General Relativity and Gravitation}, 10:\penalty0 377--393,
  April 1979.
\newblock \doi{10.1007/BF00760221}.

\bibitem[{Harte}(2012)]{Harte_2012}
A.~I. {Harte}.
\newblock Mechanics of extended masses in general relativity.
\newblock \emph{Classical and Quantum Gravity}, 29\penalty0 (5):\penalty0
  055012, feb 2012.
\newblock \doi{10.1088/0264-9381/29/5/055012}.
\newblock URL \url{https://doi.org/10.1088%2F0264-9381%2F29%2F5%2F055012}.

\bibitem[{Harte}(2015)]{harte2015}
A.~I. {Harte}.
\newblock {Motion in Classical Field Theories and the Foundations of the
  Self-force Problem}.
\newblock In D.~Puetzfeld, C.~L{\"a}mmerzahl, and B.~Schutz, editors,
  \emph{Equations of Motion in Relativistic Gravity}, pages 327--398, Cham,
  2015. Springer.
\newblock ISBN 978-3-319-18335-0.
\newblock \doi{10.1007/978-3-319-18335-0_12}.
\newblock URL \url{https://doi.org/10.1007/978-3-319-18335-0_12}.

\bibitem[{Puetzfeld} and {Obukhov}(2013{\natexlab{a}})]{dirk_2013PhLA}
D.~{Puetzfeld} and Y.~N. {Obukhov}.
\newblock {Unraveling gravity beyond Einstein with extended test bodies}.
\newblock \emph{Physics Letters A}, 377:\penalty0 2447--2449, November
  2013{\natexlab{a}}.
\newblock \doi{10.1016/j.physleta.2013.07.024}.

\bibitem[{Obukhov} and {Puetzfeld}(2014)]{Obukhov_Puetzfeld2014}
Y.~N. {Obukhov} and D.~{Puetzfeld}.
\newblock {Equations of motion in scalar-tensor theories of gravity: A
  covariant multipolar approach}.
\newblock \emph{\prd}, 90\penalty0 (10):\penalty0 104041, November 2014.
\newblock \doi{10.1103/PhysRevD.90.104041}.

\bibitem[{Puetzfeld} and {Obukhov}(2014)]{dirk_obukhov2014}
D.~{Puetzfeld} and Y.~N. {Obukhov}.
\newblock Equations of motion in metric-affine gravity: A covariant unified
  framework.
\newblock \emph{Phys. Rev. D}, 90:\penalty0 084034, Oct 2014.
\newblock \doi{10.1103/PhysRevD.90.084034}.
\newblock URL \url{https://link.aps.org/doi/10.1103/PhysRevD.90.084034}.

\bibitem[{Nutku}(1969{\natexlab{a}})]{nutku_1969}
Y.~{Nutku}.
\newblock The post-{N}ewtonian equations of hydrodynamics in the {Brans-Dicke}
  theory.
\newblock \emph{Astrophys. J.}, 155:\penalty0 999--1007, March
  1969{\natexlab{a}}.

\bibitem[{Nutku}(1969{\natexlab{b}})]{nutku_1969ApJ}
Y.~{Nutku}.
\newblock {The Energy-Momentum Complex in the Brans-Dicke Theory}.
\newblock \emph{\apj}, 158:\penalty0 991--996, December 1969{\natexlab{b}}.
\newblock \doi{10.1086/150258}.

\bibitem[{Yasskin} and {Stoeger}(1980)]{yasskin_1980PhRvD}
P.~B. {Yasskin} and W.~R. {Stoeger}.
\newblock {Propagation equations for test bodies with spin and rotation in
  theories of gravity with torsion}.
\newblock \emph{\prd}, 21:\penalty0 2081--2094, April 1980.
\newblock \doi{10.1103/PhysRevD.21.2081}.

\bibitem[{Mao} et~al.(2007){Mao}, {Tegmark}, {Guth}, and {Cabi}]{mao_2007PhRvD}
Y.~{Mao}, M.~{Tegmark}, A.~H. {Guth}, and S.~{Cabi}.
\newblock {Constraining torsion with Gravity Probe B}.
\newblock \emph{\prd}, 76\penalty0 (10):\penalty0 104029, November 2007.
\newblock \doi{10.1103/PhysRevD.76.104029}.

\bibitem[{March} et~al.(2011){March}, {Bellettini}, {Tauraso}, and
  {Dell'Agnello}]{march_2011GReGr}
R.~{March}, G.~{Bellettini}, R.~{Tauraso}, and S.~{Dell'Agnello}.
\newblock {Constraining spacetime torsion with LAGEOS}.
\newblock \emph{General Relativity and Gravitation}, 43:\penalty0 3099--3126,
  November 2011.
\newblock \doi{10.1007/s10714-011-1226-2}.

\bibitem[{Flanagan} and {Rosenthal}(2007)]{flanagan_2007PhRvD}
{\'E}.~{\'E}. {Flanagan} and E.~{Rosenthal}.
\newblock {Can Gravity Probe B usefully constrain torsion gravity theories?}
\newblock \emph{\prd}, 75\penalty0 (12):\penalty0 124016, June 2007.
\newblock \doi{10.1103/PhysRevD.75.124016}.

\bibitem[{Hehl} et~al.(2013){Hehl}, {Obukhov}, and {Puetzfeld}]{hehl_2013PhLA}
F.~W. {Hehl}, Y.~N. {Obukhov}, and D.~{Puetzfeld}.
\newblock {On Poincar{\'e} gauge theory of gravity, its equations of motion,
  and Gravity Probe B}.
\newblock \emph{Physics Letters A}, 377:\penalty0 1775--1781, October 2013.
\newblock \doi{10.1016/j.physleta.2013.04.055}.

\bibitem[{Puetzfeld} and {Obukhov}(2013{\natexlab{b}})]{dirk_2013PhRvD}
D.~{Puetzfeld} and Y.~N. {Obukhov}.
\newblock {Equations of motion in gravity theories with nonminimal coupling: A
  loophole to detect torsion macroscopically?}
\newblock \emph{\prd}, 88\penalty0 (6):\penalty0 064025, September
  2013{\natexlab{b}}.
\newblock \doi{10.1103/PhysRevD.88.064025}.

\bibitem[{Damour}(1987)]{Damour_1987book}
T.~{Damour}.
\newblock {The problem of motion in Newtonian and Einsteinian gravity.}
\newblock In S.~W. {Hawking} and W.~{Israel}, editors, \emph{Three Hundred
  Years of Gravitation}, pages 128--198. Cambridge University Press, Cambridge,
  1987.

\bibitem[{Schmidt} et~al.(1975){Schmidt}, {Walker}, and
  {Sommers}]{Schmidt_1975GReGr}
B.~{Schmidt}, M.~{Walker}, and P.~{Sommers}.
\newblock {A characterization of the Bondi-Metzner-Sachs group}.
\newblock \emph{General Relativity and Gravitation}, 6:\penalty0 489--497,
  October 1975.
\newblock \doi{10.1007/BF00762453}.

\bibitem[{Ashby} and {Bertotti}(1986)]{ashb2}
N.~{Ashby} and B.~{Bertotti}.
\newblock {Relativistic effects in local inertial frames}.
\newblock \emph{\prd}, 34:\penalty0 2246--2259, October 1986.

\bibitem[{Dubrovin} et~al.(1984){Dubrovin}, {Fomenko}, and {Novikov}]{dfn}
B.~A. {Dubrovin}, A.~T. {Fomenko}, and S.~P. {Novikov}.
\newblock \emph{{Modern Geometry - Methods and Applications.}}
\newblock Springer, New York, 1984.

\bibitem[{Arnold}(1995)]{arno}
V.~I. {Arnold}.
\newblock \emph{{Mathematical Methods of Classical Mechanics}}.
\newblock Berlin, Springer, 1995.

\bibitem[{Soffel} and {Langhans}(2013)]{Soffel_2013book}
M.~{Soffel} and R.~{Langhans}.
\newblock \emph{{Space-Time Reference Systems}}.
\newblock Springer, Berlin, 2013.
\newblock \doi{10.1007/978-3-642-30226-8}.

\bibitem[Faraoni and Gunzig(1999)]{Faraoni_1999ijtp}
V.~Faraoni and E.~Gunzig.
\newblock Einstein frame or {J}ordan frame?
\newblock \emph{International Journal of Theoretical Physics}, 38\penalty0
  (1):\penalty0 217--225, Jan 1999.
\newblock ISSN 1572-9575.
\newblock \doi{10.1023/A:1026645510351}.
\newblock URL \url{https://doi.org/10.1023/A:1026645510351}.

\bibitem[{Bhadra} et~al.(2007){Bhadra}, {Sarkar}, {Datta}, and
  {Nandi}]{Bhandra_2007MPLA}
A.~{Bhadra}, K.~{Sarkar}, D.~P. {Datta}, and K.~K. {Nandi}.
\newblock {Brans-Dicke theory: Jordan versus Einstein frame}.
\newblock \emph{Modern Physics Letters A}, 22:\penalty0 367--375, 2007.
\newblock \doi{10.1142/S021773230702261X}.

\bibitem[{Flanagan}(2004)]{Flanagan_2004cqg}
E.~{Flanagan}.
\newblock The conformal frame freedom in theories of gravitation.
\newblock \emph{Classical and Quantum Gravity}, 21\penalty0 (15):\penalty0
  3817, 2004.
\newblock URL \url{http://stacks.iop.org/0264-9381/21/i=15/a=N02}.

\bibitem[{Zschocke}(2014)]{Zschocke_2014}
S.~{Zschocke}.
\newblock {A detailed proof of the fundamental theorem of STF multipole
  expansion in linearized gravity}.
\newblock \emph{Int. J. Mod. Phys. D}, 23\penalty0 (01):\penalty0 1450003,
  2014.
\newblock \doi{10.1142/S0218271814500035}.
\newblock URL
  \url{https://www.worldscientific.com/doi/abs/10.1142/S0218271814500035}.

\bibitem[{Synge}(1964)]{syngebook}
J.~L. {Synge}.
\newblock \emph{{Relativity: The general theory}}.
\newblock Series in Physics. North-Holland, Amsterdam, 1964.

\bibitem[{Misner} et~al.(1973){Misner}, {Thorne}, and {Wheeler}]{mtw}
C.~W. {Misner}, K.~S. {Thorne}, and J.~A. {Wheeler}.
\newblock \emph{{Gravitation}}.
\newblock W.H.~Freeman, San Francisco, 1973.

\bibitem[{Weinberg}(1972)]{weinberg_book1972}
S.~{Weinberg}.
\newblock \emph{Gravitation and Cosmology}.
\newblock J. Wiley \& Sons, New York, July 1972.

\bibitem[Bekenstein(2004)]{Bekenstein2004}
J.~D. Bekenstein.
\newblock Relativistic gravitation theory for the modified {N}ewtonian dynamics
  paradigm.
\newblock \emph{\prd}, 70:\penalty0 083509, Oct 2004.
\newblock \doi{10.1103/PhysRevD.70.083509}.
\newblock URL \url{http://link.aps.org/doi/10.1103/PhysRevD.70.083509}.

\bibitem[{Jordan}(1949)]{1949Natur.164..637J}
P.~{Jordan}.
\newblock {Formation of the stars and development of the Universe}.
\newblock \emph{\nat}, 164:\penalty0 637--640, 1949.

\bibitem[{Jordan}(1959)]{1959ZPhy..157..112J}
P.~{Jordan}.
\newblock {Zum gegenw{\"a}rtigen Stand der Diracschen kosmologischen
  Hypothesen}.
\newblock \emph{Zeitschrift fur Physik}, 157:\penalty0 112--121, February 1959.
\newblock \doi{10.1007/BF01375155}.

\bibitem[{Fierz}(1956)]{1956AcHPh.29..128F}
M.~{Fierz}.
\newblock {On the physical interpretation of P. Jordan's extended theory of
  gravitation}.
\newblock \emph{Helv. Phys. Acta}, 29:\penalty0 128--134, 1956.

\bibitem[{Brans} and {Dicke}(1961)]{1961PhRv..124..925B}
C.~{Brans} and R.~H. {Dicke}.
\newblock Mach's principle and a relativistic theory of gravitation.
\newblock \emph{Physical Review}, 124:\penalty0 925--935, November 1961.
\newblock \doi{10.1103/PhysRev.124.925}.

\bibitem[{Dicke}(1962{\natexlab{a}})]{1962PhRv..125.2163D}
R.~H. {Dicke}.
\newblock {Mach's Principle and Invariance under Transformation of Units}.
\newblock \emph{Physical Review}, 125:\penalty0 2163--2167, March
  1962{\natexlab{a}}.
\newblock \doi{10.1103/PhysRev.125.2163}.

\bibitem[{Dicke}(1962{\natexlab{b}})]{1962PhRv..126.1875D}
R.~H. {Dicke}.
\newblock Long-range scalar interaction.
\newblock \emph{Physical Review}, 126:\penalty0 1875--1877, June
  1962{\natexlab{b}}.
\newblock \doi{10.1103/PhysRev.126.1875}.

\bibitem[{Dittmaier} and {Schumacher}(2013)]{Dittmaier_2013PrPNP}
S.~{Dittmaier} and M.~{Schumacher}.
\newblock {The Higgs boson in the Standard Model -- From LEP to LHC:
  Expectations, Searches, and Discovery of a Candidate}.
\newblock \emph{Progress in Particle and Nuclear Physics}, 70:\penalty0 1--54,
  May 2013.
\newblock \doi{10.1016/j.ppnp.2013.02.001}.

\bibitem[{Dehnen} et~al.(1992){Dehnen}, {Frommert}, and
  {Ghaboussi}]{dehnen_1992IJTP}
H.~{Dehnen}, H.~{Frommert}, and F.~{Ghaboussi}.
\newblock {Higgs field and a new scalar-tensor theory of gravity}.
\newblock \emph{International Journal of Theoretical Physics}, 31:\penalty0
  109--114, January 1992.
\newblock \doi{10.1007/BF00674344}.

\bibitem[{Eisenhart}(1947)]{eisen}
L.~P. {Eisenhart}.
\newblock \emph{{Differential Geometry}}.
\newblock Princeton University Press, Princeton, 1947.

\bibitem[{Damour} and {Esposito-Farese}(1993)]{damesf1993}
T.~{Damour} and G.~{Esposito-Farese}.
\newblock {Nonperturbative strong-field effects in tensor-scalar theories of
  gravitation}.
\newblock \emph{Physical Review Letters}, 70:\penalty0 2220--2223, April 1993.

\bibitem[{Will}(2011)]{Will_2011PNAS}
C.~M. {Will}.
\newblock {On the unreasonable effectiveness of the post-Newtonian
  approximation in gravitational physics}.
\newblock \emph{Proceedings of the National Academy of Science}, 108:\penalty0
  5938--5945, April 2011.
\newblock \doi{10.1073/pnas.1103127108}.

\bibitem[{Ellis} and {Uzan}(2005)]{uzan_2005AmJPh}
G.~F.~R. {Ellis} and J.-P. {Uzan}.
\newblock {c is the speed of light, isn't it?}
\newblock \emph{American Journal of Physics}, 73:\penalty0 240--247, March
  2005.
\newblock \doi{10.1119/1.1819929}.

\bibitem[{Kates} and {Kegeles}(1982)]{kake}
R.~E. {Kates} and L.~S. {Kegeles}.
\newblock {Nonanalytic terms in the slow-motion expansion of a radiating scalar
  field on a Schwarzschild background.}
\newblock \emph{\prd}, 25:\penalty0 2030--2037, April 1982.
\newblock \doi{10.1103/PhysRevD.25.2030}.

\bibitem[{Rendall}(1992{\natexlab{b}})]{rend}
A.~D. {Rendall}.
\newblock {On the definition of post-Newtonian approximations}.
\newblock \emph{Royal Society of London Proceedings Series A}, 438:\penalty0
  341--360, August 1992{\natexlab{b}}.

\bibitem[{Faraoni} et~al.(2018){Faraoni}, {Hammad}, {Cardini}, and
  {Gobeil}]{faraoni_2018PRD}
V.~{Faraoni}, F.~{Hammad}, A.~M. {Cardini}, and T.~{Gobeil}.
\newblock {Revisiting the analogue of the Jebsen-Birkhoff theorem in
  Brans-Dicke gravity}.
\newblock \emph{\prd}, 97\penalty0 (8):\penalty0 084033, April 2018.
\newblock \doi{10.1103/PhysRevD.97.084033}.

\bibitem[{Kopeikin}(1985)]{k85}
S.~M. {Kopeikin}.
\newblock General relativistic equations of binary motion for extended bodies
  with conservative corrections and radiation damping.
\newblock \emph{Soviet Astronomy}, 29:\penalty0 516--524, October 1985.

\bibitem[{Damour}(1983)]{Damour_1983grr}
T.~{Damour}.
\newblock {Gravitational radiation and the motion of compact bodies}.
\newblock In N.~{Deruelle} and T.~{Piran}, editors, \emph{Gravitational
  Radiation}, pages 59--144, Amsterdam, 1983. North-Holland.

\bibitem[{Kopeikin} and {Vlasov}(2008)]{kovl_2008}
S.~{Kopeikin} and I.~{Vlasov}.
\newblock The effacing principle in the post-{N}ewtonian celestial mechanics.
\newblock In H.~{Kleinert}, R.~T. {Jantzen}, and R.~{Ruffini}, editors,
  \emph{The 11-th MG Meeting On Recent Developments in Theoretical and
  Experimental General Relativity}, pages 2475--2477, Singapore, September
  2008. World Scientific Publishing.
\newblock \doi{10.1142/9789812834300_0437}.

\bibitem[{Mitchell} and {Will}(2007)]{mitchell_2007PhRvD}
T.~{Mitchell} and C.~M. {Will}.
\newblock {Post-Newtonian gravitational radiation and equations of motion via
  direct integration of the relaxed Einstein equations. V. Evidence for the
  strong equivalence principle to second post-Newtonian order}.
\newblock \emph{\prd}, 75\penalty0 (12):\penalty0 124025, June 2007.
\newblock \doi{10.1103/PhysRevD.75.124025}.

\bibitem[{Zharkov} and {Trubitsyn}(1978)]{Zharkov_1978book}
V.~N. {Zharkov} and V.~P. {Trubitsyn}.
\newblock \emph{{Physics of planetary interiors}}.
\newblock Astronomy and Astrophysics Series, Pachart, Tucson, 1978.

\bibitem[{Cheng}(1991)]{loven1}
Z.~{Cheng}.
\newblock {Relation between the {L}ove numbers and the {E}arth models.}
\newblock \emph{Journal Nanjing Univ}, 27:\penalty0 234--242, April 1991.

\bibitem[{Getino}(1993)]{getino}
J.~{Getino}.
\newblock {Perturbed nutations, {L}ove numbers and elastic energy of
  deformation for Earth models 1066A and 1066B}.
\newblock \emph{Zeitschrift Angewandte Mathematik und Physik}, 44:\penalty0
  998--1021, November 1993.
\newblock \doi{10.1007/BF00942762}.

\bibitem[{Yip} and {Leung}(2017)]{Yip_2017}
K.~L.~S. {Yip} and P.~T. {Leung}.
\newblock Tidal {L}ove numbers and moment-{L}ove relations of polytropic stars.
\newblock \emph{Monthly Notices of the Royal Astronomical Society},
  472\penalty0 (4):\penalty0 4965--4981, 2017.
\newblock \doi{10.1093/mnras/stx2363}.
\newblock URL \url{http://dx.doi.org/10.1093/mnras/stx2363}.

\bibitem[{Damour} and {Nordtvedt}(1993)]{1993PhRvL..70.2217D}
T.~{Damour} and K.~{Nordtvedt}.
\newblock {General relativity as a cosmological attractor of tensor-scalar
  theories}.
\newblock \emph{Physical Review Letters}, 70:\penalty0 2217--2219, April 1993.

\bibitem[{Hofmann} and {M{\"u}ller}(2018)]{LLR_2018CQGra}
F.~{Hofmann} and J.~{M{\"u}ller}.
\newblock {Relativistic tests with lunar laser ranging}.
\newblock \emph{Classical and Quantum Gravity}, 35\penalty0 (3):\penalty0
  035015, February 2018.
\newblock \doi{10.1088/1361-6382/aa8f7a}.

\bibitem[{Garc{\'{\i}}a-Berro} et~al.(2011){Garc{\'{\i}}a-Berro},
  {Lor{\'e}n-Aguilar}, {Torres}, {Althaus}, and {Isern}]{WD_2011JCAP}
E.~{Garc{\'{\i}}a-Berro}, P.~{Lor{\'e}n-Aguilar}, S.~{Torres}, L.~G. {Althaus},
  and J.~{Isern}.
\newblock {An upper limit to the secular variation of the gravitational
  constant from white dwarf stars}.
\newblock \emph{\jcap}, 5:\penalty0 021, May 2011.
\newblock \doi{10.1088/1475-7516/2011/05/021}.

\bibitem[{Pitjeva} and {Pitjev}(2013)]{Pitjeva_2013MNRAS}
E.~V. {Pitjeva} and N.~P. {Pitjev}.
\newblock {Relativistic effects and dark matter in the solar system from
  observations of planets and spacecraft}.
\newblock \emph{\mnras}, 432:\penalty0 3431--3437, July 2013.
\newblock \doi{10.1093/mnras/stt695}.

\bibitem[{Futamase} and {Schutz}(1983)]{futamase_1983PhRvD}
T.~{Futamase} and B.~F. {Schutz}.
\newblock {Newtonian and post-Newtonian approximations are asymptotic to
  general relativity}.
\newblock \emph{\prd}, 28:\penalty0 2363--2372, November 1983.
\newblock \doi{10.1103/PhysRevD.28.2363}.

\bibitem[{Gibbings}(2011)]{Gibbings_2011}
J.~C. {Gibbings}.
\newblock \emph{Dimensional Analysis}.
\newblock Springer, London, 2011.
\newblock ISBN 978-1-84996-316-9.
\newblock \doi{10.1007/978-1-84996-317-6}.

\bibitem[{Klioner} and {Soffel}(2000)]{2000PhRvD..62b4019K}
S.~A. {Klioner} and M.~H. {Soffel}.
\newblock {Relativistic celestial mechanics with PPN parameters}.
\newblock \emph{\prd}, 62\penalty0 (2):\penalty0 024019, July 2000.

\bibitem[{Dittus} et~al.(2008){Dittus}, {L{\"a}mmerzahl}, and
  {Turyshev}]{2008ASSL..349.....D}
H.~{Dittus}, C.~{L{\"a}mmerzahl}, and S.~G. {Turyshev}, editors.
\newblock \emph{{Lasers, Clocks and Drag-Free Control: Exploration of
  Relativistic Gravity in Space}}, volume 349 of \emph{Astrophysics and Space
  Science Library}, Berlin, 2008. Springer.

\bibitem[{Appourchaux} et~al.(2009){Appourchaux}, {Burston}, {Chen}, {Cruise},
  {Dittus}, {Foulon}, {Gill}, {Gizon}, {Klein}, {Klioner}, {Kopeikin},
  {Kr{\"u}ger}, {L{\"a}mmerzahl}, {Lobo}, {Luo}, {Margolis}, {Ni}, {Pat{\'o}n},
  {Peng}, {Peters}, {Rasel}, {R{\"u}diger}, {Samain}, {Selig}, {Shaul},
  {Sumner}, {Theil}, {Touboul}, {Turyshev}, {Wang}, {Wang}, {Wen}, {Wicht},
  {Wu}, {Zhang}, and {Zhao}]{astrod_2009ExA}
T.~{Appourchaux}, R.~{Burston}, Y.~{Chen}, M.~{Cruise}, H.~{Dittus},
  B.~{Foulon}, P.~{Gill}, L.~{Gizon}, H.~{Klein}, S.~{Klioner}, S.~{Kopeikin},
  H.~{Kr{\"u}ger}, C.~{L{\"a}mmerzahl}, A.~{Lobo}, X.~{Luo}, H.~{Margolis},
  W.-T. {Ni}, A.~P. {Pat{\'o}n}, Q.~{Peng}, A.~{Peters}, E.~{Rasel},
  A.~{R{\"u}diger}, {\'E}.~{Samain}, H.~{Selig}, D.~{Shaul}, T.~{Sumner},
  S.~{Theil}, P.~{Touboul}, S.~{Turyshev}, H.~{Wang}, L.~{Wang}, L.~{Wen},
  A.~{Wicht}, J.~{Wu}, X.~{Zhang}, and C.~{Zhao}.
\newblock {Astrodynamical Space Test of Relativity Using Optical Devices I
  (ASTROD I) -- A class-M fundamental physics mission proposal for Cosmic
  Vision 2015-2025}.
\newblock \emph{Experimental Astronomy}, 23:\penalty0 491--527, March 2009.
\newblock \doi{10.1007/s10686-008-9131-8}.

\bibitem[{Ciufolini}(2010)]{ciufolini_2008}
I.~{Ciufolini}.
\newblock Frame-dragging, gravitomagnetism and lunar laser ranging.
\newblock \emph{New Astronomy}, 15:\penalty0 332--337, March 2010.
\newblock \doi{10.1016/j.newast.2009.08.004}.

\bibitem[{Kopeikin} et~al.(2008){Kopeikin}, {Pavlis}, {Pavlis}, {Brumberg},
  {Escapa}, {Getino}, {Gusev}, {M{\"u}ller}, {Ni}, and
  {Petrova}]{2008AdSpR..42.1378K}
S.~M. {Kopeikin}, E.~{Pavlis}, D.~{Pavlis}, V.~A. {Brumberg}, A.~{Escapa},
  J.~{Getino}, A.~{Gusev}, J.~{M{\"u}ller}, W.-T. {Ni}, and N.~{Petrova}.
\newblock {Prospects in the orbital and rotational dynamics of the Moon with
  the advent of sub-centimeter lunar laser ranging}.
\newblock \emph{Advances in Space Research}, 42:\penalty0 1378--1390, October
  2008.
\newblock \doi{10.1016/j.asr.2008.02.014}.

\bibitem[{Dell'Agnello} et~al.(2012){Dell'Agnello}, {Maiello}, {Currie},
  {Boni}, {Berardi}, {Cantone}, {Delle Monache}, {Intaglietta}, {Lops},
  {Garattini}, {Martini}, {Patrizi}, {Porcelli}, {Tibuzzi}, {Vittori},
  {Bianco}, {Coradini}, {Dionisio}, {March}, {Bellettini}, {Tauraso}, and
  {Chandler}]{2012NIMPA}
S.~{Dell'Agnello}, M.~{Maiello}, D.~G. {Currie}, A.~{Boni}, S.~{Berardi},
  C.~{Cantone}, G.~O. {Delle Monache}, N.~{Intaglietta}, C.~{Lops},
  M.~{Garattini}, M.~{Martini}, G.~{Patrizi}, L.~{Porcelli}, M.~{Tibuzzi},
  R.~{Vittori}, G.~{Bianco}, A.~{Coradini}, C.~{Dionisio}, R.~{March},
  G.~{Bellettini}, R.~{Tauraso}, and J.~{Chandler}.
\newblock Probing general relativity and new physics with lunar laser ranging.
\newblock \emph{Nuclear Instruments and Methods in Physics Research A},
  692:\penalty0 275--279, November 2012.
\newblock \doi{10.1016/j.nima.2012.01.002}.

\bibitem[{Murphy}(2013)]{Murthy_2013RPPh}
T.~W. {Murphy}.
\newblock {Lunar laser ranging: the millimeter challenge}.
\newblock \emph{Reports on Progress in Physics}, 76\penalty0 (7):\penalty0
  076901, July 2013.
\newblock \doi{10.1088/0034-4885/76/7/076901}.

\bibitem[{Baiotti}(2016)]{Baiotti_2016JPhCS}
L.~{Baiotti}.
\newblock {Modeling of neutron-star mergers: a review while awaiting
  gravitational-wave detection}.
\newblock \emph{Journal of Physics Conference Series}, 759:\penalty0 012004,
  October 2016.
\newblock \doi{10.1088/1742-6596/759/1/012004}.
\newblock URL
  \url{http://iopscience.iop.org/article/10.1088/1742-6596/759/1/012004/pdf}.

\bibitem[{Efroimsky} and {Goldreich}(2003)]{efroim_2003JMP}
M.~{Efroimsky} and P.~{Goldreich}.
\newblock {Gauge symmetry of the N-body problem in the Hamilton-Jacobi
  approach}.
\newblock \emph{Journal of Mathematical Physics}, 44:\penalty0 5958--5977,
  December 2003.
\newblock \doi{10.1063/1.1622447}.

\bibitem[{Efroimsky} and {Goldreich}(2004)]{efroim_2004A&A}
M.~{Efroimsky} and P.~{Goldreich}.
\newblock {Gauge freedom in the N-body problem of celestial mechanics}.
\newblock \emph{\aap}, 415:\penalty0 1187--1199, March 2004.
\newblock \doi{10.1051/0004-6361:20034058}.

\bibitem[{Tulczyjew}(1959)]{tulczyjew1}
W.~{Tulczyjew}.
\newblock {Motion of multipole particles in general relativity theory}.
\newblock \emph{Acta Physica Polonica}, 18:\penalty0 393--409, 1959.

\bibitem[{Tulczyjew} and {Tulczyjew}(1962)]{tulczyjew1_1962}
B.~{Tulczyjew} and W.~{Tulczyjew}.
\newblock {On multipole formalism in general relativity}.
\newblock In \emph{{Recent Developments in General Relativity. A collection of
  papers dedicated to Leopold Infeld. }}, pages 465--472. Pergamon Press, New
  York, 1962.

\bibitem[{Papapetrou}(1951{\natexlab{c}})]{pap2}
A.~{Papapetrou}.
\newblock Equations of motion in general relativity: Ii. {The} coordinate
  condition.
\newblock \emph{Proceedings of the Physical Society A}, 64:\penalty0 302--310,
  March 1951{\natexlab{c}}.

\bibitem[{Bini} et~al.(2009){Bini}, {Cherubini}, {Geralico}, and
  {Ortolan}]{bini_2009GReGr}
D.~{Bini}, C.~{Cherubini}, A.~{Geralico}, and A.~{Ortolan}.
\newblock {Dixon's extended bodies and weak gravitational waves}.
\newblock \emph{General Relativity and Gravitation}, 41:\penalty0 105--116,
  January 2009.
\newblock \doi{10.1007/s10714-008-0657-x}.

\bibitem[{Costa} et~al.(2012){Costa}, {Herdeiro}, {Nat{\'a}rio}, and
  {Zilh{\~a}o}]{helical_2012PhRvD}
L.~F. {Costa}, C.~{Herdeiro}, J.~{Nat{\'a}rio}, and M.~{Zilh{\~a}o}.
\newblock {Mathisson's helical motions for a spinning particle: Are they
  unphysical?}
\newblock \emph{\prd}, 85\penalty0 (2):\penalty0 024001, January 2012.
\newblock \doi{10.1103/PhysRevD.85.024001}.

\bibitem[{Shilov}(1968)]{shilov_1968}
G.~E. {Shilov}.
\newblock \emph{{Generalized Functions and Partial Differential Equations:
  Mathematics and its Applications}}.
\newblock {(translated by B. Seckler}. Gordon \& Breach, Philadelphia, 1968.

\bibitem[{Gorbonos} and {Kol}(2004)]{gorbonos_2004JHEP}
D.~{Gorbonos} and B.~{Kol}.
\newblock A dialogue of multipoles: {M}atched asymptotic expansion for caged
  black holes.
\newblock \emph{Journal of High Energy Physics}, 6:\penalty0 053, June 2004.
\newblock \doi{10.1088/1126-6708/2004/06/053}.

\bibitem[{Futamase} et~al.(2008){Futamase}, {Hogan}, and
  {Itoh}]{futamase_2008PhRvD}
T.~{Futamase}, P.~A. {Hogan}, and Y.~{Itoh}.
\newblock {Equations of motion in general relativity of a small charged black
  hole}.
\newblock \emph{\prd}, 78\penalty0 (10):\penalty0 104014, November 2008.
\newblock \doi{10.1103/PhysRevD.78.104014}.

\bibitem[{Lukes-Gerakopoulos} et~al.(2014){Lukes-Gerakopoulos}, {Seyrich}, and
  {Kunst}]{Lukes_2014PhRvD}
G.~{Lukes-Gerakopoulos}, J.~{Seyrich}, and D.~{Kunst}.
\newblock {Investigating spinning test particles: Spin supplementary conditions
  and the Hamiltonian formalism}.
\newblock \emph{\prd}, 90\penalty0 (10):\penalty0 104019, November 2014.
\newblock \doi{10.1103/PhysRevD.90.104019}.

\bibitem[{Mik{\'o}czi}(2017)]{Mikoczi_2017PhRvD}
B.~{Mik{\'o}czi}.
\newblock {Spin supplementary conditions for spinning compact binaries}.
\newblock \emph{\prd}, 95\penalty0 (6):\penalty0 064023, March 2017.
\newblock \doi{10.1103/PhysRevD.95.064023}.

\bibitem[{Costa} et~al.(2018){Costa}, {Lukes-Gerakopoulos}, and
  {Semer{\'a}k}]{Costa_2018PhRvD}
L.~F.~O. {Costa}, G.~{Lukes-Gerakopoulos}, and O.~{Semer{\'a}k}.
\newblock {Spinning particles in general relativity: Momentum-velocity relation
  for the Mathisson-Pirani spin condition}.
\newblock \emph{\prd}, 97\penalty0 (8):\penalty0 084023, April 2018.
\newblock \doi{10.1103/PhysRevD.97.084023}.

\bibitem[{Fichtenholtz}(1950)]{Fichte_1950}
I.~G. {Fichtenholtz}.
\newblock {Lagrangian form of equations of motion in the second approximation
  of Einstein's theory of gravity}.
\newblock \emph{Journal of Experimental and Theoretical Physics (JETP},
  20\penalty0 (3):\penalty0 233 -- 242, 1950.

\bibitem[{Fitchett}(1983)]{Fitchett_1983MNRAS}
M.~J. {Fitchett}.
\newblock {The influence of gravitational wave momentum losses on the centre of
  mass motion of a Newtonian binary system}.
\newblock \emph{\mnras}, 203:\penalty0 1049--1062, June 1983.
\newblock \doi{10.1093/mnras/203.4.1049}.

\bibitem[{Grishchuk} and {Kopeikin}(1986)]{gk86}
L.~P. {Grishchuk} and S.~M. {Kopeikin}.
\newblock {Equations of motion for isolated bodies with relativistic
  corrections including the radiation reaction force}.
\newblock In J.~{Kovalevsky} and V.~A. {Brumberg}, editors, \emph{Relativity in
  Celestial Mechanics and Astrometry. High Precision Dynamical Theories and
  Observational Verifications}, volume 114 of \emph{IAU Symposium}, pages
  19--33, Dordrecht, 1986. Kluwer.

\bibitem[{Damour} and {Sch{\"a}fer}(1985)]{Damour_1985GReGr}
T.~{Damour} and G.~{Sch{\"a}fer}.
\newblock {Lagrangians for N point masses at the second post-Newtonian
  approximation of general relativity.}
\newblock \emph{General Relativity and Gravitation}, 17:\penalty0 879--905,
  September 1985.
\newblock \doi{10.1007/BF00773685}.

\bibitem[{Damour} et~al.(1989){Damour}, {Grishchuk}, {Kopejkin}, and
  {Sch{\"a}fer}]{Damour_1989MG5}
T.~{Damour}, L.~P. {Grishchuk}, S.~M. {Kopejkin}, and G.~{Sch{\"a}fer}.
\newblock {Higher-order relativistic dynamics of binary systems}.
\newblock In D.~G. {Blair} and M.~J. {Buckingham}, editors, \emph{The Fifth
  Marcel Grossmann Meeting. Part A.}, pages 451 -- 459, Singapore, 1989. World
  Scientific.

\bibitem[{D'Eath}(1996)]{DEath_1996book}
P.~D. {D'Eath}.
\newblock \emph{{Black holes: gravitational interactions}}.
\newblock Oxford University Press, New York, 1996.

\bibitem[{Kovalevsky} and {Mueller}(1989)]{1989rfag.conf....1K}
J.~{Kovalevsky} and I.~I. {Mueller}.
\newblock {Reference frames in astronomy and geophysics: Introduction.}
\newblock In \emph{Reference Frames in Astronomy and Geophysics}, pages 1--12,
  Amsterdam, 1989. Kluwer.

\bibitem[{Kovalevsky} and {Seidelmann}(2004)]{2004fuas.book.....K}
J.~{Kovalevsky} and P.~K. {Seidelmann}.
\newblock \emph{{Fundamentals of Astrometry}}.
\newblock Cambridge University Press, Cambridge, July 2004.

\bibitem[{Hofmann} et~al.(2010{\natexlab{a}}){Hofmann}, {M{\"u}ller}, and
  {Biskupek}]{Hofman_2010}
F.~{Hofmann}, J.~{M{\"u}ller}, and L.~{Biskupek}.
\newblock {Lunar laser ranging test of the Nordtvedt parameter and a possible
  variation in the gravitational constant}.
\newblock \emph{\aap}, 522:\penalty0 L5, November 2010{\natexlab{a}}.
\newblock \doi{10.1051/0004-6361/201015659}.

\bibitem[{Geroch}(1970)]{geroch_1970JMP}
R.~{Geroch}.
\newblock {Multipole Moments. II. {C}urved Space}.
\newblock \emph{Journal of Mathematical Physics}, 11:\penalty0 2580--2588,
  August 1970.
\newblock \doi{10.1063/1.1665427}.

\bibitem[{Hansen}(1974)]{hansen_1974JMP}
R.~O. {Hansen}.
\newblock {Multipole moments of stationary space-times}.
\newblock \emph{Journal of Mathematical Physics}, 15:\penalty0 46--52, January
  1974.
\newblock \doi{10.1063/1.1666501}.

\bibitem[{Quevedo}(1990)]{quevedo_1990ForPh}
H.~{Quevedo}.
\newblock Multipole moments in general relativity -- static and stationary
  vacuum solutions.
\newblock \emph{Fortschritte der Physik}, 38:\penalty0 733--840, 1990.
\newblock \doi{10.1002/prop.2190381002}.

\bibitem[{G{\"u}rsel}(1983)]{gursel_1983GRG}
Y.~{G{\"u}rsel}.
\newblock {Multipole moments for stationary systems: {T}he equivalence of the
  {G}eroch-{H}ansen formulation and the {T}horne formulation}.
\newblock \emph{{Gen. Rel. Grav.}}, 15:\penalty0 737--754, August 1983.
\newblock \doi{10.1007/BF01031881}.

\bibitem[{Lagerstrom}(1989)]{Lagerstrom_89}
P.~A. {Lagerstrom}.
\newblock \emph{{Matched Asymptotic Expansions: Ideas and Techniques}}.
\newblock Springer, Berlin, 1989.
\newblock URL \url{https://www.springer.com/us/book/9780387968117}.

\bibitem[{Demia{\'n}ski} and {Grishchuk}(1974)]{demgrish_1974GRG}
M.~{Demia{\'n}ski} and L.~P. {Grishchuk}.
\newblock {Note on the motion of black holes}.
\newblock \emph{General Relativity and Gravitation}, 5:\penalty0 673--679,
  December 1974.
\newblock \doi{10.1007/BF00761925}.

\bibitem[{Brumberg} and {Kopeikin}(1990)]{1990CeMDA..48...23B}
V.~A. {Brumberg} and S.~M. {Kopeikin}.
\newblock {Relativistic time scales in the solar system}.
\newblock \emph{Celestial Mechanics and Dynamical Astronomy}, 48:\penalty0
  23--44, March 1990.
\newblock \doi{10.1007/BF00050674}.

\bibitem[{Xie} and {Kopeikin}(2010)]{xie_2010AcPSl}
Y.~{Xie} and S.~{Kopeikin}.
\newblock Post-{N}ewtonian reference frames for advanced theory of the lunar
  motion and a new generation of lunar laser ranging.
\newblock \emph{Acta Physica Slovaca}, 60:\penalty0 393--495, August 2010.
\newblock URL
  \url{http://www.physics.sk/aps/pubs/2010/aps-10-04/aps-10-04.pdf}.

\bibitem[{Fisher}(1972)]{Fisher_1972AmJPh}
G.~P. {Fisher}.
\newblock {The Thomas Precession}.
\newblock \emph{American Journal of Physics}, 40:\penalty0 1772--1781, December
  1972.
\newblock \doi{10.1119/1.1987061}.

\bibitem[{Will}(2015)]{GPB_2015CQGra}
C.~M. {Will}.
\newblock {Focus issue: Gravity Probe B}.
\newblock \emph{Classical and Quantum Gravity}, 32\penalty0 (22):\penalty0
  220301, November 2015.
\newblock \doi{10.1088/0264-9381/32/22/220301}.

\bibitem[{Ciufolini} et~al.(2012){Ciufolini}, {Pavlis}, {Paolozzi}, {Ries},
  {Koenig}, {Matzner}, {Sindoni}, and {Neumayer}]{ciufolini_2012NewA}
I.~{Ciufolini}, E.~C. {Pavlis}, A.~{Paolozzi}, J.~{Ries}, R.~{Koenig},
  R.~{Matzner}, G.~{Sindoni}, and K.~H. {Neumayer}.
\newblock Phenomenology of the {Lense-Thirring} effect in the solar system:
  {M}easurement of frame-dragging with laser ranged satellites.
\newblock \emph{New Astronomy}, 17:\penalty0 341--346, April 2012.
\newblock \doi{10.1016/j.newast.2011.08.003}.

\bibitem[{Ciufolini} et~al.(2016){Ciufolini}, {Paolozzi}, {Pavlis}, {Koenig},
  {Ries}, {Gurzadyan}, {Matzner}, {Penrose}, {Sindoni}, {Paris}, {Khachatryan},
  and {Mirzoyan}]{Ciufolini_2016EPJC}
I.~{Ciufolini}, A.~{Paolozzi}, E.~C. {Pavlis}, R.~{Koenig}, J.~{Ries},
  V.~{Gurzadyan}, R.~{Matzner}, R.~{Penrose}, G.~{Sindoni}, C.~{Paris},
  H.~{Khachatryan}, and S.~{Mirzoyan}.
\newblock {A test of general relativity using the LARES and LAGEOS satellites
  and a GRACE Earth gravity model. Measurement of Earth's dragging of inertial
  frames}.
\newblock \emph{European Physical Journal C}, 76:\penalty0 120, March 2016.
\newblock \doi{10.1140/epjc/s10052-016-3961-8}.

\bibitem[{Veledina} et~al.(2013){Veledina}, {Poutanen}, and
  {Ingram}]{Veledina_2013ApJ}
A.~{Veledina}, J.~{Poutanen}, and A.~{Ingram}.
\newblock A unified {L}ense-{T}hirring precession model for optical and x-ray
  quasi-periodic oscillations in black hole binaries.
\newblock \emph{\apj}, 778:\penalty0 165, December 2013.
\newblock \doi{10.1088/0004-637X/778/2/165}.

\bibitem[{Buonanno} et~al.(2003){Buonanno}, {Chen}, and
  {Vallisneri}]{buonanna_2003PhRvD}
A.~{Buonanno}, Y.~{Chen}, and M.~{Vallisneri}.
\newblock {Detecting gravitational waves from precessing binaries of spinning
  compact objects: Adiabatic limit}.
\newblock \emph{\prd}, 67\penalty0 (10):\penalty0 104025, May 2003.
\newblock \doi{10.1103/PhysRevD.67.104025}.

\bibitem[{Buonanno} et~al.(2006){Buonanno}, {Chen}, and
  {Vallisneri}]{err_2006PhRvD}
A.~{Buonanno}, Y.~{Chen}, and M.~{Vallisneri}.
\newblock {Erratum: Detecting gravitational waves from precessing binaries of
  spinning compact objects: Adiabatic limit [Phys. Rev. D 67, 104025 (2003)]}.
\newblock \emph{\prd}, 74\penalty0 (2):\penalty0 029904, July 2006.
\newblock \doi{10.1103/PhysRevD.74.029904}.

\bibitem[{Gupta} and {Gopakumar}(2015)]{Gupta_2015CQGra}
A.~{Gupta} and A.~{Gopakumar}.
\newblock {Post-Newtonian analysis of a precessing convention for spinning
  compact binaries}.
\newblock \emph{Classical and Quantum Gravity}, 32\penalty0 (17):\penalty0
  175002, September 2015.
\newblock \doi{10.1088/0264-9381/32/17/175002}.

\bibitem[{Harry} et~al.(2016){Harry}, {Privitera}, {Boh{\'e}}, and
  {Buonanno}]{Harry_2016PhRvD}
I.~{Harry}, S.~{Privitera}, A.~{Boh{\'e}}, and A.~{Buonanno}.
\newblock {Searching for gravitational waves from compact binaries with
  precessing spins}.
\newblock \emph{\prd}, 94\penalty0 (2):\penalty0 024012, July 2016.
\newblock \doi{10.1103/PhysRevD.94.024012}.

\bibitem[{Harte}(2008{\natexlab{a}})]{harte2008_2}
A.~I. {Harte}.
\newblock {Self-forces from generalized Killing fields}.
\newblock \emph{Classical and Quantum Gravity}, 25\penalty0 (23):\penalty0
  235020, December 2008{\natexlab{a}}.
\newblock \doi{10.1088/0264-9381/25/23/235020}.

\bibitem[{Detweiler}(2011)]{Detweiler_2011mmgr}
S.~{Detweiler}.
\newblock {Elementary Development of the Gravitational Self-Force}.
\newblock In L.~{Blanchet}, A.~{Spallicci}, and B.~{Whiting}, editors,
  \emph{Mass and Motion in General Relativity}, pages 271--307. Springer,
  Berlin, 2011.
\newblock \doi{10.1007/978-90-481-3015-3_10}.

\bibitem[{Wald}(2011)]{Wald_2011mmgr}
R.~M. {Wald}.
\newblock {Introduction to Gravitational Self-Force}.
\newblock In L.~{Blanchet}, A.~{Spallicci}, and B.~{Whiting}, editors,
  \emph{Mass and Motion in General Relativity}, pages 253--262. Springer,
  Berlin, 2011.
\newblock \doi{10.1007/978-90-481-3015-3_8}.

\bibitem[Pound(2015)]{Pound_2015}
A.~Pound.
\newblock Motion of small objects in curved spacetimes: An introduction to
  gravitational self-force.
\newblock In D.~Puetzfeld, C.~L{\"a}mmerzahl, and B.~Schutz, editors,
  \emph{Equations of Motion in Relativistic Gravity}, pages 399--486, Berlin,
  2015. Springer.
\newblock \doi{10.1007/978-3-319-18335-0_13}.
\newblock URL \url{https://doi.org/10.1007/978-3-319-18335-0_13}.
\newblock http://adsabs.harvard.edu/abs/2015arXiv150606245P.

\bibitem[{Mirshekari} and {Will}(2013)]{Mirshekari_2013PhRvD87h4070M}
S.~{Mirshekari} and C.~M. {Will}.
\newblock {Compact binary systems in scalar-tensor gravity: Equations of motion
  to 2.5 post-Newtonian order}.
\newblock \emph{\prd}, 87\penalty0 (8):\penalty0 084070, April 2013.
\newblock \doi{10.1103/PhysRevD.87.084070}.

\bibitem[{Chandrasekhar} and {Esposito}(1970)]{Chandra_1970ApJ}
S.~{Chandrasekhar} and F.~P. {Esposito}.
\newblock The 2{$\frac12$}-post-{N}ewtonian equations of hydrodynamics and
  radiation reaction in general relativity.
\newblock \emph{Astrophys. J.}, 160:\penalty0 153--179, April 1970.
\newblock \doi{10.1086/150414}.

\bibitem[{Grishchuk} and {Kopeikin}(1983)]{gk_1983SvAL}
L.~P. {Grishchuk} and S.~M. {Kopeikin}.
\newblock The motion of a pair of gravitating bodies including the radiation
  reaction force.
\newblock \emph{Soviet Astronomy Letters}, 9:\penalty0 230--232, April 1983.

\bibitem[{Sch{\"a}fer}(1985)]{schaefer_1985AnPhy}
G.~{Sch{\"a}fer}.
\newblock {The gravitational quadrupole radiation-reaction force and the
  canonical formalism of ADM}.
\newblock \emph{Annals of Physics}, 161:\penalty0 81--100, April 1985.
\newblock \doi{10.1016/0003-4916(85)90337-9}.

\bibitem[{Babak, S. and Gair, J.~R. and Cole, R.~H.}(2015)]{Babak2015}
{Babak, S. and Gair, J.~R. and Cole, R.~H.}
\newblock Extreme mass ratio inspirals: Perspectives for their detection.
\newblock In D.~Puetzfeld, C.~L{\"a}mmerzahl, and B.~Schutz, editors,
  \emph{Equations of Motion in Relativistic Gravity}, pages 783--812, Cham,
  2015. Springer International Publishing.
\newblock ISBN 978-3-319-18335-0.
\newblock \doi{10.1007/978-3-319-18335-0_23}.
\newblock URL \url{https://doi.org/10.1007/978-3-319-18335-0_23}.

\bibitem[{Wardell, B.}(2015)]{wardell2015}
{Wardell, B.}
\newblock {Self-force: Computational Strategies}.
\newblock In D.~Puetzfeld, C.~L{\"a}mmerzahl, and B.~Schutz, editors,
  \emph{Equations of Motion in Relativistic Gravity}, pages 487--522, Cham,
  2015. Springer International Publishing.
\newblock ISBN 978-3-319-18335-0.
\newblock \doi{10.1007/978-3-319-18335-0_14}.
\newblock URL \url{https://doi.org/10.1007/978-3-319-18335-0_14}.

\bibitem[{Chicone} et~al.(2001){Chicone}, {Kopeikin}, {Mashhoon}, and
  {Retzloff}]{chicone_2001PhLA}
C.~{Chicone}, S.~M. {Kopeikin}, B.~{Mashhoon}, and D.~G. {Retzloff}.
\newblock {Delay equations and radiation damping}.
\newblock \emph{Physics Letters A}, 285:\penalty0 17--26, June 2001.
\newblock \doi{10.1016/S0375-9601(01)00327-9}.

\bibitem[{Detweiler} and {Whiting}(2003)]{Detweiler_2003PRD}
S.~{Detweiler} and B.~F. {Whiting}.
\newblock {Self-force via a {G}reen's function decomposition}.
\newblock \emph{\prd}, 67\penalty0 (2):\penalty0 024025, January 2003.
\newblock \doi{10.1103/PhysRevD.67.024025}.

\bibitem[{Whiting} and {Detweiler}(2003)]{Whiting_2003IJMPD}
B.~F. {Whiting} and S.~{Detweiler}.
\newblock {Radiation Reaction and the Principle of Equivalence}.
\newblock \emph{International Journal of Modern Physics D}, 12:\penalty0
  1709--1713, 2003.
\newblock \doi{10.1142/S0218271803004109}.

\bibitem[{Ni} and {Zimmermann}(1978)]{Ni_1978PRD}
W.-T. {Ni} and M.~{Zimmermann}.
\newblock {Inertial and gravitational effects in the proper reference frame of
  an accelerated, rotating observer}.
\newblock \emph{\prd}, 17:\penalty0 1473--1476, March 1978.
\newblock \doi{10.1103/PhysRevD.17.1473}.

\bibitem[{Puetzfeld} and {Obukhov}(2015)]{dirk_obukhov_2015PhRvD}
D.~{Puetzfeld} and Y.~N. {Obukhov}.
\newblock {Equivalence principle in scalar-tensor gravity}.
\newblock \emph{\prd}, 92\penalty0 (8):\penalty0 081502, October 2015.
\newblock \doi{10.1103/PhysRevD.92.081502}.

\bibitem[{Petrova}(1949)]{petrova}
N.~M. {Petrova}.
\newblock {On equations of motion and tensor of matter for a system of finite
  masses in general theory of relativity}.
\newblock \emph{Zh. Exp. Theor. Phys.}, 19:\penalty0 989--999, 1949.

\bibitem[Tolman(1930)]{PhysRev.35.875}
R.~C. Tolman.
\newblock On the use of the energy-momentum principle in general relativity.
\newblock \emph{Phys. Rev.}, 35:\penalty0 875--895, Apr 1930.
\newblock \doi{10.1103/PhysRev.35.875}.
\newblock URL \url{http://link.aps.org/doi/10.1103/PhysRev.35.875}.

\bibitem[{Dicke}(1965)]{1965AnPhy..31..235D}
R.~H. {Dicke}.
\newblock {The weak and strong principles of equivalence}.
\newblock \emph{Annals of Physics}, 31:\penalty0 235--239, January 1965.
\newblock \doi{10.1016/0003-4916(65)90239-3}.

\bibitem[{Nordtvedt}(1973)]{1973PhRvD...7.2347N}
K.~{Nordtvedt}.
\newblock Post-{N}ewtonian gravitational effects in lunar laser ranging.
\newblock \emph{\prd}, 7:\penalty0 2347--2356, April 1973.
\newblock \doi{10.1103/PhysRevD.7.2347}.

\bibitem[{Kopejkin}(1991)]{kopejkin_1991INTSA}
S.~M. {Kopejkin}.
\newblock {Relativistic reference frames in the solar system}.
\newblock \emph{Itogi Nauki i Tekhniki Seriia Astronomiia}, 41:\penalty0
  87--146, 1991.

\bibitem[{Gradshteyn} and {Ryzhik}(1965)]{gradryzh}
I.~S. {Gradshteyn} and I.~M. {Ryzhik}.
\newblock \emph{{Table of integrals, series and products, 4-th ed., edited by
  Geronimus, Yu.V. \& Tseytlin, M.Yu.}}
\newblock Academic Press, New York, 1965.
\newblock URL \url{http://adsabs.harvard.edu/abs/1965tisp.book.....G}.
\newblock First appeared in 1942 as MT15 in the Mathematical tables series of
  the National Bureau of Standards.

\bibitem[{Barker} and {O'Connell}(1975)]{Barker_Oconnell_1975PhRvD}
B.~M. {Barker} and R.~F. {O'Connell}.
\newblock {Gravitational two-body problem with arbitrary masses, spins, and
  quadrupole moments}.
\newblock \emph{\prd}, 12:\penalty0 329--335, July 1975.
\newblock \doi{10.1103/PhysRevD.12.329}.

\bibitem[{Barker} and {O'Connell}(1976)]{Barker_Oconnell_1976PhRvD}
B.~M. {Barker} and R.~F. {O'Connell}.
\newblock {Lagrangian-Hamiltonian formalism for the gravitational two-body
  problem with spin and parametrized post-Newtonian parameters gamma and beta}.
\newblock \emph{\prd}, 14:\penalty0 861--869, August 1976.
\newblock \doi{10.1103/PhysRevD.14.861}.

\bibitem[{Barker} and {O'Connell}(1987)]{Barker_Oconnell1987JMP}
B.~M. {Barker} and R.~F. {O'Connell}.
\newblock {On the completion of the post-Newtonian gravitational two-body
  problem with spin.}
\newblock \emph{Journal of Mathematical Physics}, 28:\penalty0 661--667, 1987.
\newblock \doi{10.1063/1.527600}.

\bibitem[{Futamase} and {Schutz}(1985)]{futamase_1985PhRvD}
T.~{Futamase} and B.~F. {Schutz}.
\newblock {Gravitational radiation and the validity of the far-zone quadrupole
  formula in the Newtonian limit of general relativity}.
\newblock \emph{\prd}, 32:\penalty0 2557--2565, November 1985.
\newblock \doi{10.1103/PhysRevD.32.2557}.

\bibitem[{Tolman}(1987)]{Tolman_book}
R.~C. {Tolman}.
\newblock \emph{{Relativity, thermodynamics and cosmology}}.
\newblock Dover, Mineola, 1987.

\bibitem[{Spyrou}(1978)]{spyrou_1978GReGr}
N.~{Spyrou}.
\newblock {Relativistic equations of motion of extended bodies}.
\newblock \emph{General Relativity and Gravitation}, 9:\penalty0 519--530, June
  1978.
\newblock \doi{10.1007/BF00759546}.

\bibitem[{Spyrou}(1979)]{spyrou_1979GRG}
N.~{Spyrou}.
\newblock {Relativistic effects in many-body systems of finite size, internal
  structure, and internal motions. I. 'Self-acceleration' of astrophysical
  systems}.
\newblock \emph{General Relativity and Gravitation}, 10:\penalty0 581--598, May
  1979.
\newblock \doi{10.1007/BF00757209}.

\bibitem[{Spyrou}(1981)]{spyrou_1981GRG}
N.~{Spyrou}.
\newblock {Relativistic effects in many-body systems of finite size, internal
  structure, and internal motions. II. The determination of the inertial and
  rest masses of binary stars}.
\newblock \emph{General Relativity and Gravitation}, 13:\penalty0 487--493, May
  1981.
\newblock \doi{10.1007/BF00756596}.

\bibitem[{Caporali}(1981{\natexlab{a}})]{caporali_1981_1}
A.~{Caporali}.
\newblock {A reformulation of the post-Newtonian approximation to general
  relativity. I. The metric and the local equations of motion}.
\newblock \emph{Nuovo Cimento B Serie}, 61:\penalty0 181--212, February
  1981{\natexlab{a}}.
\newblock \doi{10.1007/BF02721322}.

\bibitem[{Caporali}(1981{\natexlab{b}})]{caporali_1981_2}
A.~{Caporali}.
\newblock {A reformulation of the post-Newtonian approximation to general
  relativity. II. Post-Newtonian equations of motion for extended bodies.}
\newblock \emph{Nuovo Cimento B Serie}, 61:\penalty0 205--212, February
  1981{\natexlab{b}}.
\newblock \doi{10.1007/BF02721323}.

\bibitem[{Dallas}(1977)]{dallas_1977CeMec}
S.~S. {Dallas}.
\newblock {Equations of motion for rotating finite bodies in the extended PPN
  formalism}.
\newblock \emph{Celestial Mechanics}, 15:\penalty0 111--123, February 1977.
\newblock \doi{10.1007/BF01229052}.

\bibitem[{Vincent}(1986)]{vincent_1986CeMec}
M.~A. {Vincent}.
\newblock {The relativistic equations of motion for a satellite in orbit about
  a finite-size, rotating earth}.
\newblock \emph{Celestial Mechanics}, 39:\penalty0 15--21, May 1986.
\newblock \doi{10.1007/BF01232285}.

\bibitem[{Nordtvedt}(1994)]{nordtvedt_1994PhRvD}
K.~{Nordtvedt}.
\newblock {Gravitational equation of motion of spherical extended bodies}.
\newblock \emph{\prd}, 49:\penalty0 5165--5172, May 1994.
\newblock \doi{10.1103/PhysRevD.49.5165}.

\bibitem[{Xu} et~al.(2001){Xu}, {Wu}, and {Soffel}]{xws1}
C.~{Xu}, X.~{Wu}, and M.~{Soffel}.
\newblock {General-relativistic theory of elastic deformable astronomical
  bodies}.
\newblock \emph{\prd}, 63\penalty0 (4):\penalty0 043002, February 2001.
\newblock \doi{10.1103/PhysRevD.63.043002}.

\bibitem[{Xu} et~al.(2005){Xu}, {Wu}, and {Soffel}]{xws2}
C.~{Xu}, X.~{Wu}, and M.~{Soffel}.
\newblock {General-relativistic perturbation equations for the dynamics of
  elastic deformable astronomical bodies expanded in terms of generalized
  spherical harmonics}.
\newblock \emph{\prd}, 71\penalty0 (2):\penalty0 024030, January 2005.
\newblock \doi{10.1103/PhysRevD.71.024030}.

\bibitem[{Xu} et~al.(2003){Xu}, {Wu}, {Soffel}, and {Klioner}]{xwsk}
C.~{Xu}, X.~{Wu}, M.~{Soffel}, and S.~{Klioner}.
\newblock {Relativistic theory of elastic deformable astronomical bodies:
  Perturbation equations in rotating spherical coordinates and junction
  conditions}.
\newblock \emph{\prd}, 68\penalty0 (6):\penalty0 064009, September 2003.
\newblock \doi{10.1103/PhysRevD.68.064009}.

\bibitem[{M{\"u}ller} et~al.(2008){M{\"u}ller}, {Williams}, and
  {Turyshev}]{LLR_2008ASSL}
J.~{M{\"u}ller}, J.~G. {Williams}, and S.~G. {Turyshev}.
\newblock Lunar laser ranging contributions to relativity and geodesy.
\newblock In H.~{Dittus}, C.~{L{\"a}mmerzahl}, and S.~G. {Turyshev}, editors,
  \emph{Lasers, Clocks and Drag-Free Control: Exploration of Relativistic
  Gravity in Space}, volume 349 of \emph{Astrophysics and Space Science
  Library}, pages 457--472, Berlin, 2008. Springer.

\bibitem[{Hofmann} et~al.(2010{\natexlab{b}}){Hofmann}, {M{\"u}ller}, and
  {Biskupek}]{LLR_2010A&A}
F.~{Hofmann}, J.~{M{\"u}ller}, and L.~{Biskupek}.
\newblock {Lunar laser ranging test of the Nordtvedt parameter and a possible
  variation in the gravitational constant}.
\newblock \emph{\aap}, 522:\penalty0 L5, November 2010{\natexlab{b}}.
\newblock \doi{10.1051/0004-6361/201015659}.

\bibitem[{Fomalont} et~al.(2009){Fomalont}, {Kopeikin}, {Lanyi}, and
  {Benson}]{Fomalont_2009ApJ}
E.~{Fomalont}, S.~{Kopeikin}, G.~{Lanyi}, and J.~{Benson}.
\newblock Progress in measurements of the gravitational bending of radio waves
  using the {VLBA}.
\newblock \emph{\apj}, 699:\penalty0 1395--1402, July 2009.
\newblock \doi{10.1088/0004-637X/699/2/1395}.

\bibitem[{Bertotti} et~al.(2003){Bertotti}, {Iess}, and
  {Tortora}]{Bertotti_2003Natur}
B.~{Bertotti}, L.~{Iess}, and P.~{Tortora}.
\newblock {A test of general relativity using radio links with the Cassini
  spacecraft}.
\newblock \emph{\nat}, 425:\penalty0 374--376, September 2003.
\newblock \doi{10.1038/nature01997}.

\bibitem[{Kopeikin} et~al.(2007){Kopeikin}, {Polnarev}, {Sch{\"a}fer}, and
  {Vlasov}]{Kopeikin_2007PhLA}
S.~M. {Kopeikin}, A.~G. {Polnarev}, G.~{Sch{\"a}fer}, and I.~Y. {Vlasov}.
\newblock {Gravimagnetic effect of the barycentric motion of the Sun and
  determination of the post-Newtonian parameter {$\gamma$} in the Cassini
  experiment}.
\newblock \emph{Physics Letters A}, 367:\penalty0 276--280, July 2007.
\newblock \doi{10.1016/j.physleta.2007.03.036}.

\bibitem[{Harte}(2008{\natexlab{b}})]{harte2008_1}
A.~I. {Harte}.
\newblock {Approximate spacetime symmetries and conservation laws}.
\newblock \emph{Classical and Quantum Gravity}, 25\penalty0 (20):\penalty0
  205008, October 2008{\natexlab{b}}.
\newblock \doi{10.1088/0264-9381/25/20/205008}.

\bibitem[{Harte}(2010)]{harte2010}
A.~I. {Harte}.
\newblock {Effective stress-energy tensors, self-force and broken symmetry}.
\newblock \emph{Classical and Quantum Gravity}, 27\penalty0 (13):\penalty0
  135002, July 2010.
\newblock \doi{10.1088/0264-9381/27/13/135002}.

\bibitem[{Sauer} and {Trautman}(2008)]{sauer_2000}
T.~{Sauer} and A.~{Trautman}.
\newblock {Myron Mathisson: what little we know of his life}.
\newblock \emph{Acta Physica Polonica B Proceedings Supplement}, 1\penalty0
  (1):\penalty0 7--26, 2008.
\newblock URL
  \url{http://authors.library.caltech.edu/19170/1/Sauer2008p8776Acta_Physica_Polonica_B.pdf}.

\bibitem[Gel'fand and Shilov(1964)]{Gelfand_1964}
I.~M. Gel'fand and G.~E. Shilov.
\newblock \emph{Generalized functions. {V}ol. {I}: {P}roperties and
  operations}.
\newblock {Translated by E. Saletan}. Academic Press, New York, 1964.

\bibitem[{Kol\'a\u{r}} et~al.(1993){Kol\'a\u{r}}, {Michor}, and
  {Slov\'ak}]{Kolar_1993}
I.~{Kol\'a\u{r}}, P.~W. {Michor}, and J.~{Slov\'ak}.
\newblock \emph{{Natural operations in differential geometry}}.
\newblock Springer, Berlin, 1993.
\newblock URL \url{http://www.emis.de/monographs/KSM/}.

\bibitem[{Schouten}(1954)]{Schouten_book}
J.~A. {Schouten}.
\newblock \emph{{Ricci-Calculus: An Introduction to Tensor Analysis and Its
  Geometrical Applications}}.
\newblock {Springer}, Berlin, 1954.
\newblock see review by K. Yano at
  \url{https://projecteuclid.org/download/pdf_1/euclid.bams/1183519893}.

\bibitem[{Poisson} and {Will}(2014)]{poissonwill_book}
E.~{Poisson} and C.~M. {Will}.
\newblock \emph{{Gravity}}.
\newblock Cambridge University Press, Cambridge, UK, May 2014.

\bibitem[{Nesterov}(1999)]{nesterov_1999CQG}
A.~I. {Nesterov}.
\newblock {Riemann normal coordinates, Fermi reference system and the geodesic
  deviation equation}.
\newblock \emph{Classical and Quantum Gravity}, 16:\penalty0 465--477, February
  1999.
\newblock \doi{10.1088/0264-9381/16/2/011}.

\bibitem[{Veblen} and {Thomas}(1923)]{Veblen_1923}
O.~{Veblen} and T.~Y. {Thomas}.
\newblock The geometry of paths.
\newblock \emph{{Transactions of the American Mathematical Society}},
  25\penalty0 (4):\penalty0 551--608, 1923.
\newblock ISSN 00029947.
\newblock URL \url{http://www.jstor.org/stable/1989307}.

\bibitem[{Beiglb{\"o}ck}(1967)]{beig_1967CMaPh}
W.~{Beiglb{\"o}ck}.
\newblock {The center-of-mass in Einsteins theory of gravitation}.
\newblock \emph{Communications in Mathematical Physics}, 5:\penalty0 106--130,
  April 1967.
\newblock \doi{10.1007/BF01646841}.

\bibitem[{Bailey} and {Israel}(1980)]{bailey_1980AnPhy}
I.~{Bailey} and W.~{Israel}.
\newblock {Relativistic dynamics of extended bodies and polarized media: An
  eccentric approach}.
\newblock \emph{Annals of Physics}, 130:\penalty0 188--214, November 1980.
\newblock \doi{10.1016/0003-4916(80)90231-6}.

\bibitem[{Suen}(1986)]{1986PhRvD..34.3617S}
W.-M. {Suen}.
\newblock {Multipole moments for stationary, non-asymptotically-flat systems in
  general relativity}.
\newblock \emph{\prd}, 34:\penalty0 3617--3632, December 1986.
\newblock \doi{10.1103/PhysRevD.34.3617}.

\bibitem[{Zhang}(1986)]{zhang_1986PhRvD}
X.-H. {Zhang}.
\newblock {Multipole expansions of the general-relativistic gravitational field
  of the external universe}.
\newblock \emph{\prd}, 34:\penalty0 991--1004, August 1986.
\newblock \doi{10.1103/PhysRevD.34.991}.

\bibitem[{Pirani}(1965)]{Pirani1964}
F.~A.~E. {Pirani}.
\newblock Introduction to gravitational radiation theory.
\newblock In A.~{Trautman}, F.~A.~E. {Pirani}, and H.~{Bondi}, editors,
  \emph{Lectures on General Relativity. Vol. 1}, pages 249--373, {Englewood
  Cliffs, NJ}, 1965. {Prentice Hall}.

\bibitem[{Dixon}(2013)]{Dixon_2013}
W.~G. {Dixon}.
\newblock {Post-Newtonian approximation for isolated systems by matched
  asymptotic expansions I. General structure revisited}.
\newblock \emph{ArXiv e-prints}, November 2013.

\bibitem[{Chandrasekhar} and {Nutku}(1969)]{1969ApJ...158...55C}
S.~{Chandrasekhar} and Y.~{Nutku}.
\newblock The second post-{N}ewtonian equations of hydrodynamics in general
  relativity.
\newblock \emph{Astrophys. J.}, 158:\penalty0 55--79, October 1969.
\newblock \doi{10.1086/150171}.

\bibitem[ide(August 24, 2018)]{idempotence}
Idempotence.
\newblock Wikipedia, August 24, 2018.
\newblock URL \url{https://en.wikipedia.org/wiki/Idempotence}.

\end{thebibliography}

\end{document}